\documentclass[final,1p]{ELSarticle}

\usepackage{makeidx,showidx}
\usepackage{multind}

\usepackage{sidecap}
\usepackage{wallpaper}
\usepackage{mdframed}

\usepackage{epsf,epsfig,graphicx}
\usepackage{pdflscape}
\usepackage{rotating}
\usepackage{longtable}
\usepackage{subfigure} 
\usepackage{multirow}
\usepackage{color}
\usepackage{colortbl}
\usepackage{ctable}
\usepackage{xspace}
\usepackage{amsmath}
\usepackage{amssymb}
\usepackage{amsfonts}
\usepackage{hyperref}
\usepackage{wasysym}
\usepackage{wrapfig}
\usepackage{esint}

\usepackage{epstopdf}
\usepackage{pdfpages}
\usepackage{paralist}
\usepackage{xspace}
\numberwithin{figure}{section}
\numberwithin{figure}{subsection}
\numberwithin{table}{section}
\numberwithin{table}{subsection}

\newcommand{\npb}{Nucl. Phys. B}
\newcommand{\nuphs}{Nucl. Phys. B Proc. Supp.}
\newcommand{\plb}{Phys. Lett.  B}
\newcommand{\pla}{Phys. Lett.  A}

\newcommand{\mpo}{ }

\newcommand{\aj}{AJ}
\newcommand{\apj}{ApJ}
\newcommand{\apjl}{ApJL}

\newcommand{\aplc}{ApL\&C}
\newcommand{\apjs}{ApJS}

\newcommand{\aap}{A\&A}

\newcommand{\araa}{ARA\&A}
\newcommand{\aapr}{A\&A Rev.}
\newcommand{\aaps}{A\&AS}
\newcommand{\raa}{RAA}
\newcommand{\mnras}{MNRAS}

\newcommand{\apss}{Ap\&SS}

\newcommand{\nat}{Nature}
\newcommand{\natc}{Nature Comm.}
\newcommand{\sci}{Science}

\newcommand{\nar}{New Astronomy Review}

\newcommand{\aipconf}{AIP Conf. Proc.}

\newcommand{\procspie}{Proc. SPIE}

\newcommand{\ssr}{Space Sci. Rev.}

\newcommand{\prd}{Phys. Rev. D}

\newcommand{\prl}{Phys. Rev. Lett.}
\newcommand{\rvmp}{RvMP}
\newcommand\jcap{{JCAP}}
\newcommand{\physrep}{Phys.~Rep.}
\newcommand{\rpg}{Rep. Prog. Phys.}
\newcommand{\nima}{Nucl. Instrum. Meth. A}
\newcommand{\ap}{Astropart. Phys.}
\newcommand{\jinst}{JINST}
\newcommand{\rpp}{RPPh}
\newcommand{\exa}{ExA}
\newcommand{\ptps}{Progr. Theor. Phys. Suppl.}
\newcommand{\aph}{Astropart. Phys.}
\newcommand{\ncimc}{Nuovo Cimento C Geophys. Space Phys. C}
\newcommand{\mmsai}{MmSAI}
\newcommand{\phpl}{Phys. of Plasmas}
\newcommand{\frp}{Frontiers in Phys.}
\newcommand{\pasp}{PASP}
\newcommand{\jphg}{JPhG}
\newcommand{\jgr}{JGR}
\newcommand{\jgra}{JGRA}
\newcommand{\jgrd}{JGRD}
\newcommand{\jastp}{JASTP}
\newcommand{\epj}{EPJ}

\let\bar=\overbar

\renewcommand{\L}{{\cal L}}
\newcommand{\M}{{\cal M}}
\newcommand{\A}{{\cal A}}

\newcommand{\One}{{\bf 1}}

\newcommand{\Dslash}{\not{\hbox{\kern-4pt $D$}}}
\newcommand{\dslash}{\not{\hbox{\kern-2pt $\del$}}}

\newcommand{\msb}{{\bar{\ssstyle M \kern -1pt S}}}

\newcommand{\lsim} {\buildrel < \over {_\sim}}
\newcommand{\gsim} {\buildrel > \over {_\sim}}

\newcommand{\Fermi}{\textit{Fermi}\xspace}
 \newcommand{\g}{gamma}
\newcommand{\agile}{\textit{AGILE}\xspace}
\newcommand{\INTEGRAL}{\textit{INTEGRAL}\xspace}
\newcommand{\brem}{Bremsstrahlung\xspace}
\newcommand{\ic}{IC\xspace}
\newcommand{\ec}{EC\xspace}
\newcommand{\ssc}{SSC\xspace}

\newcommand{\Sun}{\ensuremath{\odot}}
\newcommand{\ea}{{e-ASTROGAM}\xspace}
\newcommand{\fermilat}{\emph{Fermi}-LAT\xspace}
\newcommand{\fermigbm}{\emph{Fermi}~GBM\xspace}
\newcommand{\swift}{\emph{Swift}\xspace}
\newcommand{\nustar}{\emph{NuSTAR}\xspace}
\newcommand{\pks}{PKS~1830$-$211}
\def \mtt{ }
\newcommand{\hd}{H$_2$\xspace}
\newcommand{\xco}{X$_{CO}$\xspace}

 \providecommand{\abs}[1]{\lvert#1\rvert}
\newcommand{\agam}{e-ASTROGAM\xspace}

\def\lsim{\lower.5ex\hbox{$ \buildrel < \over \sim \;$}}
\def\gsim{\lower.5ex\hbox{$ \buildrel > \over \sim \;$}}
\def\units#1{\hbox{$\,{\rm #1}$}}

\newcommand*{\vcenteredhbox}[1]{\begingroup
\setbox0=\hbox{#1}\parbox{\wd0}{\box0}\endgroup}

\makeatletter
\newcommand{\printchapterauthor}[1]{%
  {\parindent0pt%
  \normalsize\itshape#1%
  \par\nobreak\vspace*{5pt}}
  \@afterheading%
}

\makeatother

\begin{document}


\IfFileExists{CoverPDF.pdf}{\includepdf[pages=1]{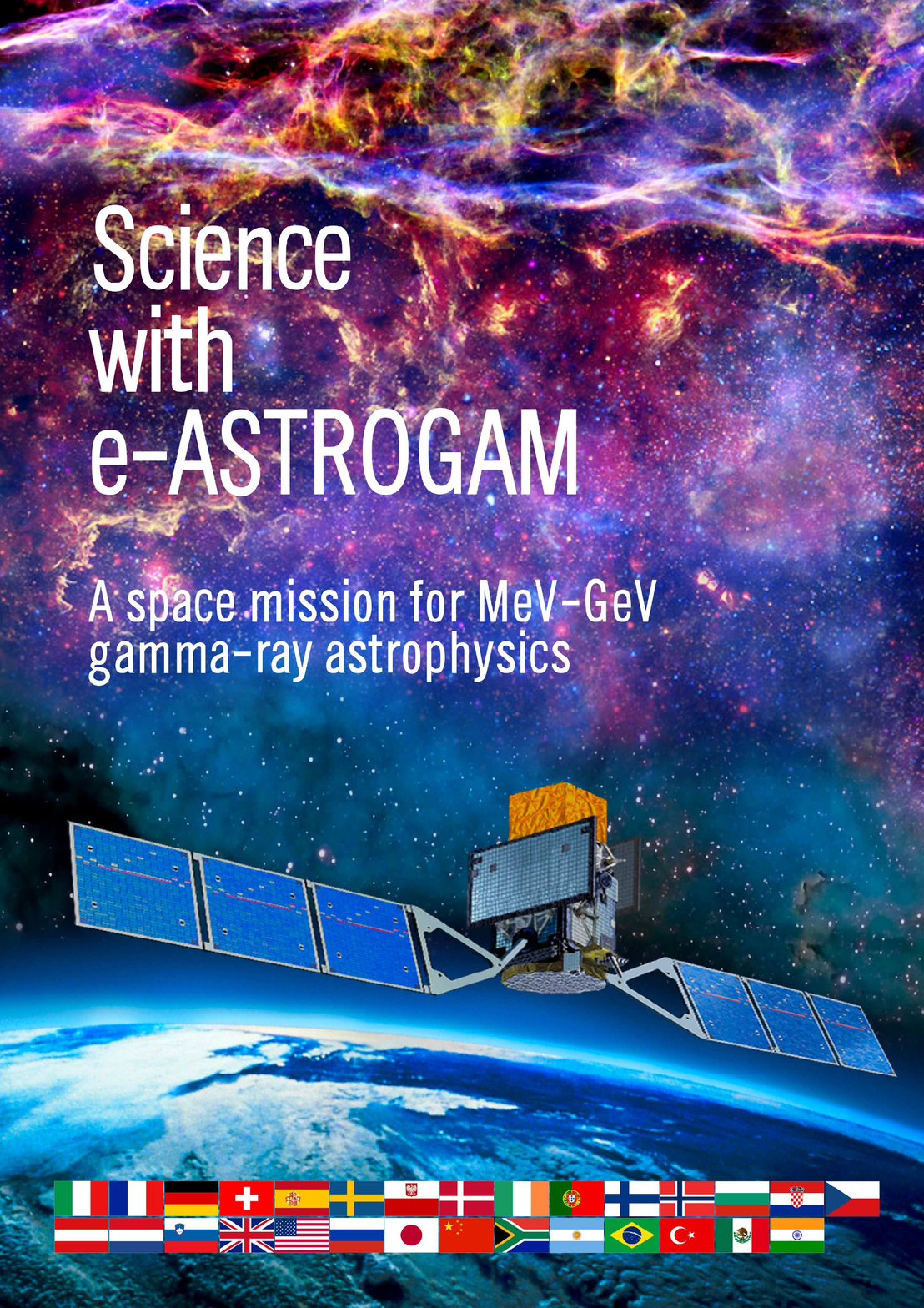}}{}

\begin{center}
\textbf{Abstract}
\end{center}
\ea (`enhanced ASTROGAM') is a breakthrough Observatory space mission, {with a detector composed by a Silicon tracker, a calorimeter, and an anticoincidence system,} dedicated to the study of the non-thermal Universe in the photon energy range from 0.3 MeV to 3 GeV {-- the lower energy limit can be pushed to energies as low as 150 keV for the tracker, and  to  30 keV for calorimetric detection}. The mission is based on an advanced space-proven detector technology, with unprecedented sensitivity, angular and energy resolution, combined with polarimetric capability.
Thanks to its performance in the MeV-GeV domain,  substantially improving its predecessors, \ea will open a new window on the {non-thermal
Universe,} making pioneering observations of the most powerful  Galactic and extragalactic  sources, elucidating the nature of their relativistic outflows and  their effects on the surroundings.
With a line sensitivity in the MeV energy range one to two orders of magnitude better than previous generation instruments, \ea will determine the origin of key isotopes fundamental {for the understanding of supernova explosion and the chemical evolution of our Galaxy}. The mission will provide unique  data of significant interest to a broad astronomical community, complementary to powerful observatories such as LIGO-Virgo-GEO600-KAGRA, SKA, ALMA, E-ELT, TMT, LSST, JWST, Athena, CTA, IceCube, KM3NeT, and LISA.



\IfFileExists{author_list.pdf}{\includepdf[pages=1-4]{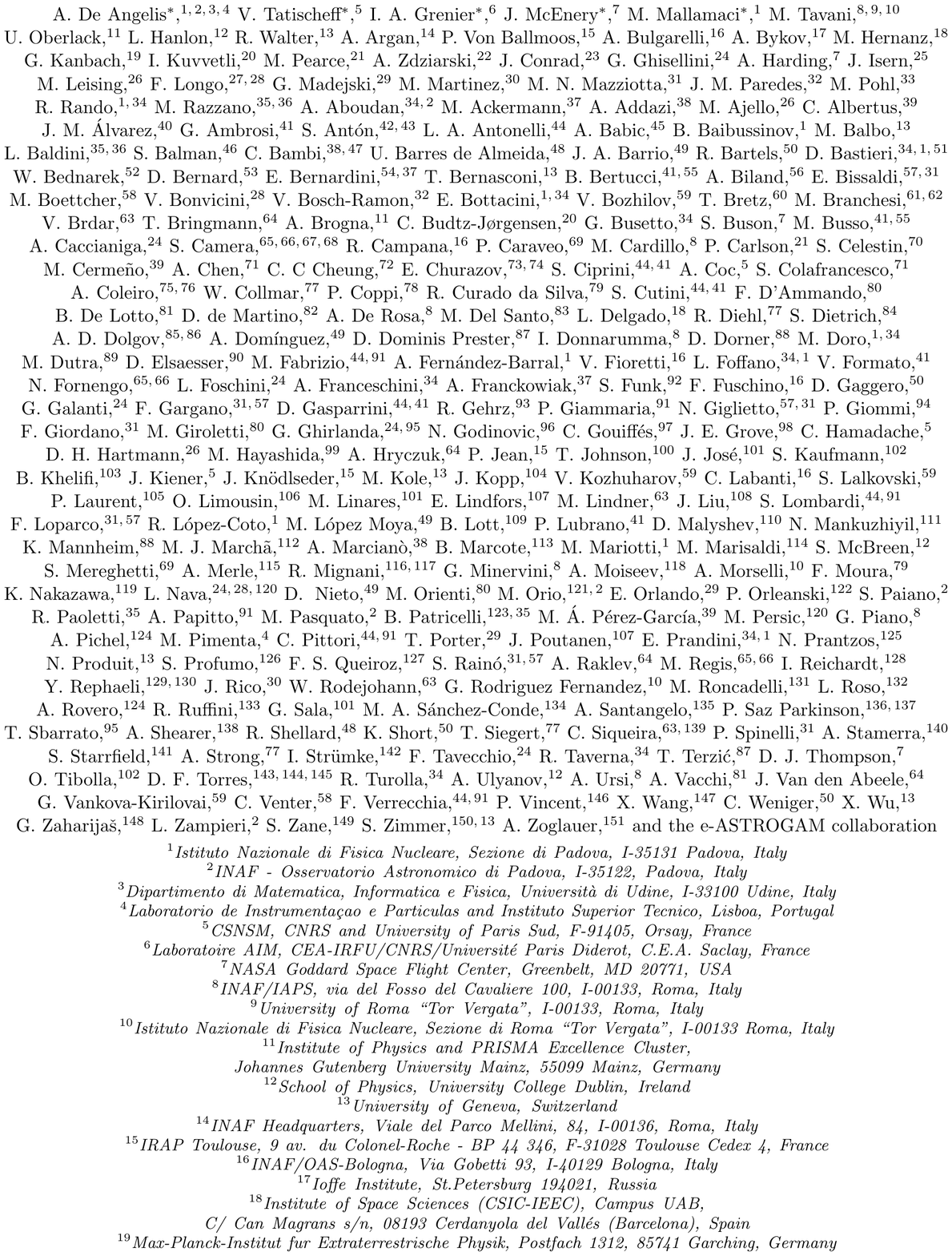}}{}

\bibliographystyle{elsarticle-num}

\tableofcontents
\newpage
\section{Introduction}
\label{intro}

{\ea \cite{fullpap,physcase} is a \g-ray mission concept proposed by a wide international community. It is conceived to operate in a maturing Gravitational Wave (GW) and multimessenger epoch, opening up entirely new and exciting synergies. The mission would provide unique and complementary data of significant interest to a broad astronomical community, in a decade of powerful observatories such as LIGO-Virgo-GEO600-KAGRA, SKA, ALMA, E-ELT, LSST, JWST, ATHENA, CTA and maybe LISA. 

The main constituents of the \ea payload are:
\begin{itemize}
\item A {\bf  Tracker} in which the cosmic \g-rays can undergo a  Compton scattering or a pair conversion, based on 56 planes of double-sided Si strip detectors, each plane with total area of $\sim$1 m$^2$;
\item A {\bf Calorimeter} to measure the energy of the secondary particles, made of an array of CsI (Tl) bars of 5$\times$5$\times$80 mm$^3$ each, with relative energy resolution of 4.5\% at 662 keV;
\item  An {\bf Anticoincidence system} (AC), composed of a standard plastic scintillator AC shielding and a Time of Flight,  to veto the 
charged particle background.
\end{itemize}

} 

The core mission science of \ea addresses three major topics of modern astrophysics.

\begin{itemize} \itemsep 0cm \topsep 0cm
\item \textbf{\em{Processes at the heart of the extreme Universe: prospects for the Astronomy of
the 2030s}}

Observations of relativistic jet and outflow sources (both in our Galaxy and 
in Active Galactic Nuclei - AGN, briefly) in the X-ray and  GeV--TeV energy ranges have shown that the  MeV--GeV band holds the key to understand the transition from the
low energy continuum to a spectral range shaped by not yet fully understood particle acceleration processes. 
{\mtt   
\ea will:
(1) identify the composition (hadronic or leptonic) of the outflows and jets, which strongly influences the environment; 
(2) identify the physical acceleration processes in these outflows and jets (e.g. diffusive shocks, magnetic field reconnection, plasma effects), that may lead to dramatically different particle energy distributions; 
(3) clarify the role of the magnetic field in powering ultra-relativistic {jets in \g-ray bursts (GRBs)}, through time-resolved polarimetry and spectroscopy.
In addition, measurements in the \ea energy band will have a big  impact on multimessenger astronomy in the 2030s. The recent discoveries of GWs emitted in the merger of two neutron stars \cite{gw170817gbm} and of a high-energy neutrino coincident with a flaring blazar \cite{txall} demonstrated that \g-ray data of high-energy transient events are crucial for making the most of multimemessenger observations. In particular, the NS-merging event generating the signal GW170817 and the corresponding short GRB detected by \Fermi GBM and INTEGRAL demonstrated that the soft gamma-ray energy range is the most appropriate electromagnetic (EM) domain for identifying the source and defining the astrophysical context of the burst event. On the other hand, joint detections of neutrinos and a X-ray/\g-ray transient sources might lead to significant associations and consequently to incontrovertible identifications of the EM counterparts of astrophysical neutrinos. e-ASTROGAM would play a fundamental role in this scenario, as addressed in more detail in Chapters \ref{chap:extreme_uni} and \ref{chap:cosmic_ray}. 

\item \textit{\textbf{The origin and impact of high-energy particles on galaxy evolution, from cosmic rays to antimatter}}

{\mtt \ea   will  resolve the outstanding issue of the origin and propagation of Low-Energy Cosmic Rays (LECRs) affecting star formation. 
It  will measure cosmic-ray diffusion in interstellar clouds
and their impact on gas dynamics; it will provide crucial diagnostics about the wind outflows and their feedback on the Galactic 
environment (e.g., Fermi bubbles, Cygnus cocoon).}
\ea will have  optimal sensitivity and energy resolution to detect  line
emissions in the keV--MeV range,
and  a variety of 
issues will be resolved, in particular: (1)  origin
of the \g-ray and positron excesses toward the Galactic
 inner regions;  (2)  determination of the astrophysical 
 sources of the local positron population from a very sensitive observation of pulsars and supernova remnants (SNRs). 
 As a consequence \ea will be able to discriminate the backgrounds to dark matter (DM) signals.

\item \textbf{\textit{Nucleosynthesis and the chemical enrichment
of our Galaxy}}

The \ea line sensitivity is more than an order of magnitude  better than previous instruments.  The deep exposure of the Galactic plane region will determine how different  isotopes are created in stars and distributed in the interstellar medium; it will also unveil the recent history of supernova explosions in the Milky Way.  
Furthermore, \ea will detect a significant number of Galactic novae and supernovae (SNe) in nearby galaxies, thus addressing fundamental issues in the explosion mechanisms of both core-collapse and thermonuclear SNe. 
The \g-ray data will provide a much better understanding of Type Ia SNe which, in turn, will allow to predict their evolution in the past, a pre-requisite for their use as standard candles for precision cosmology.}
\end{itemize}

\begin{figure}[ht]
\centering
\includegraphics[width=0.8\textwidth]{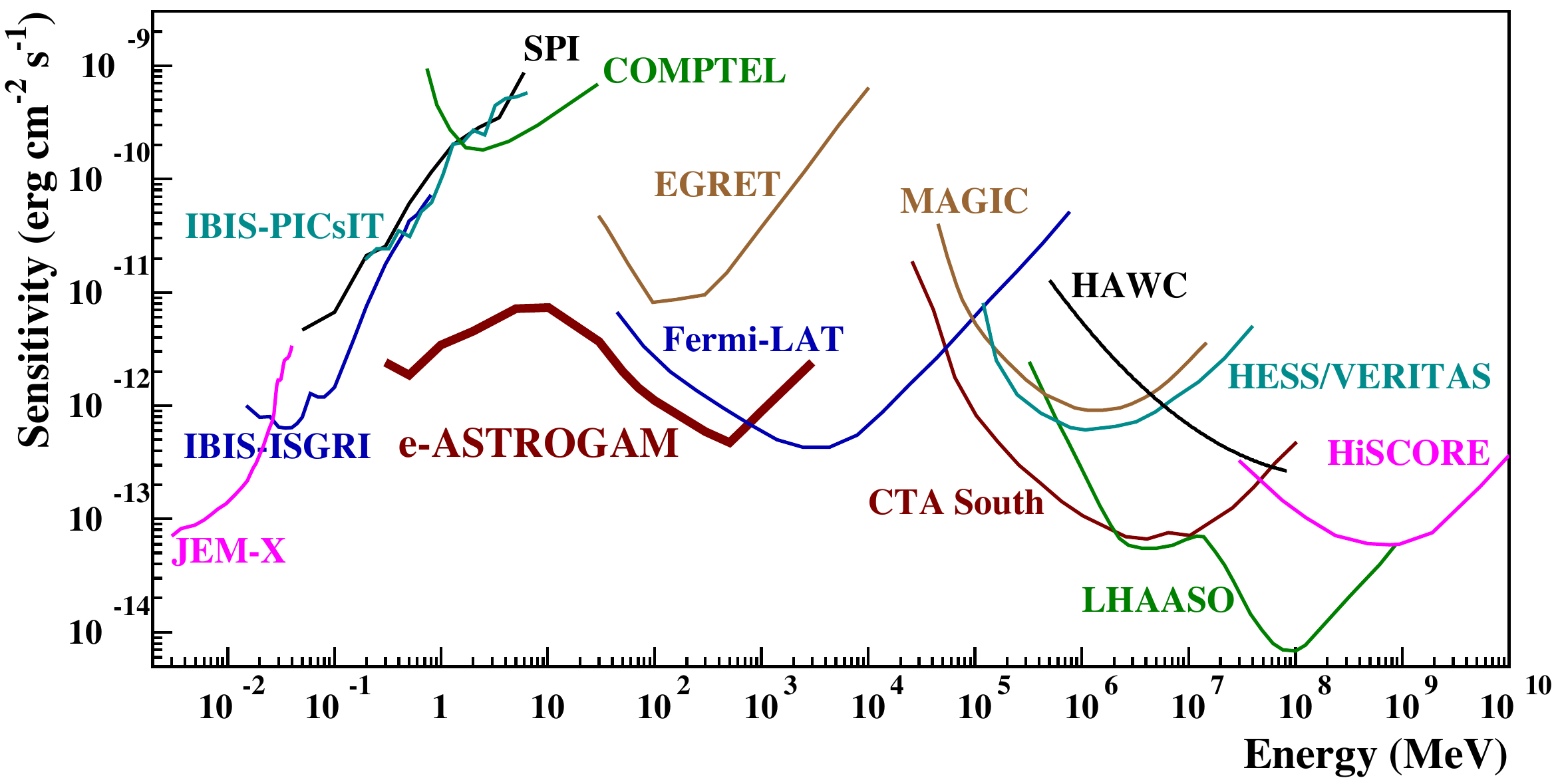}
\caption{{Point source continuum differential sensitivity of different X- and \g-ray instruments. The curves for JEM-X, IBIS (ISGRI and PICsIT), and SPI are for an effective observation time $T_{\rm obs}$~=~1~yr, which is the approximate exposure of the Galactic center region accumulated by \INTEGRAL since the beginning of the mission. The COMPTEL and EGRET sensitivities are given for {the typical observation time accumulated during the $\sim$ 9 years of the {\it CGRO} mission (see Fig. 1 in \cite{tak13}). The \fermilat sensitivity is for a high Galactic latitude source in 10 years of  observation in survey mode}. For MAGIC, VERITAS (sensitivity of H.E.S.S. is similar), and CTA, the sensitivities are given for $T_{\rm obs}$~=~50~hours. For HAWC $T_{\rm obs}$~=~5~yr, for LHAASO $T_{\rm obs}$~=~1~yr, and for HiSCORE $T_{\rm obs}$~=~100~ h. The e-ASTROGAM sensitivity is calculated at $3\sigma$ for an effective exposure of 1 year and for a source at high Galactic latitude.}
\label{fig:sensitivity}}
\end{figure}

In addition to addressing its core scientific goals, \ea will 
achieve many serendipitous discoveries (the unknown unknowns) through its combination of wide field of view (FoV) and improved sensitivity, 
measuring  in 3 years the Spectral Energy Distributions (SEDs) of thousands of Galactic and extragalactic sources,  
and providing new information on solar flares and terrestrial \g-ray flashes (TGFs). \ea will become a key contributor to multi-wavelength time-domain astronomy. 
{\mtt The mission } 
has outstanding discovery potential {\mtt as an Observatory facility that is 
open to a wide astronomical community.}\\

\ea  is designed to achieve:
\begin{itemize} \itemsep 0cm
\item Broad energy coverage (0.3 MeV to 3 GeV), with  one-two orders of magnitude improvement in continuum sensitivity in the range 0.3 MeV -- 100 MeV compared to previous instruments (the lower energy limit can be pushed to energies as low as 150 keV, albeit with rapidly degrading angular resolution, for the tracker, and to  30 keV for calorimetric detection);
\item Unprecedented performance for \g-ray lines, with, for example, a sensitivity for the 847~keV line from Type Ia SNe 70 times better than that of \INTEGRAL/SPI;
\item Large FoV ($>$2.5 sr), ideal to detect transient sources and hundreds of GRBs;
\item Pioneering polarimetric capability for both steady and transient sources;
\item Optimized source identification capability obtained by the best angular resolution achievable by state-of-the-art detectors in this energy range (about 0.15 degrees at 1 GeV);
\item Sub-millisecond trigger and alert capability for GRBs and other cosmic and terrestrial transients;
\item Combination of Compton and pair-production detection techniques allowing model-independent control on the detector systematic uncertainties.
\end{itemize}



\begin{figure}
\centering
\includegraphics[width=0.6\textwidth]{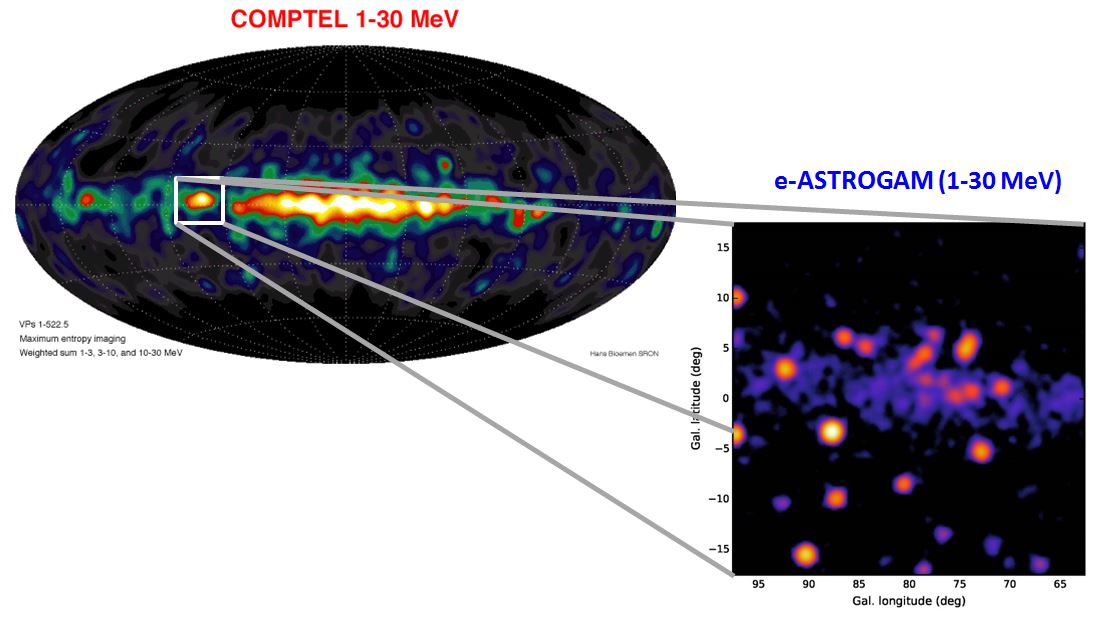}
\includegraphics[width=0.675\textwidth]{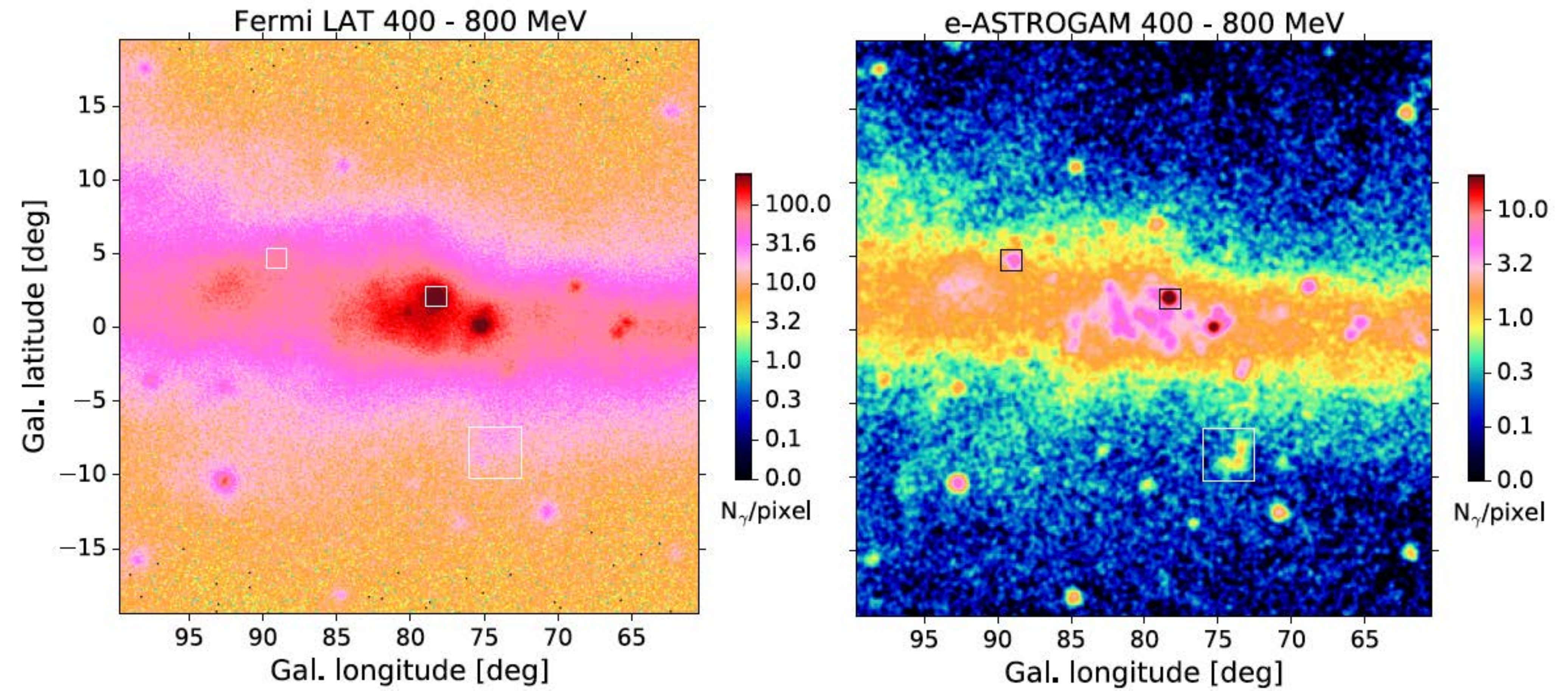}
\caption{An example of the capability of \ea to transform our 
 knowledge of the MeV-GeV sky. Upper panel: The upper left figure shows the
1-30 MeV sky as observed by COMPTEL in the 1990s; the lower right figure
  shows the simulated Cygnus region in the 1-30 MeV
energy region  from \ea. Lower panel: comparison between the view of the Cygnus region by \Fermi in 8 years (left) and that by e-ASTROGAM in one year of effective exposure (right) between 400 MeV and 800 MeV.}
\label{fig:Jurgen}
\end{figure}
\begin{figure}
\centering
\includegraphics[width=0.8\columnwidth]{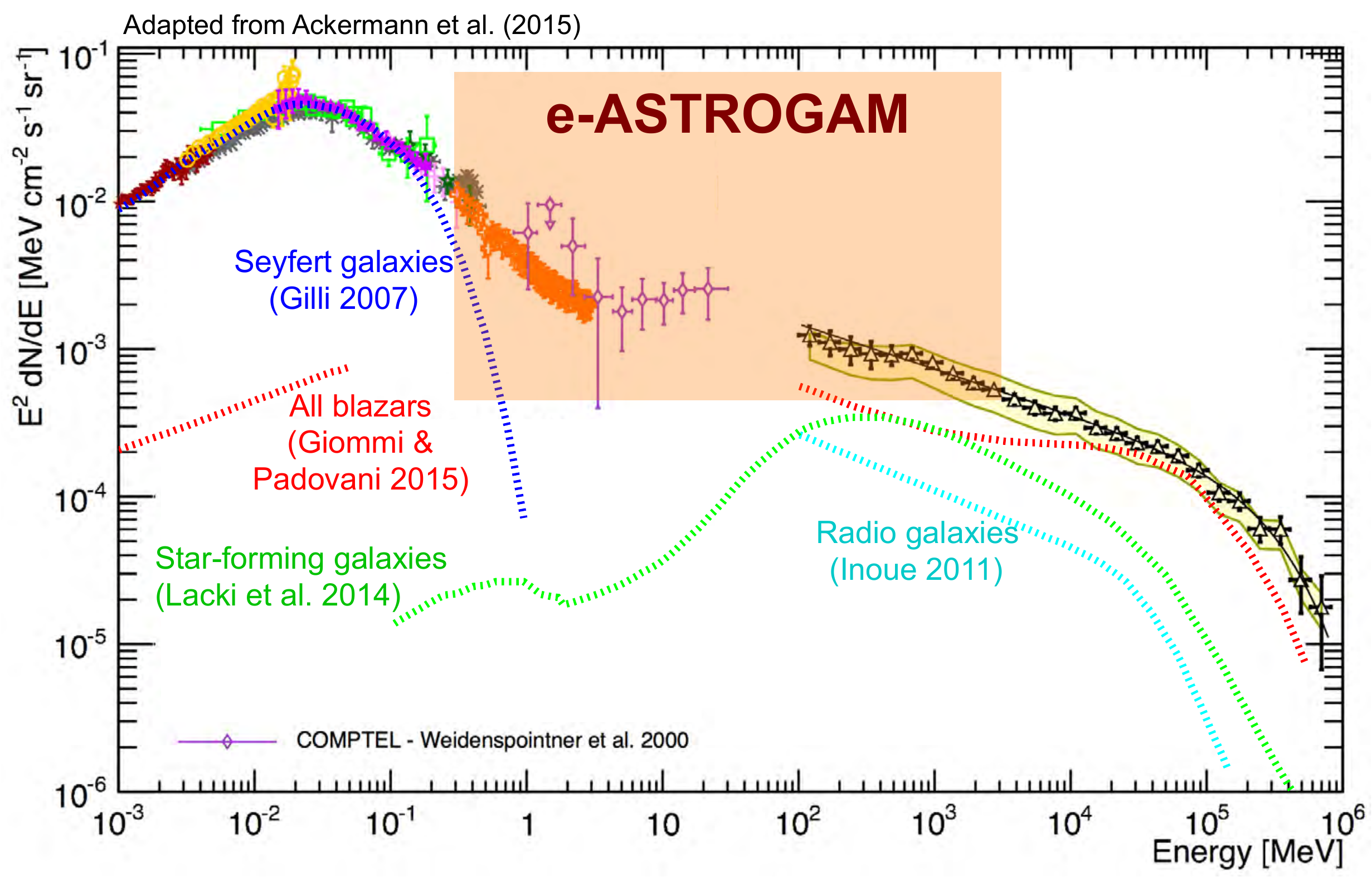}
\caption{Compilation of the measurements of the total extragalactic \g-ray intensity between 1 keV and 820 GeV \cite{FermiEGB}, {with different components from current models}; the contribution from MeV blazars is largely unknown.
  The semi-transparent band indicates the energy region in which e-ASTROGAM will dramatically improve on present knowledge.\label{fig:egb}}
\end{figure}

\ea will open the MeV region for exploration, with an improvement of one-two orders of magnitude in sensitivity (Fig.~\ref{fig:sensitivity}) compared to the current state of the art, much of which was derived from the COMPTEL instrument more than two decades ago.  It will also achieve a spectacular improvement in terms of source localization accuracy (Fig.~\ref{fig:Jurgen}) and  energy resolution, and will allow to measure the contribution to the radiation of the Universe in an unknown range (Fig. \ref{fig:egb}). 
The sensitivity of \ea will reveal the transition from nuclear processes to those involving electro- and hydro-dynamical, magnetic and gravitational interactions.

An important characteristic of \ea is its ability to accurately measure polarization in the MeV range, which is afforded by Compton interactions in the detector. The achievable Minimum Detectable Polarization (MDP) at the 99\% confidence level is 10\% for a 10 mCrab source in the 0.2-2 MeV range after 1 year of effective exposure (see Sect. \ref{sec:perfass}).
Polarization encodes information about the geometry of magnetic fields and adds a new observational pillar, in addition to the temporal and spectral, through which fundamental processes governing the MeV emission can be determined. The addition of polarimetric information will be crucial for a variety of investigations, including accreting black hole (BH) systems, magnetic field structures in jets, and the emission mechanisms of GRBs. Polarization will provide decisive insight into the presence of hadrons in extragalactic jets and the origin of Ultra-High Energy Cosmic Rays (UHECRs).

\subsection{Scientific requirements}\label{sec:scirec}

e-ASTROGAM's requirements to achieve its core science objectives, such as the angular and energy resolution, the field of view, the continuum and line sensitivity, the polarization sensitivity, {and the timing accuracy}, are summarized in Table~\ref{table:requirements}. 

\begin{table}
\centering
\caption{e-ASTROGAM scientific requirements.}
\vspace{-0.1cm}
\includegraphics[width=0.9\textwidth]{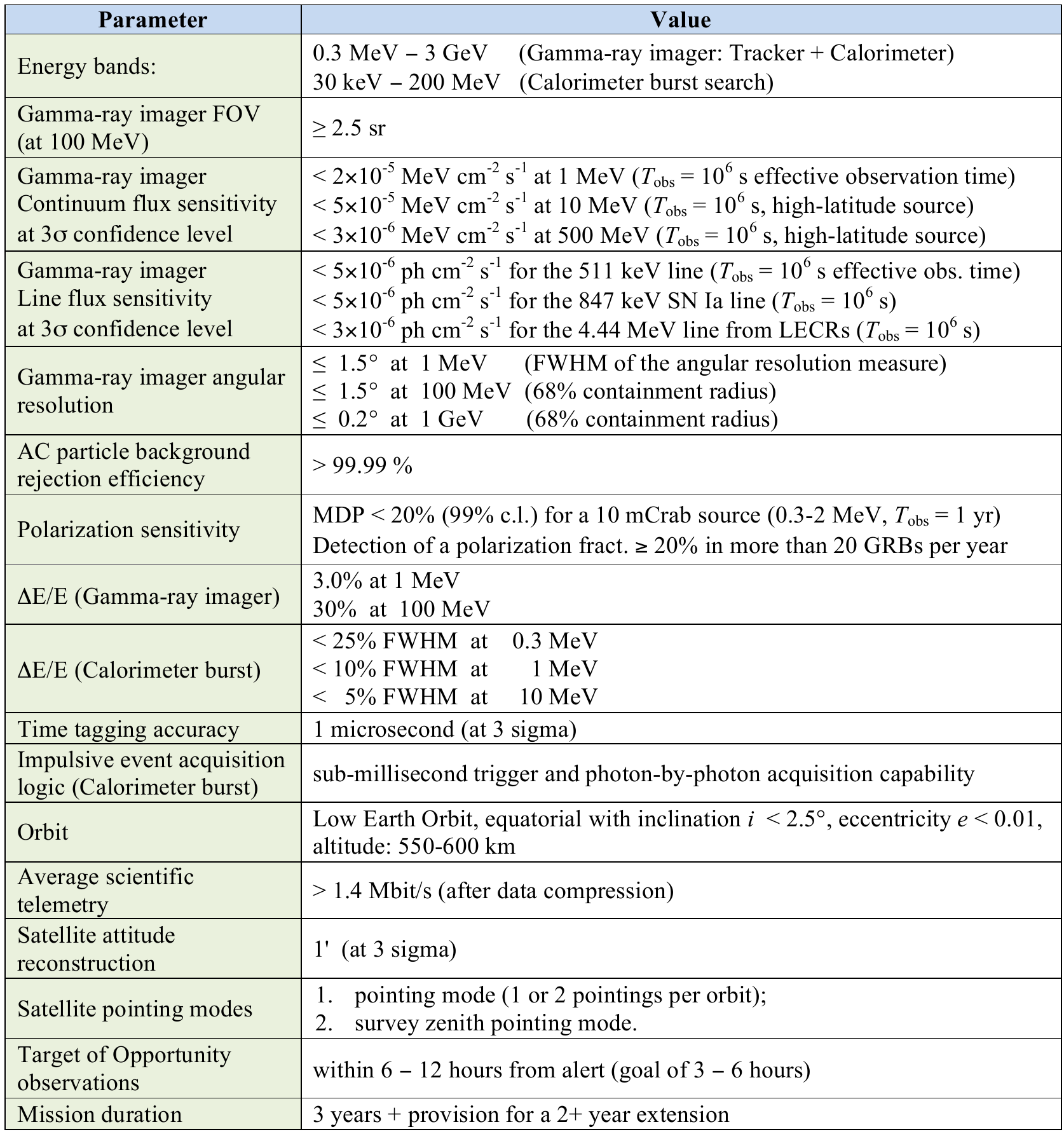}
\label{table:requirements}
\end{table}

{ 
\begin{itemize}
\item The very large spectral band covered by the telescope in the standard \g-ray acquisition mode will give a complete view of the main nonthermal processes at work in a given astrophysical object, for the first time with a single instrument. The e-ASTROGAM energy band includes the 511~keV line from $e^+e^-$ annihilation, the nuclear de-excitation lines, the characteristic spectral bump from pion decay, the typical domains of nonthermal electron bremsstrahlung and IC emission, as well as the high-energy range of synchrotron radiation. 
The designed wide energy band is particularly important for the study of blazars, GRBs, Galactic compact binaries, pulsars, as well as the physics of Cosmic Rays (CRs) in SNRs and in the Interstellar Medium (ISM). 
\item The large energy band covered by the Calorimeter in the burst search mode of data acquisition   is primarily designed for the triggering and study of GRBs. It is also well adapted to the broadband emissions of TGFs and solar flares. 
\item The wide field of view of the telescope is especially important to enable the measurement of source flux variability over a wide range of timescales both for a-priori chosen sources and in serendipitous observations. Coupled with the scanning mode of operation, this capability enables continuous monitoring of source fluxes that will greatly increase the chances of detecting correlated flux variability with other wavelengths. The designed wide field of view is particularly important for the study of blazars, GRBs, Galactic compact objects, SNe, novae, and extended emissions in the Milky Way (CRs, radioactivity). It will also enable, for example, searches of periodicity and orbital modulation in binary systems. 
\item One of the main requirements of \ea is to improve dramatically the detection sensitivity in a region of the EM spectrum, the so-called MeV domain, which is still largely unknown. The sensitivity requirement is relevant to all science drivers discussed above. Thus, the goal of detecting a significant number ($N > 5$) of SN~Ia in \g-rays after 3 years requires a sensitivity in the 847~keV line $<5 \times 10^{-6}$ ph~cm$^{-2}$~s$^{-1}$  in 1~Ms of integration time (Table~\ref{table:requirements}). 
\item Another major requirement for a future \g-ray observatory is to improve significantly the angular resolution over past and current  missions, which have been severely affected by a spatial confusion issue. Thus, the e-ASTROGAM angular resolution will be excellent in the MeV range and above a few hundreds of MeV, improving $CGRO$/COMPTEL and \fermilat by almost a factor of 4 at 1 MeV and 1 GeV, respectively. The targeted angular resolution given in Table~\ref{table:requirements} is close to the physical limits: for Compton scattering, the limit is given by the Doppler broadening induced by the velocity of the atomic electrons, while for low-energy pair production, the limit is provided by the nuclear recoil. e-ASTROGAM angular resolution will allow a number of currently unidentified \g-ray sources (e.g. 992 sources in the 3FGL catalog \cite{3FGL}) to be associated with objects identified at other wavelengths. The Galactic Center (GC) region is the most challenging case, for which the e-ASTROGAM capability will be fully employed. 
\item The polarization sensitivity of e-ASTROGAM is designed to enable measurements of the \g-ray polarization fraction in more than 20 GRBs per year (GRBs being promising candidates for highly \g-ray polarized sources, see, e.g., \cite{mcc16}). Such measurements will provide important information on the magnetization and content (leptons, hadrons, Poynting flux) of the relativistic outflows, and, in the case of GRBs at cosmological distance, will address fundamental questions of physics related to vacuum birefringence and Lorentz invariance violation (e.g., \cite{got14}). With the designed polarization sensitivity, e-ASTROGAM will also be able to study the polarimetric properties of more than 50 pulsars, magnetars, and BH systems in the Galaxy.
\item The spectral resolution of e-ASTROGAM is well adapted to the main science drivers of the mission. Thus, the main \g-ray lines produced in SN explosions or by LECRs interactions in the ISM are significantly broadened by the Doppler effect, and a FWHM resolution of 3\% at 1 MeV is adequate. In the pair production domain, an energy resolution of 30\% will be more than enough to measure accurately putative spectral breaks and cutoffs in various sources and  distinguish the characteristic pion-decay bump from leptonic emissions. 
\item The timing performance of e-ASTROGAM is mainly driven by the physics of magnetars and rotation-powered pulsars, as well as by the properties of TGFs. The targeted microsecond timing accuracy is already achieved in, e.g., the \agile mission \cite{tav09}. 
\end{itemize}
}

The e-ASTROGAM requirements reflect the dual capacity of the instrument to detect both Compton scattering events in the 0.3 (and below) -- 10 MeV range and pair-producing events in the 10 MeV -- 3 GeV energy range; a small overlap around 10 MeV allows (although in a limited energy band)   cross-calibration, thus reducing systematic uncertainties. The main instrument features of e-ASTROGAM necessary to meet the scientific requirements in Table~\ref{table:requirements}, are described in Sec. \ref{sec:perfass}.

The sensitivity performance is consistent with the requirement of an equatorial low-Earth orbit (LEO)  of altitude in the range 550~--~600 km.  Such an orbit  is preferred for a variety of reasons. It has been demonstrated to be only marginally affected by the South Atlantic Anomaly and is therefore a low-particle background orbit, ideal for high-energy observations. The orbit is practically unaffected by precipitating particles originating from solar flares, a virtue for background rejection. Finally, both ESA and ASI have satellite communication bases near the equator (Kourou and Malindi) that can be efficiently used as mission ground stations.

Table~\ref{table:requirements} also includes the most important system requirements such as the satellite attitude reconstruction, telemetry budget, and pointing capability. e-ASTROGAM is a multi-purpose astrophysics mission with the capability of a very flexible observation strategy. Two main scientific observation modes are to be managed by the Mission Operation Center (MOC):
\begin{compactitem}
\item[$\bullet$] pointing mode;
\item[$\bullet$] survey mode. 
\end{compactitem}

The pointing mode can be implemented either in a fixed inertial pointing or in the more efficient double-pointing per orbit mode. In the latter case, the e-ASTROGAM satellite is required to be able to perform two sky pointings  per orbit, lasting approximately 40 minutes each. The survey mode consists in a continuous pointing to the zenith to perform a scan of the sky at each orbit. This mode can be activated at any time in principle, and depending on the scientific prioritization and on the mission schedule foreseen by the Science Management Plan, can lead to an optimized  all-sky survey.

Requirements for the Ground Segment are standard for an observatory-class mission. Target of Opportunity observations (ToOs) are required to follow particularly important transient events that need a satellite repointing. The e-ASTROGAM mission requirement for ToO execution is within 6--12 hours, with the goal of reaching 3--6 hours. The speed of repointing depends on the torque of the reaction wheels. We expect a repointing velocity similar to \Fermi ($\sim 30$ degrees/min, which grants to have a visible object in FoV within less than 5'). 

e-ASTROGAM  does not use any consumable and could in principle be operated for a duration up to 10-20 years (well within the foreseen operation duration of 3 years
with a possible extension of two), limited mainly by orbital instabilities and by the risk of accidents. Radiation damage in LEO, with almost equatorial inclination, is negligible. As an example, the degradation of \Fermi, whose inclination implies significant crossing of the South Atlantic Anomaly, is negligible for what concerns electronics, negligible for what concerns Tracker aging, and around  1\%/year in terms of loss in light yield of the Calorimeter crystals.



Table \ref{tab:nevents} summarizes our conservative estimates of the number of sources detectable by e-ASTROGAM in 3 years, based on current knowledge and $\log N - \log S$ determinations of Galactic and extragalactic sources, {including} GRBs. It takes information from \swift-BAT 70-Month Hard X-ray survey catalog \cite{b70h}, the 4th \INTEGRAL-IBIS catalog \cite{ibisc}, and the 3rd \fermilat catalog \cite{3FGL}. Noteworthy, the latter catalog contains more than 1000 unidentified sources in the 100 MeV -- 300 GeV range with no counterparts at other wavelength, and most of them will be detected by e-ASTROGAM, in addition to  a relevant number of new unidentified sources. The discovery space of e-ASTROGAM for new sources and source classes is very large.

\vskip 2mm
\begin{table}
\begin{center}
\begin{tabular}{| l | l | l |}\hline
Type & 3 yr & New sources\\ \hline
{Total} & {3000 -- 4000} & $\sim$1800 (including GRBs)  \\
Galactic & $\sim1000$ & $\sim$400 \\
MeV blazars  & $\sim350$ & $\sim350$ \\
GeV blazars  & {1000 -- 1500} & $\sim350$ \\
Other AGN  ($< $10 MeV)& {70 -- 100} & {35 -- 50}\\
SNe  & {10 -- 15} & {10 -- 15}\\
Novae & 4 -- 6 &   4 -- 6 \\
GRBs  & $\sim$600 &  $\sim$600\\ \hline
\end{tabular}
\end{center}
\caption{Estimated number of sources {of various classes} detectable by e-ASTROGAM in 3 years. {The last column gives the number of sources not known before in any  wavelength.} \label{tab:nevents}}
\end{table}

The e-ASTROGAM mission concept aims to fill the gap in our knowledge of astronomy in the medium-energy (0.3--100 MeV) \g-ray domain \cite{deangelisbook} by increasing the number of known sources in this field by more than an order of magnitude and providing polarization information for many of them.  Between 3000 and 4000 sources are expected to be detected during the first three years of mission operation. The e-ASTROGAM \g-ray instrument inherits from its predecessors such as \agile~\cite{tav09} and \Fermi~\cite{atw09}, as well as from the MEGA prototype\cite{kan05}, but takes full advantage of recent progress in silicon detectors and readout microelectronics to achieve excellent spectral and spatial resolution by measuring the energy and 3D position of each interaction within the detectors. The e-ASTROGAM mission concept is presented at length in Ref.~\cite{fullpap}. Here, we first give an overview of the proposed observatory (Sec.~\ref{sect:ea}) and then outline the breakthrough capability of the e-ASTROGAM telescope for \g-ray polarimetric observations of some of the main targets of the mission: AGNs, GRBs, the Crab pulsar/nebula system, and microquasars.

\subsection{The e-ASTROGAM observatory}
\label{sect:ea}  

\begin{figure}
\centering
\includegraphics[width=0.75\textwidth]{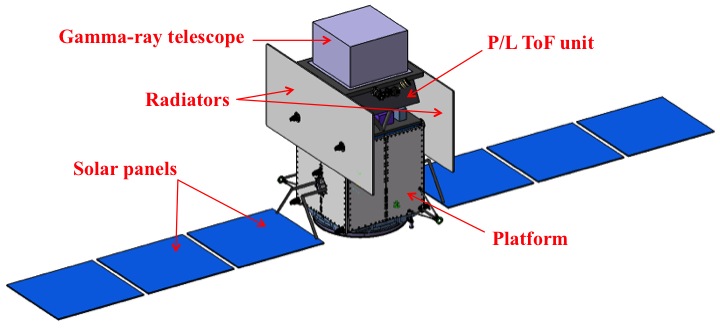}
\caption{e-ASTROGAM spacecraft with solar panels deployed.}
\label{fig:deployed}
\end{figure}

The payload of the e-ASTROGAM satellite (Fig.~\ref{fig:deployed}) consists of a single \g-ray telescope operating over more than four orders of magnitude in energy (from about 150 keV to 3 GeV) by the joint detection of photons in both the Compton (0.15 -- 30 MeV) and pair ($> 10$~MeV) energy ranges. It is attached to a mechanical structure at a distance of about 90~cm from the top of the spacecraft platform, the space between the payload and the platform being used to: (i) host a time-of-flight (ToF) unit designed to discriminate between particles coming out from the telescope and those entering the instrument from below; (ii) host several units of the payload (the back-end electronics modules, the data handling unit, and the power supply unit) and (iii) accommodate two fixed radiators of the thermal control system, each of 5.8~m$^2$ area (Fig.~\ref{fig:deployed}). This design has the advantage of significantly reducing the instrument background due to prompt and delayed \g-ray emissions from fast particle reactions with the platform materials. 

\begin{figure}
\begin{center}
\includegraphics[width=5cm]{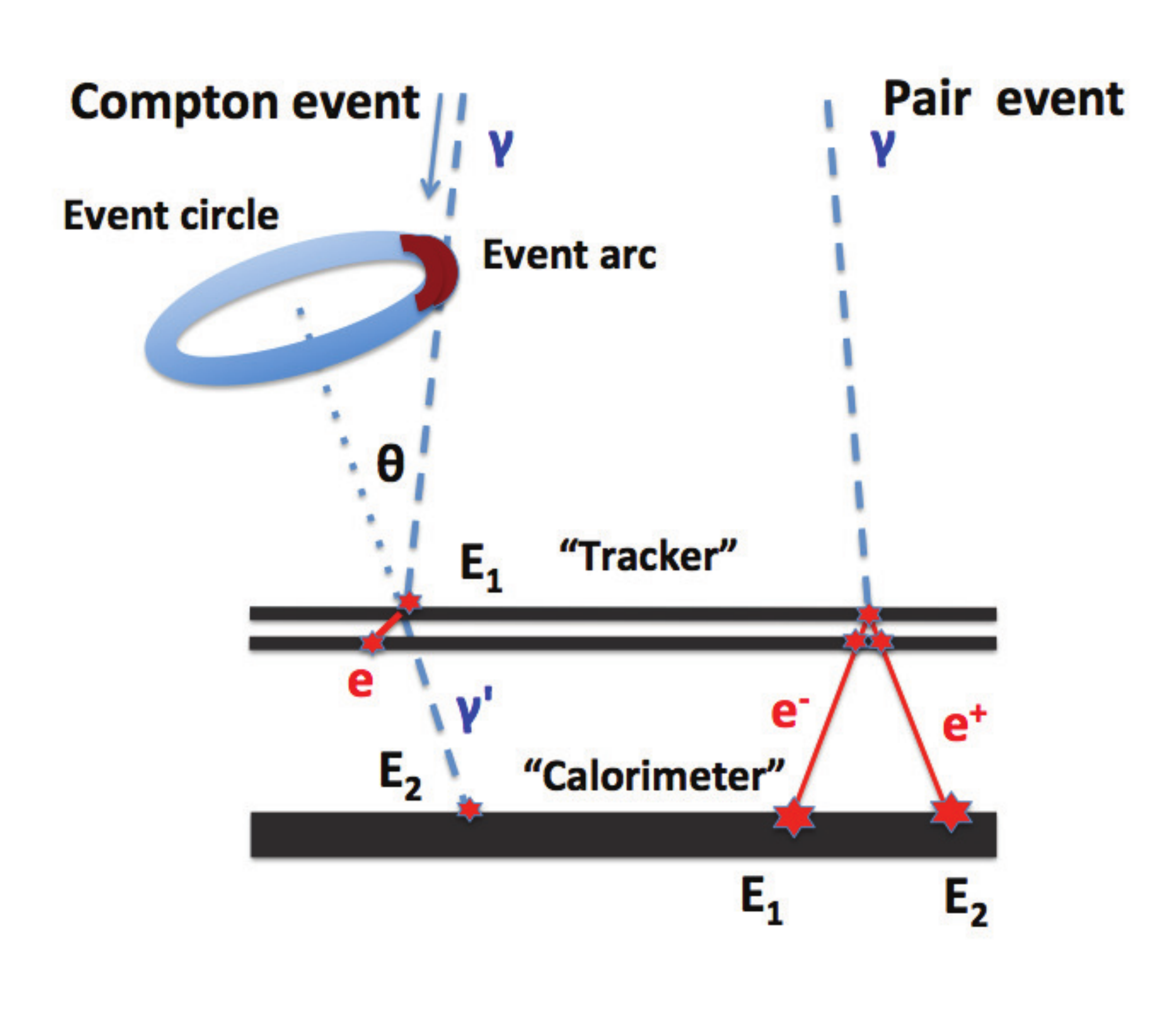}
\includegraphics[height=5cm,width=7.5cm]{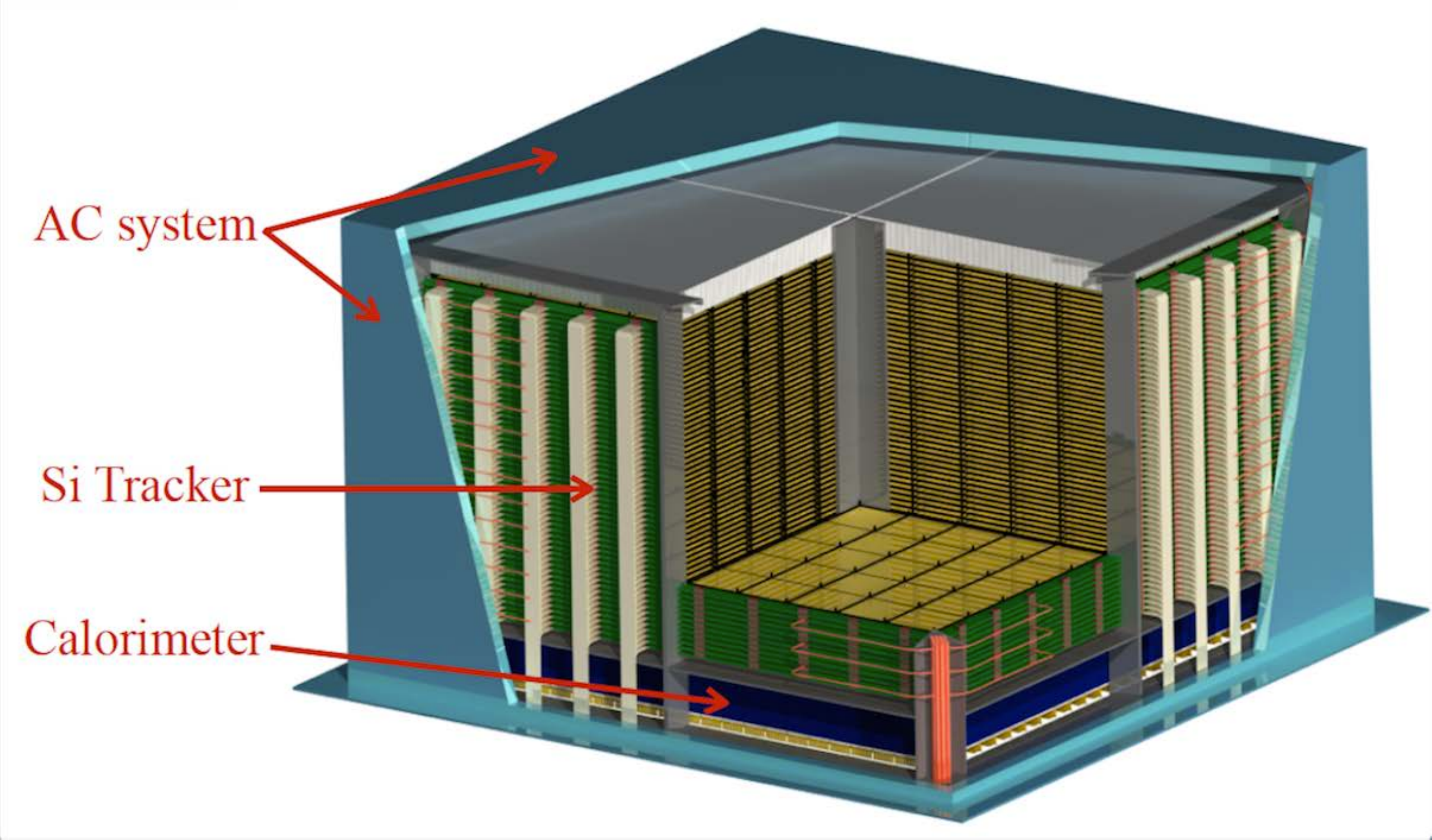}
\\
\end{center}
\caption 
{Left: Representative  topologies for a Compton event and  for a pair event. Photon tracks are shown in pale blue, dashed, and electron and/or positron tracks are in red, solid. Right: Overview of the e-ASTROGAM payload.} \label{fig:payload}
\end{figure}

The e-ASTROGAM telescope is made up of three detection systems (Fig.~\ref{fig:payload}): a silicon Tracker in which the cosmic \g-rays undergo a Compton scattering or a pair conversion (see Fig.~\ref{fig:payload} left); a Calorimeter to absorb and measure the energy of the secondary particles and an anticoincidence (AC) system to veto the prompt-reaction background induced by charged particles. The telescope has a size of 120$\times$120$\times$78 cm$^3$ and a mass of 1.2~tons (including maturity margins plus an additional margin of 20\% at system level).

The Si Tracker comprises 5600 double-sided strip detectors (DSSDs) arranged in 56 layers. It is divided in four units of 5$\times$5 DSSDs, the detectors being wire bonded strip to strip to form 2-D ladders. Each DSSD has a geometric area of 9.5$\times$9.5 cm$^2$, a thickness of 500~$\mu$m, and a strip pitch of 240~$\mu$m. The total detection area amounts to 9025 cm$^2$. Such a stacking of relatively thin detectors enables efficient tracking of the electrons and positrons produced by pair conversion, and of the recoil electrons produced by Compton scattering. The DSSD signals are read out by 860,160 independent, ultra low-noise and low-power electronics channels with self-triggering capability.

The Calorimeter is a pixelated detector made of a high-$Z$ scintillation material -- Thallium activated Cesium Iodide -- for efficient absorption of Compton scattered \g-rays and electron-positron pairs. It consists of an array of 33,856 parallelepipeds of CsI(Tl) of 8~cm length and 5$\times$5~mm$^2$ cross section, read out by silicon drift detectors (SDDs) at both ends, arranged in an array of 529 ($=23 \times 23$) elementary modules each containing 64 crystals. The depth of interaction in each crystal is measured from the difference of recorded scintillation signals at both ends. Accurately measuring the 3D position and deposited energy of each interaction is essential for a proper reconstruction of the Compton events. The Calorimeter thickness -- 8 cm of CsI(Tl) -- makes it a 4.3 radiation-length detector having an absorption probability of a 1-MeV photon on-axis of 88\%.

The third main detector of the e-ASTROGAM payload consists of an Anticoincidence system composed of two main parts: (1) a standard Anticoincidence, named Upper-AC, made of segmented panels of plastic scintillators covering the top and four lateral sides of the instrument, providing a total active area of about 5.2~m$^2$, and (2) a Time of Flight (ToF) system, aimed at rejecting the particle background produced by the platform. The Upper-AC detector is segmented in 33 plastic tiles (6 tiles per lateral side and 9 tiles for the top) coupled to silicon photomultipliers (SiPM) by optical fibers. The bottom side of the instrument is protected by the ToF unit, which is composed of two plastic scintillator layers separated by 50 cm, read out by SiPMs connected to Time Digital Converters. The required timing resolution is 300 ps. 

For best environmental conditions, the e-ASTROGAM satellite should be launched into a quasi-equatorial (inclination $i < 2.5^\circ$) LEO at a typical altitude of 550~--~600~km. The background environment in such an orbit is now well-known thanks to the Beppo-SAX~\cite{cam14} and \agile~\cite{tav09} missions. In addition, such a LEO is practically unaffected by precipitating particles originating from solar flares, a virtue for background rejection. 

Extensive simulations of the instrument performance using state-of-art numerical tools\cite{zog06,bul12} and a detailed numerical mass model of the satellite together with a thorough model for the background environment have shown that e-ASTROGAM will achieve: 
\begin{itemize}
\item Broad energy coverage ($\sim$0.15 MeV to 3 GeV), with nearly two orders of magnitude improvement of the continuum sensitivity in the range 0.3 -- 100 MeV compared to previous missions;
\item Excellent sensitivity for the detection of key \g-ray lines e.g. sensitivity for the 847~keV line from thermonuclear SNe 70 times better than that of the \INTEGRAL spectrometer (SPI);
\item Unprecedented angular resolution both in the MeV domain and above a few hundreds of MeV  i.e. improving the angular resolution of the COMPTEL telescope on board the {\it Compton Gamma Ray Observatory} ({\it CGRO}) and that of the \fermilat instrument by a factor of $\sim$4 at 5 MeV and 1~GeV, respectively (e.g. the e-ASTROGAM Point Spread Function (PSF) (68\% containment radius) at 1 GeV is 9').
\item Large field of view ($>$ 2.5 sr), ideal to detect transient Galactic and extragalactic sources, such as X-ray binaries and GRBs;
\item {Timing accuracy of 1~$\mu$s (at 3$\sigma$), ideal to study the physics of magnetars and rotation-powered pulsars, as well as the properties of TGFs;}
\item Pioneering polarimetric capability for both steady and transient sources, as illustrated in the next Section. 
\end{itemize}
%

e-ASTROGAM will be sensitive to the linear polarization of incident \g-rays over its entire bandwidth. In the Compton range, the polarization signature is reflected in the probability distribution of the azimuthal scatter angle. In the pair production domain, the polarization information is given by the distribution of azimuthal orientation of the electron-positron plane. e-ASTROGAM will have a breakthrough capacity for \g-ray polarimetry thanks to the fine 3D position resolution of both the Si Tracker and the Calorimeter, as well as the light mechanical structure of the Tracker, which is devoid of any heavy absorber in the detection volume. 

The measurement of polarization in the pair creation range, using the azimuthal orientation of the electron-positron plane, is complex and a precise evaluation of the unfolding procedures and performance requires accurate simulation and testing \cite{dbpol}. In the following, we focus on the e-ASTROGAM performance for polarimetry in the Compton domain. We discuss in particular the polarimetric capability of e-ASTROGAM for the study of AGNs, GRBs, the Crab pulsar and nebula, as well as microquasars. { e-ASTROGAM will explore for the first time the polarimetric properties of celestial sources above 1 MeV. Thus, as the mission will open a new window, it is difficult to assess what will be discovered. Anyway, we could expect to make detailed studies of jet non-thermal components observed from AGNs, stellar BHs and GRBs. We might also expect a better description of particle acceleration processes in, for example, pulsars and SNRs.}

\subsection{Instrument response}
\label{sec:perfass}

The scientific performance of the e-ASTROGAM instrument was evaluated by extensive numerical simulations with the software tools MEGAlib~\cite{zog06} and BoGEMMS (Bologna Geant4 Multi-Mission Simulator,  \cite{bul12}), together with detailed background model including the effects on the instrument response of the cosmic diffuse \g-ray radiation (both Galactic and extragalactic), the Galactic cosmic-ray protons and electrons modulated by the geomagnetic field, the secondary semi-trapped protons, electrons and positrons, as well as the atmospheric \g-rays and the secondary albedo neutrons. The environmental conditions in the quasi-equatorial (inclination $i < 2.5^\circ$) low Earth orbit (typical altitude of 550~km) of e-ASTROGAM is now well-known, thanks to the Beppo-SAX mission, which measured the radiation environment on a low-inclination ($i \sim 4^\circ$), 500 -- 600 km altitude orbit almost uninterruptedly during 1996 -- 2002 \cite{cam14} and the on-going \agile mission, which has been scanning the \g-ray sky since 2007 from a quasi-equatorial orbit at an average altitude of 535~km~\cite{tav09}.

The numerical mass model of e-ASTROGAM used to simulate the performance of the instrument includes passive material in the detector and its surroundings, true energy thresholds and energy and position measurement accuracy, as well as a roughly accurate spacecraft bus mass and position.

\subsubsection*{Angular and spectral resolution}

e-ASTROGAM will image the Universe with substantially improved angular resolution both in the MeV domain and above a few hundreds of MeV, i.e. improving the angular resolution of the {\it CGRO}/COMPTEL telescope and that of the \fermilat instrument by a factor of $\sim$4 at 1 MeV and 1 GeV, respectively.

\begin{figure}
\centering
\includegraphics[height=6cm]{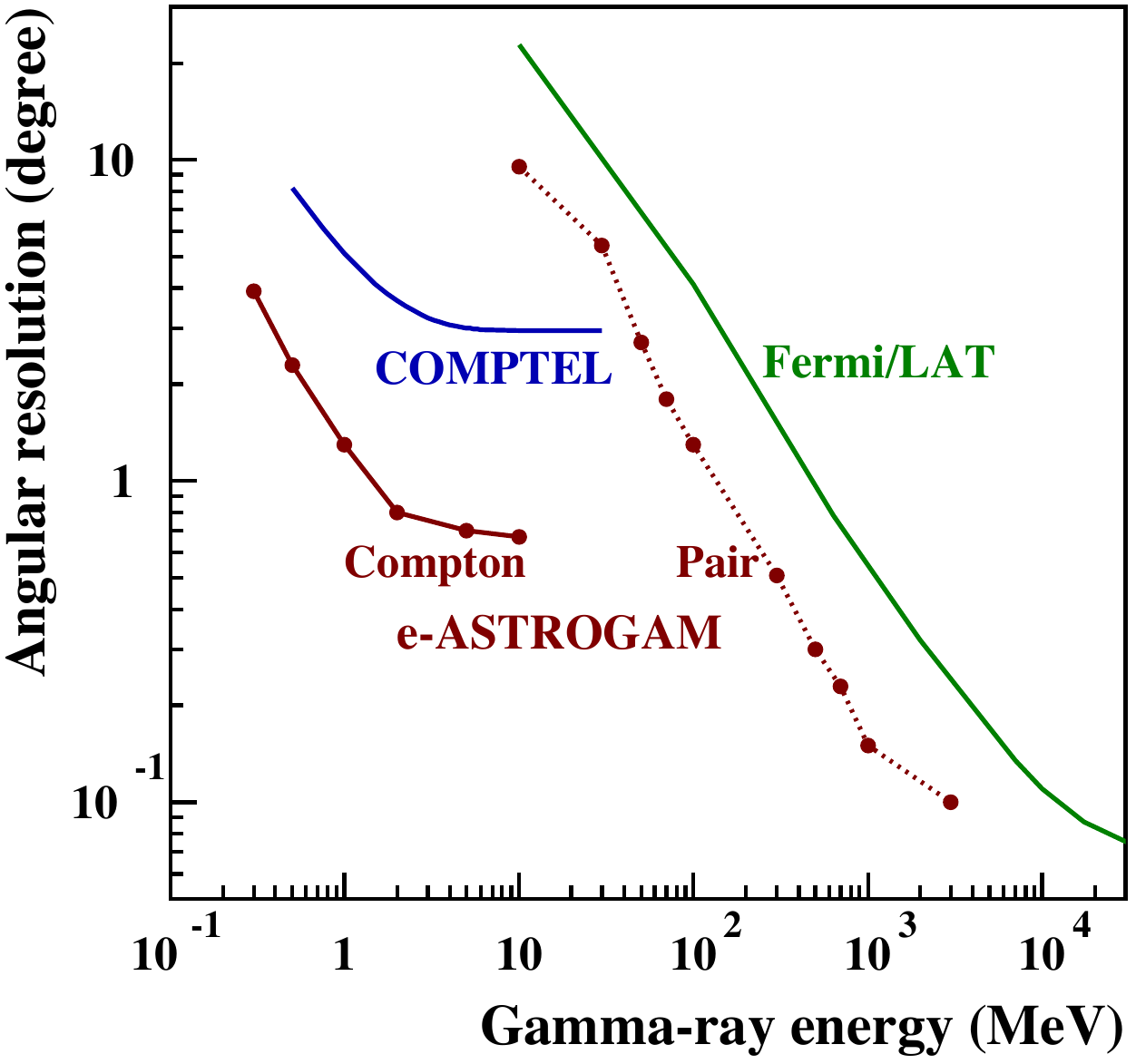}
\includegraphics[height=6cm]{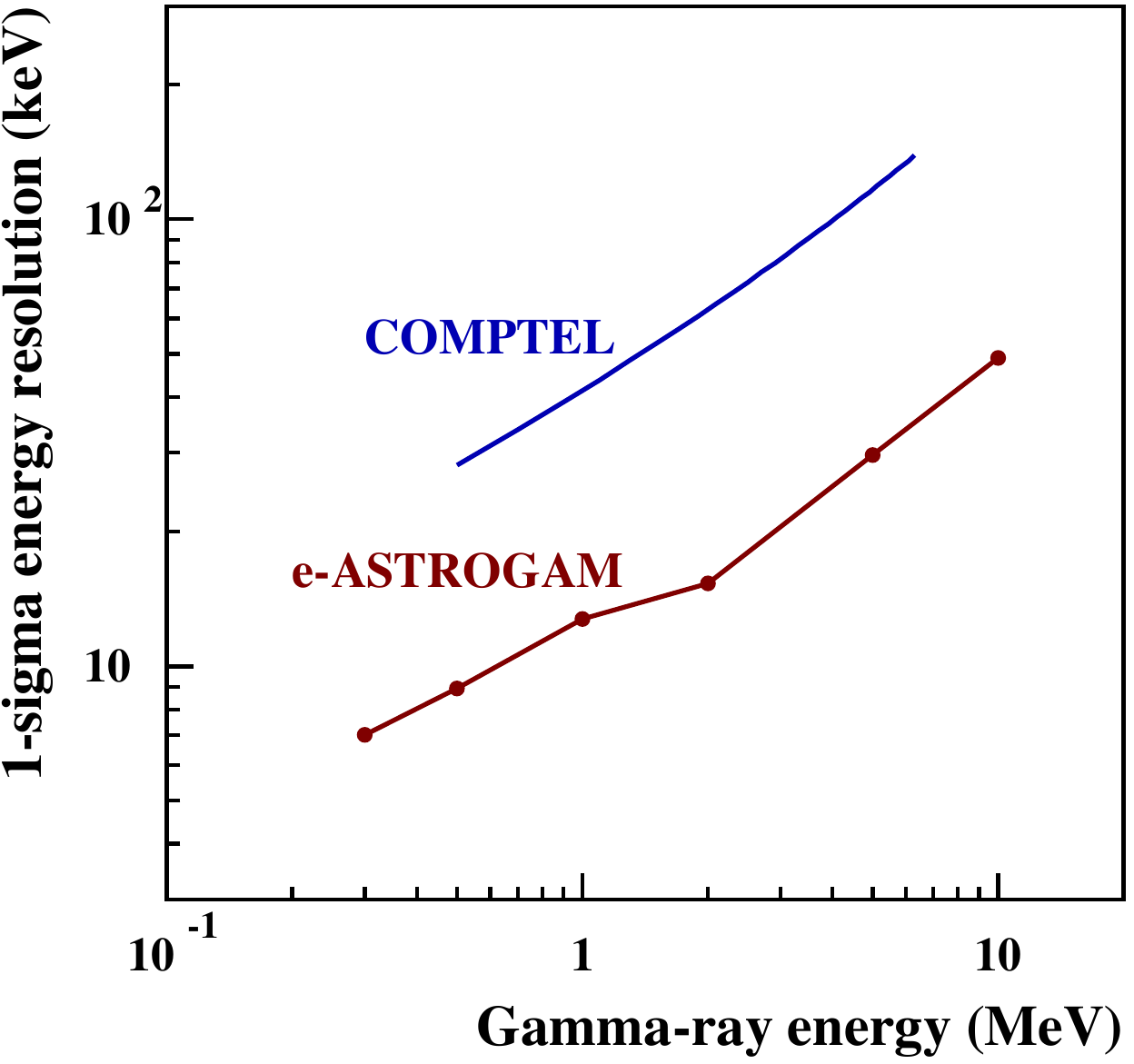}
\caption{Left: e-ASTROGAM on-axis angular resolution compared to that of COMPTEL and \fermilat. In the Compton domain, the presented performance of e-ASTROGAM and COMPTEL is the FWHM of the angular resolution measure (ARM). In the pair domain, the PSF is the 68\% containment radius for a 30$^\circ$ point source. The \fermilat PSF is from the Pass 8 analysis (release 2 version 6) and corresponds to the FRONT and PSF event type. Right: 1$\sigma$ energy resolution of COMPTEL and e-ASTROGAM in the Compton domain after event reconstruction and selection on the ARM.}
\label{fig:perf}
\end{figure}

In the pair production domain, the PSF improvement over \fermilat is due to (i) the absence of heavy converters in the Tracker, (ii) the light mechanical structure of this detector minimizing the amount of passive material within the detection volume and thus enabling a better tracking of the secondary electrons and positrons, and (iii) the analog readout of the DSSD signals allowing a finer spatial resolution of about 40~$\mu$m ($\sim$1/6 of the microstrip pitch). In the Compton domain, thanks to the fine spatial and spectral resolutions of both the Tracker and the Calorimeter, the e-ASTROGAM angular resolution will be close to the physical limit induced by the Doppler broadening due to the velocity of the target atomic electrons.

Fig.~\ref{fig:Jurgen}  shows an example of the e-ASTROGAM imaging capability in the MeV domain compared to COMPTEL. The e-ASTROGAM synthetic map of the Cygnus region was produced from the third \fermilat (3FGL) catalog of sources detected at photon energies $E_\gamma > 100$~MeV \cite{3FGL}, assuming a simple extrapolation of the measured power-law spectra to lower energies. It is clear from this example that e-ASTROGAM will substantially overcome (or eliminate in some cases) the confusion issue that severely affected the previous and current generations of \g-ray telescopes. The e-ASTROGAM imaging potential will be particularly relevant to study the various high-energy phenomena occurring in the GC region.

e-ASTROGAM will also significantly improve the energy resolution with respect to COMPTEL, e.g. by a factor of $\sim$3.2 at 1 MeV, where it will reach a 1$\sigma$ resolution of $\Delta E/E=1.3$\% (Fig.~\ref{fig:perf}). In the pair production domain above 30~MeV, the simulated spectral resolution is within 20--30\%.

\subsubsection*{Field of view}

The e-ASTROGAM field of view was evaluated from detailed simulations of the angular dependence of the sensitivity. Specifically, the width of the field of view was calculated as the half width at half maximum (HWHM) of the inverse of the sensitivity distribution as a function of the polar, off-axis angle, for a constant azimuthal angle $\phi=22.5^\circ$. In the Compton domain, the sensitivity remains high within $40^\circ$ to $50^\circ$ off-axis angle and then degrades for larger incident angles. For example, the field of view at 1~MeV amounts to 46$^\circ$ HWHM, with a fraction-of-sky coverage in zenith pointing mode of 23\%, corresponding to $\Omega = 2.9$~sr.

In the pair-production domain, the field-of-view assessment is also based on in-flight data from the \agile and \fermilat \g-ray imager detectors. With the e-ASTROGAM characteristics (size, Si plane spacing, overall geometry), the field of view is found to be $> 2.5$~sr above 10~MeV.  

\subsubsection*{Effective area and continuum sensitivity}

Improving the sensitivity in the medium-energy \g-ray domain (1--100~MeV) by one to two orders of magnitude compared to previous missions is the main requirement for the proposed e-ASTROGAM mission. Such a performance will open an entirely new window for discoveries in the high-energy Universe. Tables~\ref{table:sensitivity_Compton} and \ref{table:sensitivity_pair} present the simulated effective area and continuum sensitivity in the Compton and pair-production domains. The sensitivity below 10 MeV is largely independent of the source location (inner galaxy vs. high latitude), because the diffuse \g-ray background is not a major background component in the Compton domain.

\begin{table}
\centering
\caption{e-ASTROGAM performance in the Compton domain simulated with MEGAlib v2.26.01. The 3$\sigma$ continuum sensitivity is for the detection of a point source on axis after an observation time $T_{\rm obs}=10^6$~s.}
\vspace{0.1cm}
\includegraphics[width=0.99\textwidth]{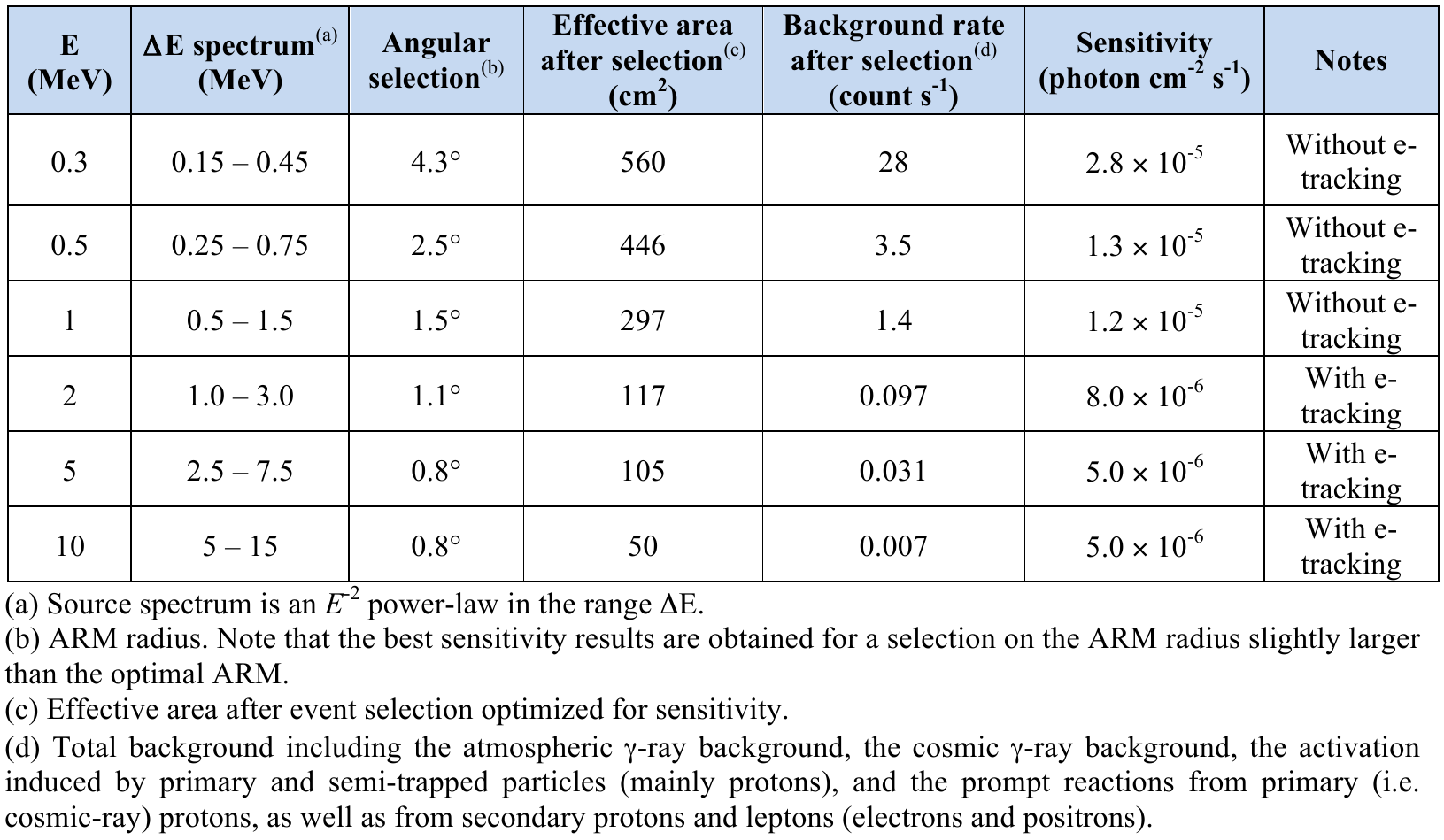}
\label{table:sensitivity_Compton}
\end{table}

\begin{table}
\centering
\caption{e-ASTROGAM performance in the pair-production domain simulated with BoGEMMS v2.0.1, together with Kalman v1.5.0 and Trigger v1.0.0. All results are for a 30$^\circ$ off-axis source and for $T_{\rm obs} = 10^6$~s. The King function used to fit the PSF, derived from the model of XMM data, is defined, e.g., in \cite{xmm04}.}
\vspace{0.1cm}
\includegraphics[width=0.99\textwidth]{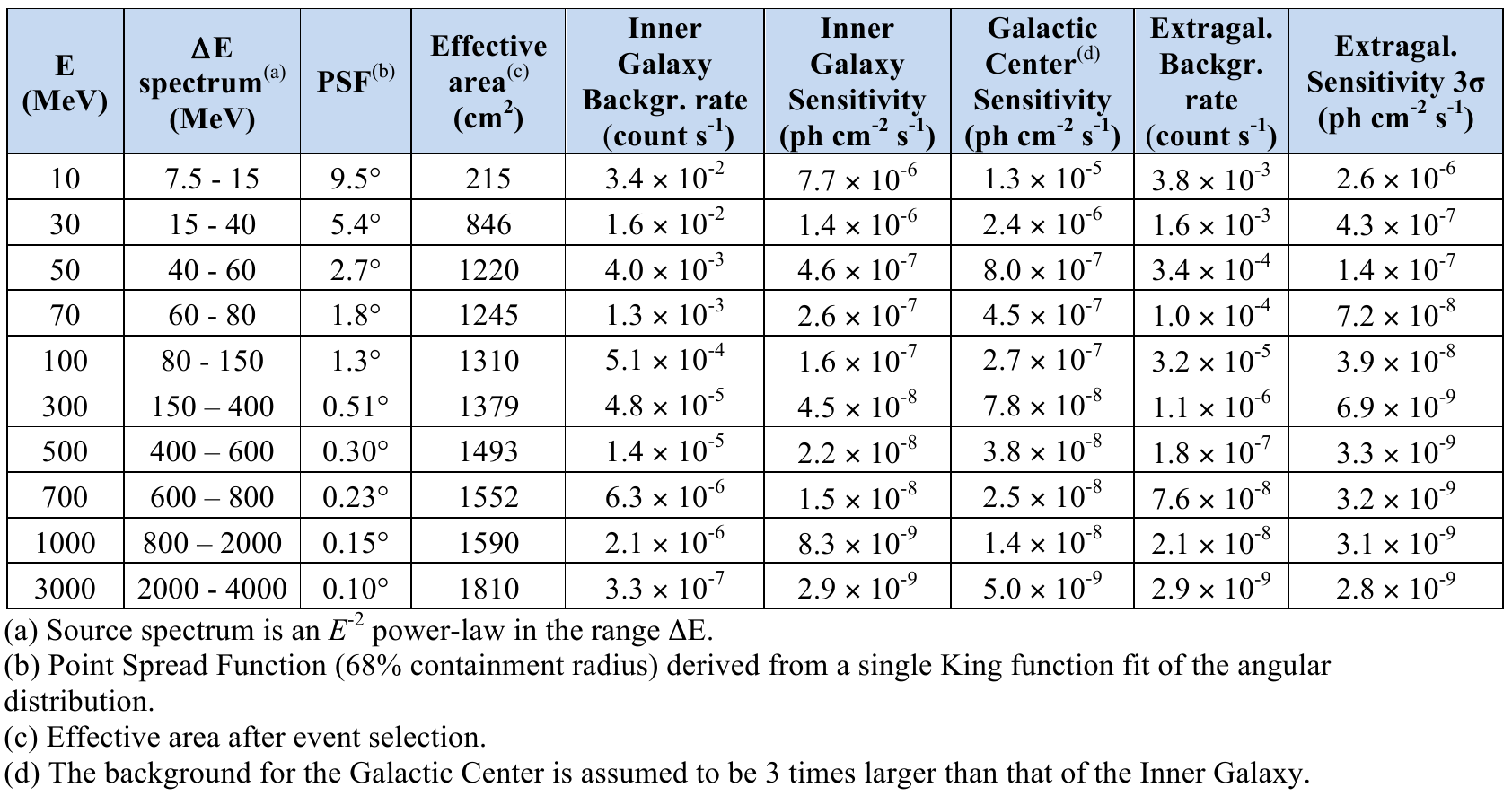}
\label{table:sensitivity_pair}
\end{table}

Fig.~\ref{fig:sensitivity}  shows the e-ASTROGAM continuum sensitivity for a 1-year effective exposure of a high Galactic latitude source. Such an effective exposure will be reached for broad regions of the sky after 3 years of operation, given the very large field of view of the instrument. We see that e-ASTROGAM would provide an important leap in sensitivity over a wide energy band, from about 200 keV to 100 MeV. At higher energies, e-ASTROGAM would also provide a new vision of the \g-ray sky thanks to its  angular resolution, which would reduce the source confusion that plagues the current \fermilat and \agile images near the Galactic plane (see, e.g., the 3FGL catalog \cite{3FGL}).

\subsubsection*{Line sensitivity}

\begin{table}
\centering
\caption{e-ASTROGAM line sensitivity (3$\sigma$ in 10$^6$ s) compared to that of \INTEGRAL/SPI\cite{roq03}.}
\vspace{0.1cm}
\includegraphics[width=0.99\textwidth]{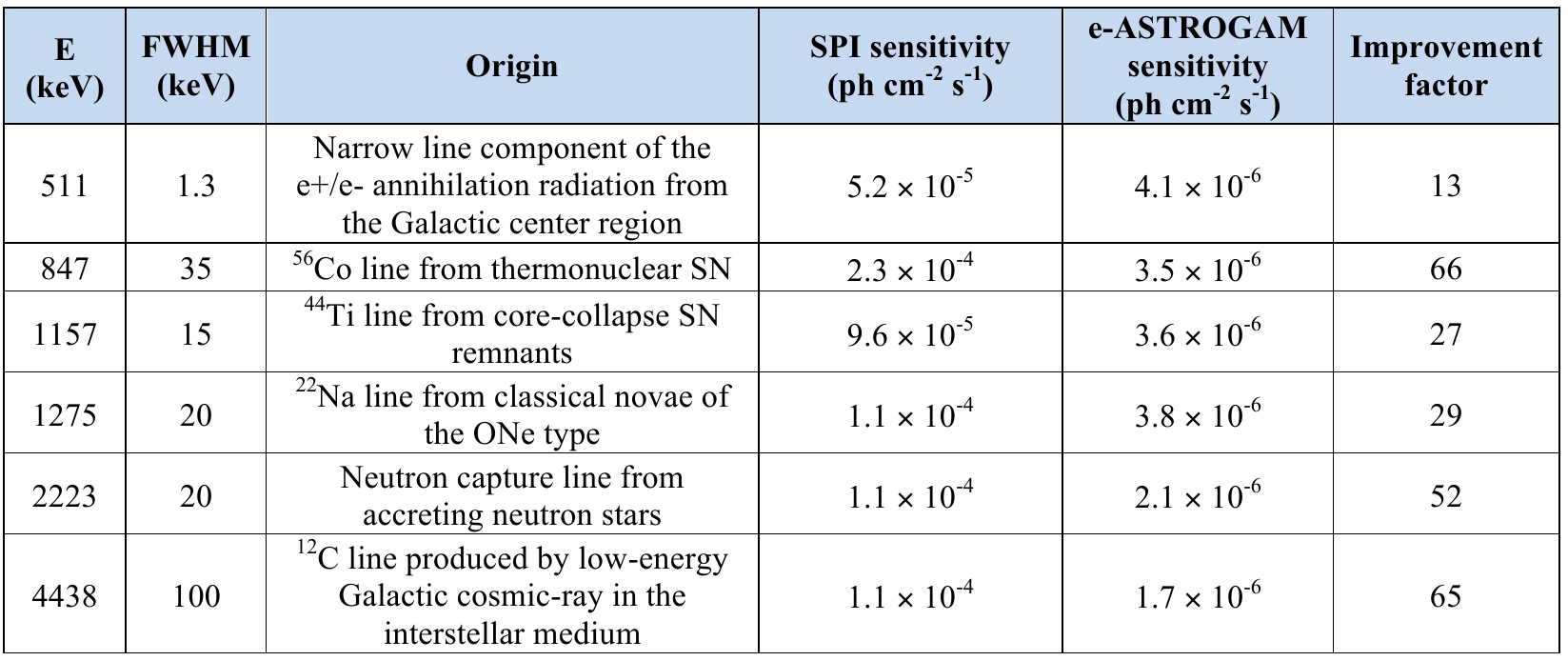}
\label{table:sensitivity_line}
\end{table}

Table~\ref{table:sensitivity_line} shows the e-ASTROGAM 3$\sigma$ sensitivity for the detection of key \g-ray lines from pointing observations, together with the sensitivity of the \INTEGRAL Spectrometer (SPI). The latter was obtained from the \INTEGRAL Observation Time Estimator (OTE) assuming 5$\times$5 dithering observations. The reported line widths are from SPI observations of the 511 and 847 keV lines (SN 2014J), and from theoretical predictions for the other lines. Noteworthy, the neutron capture line from accreting neutron stars can be significantly redshifted and broadened (FWHM between 10 and 100 keV) depending on the geometry of the mass accretion \cite{bil93}.

We see that e-ASTROGAM will achieve a major gain in sensitivity compared to SPI for all \g-ray lines, the most significant improvement being for the 847~keV line from Type Ia SNe. 

\subsubsection*{Polarization response}\label{sec:polargrb}

Both Compton scattering and pair creation partially preserve the linear polarization information of incident photons. In a Compton telescope, the polarization signature is reflected in the probability distribution of the azimuthal scattering angle. In the pair domain, the polarization information is given by the distribution of azimuthal orientation of the electron-positron plane. e-ASTROGAM will be able to perform for the first time at these energies polarization measurements thanks to the fine 3D position resolution of both the Si Tracker and the Calorimeter, as well as the light mechanical structure of the Tracker, which is devoid of any heavy absorber in the detection volume \cite{polpap}.

\begin{figure}
\centering
\includegraphics[width=\textwidth]{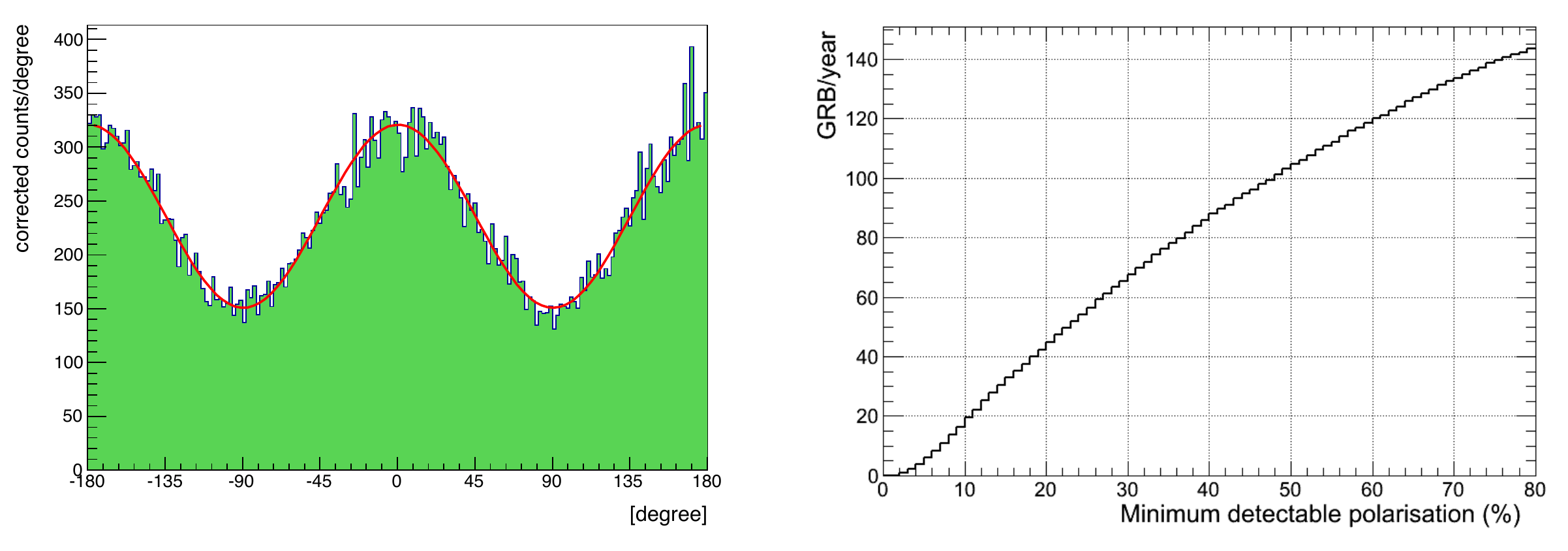}
\caption{Left: e-ASTROGAM polarization response (polarigramme) in the 0.2 -- 2 MeV range for a 100\% polarized, 10 mCrab-like source observed on axis for 10$^6$ s. The corresponding modulation is $\mu_{100}$ = 0.36.  Right: Cumulative number of GRBs to be detected by e-ASTROGAM as a function of the minimum detectable polarization at the 99\% confidence level.}
\label{fig:polarization}
\end{figure}

The left panel of Fig.~\ref{fig:polarization} shows an example of a polarigramme in the 0.2 -- 2 MeV range (i.e. in the Compton domain), simulated with MEGAlib. The calculations assume a 100\% polarized emission from a 10 mCrab-like source observed on axis. The systematic effects of instrumental origin were corrected by simulating the azimuthal response of the instrument to an unpolarized source with the same spectral distribution and position in the field of view as the polarized source. From the obtained modulation ($\mu_{100} = 0.36$), we find that at low energies (0.2 -- 2~MeV), e-ASTROGAM will be able to achieve a MDP at the 99\% confidence level as low as 0.7\% for a Crab-like source in 1~Ms (statistical uncertainties only). After one year of effective exposure of the GC region, the achievable MDP$_{99}$ for a 10~mCrab source will be 10\%. With such a performance, e-ASTROGAM will be able to study the polarimetric properties of many pulsars, magnetars, and BH systems in the Galaxy.

The right panel of Fig.~\ref{fig:polarization} shows the number of GRBs detectable by e-ASTROGAM as a function of MDP$_{99}$ in the 150--300 keV band. The total number of GRBs detected by e-ASTROGAM will be $\sim$600 in 3 years of nominal mission lifetime. Here, the GRB emission spectrum has been approximated by a typical Band function \cite{ban93} with $\alpha=-1.1$, $\beta=-2.3$, and $E_{\rm peak}=0.3$~MeV, and the response of e-ASTROGAM to linearly polarized GRBs has been simulated at several off-axis angles in the range $[0^\circ;90^\circ]$. The number of GRBs with polarization measurable with e-ASTROGAM has then been estimated as a function of polarization fraction using the 4th BATSE GRB Catalog \cite{pac99}. We see in Fig.~\ref{fig:polarization} that e-ASTROGAM should be able to detect a polarization fraction of 20\% in about 42 GRBs per year, and a polarization fraction of 10\% in $\sim$16 GRBs per year. This polarization information, combined with spectroscopy over a wide energy band, will provide unambiguous answers to fundamental questions on the sources of the GRB highly relativistic jets and the mechanisms of energy dissipation and high-energy photon emission in these extreme astrophysical phenomena.

The measurement of polarization using the azimuthal orientation of the electron-positron plane is complex and a precise evaluation of the unfolding procedures and performance requires accurate simulation and testing \cite{dbpol}. Thus, using a simplified model for pair production and multiple scattering of electrons and positrons, a MDP of $\sim$45\% at 3$\sigma$ has been estimated for the Crab Nebula in $10^6$ s in the range from 10 to 100 MeV. 

\subsection{Communicating e-ASTROGAM with science visualization}
As reported above, thanks to its unprecedented performances in the MeV-GeV domain, the e-ASTROGAM satellite will represent a focal point for the interests of different and broad scientific communities. 
In general, astronomical and astroparticle physics missions and experiments of the next future, will have to deal with the growing demand for science information and communication in the frame of the ``knowledge society''. 
The ability to communicate physics and astronomy with a wide variety of stakeholders is, however, a complex question, that is more and more interconnected to the future of the society, innovation and the future of science itself. In this view, an active communication and public outreach effort, to be planned and conducted in parallel to the technical and scientific developments of e-ASTROGAM, are expected to be very important. Public communication activities also encourage researchers to think about the big picture, avoiding to use too much details, jargon and specifics, and helping to get the essentials of their message.
The visual rendering of the rich quantity of subjects, astronomical objects and physical ambients in the MeV-GeV \g-ray Universe
is a central goal for e-ASTROGAM, unable to produce direct and fascinating images for the public like those produced by Hubble, JWST, Spitzer, Chandra or by planetary probes like Cassini. The situation is comparable to exoplanet missions where, for example, there are not direct and beautiful pictures to be shown, but dozens of pictorial artistic representations of newly discovered alien planets are disseminated around the web and the other media. The \g-ray sky at MeV-GeV energy band is also very abundant of fascinating, extreme, sources and phenomena for visual communication.
\begin{figure}[ttt!!]\label{communicscitopicsfig1}
\centering
\resizebox{\hsize}{!}{\includegraphics[width=\hsize,angle=0]{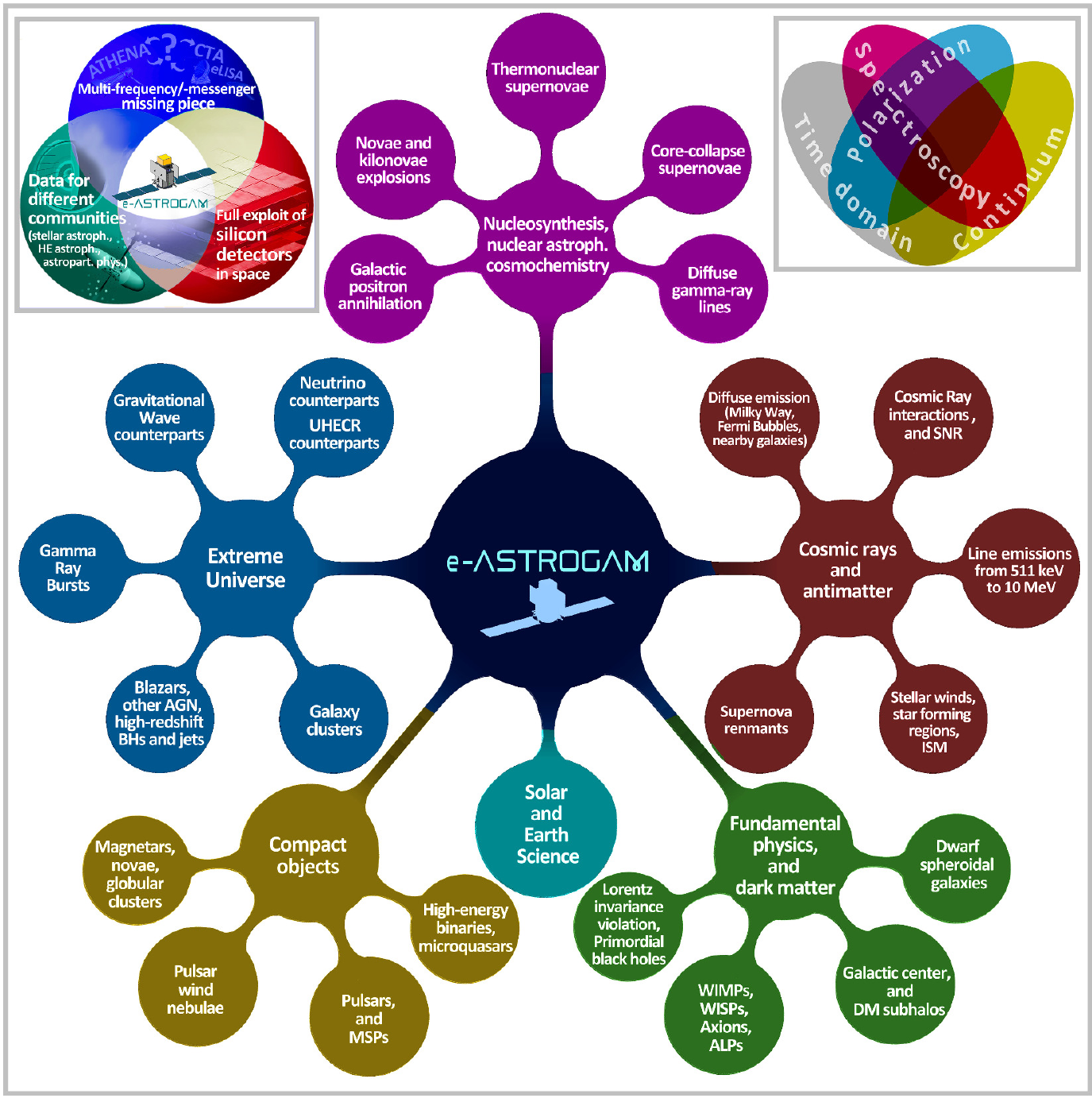}}
\vspace{-0.5cm}
\caption{An example of science visualization for e-ASTROGAM illustrating a tentative scheme for science topics and properties of the mission.}\label{science_comm}
\end{figure}
Beyond the three major ``core science'' topics, the e-ASTROGAM mission can be also conceived and communicated as a threefold mission, representing: 1) the missing, MeV-energy, piece of the multi-waveband/multimessenger puzzle placed between ATHENA and CTA; 2) the ability to provide data for different science communities as it will be addressed in this White Book (from stellar/nuclear astrophysicists to high-energy astrophysicists, to fundamental/particle physicists); 3) both an established technology with minor risks and a full use of this (double sided detectors, two-process detection) with a full capitalization of silicon detectors in space.
Beyond the three major ``core science'' topics of e-ASTROGAM, the list of supplementary, ancillary/bonus, topics can be large and summarized in Fig.~\ref{science_comm}. 
Science mapping and visual narration and conceptualization of the e-ASTROGAM instrument properties, simulations, calibrations, data analysis and scientific results, joined to the distribution of high-level data and tools for citizen science exploration, are science communication activities that can be developed in parallel to the main activities for e-ASTROGAM. Not even the most brilliant minds can keep up with today's deluge of growing scientific results. Science visualization can therefore help us to represent the landscape of what we know. A science atlas for e-ASTROGAM will better show to the public what we will know thanks to this mission. 

\subsection{Summary}

e-ASTROGAM is a concept for a \g-ray space observatory that can revolutionize the astronomy
of medium/high-energy \g-rays by increasing the number of known sources in this field by more than
an order of magnitude and providing polarization information for many of these sources -- thousands of
sources are expected to be detected during the first 3 years of operations. Furthermore, the proposed wide-field
\g-ray observatory will play a major role in the development of time-domain astronomy, and provide valuable
information for the localization and identification of GW sources.

The instrument is based on
an innovative design, which minimizes any passive material in the detector volume. The instrument performance
has been assessed through detailed simulations using state-of-the-art  tools and the results fully meet
the scientific requirements of the proposed mission. 

e-ASTROGAM will operate as an observatory open to the
international community.
The \g-ray observatory will be complementary to ground and space instruments, and multi-frequency observation
programs will be very important for the success of the mission. In particular, e-ASTROGAM will be
essential  for investigations jointly done with radio (VLA, VLBI, ALMA, SKA), optical (JWST,
E-ELT and other ground telescopes), X-ray and TeV ground instrument (ATHENA, CTA, HAWC, LHAASO and other
ground-based detectors). Special emphasis will be given to fast reaction to transients and rapid communication
of alerts. New astronomy windows of opportunity (sources of GWs, neutrinos, UHECRs) will be fully and uniquely explored.

{{\bf{Acknowledgements.}}
The contribution by P. Couzin (TAS-F), G. Cluzet (TAS-F), X. Roser (TAS-F), A. Laurens (CNES), D. Delrieu (CNES), M.-F. DelCastillo (CNES), C. Contini (CGS), P. Lattanzi (CGS), B. Morelli (CGS), A. Spalla (CGS), is acknowledged.

\newpage
\section[The extreme extragalactic universe\\
\noindent
\textnormal{\small\textnormal{Convenors:} \textit{G.~Ghisellini, L.~Hanlon, G.~Madejski, M.~Pohl, M.~Razzano}}]
{The extreme extragalactic universe}\label{chap:extreme_uni}
The universe contains objects with extreme properties than can be studied by measuring emission from particles that are accelerated near them. The emission is very intense, permitting measurements at very large distance, or redshift, when the universe was young and many galaxies still forming. In many cases, a substantial fraction of the radiated power appears in the MeV band, and so e-ASTROGAM would offer an ideal view 
of the violent processes operating close by supermassive BHs, inside the powerful explosions that we see as Gamma-ray Bursts (GRBs), and during the merger of binary neutron stars (NS). By deciphering many aspects of particle acceleration in the universe, we address why the energy distribution is so unbalanced: very few particles carry an extreme share of the available energy, and by their feedback they shape numerous cosmic objects.

GRBs are explosive events with peak emission in the MeV band. The unique capability of e-ASTROGAM to measure \g-ray polarization permits measuring the structure and amplitude of the magnetic field that shapes the acceleration and transport of particles. Lorentz-invariance violation can be searched for, and together with future gravitational wave detectors the relation between GRBs and the mergers of compact objects can be determined.

Clusters of galaxies are the largest gravitationally bound structures in the universe. In fact, they are still forming, leading to particle acceleration at structure formation shocks. Measuring their emission in the MeV band in conjunction with radio-band data lifts degeneracies in the interpretation and permits a precise study of the energy redistribution into magnetic field and accelerated particles, together with the feedback they impose on the cluster structure.

The MeV \g-ray background contains invaluable collective information about nucleosynthesis in distant SNe, DM annihilation, and supermassive BHs. The latter are also visible as AGN, and they are the most luminous persistent sources in the universe, many of which emit the bulk power in the MeV band. e-ASTROGAM can use these unique beacons to study the formation history and evolution of supermassive BHs at times when the universe had only a fraction of its current age. MeV-band observations address the energy limit to which electrons may be accelerated, the location where this happens. By studying the spectral response to changes in the activity of these objects, we can distinguish the emission from electrons from that of energetic ions. The MeV band is ideally suited for this inquiry, because emission at higher \g-ray energies may be absorbed, and the specific contribution from photo-pair-production by high-energy cosmic nuclei is a critical discriminant in the soft \g-ray band, as an analysis of the recent detection of a statistical association of a 300-TeV neutrino event with an extended \g-ray flare of the Active Galactic Nucleus TXS0506+056 shows. Finally, the MeV band carries the cascade emission of all the absorbed Very-High-Energy (VHE) \g-ray emission that is emitted in the universe, and so its study provides a unique view of its extreme particle acceleration history, including the feedback on the intergalactic medium and the magnetic-field genesis therein.

Last but not least, the MeV range is the perfect companion for multimessenger astronomy.  On top of the SED of the EM emission by TXS0506+056, mentioned before, the recent NS-NS merger generating the GW event GW170817 and the corresponding \g-ray signal detected by \fermigbm and \INTEGRAL has shown that the EM cutoff of this class of mergers is in the MeV range.
\subsection[Electromagnetic counterparts to Gravitational Wave transients in the MeV range\\
\noindent
\textit{\small{B. Patricelli, A. Stamerra, M. Razzano, V. Tatischeff, A. De Angelis, \\M. Branchesi}}]
{Electromagnetic counterparts to gravitational wave transients in the MeV range}
\paragraph*{Science questions}
The long-standing quest for the observation of GWs met with success on September 14, 2015 when the two Advanced LIGO interferometers detected the signal from the final inspiraling, merging and ring-down of a coalescing binary system formed by two stellar BHs (BBH; the event was named GW150914 \cite{LVC1}). 
After this first event, the LIGO scientific collaboration and the Virgo collaboration reported the GW detection of other three BBH mergers: GW151226 \cite{LVC2}, detected during the first observing run (O1, September 2015 - January 2016) and GW170104 \cite{LVC3} and GW170814, detected during the second observing run (O2, November 2016 - August 2017). In particular, GW170814 was the first detection made by the LIGO-Virgo network, since Advanced Virgo joined O2 on August 1, 2017. During O2, LIGO and Virgo also detected GW170817, the first signal from the coalescence of two NS 1.7 s before the \g-ray signal detected by the \fermigbm instrument.
Thanks to a dedicated follow-up campaign, EM counterparts to GW170817 were found in the visible, X-ray and radio bands \cite{mma}, marking the first multimessenger observation done with EM and GWs.
In fact, besides BBHs the most promising transient sources that emit GWs at the frequencies at which Advanced LIGO and Advanced Virgo are sensitive (20 Hz - 20 kHz) are the coalescences of binary systems NS-NS or a neutron star and a stellar mass BH (NS-BH). These sources are expected to have also an associated EM emission. Specifically, these systems are expected to be the progenitors of short Gamma-ray Bursts (GRBs): intense flashes of \g-rays lasting less than 2 s, sometimes followed by a  long lasting multi-wavelength afterglow emission (see \cite{grbreview} for a review). Furthermore, NS-NS mergers are theoretically predicted to entail significant mass ejection which interacts with the surrounding medium on timescales of years, producing a remnant in which accelerated electrons can produce \g-ray emission \cite{Takami}. The association between GW170817 and the GRB 170817A by \fermigbm \cite{gw170817gbm} supports the connection between NS-NS mergers and short GRBs.
Joint GW and EM observations are key to obtain a more complete knowledge of the sources and their environments, since they provide complementary informations. From one side, GW signals provide information about the physics of the source such as, e.g., the mass and the distance; on the other hand, the identification of the possible EM counterpart pinpoints the location of the burst, possibly identifying the host galaxy and properly defining the astrophysical context. 
Finally, the detection of the \g-ray counterpart with e-ASTROGAM will help understand if also NS-BH systems are progenitors of short GRBs and to characterize the astrophysical properties of the source. These results will also improve our knowledge of the stellar population of our Galaxy, with a particular focus on the progenitor of merging binary systems.
\paragraph*{Importance of \g-ray observations}
The search for the EM counterpart to GW transient events is challenging for several reasons. First of all, the sky localization provided by the current ground-based interferometers is in order of tens to hundreds of square degrees (see, e.g., \cite{LVC4}), therefore large field-of-view (FoV) instruments are essential to properly cover the large GW error boxes. Furthermore, within these boxes, a huge number of EM transients is expected, making it difficult a clear and univocal identification of an EM counterpart to the GW event (e.g., the number of optical transients spatially and temporally coincident with GW events is expected to be of the order of hundreds, see e.g. \cite{Nissanke}); this is somewhat mitigated at \g-ray energies, where the number of transient events is much smaller than at lower energies (for instance, the \fermigbm transient catalog comprises only a few events in an area of 100 square degrees, see \cite{Jenke}).
In the \g-ray domain, the favourite EM counterparts to NS-NS mergers are short GRBs, possibly accompanied by a thermal signal associated to the ``kilonova'' emission (see \cite{Barnes,Hotokezaka}). The EM emission from short GRBs is believed to be beamed and the observed sources are typically the on-axis ones, i.e. the ones for which the angle between the line-of-sight and the jet axis is less than the jet opening angle. However, the majority NS-NS merger events will correlate to off-axis short GRBs, as suggested by simple geometrical arguments based on the presumable small opening angle $\theta \sim 10^{\circ}$ of the jet \cite{Fong}.
Taking into consideration that the observed flux from on-axis GRBs is enhanced by beaming, off-axis GRBs flux is dramatically weaker and very sensitive \g-ray instruments are needed to reveal nearby off-axis GRBs associated to GW events.\\
From the observational point of view, the follow-up of GW170817 conducted in optical, IR and UV revealed the presence of an EM counterpart with emission consistent with a kilonova, while X-ray and radio data are interpreted as due to an off-axis afterglow emission \cite{mma}.\\
Polarization is expected if the jet launching is driven by magnetic energy and depending on the magnetic field configuration. Off-axis observations can introduce an anisotropy that enhances the degree of polarization \cite{ghisellini,granot}. In case a high-energy MeV-GeV component is observed, polarization can help to discriminate between different emission processes such as Inverse-Compton (\ic) emission of leptons (no polarization) and synchrotron polarized emission from hadrons. Measurement of the \g-ray polarization in GW triggered events could provide a new tool for the interpretation of the GW/EM emission.
\paragraph*{Expected results with e-ASTROGAM}
e-ASTROGAM fills the gap in the energy region from X-rays up to GeV and TeV \g-rays, providing a MeV \g-ray detector operating at the same time as facilities such as SKA and CTA.
e-ASTROGAM may coincide with the third generation of ground-based interferometer projects, such as the Einstein Telescope and Cosmic Explorer, with an order of magnitude increase in sensitivity (see e.g. \cite{hild,dwyer}). Furthermore, the space detector LISA will open GW observations to massive, $10^4 -10^6 M_{\Sun}$ BHs,  which could have magnetized circumbinary discs powering EM emission. 
Within the GW-sGRB paradigm, on-axis GRBs associated to GW events shall be favourably detected with e-ASTROGAM. The presence of a GW signal naturally selects nearby GRBs, thus favouring the detection of the prompt emission and possibly of the delayed afterglow. When Advanced LIGO and Advanced Virgo will operate at design sensitivity, the expected range for the detection is 200 Mpc for NS-NS mergers and $\approx 1$ Gpc for BH-NS systems \cite{LVC4}. 
Considering a maximum GRB jet opening angle of 30$^\circ$ (see, e.g., \cite{patricelli}) and taking into account the updated NS-NS merger rate estimates \cite{BNSdiscovery}, the expected detection rate of GRB prompt emission by e-ASTROGAM in coincidence with a GW detection is between $\sim$ 0.6 yr$^{-1}$ and $\sim$ 9 yr$^{-1}$ ; these numbers will double after the incorporation of KAGRA and LIGO-India into the GW network, which should happen several years before 2029. 
e-ASTROGAM will also play a key role in the multiwavelength study of GW events: in fact, its large FoV will maximize the detection probability and provide accurate sky localization ($<$ 1 sq. deg at 1 MeV), thus allowing the follow-up of the GW events by other telescopes. This capability will be crucial for the identification and the multiwavelength characterization of the GW progenitor and of its host galaxy. \\
The joint GW and EM detection rate is expected to increase if off-axis GRBs are taken into account. To verify the capability of e-ASTROGAM to detect also these sources, we estimate the minimum luminosity $L_{\rm min}$ for a short GRB to be detected at a distance equal to the horizon of Advanced LIGO at design sensitivity. We simulate a short GRB spectrum assuming the Band function, with the parameters estimated for short GRBs observed by \fermigbm \cite{nava} and different values for the luminositiy of the source; we then compare the predicted flux with the sensitivity of e-ASTROGAM in the energy range 0.2-2 MeV for an observation period of 1 s, that is 0.05 ph cm$^{-2}$ s$^{-1}$: we obtain $L_{\rm min} \sim 10^{48}$ erg/s. This value is much lower than the typical luminosity of short GRBs (see, e.g., \cite{davanzo}): this suggest that e-ASTROGAM will be able to detect also off-axis sources, with the consequent sizeable increase in the detection rates.  
e-ASTROGAM will also be able to detect events like GRB170817. This GRB is characterized by an isotropic peak luminosity $L$=1.6 $ 10^{47}$ erg/s, a luminosity distance  40 Mpc and its spectrum is well described by an exponentially cut-off power law (see \cite{GRB170817,GBMpaper}; the expected flux in the 0.2-2 MeV energy range for such an event is $\sim 0.8$ ph cm$^{-2}$ s$^{-1}$, that is above the e-ASTROGAM sensitivity (see Chapter~\ref{intro}).

e-ASTROGAM will be capable also to detect the MeV \g-ray emission associated to kilonovae, provided that the sources are located at a distance less than 10-15 Mpc, where the expected flux for $\sim 1$ MeV photons is of the order of $10^{-11} -10^{-12}$ erg cm$^{-2}$s$^{-1}$ \cite{Hotokezaka}.

Fig. \ref{fig:GW170817} illustrates the superior sensitivity of e-ASTROGAM (compared to \INTEGRAL) to detect the continuum and nuclear line emissions  expected from the kilonova following a merger event like GW170817. Kilonovae are thought to be primarily powered by the radioactive decay of r-process nuclei synthesized in the merger outflows, and e-ASTROGAM could detect the predicted \g-ray line emission \cite{Hotokezaka} up to a maximum distance of $\sim 15$~Mpc. 
Fig. \ref{fig:GRB}, top, shows that e-ASTROGAM will be more sensitive than \fermigbm over 60\% of the sky. Simulations of the large GRB database yield detection rates of about 60 short GRBs and 180 long GRBs per year in the ``Gamma-ray imager'' trigger mode of e-ASTROGAM, and these events will be localized within $\sim$2 square degrees to initiate observations at other wavelengths. Additional, softer bursts will be detected by the ``Calorimeter burst search'' mode of data acquisition (i.e. using triggers generated only by an increase of the Calorimeter count rate). The 6$\sigma$ trigger threshold in this mode is $\sim 0.05$~ph~cm$^{-2}$~s$^{-1}$ in the 100--300~keV energy range over 1~s timescale, which is an order of magnitude lower than the measured flux in the main pulse ($\Delta t =0.576$~s) of GRB~170817A, $F_\gamma$(100--300~keV)$=0.49$~ph~cm$^{-2}$~s$^{-1}$ \cite{gw170817gbm}.  Finally,  GRB170817A is likely to have been observed at an angle $\sim30^0$ from the jet axis (see also \cite{jetangle}): the on-axis energy should  be larger by a factor of 20 to 30 \cite{kim}, allowing e-ASTROGAM to track Compton events (Fig. \ref{fig:GRB}, bottom).

\begin{figure}[t!]
\begin{center}
\includegraphics[width=0.465\linewidth]{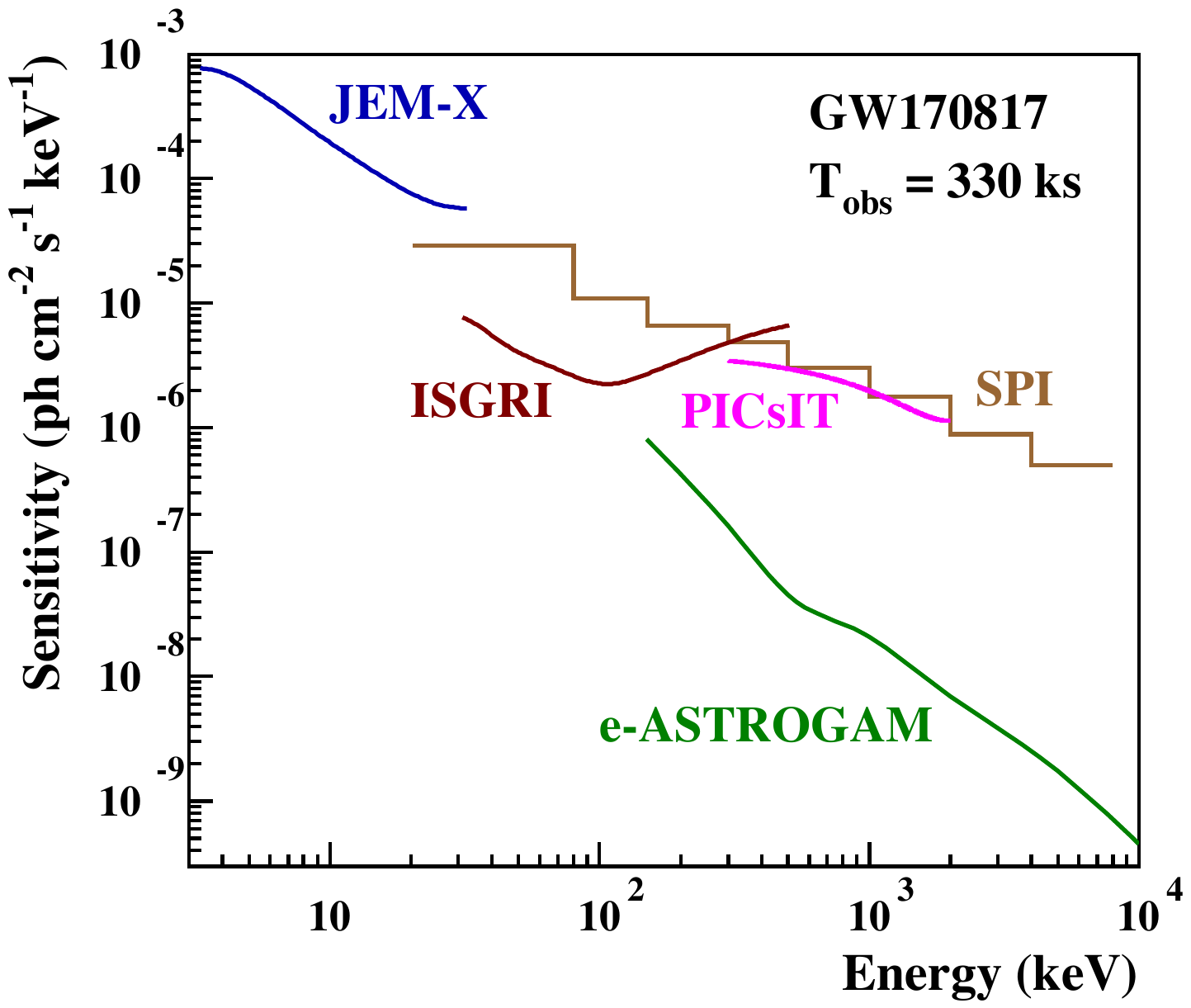}
\includegraphics[width=0.5\linewidth]{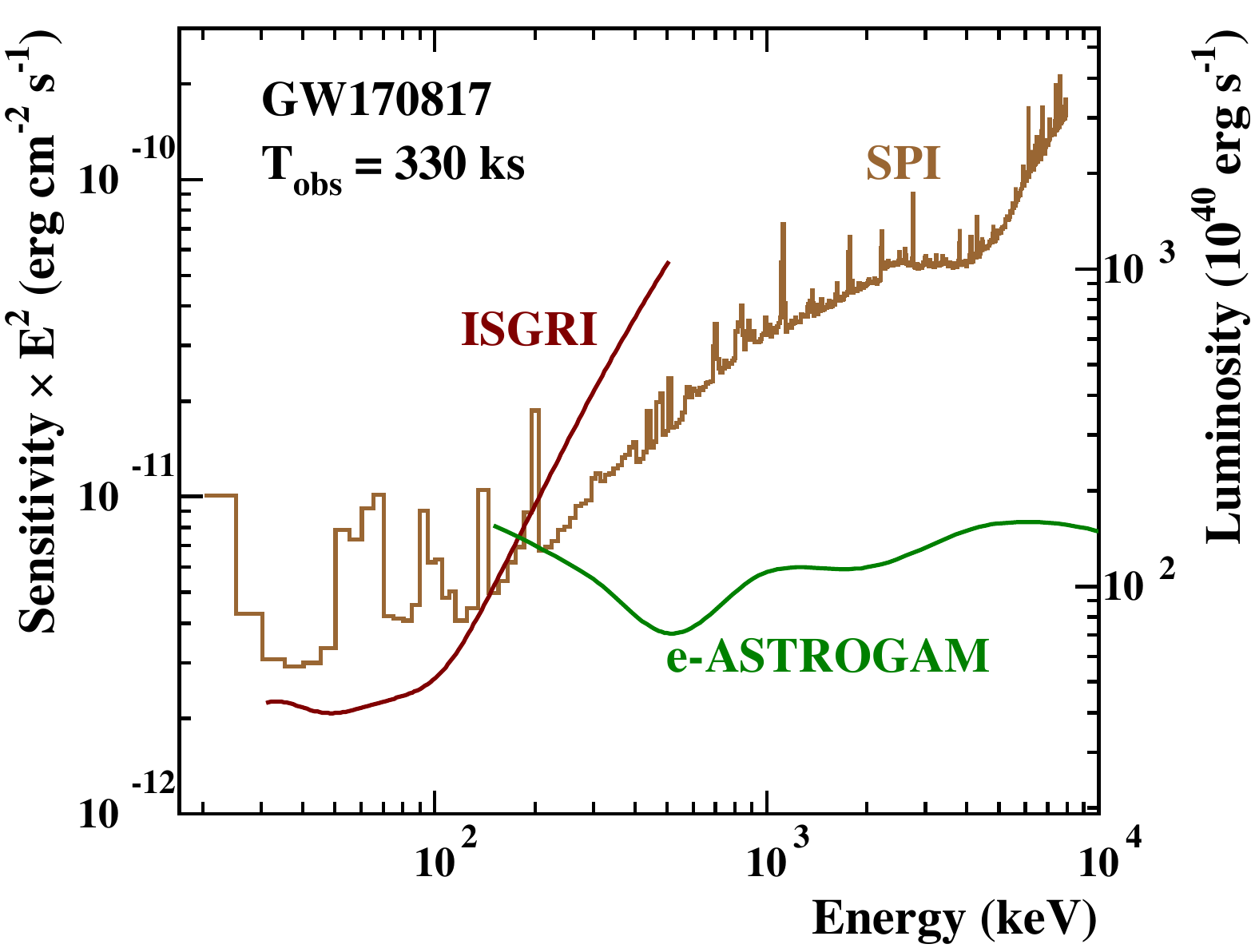}
\end{center}
\caption 
{ \label{fig:GW170817}
Left: Continuum and (b) narrow-line sensitivities reached in the \INTEGRAL targeted follow-up observation of GW170817, compared to the corresponding sensitivities of e-ASTROGAM. All sensitivities are shown for a total exposure of 330~ks. The luminosity units of the panel on the right assume a distance to the source of 40 Mpc. Adapted from Figs.~5 and 6 of Ref.~\cite{sav17}.} 
\end{figure} 

\begin{figure}[!ht]
\begin{center}
\includegraphics[width=0.43\textwidth]{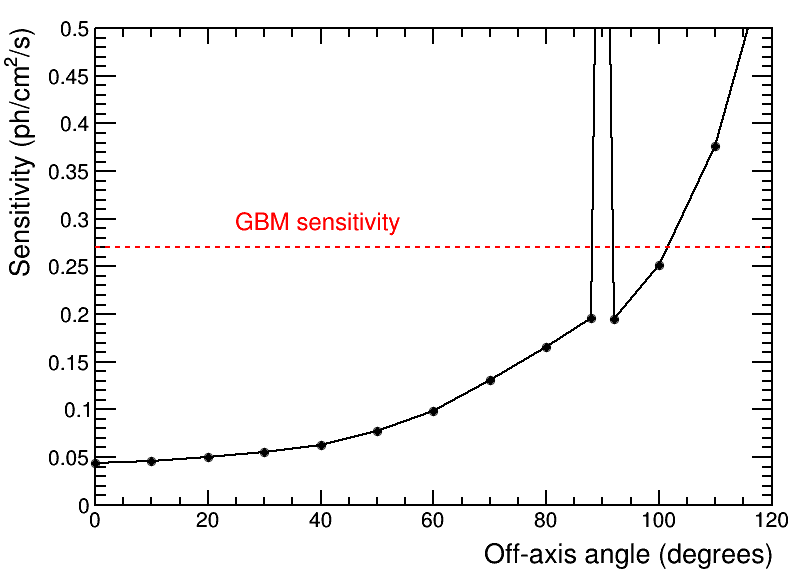}
\includegraphics[trim= 0mm 1.6cm 0cm 1.6cm, width=0.54\textwidth]{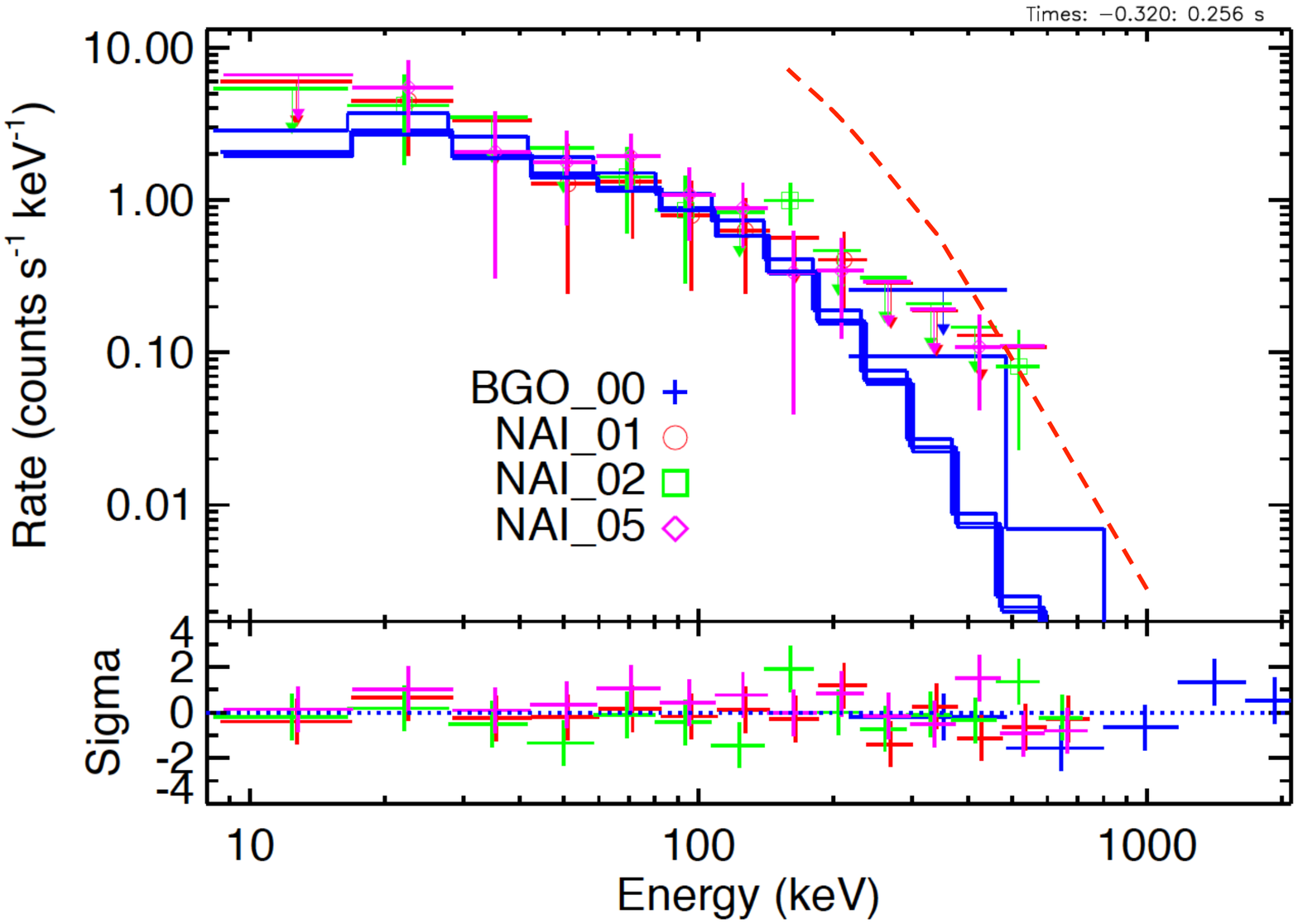}
\caption{\label{fig:GRB}
Left: 6\,$\sigma$ sensitivity of e-ASTROGAM to an average GRB on a 1 second timescale in the 0.2--2 MeV band. The sharp loss of sensitivity at 90$^{\circ}$ incidence is due to \g-rays crossing the tracker parallel to the silicon detectors. The red line gives the equivalent trigger sensitivity of \fermigbm, adapted from \cite{gbm09}. Right: the flux from the hard component of GRB170817A as recorded from \fermigbm (solid blue line), and a conservative extrapolation (20x) to an on-axis flux (dashed red line).} 
\end{center}
\end{figure} 

e-ASTROGAM will also allow to measure the polarization of the brightest events with the highest fluence, typically of the order of $10^{-4} - 10^{-5}$ erg/cm$^2$  down to the level of 10-20\%. 
The possible detection of polarization from GRB associated to GW events with e-ASTROGAM shall have a tremendous impact on the interpretation of the formation of the jet and radiation mechanisms. 

The detection of the \g-ray counterpart with e-ASTROGAM will help to understand if and which binary systems are progenitors of short GRBs and to characterize the astrophysical properties of the source. Simultaneous GW/EM emission will transform our understanding of the formation, evolution, properties and environment of different mass compact objects through cosmic history.

As a final note, also BH-NS mergers, yet undetected, might  entail electromagnetic emission with  a cutoff in the $\sim 10-20$ MeV region \cite{bhns,bhns2}.

\subsection[Synergies between neutrino telescopes and e-ASTROGAM\\
\noindent
\textit{\small{E. Bernardini, S. Buson, A. Coleiro, A. De Angelis, L. Foffano}}]{Synergies between neutrino telescopes and e-ASTROGAM}
\paragraph*{Science questions}
Neutrinos are unique probes to study high-energy cosmic sources. Contrary to CRs, they are not deflected by the magnetic fields and unlike high-energy photons, they are not absorbed by pair production via $\gamma\gamma$ interactions. Astrophysical high-energy neutrinos at TeV--PeV energies are generated by the decay of charged pions produced in inelastic photo-hadronic ($p\gamma$) or hadronuclear ($pp$) processes, involving protons $\sim20$ times more energetic than the resulting neutrinos. Photoproduction of neutrinos (and photons) via pion decay happens mainly via the $\Delta^+$ resonance: $p\gamma \rightarrow \Delta^+ \rightarrow N \pi$, and has a kinematical threshold.
The energy of the proton has to be $E_p  \gtrsim$ 350 PeV/$\epsilon$, where $\epsilon$ is the target photon energy in eV. For UV  photons, as expected in AGN jets, this translates into $E_p \gtrsim$~10 PeV, i.e., above the knee of the CR spectrum: photoproduction of neutrinos on optical/UV photons is a likely indicator of UHECR acceleration. A simultaneous emission of hadronic \g-rays is also expected from both processes. An approximate relation holds {\em{at emission}} between the spectral production rates  of neutrinos and \g-rays in hadronic production: 
\[
E_\nu^2 \frac{dN_\nu(E_\nu)}{dE_\nu} \sim \frac{3K}{4}  E^2_\gamma \frac{dN_\gamma(E_\gamma)}{dE_\gamma} \]
with $K = 1/2(2)$ {{for}} the $\gamma p \,(pp)$ process. Depending on the source optical depth, such photons may escape or further cascade, complicating  time and energy correlation between neutrinos and EM counterparts.

A diffuse flux of cosmic neutrinos has been detected by IceCube~\cite{IceCube_HESE1}, the sources of which are still unknown. Identifying those sources and their association with EM counterparts would provide unique insights into the long-standing problem of the origin of CRs~\cite{Santander2017}. Many astrophysical source classes have been suggested as responsible for the IceCube signal, like star-forming and/or star-burst galaxies, GRBs, or AGN. Galactic sources like microquasars are also expected to be emitters of astrophysical neutrinos. For a review on neutrino source candidates and multimessenger connections see e.g.~\cite{review_ahlers_halzen}.

 In conventional GRBs, the neutrino emission is expected to be in temporal coincidence with the prompt \g-ray emission. Recent results from IceCube~\cite{GRB_IC} disfavour them as the sources of the highest energy CRs and neutrinos. Such conclusions however would not apply if the central engine is surrounded by a dense material envelope,  like the shocked jets proposed in~\cite{Senno2016}. For AGN, predicted fluxes strongly vary with the assumed emission mechanisms. A recent IceCube analysis~\cite{IC_blazars} suggests that blazars contribute at most 27\% of the observed IceCube intensity. 
 
 Neutrinos could be emitted during flaring events, making simultaneous observation of neutrino and \g-ray signals mandatory to probe this scenario. Recently, the TANAMI collaboration reported that the detection of the third PeV neutrino by IceCube occurred during a major and long-lasting \g-ray ($0.1-300$ GeV) outburst of the blazar PKS~B1424-418 with a small \textit{a posteriori} chance coincidence probability of $\sim5$\%~\cite{TANAMI_ref}. While a genuine association of the PeV neutrino and the \g-ray flare seems unlikely~\cite{Gao2017}, this result illustrates well the great importance of \g-ray monitoring of high-energy sources to search for astrophysical neutrino counterparts. More recently, a candidate \g-ray precursor to a neutrino event detected by IceCube has been observed by the \agile satellite, with a 3.9$\sigma$ post-trial significance~\cite{Lucarelli2017}. 

The first  compelling evidence was recorded on September 2017, when the \fermilat and MAGIC observed enhanced \g-ray emission from a blazar positionally consistent with the neutrino IC170922A \cite{IC170922A,Tanaka2017,Mirzoyan2017}, and very recently published \cite{txall, txice, txmagic}. On September 22, 2017, IceCube detected a muon coming from the bottom of the detector through the Earth,
 produced by a neutrino of energy of $E_\nu \sim 300$ TeV.   
Promptly alerted, the \fermilat and MAGIC detected  at more than $5\sigma$ a flare from  the blazar TXS 0506 +056, at a redshift $\sim 0.34$ \cite{Paiano}, within the region of sky consistent with the 50\% 
probability region of the IceCube neutrino (about one degree in size). The MAGIC detection allowed to determine that the electromagnetic emission had a cutoff at $\sim~400$~GeV.
The simultaneous emission of gamma rays and neutrinos from the same source proves that the ``hadronic mechanism'' has been seen at work. The estimated energy of a proton producing such a high energy neutrino in 
a ``beam dump'' is:
\[
 E_p \gtrsim 20 \,E_\nu \sim 10-20 \, {\rm{PeV}} \,  
\]
an energy above the knee and well appropriate for a blazar.
This event opened the era of multimessenger astronomy with neutrinos; upgrades of IceCube are expected in the next years, and these events will become common.

\paragraph*{Importance of \g-ray observations}
One of the main challenges in neutrino astronomy is the detection of excesses of events due to astrophysical sources among background signals. To this end, directional, energy and time information are used to differentiate the signal emission from the background. Focusing on high-energy events with neutrinos vertices inside the detector volume allows to select candidates with a high probability of astrophysical origin~\cite{IceCube_HESE1}, however at the price of much lower effective area compared to through-going events. 
The ANTARES and IceCube neutrino telescopes~\cite{ANTARES_ref, IceCube_ref} operate extensive programs of real-time multi-wavelength follow-up~\cite{TAToO_ref, IC_followup_ref}. They enable to search for an EM counterpart to astrophysical neutrino candidates by generating alerts whenever an interesting neutrino event is detected (namely a significant multiplet of events, an energetic event or an event whose direction is compatible with a local galaxy \cite{TAToO_ref,GF_IC}). Broad-band data, from the radio domain to the VHE \g-ray regime, are requested as Target-of-Opportunity (ToO) observations to the partners. In particular, high-energy observations from the X-ray (keV) to the \g-ray (MeV-GeV) domains are among the most decisive if they are performed within a few hours after the neutrino trigger, since they allow for the detection of transient cataclysmic events which might involve hadronic processes. A few possible associations have been already claimed~\cite{TANAMI_ref, Lucarelli2017} and a set of serendipitous discoveries is emerging~\cite{procICRCTATOO17,icecube_ofu_sn,txall,txice}.
 Selecting only neutrino events coincident with the EM flare allows for a much better background rejection, and thus a better sensitivity. 
Such studies generally assume a correlation between X-ray/\g-ray flares and neutrino emission, and thus require light curves measured by X-ray/\g-ray instruments as an input, with the largest possible time coverage. Such studies have found so far just one source of cosmic neutrinos \cite{txall}, but  yielded already important model constraints (see e.g. \cite{GRB_IC, xrb_ANT}). 

\paragraph*{Expected results with e-ASTROGAM}
\begin{figure}[h!]
  \begin{center}
\includegraphics[scale=0.65]{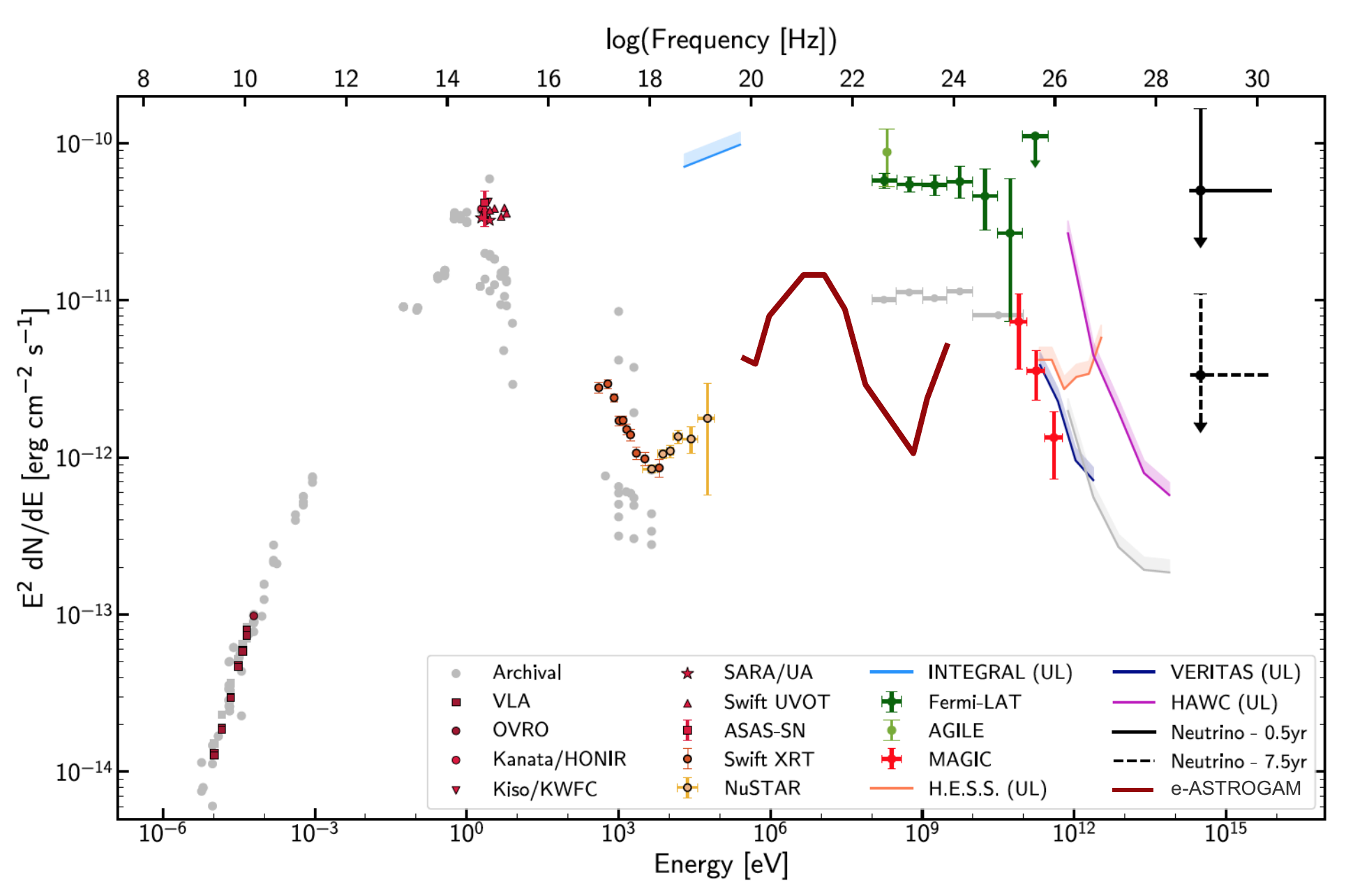}
  \end{center}
  \vspace{-5pt}
\caption{SED of the blazar TXS~0506+056, from \cite{txall}. Dark points represent data taken during the flare; grey points correspond to non-flaring states. The e-ASTROGAM expected sensitivity (solid brown) is calculated for an effective exposure of 7 days, corresponding to the observations by \fermilat and MAGIC during the flare.}
\label{fig:TXS_SED}
\end{figure}
The next generation of neutrino telescopes is currently under deployment. In the Northern hemisphere, KM3NeT will succeed ANTARES in the coming years and will greatly improve both the sensitivity to neutrino point-sources and the angular resolution ($\sim0.2^\circ$  for muon track events and $\sim1.5^\circ$ for showers). In parallel, the upgraded IceCube and IceCube-Gen2 will increase the performance of the current detector by one order of magnitude with the deployment of 120 new detection lines by the next decade. Such upgrades will enable significant improvement on the EM follow-up activities and will benefit from the multi-wavelength facilities operating at the same time. 

e-ASTROGAM can play a decisive role in this scenario. 
In Fig.~\ref{fig:TXS_SED}, the sensitivity of e-ASTROGAM in 7 days (the time in which MAGIC observed an enhanced signal associated to the blazar TXS~0506+056) is compared to the SED of TXS~0506+056 \cite{txall}. Simultaneous time-resolved multi-wavelength information of  variable objects at a daily-timescale (as TXS~0506+056) is fundamental for pinpointing the emission mechanism but is currently not feasible with \Fermi. e-ASTROGAM will have a higher sensitivity than \fermilat over the overlapping energy range. This would have allowed to resolve the state of the source associated to the IceCube neutrino IC170922A. Furthermore, the GeV energy range covers the peak of the second hump of the blazar SED, which can be dominated by \ic electron emission. e-ASTROGAM would have covered an energy range not yet dominated by \ic electron emission. There, it can be expected to be easier to single out hadronic components and constrain the efficiency as neutrino source. \color{black} The ToO capabilities of the satellite should allow for a repointing of the instrument within 6--12 hours, with the goal of reaching 3--6 hours, while its large field-of-view (FoV) will maximize the detection probability and provide an accurate sky localization. Those low-latency follow-up abilities will be important to test a potential association between high-energy neutrino candidates and various classes of transient astrophysical events, and will continue the programs currently performed with the \swift and \Fermi satellites. \\
Furthermore, thanks to its wide FoV ($>2.5$ sr at 10 MeV) in survey mode, e-ASTROGAM will detect and follow variable point-like sources (microquasars, AGN, etc). It has been recently claimed that the \g-ray transparency of astrophysical sources of neutrinos and \g-rays coming from photoproduction of pions might be severely reduced ~\cite{Murase2016}. This result consequently suggests a population of cosmic-ray accelerators invisible in GeV--TeV \g-rays but bright in the MeV domain (see e.g. \cite{Senno2016}). e-ASTROGAM will provide a good sampling of their MeV light curves that will be used to search for neutrino counterparts. \\More specifically, the typical double-humped SED of blazars peaks at MeV energy and can be explained by both hadronic and leptonic processes. In photo-hadronic models, the neutrino flux $F_\nu$ can be related to the bolometric high-energy EM flux $F_\gamma$ (integrated from 1 keV to 5 GeV) with $F_\nu \approx F_\gamma$~\cite{Krauss2014}, which makes the MeV photon flux a good proxy of the neutrino emission from blazars. Thanks to its high sensitivity in the MeV domain, e-ASTROGAM will be perfectly suited to select the best blazar candidates for a neutrino emission and will help to interpret the neutrino observations. In addition, its unique polarimetric capability will enable to reveal the structure of the magnetic field and test the presence of hadrons in relativistic jets.
e-ASTROGAM should also observe $\sim600$ GRBs during the first three years of its mission. Its sub-millisecond trigger and alert capabilities will enable to look for neutrino counterparts of GRBs in nearly real-time and will then take over from \swift, \INTEGRAL and \Fermi instruments.
 
Finally, one of the yet unanswered questions is the nature of the process generating the observed cosmic neutrinos ($p\gamma$ or $pp$ processes). If IceCube neutrinos are mainly produced by $pp$ interactions, their sources should significantly contribute to the IGRB and their flux should be consistent with the total flux. Recent studies~(see e.g. \cite{Bechtol2017}) show that $pp$ models are in tension with the IGRB, disfavoring the $pp$ origin of the cosmic neutrino flux observed by IceCube. Further understanding the contribution of different source populations to the IGRB is therefore crucial. Measurement of spectral features in the 10 -- 200 MeV range with e-ASTROGAM will help to constrain the population models of the IGRB and will consequently have an important impact on the interpretation of the multimessenger connection between \g-rays and neutrinos.   
\subsection[The physics of Gamma Ray Bursts through the polarized eyes of e-ASTROGAM\\
\noindent
\textit{\small{T.~Bernasconi, M.~Kole, N.~Produit, R.~Walter, A.~Ulyanov, S.~McBreen, L.~Hanlon, R.~Curado~da~Silva, F.~Moura}}]
{The physics of Gamma Ray Bursts through the polarized eyes of e-ASTROGAM}
\paragraph*{Science questions}
GRBs have been discovered in 1967 by the Vela satellites. BATSE detected about one bursts per day and discovered that GRBs have different and highly structured light curves \cite{Marani97} and feature an isotropic distribution, indicating their extragalactic origin \cite{Metzger97}. GRBs are the most luminous events in the Universe and the probable signature of the birth of BHs. They are classified in two categories, short ($<$2 seconds) and long. Long bursts are generally believed to be produced by the collapse of a massive star, while the short ones are linked to the merging of two compact objects like NS. The latter are particularly interesting because of the link with the recent GW detections.
GRBs have two distinct phases: the prompt and the afterglow. The prompt is an initial burst of high energy and is widely accepted to be generated by a jet forming during the gravitational collapse. The afterglow is a long-lasting multi-wavelength emission that occurs when the jet interacts with the ambient medium \cite{Gehrels04,mezaros06}. The physical origin of the high-energy \g-rays during the prompt emission of GRBs is not yet understood.  
\paragraph*{Importance of \g-ray observations}
e-ASTROGAM will be a very effective instrument to not only detect and localize GRBs, but also to measure their MeV-GeV characteristics and polarization in the prompt and afterglow phases. e-ASTROGAM will be able to study the evolution of the GRB SED, identify the various spectral components and their correlations, where GRBs have the peak of their luminosity. If the prompt emission originates from synchrotron emission of particles carried away from the central engine, variable moderate to high linear polarization is expected \cite{Fan2004,Zhang2010} and several predictions have been made:
\begin{itemize}
\item The ordered-field model assumes that an helical magnetic field is advected from the central engine and producing a highly polarized emission. The emitted photons would not be uniformly polarized, as there would be patches of different polarization over the emitting shell. Such patches would produce a polarization angle variable over time when they emit in the line of sight \cite{2003ApJ...597..998L,2003ApJ...596L..17G,0004-637X-698-2-1042}.
\item The random-field model suggests that collisionless shocks formed in the jet can produce sizable magnetic fields with random directions on plasma skin depth scales which in turn produce synchrotron emission and axisymmetric polarization angles along the line of sight. In this case, the polarization vectors will roughly cancel each-other out and the measured polarization will be small. On the other hand, if the viewing angle is off-axis, the polarization vectors do not fully cancel out and the polarization degree will be between 30 and 50\% depending on the Lorentz bulk factor \cite{2003ApJ...596L..17G,2003JCAP...10..005N,0004-637X-698-2-1042}.
\item The synchrotron model with random fields on hydrodynamic scales is very similar to the previous model except that the depth of the shock exceeds the skin depth scale. In this case, the overall polarization will remain small \cite{2041-8205-772-2-L20,Toma}.
\end{itemize}
Alternatively the \g-rays could be emitted radiatively from a photosphere (photospheric model) where they are beamed towards the expansion direction. As the polarization is produced by the last \ic scattering the linear polarization degree is correlated with the luminosity and the level of photon anisotropy. The maximal polarization degree predicted by this model is 40\% \cite{2011ApJ...737...68B,Toma}.
Other models \cite{Beloborodov:2013aa} predict that the high-energy photons are emitted by \ic scattering of the prompt MeV radiation in a thermal plasma behind the forward shock with time delays, strength and spectral shape depending on the surrounding wind density. 

All these models can produce very similar signatures of individual GRBs and a single observation is  not enough to rule out any model. However, from the correlation of the polarization degree and angle with other parameters, such as the peak energy, the physics at play can be deduced.

Finally, quantum gravity allows Lorentz invariance violation, which could be searched for using time-delays and polarization changes in the MeV range \cite{Toma,1979rpa..book.....R}.
\paragraph*{Expected results with e-ASTROGAM}
e-ASTROGAM will detect a large fraction of the GRBs and study them over the full energy range covering the prompt emission with excellent timing and energy resolutions. For very bright GRBs, it will be possible to study the variability of the polarization during the prompt phase for the first time.
Valuable information on the delay between GRBs and GWs will be obtained as well as new limits for the Lorentz invariance violation over a very wide energy range.\\
\begin{figure}[h]
\centering
\includegraphics[height=4.5cm]{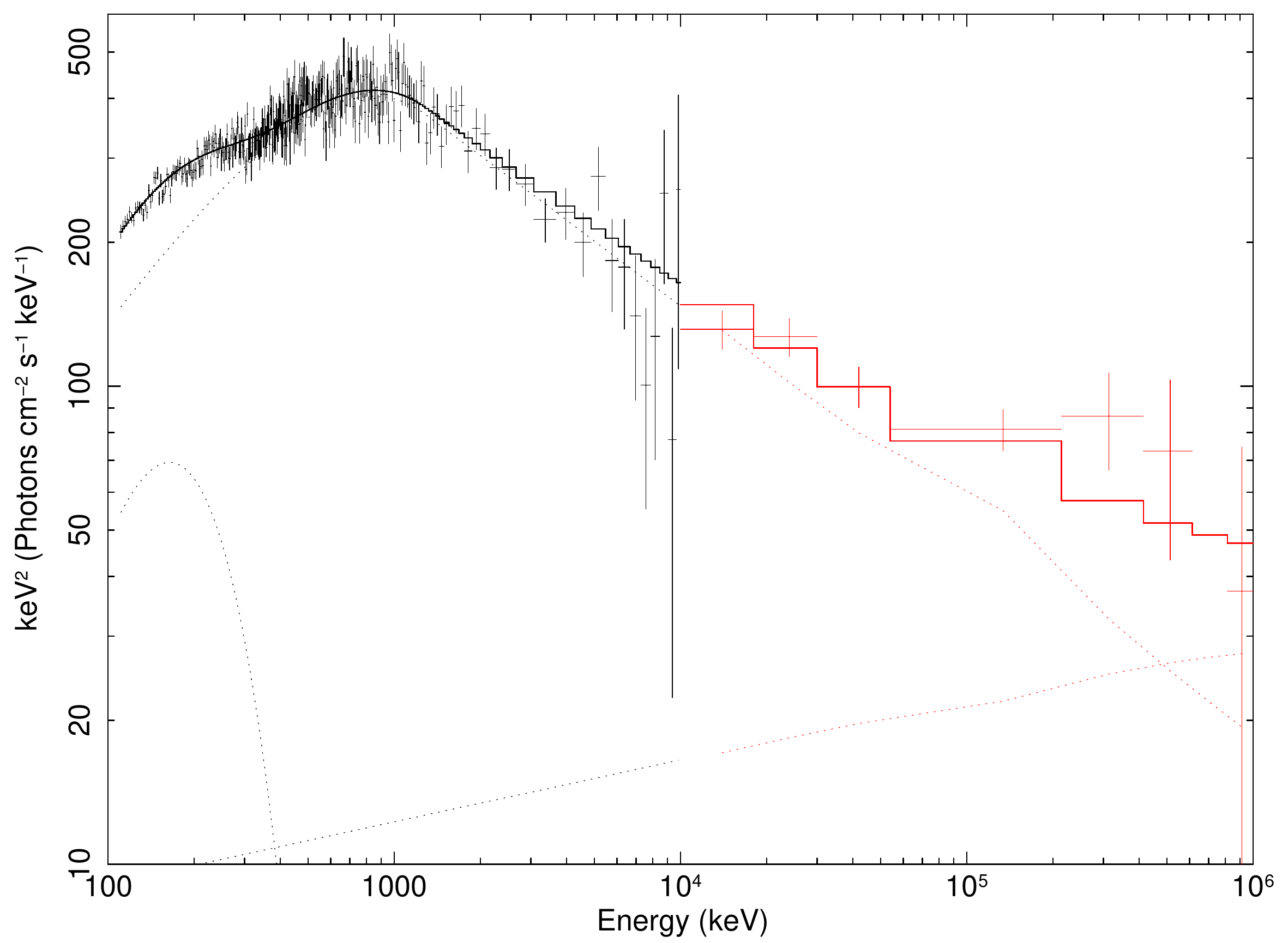}
\hspace{0.5cm}
\includegraphics[height=4.5cm, width=6cm]{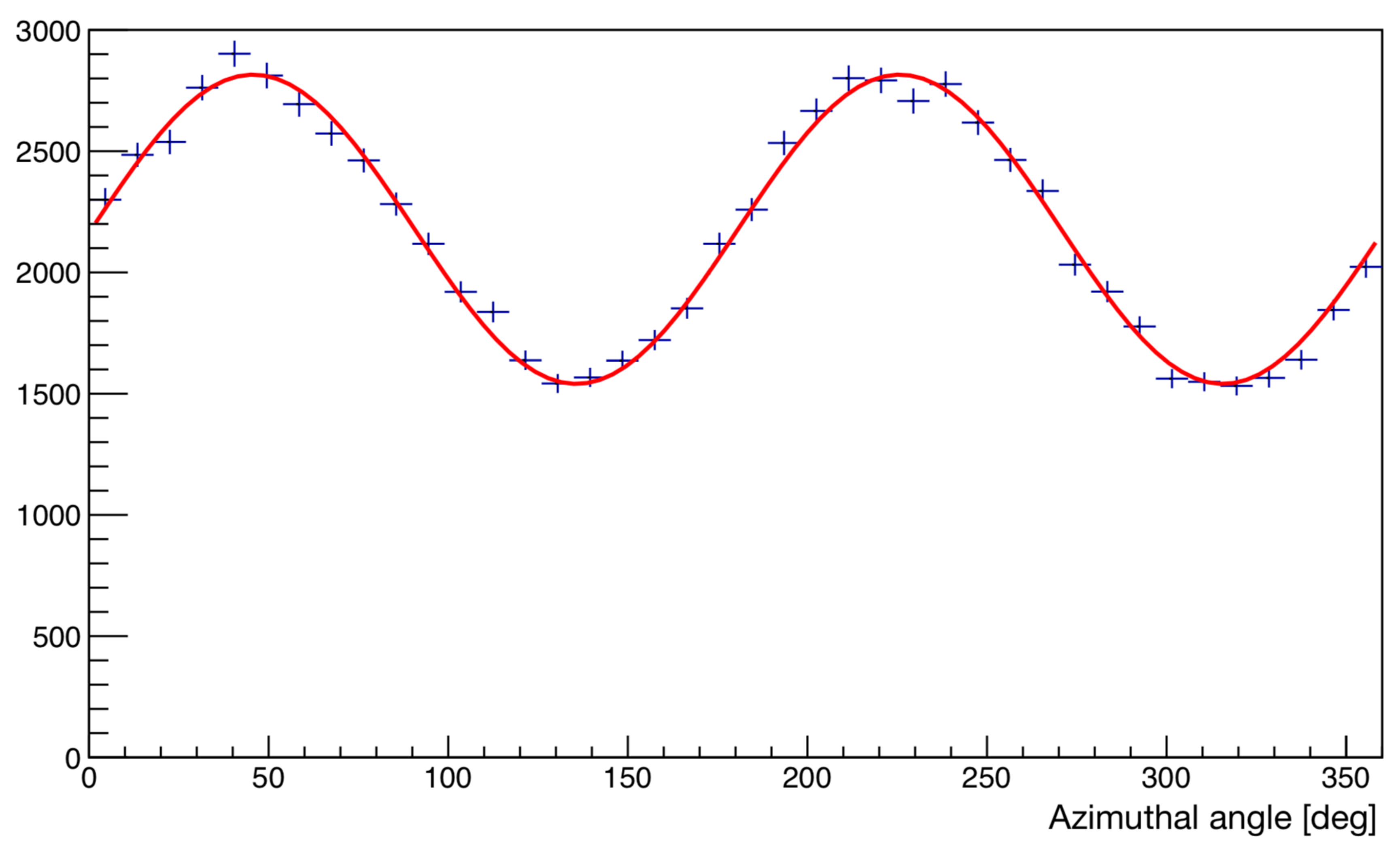}
\caption{\small{e-ASTROGAM spectra (left) and 0.1-1 MeV photon modulation with polarization angle in degrees (right) expected for a 100\% polarized GRB 080916C.}}
\label{GRB_pol:fig}
\end{figure}
To fully demonstrate the capabilities of e-ASTROGAM, we modeled the expected spectrum of the bright GRB 080916C \cite{Guiriec:2015aa} as a black-body, Band function and cutoff power-law, see Fig.~\ref{GRB_pol:fig}.
The minimum polarization which could be detected with a 99\% confidence level \cite{2010SPIE} is
\begin{equation*}
\text{MDP}=\frac{4.29}{\mu_{100}R_{\rm{src}}}\sqrt{\frac{R_{\rm{src}}+R_{\rm{bg}}}{T}}
\end{equation*}
 where $T$ is the burst duration, $R_{\rm{src}}$ and $R_{\rm{bg}}$ are the source and background count rates, and $\mu_{100}=30\%$ is the modulation of the signal for a fully polarized GRB. For GRB 080916C, the MDP ranges from 4.67\% (0.1-1 MeV) to 38.5\% (1-10 MeV). Below 1 MeV, a measurement with MDP=10\% can be obtained every 18 seconds, allowing us to probe its variability during the prompt emission (70 sec in the case of 080916C). e-ASTROGAM is expected to detect approximately 10 GRBs per year with a fluence similar to that of 080916C while approximately 1 per year will be detected with a fluence more than 10 times higher.
To further characterize the expected performance of e-ASTROGAM in polarization measurements, we simulated GRBs at several different angles with respect to the telescope axis. For each off-axis angle, the azimuth scatter distribution observed for a polarized GRB was corrected for the asymmetry of the detector acceptance, using the azimuth scatter distribution obtained for an unpolarized source. As shown in Fig.~\ref{fig:GRB_MDP}, polarization of bright bursts can be detected at very large off-axis angles.  Here, the GRB emission spectrum was approximated by a Band function \cite{ban93} with average GRB parameters of $\alpha=-1.1$, $\beta=-2.3$, and $E_{\rm peak}=0.3$\,MeV. The GRB duration was assumed to be 50\,s. 
The number of GRBs with polarization measurable with e-ASTROGAM was then estimated using the GRB fluences and durations from the Fourth BATSE GRB Catalog~\cite{pac99}. The central plot of Fig.~\ref{fig:GRB_MDP} indicates that e-ASTROGAM will be able to detect a polarization fraction as low as 20\% in about 40 GRBs per year, and a polarization fraction of 10\% will be detectable in $\sim$16 GRBs per year. 
The polarization distribution was finally established for two different models (right panel) to illustrate the model discrimination power of e-ASTROGAM.
\begin{figure}
\centering
\includegraphics[width=4cm]{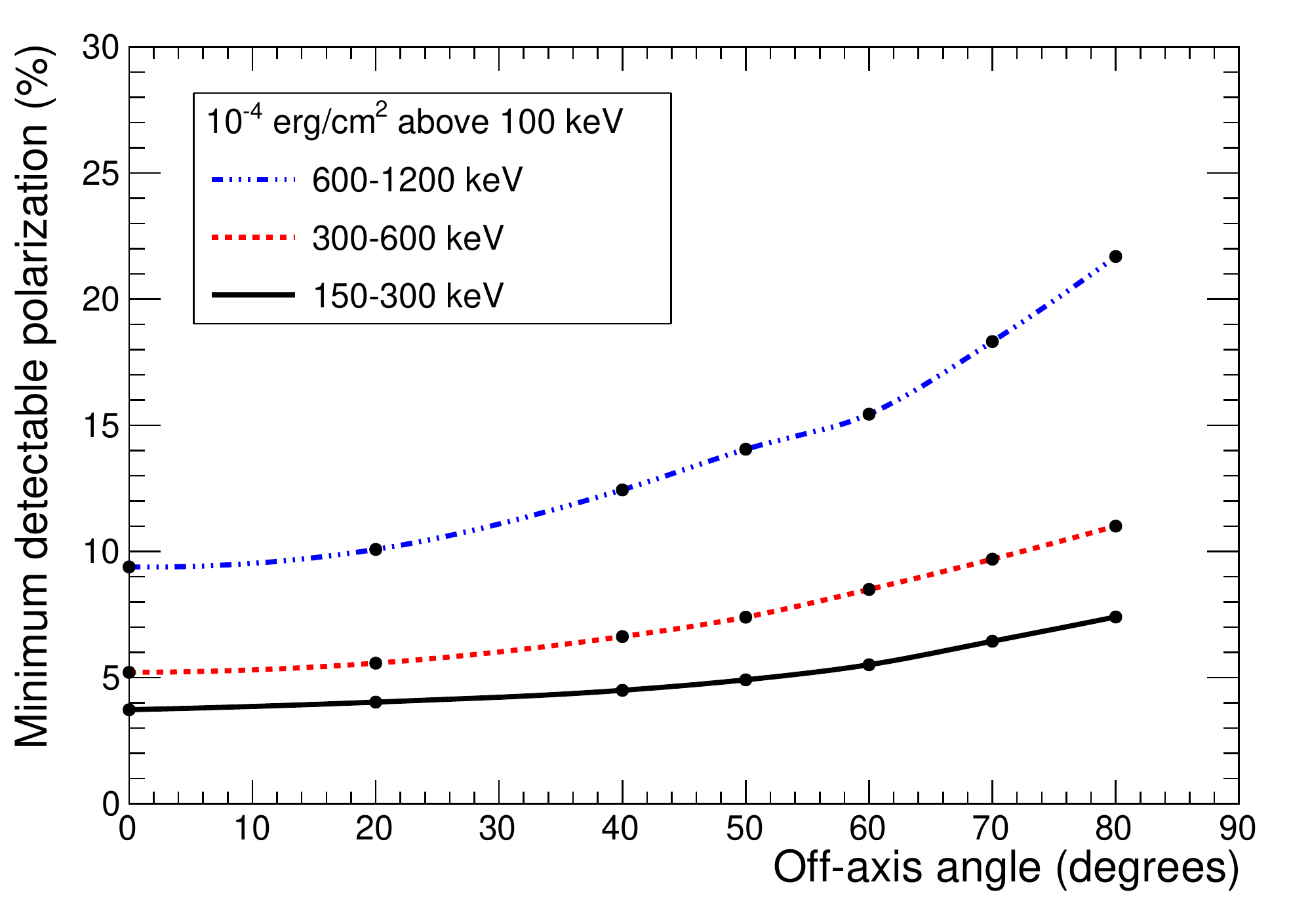}
\includegraphics[width=4cm]{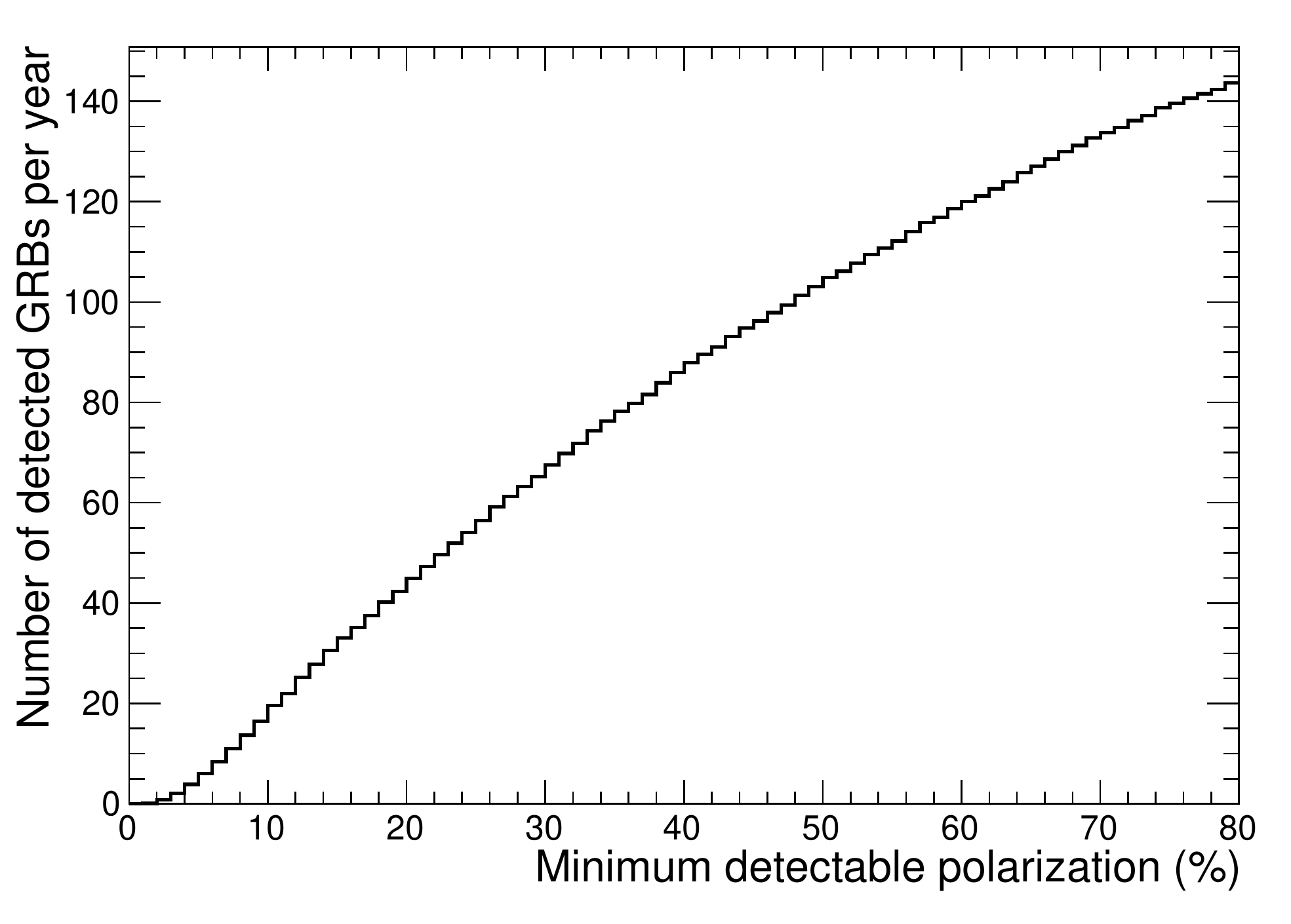}
\includegraphics[width=5cm]{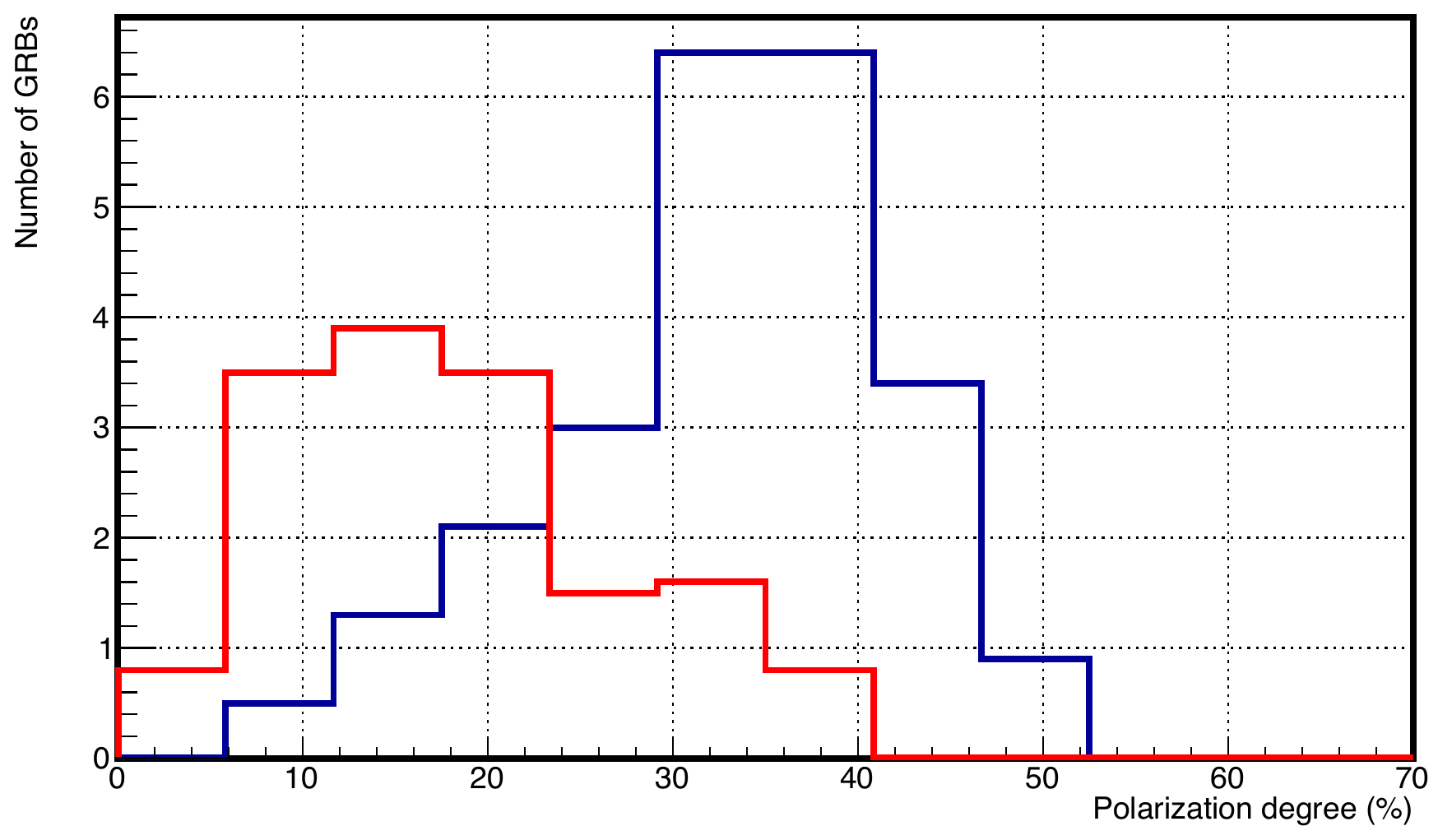}
\caption {\small{Left: Minimum polarization fraction detectable by e-ASTROGAM in three energy bands as a function of the off-axis angle.
Centre: Cumulative number of GRBs to be detected by e-ASTROGAM as a function of the MDP.
Right: Expected measured distribution of polarization degrees achieved using 1 year of data for the photospheric emission model (red) and the synchrotron with highly ordered magnetic field model (blue).}}
\label{fig:GRB_MDP}
\end{figure} 
\subsection[Understanding the Gamma Ray Burst prompt emission\\
\noindent
\textit{\small{G. Ghirlanda, L. Nava}}]
{Understanding the Gamma Ray Burst prompt emission}
\paragraph*{Science questions}
Since their discovery, GRBs have raised several questions about their origin and the nature of the physical mechanisms involved \cite{kumar15}. Both in long GRBs, produced by the core collapse of massive stars, and short GRBs, originating from the merger of two compact objects (NS--NS or NS--BH,), the central engine is most likely a compact object (BH or highly magnetized NS - magnetar) which is able to release (through neutrino and/or magnetic processes) a (isotropic equivalent) energy of 10$^{52-54}$ erg within the short duration of 0.1-100 seconds in the form of high energy keV-MeV photons. This {\it prompt} emission phase is accompanied by a long lasting (days/months) fading emission (the {\it afterglow}). 
Among the most compelling questions about GRBs is the nature of the prompt emission mechanism. Energised electrons (accelerated either by internal shocks or magnetic reconnection events) are expected to radiate via synchrotron emission \cite[e.g.]{Meszaros:1993qy}. The apparent discrepancy between the observed keV--MeV spectral shape and the expected synchrotron spectrum \cite{preece00,ghirlanda03,Frontera:2009uq,vianello09,nava11,goldstein12,sakamoto11}
seems to find a possible solution in recently published results \cite{oganesyan17}, supporting synchrotron radiation in a regime of moderately fast cooling. 
These recent findings are the results of an improved characterization of the low energy part of the prompt spectrum, namely below the $\nu F_{\nu}$ peak energy $\sim$\,300\,keV. What remains highly unexplored is the shape of the high energy part of GRB prompt emission, i.e. in the 1\,MeV -- 100\,MeV energy range.  Above the peak energy, the spectrum is expected to display a powerlaw shape $N(E)\propto E^{\beta}$, with photon index $\beta$ directly related to the power-law index describing the energy spectrum of the emitting electrons. 
The present knowledge of the prompt MeV energy range is mainly based on \fermigbm and \fermilat observations. The \fermigbm with the BGO detectors extends nominally to 40\,MeV, but the reduced effective area at such energies prevented a detailed study of the prompt emission high energy spectral tail. Fig.~\ref{GRB:f1} shows the spectral index $\beta$ of the high energy powerlaw (obtained from the \fermigbm spectral catalog  - Gruber et al. 2014\footnote{https://heasarc.gsfc.nasa.gov/W3Browse/fermi/fermigbrst.html}) versus its uncertainty. $\sim$25\% of the population has a poorly constrainted $\beta$ (rightward of the green line). 
Clear cutoff have been detected in a few cases by the LAT, sensitive down to $\sim$30\,MeV. 
\cite{vianello17} reported the existence of two remarkable cases (GRB~100724B and GRB~160509A) where the combined GBM-LAT data show that the prompt keV--MeV spectrum has an evident softening (located at 20--60\,MeV and 80--150\,MeV, respectively), well modelled by an exponential cutoff. These detections led to estimated Lorentz factors in the range $\Gamma=100-300$ for both GRBs. In other cases, the presence of a cutoff has been inferred after comparing the powerlaw extrapolation of the GBM spectrum with the lack of detection by the LAT. Using this method, \cite{LATcatalogUL12} were able to infer the presence of a cutoff only in six cases out of a sample of 288~GRBs, deriving Lorentz factors in the range $\sim$\,200 to $\sim$\,600.
\begin{figure}[h]
\center
\includegraphics[height=4.5cm]{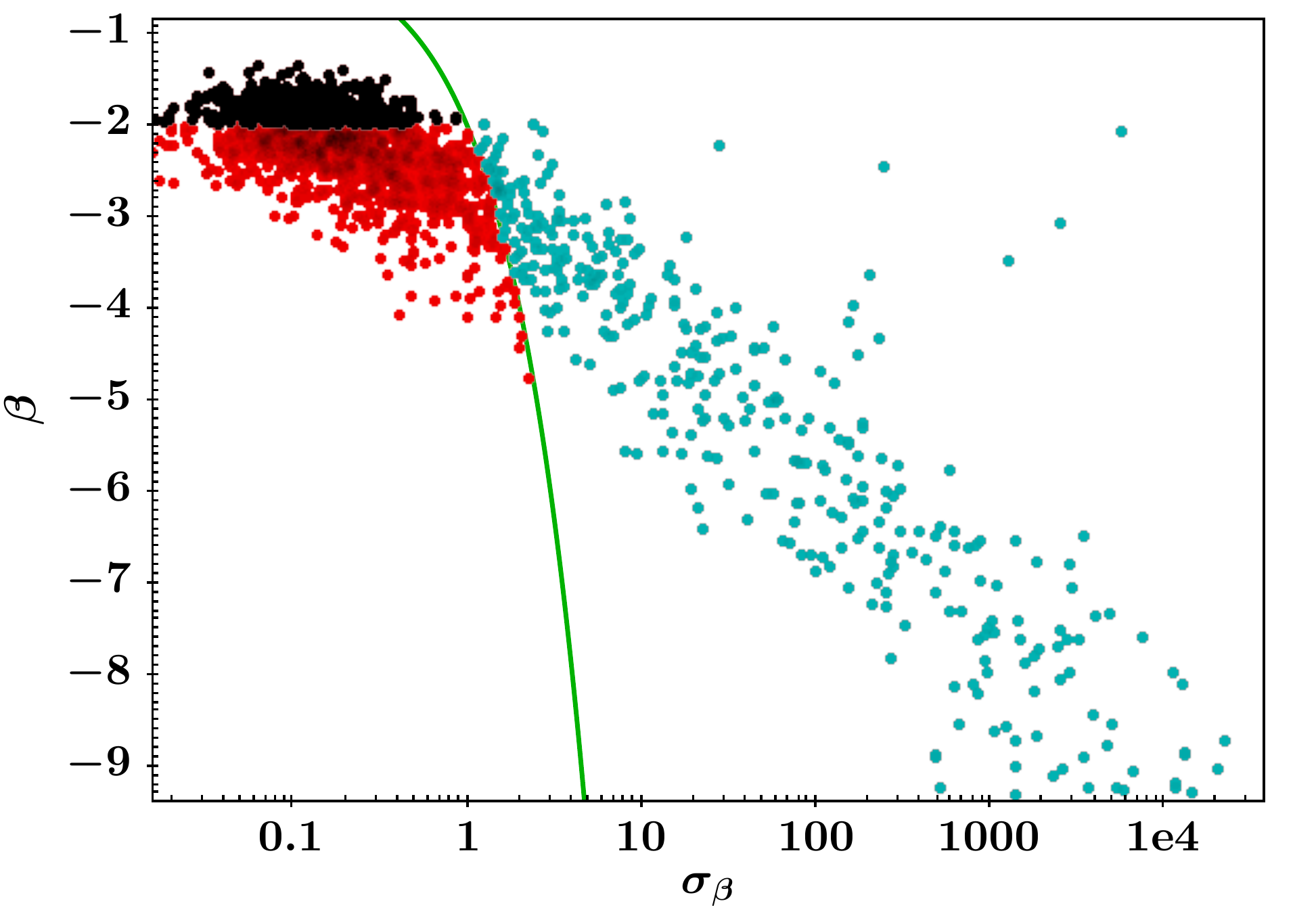}
\includegraphics[height=4.9cm]{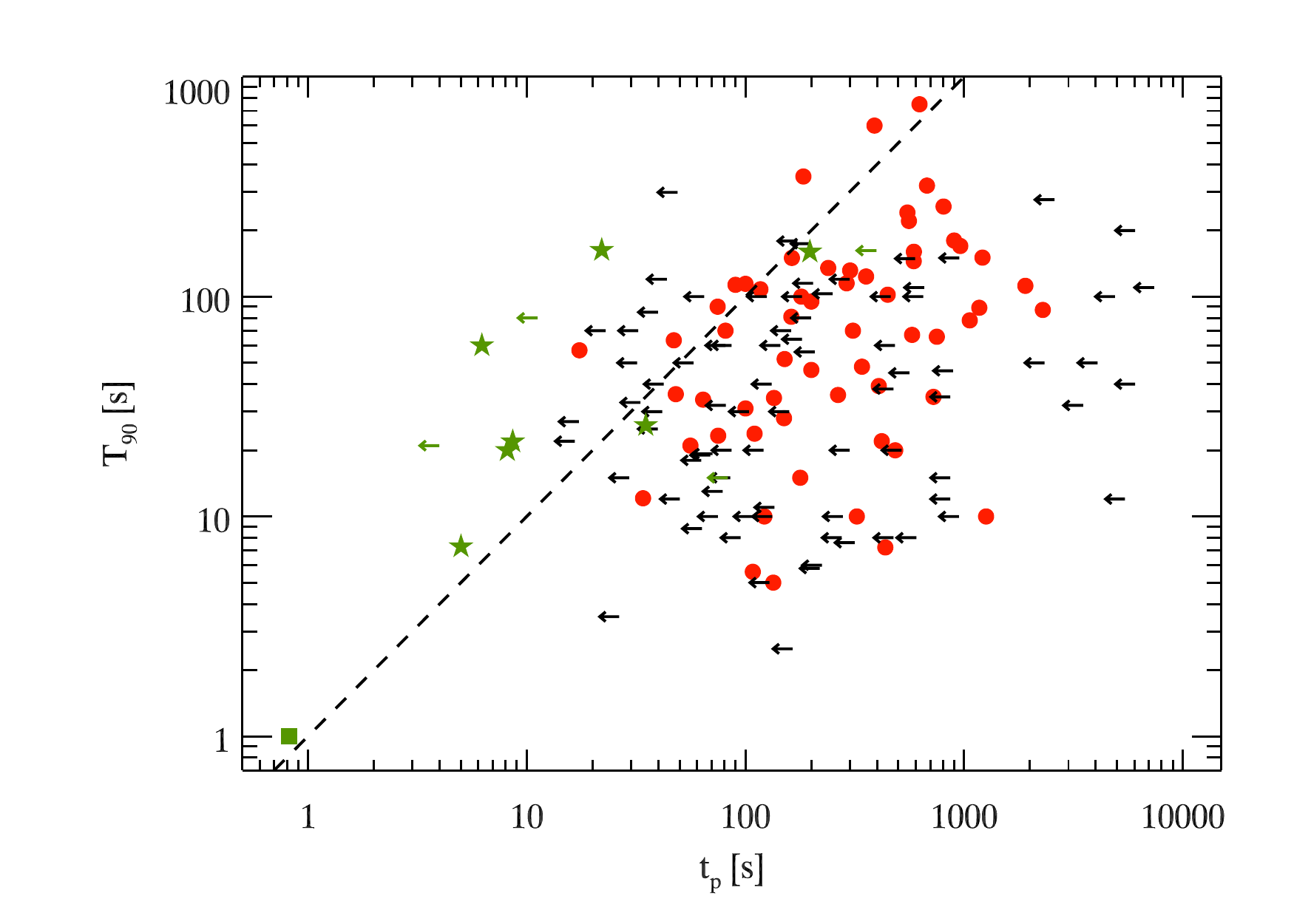}
\caption{\small{Left: high energy powerlaw spectral index $\beta$ ($N(E)\propto E^{\beta}$) versus its relative uncertainty for the population of Fermi GRBst. The green line separates (leftward) GRBs with well constrained $\beta$ (75\% of the \Fermi population) from (rightward) the GRBs with unconstrained $\beta$ (25\&).  Right: GRB duration versus time of the fireball deceleration ($t_{\rm p}$) both measured in the observer frame (Figure from Ghirlanda et al. 2017). Green symbols show the GRBs detected by LAT.}}
\label{GRB:f1}
\end{figure}
Beside the prompt emission, there is another spectral component contributing to the emission at energies above 10-50\,MeV.  The presence of this additional component was first identified by EGRET (e.g. Gonzales et al. 2004) and later confirmed by the \agile/GRID 
\cite{Giuliani:2010fj,Del-Monte:2011kx,Longo:2011yq,Giuliani:2014vn} and by the \fermilat (20\,MeV--300\,GeV) \cite{Ackermann:2014rt,Ackermann:2010ix,Ackermann:2010ys,Abdo:2009ao,Abdo:2009zm}. The LAT is detecting GRBs at an approximate rate of 14 yr$^{-1}$ \footnote{http://fermi.gsfc.nasa.gov/ssc/observations/types/grbs/lat\_grbs/table.php}. 
In most cases, however, the high-energy emission lasts much longer ($\gtrsim$ a factor 10) than the prompt, its onset is delayed by few seconds, and its spectrum is generally harder than the extrapolation of the keV--MeV component \cite{Ghisellini:2010rp,Panaitescu:2017fu}. This component is successfully interpreted as synchrotron \cite{Ghisellini:2010rp,Kumar:2010if} or \ic \cite{Beloborodov:2013aa} emission from external shocks.
The MeV--GeV luminosity after the deceleration time (i.e., after the peak) has been proven to be a robust proxy for the total energy content of the fireball, and has been used to constrain the efficiency of external shocks in accelerating electrons, the strength and configuration of the magnetic field, and the efficiency of the prompt dissipation mechanism \cite{Lemoine:2013kb,Nava:2014ye,Beniamini:2015qo}.
\paragraph*{Importance of \g-ray observations}
An improved study of the spectrum at MeV-GeV energies would reveal fundamental properties of mildly relativistic shocks, and/or acceleration in magnetic reconnection. The physics of both these acceleration processes is poorly known. The high energy powerlaw behaviour is expected to break at energies where $\gamma-\gamma$ absorption within the source becomes relevant, producing a cutoff in the high energy part of the spectrum. The exact location of the cutoff depends on the value of the bulk Lorentz factor. The detection (or the lack) of this feature can then be used to estimate (or place constraints on) the bulk Lorentz factor.
The afterglow component produces a peak in the light curve when the outflow, engulfed by the interstellar material, is substantially decelerated. This peak, observed also in MeV--GeV light curves,  allows us a direct estimate of the bulk velocity before the deceleration (i.e. the maximum velocity attained during the fireball expansion). This is a fundamental and poorly constrained parameter for the modeling of GRB emission (i.e. relativistic beaming plays a major role in GRBs). The largest $\Gamma_0$ have been measured, so far, through the peak of the GeV light curve of \fermilat GRBs \cite{Ghirlanda:2010pb,Ghirlanda:2012th,Liang:2010lr,Liang:2013ly}.
Disentangling between the two emission components (prompt and afterglow) that partially overlap in time is fundamental in order to understand the shape of the high energy part of the prompt spectrum and the properties of the high energy synchrotron afterglow spectrum. Observations in the 10\,MeV--1\,GeV range are fundamental to achieve this goal.
\paragraph*{Expected results with e-ASTROGAM}
e-ASTROGAM is going to cover a poorly explored energy range of the emission spectrum of GRBs, that receives contribution both from the prompt and from the afterglow emission. 
Presently, the 0.3\,MeV -- 100\,MeV prompt emission phase of GRBs is characterized mostly through the GBM on board \Fermi but only for the brightest events. Key questions that e-ASTROGAM will answer through systematic studies of larger samples of GRBs will be (a) whether the high energy prompt emission spectrum is a powerlaw or has a cutoff; (b) how it evolves in time (softening or hardening).
These data will provide unique opportunities to study the properties of the electron distribution (shedding light on the acceleration mechanism)
and the effect of $\gamma-\gamma$ internal absorption (shedding light on the dynamics of the outflow).  Moreover, e-ASTROGAM will allow to (c) disentangle the high energy tail of the prompt emission from the afterglow component, (d) measure the delay time with respect to the prompt keV--MeV component, (e) estimate the luminosity of the afterglow components. 
This will allow us to estimate the highest bulk Lorentz factors in long and (for the first time systematically) in short GRBs, the properties of ultra-relativistic shocks (particle acceleration efficiency, magnetic field amplification and decay),  the fireball energy content during the afterglow phase and (from the comparison with the prompt radiated energy) and the efficiency of the prompt mechanism.
\subsection[The most massive high redshift and jetted Black Holes in the universe\\
\noindent
\textit{\small{G. Ghisellini, F. Tavecchio, T. Sbarrato, S. Kaufmann, O. Tibolla}}]
{The most massive high redshift and jetted Black Holes in the universe}\label{highz_BH}
\paragraph*{Science questions}
Flat Spectrum Radio Quasars at high redshift (z$>2$) are the most persistent 
powerful hard X--ray sources in the Universe.
As such, they are well suited both to study the physics of jets and of accretion along the
cosmic history, and to be used as probes to shed light on the far Universe.
Their 15--150 keV spectrum, as seen by the Burst Alert Telescope (BAT) onboard \swift is
invariably very flat (photon spectral index $\Gamma_{\rm X}<1.5$): this, together with 
\g--ray data from \fermilat, suggests that their high energy SED peaks around 0.5--3 MeV,
where most of their EM power comes out.
All the high--$z$ FSRQ of \cite{ajello09} (10 objects at $z>2$, and 5 at $z>3$)
have a [15--55 keV] luminosity $L_{\rm X}> 2\times 10^{47}$ erg s$^{-1}$, and a bolometric one 
exceeding 10$^{48}$ erg s$^{-1}$.
The same is true for the extended sample of \cite{baumgartner13}.
Recent \nustar observations of some of these FSRQs confirmed and refined this view.
In PMN J0641$-$0320 the observed X--ray spectrum was extremely flat, with $\Gamma_{\rm X}\sim 1$,
allowing to get information on the region of the jet where most of the power comes out and on the
details of the acceleration/cooling of the emitting electrons.
In this and in other FSRQs (see Fig.~\ref{highz_BH:fig} the example of S5 0014+813 at $z=3.366$) the
optical emission is dominated by the accretion disc component (since the synchrotron
emission peaks at smaller frequencies).
Often, it is possible to observed the peak of the disc emission: once it is fitted with 
a standard disc model, we can infer the BH mass and the accretion rate with 
an uncertainty smaller than what allowed by the virial method (based on the FWHM of the broad emission lines).
{\it All} BHs in $z>2$ FSRQs detected in the hard X--rays turned out to have masses $M>10^9 M_\odot$.
Benefitting from the completeness of the SLOAN optical sample, \cite{sbarrato16} 
reconstructed the number density as a function of $z$ of massive BH with $M>10^9M_\odot$ 
and that are {\it active}, e.g. with a disc luminosity exceeding 10\% the Eddington one. 
The right panel of Fig.~\ref{highz_BH:fig} shows the corresponding number density.
For radio--quiet quasars it peaks at $z\sim 2$--2.5 and decays exponentially after the peak.
The number density of radio--loud quasars is surprisingly different. 
It peaks at $z\sim 4$. 
This result suggests that there are 2 preferred epochs of formation of massive BH, and that 
systems with jets form earlier.
Is the jet helping the mass accretion rate or is a large accretion rate required to have a jet?
\paragraph*{Importance of \g-ray observations}
The EM output of high--$z$ powerful FSRQs peaks just in the 
band of e--ASTROGAM. 
Therefore e--ASTROGAM can discover several of new sources of this kind. 
With each source we can find the BH mass, the accretion rate and the jet power.
Since the emission from these sources are beamed toward us, for each detected source
there must exist (several) other sources pointing in other directions.
Since the produced radiation is collimated within an angle $\sim 1/\Gamma$
(where $\Gamma$ is the bulk Lorentz factor) each detected source corresponds
to other $2\Gamma^2$ sources pointing elsewhere, but with the same intrinsic properties
of the detected one. 
We could start to evaluate how the number density of massive BH with jets 
behave as a function of BH mass. 
Are the BHs with -- say -- $M=10^8 M_\odot$ formed at $z=4$ or later (i.e. smaller $z$)?
\paragraph*{Expected results with e-ASTROGAM}
Powerful FSRQs are characterized by an hard ($\Gamma_{\rm X}<1.5$) spectrum.
Therefore e--ASTROGAM can find them either selecting hard spectrum sources below 1 MeV,
and then cross correlating with the radio emission, to pinpoint the arcsec position.
If no redshift is already known for the source, a spectroscopic follow--up is needed.
Alternatively, the best candidates could be selected by the upcoming X--ray surveys (i.e. by e--ROSITA)
in the 2--10 keV. 
\begin{figure}
\psfig{file=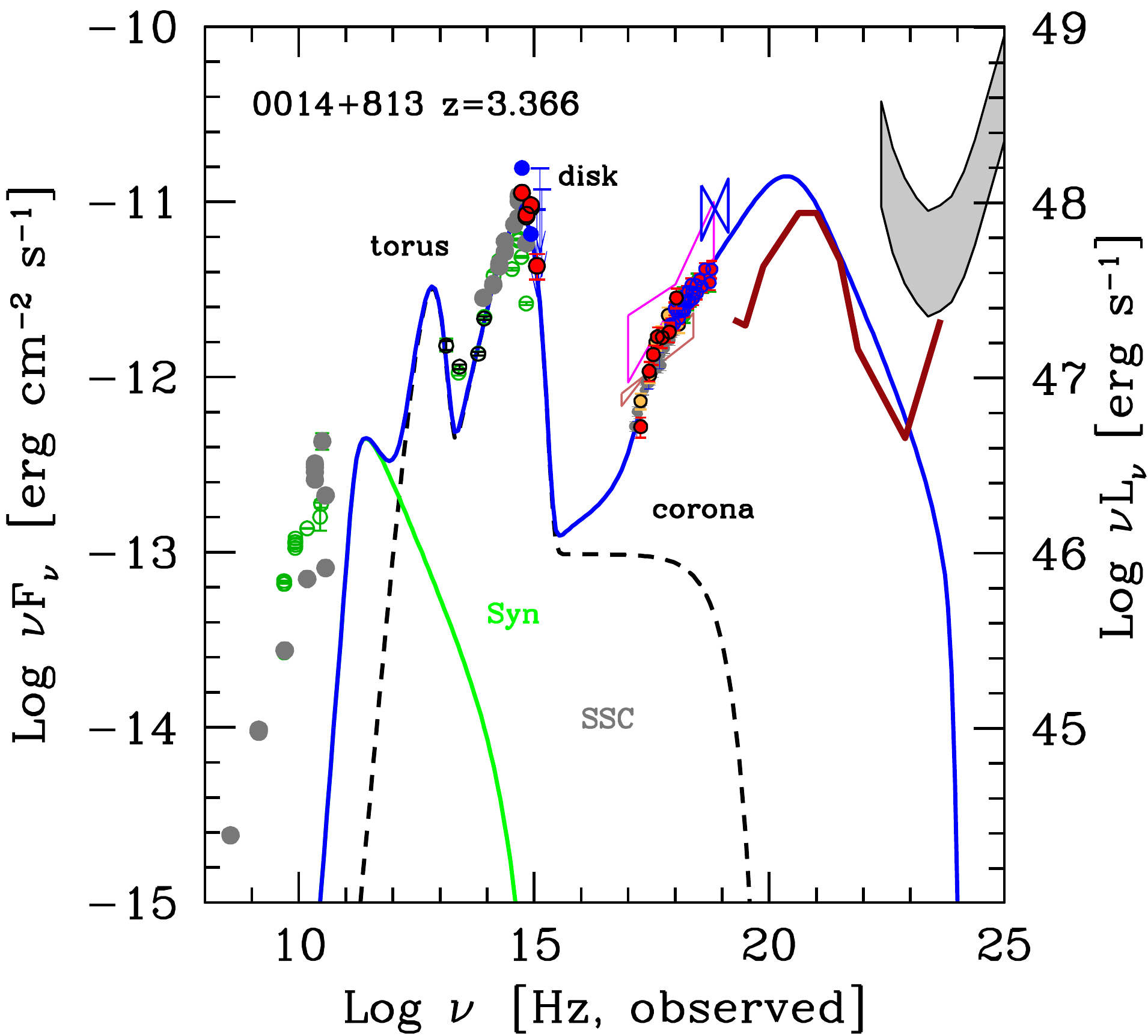,width=6.5truecm,height=6.5truecm}
\psfig{file=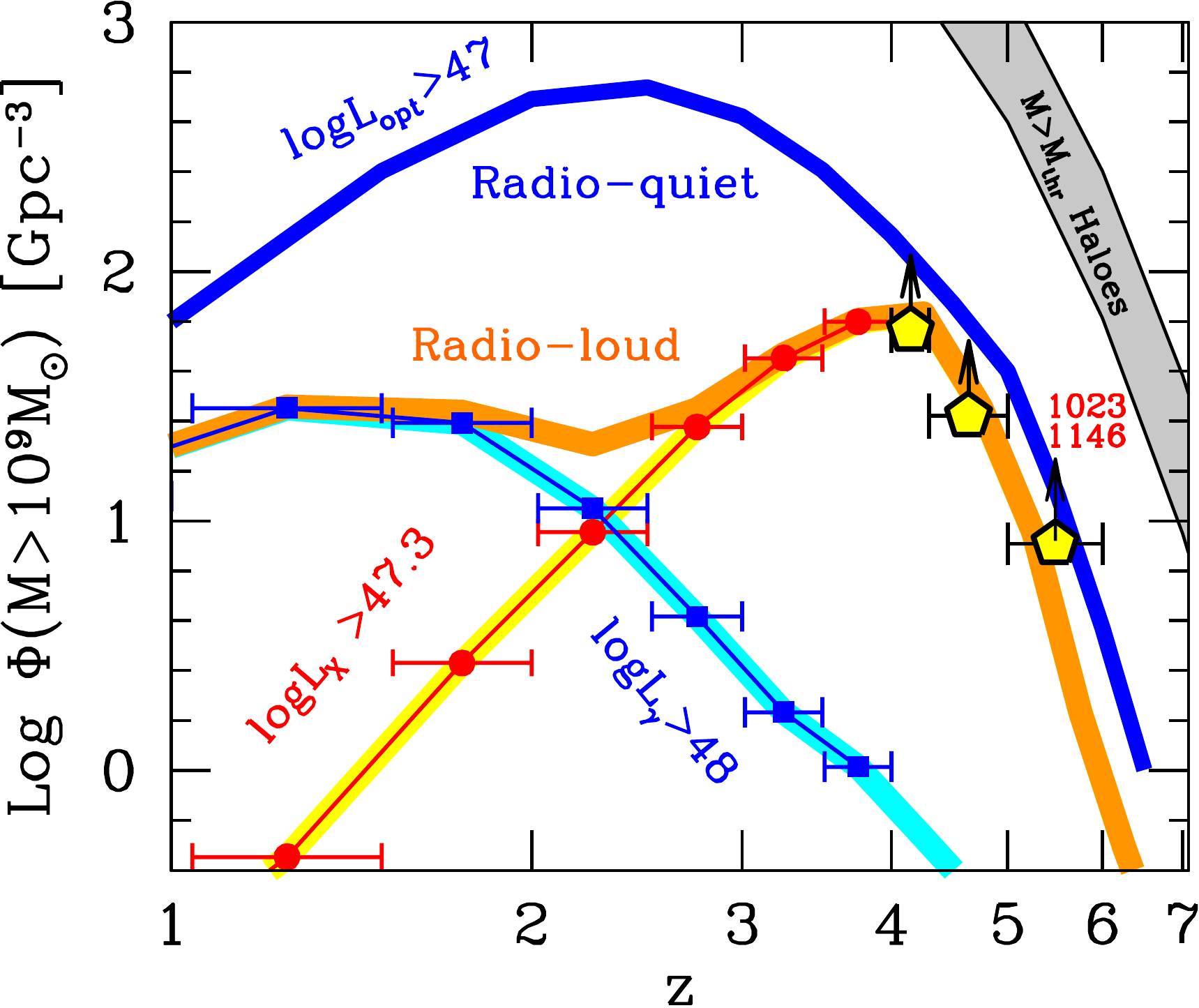,width=6.5truecm,height=6.5truecm}
\caption{\small{Left: Broad band SED of S5 0014+81 with X--ray data from \swift and \nustar.
Note the large Compton dominance, the optical peak unveiling the contribution of the accretion disc and
the fact that his FSRQs has not been detected by \Fermi.
The BH mass for this source is of the order of $M\sim 10^{10}M_\odot$ (from \cite{sbarrato16}). 
 The e-ASTROGAM sensitivity (solid brown line) is calculated for an effective exposure of 1 year (see Chap. \ref{intro}).
 Right: The number density of BHs with $M>10^9 M_\odot$
as a function of redshift. 
While massive BHs in active radio quiet
quasars (i.e. accreting at $>$10\% the Eddington rate) appear to form at $z\sim 2$, 
the ones in jetted sources appear to form earlier, at $z>4$.
Adapted from \cite{sbarrato15}.}}
\label{highz_BH:fig}
\end{figure}
Again, we have to select the hardest sources, cross correlate them 
with the radio ($>1$ mJy is enough) samples, and find the redshift if unknown.
Given the expected sensitivity of e--ASTROGAM, this second option is to be preferred, since in this case
the selected FSRQ would be a pointed target, with adequate exposure.
If the sensitivity for one year of exposure is $10^{-11}$ erg cm$^{-2}$ s$^{-1}$ at $\sim$10 MeV,
scaling with $t^{1/2}$ implies to reach a limiting flux ten times more ($10^{-10}$)
in 3.6 days of effective exposure.
At $z=1$, this corresponds to a luminosity $L_{\rm X}\sim 5\times 10^{47}$ erg s$^{-1}$.
With these new detected FSRQs we can start to refine the current ideas of the relation about the jet
and the accretion rate. 
Currently, the results (using with \Fermi blazars whose maximum redshift is $\sim$3
and mostly located at $z\sim 1$, see \cite{ghisellini14}) indicate that the jet power
is greater than the luminosity of the accretion disc.
Selecting new sources where the jet emission peaks (i.e. at $\sim$1 MeV) could 
imply to find even more dominant jets.
In turn, this impacts on our understanding of the generation process of jets itself: is it really the
Blandford--Znajek mechanism?
Or can we explain these results assuming that part of the gravitational energy of the accreting matter 
goes into amplifying the magnetic field, instead of heating the disc?
In this case we can have sub--Eddington disc luminosities with super--Eddington accretion rates.
This possibility could also explain why jetted sources have BHs that grows
at earlier epochs than in radio--quiet quasars.
\subsection[MeV blazars: understanding emission processes and blazar evolution at high-redshift\\
\noindent
\textit{\small{S.~Kaufmann, O.~Tibolla, S.~Ciprini, G.~Ghisellini, F.~Tavecchio, C.~Pittori, F.~Verrecchia, W.~Collmar}}]
{MeV blazars: understanding emission processes and blazar evolution at high-redshift}\label{MeV_bl}
\paragraph*{Science questions}
Blazars with a high luminosity at MeV energies (so-called {\bf MeV-blazars}) are the most luminous objects of their class. 
Blazars correspond to the class of AGN which are detected at a very small angle between the rotational axis of the accretion disc and the line of sight of the observer, hence in the direction of the jet. This class contains the flat spectrum radio quasars (FSRQ) and the BL Lac objects.
In the first COMPTEL source catalog and in the subsequent re-analysis of the COMPTEL database (0.75-30 MeV), evidence for MeV emission of several blazars both in the lower ($<3$ MeV) and upper ($>3$ MeV) COMPTEL energy bands were reported \cite{Schoenfelder2000}, \cite{Collmar2006}.
Only few MeV-blazars have been detected so far (\cite{Bloom1996},\cite{Collmar2006},\cite{Sambruna2006},\cite{Ajello2009},\cite{Ajello2016},\cite{Sbarrato2013},\cite{Tagliaferri2015}). These very luminous objects are mostly found at high redshifts (z$>$2) \cite{Ajello2016}, they are thought to be fueled by super-massive BH accretion ($M \geq 10^9 \rm{M_\odot}$) \cite{Ghisellini2010}, and they have luminous accretion disc photon fields \cite{Ghisellini2017}.  At high luminosities and redshift, the accretion disc is expected to become visible in FSRQs, which could testify a sequence in physical parameters and in the dominance of the \ic emission.\\
\\
{\it External photon field: BLR or torus:} \\
The SED represents the clear Compton dominance of the MeV blazars (see Fig.~\ref{MeVbl:fig1}), in which the ratio of the \ic to synchrotron luminosity is of the order of $\sim 100$. The optical and UV radiation is dominated by the thermal emission from the accretion disc \cite{Sambruna2006}, which is strengthen by the fact of lack of variability in these bands.
As described by e.g. \cite{Ajello2016}, an external photon field, in addition to the photons produced by the synchrotron, is needed to account for the high \ic flux, since the synchrotron Self-Compton (\ssc) model would produce a much less luminous Compton peak. There are two favorable locations for this external photon field which could yield to this large Compton dominance (see \cite{Ghisellini2009}, \cite{Sikora2009}): the broad line region (BLR) and the torus region. In both locations, the ratio between radiation and magnetic energy density are large enough to explain the Compton dominance (\cite{Ajello2016}). The size of the emitting region is a good indicator to distinguish between the two options for the  location of the photon field responsible for the Compton dominance. The size can be identified by the variability time scale of the X-ray and \g-ray emission, e.g. day time scale for the BLR and five times longer for the torus option (\cite{Ajello2016}).

\noindent {\it Understanding cause of violent outbursts:} \\
As \cite{Sambruna2006} stated, one need to understand the cause of the violent outbursts at hard X-rays, which are expected to be detected as well in the MeV range and what is their duty cycle. 
The time scale of the variability gives an estimation of the size of the emission region and hence clues about the most reasonable external photon fields.

\noindent {\it High-redshifts studies of blazars:} \\
The very luminous objects are mostly found at high redshifts (z$>$2-3)(\cite{Ajello2016}). Therefore they are the best cases to study the redshift evolution of blazars. 
As mentioned in \cite{Ajello2016}, the Burst Alert Telescope (BAT) on-board \swift detected 26 FSRQ of which $\sim 40 \%$ are located at z$>2$ and \fermilat 
instead detected $>400$ FSRQ of which only $\sim 12\%$ are with $z>2$. 
Although currently the number of detected MeV blazars is very small,  they enlarge the redshift range, e.g. one MeV blazar is detected at redshift of 5.3 (\cite{Sbarrato2013}).

\noindent {\it MeV background:} \\
 Ajello et al \cite{Ajello2009} pointed out, that MeV blazars can contribute to the MeV background. Moreover, the mass density of massive BHs might be constrained by the measurements of MeV blazars \cite{Ghisellini2010}.
\begin{figure}[h]
\includegraphics[width=0.47\textwidth]{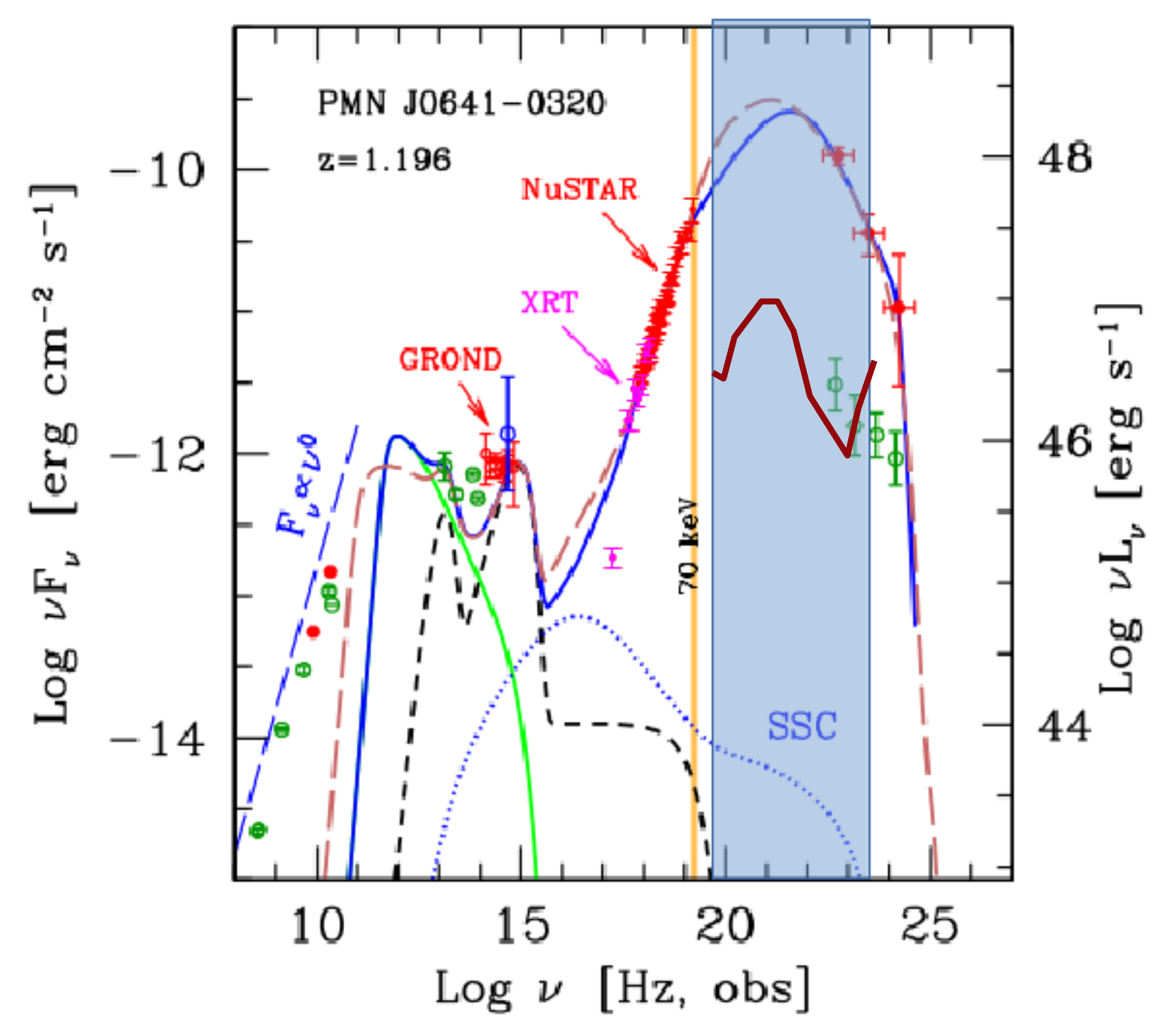}
\includegraphics[width=0.457\textwidth]{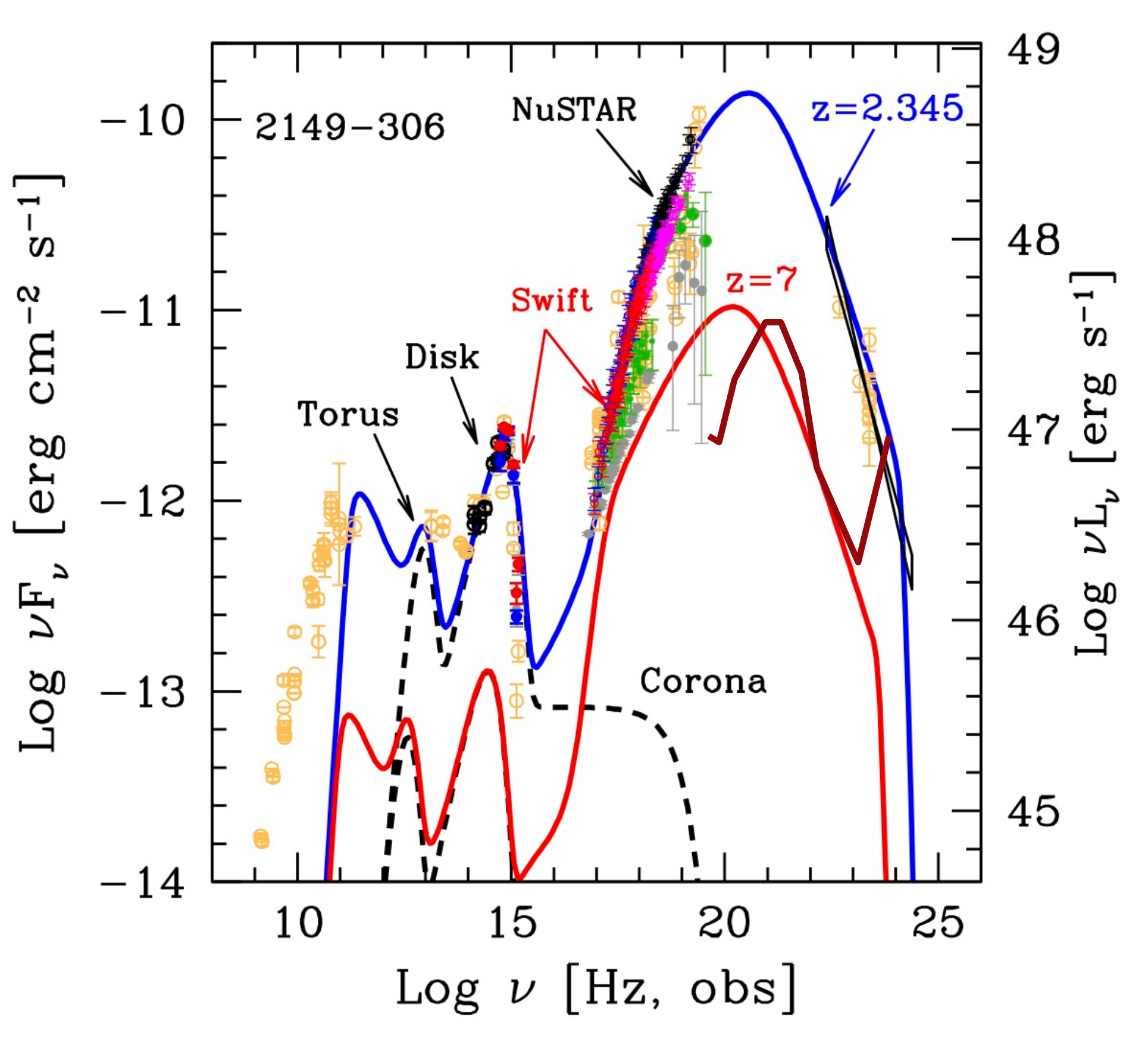}
\caption{\small{Left: SED and model of the MeV blazar PMN J0641-0320 taken from \cite{Ajello2016}. Observations with quasi-simultaneous observations by GROND, \swift, \nustar and \fermilat (in red). The black dashed curve represents to contribution of the torus, accretion disc and X-ray corona. The solid green line shows the synchrotron emission. The solid blue line represents a model with external photon field in the BLR, while the dashed brown model takes into account a photon field of the region between the BLR and the torus.  
In blue the energy range of e-ASTROGAM is marked to illustrate the coverage of the MeV energy where we have a gap of observations due to the lack of an instrument like e-ASTROGAM. Its expected sensitivity is shown (solid brown line) for an effective exposure of 7 days, corresponding to the integration time of \fermilat data.
Right:
SED and model of the MeV blazar PKS~2149$-$306   (redshift of z=2.345), which was observed by \nustar together with multi-wavelength instruments and discussed in \cite{Tagliaferri2015}.
Additionally, a model, illustrating how such source might appear at a redshift of z=7, was added. 
Hence, e-ASTROGAM will have the sensitivity (solid brown line, for 1 year of effective exposure) to detect this kind of source even at a redshift of z=7.}}
\label{MeVbl:fig1}
\end{figure}
\paragraph*{Importance of \g-ray observations}
The physical parameters of MeV blazar jets need to be studied in detail and only higher statistics of MeV bright blazars can give a good parameter space to describe this class of very luminous objects. Moreover, very few high-redshift blazars could yet be studied at MeV and GeV energies. 
The detailed modeling of the Compton dominated SED is crucial to identify the underlying physical properties of the MeV blazars. Moreover, the measurement of the time scale of flux variability in the \g-rays together with X-ray observations (e.g. e-ROSITA heritage, ATHENA and further future X-ray monitoring satellites), gives the indication about the location of the external photon fields necessary to explain the Compton emission. In addition cross-correlation studies with mm/infrared data (for example from ALMA, JWST, WFIRST) will also be very important.
The redshift distribution of the MeV blazars can reach much higher redshift than the GeV detected blazars due to the absorption by the extragalactic background light (EBL) at \g-ray energies. The study of high-redshift blazars at MeV energies gives a very detailed information about the source intrinsic spectra. This intrinsic spectra are important to verify the current EBL model predictions for TeV blazars. 
\paragraph*{Expected results with e-ASTROGAM}
The hard continuum spectrum at hard X-rays and the peak (in $\nu F_\nu$ presentation) of the Compton component at MeV energies makes them a wonderful target for e-ASTROGAM observations, especially due to the current lack of data in the 100keV-100 MeV energy band. 
A numerous detection of MeV blazars are expected with the covered broad energy range from 0.3 MeV to 3 GeV and its planned high sensitivity of e-ASTROGAM, which, e.g. in the range 0.3-100 MeV will be one to two orders of magnitude better than that of previous instruments (see Chapter~\ref{intro}).
This will give rise to a more detailed study of the underlying emission processes and to identify the characteristic parameters for the general class of blazars.
Based on the number of FSRQ mentioned in the \fermilat catalog 3FGL \cite{3FGL} for which the \g-ray spectrum can be described with a photon index of $\Gamma>2$ and which hence could be good candidates for high luminosity at MeV energies, we expect that at least more than 450 blazars (a conservative estimation) will be detected with a high flux in the MeV range. 
Gamma-ray observations in the MeV energy range are important to increase the number statistics of the MeV blazars to verify, if Compton dominance is a general characteristics of them and hence that external photon fields are necessary to explain their high luminosity at MeV energies.
A combination and interplay of external-jet infrared photon field Comptonization and in-jet \ssc mechanism, both producing \g-rays, can be unveiled and well studied only with sensitive observations in the MeV regime.
The 0.1-100 MeV region is a new discovery window for the possible emergence of multi-component and multi-process \g-ray signatures observable in this poorly known portion of the blazars SED.
The increased number of detections, based on e-ASTROGAM, will enlarge the redshift distribution of blazars up to highest redshifts, which is very important for the study of the evolution models. 
Blazar with highest redshift, even up to z=7, are expected to be detected with e-ASTROGAM, as illustrated in Fig.~\ref{MeVbl:fig1}. 
Therefore e-ASTROGAM will be crucial to study the evolution of blazars.
MeV-blazars have their peak emission in the high sensitivity range of e-ASTROGAM, which will detect hundreds of these sources up to high redshifts, revolutionizing our understanding of blazar emission processes and evolution.

\subsection[Unraveling Active Galactic Nuclei using Time-resolved Spectral Energy Distributions\\
\noindent
\textit{\small{D. Dorner, T. Bretz}}]
{Unraveling Active Galactic Nuclei using Time-resolved Spectral Energy Distributions}
\paragraph*{Science questions}
The importance of studying the SED of AGN with e-ASTROGAM and the relevance of its all-sky survey have already been highlighted in this work. 
For AGN, one of the most fundamental and still open questions is the origin of the high energy emission, i.e.\ the identification of the processes in the central engine responsible for the highest energy photons. 
While stationary SEDs can be explained with a variety of models, some of the most intriguing variability features are still not understood. Leptonic models predict simultaneous flux increases in the low energy and high energy peak, while lepto-hadronic models can accommodate more complex variability patterns depending on the dominant process responsible for the \g-ray emission. 
\begin{figure}
\centering
\includegraphics[width=0.5\textwidth]{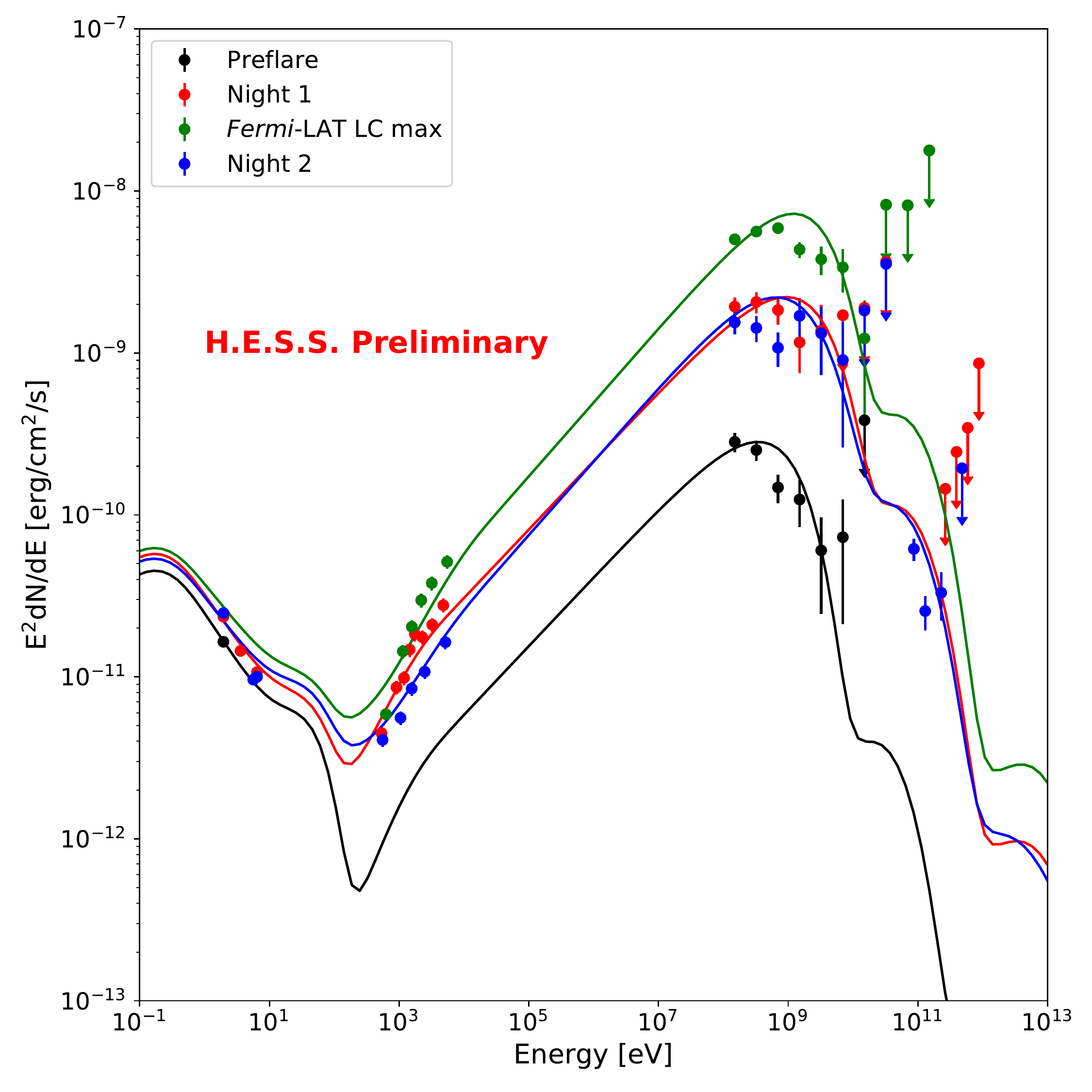}
\caption{\small{Multi-wavelength SED (\swift-XRT, \fermilat and H.E.S.S.) for different flux states superimposed with fits of an hadronic model. From \cite{Romoli}.}}
\label{AGN_time:fig1}
\end{figure}
Bright blazars, such as Markarian 421 and Markarian 501, are well studied in different energy bands (e.g.~\cite{Abdo2011a,Acciari2011,Abdo2011b,Fossati2008,Pian2013,Lichti2008}). Their quiescent-state SEDs are well described by leptonic or hadronic models. Also the broadband SEDs of individual high-states can be explained tuning the parameters of the models. While usually the different flux states are studied in detail but individually, the temporal evolution is rarely considered. Fig.~\ref{AGN_time:fig1} shows an example of SEDs of the flat spectrum radio quasar 3C\,279 in different flux states. While snapshot SEDs can be explained by a variety of models, their temporal evolution challenges stationary models. Only few first approaches, as the one shown, feature time-dependent modeling. To overcome the sparse sampling at VHE, the \g-ray telescope FACT~\cite{Anderhub2013} is monitoring bright TeV blazars with an excellent temporal coverage allowing for time-resolved SEDs~\cite{Dorner2017}. 
In the framework of \ssc models, a quadratic dependence between the synchrotron- and the \ic flux is predicted. Apart from effects due to the shift of the peak-frequency, this simple correlation is expected. The continuous gain during the assumed Fermi-I acceleration will produce a time lag between lower and higher energy photons (hard lag) in each hump. The ratio between the acceleration timescales of electrons and protons is expected to produce a very clear time lag between the two synchrotron components (e.g.~\cite{Spanier2012}). 
For blazars, another peculiar phenomenon challenging especially leptonic models are orphan flares, i.e.\ outbursts in \g-rays not accompanied by a low energy counterpart. Continuous monitoring of the SED is needed to allow to distinguish such events from time lags and from changes in the spectral shape. 
Also periodic modulations of the \g-ray emission have been derived from a number of models of the core regions of blazars. Their observation would put constraints on the possible intrinsic source processes. An example was found in a multi-wavelength campaign ~\cite{Ackermann2015}, which combined \g-ray measurements from \fermilat with data from optical- and radio-waveband long-term monitoring. It revealed a possible quasi-periodic oscillation in PG\,1553+113 on a time-scale of about two years. The paper also lists numerous proposed models for periodic emission such as binary BH systems (e.g.~\cite{Begelmann1980,Komossa2003}), accretion flow instabilities (e.g.~\cite{Honma1992,Ackermann2015}) or helical jet motion~\cite{Rieger2004,Ackermann2015}.
\paragraph*{Importance of \g-ray observations}
To draw conclusions on the mechanisms in the central engine of AGN not only the spectral but also the temporal coverage of e-ASTROGAM is important.  
Depending on the position of the high energy peak of the source, the e-ASTROGAM observations will allow to probe different ranges of the high energy part of the SED. Although, for blazars the time resolution will be limited since measurements take place in the gap between high- and low-energy bump where fluxes are low, e-ASTROGAM's unprecedented sensitivity will allow for a time-resolution good enough to yield additional model constraints. For other AGN with lower peak position, the studies can be carried out with better timing resolution. In the context of multi-wavelength studies, unprecedented time-resolved SEDs can be studied and allow to constrain models and draw conclusions on the dominating emission process. Measurements of different classes of AGN can be compared. In this way, e-ASTROGAM provides an essential contribution to the measurement and understanding of the high-energy peak. 
After the \fermilat era, e-ASTROGAM will be the only instrument monitoring the non-thermal sky not only in space but also in time. This fits very well with future monitoring programs of the planned Cherenkov Telescope Array (CTA). Although targeted on single sources, CTA will extend the e-ASTROGAM measurements to higher energies. Only together, both measurements will allow to study time-resolved SEDs with unprecedented sensitivity and precision. 
\paragraph*{Expected results with e-ASTROGAM}
The time evolution of all measured sources is a natural by-product of the proposed cataloging of the MeV sky. 
New MeV sources will be detected extending and complementing the catalog of AGN. For unidentified sources, only e-ASTROGAM will be able to provide the crucial time evolution and spectral information helping to classify them. In case an association at other wavelengths is found, e-ASTROGAM will at that time be the only instrument available to provide time-resolved spectra of a large number of sources simultaneously.
With this valuable information, numerous models on periodicity or acceleration processes can be tested and excluded or further constrained. Thus, the existence of the time evolution from the only all-sky survey instrument available in the \ic regime and the only MeV instrument available will render exceptionally useful.
e-ASTROGAM will provide an essential contribution to the multi-wavelength picture of AGN. Covering a large energy range in \g-rays, it fills a gap in the SEDs which is important to constrain the models. While snapshots of SEDs can be explained with a variety of models, studying the temporal evolution will allow to further constrain models by enforcing a smooth evolution of the model parameters with time or comparing it to time-dependent models. The continuous coverage will also allow to search for orphan flares from others sources than blazars and allow to distinguish these special flares from time lags between the low and high energy peak and from changes in the spectral shape.
With increased source statistics from the all-sky survey, it is not needed anymore to generalize the result obtained from a single source, but population studies allow for a wider and more general picture. 
\subsection[Extreme blazars: testing the limit of particle acceleration in the jet\\
\noindent
\textit{\small{E.~Prandini, E.~Bottacini, L.~Foffano, M.~Mariotti, S.~Paiano, U.~Barres de Almeida, F.~D'Ammando}}]
{Extreme blazars: 
 testing the limit of particle acceleration in the jet}
\paragraph*{Science questions}
Blazars are supermassive BHs accreting material and ejecting part of it in a jet closely aligned to the line of sight of the observer. They are the most powerful, persistent accelerators known in the Universe. The blazar SED is dominated by the jet emission and it encodes the particle acceleration. The SED is, in fact, characterized by a low frequency peak (from 10$^{12}$ to $>$ 10$^{18}$ Hz), due to synchrotron radiation emitted by ultra-relativistic electrons and a second peak at higher frequencies ($>$ 10$^{21}$ Hz). The nature of this second peak is still largely discussed, in particular a debated issue is the contribution of hadrons in addition to leptons to the \ic emission, as discussed for example in \cite{2013ApJ...768...54B}.
Blazars are further divided into flat spectrum radio quasars (FSRQs) and BL Lac objects (BL Lacs), depending on the characteristics of their optical spectrum and accretion regime. 
Very remarkably, the analysis of the SEDs revealed that blazars display an anti-correlation between the  bolometric luminosity and the location of the synchrotron peak (the so called blazar sequence \cite{1998MNRAS.299..433F,Ghisellini2017}
). FSRQs display the synchrotron peak at low frequencies (IR - optical) while BL Lacs feature a lower luminosity and the peak shifted to higher frequencies. The subclasses of low/intermediate/high synchrotron peaked BL Lac objects (LBL, IBL, and HBL respectively) reflect this behaviour.
In the MeV domain, the blazar SED may feature the second peak (FSRQs), the valley between the two peaks (LBL and IBL) or even part of the synchrotron peak (HBL).
The extreme blazars \cite{2001AA...371..512C} are BL Lac objects  characterized by a synchrotron peak located at energies exceding the hard X-ray band, and therefore not well constrained yet, and by the extreme hardness of the spectrum in the GeV to TeV energy range.
The analysis of a number of these extreme blazars raised the question about the limit of particles acceleration in the blazar jets. Moreover, in some extreme blazars, like 1ES~0229+200, the high-energy part of the SED seems to show evidence of a non-negligible hadronic component in the jet. A precise, complete sampling of the SED is therefore necessary to fully characterize it and disentangle between the leptonic and hadronic contributions.
We propose to measure with e-ASTROGAM the missing part of the SED for the most luminous extreme blazars known.
With this measure we aim at answering  the following questions: How do the synchrotron peak and the second peak connect in such extreme objects? What is the maximum energy reached by electrons in the jet of blazars?  Is the SED obtained in agreement with the standard model of particle acceleration in the blazar jet? What is the hadronic contribution to the overall power emitted in the jet of an extreme blazar? Is there any additional, unexpected component in the spectrum of extreme blazars at MeV energies?
\paragraph*{Importance of \g-ray observations}
\begin{figure}
\centering
\includegraphics[width=0.68\textwidth]{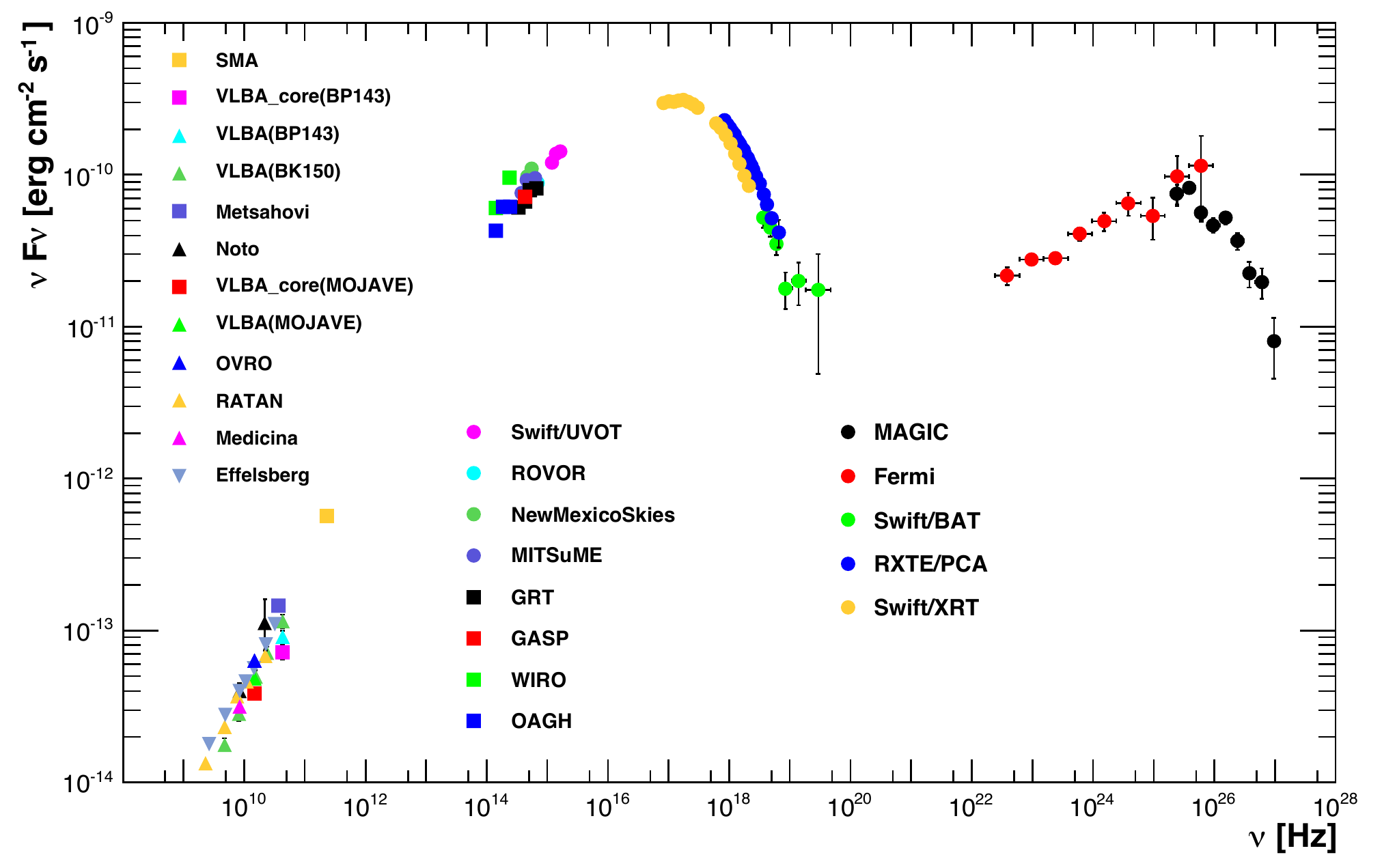}
\caption{\small{Overall SED of the TeV blazar Mkn\,421. Multi-wavelength observations of the source emission, from radio to TeV energies, allowed for an accurate measurement of the acceleration mechanisms of the electrons in the jet  \cite{2011ApJ...736..131A}. In this HBL object the synchrotron emission lies in the optical/soft X-ray energy range. Therefore, the synchrotron peak of the source could be precisely determined and modeled.}}
\label{fig:mkn421}
\end{figure}
A large number of blazars has been detected at energies above 100\,MeV  by current generation of \g-ray satellites (\fermilat and \agile). 
The last \fermilat AGN catalogue released is the 3LAC \cite{3lac} 
containing 1591 sources mainly belonging to three categories, which are blazars of uncertain type, FSRQs, and BL Lacs. Of the latter two categories, only a minor fraction emits up to the highest energies (E $>$ 100 GeV). The TeV catalog counts, in mid 2017, $\sim$ 70 sources, mainly HBLs. This drop in the number of sources is mainly due to the fact that only the most powerful and nearby objects  reach such high energies with a sufficient flux to be detected by current generation of instruments.
Moreover, the most sensitive telescopes covering this energy range operate in pointing mode and feature a relatively small field of view of few degrees. The Cherenkov telescope array (CTA) will start  operations in a few years from now and is expected to detect hundreds of blazars above 100\,GeV, thanks to its unprecedented sensitivity and to a larger field of view. 
At energies below 100\,MeV, the number of blazars with a significant \g-ray emission detected by COMPTEL, in orbit from 1991 to 2000, is very low in comparison to the sources reported in the 3LAC (only $\sim$1\%). The reason is twofold: the potential of source detection  of COMPTEL was quite poor in comparison to that of  \fermilat, due to its low sensitivity (only down to 10\% of the Crab Nebula flux, while it is below 0.5\% in case of \fermilat. Moreover,  most of the BL Lacs are expected to have a dip of the emission at these energies, due to the transition from the synchrotron emission to the \ic one. Interestingly, some objects (FSRQs and extreme blazars) presumably emits a large fraction of their power in this band which is still largely undetected.
In the last decade, an increasing number of blazars has been intensively studied at different bands, from radio to VHE \g-ays. The characterization of the SED over more than a decade in energy allowed very detailed studies of the physical conditions responsible for the emission. 
In general, a precise characterization of the first peak of the SED, the synchrotron peak, allows us to determine the electron acceleration in the jet, while the study of the high-energy emission helps to constrain the possible hadronic component in the jet or the presence of external radiation fields. An example is Mkn 421, whose SED collected during a multi-wavelength campaign carried out in 2009 is displayed in Fig.~\ref{fig:mkn421}.  For this HBL object the synchrotron emission lies in the optical/soft X-ray energy range. Therefore, the synchrotron peak of the source could be precisely determined and modeled. The overall SED from Mkn~421 including the second peak is well modelled by a standard, \ssc model, where no additional contribution (e.g. hadronic emission or external radiation fields) is needed.
Another deeply studied blazar is  the BL Lac object 1ES~0229+200, located at redshift 0.14. It is one of the few extreme HBLs detected at TeV energies \cite{2007AA...475L...9A}. The SED of 1ES~0229+200 is displayed in Fig~\ref{fig:1es0229}, obtained from the SSDC website\footnote{https://tools.asdc.asi.it/SED}.
\begin{figure}
\centering
\includegraphics[width=0.6\textwidth]{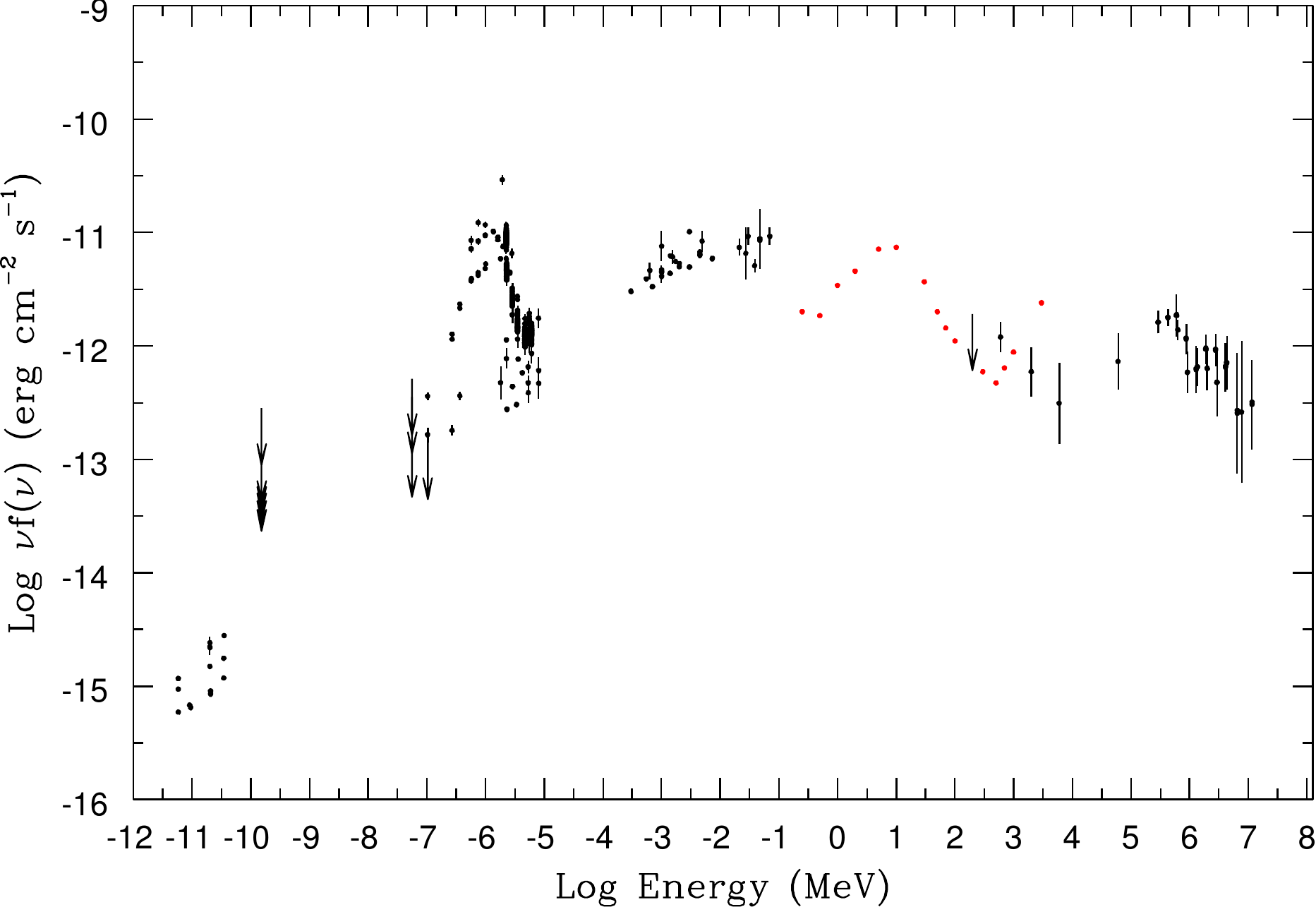}
\caption{\small{SED of the extreme blazar 1ES~0229+200 (black markers). In this case, the peak of the synchrotron emission is not well determined due to the lack of measurements at frequencies above 100\,keV. The 1-year sensitivity of e-ASTROGAM is plotted with red markers. Adapted from SSDC website.}} 
\label{fig:1es0229}
\end{figure}
From its SED we can conclude that the X-ray emission is detected up to $\sim$100\,keV without any significant cut-off \cite{2011AA...534A.130K}, meaning that the synchrotron peak is located at extremely high frequencies.
Moreover, the luminosity of the source is orders of magnitude below that of Mkn~421, as foreseen by the blazar sequence (extreme = faint). Finally, once corrected for the effect of absorption due to the extragalactic background light (EBL), the VHE spectrum of 1ES~0229+200 is very hard at TeV energies, which is in tension with the classical, leptonic model of blazar emission.

The last point had a great relevance for the astrophysical community. The hard spectrum was used to set new constraints on the EBL itself in the IR regime \cite{2007AA...475L...9A}, and to determine an upper limit on the intergalactic magnetic field \cite{2011AA...529A.144T, 2015ApJ...814...20F}. Several authors proposed a hadronic origin for the peculiar TeV spectrum \cite{2015MNRAS.448..910C,2015arXiv151205080T}, suggesting that they could be responsible for a significant neutrino emission and could also be the site of UHECR acceleration.

To conclude, extreme blazars have extended by nearly two orders of magnitude to higher energies the known range of both synchrotron and \ic peak frequencies, disclosing a new region of possible physical parameters that can give us new and valuable insights on the particle acceleration mechanism.
In  order  to  explain  such  a  shift  in  the  SED peaks,  the  minimum Lorentz factor of the electron energy distribution and the magnetic field intensity should both have significantly different values than those inferred in classical TeV BL Lacs.
A better knowledge of the MeV spectrum, being related to the synchrotron emission of the electrons,  could be of particular relevance to constrain the leptonic component of the emission.
\paragraph*{Expected results with e-ASTROGAM}
We propose to carry out multi-wavelength observation campaigns on a sample of extreme blazars including for the first time e-ASTROGAM observations. 
The target list can be extracted from \cite{2015MNRAS.451..611B}, which collects the most promising extreme blazars known to date. In addition, we propose to include Mkn~501 during flaring states, since this source usually features an extreme behaviour in such circumstances.
Goal of the campaign is to achieve the most accurate and complete characterization of the SED of a sample of extreme blazars.
In particular, the goals of e-ASTROGAM observations are the measurement of the flux level at MeV energies and, possibly, the spectral slope. This will allow, for the first time, to determine the location of the synchrotron peak of these extreme and intriguing sources and to build an almost continuous broad-band SED from radio to TeV energies.
\subsection[Gravitationally lensed MeV blazars\\
\noindent
\textit{\small{S.~Ciprini, C.~Pittori, C.~C~Cheung, S.~Buson, F.~Verrecchia, D.~Gasparrini, S.~Cutini}}]		
{Gravitationally lensed MeV blazars}
\paragraph*{Science questions}
Blazars, namely BL Lac objects and flat-spectrum radio quasars (FSRQs), are a small but important fraction of the entire population of AGN (e.g. \cite{2017A&ARv..25....2P}).
For blazar populations which are typically distributed at larger distances, such as the FSRQs,
	the sub-GeV and MeV \g-ray emission can dominate the EM radiative bolometric power. This makes them optimal probes of the distant and young Universe (Sec. \ref{MeV_bl}, \ref{highz_BH} and \cite{ghisellini14, 2017ApJ...837L...5A}), 
	and targets for astrophysical ``tomography'' in the MeV regime.
The sky in the 0.2-30 MeV energy region is, however, insufficiently covered with only a few tens of steady sources detected so far.

In parallel, strong gravitational lensing of EM radiation from distant sources (predicted in Einstein's theory of General Relativity \cite{1936Sci....84..506E}), has been discovered and studied in hundreds of radio/optical lens systems, since the first detection of multiple images of SBS 0957+561 \cite{1979ApJ...233L..43W}. When the distant source, the lensing galaxy and the observer are aligned, a circle, known as the Einstein ring, may be formed \cite{schneider92}.

An example of (spatially unresolved) strong-lensing is the case of the powerful, MeV-peaked FSRQ, \pks\ ($z=2.507$, routinely detected in GeV band by \agile and \Fermi, Fig.~\ref{gravlens:fig} and \cite{donnarumma11, 2015ApJ...799..143A}). \pks\ is the brightest strong lens in the sky at cm, hard X-ray, MeV \g-ray energies, and detected already by COMPTEL \cite{Collmar06} in 0.75$-$30 MeV band.
The the two lines of sight to this object have been used in the past also as a cosmological probe \cite{blandford92}. S3 0218+35 (lens B0218+357, $z=0.6847$) is another GeV lensed blazar detected by \Fermi (and by MAGIC at $E>100$ GeV, \cite{2016A&A...595A..98A}), representing the smallest-separation lens known. For S3 0218+35 the first \g-ray delay measurement was possible thanks to \fermilat data. This opened the possibility of delay measurements for other distant lensed \g-ray FSRQs. In the MeV regime, the largest amplitude for flares and variability patterns occurs, enriching the statistics in strong-lensing/microlensing \g-ray temporal features.

How we can substantially improve the spatial resolution of the central engine and identify the sizes and locations of \g-ray emission regions from distant sources? How independent \g-ray delay measurements and radio-delays are related in strong macro-lensing? Which is the role of micro/milli lensing (a view into astrophysical emission regions or a probe for DM substructure and subhalos)? Can high-redshift, lensed MeV blazars, help us in the detection of cosmic neutrinos from the distant Universe? Are, at the end, distant gravitationally lensed MeV blazars a potential and unexplored gold-mine for multimessenger and fundamental physics?\\
\begin{figure}
\includegraphics[width=0.5\textwidth]{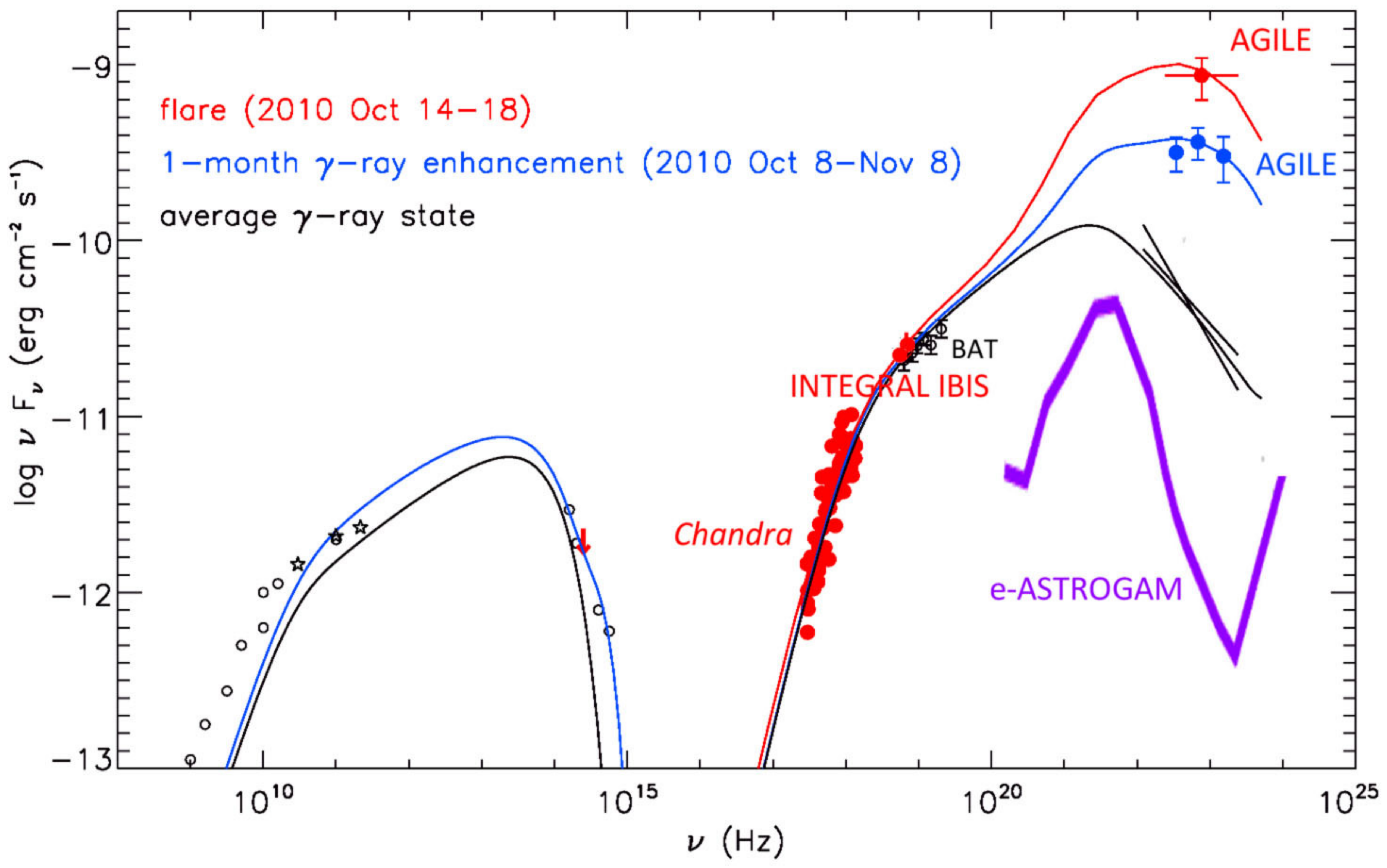}
\includegraphics[width=0.5\textwidth]{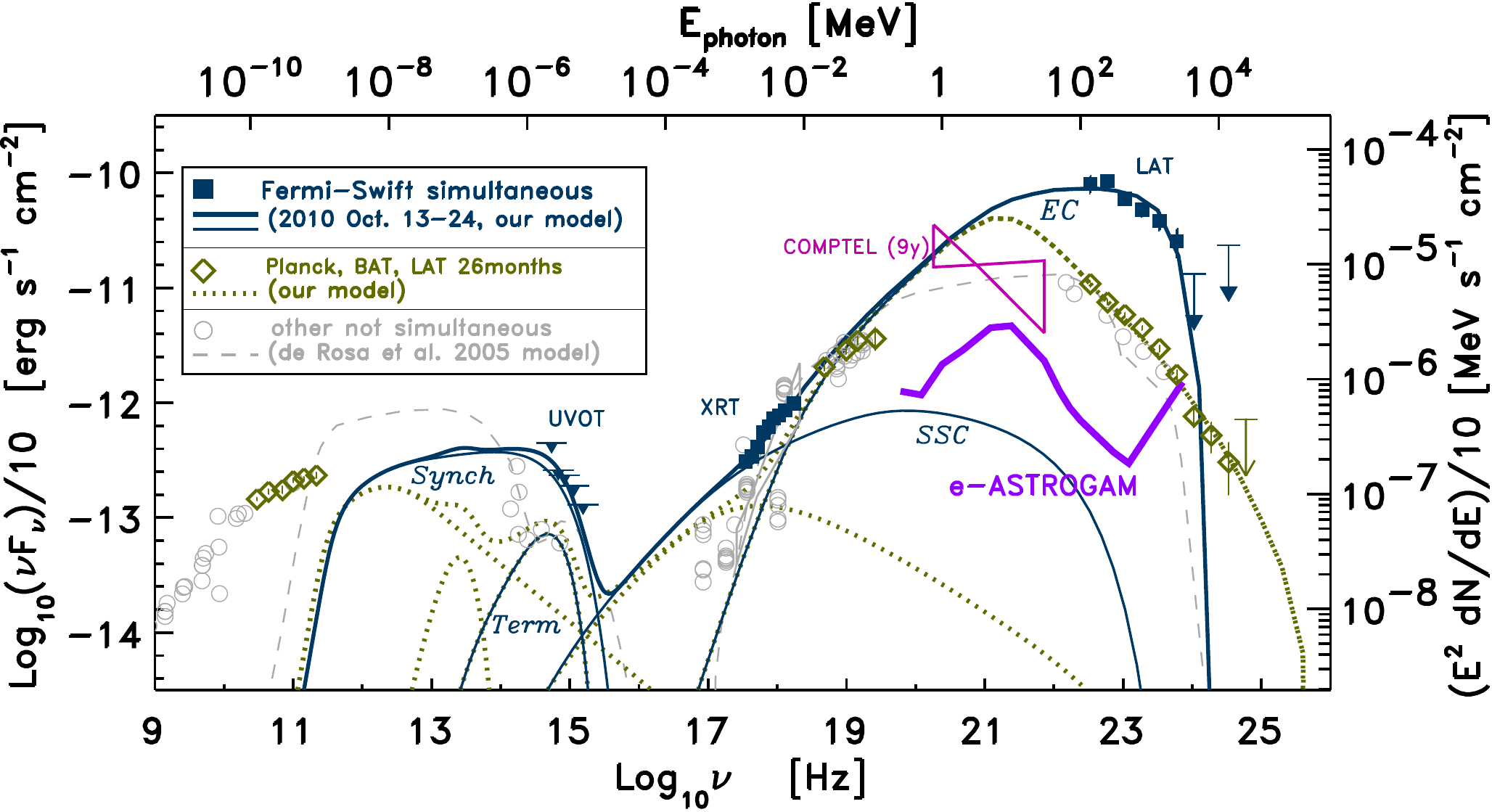}
\caption{\small{Left: Observed SED of the gravitationally lensed flat spectrum radio quasar PKS~1830$-$211 ($z=2.507$), built with archival data and Oct.-Nov. 2010 simultaneous data from the multi-wavelength campaign led by \agile (models magnified by a factor of 10$\times$ for the lensing). Adapted from \cite{donnarumma11}.
Right: observed SED of PKS~1830$-$211 built with simultaneous LAT and \swift XRT and UVOT data, averaged over the Oct. 13-24, 2010 period of the multi-frequency campaign led by \Fermi. Past 26-month LAT, 58-month BAT, \textit{Planck} ERCSC, Gemini-N, \textit{Hubble}-ST, \textit{Chandra} and \INTEGRAL IBIS \cite{2005A&A...438..121D}, COMPTEL and EGRET data are also reported (corrected for lensing by a factor of 10$\times$). Adapted from \cite{2015ApJ...799..143A}. This study is an example of a possible future confluence in e-ASTROGAM of synergetic legacy science from teams of \Fermi, \agile and \INTEGRAL missions.}}\label{gravlens:fig}
\end{figure}
\paragraph*{Importance of \g-ray observations}
A gravitational lens magnify the radiation emitted from a distant blazars and produce time delays between the diffraction mirage images, with delays depending on the position of the emitting regions in the source plane. Time delays in AGN/galaxy-scale lenses typically range from hours to weeks. The possibility to obtain independent \g-ray delay measurements from strong macro-lensing, and to derive accurate measurements of the projected size of the \g-ray emission regions in central engine and the jet, disentangling micro-lensing temporal features, was attested for S3 0218+35 \cite{2014ApJ...782L..14C} and \pks\ \cite{2015ApJ...809..100B, 2013A&A...558A.123M,  2015NatPh..11..664N}. The evidence for micro/milli-lensing effects in strong lensed quasars is increasing in general. These can introduce a variability in the flux ratio of the two images, in addition to an intrinsic energy-dependent source structure and the different region sizes, resulting in a ``chromatic'' spectral variability \cite{2015ApJ...799..143A, donnarumma11, 2013A&A...558A.123M}.
The study of variability of gravitationally lensed blazars emitting in the 0.2 MeV$-$3 GeV band, can open interesting perspectives:\\
$\bullet $ MeV data are important to understand blazar particle acceleration and emission processes, the combination and interplay of different leptonic \ic mechanisms (\ssc, BLR, torus, diffuse dust photon fields) or hadronic emission processes (photopion, e.m. cascades, proton synchrotron, Bethe-Heitler).\\
$\bullet $ MeV temporal/spectral variability produced by unresolved lensing of distant FSRQs is able to probe the central engine and jet structures and the origin of the HE emission, this also in synergy with facilities like SKA, ALMA (\cite{2013A&A...558A.123M}), LSST and Euclid.\\
$\bullet $ MeV data, placed around the emission peak with more pronounced variability and flares, enhance
the detection of temporally delayed events and micro-lensing signals. \fermilat already observed common 1-day GeV flares by a factor 3 to 10 compared to few-10\% increases in mm/radio bands.\\
$\bullet $ More, small separation, lenses that cannot be resolved, can be discovered in MeV band thanks to measured delays. This is also relevant for unidentified \fermilat point sources. \\
$\bullet $ Gravitational lensing might help to enhance the sensitivity to cosmic neutrinos emitted by hadronic-dominated \g-ray FSRQs that are typically placed at much larger distances with respect to other expected neutrino sources. 
The neutrino signal magnification by astrophysical lenses is of much interest for the next large-scale neutrino detectors. Lens multiple paths might induce also neutrino quantum interference and oscillations \cite{2004PhRvD..69f3008C}.\\
$\bullet $ Pseudoscalar axion-like particles (ALPs) generically couple to two photons, giving rise to possible oscillations with \g-ray photons emitted by a FSRQ in the intergalactic/intervening galaxy magnetic fields. Strong lensing of a background MeV FSRQ has some, speculative, possibility to enhance the flux of non-isotropic/streaming ALPs. Anomalies in the flux ratios of lensed images are foreseen by some DM theories. Time-variable lenses are also probes on the behavior of DM substructure in the intervening galaxy halo.\\
$\bullet $ Depending from particle properties, cosmological parameters, masses and separations of elements in the lensing system, differential arrival times of multimessenger particles (\g-ray photons, massive-neutrinos, GWs, even massive axions and gravitons) are expected. Multi-messenger detections of different time delays from a lensed MeV FSRQ would be an unexplored fundamental physics phenomenon.
\paragraph*{Expected results with e-ASTROGAM}
The capability of e-ASTROGAM 
to obtain independent \g-ray delay measurements from unresolved strong macro-lensing, and to identify variability features related to micro-lensing, in the case of MeV blazars, will be already a very useful goal. e-ASTROGAM is expected to discover several new high-redshift FSRQs undetected by the \fermilat because of GeV cutoffs, and to see many MeV \g-ray flares, including those from lensed FSRQs.  In addition, space-borne wide field imaging observatories, such as ESA's \textit{Euclid} space telescope, would soon produce hundreds of new useful strong lenses to be searched for a MeV detection.  Time-series and spectral analysis of \g-ray variability, combined with the properties of the lens from radio observations (SKA, ALMA, etc.) or IR/optical observations (LSST, \textit{Euclid}, JWST, etc.) can yield an improvement in spatial resolution at \g-ray energies by a factor of $10^{4}$ \cite{2015ApJ...809..100B, 2015NatPh..11..664N}.
Multi-messenger studies using FSRQ sources with candidate hadronic processes, will also be potentially opened by e-ASTROGAM, in conjunction with the foreseen large scale neutrino array experiments (KM3NeT and other). The lens magnification of the neutrino flux is expected to be equal to that of \g-ray photon flux, and this could drive to the measure the intrinsic neutrino luminosity of powerful MeV-GeV FSRQs. MeV \g-ray lensed blazar might also be of interest for, speculative, hypotheses in multimessenger and fundamental physics.
\subsection[Narrow-Line Seyfert 1 galaxies: high accretion rates and low Black Hole masses\\
\noindent
\textit{\small{S. Kaufmann, O. Tibolla, L. Foschini}}]
{Narrow-Line Seyfert 1 galaxies: high accretion rates and low Black Hole masses}
\paragraph*{Science questions}
Radio-loud Narrow-Line Seyfert 1 galaxies have been established as a new class of \g-ray emitting AGN with relatively low BH masses, but near-Eddington accretion rates.
The mass of the central BH is much smaller ($10^6 - 10^8 M_\odot$, e.g. \cite{Boroson2002}) and the accretion rate much higher than those estimated for the class of blazars (see \cite{Foschini2013} for a review). 
Narrow-Line Seyfert 1 (NLSy1) galaxies are characterized by broad permitted and narrow forbidden lines in their optical spectra, classifying them as Seyfert 1 galaxies. 
However, the permitted lines are narrower than usual with FWHM(H$\beta$) $< 2000 \rm{km s^{-1}}$, the ratio of [O III] to H$\beta$ is smaller than 3, and a bump due to Fe II exist (see, e.g. \cite{Pogge2000} for a review).
A larger study by \cite{Zhou2006} based on SDSS Data Release 3 identified a sample of 2011 NLSy1 galaxies.
Only a small fraction of NLSy1 galaxies are radio loud ($S_{4.85 \rm{GHz}}/S_{440 \rm{nm}}>10$.), e.g. $7\%$ in the study of \cite{Komossa2006}.

\begin{figure}
\centering
\includegraphics[width=0.49\textwidth]{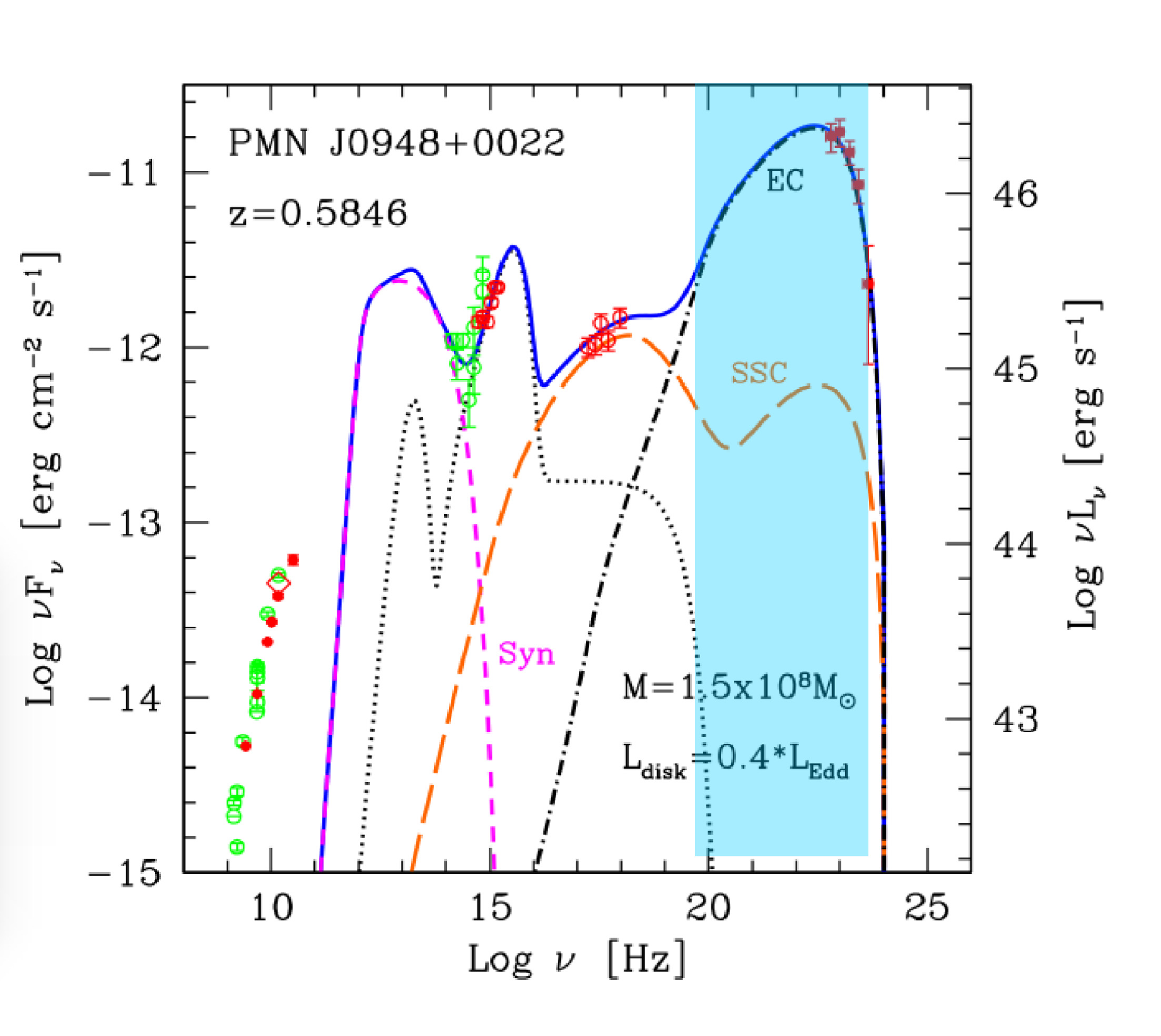}
\includegraphics[width=0.49\textwidth]{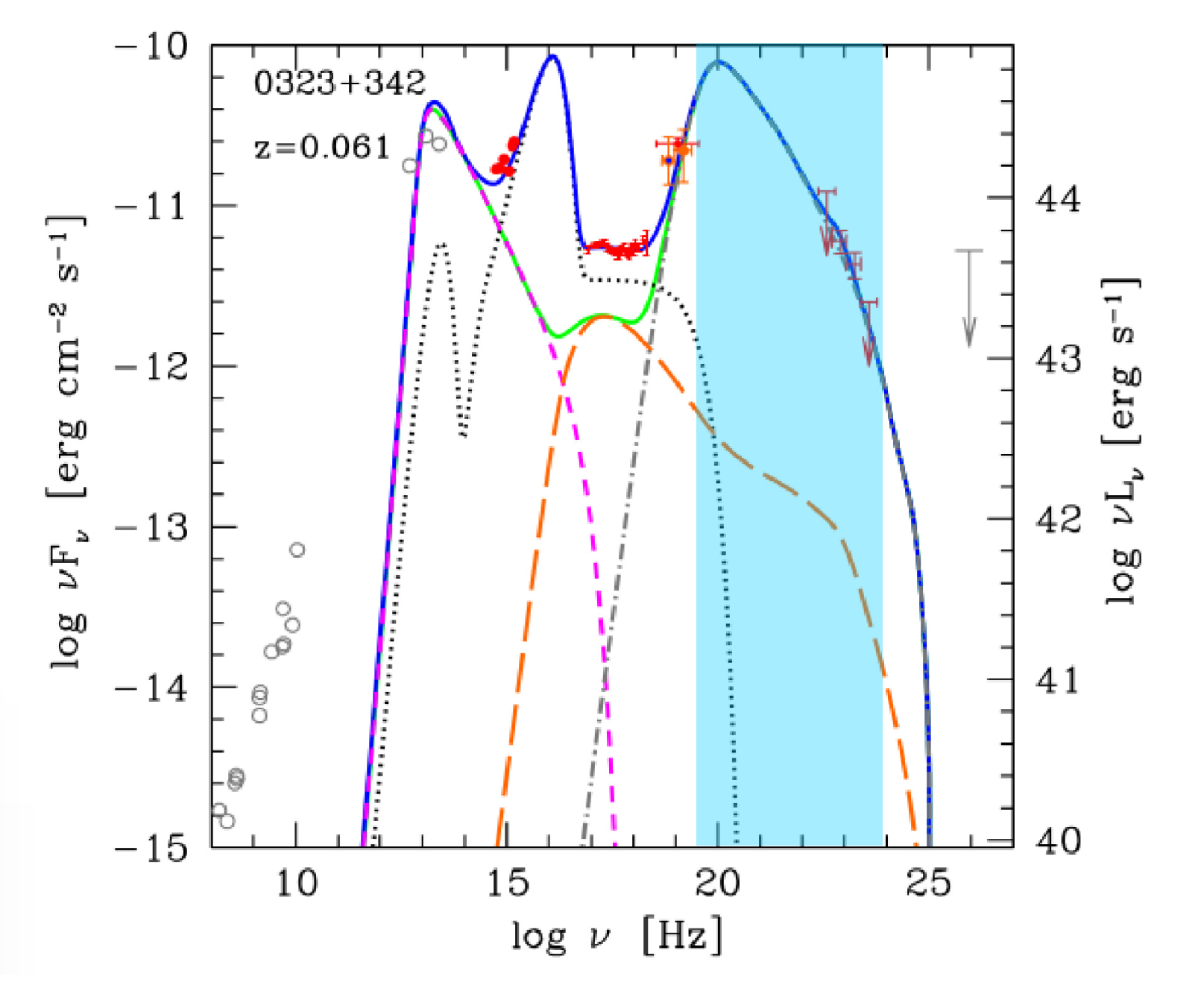}
\caption{\small{Left: SED of PMN J0948+0022 taken from \cite{Abdo2009a}. The dotted line shows the contribution from the infrared torus and accretion disc. The \ssc is shown in dashed and the \ic from external photon fields (\ec) is shown with a dot-dashed line. The blue area represents the energy range of \ea.
 Right: 
1H 0323+342 is the closest among the \g-ray emitting NLSy1. The SED is taken from \cite{Abdo2009}. 
The BH mass is assumed to be $M_{BH} \sim 10^7 M_\odot$ and accretion disc luminosity of $L_{\rm{disc}} = 0.9  L_{\rm{Edd}}$ (\cite{Abdo2009}).
Based on its characteristics in optical/X-rays, 1H 0323+342 seem to represent a transitional case in which the inner disc heats up and blows up to a torus configuration, as the accretion rate goes down (\cite{Tibolla2013}).}}
\label{NLSy1:fig1}
\end{figure}

The detection of high-energy \g-rays and its variability (\cite{Abdo2009a},\cite{Abdo2009}) confirmed the existence of powerful relativistic jets in radio-loud NLSy1 galaxies (see \cite{Foschini2014} for a review), which therefore can be now named as  jetted NLS1, according to the classification recently proposed by Padovani \cite{Padovani2017}. \\
\\
{\it MeV peaked emission - high energetic jet:} \\
An important feature, that NLSy1 galaxies have in common with the other class of jetted AGN (e.g. FSRQ), is the MeV peaked spectral emission, which should be studied in more detail.
Fig.~\ref{NLSy1:fig1} illustrate two prominent examples of radio-loud NLSy1 galaxies with detected GeV emission. 
The keV-GeV peak (in $\nu F_\nu$ presentation) is generally described by the \ssc emission from the highly energetic jet and the External Compton (\ec) emission in which the relativistic electrons interact with a photon field close to the jet (generally from the broad-line region). The \ec component is generally necessary to describe the detected GeV \g-ray emission. It is still needed to understand the different contributions of the \ssc and \ec in the high energy band, which are currently difficult to establish precisely. Measurements of the polarization will help to distinguish between the \ssc (polarized) and the \ec (un-polarized) emission. \\
\\
{\it High accretion rate and low BH mass:}\\
As can be seen clearly in Fig.~\ref{NLSy1:fig2}, the radio-loud NLSy1 galaxies have much lower BH masses than the class of FSRQ and BL Lac objects (class of object observed in the direction of the high energetic jet). In addition, the accretion rate is very high, comparable to the ones of FSRQs. 
The important idea was established by \cite{Foschini2017}, that the sequence of NLSy1 galaxies to FSRQ to BL Lac objects, going from small-mass BHs with high accretion rate to large-mass BHs and low accretion rate, could describe the cosmological evolution of the same type of object. Hence, the NLSy1 galaxies represent the young state with low BH masses and their study will give the opportunity to understand better the cosmological evolution of AGN.\\
\\
{\it Comparable characteristics to X-ray binaries:} \\
One hot topic of discussion is the simultaneous existence of the jet and a very high accretion rate. The investigation about the flux variability will give more insight in this question. As shown in \cite{Abdo2009a},\cite{Tibolla2013}, the \g-ray emission of NLSy1 galaxies is variable. Hence, the jet may be formed accompanying with relatively weak soft X-ray, as was commonly seen in X-ray binaries. 
\paragraph*{Importance of \g-ray observations}
Based on survey studies by \cite{Foschini2015} with multi wavelength spectral studies of jetted NLSy1 galaxies, a peak of the \g-ray emission in the MeV energy band is expected, as in the other jetted AGN.
There is a current lack of data in the 100 keV - 100 MeV energy band, in which luminous emission is expected from all jetted NLSy1 galaxies.
As can be seen in Fig.~\ref{NLSy1:fig1}, the SED of NLSy1 galaxies is rather complex. In the energy range from keV to MeV, the dominant emission process seems to be the \ssc emission from the high energetic jet and the \ec emission in which the relativistic electrons interact with the BLR photon field close to the jet. The \ec component is necessary to describe the detected GeV \g-ray emission. Constraints on the model can be obtained with \g-ray observations on jetted NLSy1 galaxies which will provide a good coverage of the current gap in the MeV energy band. The currently not well determined ratio between the \ssc and \ec components can be defined more precisely with such measurements.
By using polarization measurements, it will be possible to disentangle \ssc (polarized) from \ec (not polarized).
Especially, it is very important to fix the \ssc contribution to be able to estimate the strength of the magnetic field.

\begin{figure}
\centering
\includegraphics[width=0.61\textwidth]{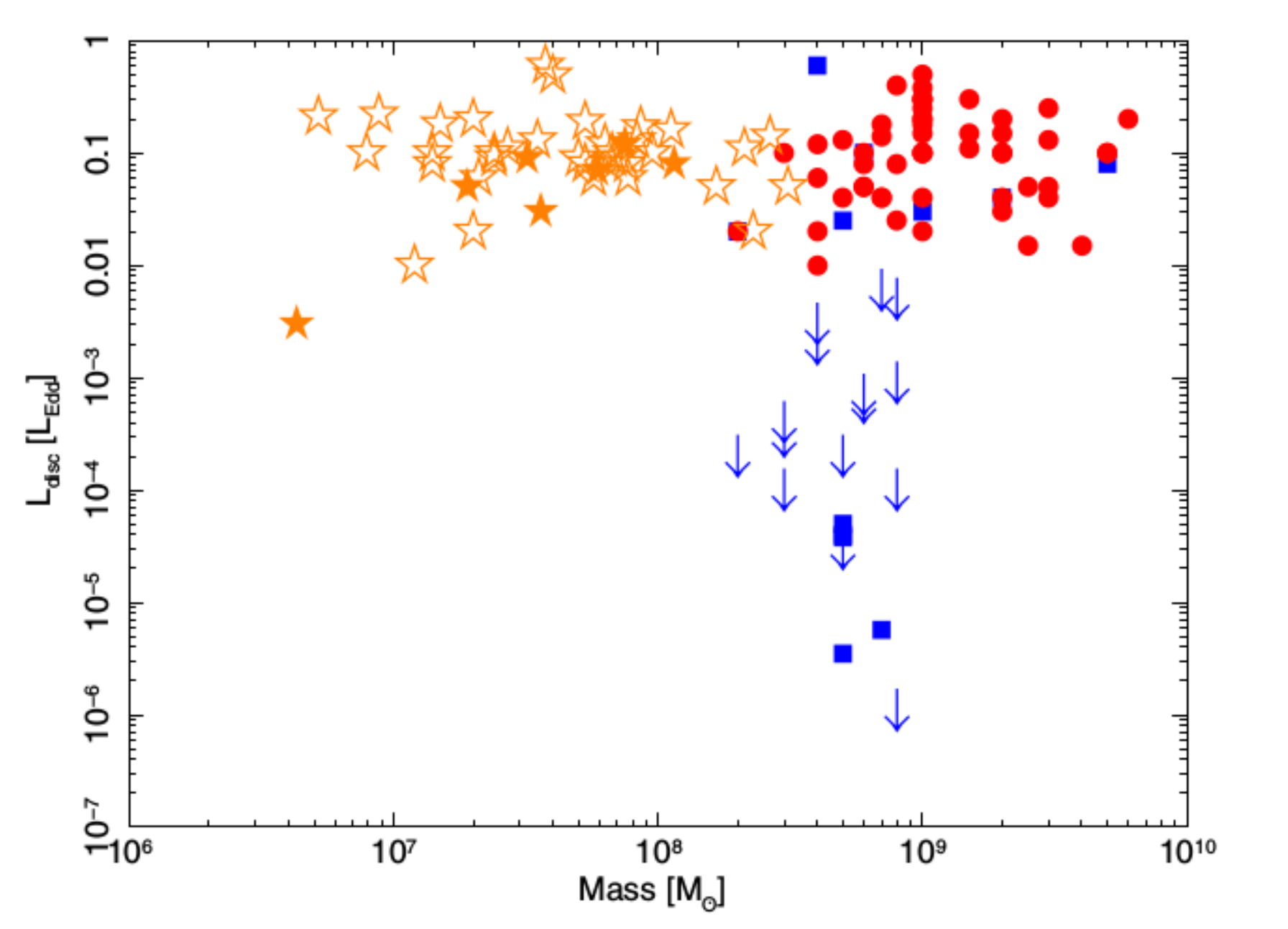}
\caption{\small{Accretion disc luminosity in Eddington units versus the mass of the central BH, taken from the survey study of \cite{Foschini2015}. The orange stars represent the characteristics of the radio-loud NLSy1 galaxies. The red circles show the FSRQs and the blue squares and arrows the BL Lac objects.}}
\label{NLSy1:fig2}
\end{figure}

Due to the variability in X-rays with changing spectral behavior, it will be important to observe simultaneously in the \g-ray and lower energy range (e.g. monitoring observations expected from eROSITA or triggered, pointed observations with current X-ray satellites).
The measurement of the time scale of the flux variability in the X-ray and \g-ray range will also give indications about the location of the external photon field responsible for the IC emission.

Another point is the study of \g-ray emission from the parent population of beamed NLS1. A steeper \g-ray spectrum is expected  \cite{Liao2015}, and therefore the detection below 100 MeV could be an asset with respect to \fermilat.
\paragraph*{Expected results with e-ASTROGAM}
The peak (in $\nu F_\nu$ presentation) of the Compton component at MeV energies makes jetted NLSy1 galaxies a wonderful target for e-ASTROGAM observations, especially due to the current lack of data in the 100 keV-100 MeV energy band. 
A large number of detections are expected with the covered broad energy range from 0.3 MeV to 3 GeV and its planned high sensitivity of e-ASTROGAM, which, e.g. in the range 0.3-100 MeV will be one to two orders of magnitude better than that of previous instruments (see Chapter~\ref{intro}). This will give rise to a more detailed study of the underlying emission processes and to identify the characteristic parameters.

Based on the sample of radio-loud NLSy1 galaxies by \cite{Foschini2015}, in which the SED of 42 NLSy1 galaxies have been studied in detail, we expect a large number of 
MeV peaked NLSy1 galaxies (based on the spectral characteristics in the X-ray regime) to be easily studied with \ea.

Berton et al. \cite{Berton2015} performed simulations indicating that SKA will detect thousands of jetted NLSy1 for which a multi-wavelength coverage will be required (and \Fermi will likely be no more available, and CTA has a too high low-energy threshold).
\subsection[Misaligned Active Galactic Nuclei\\
\noindent
\textit{\small{F.~D'Ammando, M. Orienti, M. Giroletti, A. De Rosa}}]
{Misaligned Active Galactic Nuclei
}
\paragraph*{Science questions}
Relativistic jets are one of the most spectacular manifestations of the release of energy by the super-massive BH at the center
of AGN. When the jet axis is closely aligned with our
line of sight, the rest-frame radiation is strongly amplified due to Doppler
boosting effects. A large fraction of their energy output is observed at high
energies, giving rise to the blazar phenomenon \cite{urry95}. Radio galaxies are viewed at
larger angles than blazars, with less severe boosting effects. Based on their
radio morphology and power, radio galaxies are classified as Fanaroff-Riley
type I (FR I) and type II (FR II) \cite{fanaroff74}. FR I are
characterized by a two-sided, low-power decelerating jet whose intensity falls
away from the nucleus. FR II show one-sided, powerful and
collimated relativistic jet leading to well-defined lobes with prominent hot
spots at their edge. According to the Unified model of AGN proposed by
\cite{urry95}, FR I and FR II are the non-aligned (to the
observer viewing angle) parent populations of the low-power (BL Lac objects)
and high-power (flat spectrum radio quasar, FSRQ) blazars, respectively. 

Misaligned AGN (MAGN), including radio galaxies and steep-spectrum radio quasars
(SSRQ; i.e. lobe-dominated radio quasars), have been discovered as a new class
of \g-ray emitting AGN by the \Fermi satellite \cite{abdo10}. The
SED of MAGN seems to resemble that of blazars,
the only difference being a smaller Doppler factor. The high-energy
emission of MAGN in the leptonic scenario is usually due to \ic scattering of seed
photons internal or external to the jet, with the \ssc process proposed as the main mechanism for FR I and the \ec process
for FR II. This difference should be related to the different environment of
optical/UV photons produced by the accretion disc in the two classes. Hadronic models have been proposed to contribute to the
high-energy emission of radio galaxies, suggesting these as high-energy
neutrinos sources \cite{becker14}. 

In the leptonic scenario the SED of FR I is generally well fit with bulk Lorentz factor $\Gamma$ $\sim$ 2--3, much lower that the typical
values of BL Lacs ($\Gamma$ = 10--20) \cite{aleksic14}. Different scenarios involving a gradient of velocity in the jet have been proposed to explain this discrepancy, in particular a spine-layer jet \cite{ghisellini05} or a decelerating jet \cite{georganopoulos03}. Theoretical arguments and numerical simulations suggest that jets in AGN are not uniform outflows, but are characterized by a transverse velocity structure composed of a fast central part, the spine, surrounded by a slower layer \cite{ferrari98}. The regions with different speeds would interact through their radiation fields, relativistically boosted in the different frames. Such interaction leads to the enhancement of the IC emission of the two zones. Another consequence of the radiative coupling is the progressive deceleration of the spine. Since the layer is expected to have lower bulk Lorentz factors than the spine, its less beamed emission can be detected even when the jet is misaligned with respect to us, as in the case of the radio galaxies. A strong support to the existence of a stratified jet structure comes from the observation of a limb-brightened structure in the FR I NGC 1275 \cite{nagai14} and M87 \cite{hada13}, as well as in the MAGN PKS~0521$-$36 \cite{dammando15}. The spine-layer model was applied to the SED of NGC 1275 and M87, finding a good agreement with the data \cite{tavecchio08,tavecchio14}. A structure of this type has been detected also in the TeV BL Lacs Mrk 501 \cite{giroletti04} and Mrk 421 \cite{lico12}, in agreement with the unification of BL Lacs with FR I radio galaxies, their parent population \cite{chiaberge00}. It is not clear whether the most powerful jets (FSRQ and FR II) have analogues structures. A difference in the jet structure between powerful and weak sources could be related either to a different environment enshrouding the jet (e.g. gas density and temperature) or to intrinsic jet properties, causing the weak jets to be more prone to instabilities. 

One key issue of the high-energy study of MAGN is the localization of the \g-ray emitting region. There is observational evidence
supporting either a location close to the super-massive BH, on sub-pc scale, or a site at
few parsecs from the central engine downstream along the jet. The detection of
\g-ray emission from the radio lobes of Cen A and Fornax A have
demonstrated that the inner jet is not the only region responsible for the GeV
emission, involving \ic scattering of Cosmic Microwave Background (CMB) photons
and also hadronic processes \cite{abdo10b, ackermann16}. The nearby radio
galaxy M87 offers a unique opportunity for exploring the connection between
\g-ray production and the jet formation at an unprecedented linear
resolution. However, the origin and location of the \g-rays even in this
source is still elusive. Based on previous radio/TeV correlation events, the
unresolved jet base (radio core) \cite{hada14} or the peculiar knot HST-1 at
120 pc from the nucleus are proposed as candidate site(s) of \g-ray
production \cite{cheung07}. FR II are detected in the GeV regime mainly during
flaring periods, as observed in 3C 111 and 3C 120. A correlation between a
\g-ray flare and the ejection of a new jet component has been observed
in these sources, suggesting a \g-ray emitting region at sub-pc distance
from the super-massive BH \cite{grandi12,tanaka15}. 
\paragraph*{Importance of \g-ray observations}
Only three FR I have been tentatively detected in \g-rays by EGRET. With the advent of the Large Area Telescope (LAT) on board the \Fermi satellite the number of MAGN detected in \g-rays increased to 21: 11 FR I, 3 FR II, and 7 SSRQ \cite{3lac}. Recently, a \g-ray source was associated to an FR 0, a new class of radio galaxies with similar nuclear properties of FR I but lacking extended radio emission \cite{grandi16}. The low number of \g-ray emitting MAGN detected so far leaves a discovery space for the high-energy emission of this class of object.

The MeV regime is still an almost unexplored window for studying AGN. Only a handful of sources have been detected at MeV, with only Cen A among radio galaxies \cite{collmar99}. The \g-ray spectrum of MAGN detected by \fermilat is usually soft ($\Gamma$ $>$ 2; Fig.~\ref{magn:fig1}, left panel), indicating an high-energy peak at MeV; therefore information in the MeV regime is crucial for characterizing the broad band SED of these sources and set tight constraints on the emission mechanisms at work and the jet parameters (Fig.~\ref{magn:fig1}, right panel). Moreover, being the high-energy emission peaked at MeV energies, observations in this band will be important for discovering many new \g-ray emitting MAGN.
\begin{figure}
\centering
\includegraphics[height=5.2cm]{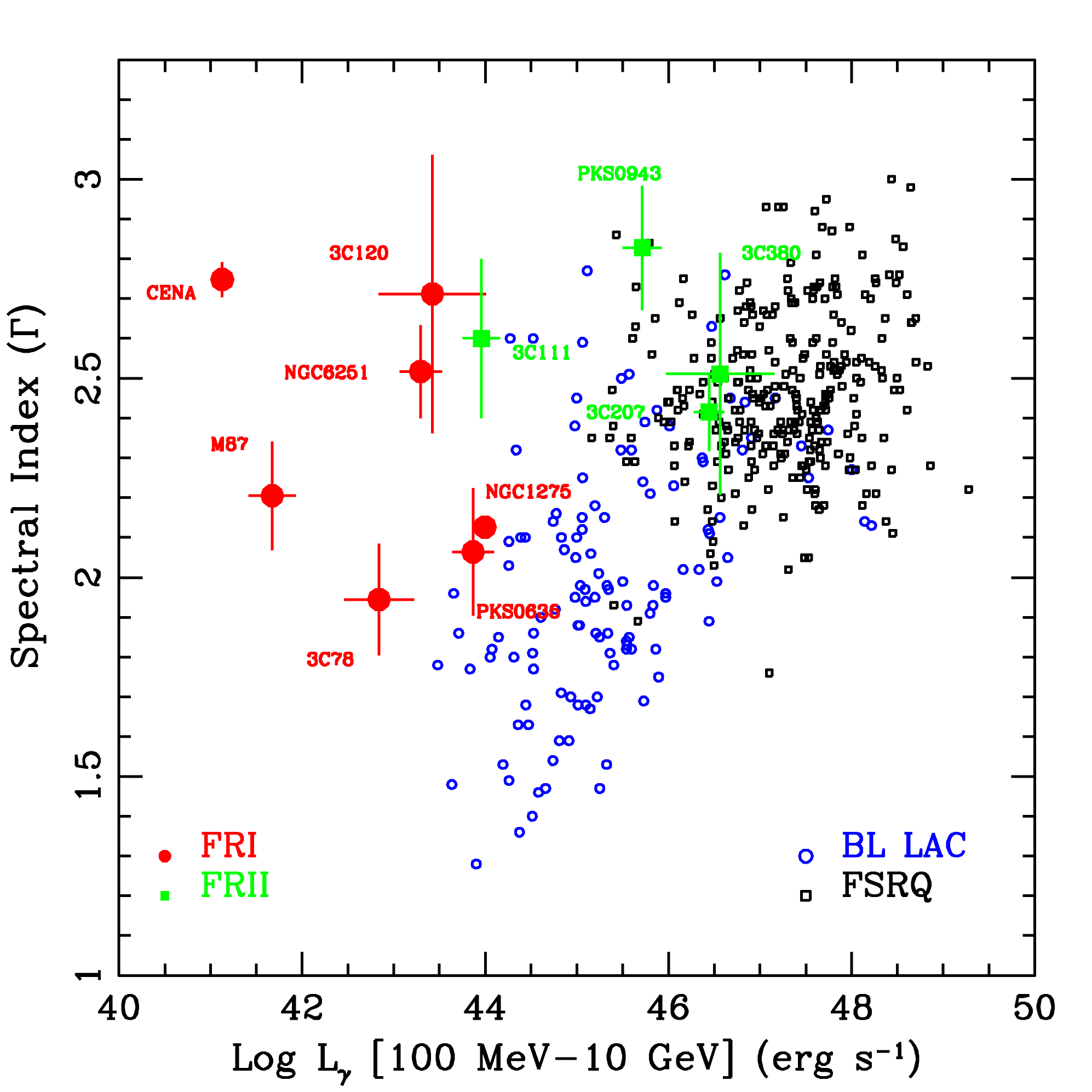}
\includegraphics[height=4.7cm]{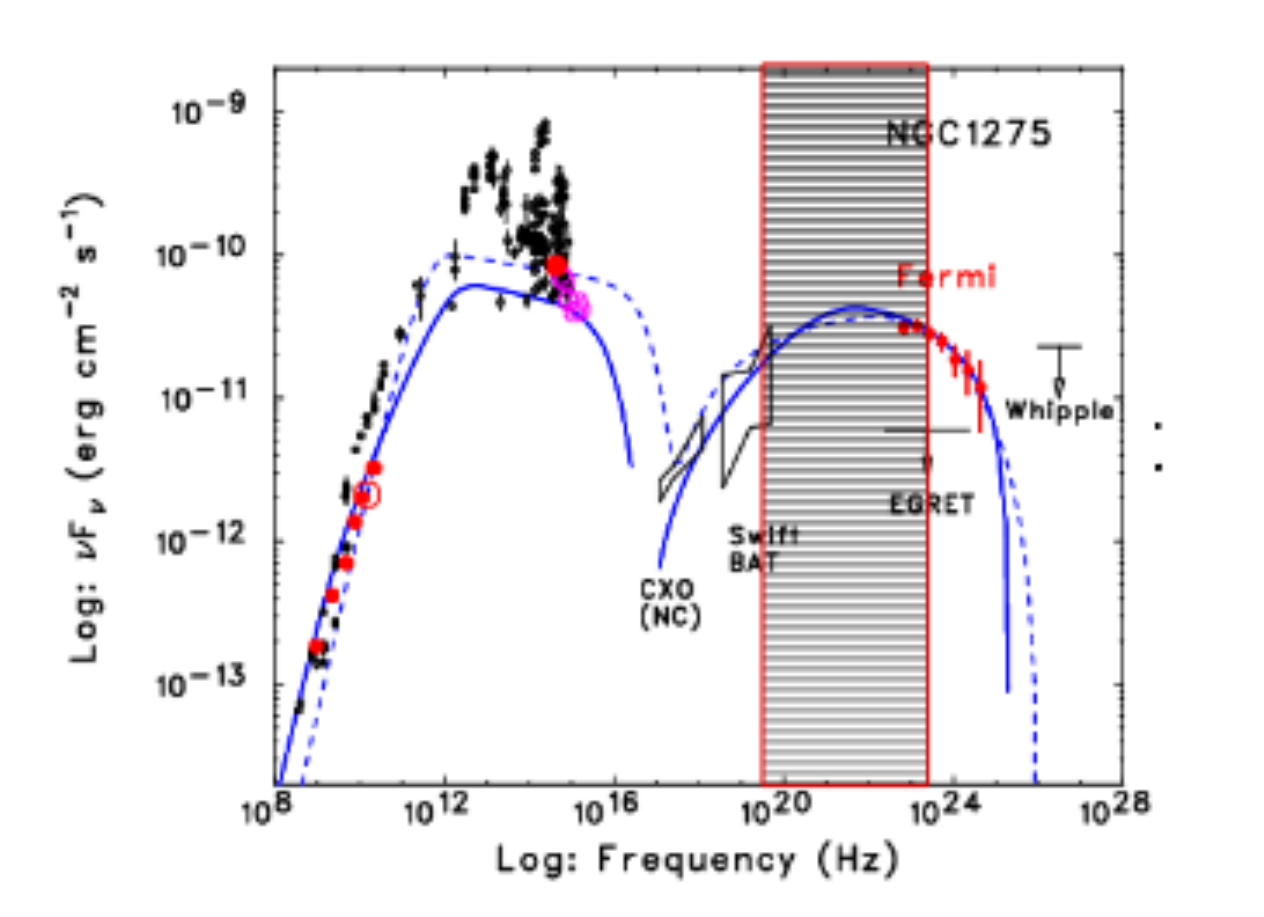}
\caption{\small{Left: Photon index vs. \g-ray luminosity of MAGN and blazars detected by \fermilat \cite{abdo10}. Right: SED of NGC 1275 from radio to TeV taken from \cite{abdo09}. The shaded area represents the energy range covered by e-ASTROGAM.}}
\label{magn:fig1}
\end{figure}
\paragraph*{Expected results with e-ASTROGAM}
Radio galaxies represent the larger population of radio-loud AGN. The energy flux in the range 100 MeV -- 100 GeV of the MAGN detected by \fermilat in the first 4 years of operation ranges between 2.2$\times$10$^{-12}$--2.0$\times$10$^{-10}$ erg cm$^{-2}$ s$^{-1}$
\cite{3lac}. These values are above the 3-$\sigma$ e-ASTROGAM sensitivity for one year of exposure in the 100 MeV--1 GeV range,
5$\times$10$^{-13}$--10$^{-12}$ erg cm$^{-2}$ s$^{-1}$. With the increase of the e-ASTROGAM exposure time throughout the whole mission and considering that the IC peak of these sources lies in the MeV regime, a large number of MAGN are expected to be detected with the covered broad energy range and sensitivity of e-ASTROGAM. MAGN can be also detected during high activity states with dedicated e-ASTROGAM pointing observations of 500 ks. 

Thanks to the increasing number of MAGN detected at high energies and to the opportunity to detect simultaneously the hard X-ray, MeV and soft GeV emission components of different origin, we will be able to study in detail the emission processes at work (i.e. disc vs jet components, \ssc vs \ec process), the location of the \g-ray emission region, and the jet parameters for galaxies with different radio morphologies and power. Polarization measurements will help to distinguish between the \ssc process (polarized) and the \ec process (unpolarized).
Owing to the unprecedented sensitivity in the 0.3--100 MeV energy range it will be possible to discriminate between one-zone and spine-layer model, leptonic and hadronic emission components for modelling the SED of radio galaxies.
Deep observations of the lobes of Cen A and Fornax A with e-ASTROGAM would allow a better measurement of the CMB that should be the main seed photon field for the IC mechanism that produces their \g-ray emission.

The increasing number of MAGN will allow us also to estimate the contribution of these sources to the extragalactic \g-ray background in the largely unexplored 0.3--100 MeV energy range. The e-ASTROGAM spectral sensitivity will allow population analysis of MAGN through variability studies coordinated with the forthcoming facilities such as SKA, JWST, Athena, and CTA that will cover the whole EM spectrum. 
\subsection[Chasing the lowest frequency peaked synchrotron emitters\\
\noindent
\textit{\small{S. Ant\'on, A. Caccianiga, M. J.  March\~a}}]
{Chasing the lowest frequency peaked synchrotron emitters
}
\paragraph*{Science questions}
One of the major topics in the study of AGN is related to the formation and the physics of the relativistic jets that are typically observed in radio-loud AGN and even in some of the €œso-called radio-quiet AGN. Currently one of the most effective ways to
investigate this issue is by studying the blazar class, i.e. the subset of AGN that are supposed to be oriented in such a way that the relativistic jet and the observer are closely aligned: in this particular condition, the non-thermal emission produced within the jet is relativistically boosted and dominates the entire nuclear SED, offering the best arrangement to infer the jet properties (for a recent review see e.g. \cite{2017A&ARv..25....2P}).\\

\noindent Most of the investigation of blazar phenomena has been based on the brightest and most powerful objects, which may not be representative of the population as a whole. The fainter blazar population has been studied  by selecting samples of low-luminosity radio-loud sources, like the 200-mJy sample or the CBS  (\cite{anton2004},\cite{marcha2001},\cite{marcha1996}). These samples contain a class of objects having milliarcsecond VLBI core-jet structure of the type found in superluminal radio sources and sharing other blazar characteristics (e.g. high levels of radio polarisation) but showing an
optical spectrum dominated by the host galaxy light. They are very-low frequency peaked objects (\cite{anton2005},\cite{caccianiga2004}), and, as explained below, they are expected to be strong MeV emitters. Therefore, these sources are excellent e-ASTROGAM candidates and may represent a relevant fraction of the MeV sky.
\paragraph*{Importance of \g-ray observations}
The SEDs of blazars have long been modelled by two broad components: one extending from radio frequencies to the IR/optical or even to the X-rays, attributed to synchrotron emission, and another one extending to the \g-rays, which is associated to the \ic process. In units of  $\nu F_{\nu}$ the two components have roughly a 2-humped structure, different objects having the maximum of emission at different frequencies. The broad range of frequencies at which the SEDs peak has a great impact on the observational properties of these objects: the 
sources with the synchrotron emission peaking in the optical band are more easily recognised as blazars thanks to their extreme optical properties (like a polarised and variable featureless continuum). On the contrary, blazars with the synchrotron emission peaking in the infrared band may be hardly recognised since their optical emission can be easily overwhelmed by the light from the host galaxy, particularly in low-luminosity sources. This means that sources with a synchrotron peak falling at very low-frequencies and with relatively low-luminosities could have been systematically overlooked in current surveys. This potential incompleteness can have a profound impact on several open issues about blazars like the shape of their radio luminosity function at low powers or their actual contribution to the \g-ray background.\\

\noindent The possibility of observing blazars in the MeV range is of fundamental importance from this point of view since, as for those objects that have their
synchrotron peak in the infrared band, ie,  very-low frequency peaked blazars (VLBL, $\nu_{\mbox{{\tiny peak}}}^{\mbox{{\tiny sync}}}\ \sim 10 ^{12} -10 ^{13}$ Hz), the IC bump is expected to peak at MeV energies, considering the almost fixed relative ratio of $\sim$10$^7$ between synchrotron and IC peak frequencies (\cite{1998MNRAS.299..433F}). Therefore, an (optically) unrecognised population of blazars should clearly emerge in e-ASTROGAM observations. The existence of such a population of sources has a great impact on our current knowledge of the blazar population and it may be also relevant for the understanding of the \g-ray background.\\

\noindent In the past years we have worked on two radio surveys (200-mJy \cite{anton2004}, \cite{marcha1996} and CBS \cite{marcha2001}) specifically aimed at selecting low-power blazars and we have found that a significant fraction of objects (~20\%) has an optical spectrum dominated by the host-galaxy light. At the same time, high resolution radio data (VLBI, see \cite{bondi2004}) show in many of these sources a core-jet morphology strongly supporting their blazar nature. The analysis of the SED of these low-power blazars indicates that they are likely VLBL objects with a peak falling between 10$^{12}$ and 10$^{13}$ Hz. If these objects are really VLBLs we expect that many of them will be detected by e-ASTROGAM.
\begin{figure}
\includegraphics[height=5cm]{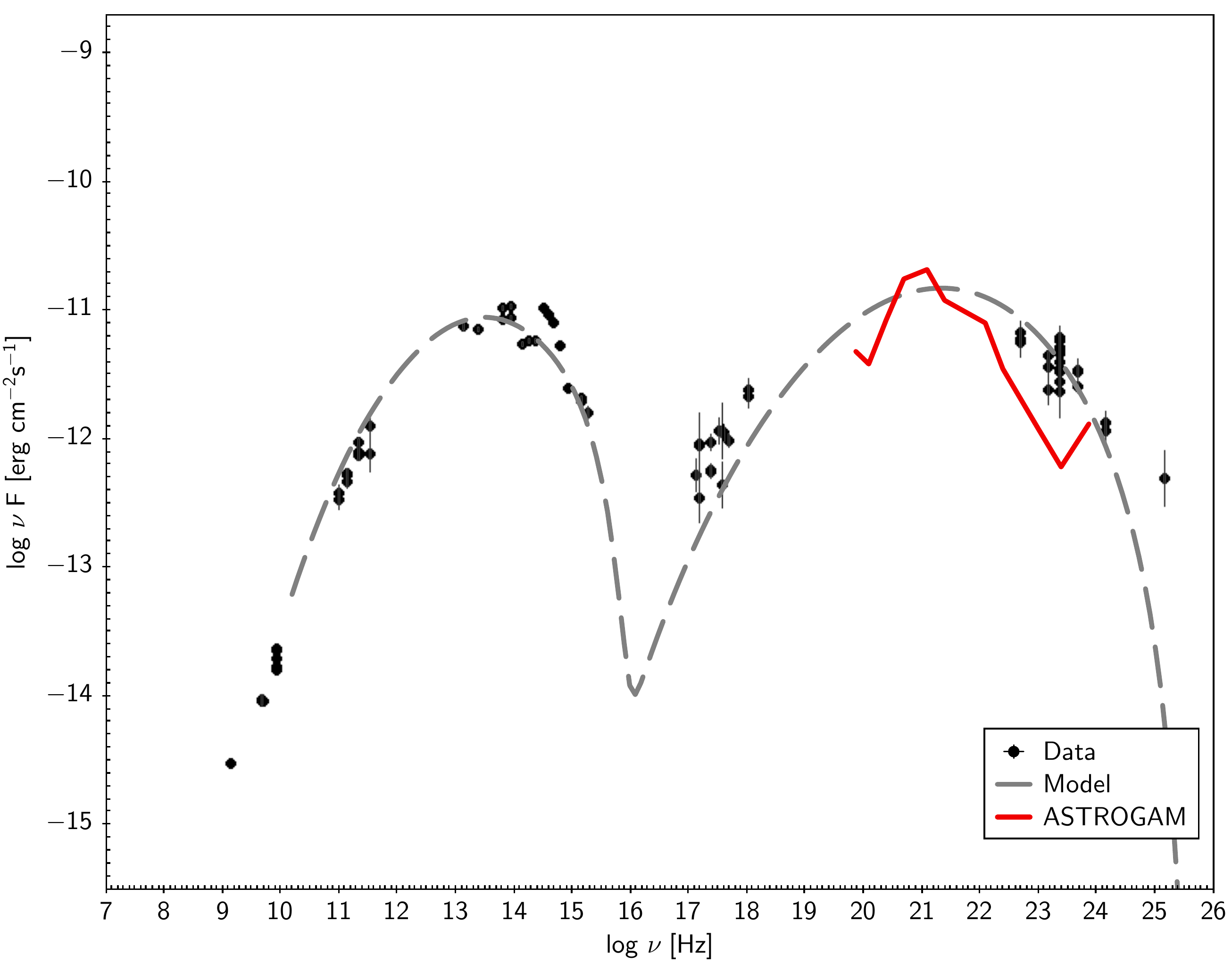}
\hspace{0.5cm}
\includegraphics[height=5cm]{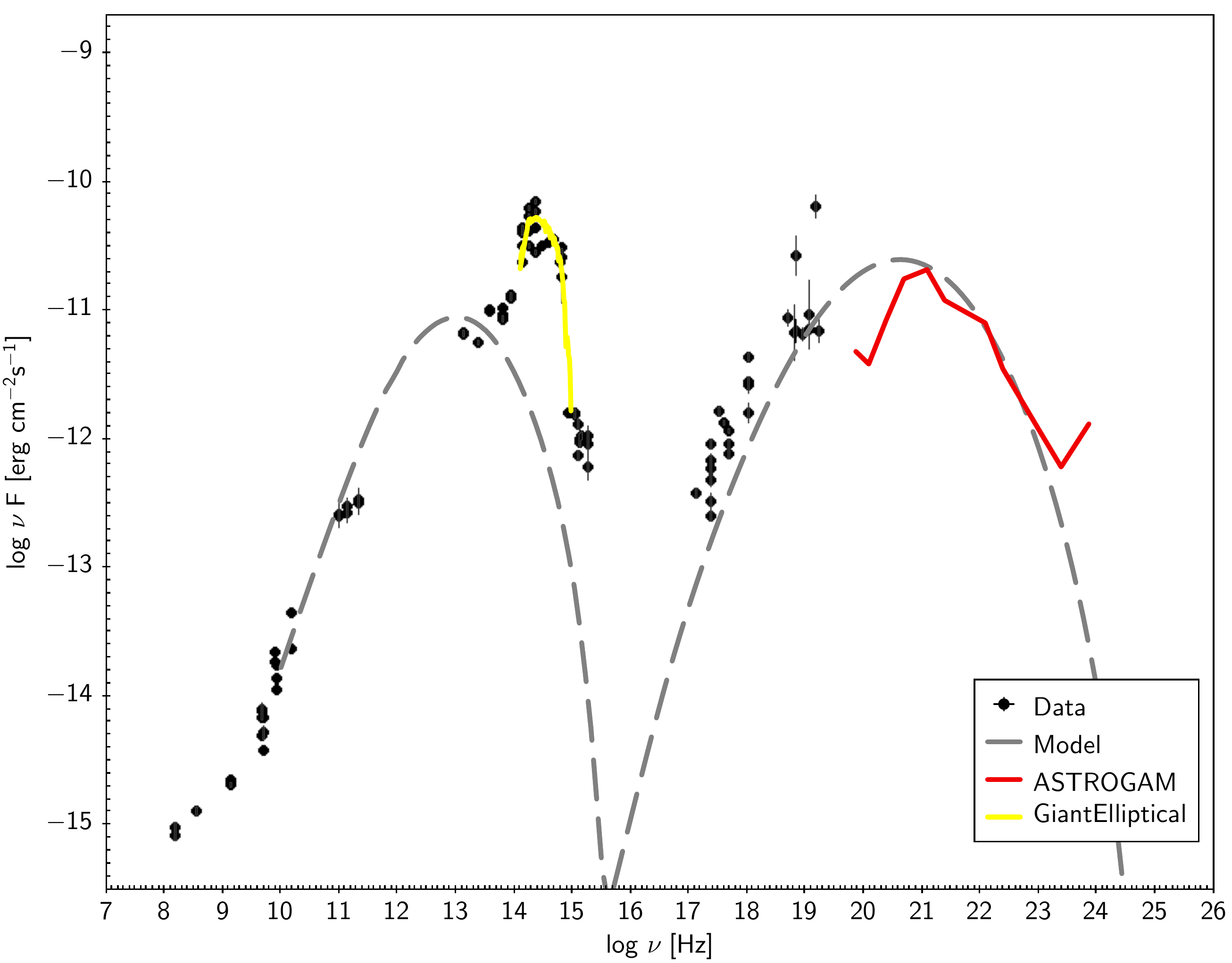}
\caption{\small{SEDs 1647+495 (left) and 1719+485 (right), that comprise multi-wavelength from the radio 
to the \g-ray bands. The objects are representative of the class of low
luminosity radio loud VLBL:  (1) have the peak of synchrotron emission at
low frequencies ( $\nu_{\mbox{{\tiny peak}}}^{\mbox{{\tiny sync}}} <10^{14}$ Hz),  and  for that reason their blazar
nature appear disguised in the optical band, particularly when in the presence of a prominent host galaxy component; (2) they
are core-jet VLBI sources, which is in general a good indication of alignment
between the  jet and the observer; (3) a fraction of these objects are \g-ray
sources. The dashed gray line represents the emission from
\ssc models, see text for details. The red line shows  e-ASTROGAM sensitivity curve and in the case of 1719+485 an elliptical galaxy template
is also shown in yellow. Data, templates and models (which are based on \cite{tramacere2009} and  \cite{mannuci2001}) were obtained from SSDC website.}} 
\label{low_sync:fig1}
\end{figure}
\paragraph*{Expected results with e-ASTROGAM}
Fig.~\ref{low_sync:fig1} presents the SEDs of two low power jet objects taken from the 200-mJy and CBS samples, of the type described above. Both the multi-frequency data and superimposed models were obtained from SSDC SED builder Tool. Gray dashed lines represent the \ssc emission models, where a log parabola electronic distribution plus a synchrotron self-absorption component were chosen, the free parameters consistent with those reported in the literature (eg. \cite{ghisellini05}). Note that the presented \ssc curves
 are just an  estimate to valuate how feasible is the MeV detection in these objects, and for that reason e-ASTROGAM sensitivity curve for 1-year
 exposure is also shown. According to our estimates these low luminosity
 radio loud objects should be detected by e-ASTROGAM. The detection of these sources at MeV energies will shed light on the poorly studied population of low-power blazars.
\subsection[Estimation of magnetic-to-particle energy density ratio of BL Lac objects\\
\noindent
\textit{\small{N. Mankuzhiyil}}]
{Estimation of magnetic-to-particle energy density ratio of BL Lac objects}
\paragraph*{Science questions}
As already pointed out, the  overall  radio to \g-rays Spectral  Energy Distribution (SED)  of  BL Lacs  displays  two  broad non-thermal  continuum  peaks.  The low energy peak is thought to arise from the synchrotron process, while the  leptonic  models, which are popular models used in literature,  suggest  that  the  second  peak  forms out of \ic emission. If the low energy photons which undergo the \ic process are the synchrotron photons, the process is known as the \ssc emission \cite{Maraschi92}. In the current scenario, \ssc models can satisfactorily reproduce the observed flux of blazars in optical-to-\g-rays broad band window.
The electron energy distribution  responsible for the non-thermal emission can be represented by a double power-law
\begin{equation*}
N(\gamma)= \begin{matrix}
{\mathrm{K \gamma^{-n_{1}}}} \,\,\,\,\,\,\,\,\,\,\,\,\,\,\,\,\,\,;    {\mathrm{\gamma_{min} < \gamma < \gamma_{br}}}
\\ 
 {\mathrm{K \gamma_{br}^{n_{2}-n_{1} } \gamma^{-n_{2}}}} \,\,; {\mathrm{\gamma_{br} < \gamma < \gamma_{max}}}
\end{matrix}
\end{equation*}
where $\mathrm{\gamma_{min}}$, $\mathrm{\gamma_{\rm br}}$, and $\mathrm{\gamma_{max}}$ are the lowest, break, and highest Lorentz factors of the electron energy distribution, K is the normalization constant, and $\mathrm{n_1}$ and $\mathrm{n_2}$ are, respectively, the slopes below and above the break.
 The kinetic energy density of relativistic electrons can be estimated as 
\begin{equation*}
\mathrm{U_e =  m_{e}c^{2} \int_{\gamma_{min}}^{\gamma_{max}} N(\gamma) (\gamma -1 )d\gamma \simeq  m_{e}c^{2} N <\gamma>}
\end{equation*}
where, N is the integrated electron density. If $\mathrm{n_{1}\approx 2}$, the average Lorentz factor of the particle can be written as
\begin{equation*}
<\gamma>  \begin{matrix}
{\mathrm{\simeq \gamma_{min}\, ln(\gamma_{br}/\gamma_{min})}}
\end{matrix}
\end{equation*}
\begin{figure}
\centering
	\includegraphics[trim= 0mm 6cm 0cm 12cm, width=0.7\textwidth,clip=t,angle=0.,scale=0.55]{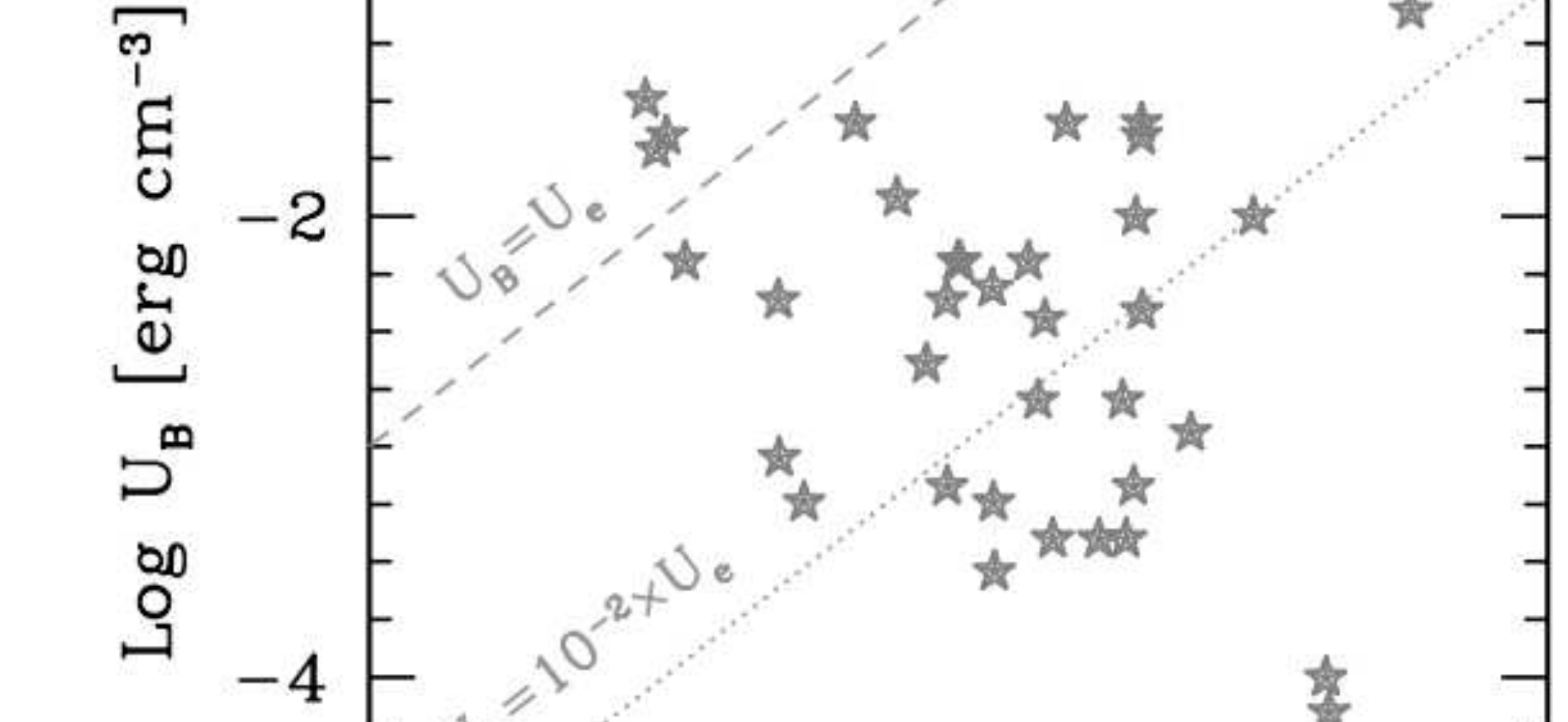}
	\hspace{1cm}
	\includegraphics[trim= 0mm 6cm 0cm 10cm, width=0.7\textwidth,clip=t,angle=0.,scale=0.57]{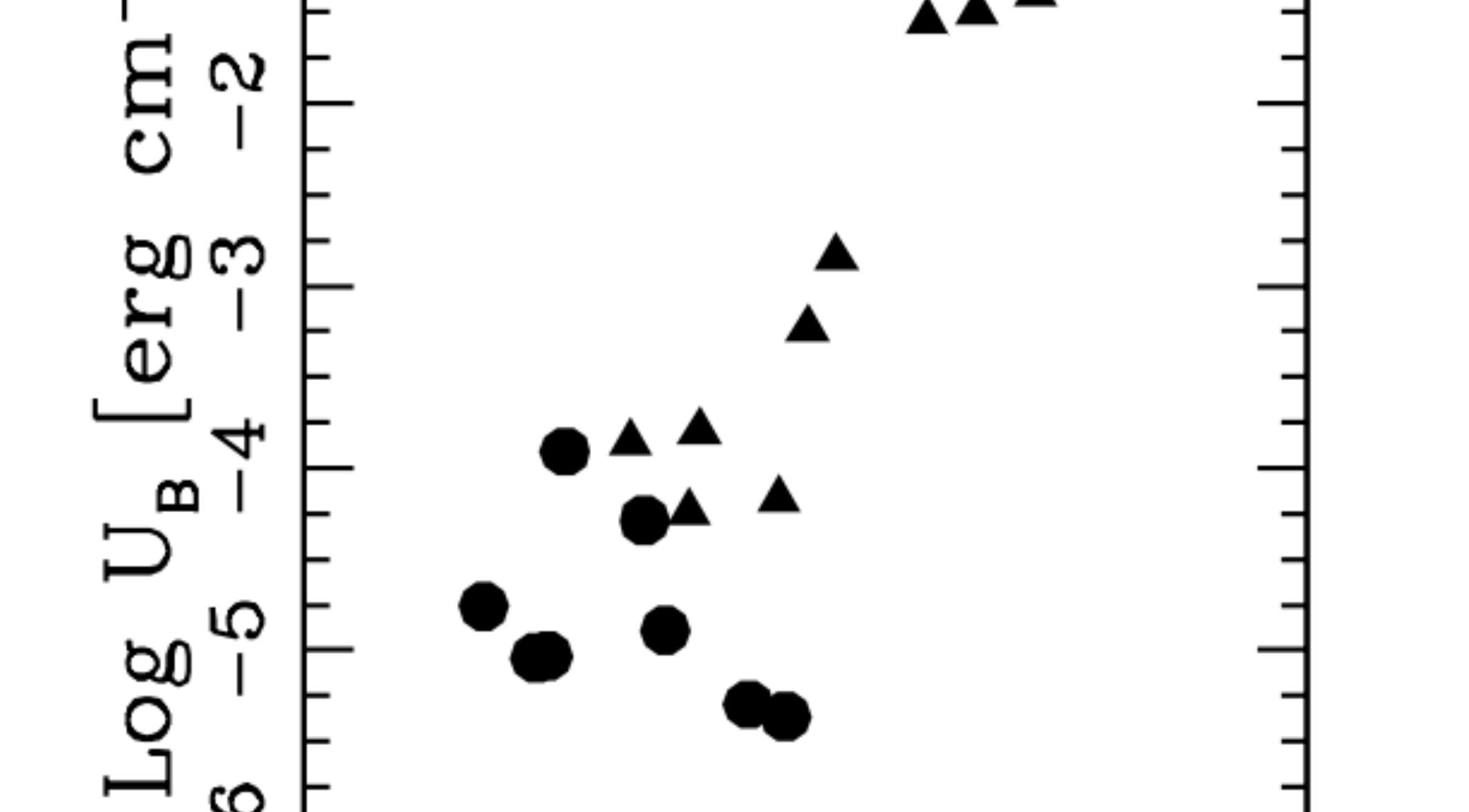}
	\vspace{2cm}
	\caption{\small{Left:  Magnetic energy density and relativistic electron density estimated from SEDs of 45 BL Lac objects \cite{Tavecchio16}.  Right:  Same, for Mrk\,421 (triangles) and Mrk\,501 (circles) in different activity states \cite{Mankuzhiyil12}.}}
\label{BLLacs:fig1}
\end{figure}
Therefore, $\gamma_{\rm{min}}$ plays the major role in estimating $U_{\rm{e}}/U_{\rm{b}}$ ratio of the jet, where  $U_{\rm{b}}$ is the magnetic energy density. 
The present estimates of $U_{\rm{e}}/U_{\rm{b}}$ hint that, the situation is far from particles-field equilibrium (except for a few sources), with electrons dominating over the field by orders of magnitude \cite{Mankuzhiyil12, Tavecchio16, Kino02}, see Fig.~\ref{BLLacs:fig1}
\paragraph*{Importance of \g-ray observations}
The detection of BL Lac objects in \g-rays have been significantly increased after the launch of \fermilat. The third \fermilat  catalog of AGN \cite{3lac} lists $\sim$ 600 BL Lac objects. Due to the relatively lower sensitivities of the current generation Cherenkov Telescopes, and the $\gamma$-$\gamma$ attenuation from the Extragalactic Background Light (which is significant for high redshift sources; z $\gtrsim$ 0.5), the number of BL Lacs detected in the \g-ray VHE range is significantly lower ($\sim$ 50). However, the recent advancement in the simultaneous multi-wavelength campaigns on BL Lacs have significantly improved our understanding on the jet energetics. \\ 

The non-thermal emission parameters of blazars are inferred from the observational quantities, like the peak frequencies (together with their peak luminosities) of the synchrotron and \ic peaks, spectral slopes, flux variability of the source etc. However, the spectral information at the rising part of the synchrotron or \ssc peak is essential to constrain $\gamma_{\rm{min}}$. As a demonstration to the current observational scenario, we show an SED of Mrk 421 \cite{mrk421mwl2011}, Fig.~\ref{mrk421mwl},
averaged over the observations taken during the multi-frequency campaign from 2009 January 19 to 2009 June 1 . The shaded area corresponds to the frequency band to constrain $\gamma_{\rm{min}}$.  Even though the rising part of the synchrotron peak falls at the radio band (in which, a wide coverage of observation is accessible), the emission at this frequency band is self absorbed, and significantly dominated by the emission from the extended region of the jet.  That would in turn make the rising part of the \ssc peak as the unique band to probe $\gamma_{\rm{min}}$. 
\begin{figure}
\vspace{0cm}
\centering
\includegraphics[width=0.6\textwidth]{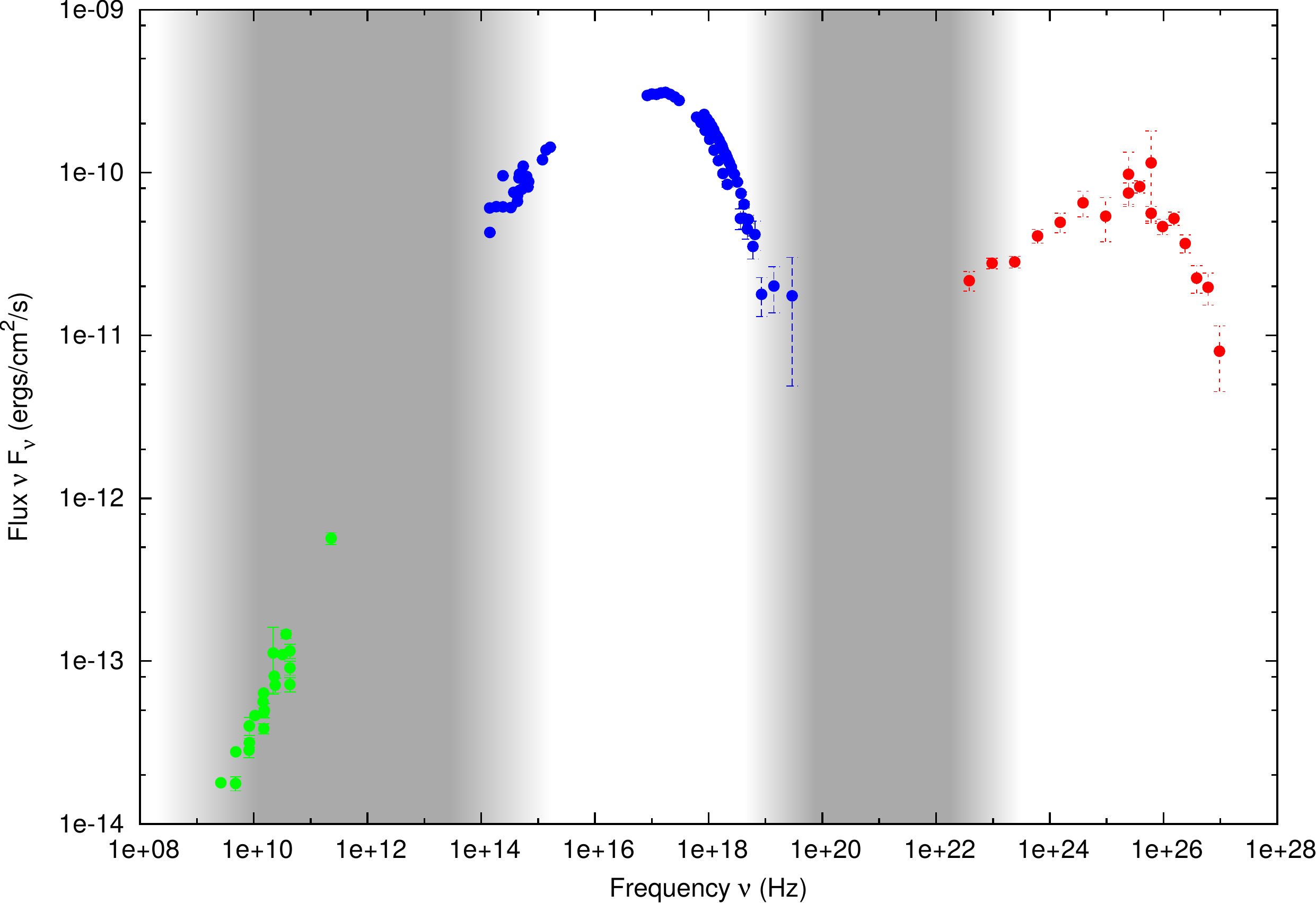}
\caption{\small{SED of Mrk\,421 measured in muti-wavelength campaigns including \Fermi and MAGIC telescopes \cite{mrk421mwl2011}. The green circles corresponds to the radio emission from an extended region of the jet. The blue and red circles denote synchrotron and \ssc emission respectively. The shaded area corresponds to the most appropriate frequency band, in order to  constrain $\gamma_{\rm{min}}$.}}
\label{mrk421mwl}
\end{figure}
\paragraph*{Expected results with e-ASTROGAM}
\begin{figure}
\vspace{3cm}
\includegraphics[trim= 0mm 6cm 0cm 12cm, width=0.8\textwidth,clip=t,angle=0.,scale=0.7]{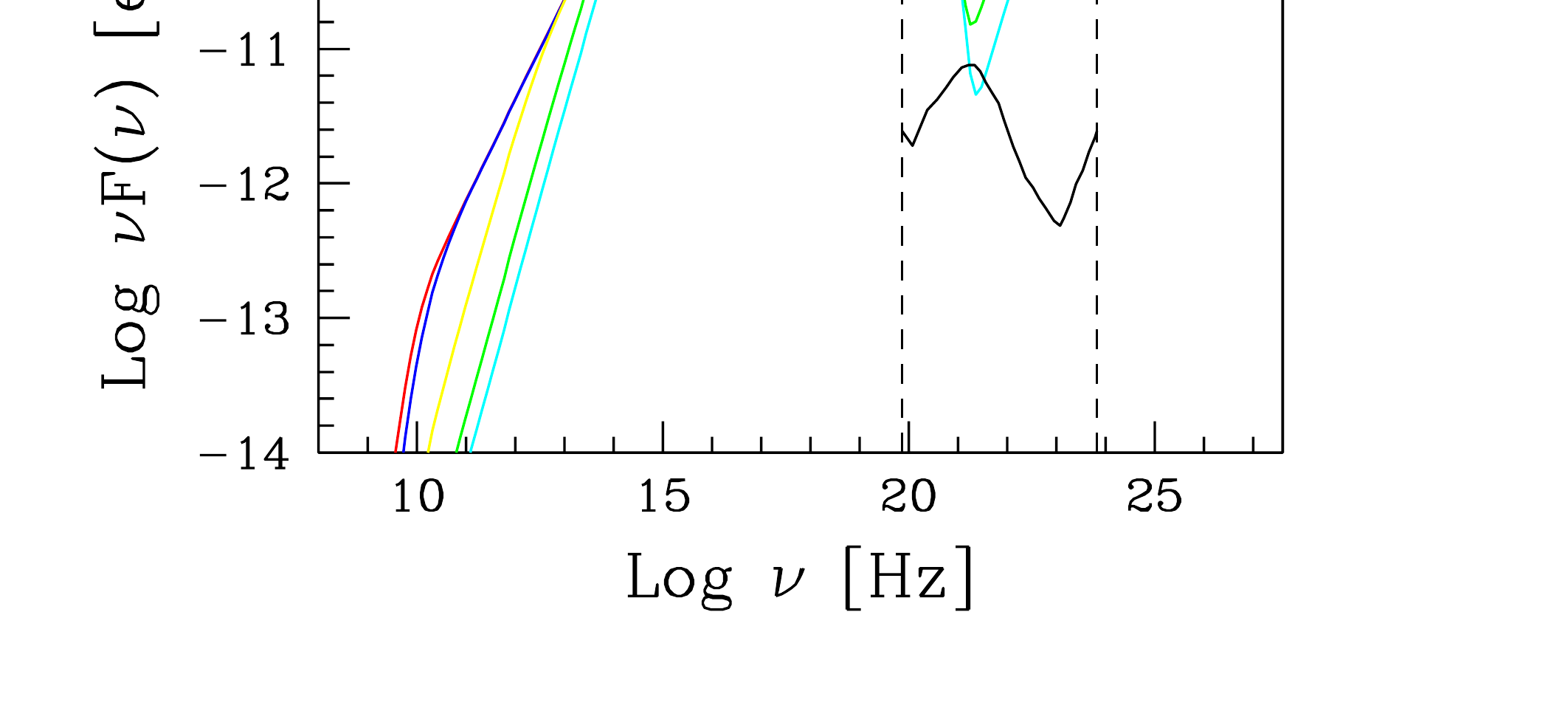}
\hspace{-2cm}
\includegraphics[trim= 0mm 6cm 0cm 10cm, width=0.8\textwidth,clip=t,angle=0.,scale=0.7]{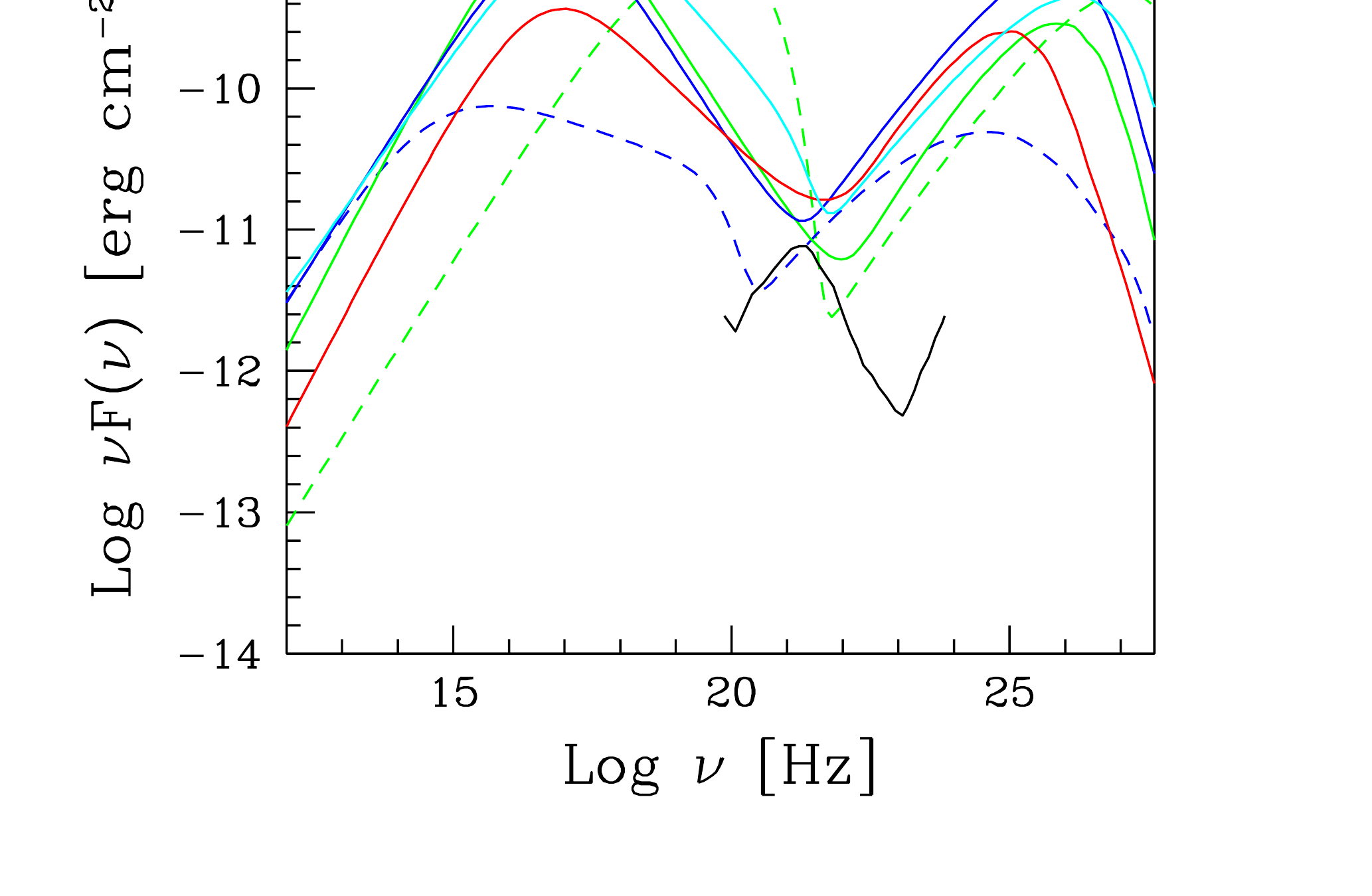}
\caption{\small{Left: \ssc emission models produced by varying  $\gamma_{\rm{min}}$, while keeping other emission parameters as constant. Red ($\gamma_{\rm{min}}$=1), blue ($\gamma_{\rm{min}}$=100), yellow ($\gamma_{\rm{min}}$=1000), green ($\gamma_{\rm{min}}$=5000), and cyan ($\gamma_{\rm{min}}$=10000) model curves show a significant difference at the e-ASTROGAM range (dashed vertical lines), and the radio region (where the emission from the extended region dominate). The black line corresponds to the sensitivity of e-ASTROGAM. Right: \ssc model curves considering the SEDs of Mrk 501 (continuous line \cite{Mankuzhiyil12}) and Mrk 421 (dotted line \cite{Mankuzhiyil11}), which can be detected by e-ASTROGAM. The different colors indicate different activity levels.}}
\label{sensg1}
\end{figure}
We compare the fitted \ssc models for Mrk\,421 \cite{Mankuzhiyil11} and Mrk\,501 \cite{Mankuzhiyil12}  with the sensitivity of e-ASTROGAM in Fig.~\ref{sensg1} (right). The predicted flux of these sources in the e-ASTROGAM range show the feasibility of detection of such sources. 
 In Fig.~\ref{sensg1} (left) we show the behavior of \ssc model curve as a function of $\gamma_{\rm{min}}$. It is very evident that the variation in $\gamma_{\rm{min}}$ is clearly reflected in the frequency band of e-ASTROGAM. Hence,  e-ASTROGAM observations, together with simultaneous multi-wavelength observation of optical to VHE instruments can provide a robust limit to the non-thermal emission parameters, especially $\gamma_{\rm{min}}$.
This would in turn increase the precision of the current $U_{e}/U_{b}$ estimations, in which 
the value of $\gamma_{\rm{min}}$ arbitrarily chosen from $\sim $1 to $10^{4}$ \cite{Mankuzhiyil11, Mankuzhiyil12, Tavecchio10}. 
Hence, the observation of BL Lac objects using e-ASTROGAM will address the energetics of jets, which is one of the most fundamental questions on blazars.
\subsection[On The Origin of the Extragalactic MeV Background\\
\noindent
\textit{\small{M. Ajello, D. Hartmann, M. Ackermann}}]
{On The Origin of the Extragalactic MeV Background}
\paragraph*{Science questions}
The origin of the MeV background, in the $\sim$0.2-100\,MeV gap region, remains a long-standing issue in
astrophysics. The first measurements by the APOLLO 15/16 missions
\cite{trombka77} displayed an intriguing `MeV bump' that was not later
confirmed  by HEAO-4 , SMM and COMPTEL
\cite{kinzer97,watanabe99,weidenspointner00}. These latter missions
characterized the MeV background spectrum as
a power-law extension of the cosmic X-ray background (up to $\sim$3\,MeV)
\cite{ajello08}. Up to this day there is no clear understanding of
which source population, or emission mechanism, may account for the
intensity of the MeV background.
\paragraph*{Importance of \g-ray observations}
 DM annihilation \cite{ahn05}, non-thermal emission from Seyfert
galaxies \cite{inoue08}, nuclear decays 
from Type Ia SNe \cite{clayton75,ruiz16}, and emission from blazars \cite{ajello09} and
radio-galaxies \cite{inoue11} are among the candidates that were considered to explain part or the totality of the MeV background. 
Blazars, radio-galaxies, and type Ia SNe have been detected at
MeV energies and as such their contribution to the MeV background is
guaranteed. On the other hand, the contribution from the putative DM interaction or the non-thermal emission of Seyfert galaxies is
less secure. The latter is however worth of attention because by invoking the
presence of non-thermal electrons in AGN coronae, it makes
radio-quiet AGN a population able to account for both the X-ray and
MeV backgrounds, justifying at the same time the power-law shape of the
low-energy part of the  MeV
background. However, the $<3$\,MeV part of the MeV background spectrum can
be accounted for by the emission of extremely powerful blazars, which
are easily detected in the hard X-ray range and display very hard
power-law spectra \cite{ajello09}. The most interesting aspect is that
in order to connect the X-ray and the \g-ray (i.e. GeV)
background, the spectrum of the MeV background must harden at around
40-60\,MeV (see Fig.~\ref{fig:mev}). This
implies that either at least two source classes are major contributors
to the MeV background or that another source class that exhibits a
spectral bump needs to be considered. Star-forming galaxies, whose MeV
to GeV emission is powered by CRs, may be this additional
population \cite{lacki14}.
\paragraph*{Expected results with e-ASTROGAM}
e-ASTROGAM will provide a new, accurate, measurement of the MeV
background at $>300$\,keV and up to a {\it few} GeV providing good
overlap with the X-ray and the \g-ray backgrounds. At the same time
e-ASTROGAM will detect thousands of sources providing direct
insight into which populations can explain the MeV background. The
measurements of luminosity functions (for example, for populations of
blazars, star-forming and radio galaxies) will provide direct
prediction of the contributions of those source classes to the background.

The measurement of the MeV background will require careful modeling of
the Galactic diffuse emission and of the instrumental background. The
former can be achieved using predictions of Galactic CR
propagation models \cite{strong98} tuned to fit the e-ASTROGAM data, while the latter
will require detailed Monte Carlo simulations and an event selection
that minimizes non celestial signal. 

Thanks to its excellent point-source detection sensitivity, e-ASTROGAM
will detect hundreds of sources. Spectroscopic campaigns will be
needed to determine their redshift and ultimately their luminosity
function. For the unresolved component of the MeV background, both a
stacking analysis and the analysis of the angular fluctuations \cite{inoue13} of the
background will be able to provide further insight into its origin.

\begin{figure}[ht!]
\centering
  	 \includegraphics[scale=0.6]{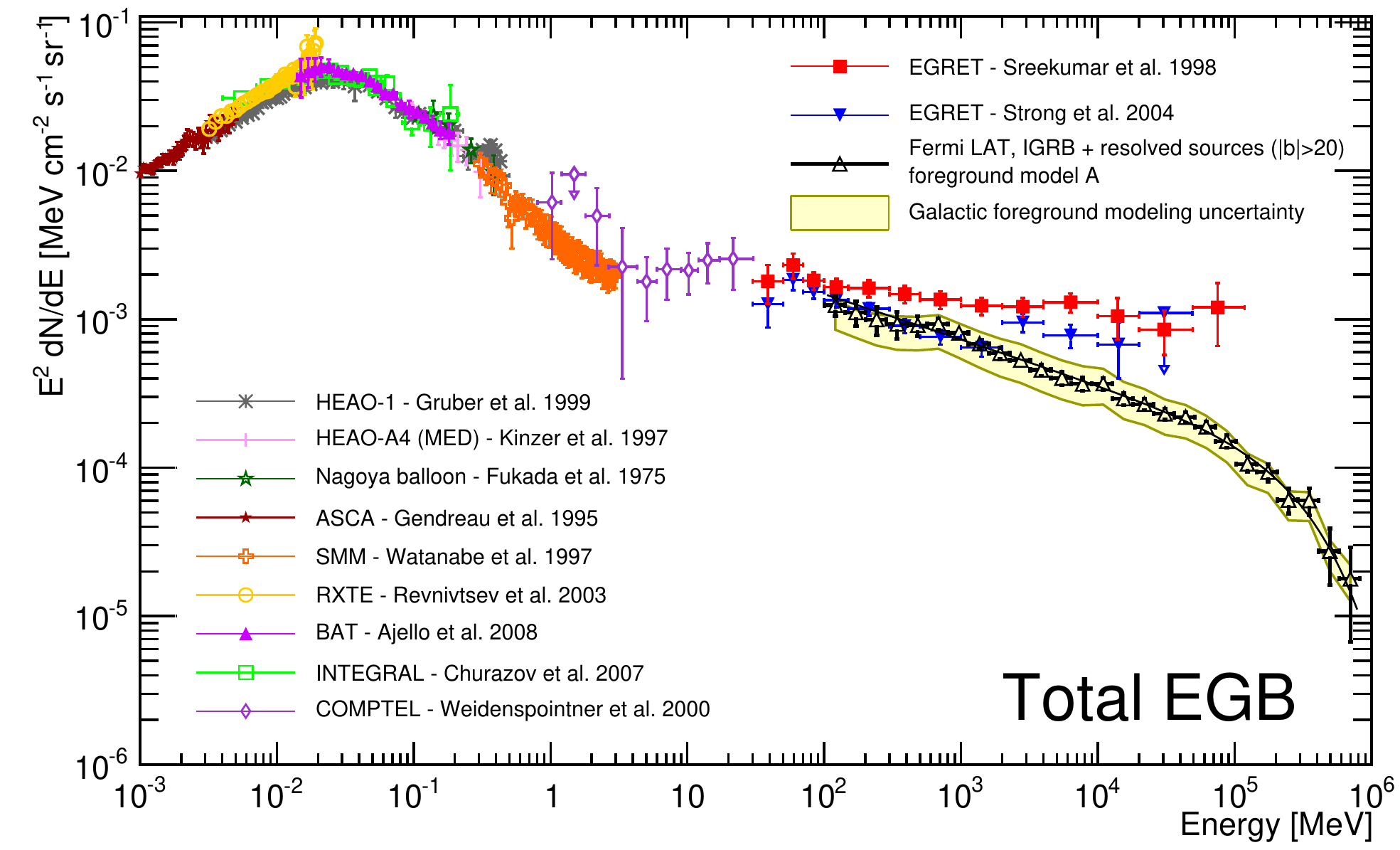}
  \caption{\small{Spectrum of the high-energy background from X-ray to TeV
    \g-rays. Adapted from \cite{ackermann15:799}.}}
\label{fig:mev}
\end{figure}
\subsection[Observations of Galaxy Clusters\\
\noindent
\textit{\small{Y. Rephaeli, S. Zimmer}}]{Observations of Galaxy Clusters} 
\paragraph*{Science questions}
Clusters of galaxies are important intrinsically, in the study of growth and evolution of the large scale structures in the Universe, and for understanding phenomena on intergalactic and cosmological scales. As in galaxies, non-thermal processes in clusters can provide essential insight on the origin and properties of energetic particles and magnetic fields. 
Evidence for relativistic electrons in intracluster space has so far been limited to measurements of extended regions of radio emission, `halos' and relics, which have already been observed in many galaxy clusters \cite[e.g.,][]{Ferrari:2008}. The radiative yield of energetic radio-emitting electrons is expected to extend to the X- and \g-ray region by Compton scattering off the CMB; searches for non-thermal X-ray emission below 100 keV were inconclusive \cite[e.g.,][]{Rephaeli:2008}. Analysis of \fermilat measurements of 50 clusters resulted in an upper limit on the mean emission above 
$\sim500$ MeV \cite{Ackermann:2013}. Improved analysis of an expanded LAT dataset (extending to $\sim100\,\mathrm{MeV}$ on the nearby Coma cluster) also resulted in an upper bound \cite{Ackermann:2016}, which yields a lower bound on the mean strength of the magnetic field. 
Quantitative description of energetic electrons and protons in clusters requires modeling of the spectro-spatial distribution of their sources, propagation mode,  and energy losses by all relevant processes. Considerations of typical energy loss times indicate that primary electrons diffusing out of sources in the cluster inner core cannot account for the observed, relatively large size ($\sim 1$ Mpc) of radio `halos', leading to the realization that energetic protons may play a major role in accounting for the radio emission through their yields of secondary electrons, produced in charged pion decays following interactions of the energetic protons with protons in the gas. Moreover, energetic electrons and protons could be turbulently re-accelerated \cite[e.g.,][]{Brunetti:2014}. These key considerations provide strong motivation to continue the search for cluster hard X and \g-ray emission. Detection of significant emission in these bands, above that from the cluster galaxies, can potentially yield first direct quantitative information on energetic electrons and protons in intergalactic space, thereby advancing our knowledge on their origin. \\The motivation to search for cluster X-and-\g-ray emission is further enhanced by the possibility that particle Dark Matter candidates may have decay and annihilation channels in these spectral regions \cite[e.g.][]{Conrad:2015}. Rich nearby clusters would clearly be prime targets for observation of such Dark Matter signatures. Here we consider only the non-thermal origin of cluster high-energy emission which is directly related to cluster radio `halos'.
For specific estimates of the feasibility of detecting cluster \g-ray emission with e-ASTROGAM, we select the rich nearby Coma cluster, whose radio `halo' has been well mapped spectrally, and partly also spatially. The presence of two powerful radio galaxies in the Coma core, in addition many star-forming galaxies (SFGs) distributed across the cluster, imply that there could be a significant distribution of energetic electrons and protons that diffuse out of these galaxies. As such, Coma has been the target of many X- and \g-ray satellites and a planned target of the upcoming Cherenkov Telescope Array\cite{2017arXiv170907997C}.
\paragraph*{Importance of \g-ray observations}
Since \g-ray emission from galaxies in the Coma cluster cannot be adequately resolved, the total emission from all relevant galaxies has to be properly accounted for in order to reliably determine emission that originates outside the cluster galaxies. These galaxies are also sources of energetic electrons and protons, so their numbers and spatial distribution need to be specified when modeling the total cluster \cite{Rephaeli:2016} in 
a treatment based on a diffusion model for energetic electrons and its predicted emission \cite{Rephaeli:2016}. Both electrons and protons originate in SFGs, whose spatial distribution is known to be very extended, whereas it was (conservatively) assumed that only electrons diffuse out of the dominant two central radio galaxies, with escape rates scaled to the reasonably well determined Galactic rates. Coulomb, Compton, and synchrotron processes couple the particles to the magnetized gas, with the gas density and magnetic field described by realistic spatial profiles. 

Predicted spectra and spatial profiles of radio emission from primary and secondary electrons in the above treatment were found to be roughly consistent with current `halo' measurements within the central $\sim 0.5$ Mpc radial region, but the emission level was significantly 
lower outside this region. Non-thermal X-ray emission is largely by Compton scattering of electrons from radio galaxies off the CMB, whereas \g-ray emission is primarily from the decay of neutral pions produced in interactions of protons from SFGs with protons in the 
gas. Whereas a Compton spectral component has a simple power-law form, the 
$\pi^{0}$ decay component has the characteristic bell-shape form around the peak at 
$\sim 70$ MeV, whose presence is diagnostically important in determining the nature 
of the dominant emission process at O($100$) MeV. Since this estimate of \g-ray emission is based on energetic particle populations deduced directly from radio emission from SFGs and the two dominant radio galaxies, and the fact that the predicted level of radio `halo' emission falls below the observed level, it constitutes a conservative minimal total emission. 
If energetic particles are efficiently re-accelerated during the few Gyr cluster merger 
phase, then their distribution would be boosted \cite[e.g.][]{Brunetti:2014} beyond the 
levels predicted in the diffusion model discussed above \cite{Rephaeli:2016}. Theoretical treatment of the re-acceleration process, and the spectral features of the particle distribution obviously depend strongly on various dynamical and gas parameters. In particular, two additional degrees of freedom are the typical re-acceleration time, and the duration of the 
re-acceleration period.
With nearly a decade of observations, \fermilat has provided important insight on cluster \g-ray emission \cite[e.g.][]{Ackermann:2013,
Ackermann:2016, Ackermann:2010}. As of yet there has not been a statistically significant detection of 
extended cluster emission, neither by the LAT, nor by imaging Cherenkov telescopes, implying that the ratio energetic particle pressure to thermal gas pressure is lower than $\sim1\%$ \cite{Ackermann:2013}. Recent work indicates that current observations of the Coma cluster \cite{Ackermann:2016} are sufficiently deep to probe a meaningful part of the parameter space of the main viable models \cite{Brunetti:2017}, and if \Fermi observations continue, the expanding database will likely lead to a detection of Coma at the  $3-5\sigma$ significance level. This will have important ramifications for essentially all currently viable models for \g-ray emission.
\paragraph*{Expected results with e-ASTROGAM}
\begin{figure}[t]
\begin{center}
\includegraphics[width=0.65\textwidth]{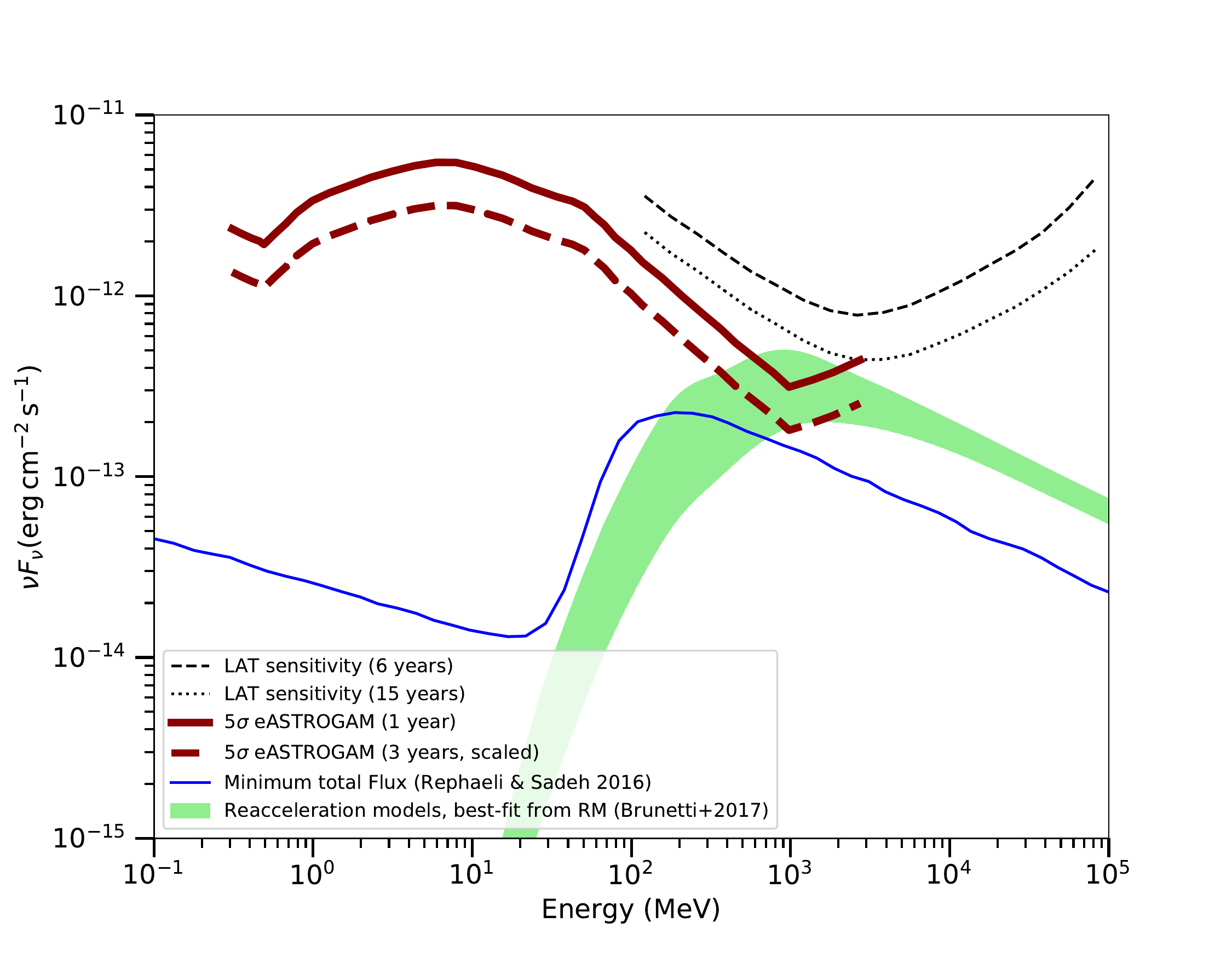}
\caption{\small{Predicted \g-ray spectra for viable models of
energetic electrons and protons in clusters, based on various cluster dynamical, intracluster
gas, and magnetic field properties. The blue curve shows the minimal level of the total
emission predicted in the model based on the assumption that particles originate in
star-forming galaxies and the two dominant radio galaxies in the cluster core
(without re-acceleration) \cite{Rephaeli:2016}. Predicted emission in turbulent
re-acceleration models assuming various values of the acceleration time and duration of the acceleration period \cite{Brunetti:2017} are shown by the green-colored region. Black lines correspond to the sensitivity of \fermilat (solid for 6 years, dashed for 15 years, respectively). The bold red curves are the same as in 
Chapter~\ref{intro}, for 1 year (solid) and 3 years (dashed). The figure is based in part on a similar figure in \cite{Brunetti:2017}.}}
\label{Cl_galaxies:fig}
\end{center} 
\end{figure}
We summarize the current observational status and the expected capability of 
e-ASTROGAM 
in Fig.~\ref{Cl_galaxies:fig}. The projected 
sensitivity of the LAT for 15 year observation time is based on the published 6 year 
likelihood analysis. These sensitivity curves require a value of the likelihood test statistic above 25 (with at least 3 photons attributed to cluster emission), and with Galactic 
foreground emission serving as proxy for the background estimate.\footnote{The 
LAT sensitivity curves were computed assuming the same analysis choices (binning, 
energy range) as in \cite{Ackermann:2016}.} Estimated e-ASTROGAM sensitivity 
curves 
are based on observation periods of 1 and 3 years, respectively.
The estimated minimal emission level \cite{Rephaeli:2016} includes also the emission 
from SFGs and the two dominant radio galaxies; it consists largely of a 
Compton component below $\sim 20$ MeV, and $\pi^{0}$ decay yield at higher energies. Enhanced particle distribution that is implied by the observed radio `halo' emission would result in a higher level of extended emission, exemplified here by predicted spectra of re-acceleration models \cite{Brunetti:2017}. These models do not include the lower energy Compton component, nor emission from the cluster galaxies; the width of the region reflects assumed parameter ranges, including the central value of the magnetic field deduced \cite{Bonafede:2010} from Faraday Rotation (FR) measurements, $B_{0}= 4.7\, \mu$G. The predicted range would be higher if a lower value is assumed for $B_{0}$, not an unlikely possibility given the substantial uncertainty in the analysis of cluster FR measurements. As evident from this figure, 1-year observations with e-ASTROGAM will already allow probing a number of models with magnetic fields that are within a factor of $\sim 2$ from the above value of $B_{0}$, with substantially improved diagnostic power expected over the lifetime of 
the e-ASTROGAM mission. 

\newpage
\section[Cosmic-ray interactions\\
\noindent
\textnormal{\small\textnormal{Convenors:} \textit{I. A. Grenier, A. Bykov}}]{Cosmic-ray interactions}\label{chap:cosmic_ray}
A clear understanding of the origin and evolution of CRs is still missing despite one century of impressive observational discoveries and theoretical progress. Understanding their origin is an interdisciplinary problem involving fundamental plasma physics, to describe the diffusive shock acceleration process, as well as astrophysical and particle-physics diagnostics, to characterize the particle properties and the local conditions in the acceleration zones. While we still lack a reliable explanation for the existence of CRs near and beyond PeV energies in the Milky Way and beyond EeV energies in the extragalactic space, we also hardly know the Galactic population of low-energy CRs, with energies below a few GeV nucleon$^{-1}$. We still need information on their sources and injection spectra into the interstellar medium, on their transport properties and flux distribution at all interstellar scales in the Galaxy, and on their impact on the overall evolution of the interstellar medium and on the dynamics of Galactic outflows and winds. The performance of the e-ASTROGAM mission would provide unique results in a number of important CR issues. 

Sensitive observations of a set of CR sources, as young SNRs, across the bandwidth of e-ASTROGAM would allow for the first time to distinguish the emission produced by the interactions of CR nuclei with the ambient gas and the non-thermal emission from CR electrons. Combined with high-resolution radio and X-ray observations of the remnants, the e-ASTROGAM data would provide information on CR injection into the acceleration process, on the structure of magnetic fields inside the remnants, and on the spectrum of CRs freshly released into surrounding clouds.

\fermilat could resolve only one case of CR activity in a Galactic superbubble to study the collective effects of multiple supernovae and powerful winds of young massive stars. The improved angular resolution of e-ASTROGAM would provide more study cases, individually as well as collectively in the inner Galaxy, which would help to probe the interplay between CRs and the turbulent medium of star-forming regions during the early steps of their Galactic voyage. Individual massive binary stars like $\eta$ Carinae, which is the most luminous massive binary system in the Galaxy and the likely progenitor of the next Galactic supernova, are promising candidates to study particle acceleration by their powerful winds. Following their time variability from radio to e-ASTROGAM energies can provide key diagnostics on the acceleration efficiency.

CR nuclei of energies below a few GeV nucleon$^{-1}$ contain the bulk energy density of the Galactic CRs. They are the main source of ionization and heating in the highly obscured star-forming clouds that are well screened from UV radiation. At the same time they are the source of free energy and pressure gradients to support large-scale magnetohydrodynamic (MHD) outflows and Galactic winds that control the overall evolution of a galaxy. e-ASTROGAM observations of the inner Galaxy would provide the first nuclear spectroscopic data on the LECR population. The energy coverage of the telescope would also allow a precise separation of the CR nuclei and electron/positron populations (and spectra) across the Galaxy. The higher-resolution images would shed light on the degree of correlation between the CR distributions and stellar activity, at the scale of cloud complexes up to that of spiral arms, in order to better constrain the diffusion properties of CRs in a galaxy. 

Last, but not least, maps of the total interstellar gas mass inferred from CRs and the GeV data from e-ASTROGAM at a resolution of 9' would serve a broad interstellar community to improve the calibration of gas tracers (radio and dust tracers) in a large variety of cloud states.

\subsection[Cosmic rays and Supernova Remnants at MeV energies\\
\noindent
\textit{\small{M. Cardillo, M. Pohl, S. Kaufmann, O. Tibolla}}]{Cosmic rays and Supernova Remnants at MeV energies}
\paragraph*{Science questions}
CRs are VHE particles (mainly protons and He nuclei) with an energy spectrum extended up to $E\sim10^{20}$ eV and a Galactic component likely accelerated at the shocks of Supernova Remnants (SNRs) \cite{ginzburg61}, persistent sources of non-thermal radiation that can be resolved in nearly all wavebands \cite{2008ARA&A..46...89R}. 
There is evidence of hadronic CRs in middle-aged SNRs, based on their \g-ray emission spectra; they are characterized by the ``pion bump''. This is a typical feature of hadronic \g-ray spectra that reflects the kinematic impact of the rest mass of the neutral pion generating the \g-ray photons and permits distinguishing pion-decay emission from electron \brem or \ic radiation. It is unclear, however, what fraction of these particles is freshly accelerated and not re-accelerated.
Studying directly accelerated particles is fundamental for finding the sources of CRs, and SNRs are ideal systems to observe on account of their persistence and resolvability. The insights on the micro-physics of particle acceleration can be extrapolated to other outflow systems, where the process operates as well, but in which observations as detailed as those of SNRs are not possible.
Consequently, here we highlight some aspects of inquiry with e-ASTROGAM:\\[1ex]
1. Direct proof of the presence of freshly accelerated (and not re-accelerated) CRs at SNR shocks through the detection of the ``pion bump'' in young sources.\\[0.3ex]
2. Search for non-thermal \brem from energetic electrons for correlating with radio synchrotron emission and determining environmental parameters such as the level of magnetic-field amplification driven by CRs.\\[0.3ex]
3. Search for nuclear de-excitation lines to infer the elemental composition of CRs at their acceleration site and to determine the SNR environment that is most conducive to particle acceleration.\\[0.3ex]
4. Measure the extent of re-acceleration of Galactic CRs at the shock fronts of SNRs and its impact on the elemental composition of CRs on Earth.
\paragraph*{Importance of \g-ray observations}
{AGILE and \fermilat detected for the first time \g-ray emission below $E\sim200$ MeV from two very bright SNRs, W44 and IC 443 \cite{giuliani11_W44,ackermann13_W44,cardillo14_W44}, later followed by another middle-aged SNR, W51C \cite{jogler16}. The measurement of the specific shape of the hadronic \g-ray spectrum, the so-called ``pion bump'', was claimed to be direct proof of the acceleration of CR nuclei at the shock of SNRs. The issue is not so simple though. First of all, the CR spectra needed to reproduce the hadronic \g-ray spectra from these remnants are far from those than one would expect on theoretical grounds \cite{2011NatCo...2E.194M}, and active research targets the relation between the instantaneous particle spectra and that of CRs released over the entire lifetime of the remnant \cite{2017arXiv170702744H}. Then, the two SNRs with confirmed ``pion bump'' are middle-aged ($t_{age}>10^{4}$ yrs) and consequently have slow shocks with $v_{sh}\sim100$ km/s \cite{reach00_W44}. The CR acceleration efficiency is strictly correlated with the shock velocity and should be low at shocks that slow. It may be that for older SNRs re-acceleration of pre-existing Galactic CR dominates over acceleration of low-energy particles \cite{uchiyama10_W44,lee15_W44,cardillo16_W44} . 
As the composition of Galactic CRs includes elements that are not abundant in the interstellar medium, a significant re-acceleration of CRs would modify the required source abundances and would have strong impact on our understanding of the propagation history of CRs in the Galaxy. We need spatially resolved studies of older SNRs with slow shocks to infer the role of CR re-acceleration in late phases of SNR evolution, which require an excellent angular resolution and high continuum sensitivity.
Simulations suggest that re-acceleration of Galactic CRs is at most a secondary process in young SNRs \cite{2012APh....35..300T,2013A&A...552A.102T,2015A&A...574A..43P}, and primary particles would dominate. The detection below 200 MeV of \brem emission from primary electrons would offer invaluable insights. Measuring the low-frequency radio synchrotron emission of electrons of similar energy, with e.g. LOFAR or SKA, provides a direct measure of the strength of the turbulently amplified magnetic field \cite{1980MNRAS.191..855C}, arguably the most critical ingredient in particle acceleration theory \cite{1987PhR...154....1B}. This measurement would also remove the degeneracy in the interpretation of TeV-band \g-ray emission. At the same time, we could measure the electron/ion ratio in CRs at the source, which would significantly advance our understanding of the injection processes into diffusive shock acceleration. It is evident that an improved low-energy sensitivity would also be very useful for the study of young SNRs.
Several de-excitation lines will be visible in e-ASTROGAM energy range. Supernovae often expand into enriched material provided by the progenitor wind or nearby earlier supernovae. Heavy elements among the accelerated particles and in the ambient medium will collide and eventually radiate nuclear de-excitation lines that are characteristic of the element, thus allowing abundance tomography. 
From the quasi-spontaneously de-excitation, unique features arise due to C and O lines in the 4-6 MeV band, while the lines induced by the Ne-Fe group will dominate in the 1-3 MeV band. For the historic SNR Cas~A we can estimate the line flux in the 4.4~MeV line from $^{12}$C and use that as a proxy for all the other lines. This particular supernova expands into the wind zone of a red supergiant that is not rich in heavy elements, and so it is the particle acceleration at the reverse shock running into the ejecta of the supernova explosion that provides a detectable line flux on the order $10^{-10}\ \mathrm{erg\,cm^{-2}\,s^{-1}}$, which is well above the sensitivity limit of e-ASTROGAM.
Combined with isotopic abundance measurements performed near Earth, in particular those of unstable isotopes such as $^{60}$Fe \cite{2016Sci...352..677B}, line observations provide direct insight into the environment in which CRs are accelerated. 
A high energy resolution is needed for studies of nuclear de-excitation lines.
\paragraph*{Expected results with e-ASTROGAM}
It is easy to understand the importance of an instrument like e-ASTROGAM. Its sensitivity in the range $0.3-100$ MeV will be one to two orders of magnitude better than that of previous instruments. 
As shown in Fig.~\ref{lines_casa}, e-ASTROGAM  should detect many SNRs within one year of operation.  The angular resolution offered by e-ASTROGAM is unprecedented, reaching 0.15$^\circ$ at 1 GeV, as highlighted in Chapter \ref{intro}, which will be decisive for resolving sources and avoiding source confusion in the Galactic plane.
\mpo{The expected results with e-ASTROGAM include:\\[1ex]
1. Observation of \g-ray emission below $200$ MeV from known young SNRs, like Cas A or Tycho, and from yet undetected young SNRs, which is expected to come from freshly accelerated CRs on account of the high shock speed in these sources.\\
2. Measurement of electron \brem below $100$~MeV from a number of SNRs. For a magnetic-field strength of $250\ \mu$G, we expect a \brem flux from Cas A of $E\times F(E)\simeq 10^{-11} \ \mathrm{erg\,cm^{-2}\,s^{-1}}$, which is more than twice the one-year sensitivity level of e-ASTROGAM. This measurement would be decisive in determining the magnetic-field strength and in the separation of leptonic and hadronic contributions to the \g-ray emission.\\[0.3ex] 
3. Detection of nuclear de-excitation lines from a number of SNRs. The C and O lines from Cas A should stand out clearly and would likely constitute one of the early breakthrough results with e-ASTROGAM. Moreover, we will be able to measure element abundances by studying line ratios. In fact, we may derive the spallation rate of heavy nuclei measuring their impact on the abundance of lighter elements and providing a new estimate of their primordial abundances.\\[0.3ex]
4. Distinction of the \g-ray emission from the remnant from that of nearby molecular clouds that are illuminated with freshly accelerated CRs. For older SNRs such as W44 or IC443, this measurement will permit the study of re-acceleration of existing CRs which, if efficient, would change our understanding of CR physics and would also have an impact on indirect searches for DM using CR annihilation products.}
\begin{figure}
 \centering
 {\includegraphics[scale=0.56]{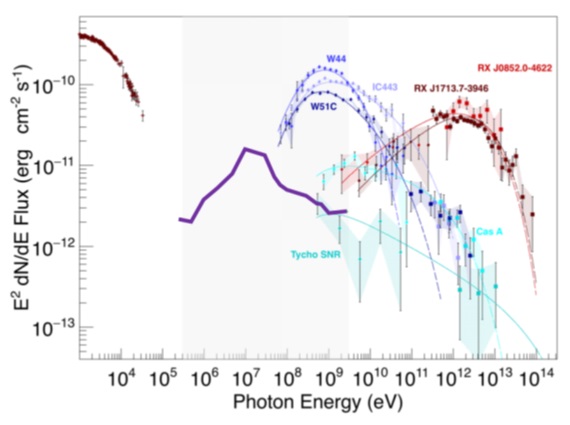}}
\caption{e-ASTROGAM sensitivity for 1-year exposure (thick purple line) compared to typical \g-ray energy spectra for several SNRs; young SNRs ($<$1000
yrs) are shown in green.} 
 \label{lines_casa}
\end{figure}
\subsection[Cosmic-ray acceleration in stellar wind collisions\\
\noindent
\textit{\small{R. Walter, M. Balbo}}]
{Cosmic-ray acceleration in stellar wind collisions}
\paragraph*{Science questions}
Diffusion of Galactic CRs leads to particle energy densities dominating the pressure in the central regions of galaxies. This pressure might be sufficient to generate Galactic winds and central outflows \cite{2016ApJ...824L..30P,2016ApJ...827L..29S}. These Galactic CRs are likely produced through Fermi acceleration processed in SNR shocks and in other exotic sources. Identifying the different contributors to CR acceleration in  galaxies is fundamental to understand Galactic processes, how Fermi acceleration works in various environments, and the feed-back between CR acceleration, Galactic magnetic fields and the dynamics of the interstellar medium. Gamma-ray observation are particularly enlightening as they are the main signature of particle acceleration, free of the pollution from thermal processes. Variable sources are interesting targets to study particle shock acceleration as the correlated observations in various energy bands provide key signatures of the physical processes at play and allow to understand how particle acceleration takes place and the luminosity of the source in the different particle species.
\paragraph*{Importance of \g-ray observations}
Particle acceleration in stellar wind collisions can be particularly well studied in $\eta$ Carinae, the most luminous massive colliding wind binary system of our Galaxy and the first one to have been detected at VHE without hosting a compact object. The relative separation of the two stars varies by a factor $\sim20$, reaching its minimum at periastron, when the two objects pass within a few AU of each other (the radius of the primary star is estimated as 0.5 AU). In these extreme conditions their supersonic winds interact forming a colliding wind region of hot shocked gas where charged particles can be accelerated via diffusive shock acceleration up to high energies. 
\cite{2017arXiv170502706B} shows the analysis of the \fermilat data of $\eta$ Carinae (Fig.~\ref{fig:simul}).
The low-energy and high-energy \g-ray light curves probe radiation from electrons and ions, respectively. 
The 0.3-10 GeV flux varied similarly for the two periastrons and, when combined, a higher resolution light curve could be obtained. Instead the 10-300 GeV flux peaked during the 2009 periastron, decreased slightly towards apastron and did not increase again during the 2014 periastron.  A calculation has been done for the maximum energies that could be reached by electrons and ions in every cells of the hydrodynamic simulations presented by \cite{2011ApJ...726..105P}, assuming a dipolar magnetic field at the surface of the primary star. As expected, most of the shock power is released on both sides of the wind collision zone and in the cells downstream of the wind-collision region \cite{2006ApJ...644.1118R}. The photon-photon opacity could also be estimated as $<10^{-2}$, excluding a significant effect on the observed GeV spectrum. Fig.~\ref{fig:simul} shows the X and \g-ray light curves predicted by the simulations for a magnetic field of 500 G and assuming that 1.5\% and 2.4\% of the mechanical energy is used to respectively accelerate electrons and protons. To ease the comparison between observations and simulations, the results of the latter were binned in the same way as the data.
\begin{figure}[]
\includegraphics[width=0.45\textwidth]{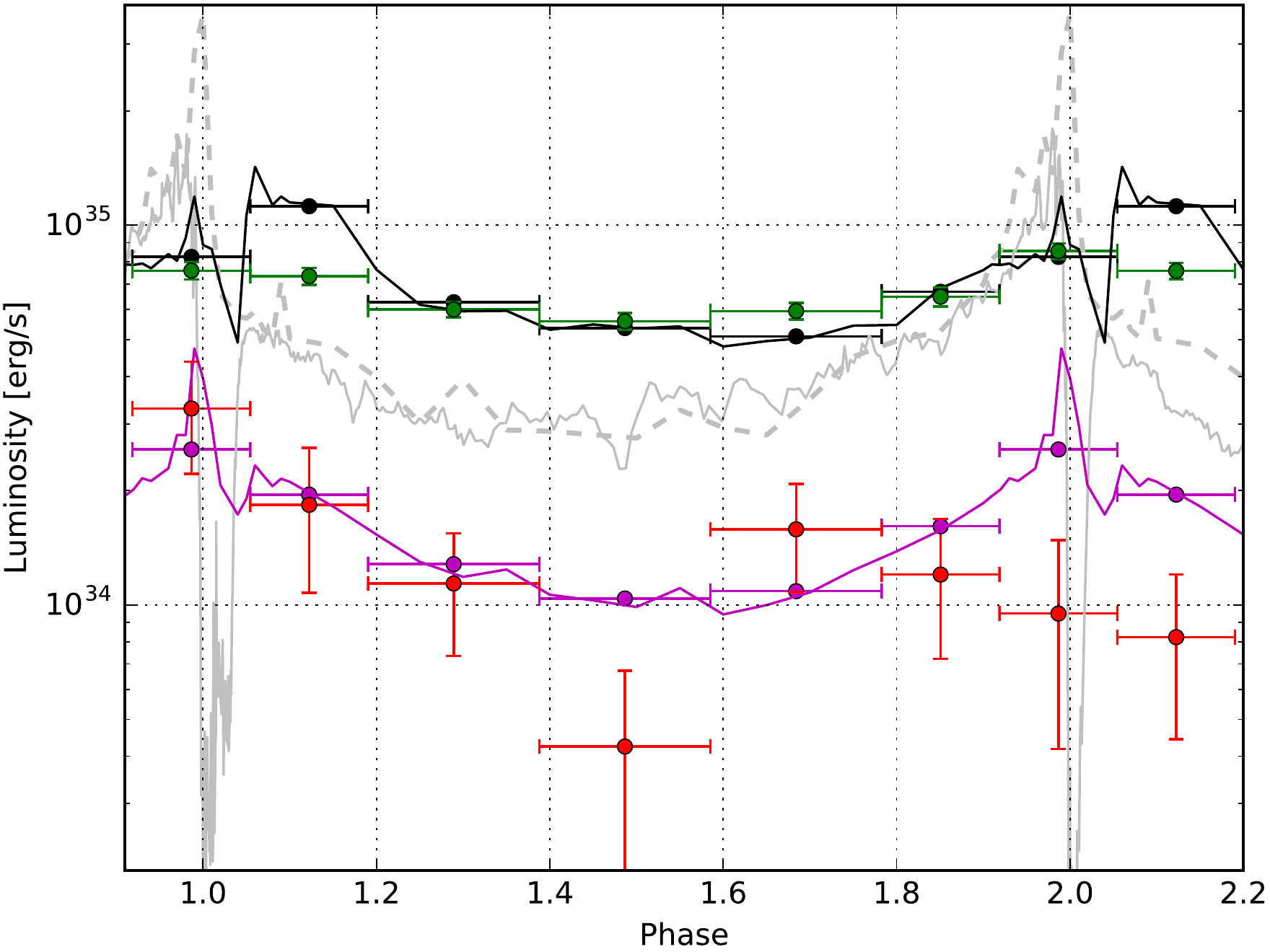}\hfill
\includegraphics[width=0.45\textwidth]{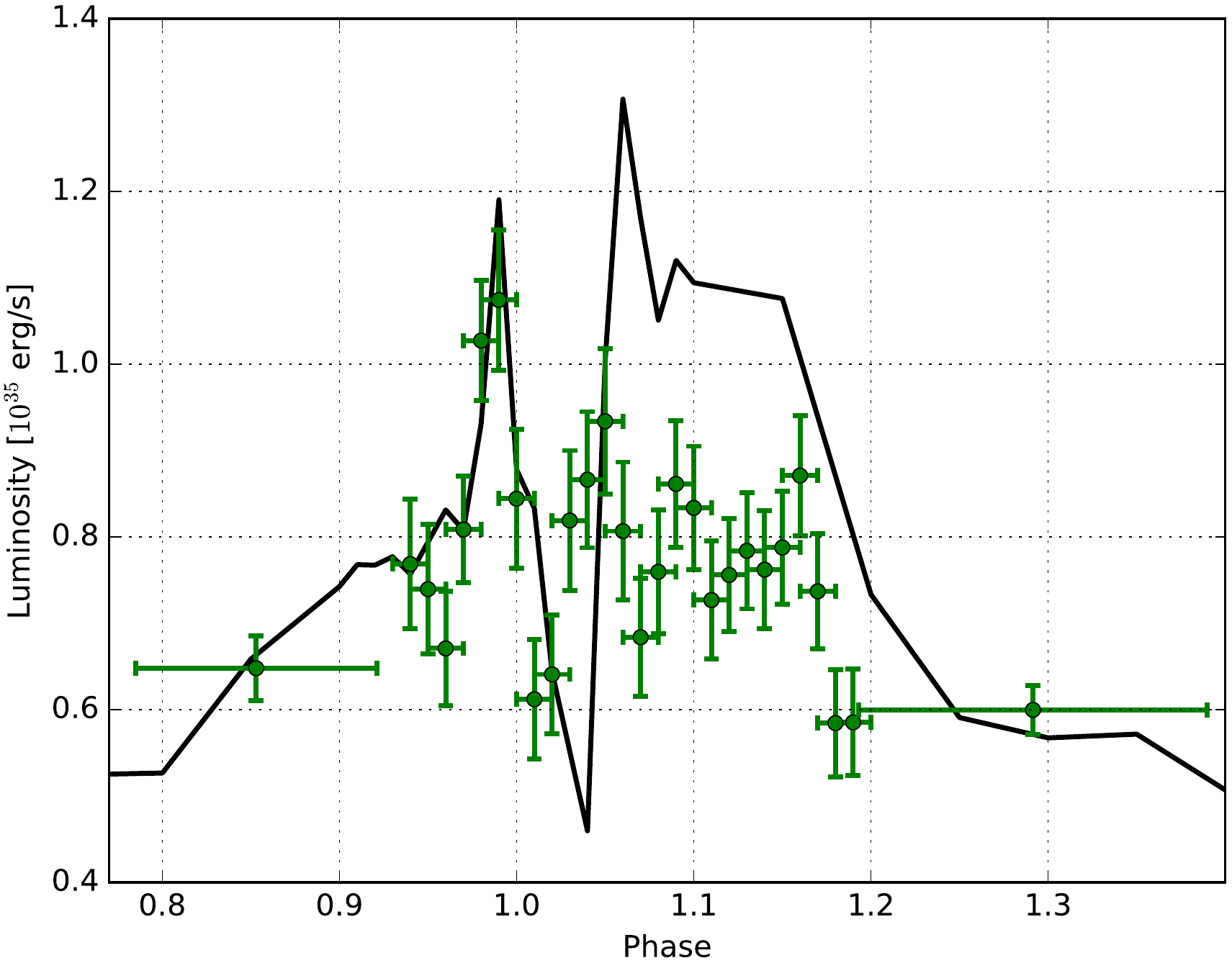}
\caption{\small Left: Simulated and observed X-ray and \g-ray light curves of $\eta$ Carinae through two periastron passages. The green and red points show the observed \fermilat light curves at low (0.3-10 GeV) and high (10-300 GeV) energies. The dim grey light curves show the observed (continuous) and predicted (dash, without obscuration) thermal X-ray light curves. The black and purple lines and bins show the predicted \ic and neutral pion decay light curves. Right: A merged \fermilat analysis (0.3-10 GeV) of the two periastrons for narrower time bins.}\label{fig:simul}
\end{figure}
Electron cooling, through \ic scattering, is very efficient and the corresponding \g-rays are expected to peak just before periastron. Both the observed (0.3-10 GeV) LAT light curve and the predicted \ic emission show a broad peak extending on both sides of periastron, as expected from the evolving shock geometry. A secondary \ic peak could be expected above phase 1.05 although its spectral shape could be very different as the UV seed thermal photons will have lower density when compared to the location of the primary shock close to the center of the system. The combined light curve is very similar to the prediction of the simulation for the \ic luminosity (Fig.~\ref{fig:simul}, right). The only notable exception is that the observed second broad peak is slightly shifted towards earlier phases and has a lower luminosity when compared to the simulation. 
This could be related to the assumed eccentricity in the simulation $(\epsilon=0.9)$, which is not well constrained observationally \cite{2000ApJ...528L.101D,2001ApJ...547.1034C} and that has an important effect on the inner shock geometry. Turbulence and instabilities can also play a key role in the wind geometry and shock conditions that could trigger differences in electron and ion distributions/emission from one periastron to another. 
The distribution of $\gamma_e$, weighted by the emissivity, is relatively smooth and the expected photon distribution is very smooth. The difference of the electron spectral shape on both sides of the wind collision zone cannot explain the two components of the \g-ray emission as suggested by \cite{2011A&A...530A..49B}.
The situation is different for ions. Unless the magnetic field would be very strong ($>$ kG) hadronic interactions mostly take place close to the center and a single peak of neutral pion decay is expected before periastron.
The simulated pion induced \g-ray light curve shows a single peak of emission centred at periastron, in good agreement with the observations of the first periastron. The results of the observations of the second periastron are different. It has been suggested that the change of the X-ray emission after that periastron was the signature of a change of the wind geometry, possibly because of cooling instabilities. A stronger disruption or clumpier wind after the second periastron could perhaps induce a decrease of the average wind density and explain that less hadronic interactions and less thermal emission took place, without affecting  much \ic emission.
The better sensitivity of e-ASTROGAM 
will allow us to study periastron-to-periastron variability in more details. 
Ions could be accelerated up to $10^{15}$ eV around periastron and reach $10^{14}$ eV on average. $\eta$ Carinae can therefore probably accelerate particles close to the knee of the CR spectrum. 
Gamma-ray observations can probe the magnetic field and shock acceleration in details, however the quality of the current data below 100 MeV and above 1 GeV does not yet provide enough information to test hydrodynamical models including detailed radiation transfer (\ic, pion emission, photo-absorption). More sensitive \g-ray observations will provide a wealth of information and allow us to test the conditions and the physics of the shocks at a high level of details, making of $\eta$ Carinae a perfect laboratory to study particle acceleration in wind collisions. $\eta$ Carinae could yield to $10^{48-49}$ erg of CR acceleration, a number close to the expectation for an average supernova remnant \cite{2016APh....81....1B}. 
\begin{figure}
\begin{center}
\includegraphics[width=0.55\textwidth]{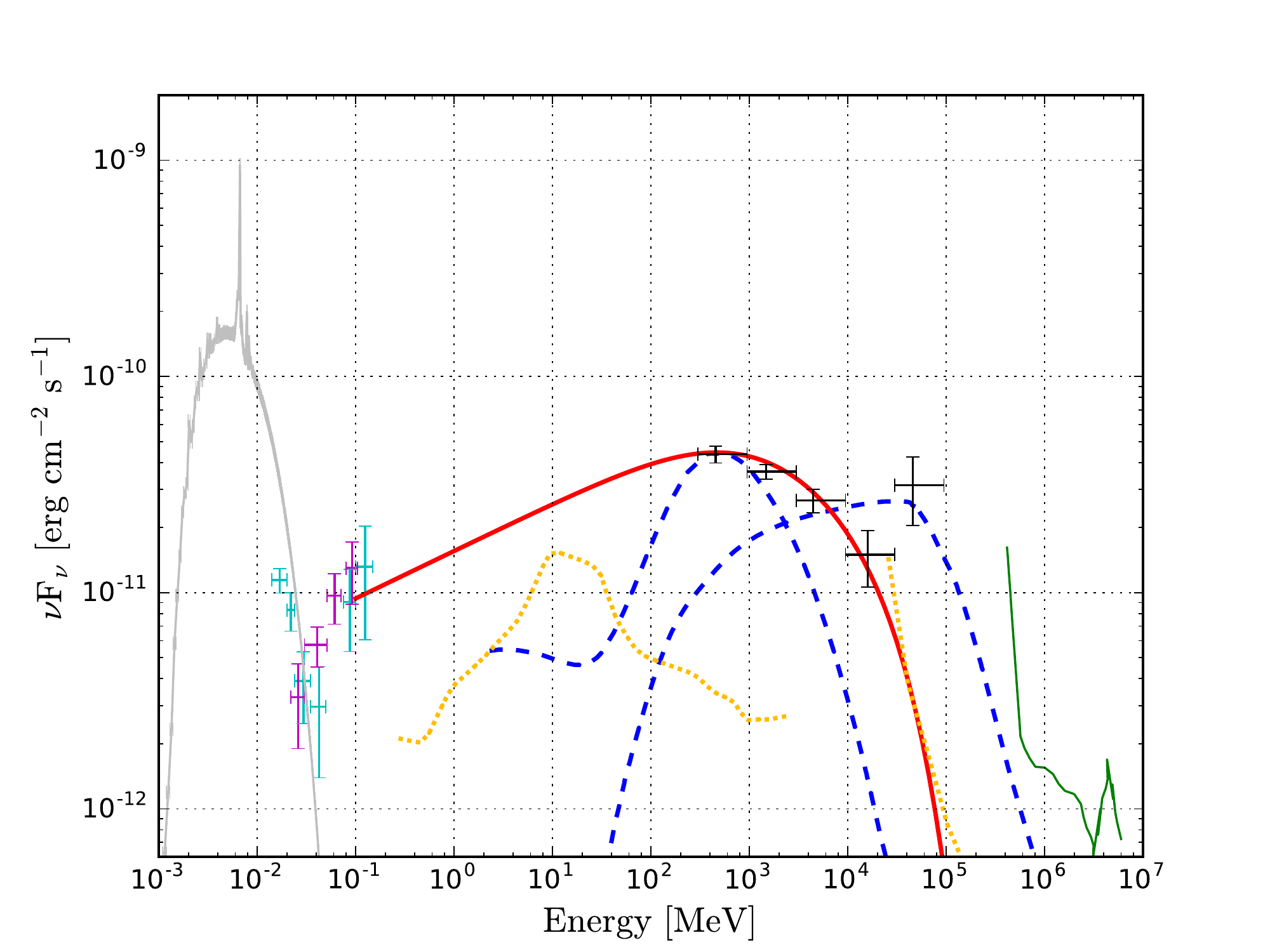}
\caption{\small SED of $\eta$ Carinae from 1 keV to 10 GeV. The data are from \nustar (grey), \swift/BAT (cyan), \INTEGRAL (purple), \fermilat (black) and the upper limits from HESS (green). The predictions are from mostly hadronic (dashed blue line) and lepto-hadronic (red line for the leptonic part) models. The sensitivity curves of e-ASTROGAM and CTA are also indicated (dotted yellow lines).}\label{Fermi_acc:sed}
\end{center}
\end{figure}
\paragraph*{Expected results with e-ASTROGAM}
The SED of $\eta$ Carinae features an excess of emission at hard X-rays, beyond the extrapolation of the thermal emission \cite{Panagiotou2017} that should connect to the \Fermi spectrum in a yet unknown manner.
In the previous section, we have presented a model where electrons and protons are accelerated (as initially proposed by \cite{1993ApJ...402..271E}). The fraction of the shock mechanical luminosity accelerating electrons appears to be slightly smaller than the one that accelerates protons. These results contrast with the efficiencies derived in the context of SNR from the latest particle-in-cell simulations \cite{2015PhRvL.114h5003P}, involving low magnetic fields, radiation and particle densities and favouring acceleration of ions.
Purely hadronic acceleration has been proposed \cite{2015MNRAS.449L.132O} to explain the GeV spectrum of $\eta$ Carinae. In that case, the two spectral components are related to the different hadron interaction times observed on the two sides of the wind separation surface, largely because of the contrast in density and magnetic field. In the simulations included in \cite{2017arXiv170502706B}, this effect is smoothed by the many zones of the model, each characterized by different conditions. Even if the shock on the companion side does contribute more at high energies, the resulting pion decay spectrum does not feature two distinct components.
e-ASTROGAM 
will easily discriminate between the lepto-hadronic and the hadronic models for the \g-ray emission as the \ic leptonic emission of the former would be much stronger than predicted by the latter (Fig.~\ref{Fermi_acc:sed}). e-ASTROGAM can therefore decide which is the model likely to explain the high energy emission of $\eta$ Carinae and strongly constrain the acceleration physics (through the hadronic over leptonic luminosity ratio) in more extreme conditions than found in SNR.
\subsection[Cosmic-ray production in star-forming regions\\
\noindent
\textit{\small{I. A. Grenier, A. Bykov, E. Orlando, A. Strong}}]{Cosmic-ray production in star-forming regions}
\paragraph*{Science questions}
Understanding the complex interplay between stars, gas, and CRs in star-forming regions is of fundamental importance for astrophysics. Multi-wavelength studies of star-forming clouds in the Galaxy and of extreme examples of massive stellar clusters in the Large Magellanic cloud and in starburst galaxies have revealed a wealth of information on the physics of star formation and on the radiation impact of massive stars on their parent cloud. Yet, little is known about the activity of such sites in terms of CR production, nor on their ability to confine and modify Galactic CRs as they diffuse through those turbulent sites. Recent data have provided a wealth of details on local CRs, from direct spectral measurements in and near the heliosphere, to remote \g-ray observations in interstellar clouds within a few hundred parsecs \cite{Grenier15}. Yet, we lack a global and resolved description of the CR distribution in the Milky Way and we don't know how much of an imprint star-forming regions leave on this distribution. This imprint must be significant, in spectrum and in composition. On the one hand, \fermilat observations have detected a cocoon of anomalously hard CRs in the Cygnus X superbubble that has been blown by multiple OB associations \cite{AckermannSB11}. On the other hand, ACE abundance measurements of heavy CR nuclei indicate that 20\% of the local CRs come from massive-star outflows and ejecta, the rest having been probably swept up from the interstellar medium (ISM) by the supernova shock waves that have accelerated them \cite{Murphy16}. Massive stars are clustered in space and time, so are their massive supersonic winds and the ensuing core-collapse supernovae. Thus what happens to CRs freshly escaping from their accelerators? Are they confined for some time and potentially reaccelerated in the highly turbulent medium of star-forming regions? What impact do they have on the surrounding ISM? Our views on the diffusion properties of Galactic CRs have largely been inferred locally. Could they be significantly biased by our viewpoint inside the Local Bubble and in the Gould Belt with its numerous OB associations \cite{Grenier00}? The recent detection of radioactive $^{60}$Fe in the local CRs indeed implies that the time required for acceleration and transport to the Solar System does not greatly exceed 2.6 Myr and that the supernova source of $^{60}$Fe lied within 1 kpc \cite{Binns17}.
\begin{figure}
\includegraphics[width=0.55\textwidth]{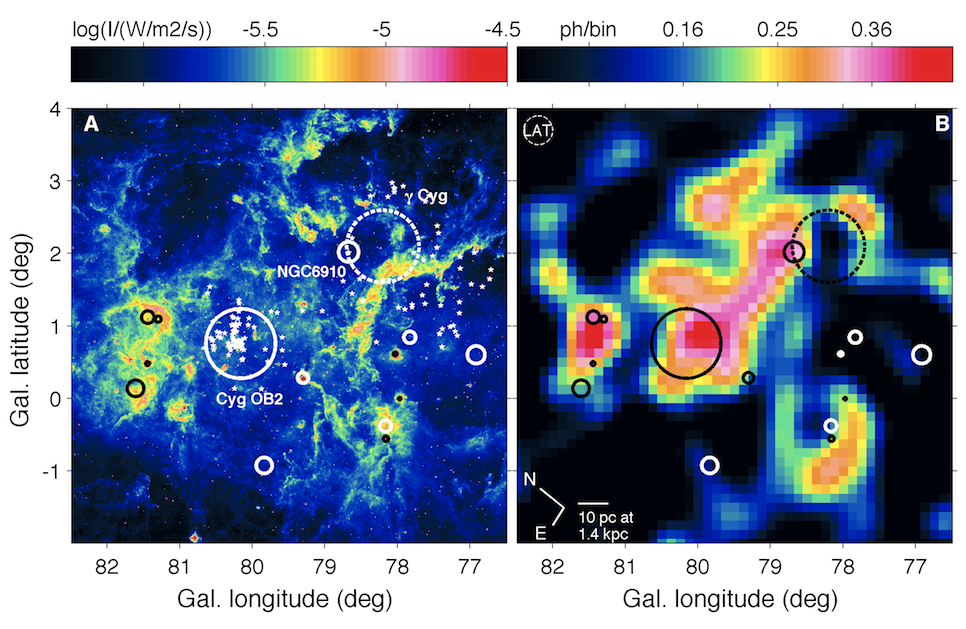}
\includegraphics[width=0.44\textwidth]{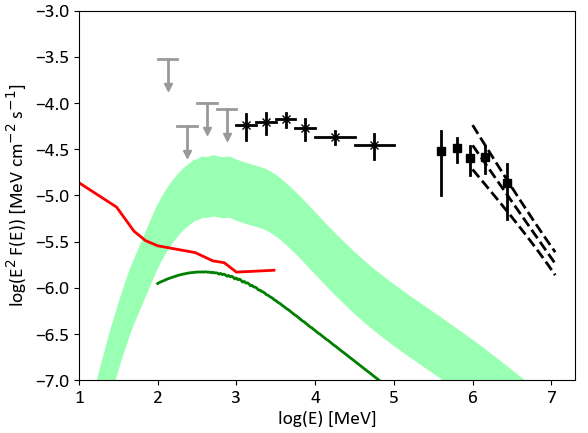}
\caption{Left:  8 $\mu$m intensity map of
the Cygnus X region from \textit{MSX} showing the heated contours of the superbubble. OB stars (white stars), OB associations (white circles), and the supernova remnant  $\gamma$ Cygni  (dashed circle) are overlaid  \cite{AckermannSB11}.
Center: \fermilat photon count map of the
same region in the 10-100 GeV band. The 50-pc-wide excess coincident with the bubble signals a cocoon of freshly accelerated CRs \cite{AckermannSB11}. 
Right: Energy spectrum of the Cygnus cocoon emission as detected
by \fermilat (crosses), \textit{ARGO-YBG} (squares), and MILAGRO (dashed lines). The curves show model expectations from normal Galactic CRs spreading the ionized gas (green band) or up-scattering the stellar and interstellar light fields (green curve). The sensitivity of e-ASTROGAM (red curve) is given for one year of effective exposure in the Galactic disc. Adapted from \cite{Grenier15,AckermannSB11}.}\label{SB:spectra}
\end{figure}
\paragraph*{Importance of \g-ray observations}
Shock waves from supernovae and from multiple powerful winds of early-type stars were suggested as favorable sites of CR acceleration in rich stellar clusters (see \cite{cm83,Bykov2014} for review). Those winds and supernova remnants blow extended superbubbles over a time scale of ${\sim} 10$~Myr \cite{Krause14}. The bubbles are filled with hot X-ray emitting gas where numerous weak and strong shocks can amplify the turbulent magnetic fields. The efficiency of the ensemble of MHD shocks to transfer kinetic power to accelerate CRs and/or re-accelerate passing-by Galactic CRs may exceed 10\%, so superbubbles can substantially modify the CR spectra over a period of 10 Myr. Non-linear modelling predicted the time-asymptotic spectra to be a power law with an index close to 2  in the MeV-TeV regime \cite{Bykov2001,Ferrand2010}. This is consistent with the \g-ray spectrum recorded in the Cygnus cocoon (shown in Fig.~\ref{SB:spectra}) if the \g-rays are mainly produced in inelastic collisions of CR nuclei with the ambient gas. CR electrons can also up-scatter the stellar and dust radiation fields to \g-rays, but they should contribute less to the total cocoon emission (see Fig.~\ref{SB:spectra}). 

Gamma-ray observations provide key probes of the high-energy particle content of superbubbles, both in nuclei and in electrons. X-ray observations can probe the diffuse synchrotron emission from the highest-energy ($>$TeV) electrons if they don't rapidly cool or escape the superbubble. The large magnetic fields ($\sim 20\,\mu$G in the Cygnus X cocoon \cite{AckermannSB11}) yield detectable fluxes for the current X-ray telescopes, but the detection of diffuse non-thermal X rays is challenging toward these hot and complex regions and the small fields of view of the instruments are ill adapted \cite{Mizuno15}. The production of 10-100 TeV neutrinos in CR interactions with gas was estimated to be barely visible with the IceCube Observatory \cite{2017PhRvD..96d3011Y}. Jointly with GeV-TeV \g-ray observations they would inform us on the acceleration efficiency and maximum CR energy attainable in such sites, whereas \g-ray observations below 1 GeV are the only means to reveal the bulk of the CR population filling a superbubble and to measure the CR diffusion lengths inside the bubble.
\paragraph*{Expected results with e-ASTROGAM}
The Cygnus X region, located at an estimated distance of 1.4 kpc \cite{2012A&A...539A..79R}, is a prime target for resolving details of the high-energy activity of superbubbles. The extended region, about 4$^{\circ}$ in size, contains several thousand OB stars 
and it holds a few million solar masses of gas for collisions with CR nuclei \cite{AckermannSB11}. The flux of $(5.8 \pm 0.9)\times 10^{-8}$ cm$^{-2}$ s$^{-1}$ detected from the hard cocoon in the 1 to 100~GeV band corresponds to a luminosity of $(9 \pm 2)\times 10^{34} (D/1.4\,\text{pc})^2$ erg s$^{-1}$, which is below one per cent of the kinetic power of the stellar winds in Cygnus OB2. Fig.~\ref{SB:spectra} shows that the cocoon emission is easily detectable by e-ASTROGAM. Yet, several other GeV and TeV sources have been found in this crowded direction as we look tangentially down the Local Spiral Arm. One has been identified with the pulsar PSR J2032+4127 and its wind nebula, another with the extended $\gamma$ Cygni supernova remnant and its associated pulsar PSR J2021+4026. The improved angular resolution of e-ASTROGAM will therefore be crucial to separate the diffuse cocoon emission from these sources and from the rest of the interstellar Galactic background. Spatial confusion prevented the cocoon detection below 1 GeV with \fermilat, despite its brightness. The performance of e-ASTROGAM will be key to reliably extend the cocoon spectrum below 1 GeV in order to estimate the energy distribution of the bulk of the CR nuclei, to estimate the CR pressure inside the bubble, to separate the emissions from CR electrons and nuclei, and to search for spectral variations across the bubble that would serve to test possible acceleration scenarios, by individual sources or by the collective action of wind and supernova shock waves. A refined morphology of the GeV cocoon will help capture its diffuse counterpart at TeV energies to study the cut-off energy of the particles since the extension of the cocoon spectrum beyond 100 GeV (shown in Fig.~\ref{SB:spectra}) is still unclear in the latest data \cite{Bird17,Hona17}. 

Another extended \fermilat source with a hard $E^{-2.1\pm 0.2}$ spectrum toward G25.0+0.0 has recently been proposed as a second case of a \g-ray emitting star-forming region in the Milky Way  \cite{Katsuta17}. It may be associated with a candidate OB association G25.18+0.26, comparable to Cygnus OB2 in mass, but at a larger distance of 6 to 8~kpc. If so, the \g-ray luminosity would be about 10 times larger than that of the Cygnus cocoon, reflecting the 9 times larger volume and/or mass of the emitting region. There again, severe confusion limits the identification of the origin of the extended emission and the improved performance of e-ASTROGAM will open new avenues for studies.

Younger OB associations, where no supernova explosion has occurred yet, may also impart a fraction of the kinetic energy of their strong supersonic stellar winds to CR acceleration. Nearby OB associations, such as NGC 2244 in the Rosette nebula and NGC 1976 in the Orion nebula, have been proposed as test beds \cite{Maurin16}. They can be detected by e-ASTROGAM below 3 GeV if a few per cent of the stellar-wind powers are supplied to CRs.

Despite the long observational and theoretical efforts to identify and study CR acceleration in supernova remnants, a number of fundamental questions remain unanswered about the acceleration efficiency and the time-dependent spectrum of the escaping particles. The detection of the high-energy activity of turbulent bubbles blown by stellar clusters adds another level of complexity between the individual CR sources and the large-scale distribution of CRs in the Galaxy. It needs to be addressed by resolving the MeV to TeV emission of active star-forming regions, by comparing them at different stages of evolution and for different cluster masses, and by uncovering new examples in the Galaxy (e.g. Westerlund 1 or 2) or in the Large Magellanic Cloud (e.g. 30 Doradus). An instrument such as e-ASTROGAM, in synergy with the HAWC and CTA observatories at TeV energies and with e-ROSITA in X-rays, will be pivotal to make progress.
\subsection[Understanding the nature of the \g-ray emission from the \Fermi Bubbles\\
\noindent
\textit{\small{D. Malyshev, A. Franckowiak, I. A. Grenier}}
]{Understanding the nature of the \g-ray emission from the \Fermi Bubbles}
\paragraph*{Science questions}
The \Fermi Bubbles (FB) are one of the most spectacular and unexpected discoveries based on the \fermilat data
\cite{Su:2010qj, Fermi-LAT:2014sfa}.
The FB extend to $55^\circ$ above and below the Galactic center. 
There exist lobes in other galaxies with similar shape and size as the FB.
The origin of these lobes is attributed either to jets from the supermassive black holes at
the centers of the galaxies (AGN scenario) or a period of starburst activity which results in a combined wind
from supernovae explosions of massive stars (starburst scenario) \cite{Sharp:2010aj}.
The lobes in other galaxies are usually too distant to be resolved by \g-ray telescopes.
Thus, the study of the FB provides a unique opportunity to test, using \g-ray data,
predictions of computer simulations of the evolution of jets from supermassive black holes 
\cite{Guo:2011eg, Yang:2012fy}, winds from supernova explosions, or CR-driven winds \cite{Wiener:2016zcr}.
The problem is that, in spite of the fact that the FB were discovered more than seven years ago, 
the formation mechanism of the FB (AGN vs starburst) is still unknown.
\paragraph*{Importance of \g-ray observations}
The \g-ray emission from the FB can be produced either by \ic scattering 
of high-energy electrons and positrons off starlight, infrared, and cosmic microwave background photons
(leptonic model),
or by interactions of CR nuclei with gas (hadronic model).
The leptonic \g-ray emission model is characteristic for the AGN-type scenario of the FB formation,
while the hadronic model of \g-ray emission is more likely in the starburst scenario.
As a result, understanding the \g-ray emission mechanism can uncover the
formation process of the FB.
The \g-rays in the hadronic model are produced as a result of the cascade of hadronic interactions
mostly from the decay of $\pi^0$ mesons.
The spectrum of these ``primary'' \g-rays has a characteristic cutoff below 100 MeV due to 
the mass of the $\pi^0$ meson.
This cutoff is usually used to distinguish the hadronic \g-ray production from
the leptonic IC model, which should not have a cutoff below 100 MeV.
The problem is that there are secondary electrons and positrons produced in the hadronic cascades alongside the \g-rays.
These electrons and positrons propagate and create ``secondary'' \g-rays via IC scattering.
In case of the FB, the secondary IC emission can dominate the spectrum below 100 MeV (Fig.~\ref{fig:bubbles_emission}), 
which results in the absence of the $\pi^0$ cutoff \cite{Fermi-LAT:2014sfa}.
\begin{figure}[h]
  \centering 
  \includegraphics[width=0.7\linewidth]{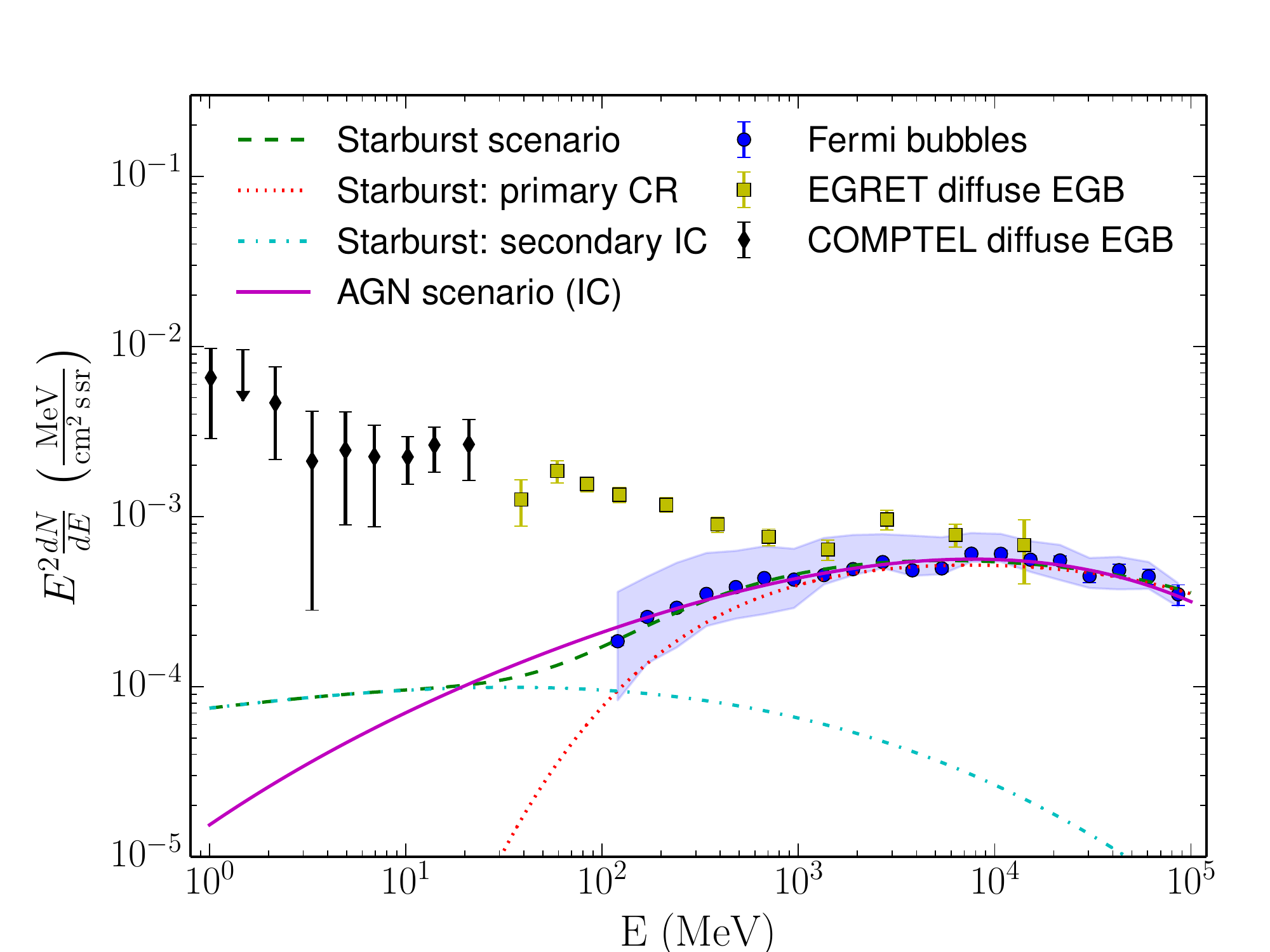}
  \caption{ 
  \Fermi Bubbles spectrum compared to the AGN (leptonic IC, solid purple line) 
  and starburst (hadronic, dashed green line) models of the \g-ray emission \cite{Fermi-LAT:2014sfa}.
  The shaded band shows the systematic uncertainty on the FB spectrum \cite{Fermi-LAT:2014sfa}.
  The hadronic model includes the primary emission (dotted red line) of \g-rays and the \g-rays produced in IC interactions
  of secondary electrons and positrons (dash-dotted cyan line).
  The secondary component of \g-ray emission in the hadronic model is significantly softer than the primary component 
  which results in a break around 30 MeV and a significant difference between the hadronic and leptonic models of \g-ray 
  emission around a few MeV.
  For comparison, we also plot the diffuse extragalactic \g-ray background (EGB) fluxes measured by the EGRET~\cite{Strong:2004ry}
  and COMPTEL~\cite{2000AIPC..510..467W} experiments.}\label{fig:bubbles_emission}
\end{figure}
As one can see from Fig.~\ref{fig:bubbles_emission}, it is very hard to distinguish the leptonic and the hadronic models based on observations above 10 MeV due to a contribution from the secondary IC emission.
However, the secondary leptons have a spectrum that is softer by $E^{-1}$ than the spectrum of the primary protons 
because of the energy losses via IC scattering and synchrotron radiation.
As a result, the secondary IC spectrum is softer than the IC spectrum in the leptonic model or the primary \g-ray spectrum in the hadronic model.
Below a few tens of MeV, the soft IC component dominates the \g-ray emission in the hadronic model which results in a break in the \g-ray spectrum around 30 MeV, while in the leptonic model the spectrum is expected to be featureless.
The presence (absence) of the break can be used to confirm the hadronic (leptonic) model of the \g-ray emission.
\paragraph*{Expected results with e-ASTROGAM}
The main backgrounds at high latitudes include resolved point sources and the diffuse extragalactic \g-ray background (EGB) coming from unresolved point sources and truly diffuse background.
In order to estimate the e-ASTROGAM sensitivity to distinguish the models of \g-ray emission in the FB, 
we compare the expected flux from the FB to the diffuse EGB at 2 MeV.
The expected e-ASTROGAM effective area at this energy in the Compton regime is $ \approx 117\; {\rm  cm^2}$, as highlighted in Chapter \ref{intro}. 
If we take into account that the effective coverage is expected to be about 23\%, 
then the exposure after one year of observations can be estimated as $\approx 8.5 \times 10^8\; {\rm  cm^2}$.
The area of the bubbles is $\approx 1$ sr \cite{Fermi-LAT:2014sfa}.
For an energy bin with a width of 1 MeV,
the number of signal counts around 2 MeV in the hadronic model after one year of observations is $\sim 3 \times 10^4$,
while the number of background photons from the diffuse EGB is expected at a level of $10^6$. 
Thus, the signal to noise ratio is expected to be at the level of 10 or more while the fractional signal is about 3\%.
Consequently, e-ASTROGAM will be sensitive to detect the difference between the AGN (leptonic) and starburst (hadronic) models of the FB already after one year of observations.
The possibility to distinguish the flux from the FB and the diffuse EGB at a few percent level is based
on the fact that the FB have a well defined shape with sharp edges while the diffuse EGB is expected to be approximately isotropic.
The improved angular resolution of e-ASTROGAM relative to the PSF of the \fermilat will be essential in 
the derivation of the shape of the FB at energies below 1 GeV, where the \fermilat measurement suffers from a significant
systematic uncertainty (Fig.~\ref{fig:bubbles_emission}).
\subsection[De-excitation nuclear \g-ray line emission from low-energy cosmic rays\\
\noindent
\textit{\small{V. Tatischeff, J, Kiener, I. Grenier, A. Strong}}
]{De-excitation nuclear \g-ray line emission from low-energy cosmic rays}
\paragraph*{Science questions}
LECRs of kinetic energies $\lsim$1~GeV~nucleon$^{-1}$ are thought to be a major player in the process of star formation. They are a primary source of ionization of heavily shielded, dense molecular clouds and the resulting ionization fraction conditions  the coupling of the gas with the ambient magnetic field in these regions. LECRs also represent an important source of heating that contributes to hold molecular cores in equilibrium against gravitational collapse. In addition, LECRs play a central role in astrochemistry by initiating a rich ion-neutral chemistry within the cold neutral medium of ISM. Furthermore, LECRs are thought to drive large-scale magnetohydrodynamic turbulence and cause amplification of magnetic field in the ISM, and also provide critical pressure support in starburst regions to launch Galactic winds into the halo (see \cite{Grenier15} and references therein, as well as \cite{2016ApJ...824L..30P}). 
LECRs are thought to be a fundamental component of the ISM. Their energy density compares to that of the interstellar gas, magnetic field or stellar radiation. Yet, their composition and flux are poorly known. The {\it Voyager 1} spacecraft has recently provided valuable measurements of the local energy spectra of Galactic CR nuclei down to 3 MeV~nucleon$^{-1}$ and electrons down to 2.7 MeV~nucleon$^{-1}$ beyond the heliopause (LECRs are severely depleted inside the Solar System because of the solar wind). But the total CR ionization rate of atomic hydrogen resulting from the measured spectra, $\zeta_{\rm H}=(1.51 - 1.64)\times10^{-17}$~s$^{-1}$, is a factor $>10$ lower than the average CR ionization rate of $\zeta_{\rm H}=1.78\times10^{-16}$~s$^{-1}$ \cite{ind15} measured in clouds across the Galactic disc using line observations of ionized molecules by Herschel (see also \cite{neu17}). The difference suggests that LECRs are relatively less abundant in the local ISM than elsewhere in the Galaxy. Observations of  H$_3^+$  in diffuse molecular clouds show indeed that the density of LECRs can strongly vary from one region to another in the Galactic disk, and, in particular, that the LECR flux can be significantly higher than the average value in diffuse molecular gas residing near a site of CR acceleration such as a supernova remnant (SNR) \cite{ind10,ind12}. Measurements of the DCO$^+$/HCO$^+$ abundance ratio have shown that the CR ionization rate can also be very high ($\gsim 100$ times the standard value) in dense molecular clouds close to SNRs \cite{cec11,vau14}. 
Various astrophysical sources could produce significant amounts of LECRs in the Galaxy besides supernova remnants (SNRs), e.g., OB associations \cite{mon79,par04}, compact objects such as microquasars \cite{hei02}, cataclysmic variables \cite{kam18}, and normal stars producing astropheric anomalous CRs \cite{sch08}. The observed quasi-linear increase of the Be abundances measured in stellar atmospheres with the star metallicity provides an independent argument for the existence of a significant component of LECR nuclei in the Galaxy, in addition to the standard CRs thought to be produced by diffusive shock acceleration in SNRs (see \cite{tat11} and references therein). Obviously, our knowledge of the production pathways and transport properties of LECRs in our Galaxy is very rudimentary.
\paragraph*{Importance of \g-ray observations}
MeV \g-ray astronomy is the only direct way of studying the various effects of sub-GeV CR nuclei in the ISM. In the GeV range, the diffuse Galactic emission is dominated by $\pi^0$-decay \g-rays from the interaction of CR nuclei (mostly protons) with interstellar matter, and observations in this domain probe CR spectra above about 1~GeV per nucleon only. Nevertheless, \fermilat observations of the diffuse Galactic emission above $E_\gamma=100$~MeV put stringent constraints on the CR origin and propagation (see, e.g., \cite{AckermannSB11,ack12a,ack12b,cas15}). See Sec. \ref{Sec:gammaISM}  for more details on the CR contributions to the multiwavelength spectrum of the inner Galaxy.  
A very promising way to study CR nuclei below the kinetic energy threshold for production of neutral pions would be to detect characteristic \g-ray lines in the $0.1 - 10$~MeV range produced by nuclear collisions of CRs with interstellar matter. The most intense lines are expected to be the same as those frequently observed from strong solar flares, i.e. lines from the de-excitation of the first nuclear levels in $^{12}$C, $^{16}$O, $^{20}$Ne, $^{24}$Mg, $^{28}$Si, and $^{56}$Fe \cite{ram79}. Strong narrow lines are produced by excitation of abundant heavy nuclei of the ISM by CR protons and alpha particles of kinetic energies between a few MeV and a few hundred MeV. The total nuclear line emission is also composed of broad lines produced by interaction of CR heavy ions with ambient H and He, and of thousands of weaker lines that together form a quasi-continuum in the range $E_\gamma \sim 0.1 - 10$~MeV \cite{ben13}. Some of the prominent narrow lines may exhibit a very narrow component from interactions in interstellar dust grains, where the recoiling excited nucleus can be stopped before the \g-ray emission \cite{tat04}. The most promising of such lines are from $^{56}$Fe, $^{24}$Mg, $^{28}$Si and $^{16}$O
\paragraph*{Expected results with e-ASTROGAM}
\begin{figure}
\centering
\includegraphics[width=0.6\textwidth]{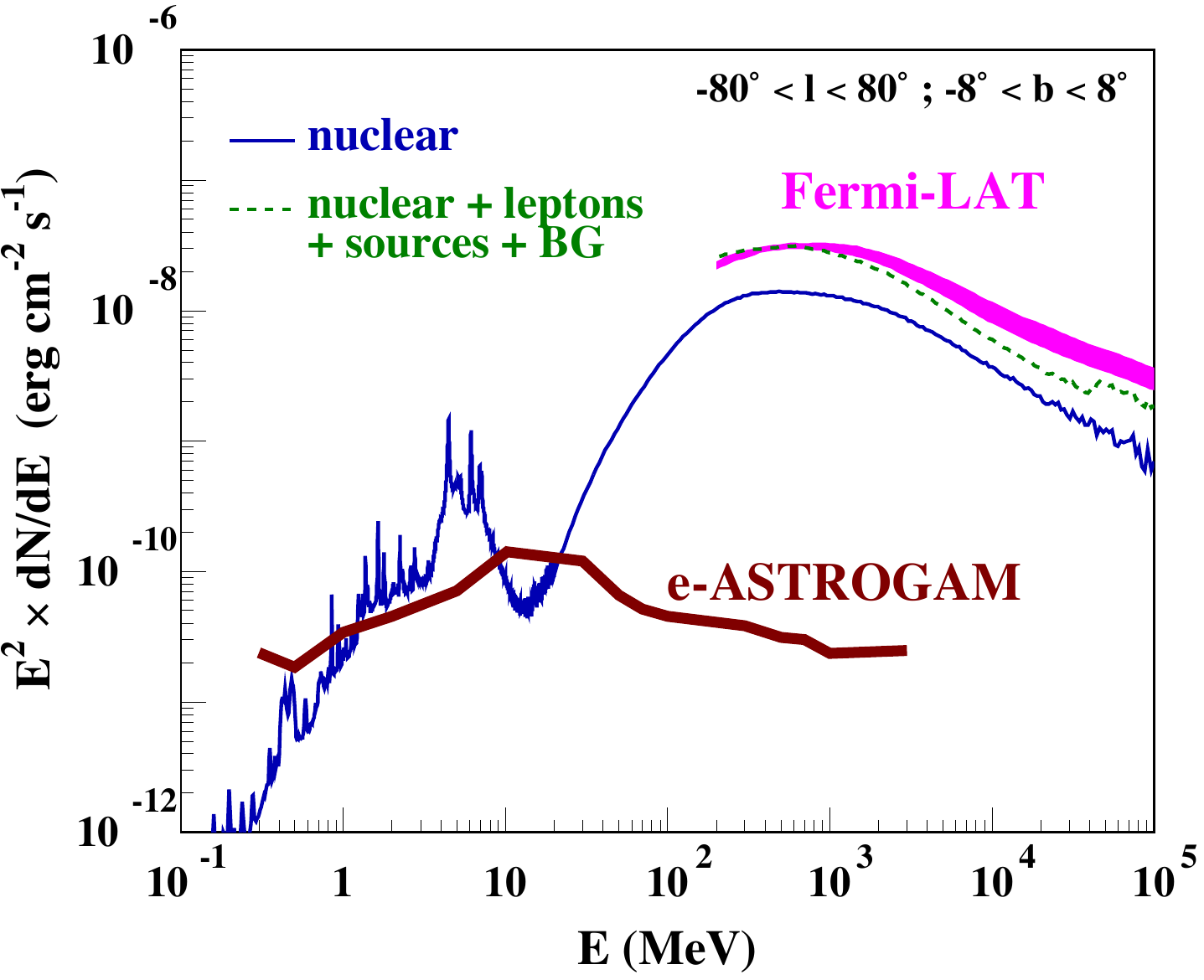}
\caption{Predicted \g-ray emission due to nuclear interactions of CRs in the inner Galaxy (longitude $-80^\circ \leq l \leq 80^\circ$ and latitude $-8^\circ \leq b \leq 8^\circ$). The \g-ray line emission below 10 MeV is due to LECRs, whose properties in the ISM have been adjusted such that the mean CR ionization rate deduced from H$_3^+$ observations and the  \fermilat data (magenta band) at 1 GeV are simultaneously reproduced (adapted from \cite{ben13}). The dashed green line shows the total calculated emission when
adding leptonic contributions, point sources and extragalactic
\g-ray background that were taken from \cite{ack12b}.The 1-year sensitivity of \ea (for Galactic background) is superimposed.}
\label{leCR:fig1}
\vspace{0pt}
\end{figure}
Fig.~\ref{leCR:fig1} shows a calculated \g-ray emission spectrum from CRs in the inner Galaxy containing a low-energy component that would account for the observed mean ionization rate of diffuse molecular clouds. A future observation of this emission would be the clearest proof of an important LECR component in the Galaxy and probably the only possible means to determine its composition, spectral and spatial distribution. A particularly promising feature of the predicted \g-ray spectrum is the characteristic bump in the range $E_\gamma = 3 - 10$~MeV, which is produced by several strong lines of $^{12}$C and $^{16}$O. The calculated flux in this band integrated over the inner Galaxy ($\mid$$l$$\mid \leq 80^\circ$; $\mid$$b$$\mid \leq 8^\circ$) amounts to $7 \times 10^{-5}$~cm$^{-2}$~s$^{-1}$, which is well above the predicted sensitivity of \ea after one year of effective exposure of such a spatially extended emission, $S_{3\sigma}=1.1 \times 10^{-5}$~cm$^{-2}$~s$^{-1}$. 
\subsection[Gamma rays from the interstellar medium: probing cosmic rays throughout the Galaxy\\
\noindent
\textit{\small{E. Orlando, A. Strong, I. A. Grenier, A. Bykov}}
]{Gamma rays from the interstellar medium: probing cosmic rays throughout the Galaxy}\label{Sec:gammaISM}
\paragraph*{Science questions}
The Milky Way is an intense source of \g-rays. These photons originate mainly from the interactions of CRs with the gas in the interstellar medium (ISM) and with the interstellar radiation field (ISRF), via leptonic (\brem and \ic scattering) and hadronic (pion decay) processes. 
Observations of this \g-ray interstellar emission have been widely used to study the large-scale distribution and spectrum of CRs, and to understand CR propagation and interactions in the Galaxy. This is often done by comparing \g-ray observations with propagation models and direct CR measurements in or near the Solar System.
A recent extensive review of this topic can be found in \cite{Grenier15}, where it is underlined how CRs is a piece of the puzzle to understand Galaxy formation and evolution. 
Our knowledge regarding the Galactic distribution of CRs, if they concentrate along spiral arms or in the Central Molecular Zone, and regarding the influence of Galactic winds and the possible anisotropy of the diffusion properties are still very limited.
Recent simulations \cite{2016ApJ...824L..30P} have showed that the transport properties have important consequences on the evolution of a galactic disc. 
While with an isotropic diffusion CRs quickly diffuse out of the disk, with an anisotropic diffusion they remain long enough in the disk to impact the gas dynamics and magnetic-field growth. Hence dynamical effects of CRs on the ISM should be investigated.
Over the past decade, many detailed studies on CRs and on the induced interstellar \g-ray emission have been performed thanks to the \fermilat and AGILE missions, and to the improved precision of the direct CR measurements.
However, these data are deeply challenging our knowledge of CRs, requiring a broader energy coverage and a better angular resolution for \g-ray instruments in order to distinguish the different emission processes and solve many open questions. Among them, the \fermilat data have confirmed that the CR distribution only weakly declines in flux and mildly softens from the inner Galaxy to its outskirts \cite{IEM}, at variance with the expectations from the distribution of potential CR sources and uniform diffusion properties. 
Possible solutions include a large halo size of order 10 kpc, additional gas or CR sources in the outer Galaxy, diffusion coefficients linked to the CR source rate and spiral arm structure, differential motions of CR sources and target gas due to the dynamics of spiral arms, and non-linear transport properties with CRs being advected by and scattering off self-generated Alfven waves (see \cite{Grenier15,gaggero15,recchia17,nava17} for review). Testing those ideas and realistic transport models runs up against the limited angular resolution of the \g-ray data that rapidly degrades spatial and spectral contrasts in the diffuse emission from CRs and that adds confusion with unresolved point sources unrelated to CR activity.
The main science questions e-ASTROGAM will address are: investigating the distribution of CR sources, understanding CR propagation in the Galaxy, and describing their density and spectral variation over the Galaxy.
\paragraph*{Importance of \g-ray observations}
While direct CR measurements with balloons and satellites inform us about the local CR spectrum in great detail, only observations of the interstellar emission in \g-rays reveal the large-scale distribution and spectrum of CRs, and help in understanding CR propagation and interactions in the Galaxy. \\
\fermilat and AGILE have provided a detailed view of the \g-ray sky in the  range above 100 MeV,
 which can extend down to about 30 MeV with the latest \fermilat event reconstruction ``Pass 8'', but with an angular resolution $>9^{\circ}$. At lower MeV energies, our overall view of the diffuse emission in the Galaxy is very limited\footnote{See the contribution `COMPTEL Heritage Data Project' in this White Book for more details on the COMPTEL MeV sky} \cite{strong1999} and of the few thousand sources known at GeV energies, only about 20 have been detected in the 1-30 MeV range by GRO/COMPTEL. \\
The diffuse hard-X-ray spectrum in the inner Galaxy has been derived up to MeV energies with the SPI coded-mask telescope on board \INTEGRAL \cite{Bouchet2011}  and with COMPTEL. 
This diffuse emission has recently been compared with updated propagation models based on the latest CR measurements such as AMS02 and Voyager~1, and constrained by observations of the radio-synchrotron emission \cite{O2017}. 
The results show that the hard X-ray intensity of the Galactic ridge is a factor of $\sim$3 above the expectations.
An increased CR electron density or a more intense ISRF in the central regions could explain the excess emission. Another explanation could be the contribution of unresolved soft \g-ray point sources, which e-ASTROGAM could resolve.
Fig.~\ref{IS_CR:spectra} shows the multiwavelength spectrum from \cite{strong11,Grenier15} combining SPI, COMPTEL, and \fermilat data, together with the spectrum expected for the separate components of the interstellar emission for a standard model. 
In addition to the hadronic gas-related emission, which peaks at GeV energies, below 100 MeV most of the interstellar emission comes from the \ic scattering of CR electrons on the ISRF and cosmic microwave background, and from the \brem emission due to CR electrons interactions with gas.
As shown in the figure, the IC component is believed to be the dominant interstellar component below a few tens of MeV \cite{O2017}.
\begin{figure}[h!]
\center
\includegraphics[width=0.6\textwidth,angle=0.]{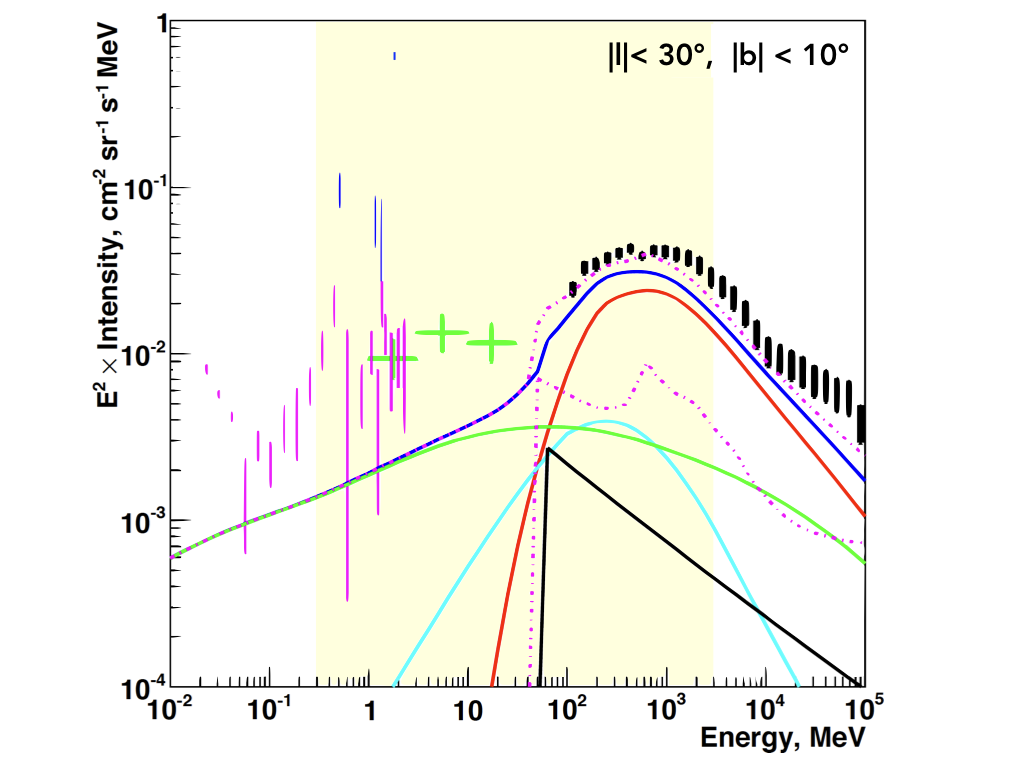}
\caption{Spectrum of the inner Galaxy from \cite{strong11,Grenier15}, including data from \INTEGRAL/SPI (magenta and blue bars), COMPTEL (green crosses)  and \fermilat (black bars).
 The components are: pion decay (red line), \ic (green line), \brem (cyan line),  total (blue line), isotropic (black line), detected sources (magenta lower dashed line), detected sources plus total (magenta upper dashed line). The spectral coverage of e-ASTROGAM is highlighted in yellow. Its extended-source sensitivity for one year of observations based on simulations for the inner Galaxy is of the order of a few 10$^{-5}$ cm$^{-2}$s$^{-1}$sr$^{-1}$MeV below a few MeV, increasing to 10$^{-4}$ cm$^{-2}$s$^{-1}$sr$^{-1}$MeV around 10 MeV, and decreasing again to few 10$^{-5}$ cm$^{-2}$s$^{-1}$sr$^{-1}$MeV above 30 MeV. This is a factor of $\sim$~30~--~10$^3$ below the interstellar intensity depending on the energy. The interstellar propagation model shown here is an example of standard models.
 } 
\label{IS_CR:spectra}
\end{figure}
\paragraph*{Expected results with e-ASTROGAM}
Since e-ASTROGAM will extend below the maximum of the pion-decay peak at 67.5 MeV, it will for the first time allow us to fully resolve the pion-decay signature to precisely separate emissions from CR nuclei and electrons.
The energy coverage of e-ASTROGAM  is  also well  suited  to reveal the spatial and spectral distributions of the IC emission in the Galaxy. This is crucial since this emission spans the entire \g-ray domain, up to TeV energies, and we can presently only rely on uncertain model predictions and on \g-ray observations above tens of GeV to subtract this pervasive component from the other sources of diffuse emissions in order to study the propagation of CRs, the extragalactic diffuse \g-ray background, or potential dark-matter annihilation signals around the Galactic centre.
e-ASTROGAM will also uncover, from the \brem and the IC emission, the distribution of CR electrons in the Galaxy down to below GeV energies. 
Because electrons are affected by energy losses more strongly than protons and heavier nuclei, they remain much closer to their sources and they better sample CR inhomogeneities, hence the importance of mapping CR electrons to constrain the large-scale distribution of CR sources.\\
The difficulty of extending \fermilat analyses below 100 MeV where the leptonic component dominates is due to the relatively large PSF and energy dispersion at those energies. With its improved PSF and energy resolution e-ASTROGAM will be finally able to access those energies that have never been studied after the COMPTEL era  to provide essential information on the bulk of CRs. 
\subsection[Probing the interplay between cosmic rays and the interstellar medium\\
\noindent
\textit{\small{I. Grenier}}
]{Probing the interplay between cosmic rays and the interstellar medium}
\paragraph*{Science questions}
The interstellar medium (ISM) is filled with gas, magnetic fields, dust, light, and CRs. The ever-changing structure of this medium controls the efficiency of star formation and the evolution of galaxies. CRs play a key role in this evolution \cite{Grenier15} as they heat and ionize the star-forming clouds and they initiate a rich network of chemical reactions (leading to gas coolants). They provide pressure support to launch strong galactic winds and regulate the gas transfer in and out of a galaxy \cite{Zweibel13,2016ApJ...824L..30P}. They influence the growth of magnetic fields by supporting gas outflows  \cite{2016ApJ...824L..30P}. 
These stimuli are driven by CRs with GeV and sub-GeV energies. Such CRs abound, but they are poorly known. \textit{Voyager 1} has measured their spectrum just outside the heliosphere \cite{Stone13}, but little is known 
elsewhere in the Milky Way. We lack observational constraints on their spatial distribution, on the degree of anisotropy in their diffusion, on the heterogeneity of their properties on the scale of star-forming regions, on their penetration inside the dense gas, and on their feedback on the multi-phase structure of clouds. These are central questions to be answered primarily in \g rays in order to better understand the CR feedback on galaxy evolution.

Accurate measurements of the gas mass at all scales are also pivotal in understanding galaxy evolution and in connecting the mass distributions of stars and of their parental clouds. The gas exists in several phases according to the conditions of pressure, heating, cooling, ionization, and screening from stellar UV radiation. The phases are interleaved in turbulent, fractal structures \cite{Hennebelle12}. By producing \g rays in their interactions with the gas, CRs expose the total gas to view, regardless of its thermodynamical and chemical state. The full \g-ray census of the gas mass provides important insight into the use of other gas tracers.
Most of the mass resides in the neutral gas at medium densities (0.1-$10^3$~cm$^{-3}$), in atomic and molecular forms that are commonly traced by HI (21 cm) and CO (2.6 mm) lines. One critical challenge is to detect the ``Dark" Neutral Medium (DNM) that lies at the H-\hd interface. By gathering optically-thick HI and CO-dark \hd, the DNM easily escapes observations even though it is ubiquitous and massive \cite{Grenier05,Planck15,Pineda13,Remy18a}. A second challenge is to evaluate \hd masses as we cannot directly detect cold \hd molecules. The \xco factor relates integrated CO line intensities to \hd column densities and the challenge is to estimate the \xco ratios in a variety of molecular clouds more or less susceptible to UV radiation \cite{Smith14}. A third challenge is to quantify how dust grains evolve across gas phases. The grains are well mixed with all forms of gas, but their emission cross section and, to a lesser extent, their specific reddening, have been found to gradually, but markedly change with increasing gas density (see \cite{Remy17,Remy18b} for review). Infrared dust emission being the prime gas tracer in distant galaxies, quantifying how dust properties vary per gas nucleon in the ISM is of paramount importance to interpret galaxy evolution. The total-gas tracing capability of CRs provides decisive information to progress on these three fronts.
\paragraph*{Importance of \g-ray observations}
Gamma-rays are produced by CR nuclei in inelastic collisions with gas nuclei (pion decays), by CR electrons in \brem radiation in the gas, or by CR electrons up-scattering the interstellar radiation fields (\ic scattering). Characterizing the ``pion bump" near 70 MeV gives access to the low-energy turnover in CR momentum spectrum near and below one GeV, with the advantage over direct nuclear line detections of a larger continuum emissivity which allows detection throughout the Milky Way and for a large range of cloud masses ($>10^3$ M$_{\odot}$, depending on distance). Observations at energies below the pion bump give access to the lowest energy CR electrons that heat and ionize the gas, to complement the higher-energy observations of the bulk of the CR nuclei that provide pressure support. 

Tracing the gas with CR nuclei relies on the assumption of a uniform CR flux through the phases of a given cloud complex, and on the measurement of the \g-ray emissivity spectrum per gas nucleon in the warm atomic part of the complex where the gas mass can be inferred from HI line emission. Since CR concentration or exclusion processes in a cloud become significant at momenta below 1 GeV \cite{Schlickeiser16}, higher-energy CR nuclei emitting above the pion bump can be used to measure the total gas for ISM studies. Their large diffusion lengths \cite{Zweibel13} and the uniformity of the GeV \g-ray spectra seen across the gas phases of nearby clouds \cite{Grenier15} give strong weight to this method. 
\begin{figure}
\includegraphics[width=0.5\textwidth]{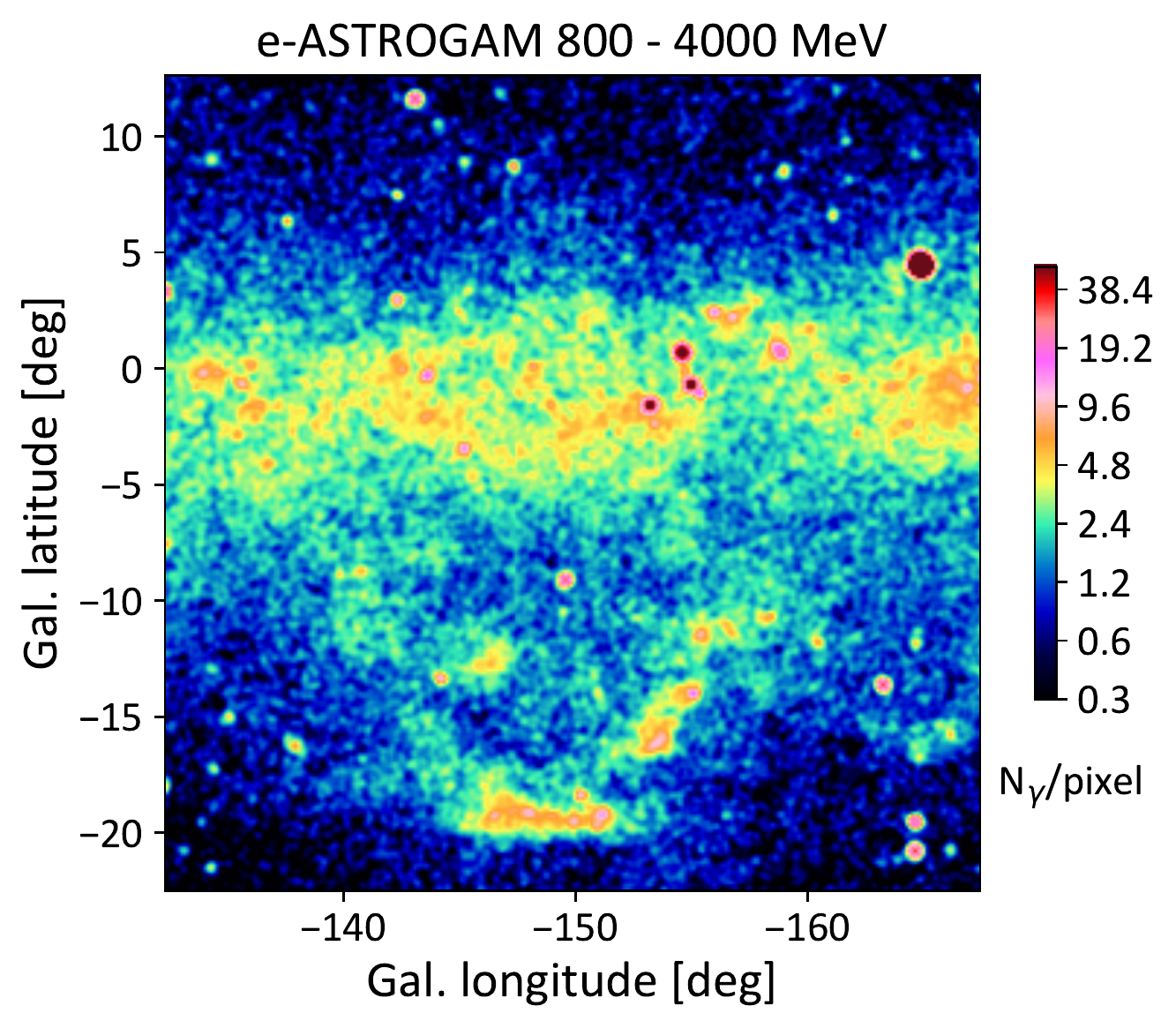}
\includegraphics[width=0.5\textwidth]{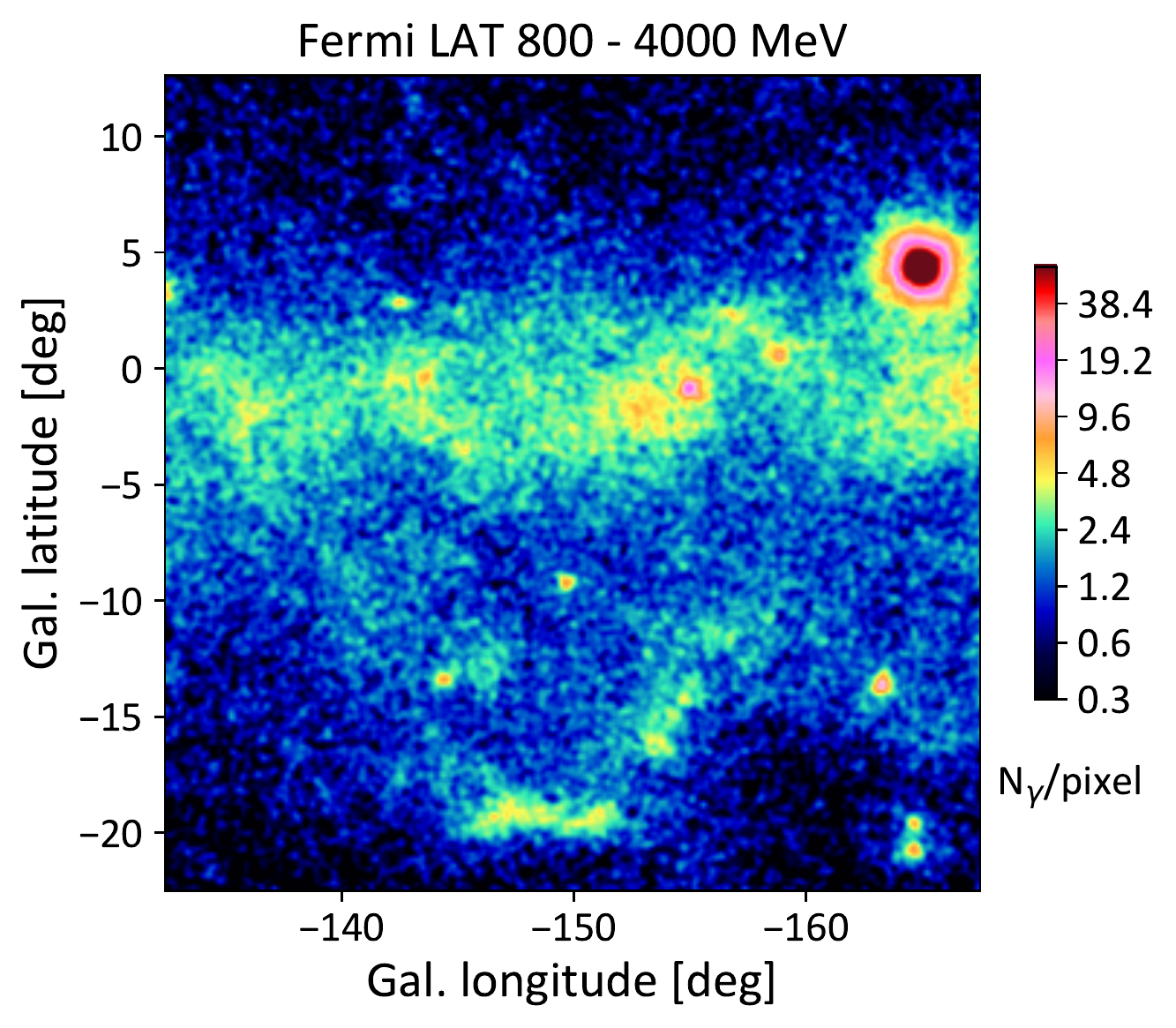}
\includegraphics[width=0.5\textwidth]{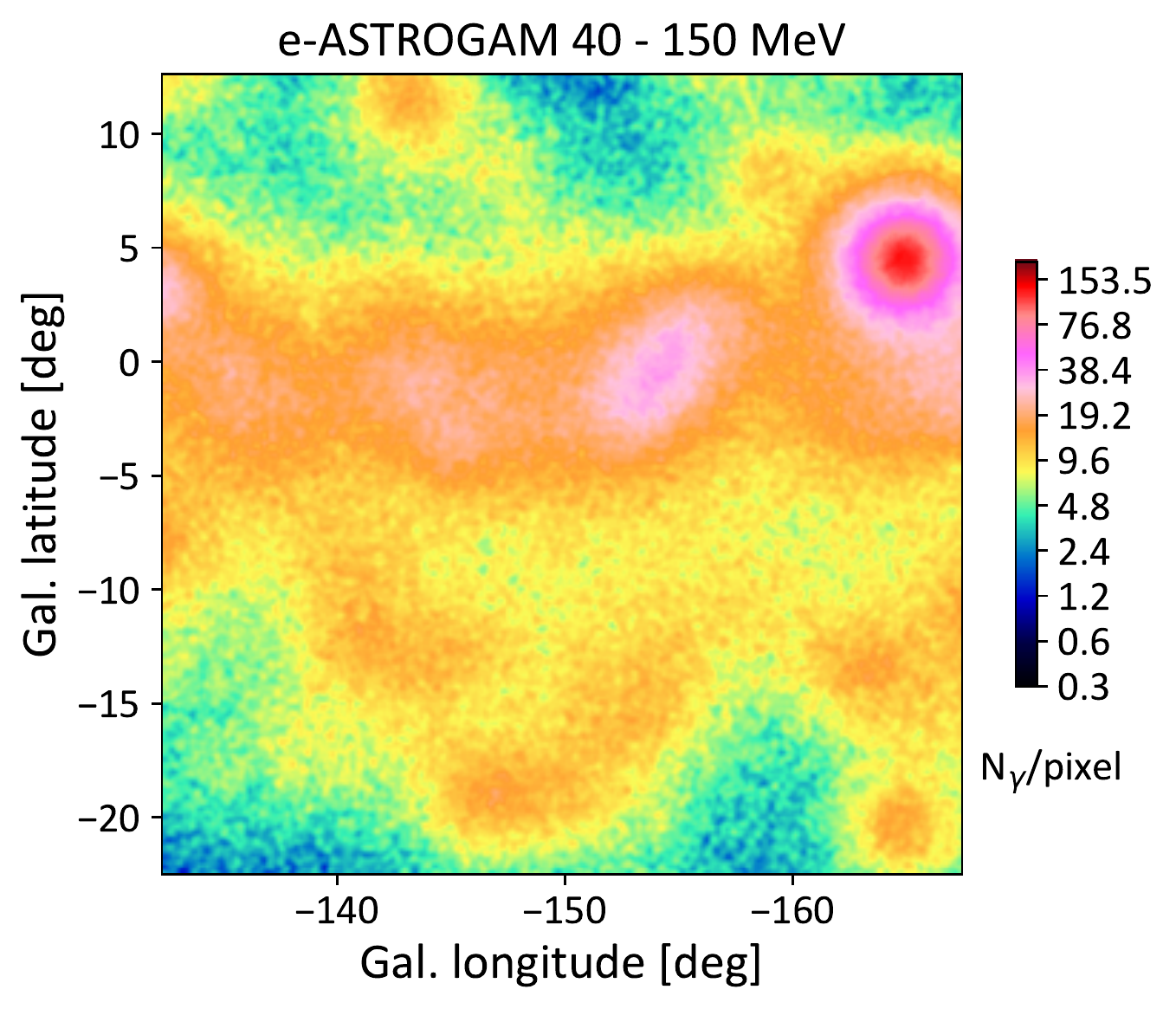}
\includegraphics[width=0.5\textwidth]{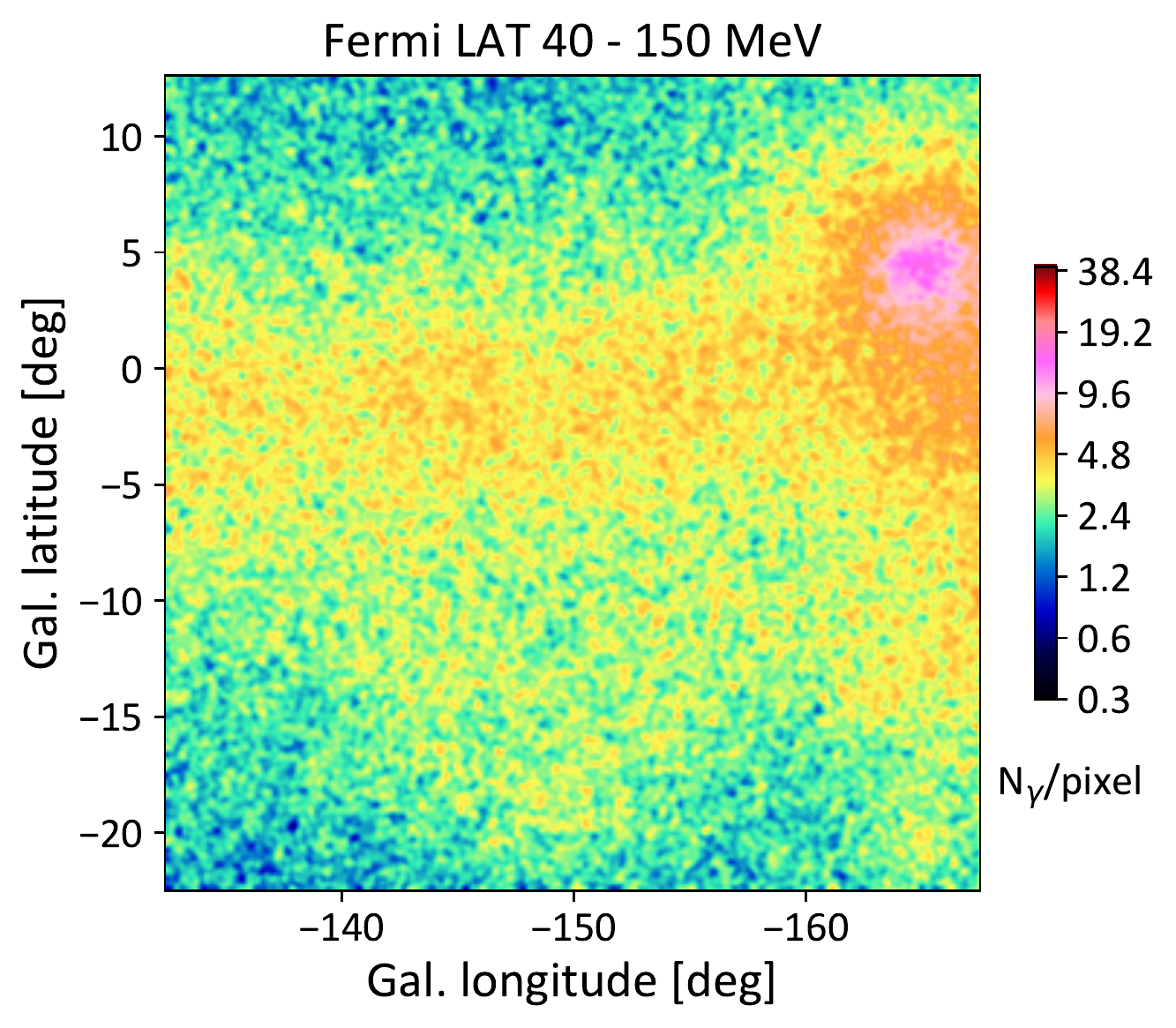}
\caption{\small Photon count maps of the Galactic disc and Orion clouds in the 0.8-4 GeV (upper panels) and 40-150 MeV (lower panels) energy bands, simulated for e-ASTROGAM for one year of effective exposure (left panels) and compared to four years of \fermilat data in scanning mode (right panels).}
\label{CR_ISM:fig1}
\end{figure}
\paragraph*{Expected results with e-ASTROGAM}
Fig.~\ref{CR_ISM:fig1} illustrates that e-ASTROGAM can easily detect and resolve the 10 MeV to GeV emission from the Galactic ISM, with evident benefits over the current \fermilat data. One can better resolve local cloud structures above 100 MeV to probe the penetration and pressure of GeV CR nuclei down to the 0.5-pc scale of dense molecular cores. The improved sensitivity enables comparisons of the CR content of tenuous cirrus clouds and of  massive clouds with stronger and more entangled magnetic fields. Measurements of the \brem intensity around 50 MeV allow firm derivations of the CR electron spectra at very low momenta at the 10-pc scale of the DNM and faint CO cloud envelopes where CRs take over other agents to heat and ionize the gas. Access to the \ic emission below 10 MeV from the same electrons that produce the radio and microwave synchrotron radiation provides tight constraints on the magnetic field strength and on the CR electron distribution inside clouds.

The gain in sensitivity and in spatial separation of the different phases of a cloud should enable the first estimation of \xco gradients across the molecular parts to shed light on the relative efficiencies of the formation and photodissociation of CO molecules as the \hd gas becomes denser \cite{Bertram16,Remy17}. Gauging the importance of these gradients is essential to determine reliable \hd masses in Galactic and extragalactic clouds. They cannot be explored with dust emission because of the strong evolution of the grains with gas density. 
e-ASTROGAM will enable studies of a variety of clouds within a couple of kiloparsecs to shed light on \xco gradients within clouds and on \xco trends with cloud state. Tighter constraints on the mass hidden in the DNM interface will bring clues to its apparent scaling with the \hd mass present in the CO-bright parts \cite{Grenier05,Remy18a}. 

The recent finding of a gradual, 4 to 6-fold rise in dust emission cross section with increasing gas density \cite{Remy17} limits the use of dust emission as a gas tracer. The improved angular resolution of e-ASTROGAM at GeV energies will allow to follow dust evolution per gas nucleon to smaller scales in the dense molecular cores where grain evolution should be stronger \cite{Kohler15}. 

At larger scales, the superior resolving power of e-ASTROGAM will be central to study remote clouds to explore the large-scale gradients in \xco ratios, DNM abundances, and in dust properties due to metallicity changes across the Milky Way.
It will also open the way to study differences in \xco ratios and in DNM abundances between clouds compressed in spiral arms and clouds sheared after their passage through an arm \cite{Smith14}. It will shed light on the 2 to 3-fold difference found between \xco values measured at parsec scales locally and at kiloparsec scales in spiral arms \cite{Remy17}. 
Explorations of clouds in extreme environments will leap forward, for instance in the Central Molecular Zone or in starburst regions where the enhanced magnetic fields, intense stellar radiation fields, high levels of turbulence and shearing, and large CR fluxes should modify the cloud states. 

Difficulties rest in that all gas tracers are non-linear and suffer from spatial confusion inside a cloud complex and along the line of sight. The improved angular resolution of e-ASTROGAM at GeV energies and its coverage extending down to MeV energies will bring a wealth of new information on the subtle interplay between CRs and the ISM. 

\newpage
\section[Fundamental physics\\
\noindent
\textnormal{\small\textnormal{Convenors:} \textit{J.~Conrad, M.~Martinez, U.~Oberlack}}]
{Fundamental physics}
The topic of fundamental physics in the context of high-energy astrophysics is often related to fundamental symmetries of nature which can be studied over cosmological distances, at high energies and in extreme environments. \\
Gamma-rays as a probe have been used for a variety of subjects in fundamental physics, the  most studied question for gamma-ray observations in general and for e-ASTROGAM in particular being the quest for DM. The exploration of topics in fundamental physics that can be addressed with e-ASTROGAM is gaining momentum and the present report gives a snapshot of current ideas: axion-like particles and primordial black-holes as well as possible observations elucidating the question of matter-antimatter asymmetry and, last but not least, different aspects of searches for DM particles with some focus on small masses, on which this introduction is focused.\\
The existence of DM is by now established beyond reasonable doubt, see e.g. \cite{bergstrom12,Ade:2015xua}, however its nature is one of the most pressing questions in science today.  One of the most popular DM candidates are weakly interacting massive particles (WIMPs), with masses and coupling strengths at the electroweak scale. Besides the fact that many of these are theoretically very well motivated, such as the supersymmetric neutralino \cite{Jungman:1995df}, an attractive feature of this class of candidates is that the observed DM abundance today can straight-forwardly be explained by the thermal production of WIMPs in the early universe. WIMPs are searched for by a variety of techniques:  directly by placing sensitive detectors in underground locations with the aim to detect WIMP-induced nuclear recoils and indirectly by detecting the secondary products of WIMP annihilation or decay.
WIMP candidates can also be produced at the Large Hadron Collider (LHC) by proton-proton collisions, which then would need to be confirmed by astrophysical observations.  The latest LHC results, based on almost 40 fb$^ {-1}$ of data at $\sqrt{s} $= 13 TeV (e.g. \cite{Aaboud:2017yqz}) did not reveal any sign of WIMP DM, in indirect detection the Fermi Large Area Telescope managed to push the sensitivity below the canonical thermal WIMP cross-section for WIMPs in the mass range from about 5 to 100 GeV without firmly confirmed detection. There is, however, significant remaining uncertainty, e.g.,  on DM distribution, which motivate further searches. Direct detection, mainly lead by deep underground liquid xenon time projection chambers, has improved sensitivity by two orders of magnitude in the last decade without any DM evidence, see e.g. 
\cite{Liu:2017drf,Aprile:2017iyp}. \\While clearly it is too early to abandon the WIMP paradigm, especially in the view of experimental programs in the next five years, the community has started to shift focus to alternative models for DM. \\One particularly interesting, and experimentally largely unexplored region is DM masses at or below the GeV scale. For example, it was pointed out that thermal production is also an attractive option for smaller DM masses \cite{Feng:2008ya}. Other relevant DM models with (sub-)GeV masses include light gravitino DM~\cite{Takayama:2000uz} inelastic DM \cite{TuckerSmith:2001hy}, light scalar DM \cite{Boehm:2003hm} or secluded DM \cite{Pospelov:2007mp}. Recently, an anomaly in the absorption profile at 78 MHz in the sky-averaged spectrum \cite{Bowman2018} has been interpreted as an excess cooling of the cosmic gas induced by its interaction with DM particles of mass lighter than few GeV \cite{Barkana2018}.\\
In the interest of avoiding duplication, we will introduce some common concepts and notation for the indirect search for DM with e-ASTROGAM. \\
The differential (Eq.\,\ref{e:dm:flux:diff}) or integrated (Eq.\,\ref{e:dm:flux}) flux of gamma-rays resulting from DM annihilation in celestial sources is given by:
\begin{equation}\label{e:dm:flux:diff}
\frac{d\Phi_{\gamma}}{dE_{\gamma} d\Omega} = \frac{a \langle \sigma v \rangle J}{4\pi m_{\chi}^{2}}\frac{dN_{\gamma}}{dE_{\gamma}}
\end{equation}
\begin{equation}\label{e:dm:flux}
\Phi_{\gamma}(\Delta\Omega) = \frac{a}{4\pi}\frac{\langle \sigma v \rangle}{m_{\chi}^{2}}\int^{E_{max}}_{E_{min}}\frac{dN_{\gamma}}{dE_{\gamma}}dE_{\gamma} \times \bar{J} \, \Delta\Omega
\end{equation}
\begin{equation}\label{e:dm:dNdE}
\frac{dN_{\gamma}}{dE_{\gamma}} = \sum_f{B_f\frac{dN_{\gamma}^f}{dE_{\gamma}}} 
\end{equation}
where $a$ denotes a numerical factor that equals either $a=1/2$ if $\chi$ is a Majorana particle (e.g., for a neutralino WIMP, with $\chi \chi \rightarrow f\bar{f}, \ldots$) or $a=1/4$ if $\chi$ is a Dirac particle ($\chi \bar{\chi} \rightarrow f\bar{f},\ldots$). $\langle \sigma v \rangle$ is the self-annihilation cross-section averaged over the local relative velocity of DM particles in the observed object and, for thermal relics, is compared to the thermally averaged self-annihilation cross-section times velocity in the early universe of approximately $3 \times 10^{-26} \text{cm}^{3} \text{s}^{-1}$\,\cite{Steigman:2012nb}. $m_{\chi}$ is the DM particle mass, $E_{min}$ and $E_{max}$ are the energy limits for the measurement and $\frac{dN_{\gamma}}{dE_{\gamma}}$ is the energy spectrum of the gammas produced in the annihilation (see, e.g., \cite{Cirelli:2010xx}), which depends on the DM model, defining its coupling to Standard Model (SM) particles, as well as the DM mass. The products of DM annihilation are thought to come from decay and/or hadronization of the primary SM particles produced in the annihilation: quark-antiquark, lepton and boson, etc., and each channel is expected to have its own branching ratio $B_f$ with photon yield per annihilation of $\frac{dN_{\gamma}^f}{dE_{\gamma}}$. Examples of DM annihilation spectra are discussed in Sec.~\ref{s:fp:addazi}, \ref{s:fp:brdar},\ref{s:fp:bringmann}, \ref{s:fp:bartels}, \ref{s:fp:vankova}.\\
The astrophysical $J$-factor is the integral of the squared DM density profile along the line of sight to the target, typically dominated by the DM density of the target itself, and often contains substantial uncertainty. Sometimes $J$ is defined as integrated or averaged over an aperture angle. Here we use the notation:
\begin{eqnarray}\label{e:dm:J-factor:ann}
 J &=& \int_{l.o.s.}\rho^{2}_{DM}(\boldmath{r})\,\text{d}s \\
 \bar{J} &=& \frac{1}{\Delta\Omega}\int_{\Delta\Omega}\int_{l.o.s.}\rho^{2}_{DM}(\boldmath{r})\,\text{d}s\,\text{d}\Omega^{'}. \nonumber
\end{eqnarray} 
For models of DM particles decaying on time scales of the Hubble time, the expected flux is modified to 
\begin{equation}\label{e:dm:flux:dec}
  \Phi_{dec}(\Delta\Omega)=\frac{1}{4\pi}\frac{1}{m_{\chi}\tau_{\chi}}\int^{E_{max}}_{E_{min}}   \frac{dN_{\gamma}}{dE_{\gamma}}dE_{\gamma} \times \bar{J}_{dec} \, \Delta\Omega,
\end{equation}
with a modified $J$-factor integrated only linearly over the DM density:
\begin{equation}\label{e:dm:J-factor:dec}
 \bar{J}_{dec} \, \Delta\Omega = \int_{\Delta\Omega}\int_{l.o.s.}\rho_{DM}(r)\,\text{d}s\,\text{d}\Omega^{'}.
\end{equation} 
Targets for searches for DM are commonly those of enhanced DM density: the Milky Way galaxy, including  the  GC, dwarf galaxies and groups of galaxies, as well as galaxy clusters. The GC is by orders of magnitude the largest potential source of signal from DM annihilation. Dwarf spheroidal galaxies provide the cleanest target with the potential to derive the DM distribution from spectral velocities and are (unlike the GC) essentially free from conventional sources or diffuse backgrounds that could hamper an identification of DM induced signal. Galaxy clusters are potential targets if a substantial fraction of DM is in substructures. Diffuse backgrounds, such as the Galactic and extragalactic backgrounds, are promising targets, especially exploiting angular autocorrelation or in cross-correlation with other wavelengths, like for example with galaxy catalogues. For a more detailed review of challenges and opportunities of different gamma-ray signatures and techniques, see e.g. \cite{Conrad:2015bsa,Gaskins:2016cha}.
\subsection[Limiting MeV-ish dark matter decays: light WIMPs, dark photons, majorons\\
\noindent
\textit{\small{A. Addazi, D. Bastieri, A. Marcian\`{o}}}]{Limiting MeV-ish dark matter decays: light WIMPs, dark photons, majorons}
\label{s:fp:addazi}
\paragraph*{Science questions}
The lack of evidence of WIMPs in the canonical mass range motivates the pursuit of new experimental constraints in order to test non-standard candidates of DM. For example, instead of considering masses of $10\, {\rm GeV}\div 1\, {\rm TeV}$, it is still an open and viable possibility to have lighter DM particles composing the halo. In particular, the mass window $1\div 100\, {\rm MeV}$ requires new kinds of direct and indirect detection experiments with respect to the current ones. 
Here we suggest to test MeV-ish DM decays with e-ASTROGAM. The idea is not only to use the e-ASTROGAM data to probe standard astrophysical objects, but also to obtain useful information in understanding particle physics. The presence of MeV-ish DM is highly motivated within the context of many different extensions of the Standard Model. 
For instance, within the WIMP paradigm one can consider mechanisms for the genesis of non-thermal DM that favor lighter WIMP candidates than the thermal WIMP miracle ones. If WIMPs are indeed MeV-ish, they can decay into light SM particles, and in particular into photons.  Another possible model which may be tested is the massive dark photon model. We also mention here the possibility to test majoron DM, which naturally favors light particles while explaining neutrino mass generation. Finally, the presence of MeV-ish DM can be related to dark first order phase transitions that produce a stochastic gravitational waves background. This is a novel multi-messenger approach to address new physics by comparing gamma-rays observations with gravitational radiation \cite{Addazi:2016fbj,Addazi:2017gpt,Addazi:2017oge}. 

\paragraph*{Importance of gamma-ray observations}
Depending on the mass of the DM particle, WIMPs can annihilate into several different channels: a photon pair $\chi\chi \rightarrow \gamma\gamma$, a neutral pion and photon $\chi \chi \rightarrow \pi^{0}\gamma$, a neutral pions pair $\chi\chi \rightarrow \pi^{0}\pi^{0}$, light lepton-antilepton pairs (electron, muons, neutrinos) $\chi \chi \rightarrow l\bar{l}$ and more complicated cascade annihilations. The primary component is constituted by all the photons in the final state directly arising from annihilation of WIMPs \cite{LWIMP1,LWIMP2,Bartels:2017dpb,squared}. 
In principle the annihilation of light WIMPs can then be detected. The expected flux grows as the square of the energy density, i.e. a higher signal is expected in places with the highest DM density. For instance, in the GC the density profile roughly grows as a power law $\rho(r)\sim r^{-\gamma}$, with $\gamma$ a fit parameter.
The greatest uncertainty is contained in the $J$-factor of Eq. \,\ref{e:dm:J-factor:ann}, here defined as $J=\int_{l.o.s}\,d s \,\rho^{2}(s,\theta)$, where $s$ is the distance along the line of sight and $\theta$ is the l.o.s. angle. This, in turn, is defined by the relation $r^{2}=s^{2}+R_{0}^{2}-2sR_{0}\cos \theta$, in which $R_{0}\sim 8\,{\rm kpc}$ represents the solar distance from the GC. 
The density profile is affected by many astrophysical uncertainties, and is usually parametrized in an analytic form as 
\begin{equation}
\rho(r)=\frac{\rho_{0}}{(r/R)^{\gamma}[1+(r/R)^{\alpha}]^{(\beta-\gamma)/\alpha}}\, ,
\end{equation}
where $\alpha,\beta,\gamma$ are model parameters, $R$ is the characteristic length scale and $\rho_{0}$ is the local DM density, approximately $0.4\, {\rm GeV}\,{\rm cm}^{-3}$. For instance, possible models are the Navarro-Frenk-White (NFW) model~\cite{1996ApJ...462..563N} $(\alpha=1.0,\beta=3.0,\gamma=1.0,R=20\,{\rm kpc})$, the Moore model~\cite{1998ApJ...499L...5M} $(\alpha=1.5,\beta=3.0,\gamma=1.5,R=28\,{\rm kpc})$ and the isothermal halo model $(\alpha=2.0,\beta=2.0,\gamma=0,R=3.5\,{\rm kpc})$.
For primary $\chi\chi \rightarrow \gamma\gamma$ annihilation, the photon spectrum is expected to be a spike in the flux spectrum, namely ${dN_{\gamma}}/{dE}=2\delta(E-m_{\chi}).$ For primary $\chi\chi \rightarrow \pi^{0}\pi^{0}$, the chiral anomaly induces the subsequent decays 
$\pi^{0}\rightarrow \gamma\gamma$, i.e. the flux spectrum can be modeled as a box-like distribution \cite{squared},
$$\frac{dN_{\gamma}}{dE}=\frac{4}{E_{+}-E_{-}}\Theta(E_{+}-E)\Theta(E-E_{-})\,, \qquad {\rm with} \qquad
E_{\pm}=\frac{m_{\chi}}{2}\left(1\pm \sqrt{1-\frac{m_{\pi}^{2}}{m_{\chi}^{2}}}\right)\, .$$
The $\chi\chi\rightarrow \pi_{0}\gamma$ decay's contribution has to appear out of the spectrum as a spike that is over-imposed on the box spectrum contribution \cite{squared} to the photons flux spectrum, namely
$$\frac{dN_{\gamma}}{dE}=\delta(E-E_{0})+\frac{2}{\Delta E}\Theta(E_{+}'-E)\Theta(E-E_{-}')\, , $$
where 
$$E_{0}=m_{\chi}-\frac{m_{\pi_{0}}^{2}}{4m_{\chi}},\,\,\,\Delta E'=m_{\chi}-\frac{m_{\pi_{0}}^{2}}{4m_{\chi}} \qquad {\rm and} \qquad E_{\pm}'=\frac{m_{\chi}}{2}\left[\left(1+\frac{m_{\pi^{0}}^{2}}{4m_{\chi}^{2}}\right)\pm \left(1-\frac{m_{\pi^{0}}^{2}}{4m_{\chi}^{2}}\right)\right]\, .$$
The estimate of the secondary emission processes requires a much more involved numerical analysis.  This is the case of $\chi\chi\rightarrow e^{+}e^{-}$ processes, in which we can have a large contribution from Bremsstrahlung emission. 

\paragraph*{Expected results with e-ASTROGAM}
The {\it dark photon} model extends the Standard Model (SM) as to encode an extra dark gauge sector. In the minimal model, just an extra $U(1)_{X}$ gauge group is added $G_{SM}\times U_{X}(1)$ \cite{DF1,DF2}. An interesting case consists in a minimal particle spectrum $(s,\chi, A'_{\mu})$, where $s$ is a scalar singlet, $\chi$ is a fermion charged with respect to the extra $U(1)_{X}$, while $A'$ is the dark photon. The dark photon can become massive thanks to a spontaneous symmetry breaking induced by the scalar singlet \cite{ArkaniHamed:2008qn}. In this scenario, fermions are thought as DM particles and their masses can be generated by Yukawa terms involving the singlet. A renormalizable gauge portal among DM and the standard model particles is the so dubbed {\it kinetic mixing term}, $-\epsilon F^{\mu\nu}_{(Y)}F_{\mu\nu}^{(X)}$, which mixes the SM hypercharge with the dark photon. This allows an EM-like annihilation process of dark fermions into SM particles. For instance, if we assume dark particles to be lighter than the electrons, the cross-section for this kind of processes reads
$$\sigma(\chi\chi\rightarrow \gamma\gamma)\, v=\frac{\pi \alpha'^{2}}{m_{\chi}^{2}}=\left(6.5\times 10^{-4}\,{\rm pb}\right)
\frac{\epsilon_{5}^{4}}{m_{keV}^{2}}\, ,$$
where $m_{keV}=m_{X}/keV$, $\alpha'=e'^{2}/4\pi=\epsilon^{2}\alpha$ and $\epsilon_{5}=10^{5}\epsilon$. Other more complicated processes from Bremsstrahlung emissions can be envisaged, involving a more sophisticated analysis, which is in preparation. \\
The {\it majoron} is the Nambu-Goldstone boson of a global lepton symmetry that generates a Majorana mass for the neutrino. It can be very long-living, if in the KeV-MeV mass spectrum range, hence providing a natural candidate for DM. At the upper end of this mass scale, it could produce primary gamma-ray emission detectable by e-ASTROGAM. In various model of neutrino mass generated with a spontaneous 
symmetry breaking of the global lepton number symmetry, majorons are coupled with photons with a dimension 5 operator like 
$$
g_{J \gamma\gamma} J\epsilon^{\nu\mu\rho\sigma}F_{\nu\mu}F_{\rho\sigma}\, ,
$$
in which $[g_{J\gamma\gamma}]=M^{-1}$ is a negative dimensional coupling and $J$ the majorons' field. This means that majorons have to decay radiatively into two photons (compare equation~\ref{e:dm:flux:dec}), each one carrying an energy $E_{\gamma}\simeq m_{J}/2$ --- the decay can be studied as if it were happening in the DM rest frame with very good approximation. In order to recover the CMB constrains, the decay rate of the majoron must be $\Gamma_{J}<\zeta \times 2.4\times 10^{-25}\, {\rm s}^{-1}\, ,$ where $\zeta$ is the inverse efficiency factor that describes how much decay energy is deposited on baryons. This opens a pathway to test long-living metastable majorons from indirect detection in the KeV-MeV region.

\subsection[MeV dark matter complementarity\\
\noindent
\textit{\small{M. Dutra, M. Lindner, S. Profumo, F. S. Queiroz, W. Rodejohann, C. Siqueira}}
]{MeV dark matter complementarity}
\label{s:fp:dutra}
\paragraph*{Science questions}
We have learned from the success of the Big Bang Nucleosynthesis (BBN) and the Cosmic Microwave Background (CMB) that the particles of the Standard Model were once in thermal equilibrium. Therefore, it is quite plausible to assume that the DM particle also belonged to a thermal history. That said, the detection of a gamma-ray signal would constitute a strong evidence for thermal production of DM in the early universe. In this context, WIMPs are regarded as predictive and the most compelling DM candidates, but they have been severely constrained by data due to non-detection of any signals. However, the reason behind WIMPs predictivity is the thermal production. In the GeV-TeV energy range several gamma-ray experiments have played an important role, but in the MeV or sub-GeV scale, there is a notorious lack of experimental results. Thus, the e-ASTROGAM mission is much needed. It will allow us to test the thermal production of DM at the MeV scale, as well as many other interesting scenarios.
\paragraph*{Importance of gamma-ray observations}
MeV DM is hardly detectable at colliders, but can still leave traces at direct and indirect detection experiments. The experimental limits from direct detection experiments are not very stringent and probe a different observable, namely the DM scattering cross section off electrons or nuclei. The existing limits from the CMB are rather restrictive \cite{Slatyer2016}. Such limits rely on the fact that DM annihilations at early times, between the period of recombination and reionization, could inject EM radiation in the intergalactic medium. This process may significantly modify the CMB power spectrum, precisely measured by the Planck satellite, leading up to strong bounds only for DM masses above $10$~MeV. Therefore, in light of the difficulty of probing DM particles below $10$~MeV, an instrument capable of observing gamma-rays at low energies is paramount to test DM models and potentially discover MeV DM. 

In Fig.\ref{fig1:profumo}, we show the expected sensitivity of the e-ASTROGAM mission to DM annihilations into electron-positron pairs compared to the existing one from the Planck satellite. In Fig.\ref{fig1:profumo}, $m_{\chi}$ is the DM mass. This expected sensitivity of the e-ASTROGAM mission to DM annihilation was derived in ~\cite{Bartels:2017dpb}, where the region of interest was chosen to be the GC. A local DM density of $0.4$ GeV/cm$^3$, a Navarro-Frenk-White density profile and systematic uncertainties similar to those present at the Fermi-LAT telescope were crucial assumptions in the study. Based on these assumptions, one can clearly notice from Fig.\ref{fig1:profumo} that e-ASTROGAM constitutes a complementary and independent probe for DM masses above $10$~MeV and a discovery machine for smaller masses.
\begin{SCfigure}
\includegraphics[width=0.48\textwidth]{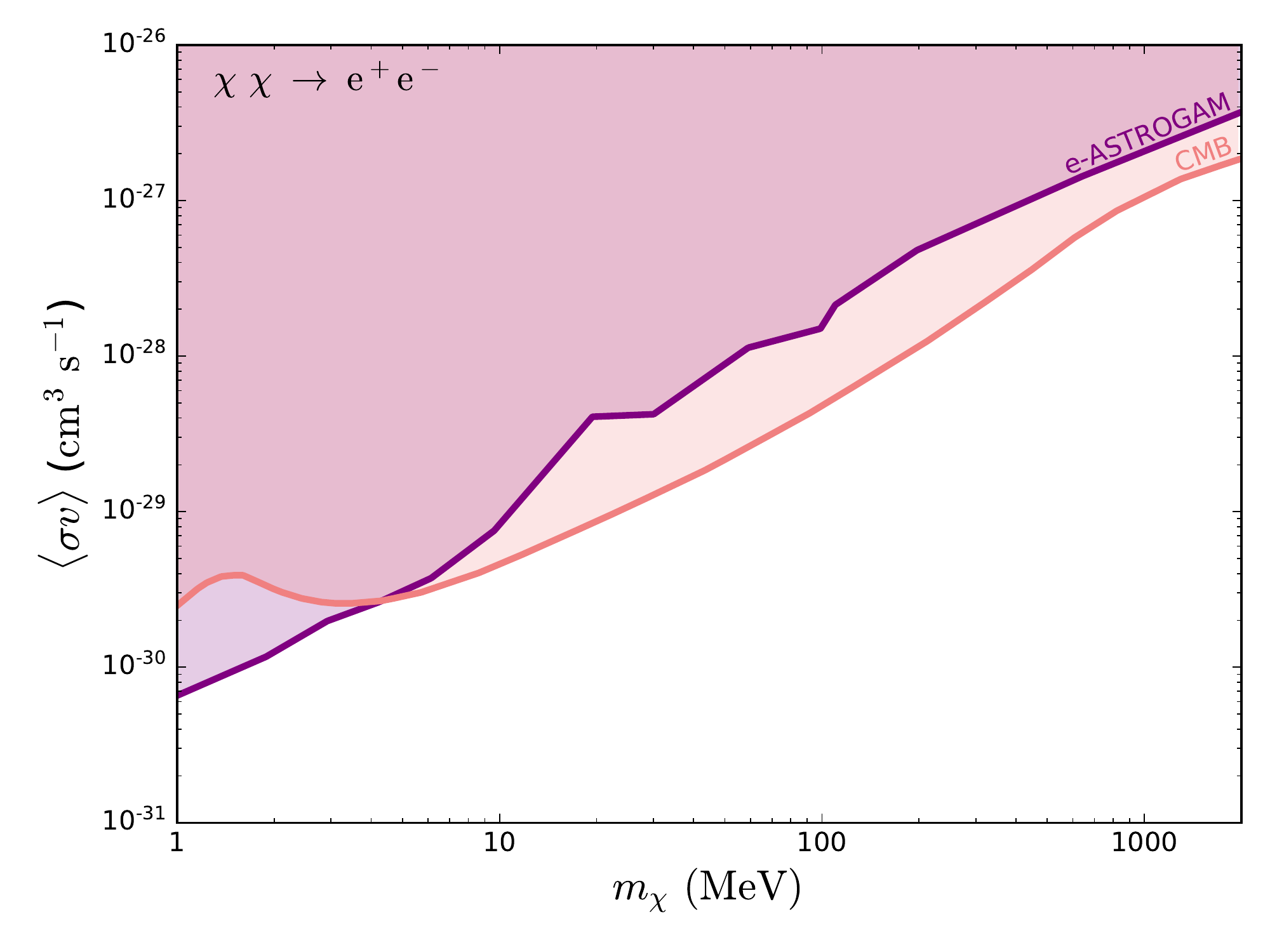}
\caption{\small
Model independent limits on the DM annihilation cross section as a function of the DM mass, $m_{\chi}$, based on the existing CMB data (red curve) \cite{Slatyer2016} and the e-ASTROGAM mission (purple curve) \cite{Bartels:2017dpb}. It is visible that e-ASTROGAM will furnish limits competetive to those from the CMB and have the potential to discover DM below $\sim 10$~MeV. See \cite{Dutra:2018gmv} for details.}
\vspace{-10pt}
\end{SCfigure}\label{fig1:profumo}
\paragraph*{Expected results with e-ASTROGAM}
Since e-ASTROGAM will be particularly sensitive to DM masses at the MeV scale, we illustrate its importance to MeV DM in the popular dark photon model by putting the results in perspective with several other existing bounds from accelerators, colliders and direct detection experiments. Assuming the DM particle to be Dirac fermion that interacts with the standard model particles via dark photon, $A^\prime$, which features a kinetic mixing with the photon, the Lagrangian that describes this model reads \cite{Dutra:2018gmv},
\begin{eqnarray}
\mathcal{L} \supset && -\frac{1}{4}F_{\mu \nu}F^{\mu \nu} -\frac{1}{4}F^\prime_{\mu \nu}F^{\prime\mu \nu} + \frac{1}{2} M_{A^\prime}^2 A'^2 \nonumber\\
&& + \sum_i \overline{f}_i(-e q_{f_i} \partial_\mu A^\mu - \varepsilon e q_{f_i}  \partial_\mu A^{\mu \prime}-m_{f_i})f_i \nonumber\\
&& + \overline{\chi}(-g_D  \partial_\mu A^{\mu \prime} - m_\chi)\chi\,,
\end{eqnarray}
where $m_{f_i}$ and $M_{A^\prime}$ are the standard model fermion and dark photon masses, respectively, $F^{\mu \nu}$ and $F^{\prime\mu \nu}$ are the fields strength tensors of the photon $A$ and of the dark photon $A^\prime$, $g_D$ is the coupling constant between the dark photon and the DM, and $\varepsilon e$ the dark photon interaction strength to the standard model fermion of charge $q_{f_i}$.

By solving the Boltzmann equation we can derive the DM relic density and draw the region of parameter space that yields the correct relic density as represented by the turquoise solid curve in Fig.\ref{fig2:profumo} for $g_D=0.1$ (left-panel), and $g_D=1$ (right-panel). Moreover, we can compute the DM-electron scattering cross section and compare with existing limits from XENON collaboration  \cite{Essig:2015cda} to obtain the red hatched exclusion region in Fig.\ref{fig2:profumo}. Existing (projected) limits based on a multitude of accelerator as well as collider searches for dark photons are drawn with solid (dashed) lines \cite{Alexander:2016aln}. The e-ASTROGAM projected exclusion region lies in the hatched purple region. It is remarkable that e-ASTROGAM will be able to fully test the model for $g_D=0.1$ and leave only a small window for $g_D=1$ and $M_{A^\prime} =10$~MeV where one can successfully accommodate a thermal Dirac fermion as DM. Hence, e-ASTROGAM will play a crucial role in the search for MeV DM and provide an orthogonal and independent probe for dark photons masses above $\sim 10$~MeV.
\begin{figure}[h]
  \includegraphics[width=0.48\textwidth]{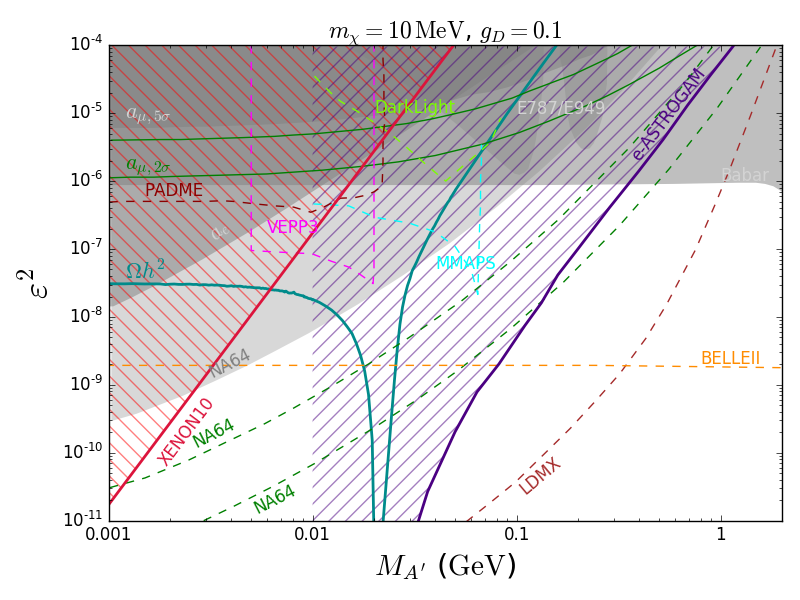}
  \includegraphics[width=0.48\textwidth]{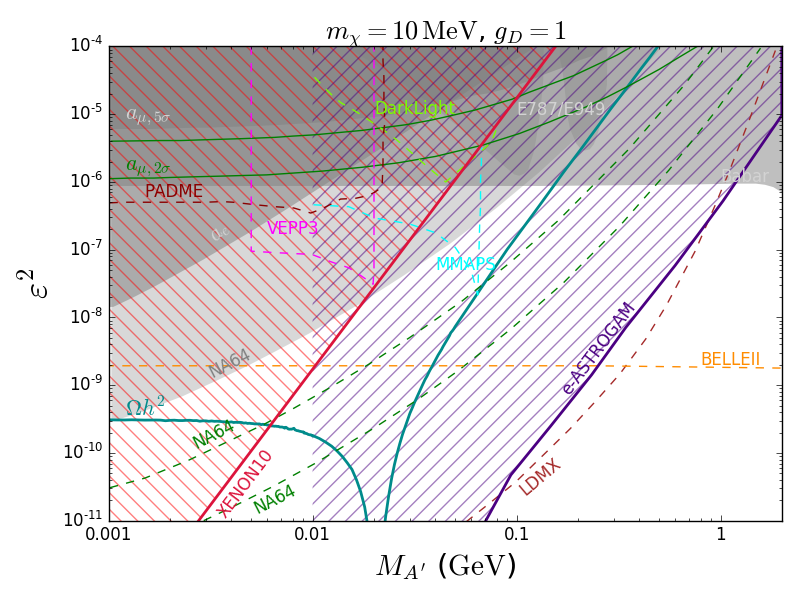}
  \caption{\small 
 {\it MeV DM complementarity:} Bounds on the plane $\varepsilon^2$ \textit{versus} dark photon mass. Direct detection and e-ASTROGAM exclusion regions are shown in red and purple hatched regions respectively. The correct relic density curve with turquoise lines for DM mass $m_{\rm \chi}=10$~MeV and two different values for the dark coupling $g_{D}$, $g_{D}=0.1$ (left panel) and $g_{D}=1$ (right panel). Existing (projected) bounds resulted from dark photon searches are displayed with gray regions (dashed lines).
  \label{fig2:profumo}
  }
\end{figure}
In summary, the e-ASTROGAM mission will be able to almost fully probe the thermal DM production mechanism in one of the most popular examples of MeV DM, a Dirac fermion, in the context of the extensively explored dark photon portal. Therefore, e-ASTROGAM will be paramount to assess unexplored MeV DM models and the thermal production of DM at the MeV scale.
\subsection[Decay or annihilation of non-thermally produced dark matter\\
\noindent
\textit{\small{V. Brdar, J. Kopp, J. Liu, A. Merle, X. Wang}}]{Decay or annihilation of non-thermally produced dark matter}
\label{s:fp:brdar}
\paragraph*{Science questions}
Searches for DM have traditionally focused on
particles around the electroweak scale, where many theoretically well motivated
DM candidates have been proposed. As these scenarios are coming under
pressure from the LHC and from direct and indirect DM searches,
scenarios with much lighter DM are entering the spotlight. Of particular
interest is the mass range from $\sim 100$\,keV
to 1\,GeV. DM particles in this range are still heavy enough
to act as Cold DM, even if the original production mechanism was non-thermal.\footnote{The intuitive picture is that, independent of the shape of the initial velocity spectrum, sufficiently heavy DM particles will cool down fast, thereby shifting all particle velocities to a value close to zero. Thus, no matter what the shape of the spectrum was originally, these DM particles could always be approximated as being essentially at rest.} However, unfortunately their masses are below the detection threshold
of typical searches for DM--nucleus scattering. 

In the early Universe, sub-GeV DM particles could in principle be produced via thermal
freeze-out. However, in many scenarios of this type, in particular those with
$s$-wave annihilation, the required DM annihilation cross sections of order
$\text{few} \times 10^{-26}\,\text{cm$^3$/sec}$~\cite{Steigman:2012nb}
is in conflict with gamma ray limits~\cite{squared} and with limits on
additional energy injection into the primordial plasma around the time of
recombination~\cite{Madhavacheril:2013cna}.
This leaves out-of-equilibrium freeze-in as a viable production mechanism~\cite{Hall:2009bx}.
In the following, we will focus on scenarios
of the latter type. Freeze-in can occur for instance
through a ``Higgs portal'' coupling of the form
\begin{align}
  \mathcal{L}_\text{Higgs-portal} = \lambda (\phi \phi)(H^\dag H)
\end{align}
between a new scalar $\phi$ and the Standard Model Higgs field $H$.
Here, $\lambda$ is a small coupling constant. $\phi$ can either be the
DM particle itself or a heavier dark sector
particle that decays or annihilates to DM at a later time (see for
instance~\cite{Merle:2013wta}).
Alternative freeze-in scenarios include $\phi$ couplings to additional new
particles, or
freeze-in through a higher-dimensional coupling such as
\begin{align}
  \mathcal{L}_\text{5d} = \tfrac{\alpha}{4 \pi \Lambda} \phi F_{\mu\nu} F^{\mu\nu} \,,
\end{align}
where $F^{\mu\nu}$ is the photon field strength tensor and $\alpha$ is
the EM fine structure constant. See Sec.~\ref{s:fp:bartels} and
ref.~\cite{Bartels:2017dpb} for a more phenomenological,
less model-dependent discussion.
All production mechanisms of MeV--GeV scale DM require the couplings between
the dark and visible sectors to be extremely weak to explain the observed
DM abundance, making direct detection and production of DM particles in
experiments at particle accelerators challenging.
It is therefore likely that such DM particles would have
escaped detection so far, and it is crucial to close this gap.

\paragraph*{Importance of gamma-ray observations}
Due to the difficulty of detecting DM particles at the MeV--GeV scale
using other means, indirect astrophysical searches are of primary interest
for them, even more so than for heavier DM.
When MeV--GeV scale DM particles decay or annihilate to Standard
Model particles, they typically leave signatures in the gamma ray sky at
precisely the right energies for e-ASTROGAM to play out its stengths.
Only few decay or annihilation channels
are available for such light DM particles: below the electron threshold
at $\sim 1$\,MeV, only decay or annihilation to photons or neutrinos
is possible. Given the small neutrino interaction rate, searches in gamma
rays are most promising in practice. At somewhat larger masses, the secondary
gamma rays radiated in decay or annihilation to electrons/positrons, muons,
or light mesons offer the most promising target for indirect searches (Sec. \ref{s:fp:bringmann}, \ref{s:fp:bartels}) and references
\cite{Bartels:2017dpb,Bringmann:2016axu}.

\begin{figure}
  \begin{tabular}{cc}
    \includegraphics[width=0.48\textwidth]{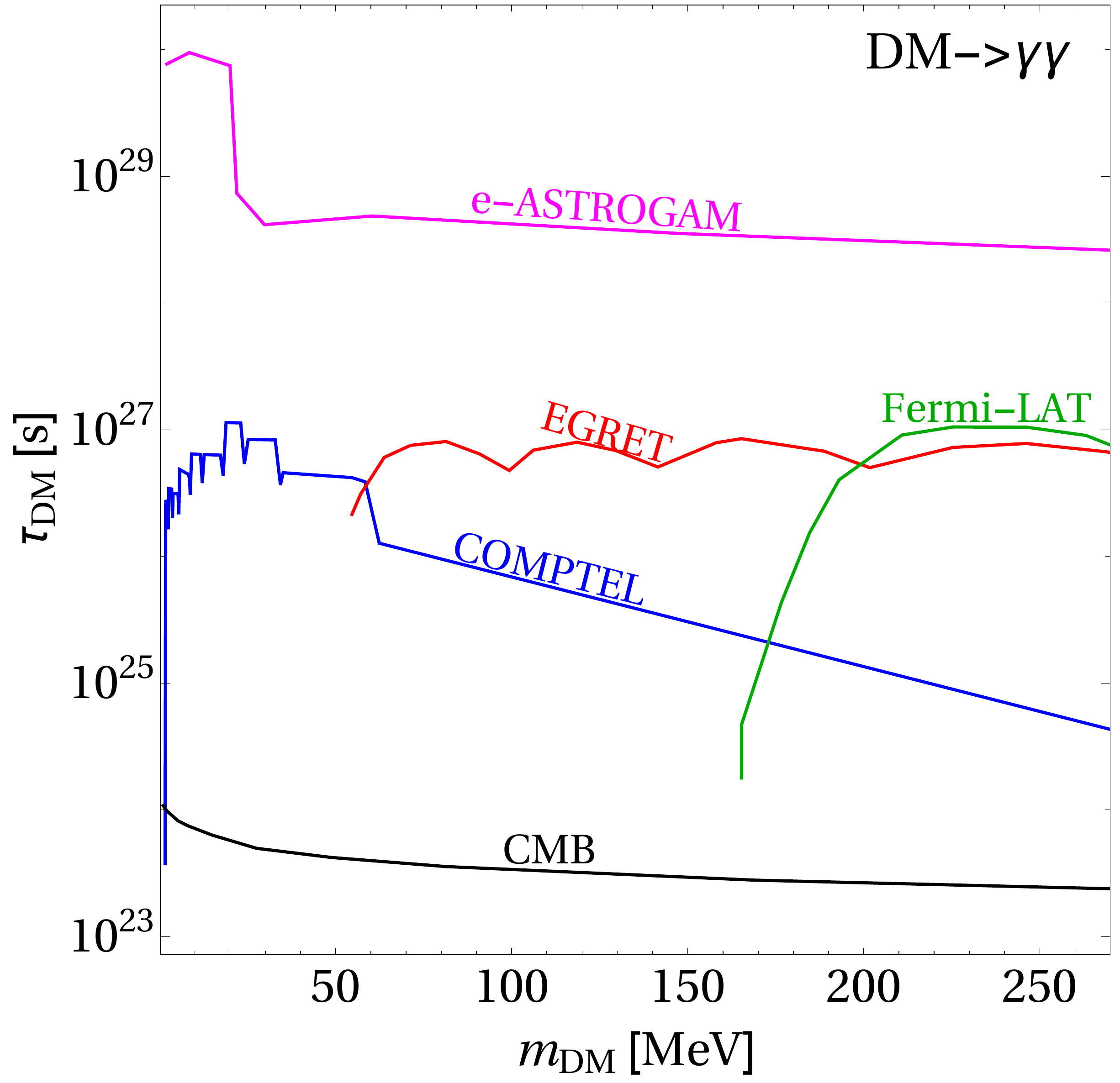} &
    \includegraphics[width=0.48\textwidth]{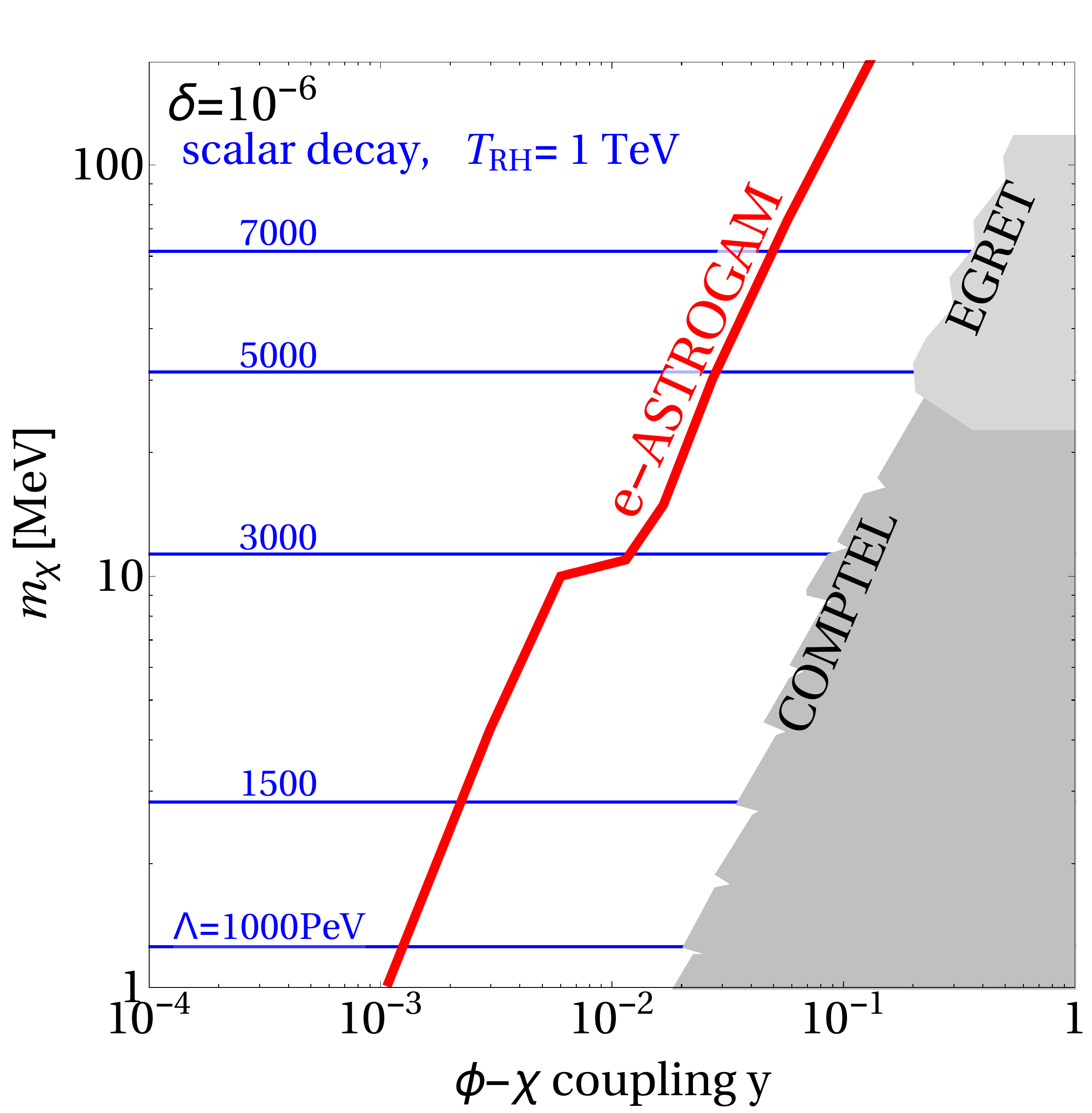}
  \end{tabular}
  \caption{\small Left: Expected sensitivity of e-ASTROGAM to
    DM decay to photons (obtained by converting the limits on
    DM annihilation presented in \cite{Bartels:2017dpb} and Sec.~\ref{s:fp:bartels}),
    compared to existing constraints (based on \cite{squared}).
    Right: Expected e-ASTROGAM constraints on the parameter
    space of the DM model from Ref.~\cite{Brdar:2017wgy}.
    }
  \label{fig:sensitivitydm}
\end{figure}

The three main classes of gamma ray signatures expected from MeV--GeV scale DM
are
\begin{enumerate}
  \item {\bf Continuum photons from final state radiation.}
    If DM decays or annihilates to charged final state particles, the radiative
    production of photons from the final state leads to peaked
    spectra at energies somewhat below the DM mass ~\cite{Bartels:2017dpb}.
    
  \item {\bf Mono-energetic photons.}
    There is a multitude of particle physics scenarios predicting this signature.
    The simplest
    example is perhaps a fermionic DM candidate $\chi$ (``sterile neutrino'')
    mixing with neutrinos. If DM
    is a fermion that does not carry gauge charges, the corresponding coupling
    $y \bar{L} (i \sigma_2 H^*) \chi$ (with $L$ a SM lepton doublet and
    $\sigma_2$ a Pauli matrix) is not forbidden by any symmetry. It is therefore
    generically expected to be present and leads to the decay
    $\chi \to \nu\gamma$ via a $W$--charged lepton loop. For scalar
    or pseudoscalar DM $\phi$, direct decay to photons may be possible via an
    effective coupling of the form $\tfrac{\alpha}{4\pi\Lambda} \phi F_{\mu\nu}
    F^{\mu\nu}$. Such a coupling will be induced for instance if DM couples
    to super-heavy charged particles. In fact, the decay rate
    \begin{align}
      \Gamma_{\phi \to \gamma\gamma}
        = 2.4 \times 10^{24}\,\text{sec} \times
              \bigg(\frac{\text{MeV}}{m_\phi} \bigg)^3
              \bigg(\frac{\Lambda}{10^{16}\,\text{GeV}} \bigg)^2
    \end{align}
    suggests that in particular
    DM couplings to particles around the Grand Unification Scale --- where
    we would generically expect such couplings --- are of interest here.
    
  \item {\bf Box-shaped spectra.}
    If DM decays or annihilates to neutral pions, or to new
    intermediate particles that decay onward to photon, the expected
    gamma ray spectrum is box-shaped. For instance, in Ref.~\cite{Brdar:2017wgy},
    a simple and successful scenario has been presented in which fermionic DM
    $\chi$ annihilates to a scalar $\phi$ that is long-lived, but
    eventually decays to photons.
    Note that, if $\chi$ and $\phi$ are nearly degenerate in this scenario,
    the box-shaped spectrum reduces again to a monochromatic one.
    Near mass-degeneracy
    of $\chi$ and $\phi$ could be understood for instance if nature is fundamentally
    supersymmetric and the two particles are members of the same supermultiplet.
    An interesting aspect of scenarios with long-lived intermediate
    particles, which travel over astrophysical distance scales before decaying,
    is that the morphology of the gamma ray signal may not directly
    trace the DM distribution in the observation target. Rather, it will
    be smeared out compared to the DM distribution.
\end{enumerate}

\paragraph*{Expected results with e-ASTROGAM}
With its superior sensitivity to gamma ray signals at MeV--GeV energies,
e-ASTROGAM will significantly extend the sensitivity to DM particles at this
mass scale. The mission thus has the potential to play a similarly transformative
role as Fermi has played for DM at larger mass scales.
Across all decay or annihilation final states, an improvement of the sensitivity
by several orders of magnitude is expected compared to current constraints,
as shown in detail in Sec.~\ref{s:fp:bringmann}, \ref{s:fp:bartels} and Refs.~\cite{Bartels:2017dpb,Bringmann:2016axu}).
In Fig.~\ref{fig:sensitivitydm}, we illustrate this
for two test cases: decaying scalar DM (left panel) and two-step
annihilation $\bar\chi\chi \to \phi\phi \to 4\gamma$ in the context
of the model presented in Ref.~\cite{Brdar:2017wgy}.

\subsection[Smoking gun dark matter signatures in the MeV range\\
\noindent
\textit{\small{T. Bringmann, A. Hryczuk, A. Raklev, I. Str\"umke, J. Van den Abeele}}]{Smoking gun dark matter signatures in the MeV range}
\label{s:fp:bringmann}
\paragraph*{Science questions}

Among the most favourite DM candidates are WIMPs, with masses 
and coupling strengths at the electroweak scale. Besides the fact that many of these are theoretically very 
well motivated, such as the supersymmetric neutralino \cite{Jungman:1995df}, an attractive feature of this 
class of candidates is that the observed DM abundance today can straight-forwardly be explained by the 
thermal production of WIMPs in the early universe. In recent years however -- triggered not the least by the 
lasting  absence of any undisputed WIMP signals, despite immense experimental efforts -- the focus of the 
community has started to shift beyond WIMPs as the main DM paradigm.

For example, it was pointed out that thermal production is also an attractive option for smaller DM 
masses \cite{Feng:2008ya}. Other relevant DM models with (sub-)GeV masses include light 
gravitino DM~\cite{Takayama:2000uz},
inelastic DM \cite{TuckerSmith:2001hy}, light scalar DM \cite{Boehm:2003hm} or secluded 
DM \cite{Pospelov:2007mp}. Models in this mass range have received significant interest because they 
could have easily escaped the 
ever more stringent constraints from direct DM detection experiments (for a suggestion of how to overcome 
the lack of sensitivity of traditional methods in this mass range, see e.g.~Ref.~\cite{Essig:2011nj}). From the 
indirect detection perspective, an intriguing feature of such models is furthermore that the center-of-mass 
energy, and hence the energy of final state quarks, is at the same mass scale as standard model hadronic 
states. As we argue in this contribution, this can lead to a potentially rich phenomenology in MeV gamma 
rays that may allow to draw far-reaching conclusions about the nature of the DM particles and the 
underlying theory.

\paragraph*{Importance of gamma-ray observations}
Gamma rays from both decaying and annihilating DM have sometimes been argued to be the {\it golden 
channel} of indirect DM searches \cite{Bringmann:2012ez}  because they directly point back to their 
sources and hence provide the potentially most accurate way to probe the astronomically observed DM 
distribution {\it in situ}. Furthermore, they may carry distinct spectral features that can both act as 
`smoking gun' signals for the particle nature of DM and convey further detailed information about the 
nature of these particles.

Motivated by the WIMP case, the main focus has traditionally been on spectral features in the 
100\,GeV -- TeV range, with relevant limits presented e.g.~in Ref.~\cite{Ackermann:2015lka}; also exotic 
line contributions in the keV range have been scrutinized in detail, where a signal could be expected 
from decaying sterile neutrino DM \cite{keVline}. Here, we point out that also the largely neglected 
MeV range is very well motivated in this respect (for earlier work, see 
Refs.~\cite{oldMeVfeatures,Bringmann:2016axu}), and hence ideally suited for searches with e-ASTROGAM.

In fact, gamma-ray and cosmic microwave background observations already put significant constraints on light DM 
candidates, and e-ASTROGAM would imply an additional boost in sensitivity (Sec.\,\ref{s:fp:dutra}, \ref{s:fp:bartels}, \ref{s:fp:antonelli}). 
As we show here, hadronic final states from DM decay or annihilation could furthermore lead to a 
plethora of potential smoking-gun signatures for a DM signal in MeV gamma rays that only a dedicated 
mission like \mbox{e-ASTROGAM} may be able to detect.

\paragraph*{Expected results with e-ASTROGAM}

\begin{figure}[t]
\includegraphics[width=0.49\textwidth]{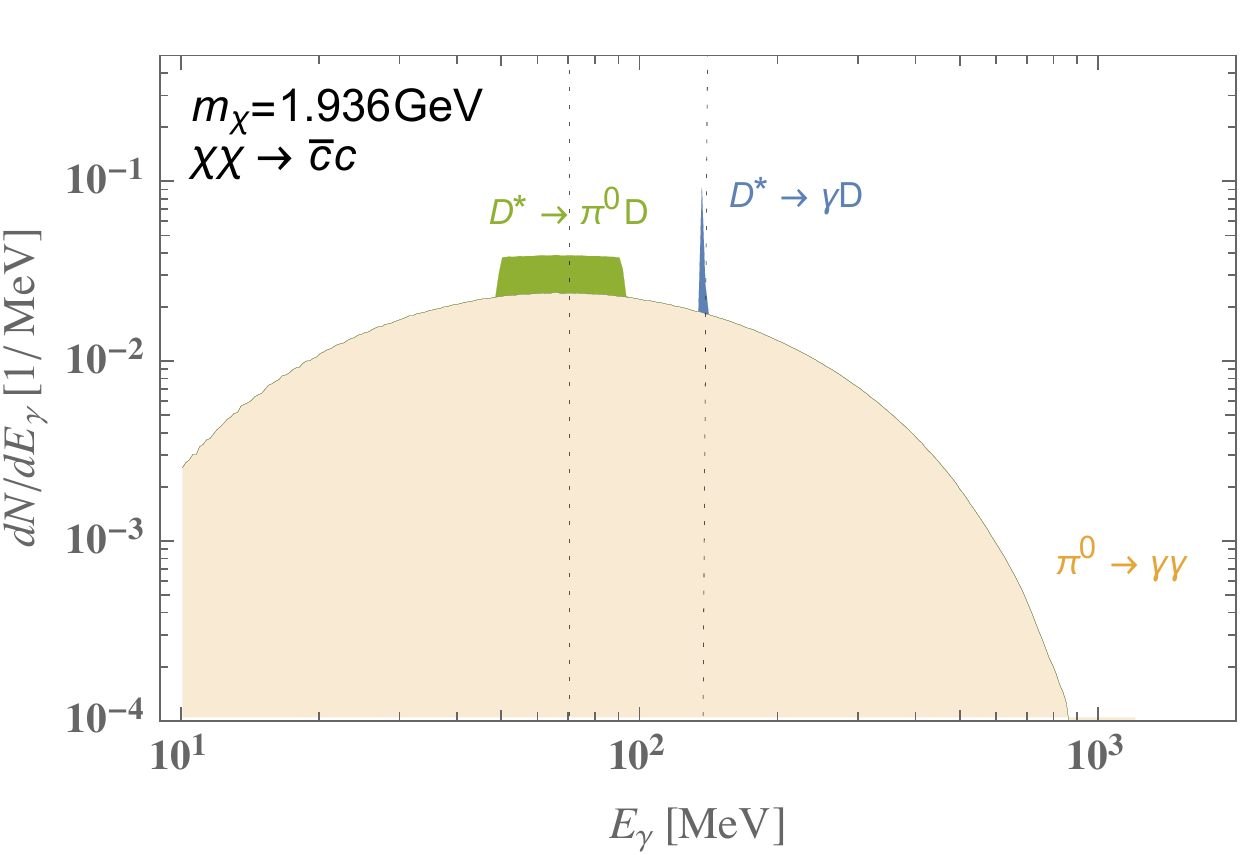}
\includegraphics[width=0.51\textwidth]{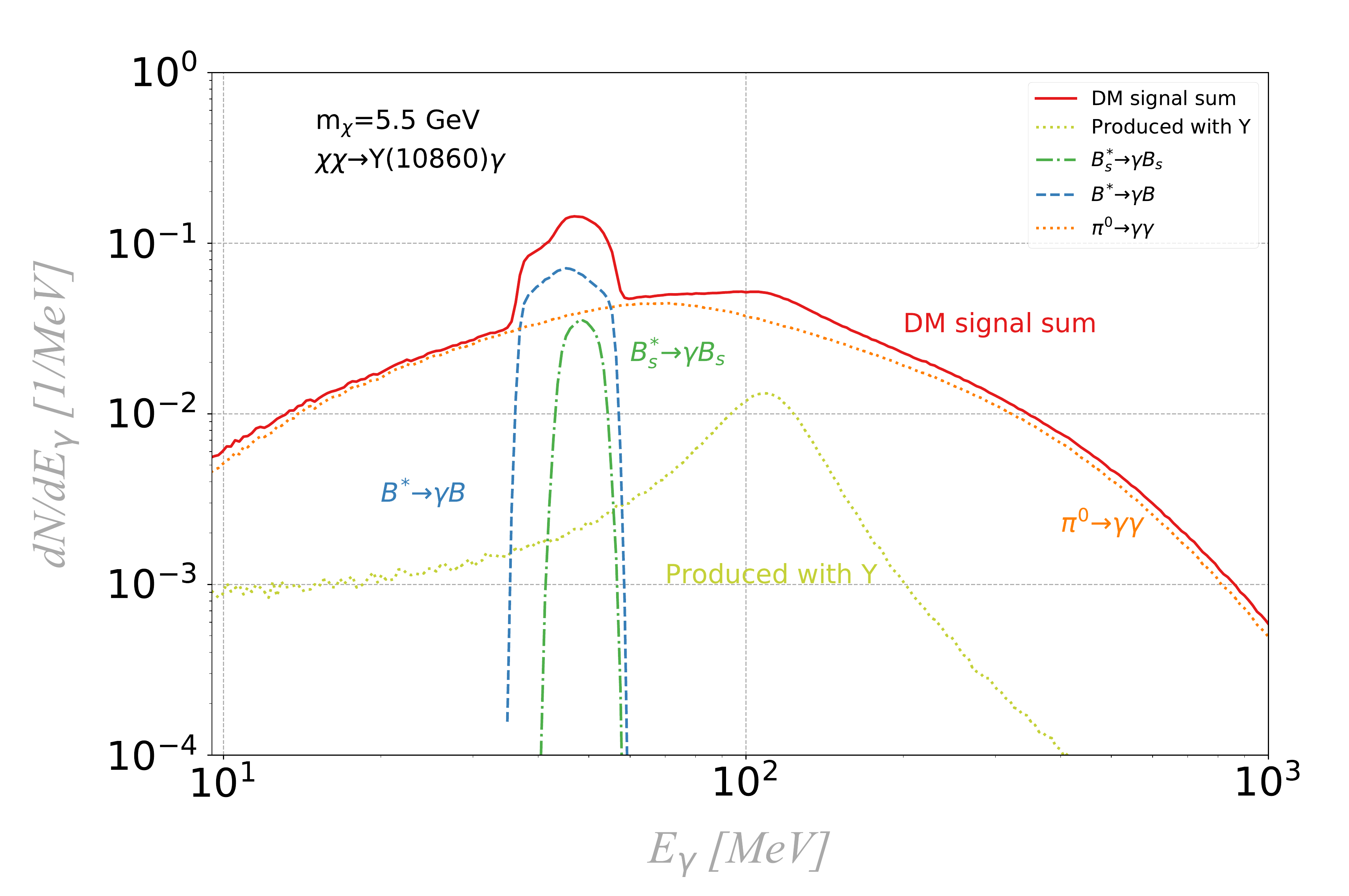}
\caption{\small Left:  Example of the expected gamma-ray spectrum for DM annihilation into 
charm quarks, with a DM mass $m_\chi$ just above the kinematic threshold to produce $D$-mesons. 
The sharp spectral features result from the indicated meson transitions, while the background is mostly 
due to $\pi^0\to\gamma\gamma$. For more details, see Ref.~\cite{Bringmann:2016axu}. 
Right: Gamma-ray spectrum from DM annihilation through the quarkonium channel $\chi\chi\to \Upsilon(10860) \gamma$. The three visible spectral features are due to two different meson transitions and the photon produced in conjunction with the quarkonium. For more details, see Ref.~\cite{smokinggun2}.
\label{fig:spectra}
}
\end{figure}

Among the various processes that could potentially lead to spectral features in MeV gamma rays
(see also Ref.~\cite{Bringmann:2016axu} for an overview, and Sec. \ref{s:fp:brdar} for further examples), 
we will focus here on standard model
meson transitions and quarkonium resonances. We consider a center-of-mass energy of the annihilating 
DM pair, or DM mass in the case of decaying DM, that is close to the threshold for the production of 
(excited) heavy mesons. The de-excitation of excited meson states in the final state, via the emission of a 
photon or neutral pion, will then generate 
box-like signatures (which in the case of photon emission can be almost monochromatic).

For illustration, we show in Fig.~\ref{fig:spectra} an example where DM is assumed to annihilate 
dominantly into $\bar cc$ pairs. In this example, both types of de-excitation processes lead to spectral 
features that are clearly visible above the standard `background' part of the signal, resulting 
from decaying neutral pions that are copiously produced in fragmentations and decays of heavier mesons. 
Implementing a realistic modelling of the expected astrophysical 
background, we have shown that the sensitivity of e-ASTROGAM to this DM annihilation channel 
improves by a factor of up to about 2 by taking into account these spectral features, compared to using 
the standard pion bump as a signal template \cite{Bringmann:2016axu}.  For $\bar bb$ final states, the 
effect can be twice as large. We note that the exact form and location of these spectral features are very 
specific for each final state. This allows, in principle, a highly accurate reconstruction not only of the DM 
mass but also of the branching ratios for the DM decay or annihilation channels. 

The possibility of MeV gamma-ray features from annihilation into heavy meson pairs also raises the issue of 
contributions from quarkonia. Either through the process $\chi\chi\to (\bar QQ) \gamma$, where gamma-rays are 
produced both directly and through subsequent decay into (excited) heavy mesons, or heavy-meson production 
enhanced by a quarkonium resonance $\chi\chi\to (\bar QQ) \to M_A M_B$, where $M_A$ and $M_B$ are two 
heavy mesons with radiative decays. An example of the resulting expected 
spectrum for a DM mass $m_\chi=5.5$ GeV and the channel $\chi\chi\to \Upsilon(10860) \gamma$ is 
shown in Fig.~\ref{fig:spectra} (right). Here structures from three processes, $B^*\to\gamma B$, 
$B_s^*\to\gamma B_s$, and direct production in the annihilation, can all be identified. Notably, such a signal 
would also exist in the annihilation of {\it sub}-GeV DM into light quarkonium states, 
e.g.\ $\chi\chi\to \eta^{(\prime)}\gamma$, with subsequent decay of the $\eta^{(\prime)}$ into photon pairs.
Furthermore, it is well known experimentally that for heavy-meson production at $e^+e^-$ colliders, quarkonium 
resonances can be dominant near threshold \cite{Aubert:2008ab}. We have explored DM annihilation 
through the related vector currents $\chi\Gamma^\mu\chi \bar Q \gamma_\mu Q$. Using collider data as 
input to our model we observe significant enhancement of the MeV features \cite{smokinggun2} due to 
these resonances. We also find that the existence and dominance of different processes is highly 
dependent on the structure of the DM-quark interaction and the nature of the DM particle, e.g.\ as 
seen in the well-known suppression of the vector current for Majorana or scalar DM~\cite{oldMeVfeatures}.

In conclusion, the sensitivity gap in the MeV range explored by e-ASTROGAM is a window of 
opportunity to detect new physics -- not only by confirming the particle nature of DM, but with
the additional potential of closing in on some of its detailed properties, like the DM particle's mass,  
its branching ratios to quark final states and, to some degree, its underlying interaction structure.
\subsection[Sub-GeV dark matter searches \\
\noindent
\textit{\small{R. Bartels, D. Gaggero, C. Weniger, J. Rico, M. Martinez}}]{Sub-GeV dark matter searches}
\label{s:fp:bartels}
\paragraph*{Science questions}
Due to non-detection of any WIMP signal in various experiments, the attention of the community is shifting beyond the WIMP paradigm. As such, MeV (or sub-GeV) DM models have increasingly attracted attention \cite{LWIMP1}. Examples of MeV DM models include self-interacting DM \cite{Boehm:2003hm, Pospelov:2007mp}, 'cannibal' DM~\cite{Pappadopulo2016} and strongly-interacting DM \cite{Hochberg2014}. 
For further models and a discussion on the early-universe production mechanism of sub-GeV DM see Sec.~\ref{s:fp:addazi} and \ref{s:fp:brdar}. 

\paragraph*{Importance of gamma-ray observations}
Indirect detection of DM includes the search for gamma-rays from decaying or annihilating DM. In particular, DM could potentially produce sharp spectral features (see below for further details) that are considered to be a smoking gun of DM. A particularly sensitive target in case of annihilating DM is the GC, since the annihilation luminosity scales with the DM density squared, which is expected to be highest at the center of galaxies \cite{Bringmann:2012ez}. Searches for monochromatic lines from DM towards the GC have been performed by the Fermi--LAT ($>200\mathrm{\, MeV}$) and H.E.S.S. ($> 200\mathrm{\, GeV}$) \cite{Ackermann:2015lka,Abdalla:2016olq}.  On the other hand, dwarf spheroidal (dSph) satellite galaxies of the Milky Way provide clean observational targets, devoid of any astrophysical background that could potentially outshine a DM-induced signal. Considered individually, each dSph would be much less luminous than the GC, but this is partially overcome by the fact that dSphs are numerous (and still being discovered by running optical surveys \cite{Bechtol:2015}). Analyzed collectively, dSphs provide competitive and robust sensitivity for DM searches. Searches for monochromatic lines and other spectral features from DM towards the Segue 1 dSph have been performed by MAGIC ($> 100\mathrm{\, GeV}$) \cite{Aleksic:2014}.
The sensitivity of current gamma--ray experiments in the MeV regime, and therefore the constraint on DM with masses $\lesssim 1\mathrm{\,GeV}$, is lacking. Current bounds from diffuse gamma-rays are given in \cite{Essig:2013goa, squared}.
In addition, MeV--DM is difficult to detect via other probes:
detecting DM through the measurement of the local CR flux is impaired by solar modulation and underground direct detection experiments are insensitive due to the small recoil energies. But, gamma-rays are expected for most of the annihilation channels and can provide a potentially powerful probe.

\paragraph*{Expected results with e-ASTROGAM}
e-ASTROGAM will be particularly sensitive to spectral features due to the annihilation of sub-GeV DM. 
In addition, Ref.~\cite{Bringmann:2016axu} and Sec.~\ref{s:fp:bringmann} showed that annihilation of slightly heavier DM can produce excited meson states which also lead to spectral features in MeV \g-rays.
For the DM with $MeV$ masses only a limited number of kinematically-allowed final states exist. 
For large enough masses, DM can potentially annihilate into pions or muons.
Below the mass of the muon and pion, the only possible final states are into electrons or photons. Neutrinos are
also possible, but this does not lead to a gamma-ray signal.
\begin{SCfigure}
\centering
\includegraphics[width=0.48\textwidth]{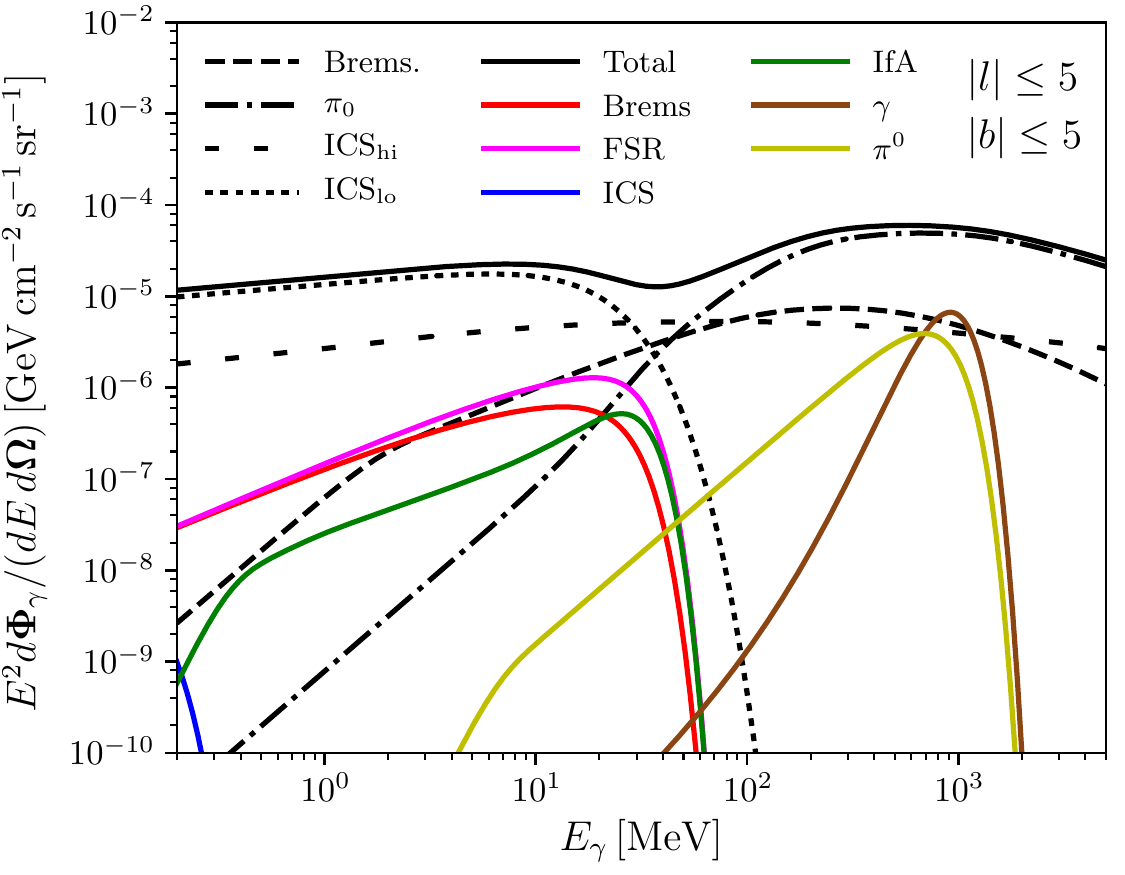}
\caption{
Total background emission in the inner $10^\circ\times10^\circ$ (black, solid)
and broken down into subcomponents (interrupted).
Colored lines show spectral features from primary and secondary
emission due to DM annihilation, convolved with an energy resolution of $\Delta E / E = 0.3$. The yellow (box) and brown lines (monochromatic photon) are for $\chi\chi\rightarrow \pi^0\gamma$ with $m_\chi=800\mathrm{\,MeV}$. 
The red, magenta, blue and green line correspond to the emission for an $m_\chi=30\mathrm{\,MeV}$ DM particle annihilating to $e^+e^-$. 
In both cases  $\left<\sigma v\right>=10^{-28}\mathrm{\,cm^3\,s^{-1}}$.
See Ref.~\cite{Bartels:2017dpb} for 
details.}
\label{fig1:bartels}
\vspace{-10pt}
\end{SCfigure}
In Fig.~\ref{fig1:bartels} we show the gamma-ray signal 
from the inner-Galaxy for two characteristic annihilation channels.
The yellow and brown line are due to DM with a mass ($m_\chi=800\mathrm{\,MeV}$) annihilating into a pion and a photon, 
$\chi\chi\rightarrow \pi^0\gamma$, leading to a box feature and a monochromatic line, respectively.
This emission is prompt, and thus traces the DM distribution exactly.
We refer to Sec. \ref{s:fp:brdar} for a detailed discussion of the generation of the various prompt-emission features.
In addition, we show the spectrum
resulting from a $m_\chi=30\mathrm{\,MeV}$ DM particle annihilating through $\chi\chi\rightarrow e^+e^-$.
This leads to a prompt signal from final-state radiation (FSR), and additional secondary signals from the injected electrons and positrons, which can have a characteristic spectrum as well. 
Only prompt emission is expected for dSphs, however, for low DM masses secondaries can
be important in the GC.
\begin{figure}[ht]
  \includegraphics[width=0.48\textwidth]{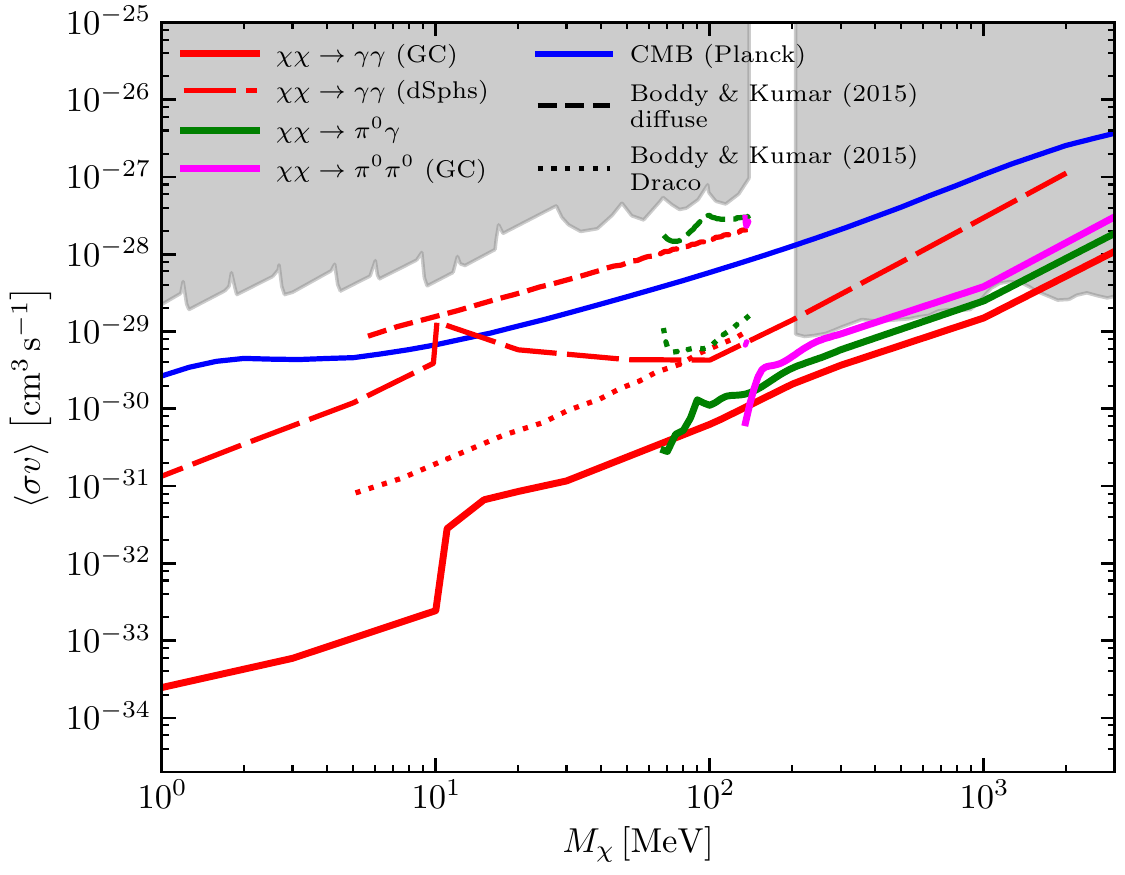}
  \includegraphics[width=0.48\textwidth]{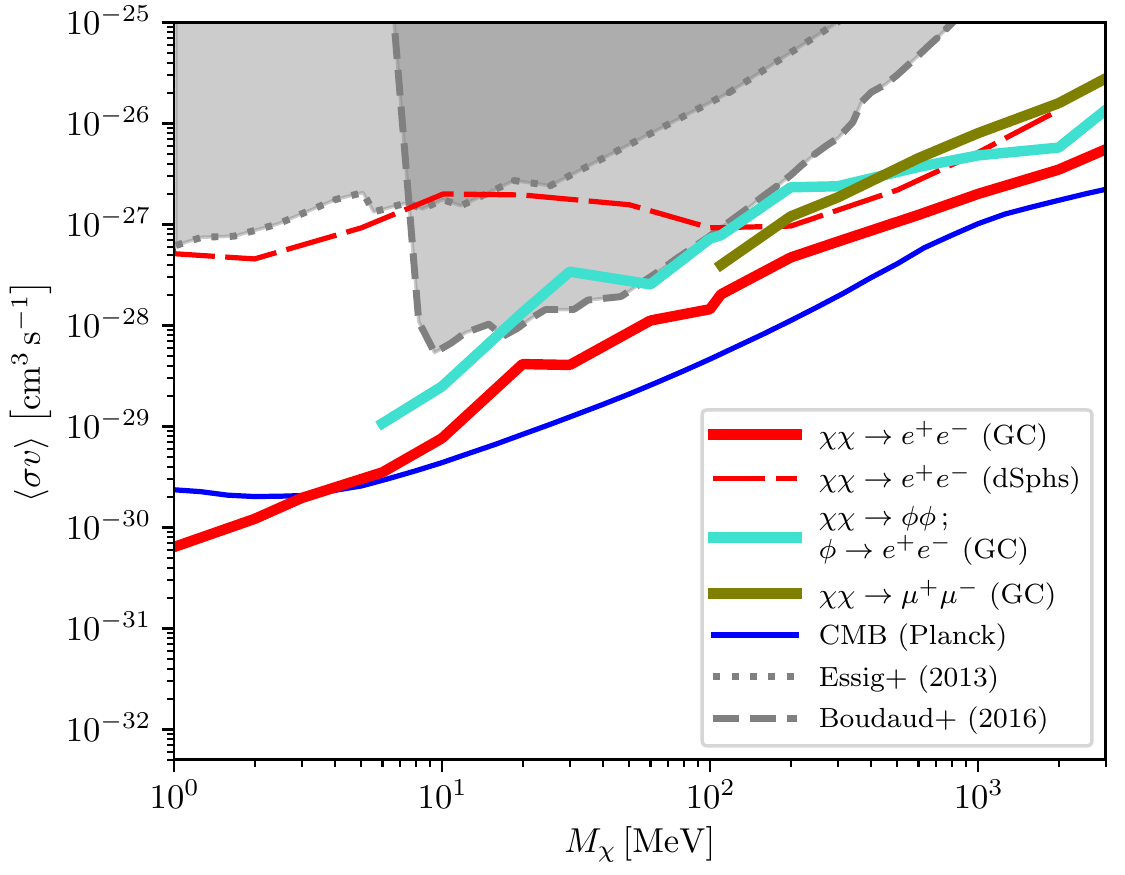}
  \caption{\small 
  Projected 95\% CL upper-limits on DM annihilating to various final states.
  Adapted from Ref.~\cite{Bartels:2017dpb}. Projections are compared to 
  current limits from the CMB (blue, \cite{Ade:2015xua}.
  Left:
  Projections from \cite{Bartels:2017dpb} DM annihilation into $\gamma\gamma$
  (red, solid=GC, long-dashed=dSphs), $\pi^0\gamma$ (green) and $\pi^0\pi^0$ (magenta).
  The CMB constraints are for
  $\chi\chi\rightarrow\gamma\gamma$
  \cite{Ade:2015xua, Slatyer2016}. 
  For
  the same channel we show \g-ray limits derived
  \cite{squared} and \cite{Ackermann:2015lka} in grey.
      Right:
  Projected 95\% CL upper-limit on \g-ray
  emission from DM annihilating to $\mathrm{e^{+}e^{-}}$.
  Results are for the total DM spectrum from the
  three reference leptonic cases: direct annihilation 
  (red, solid=GC, long-dashed=dSphs)),
  cascade channel (turquoise) and the muon channel (olive).
  The blue solid line shows the CMB limits on
  DM $s$-wave annihilation into $e^+e^-$
  from Planck 
  \cite{Ade:2015xua, Slatyer2016}.
  In addition we show in light-grey the limits for 
  $\chi\chi\rightarrow e^+e^-$ 
  from Voyager 
  \cite{Boudaud2016} and current 
  limits from diffuse emission 
  \g-rays \cite{Essig:2013goa}.
}
\end{figure}  \label{fig2:bartels}

We study e-ASTROGAM detectability of sub-GeV DM-induced gamma-ray signals from the GC and dSphs (see \cite{Gonzalez-Morales2017} for another recent study). For the GC we include prompt and secondary emissions from annihilation into photons, neutral pions, muons or electrons (for the latter, either directly or via a cascade) \cite{Bartels:2017dpb}. Projected limits on the annihilation cross-section are derived using Fisher forecasting \cite{Edwards:2017mnf}, taking into account astrophysical backgrounds and both long-- and short--range systematics in energy. For dSphs, only prompt emission from annihilation into photons and electrons are included at this stage, which already provides a useful comparison between the results expected from the two types of targets. In this case, limits are obtained from a maximum likelihood analysis \cite{Aleksic:2012} and include all dSphs listed in Table 1 of Ref.~\cite{Ahnen:2016}, taking into account the uncertainty in the gamma-ray luminosity from each of them.

The results are shown in Fig.~\ref{fig2:bartels}. In the left panel projected-limits are shown for final states involving neutral pions and/or monochromatic photons. Existing constraints from diffuse gamma-rays and the CMB are shown as shaded grey areas and a blue line respectively. e-ASTROGAM observations of the inner-Galaxy can produce significantly stronger constraints below $\mathcal{O}(\mathrm{few}\times 100\mathrm{\,MeV})$. Probing cross-sections close to what is expected for thermal DM with a $p$--wave cross-section. The right panel displays the limits that can be obtained for final states involving leptons. Again, e-ASTROGAM will improve on current bounds, both from diffuse gamma-rays and measurements of the local electron and positron flux. Current CMB limits for the $e^+e^-$ state are stronger, but only apply to $s$--wave annihilating DM. In case of $p$--wave annihilating DM e-ASTROGAM will provide the best constraints.

In conclusion, e-ASTROGAM will be able to place very competitive, and sometimes the strongest, constraints on the DM self-annihilation cross-section for sub-GeV DM by observing the inner-Galaxy and dSph satellites of the Milky Way.
\subsection[Synergy with optical observations for indirect dark matter searches\\
\noindent
\textit{\small{L. A. Antonelli, M. Fabrizio, P. Giammaria, S. Lombardi}}]{Synergy with optical observations for indirect dark matter searches}
\label{s:fp:antonelli}
\paragraph*{Science questions}

In the last years, due to the lack of clear evidence of DM signal in all current complementary experiments (i.e. colliders, direct and indirect searches~\cite{Cahil-Rowley14}), scenarios beyond the WIMP paradigm are also getting increasing attention. Among them, MeV (or sub-GeV) DM models, such as self-interacting DM, "cannibal" DM, and strongly-interacting DM (see Sec.~\ref{s:fp:bartels} and \ref{s:fp:addazi}) for further models and details in the context of
e-ASTROGAM) are currently widely considered. In all these scenarios, gamma rays in the energy range where e-ASTROGAM will operate are expected to be produced.

In the local Universe, the GC and the dwarf spheroidal satellite galaxies (dSphs) are among the most promising targets for indirect DM searches. 
The dSphs are believed to be the smallest (size $\sim$1~kpc) and faintest ($10^2$~--~$10^8$~$L_{\odot}$) astronomical objects whose dynamic is dominated by DM, with mass-to-light ratios $\frac{M}{L} \sim 1000 \frac{M_\odot}{L\odot}$ for the ultra-faint (UF) ones. Despite a typically lower DM-induced gamma-ray flux compared to the expected one from the GC, the dSphs satellites of the Milky Way (MW) represent particularly clean targets for DM searches in the gamma-ray energy domain, due to their proximity (from few tens of kpc up to few hundreds of kpc)
and their generally negligible intrinsic gamma-ray emission from standard astrophysical sources [365]. At the same time, most dSphs are located at intermediate or high galactic latitudes where Galactic foregrounds are suppressed. 

In addition, it has become increasingly clear over the last two years that the census of Local Group satellites is very incomplete. Moreover, the history itself of dSphs discovery has already shown amazing big steps thanks to the mployment of instruments able to realize ever deeper photometric and astrometric scans of the sky~\cite{Walker12}. Hence, the new generation of sky surveys (Pan-STARRS~\cite{Chambers16}, DES~\cite{DES2Y}, GAIA ~\cite{GAIA15}, LSST~\cite{Juric15}, etc.)~$-$~already operating and/or upcoming~$-$~are bringing new discoveries\footnote{In 2015, they enabled the discovery of more than 20 new Milky Way satellites having morphological characteristics similar to the known DM-dominated dSphs.}. These surveys are indeed extending the knowledge of possible sites of large DM concentrations and a detailed study should be made to continuously select the best targets.

\paragraph*{Importance of gamma-ray observations}
Satellite dSphs have been under the eyepiece of many telescopes of different telescope classes for the last twenty years. Optical telescopes are devoted to investigate their dynamics. However, to probe their DM content, instruments working in the high-energy (from MeV up to TeV) band are believed to be well suited to shed light. The reason is directly related to the mass of the DM particles expected within well-motivated theoretical scenarios. In this regard, the MeV-GeV band could be the crucial regime to understand the low energy continuum spectrum expected from DM annihilation/decay processes.

In addition to this, the sensitivity of e-ASTROGAM in the MeV-GeV domain could enable a further characterization of dSphs, allowing for possible discovery of a new class of gamma-ray emitters such as millisecond pulsars, still undetected in these galaxies at higher energy (GeV-TeV domain). Studies to estimate the GeV emission of millisecond pulsars in dSphs have been recently performed in order to evaluate the impact of their emission in the DM search~\cite{Winter17}. 
\paragraph*{Importance of optical observations}
The DM density profile of the target of interest is a crucial point in the indirect DM search. Mass models are most commonly derived by exploiting the stellar population as a dynamical tracer of the underlying gravitational potential well (and hence of the dominant mass component, namely the DM mass profile).
The dynamical mass of a dSph is estimated by quantifying the stellar velocity dispersion ($\sigma_v$).  Due to the lack of deep photometric and spectroscopic data of several ultra-faint dSphs~$-$~the most promising DM search target among satellite galaxies~$-$~current studies suffer from great uncertainties in M/L estimation, and even in target selection.
In order to identify the best targets among MW dSphs' population multi-epoch photometric and spectroscopic observations have to be performed. These studies allow a better constraining of the astrophysical properties required to infer the DM content estimation (total luminosity, presence of binary systems, kinematics of member stars, ...). Optical studies devoted to dSphs have been already carried out e.g. with the Large Binocular Telescope (LBT)~\cite{Fabrizio14} and the Very Large Telescope (VLT)~\cite{Fabrizio16}. In addition, new GAIA releases are expected to both discover new dSph candidates and improve the dSphs' luminosity estimation.

\paragraph*{Expected results with e-ASTROGAM}
In Sec.~\ref{s:fp:bartels} and \ref{s:fp:morselli}, preliminary prospects for e-ASTROGAM observations of the inner Galaxy and dSphs are provided. In particular, in Sec.~\ref{s:fp:bartels} it is shown that e-ASTROGAM will be particularly sensitive to spectral features due to the annihilation of sub-GeV DM and be able to place for those models very competitive constraints on the DM self-annihilation cross-section. Similar conclusions are also achieved in Sec.~\ref{s:fp:morselli}, considering the observations of two dSphs, Draco and Ursa Minor.

In order to reduce systematics associated to already-known DM targets and establish new ones, optical observations are of major importance. Therefore, a synergy between gamma-ray observations with e-ASTROGAM and optical surveys is expected to strengthen the overall DM scientific case.
\subsection[WIMP annihilation in dwarf spheroidal galaxies\\
\noindent
\textit{\small{A. Morselli, G. Rodriguez Fernandez}}]{WIMP annihilation in dwarf spheroidal galaxies}
\label{s:fp:morselli}
\paragraph*{Science questions}
WIMPs are one of the most promising particle DM candidates, and typically can self-annihilate and generate gamma rays~\cite{Bertone2005, Bergstrom:2000}. If WIMPs are produced thermally in the early Universe then the self-annihilation cross-section has a natural value of approximately $3 \times 10^{-26} \text{cm}^{3} \text{s}^{-1}$~\cite{Steigman:2012nb}. 

WIMP models, such as the supersymmetric neutralino, provide predictions for the
gamma-ray energy spectra from annihilations, which are crucial inputs, together with
the DM distribution in the observed target, for estimating the sensitivity of
indirect searches~\cite{Martin:1998a}. The present study provides preliminary comparative
expectations from indirect DM searches with the planned mission e-ASTROGAM,
taking into account continuum gamma-ray signatures coming from typical DM annihilation
channels. The indirect DM search with e-ASTROGAM has many possible
astrophysical targets with different advantages and disadvantages.

The total mass of DM in the Galactic halo together with its proximity to Earth make it the most promising source for DM searches
and  the  perspective for e-ASTROGAM  is described in  Sec.~\ref{s:fp:bartels}. However its proximity means that the source is diffuse and signal and background separation is problematic. The limits from the GC in principle are stronger but the 
limits from spheroidal galaxies are much less dependent on uncertainties like the halo distribution,
other astrophysical signals and backgrounds. 
A detection from spheroidal galaxies will be a smoking gun for the discovery of DM.

\paragraph*{Importance of gamma-ray observations}
Indirect detection of DM annihilations through gamma rays has attracted much interest due to several unique properties of gamma rays. First of all, they do not scatter appreciably during their travel through the Galaxy, but rather point back to the site where the annihilation took place. Also, absorption can generally be neglected, as the cross-section for scattering on electrons and nuclei for MeV to TeV photons is small. This means that one may use properties of the energy distribution resulting from these processes to separate a signal from astrophysical foreground or backgrounds. And, as the EM cross-section of gamma rays is so much higher than the weak interaction cross-section for neutrinos, they are relatively easy to detect.

This is particularly true for the possible signals coming from dwarf spheroidal galaxies because it could give a clear and unambiguous detection of DM. Neither astrophysical gamma-ray sources (supernova remnants, pulsar wind nebulae,...) nor gas acting as target material for CRs have been observed in these systems.

\begin{figure}
\centering
\includegraphics[width=0.47\linewidth]{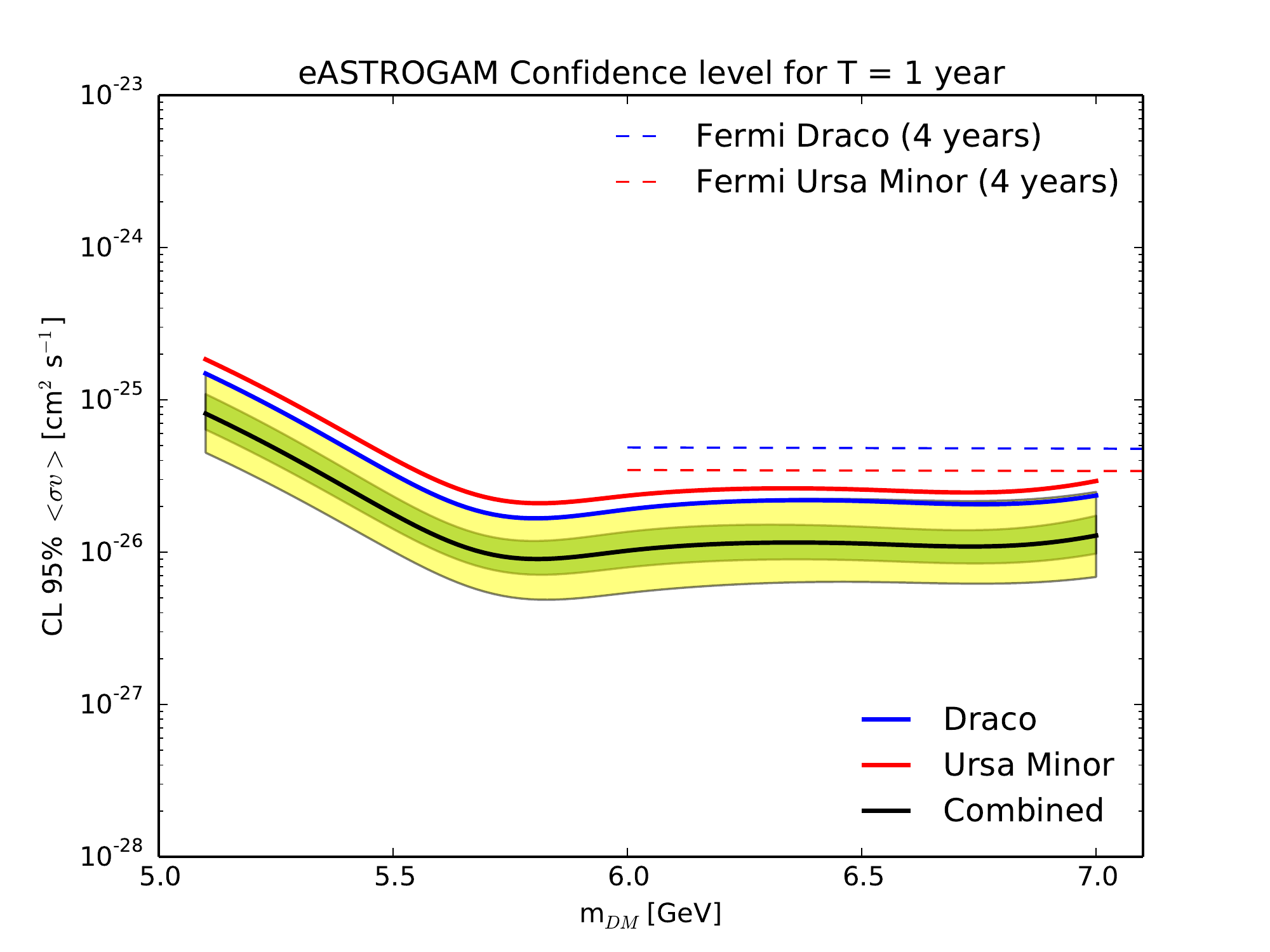}
\caption{\small  Sensitivity for $<\sigma v>$ from observation of the classical dwarf galaxy Draco and Ursa Minor for self-annihilation channel $b\bar{b}$.  
\label{fig1:morselli}
}
\end{figure}

\paragraph*{Expected results with e-ASTROGAM}
The indirect detection experiments aim at searching for a flux of annihilation products created in astrophysical environments where DM annihilation may be occurring at an appreciable rate \cite{FermiLimits}. In particular, e-ASTROGAM will look for gamma-rays from WIMPs in the mass range $\sim$0.3 MeV up to $\sim$3000 MeV. 
As an example, figure \ref{fig1:morselli} shows the expected flux for two self-annihilation channels in comparison with the e-ASTROGAM sensitivity for 1 year.

\subsection[High Galactic latitude, unassociated gamma-ray sources: uncovering dark matter subhalos in the MeV band\\
\noindent
\textit{\small{D. Nieto, J. A. Barrio, M. A. S\'anchez-Conde}}]{High Galactic latitude, unassociated gamma-ray sources: uncovering dark matter subhalos in the MeV band}
\label{s:fp:nieto}
\paragraph*{Science questions}

High-resolution N-body simulations of Milky-Way-like galaxies have revealed that the distribution of DM in this type of objects is far from smooth, rather exhibiting a wealth of substructures, or subhalos, at all spatially-resolved mass scales~\cite{Springel:2008b,Diemand:2009a,Garrison-Kimmel:2013eoa}. 
It is believed that the most massive of these subhalos host the satellite galaxies we observe today, while there should be a large population of subhalos not massive enough to capture gas and/or stars at all. The effective lack of baryonic gas renders star formation unlikely in these small subhalos, making them virtually invisible.
Yet, in models where the DM particle self-annihilates or decays into standard model products, some of these DM subhalos might be located sufficiently close to Earth as to produce detectable signals. Indeed, these objects are expected to possess very dense DM cores. Therefore, they are probably not only resilient to the strong tidal forces they are subject to in the inner Galactic regions, but also potentially yielding very high annihilation fluxes at Earth.
\begin{wrapfigure}{R}{0.63\textwidth}
\centering
\includegraphics[width=0.6\textwidth,clip=true,trim= 0 0 0 0]{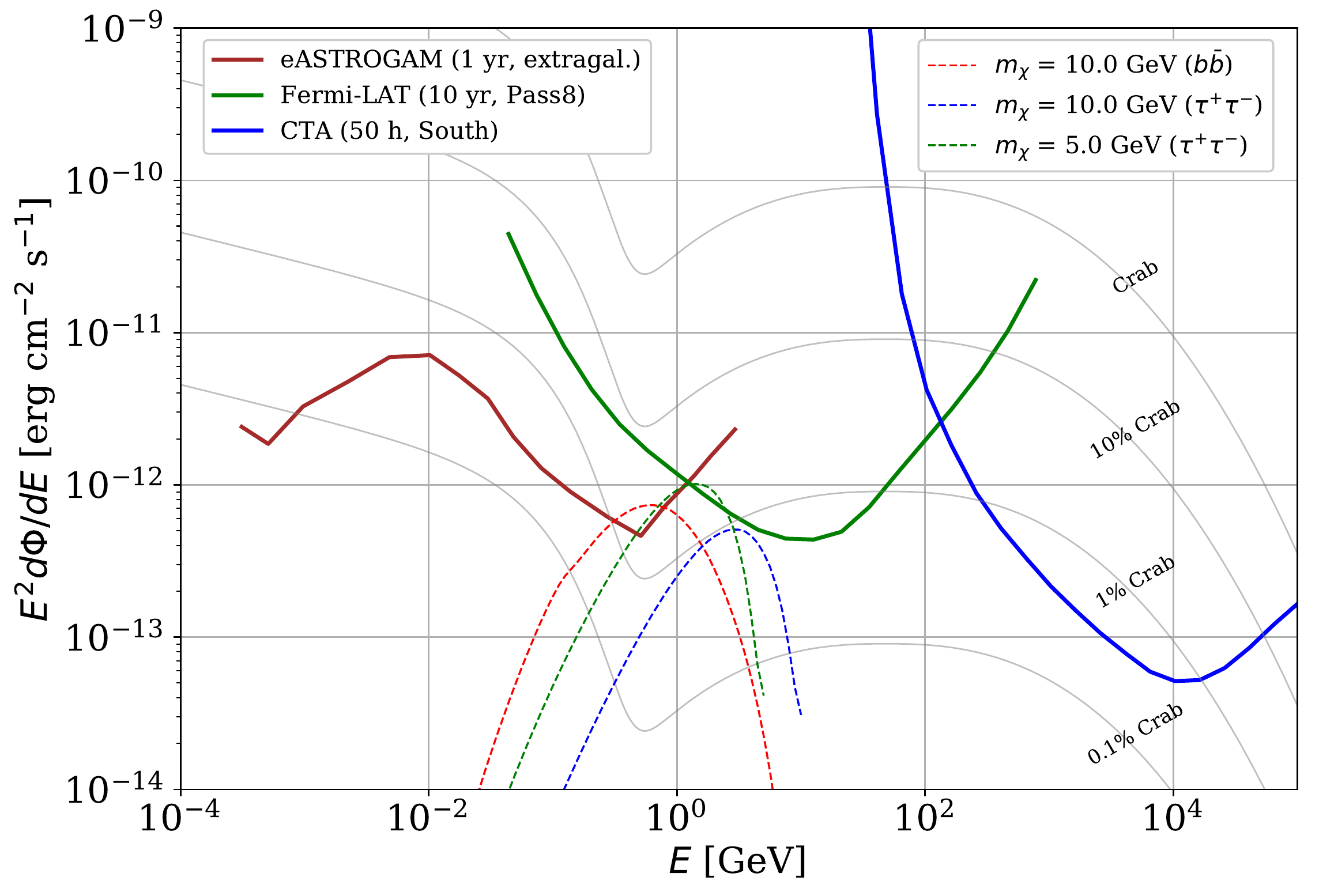}
\caption{e-ASTROGAM differential sensitivity (solid brown line) to point-like WIMP DM subhalos, compared to those of the \emph{Fermi}-LAT (green) and the future CTA (blue). Dashed lines represent examples of DM subhalo energy fluxes for three DM particle models. A $J$-factor value of $1.5\times10^{19}$ GeV$^2$ cm$^{-5}$ and a $\langle\sigma_{ann}v\rangle$ value of $2.2\times 10^{-26}$ cm$^3$ s$^{-1}$ have been assumed. See text for further details.}
\vspace{-0cm}
\label{fig:point-like-sensitivity}
\end{wrapfigure}

\paragraph*{Importance of gamma-ray observations}
Should the DM particle mass be on the MeV~\cite{Gonzalez-Morales2017,Bartels:2017dpb} or GeV scale~\cite{Bertone2005}, annihilation or decay signals from DM subhalos could be potentially detected by telescopes sensitive to these energies. Since dark-matter-induced gamma-ray emission is expected to be constant, subhalos could then appear in all-sky surveys sensitive at gamma-ray energies. Depending on the proximity of those subhalos to Earth, they might show up as point-like or extended sources in such surveys.
The search for DM subhalos in the GeV gamma-ray band has a long record: the \emph{Fermi}-LAT Collaboration has thoroughly searched their data for potential point-like subhalos \cite{2012ApJ...747..121A}, and searches for candidates among the unassociated sources in the different LAT catalogs have been conducted, e.g.,~\cite{Zechlin:2012b,Bertoni:2015mla}. Currently, there are two intriguing candidates, not only showing a lack of counterparts at other wavelengths and spectra compatible with the hypothesis of annihilating DM, but also showing spatial extension in LAT data~\cite{Bertoni:2016hoh,PhysRevD.95.102001}. Higher angular resolution experiments sensitive to gamma rays may be able to shed light on the actual morphology of the sources, resolving the standing ambiguity between the hypothesis of an extension originated by unresolved multiple sources or by the distribution of DM in a nearby subhalo~\cite{Chou:2017wrw}.
These searches for subhalo candidates in the GeV band have also been complemented by searches in the TeV energy regime by the current generation of imaging atmospheric Cherenkov telescopes. TeV subhalo searches base their strategy on follow up observations of subhalo candidates in \emph{Fermi}-LAT catalogs that are spectrally compatible with DM particle masses in the several hundred GeV to multi TeV range~\cite{Nieto:2011c}. Dedicated observations of DM subhalo candidates have been reported by both MAGIC~\cite{Nieto:2011e} and VERITAS~\cite{Nieto:2015hca} Cherenkov telescopes.

\paragraph*{Expected results with e-ASTROGAM}
e-ASTROGAM will be able to contribute to the search of both point-like and extended DM subhalos due to its large field-of-view, sensitivity in the MeV to few GeV range, and improved angular resolution below 1 GeV compared to Fermi-LAT. Indeed, the exceptional sensitivity in the whole MeV energy range will naturally allow to test DM models with particle masses in the same range~\cite{Gonzalez-Morales2017,Bartels:2017dpb}, as well as low-mass weakly-interacting massive particle (WIMP) models ~\cite{Bertone2005} in the mass range around $\sim$1-20 GeV. WIMP DM models with particle masses beyond the e-ASTROGAM upper energy threshold can be within the reach of the instrument, since a substantial fraction of the annihilation photon yield for GeV-mass DM particles would be deposited in the MeV range and resolved by e-ASTROGAM up to a few GeV. Fig.\ref{fig:point-like-sensitivity} shows the expected e-ASTROGAM differential sensitivity to point-like DM subhalos. Also shown are the energy fluxes corresponding to several DM models, obtained following eq.\,\ref{e:dm:flux:diff} for the annihilation flux in in its energy differential form.

For the calculation of the energy fluxes in Fig.~\ref{fig:point-like-sensitivity}, a reference $J$-factor of $1.5\times10^{19}$ GeV$^2$ cm$^{-5}$ is assumed, which corresponds to that of \emph{Segue 1} dwarf spheroidal galaxy~\cite{bonnivard:2016}, along with a value for $\langle\sigma_{ann}v\rangle$ of $2.2\times 10^{-26}$ cm$^3$ s$^{-1}$~\cite{Steigman:2012nb}. We present three different models in which a WIMP DM particle completely annihilates either to a $\tau^+ \tau^-$ pair or to a $b \bar{b}$ pair. The corresponding $\frac{dN_{\gamma}^i}{dE_{\gamma}}$ spectra are obtained from \cite{Cirelli:2010xx}.
Additionally, e-ASTROGAM's improved angular resolution with respect to past gamma-ray missions is of remarkable importance to search for DM subhalos. There are at least two strong arguments supporting the latter statement: first, a more precise source localization and a smaller containment region will help with source association, especially for those cases where multiple counterparts currently coexist within the source containment region derived from previous missions. This will allow a cleaner sample of unassociated sources for point-like DM subhalo search studies. Second, as previously mentioned, a better definition of source spatial morphology can be used as a handle to tell extended DM subhalos from conventional unresolved multiple sources. Fig.\,\ref{fig:extended-ability} depicts a simulation result extracted from~\cite{Chou:2017wrw} showing how e-ASTROGAM can successfully resolve an extended source (with $\sigma_{68} = 0.25^\circ$) from a pair of point sources (separated by 0.28$^\circ$), as opposed to  \emph{Fermi}-LAT. As a matter of fact, e-ASTROGAM will surely enlarge the population of high Galactic latitude, unassociated sources in the gamma-ray band, thus increasing the likelihood of discovery of DM subhalos. 
\begin{figure}[t]
\vspace{-0.2cm}
\centering
\includegraphics[width=0.49\textwidth,clip=true,trim= 10 190 10 0]{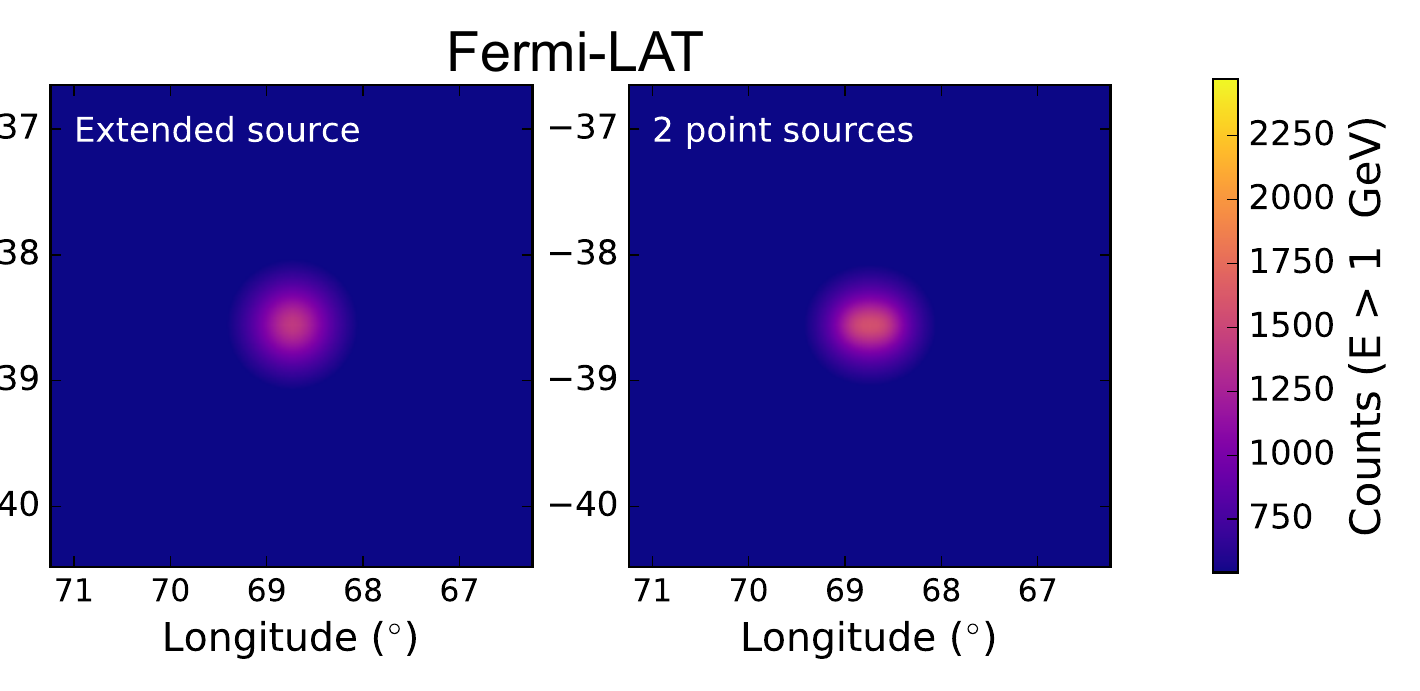}
\includegraphics[width=0.49\textwidth,clip=true,trim= 10 0 10 200]{DM_subhalo/LAT-eASTROGAM_extended-ability.pdf}
\caption{Comparison between the \emph{Fermi}-LAT and e-ASTROGAM abilities to resolve between an extended source and two nearby point sources (extracted from~\cite{Chou:2017wrw}). See text for details.}
\vspace{-0.5cm}
\label{fig:extended-ability}
\end{figure}

Finally, it is worth mentioning that the synergy in this field between current generation space-borne and ground-based gamma-ray telescopes could be extended in the future through the complementarity of the e-ASTROGAM mission and the upcoming Cherenkov Telescope Array~\cite{2013APh....43..189D}.

\subsection[All-sky mapping in the 100 MeV region in search for point-like dark matter sources\\
\noindent
\textit{\small{G. Vankova-Kirilova, V. Bozhilov, V. Kozhuharov, S. Lalkovski}}]{All-sky mapping in the 100 MeV region in search for point-like dark matter sources}
\label{s:fp:vankova}
\paragraph*{Science questions}
The nature of DM and the mechanisms leading to its
creation are among the biggest open questions in modern
physics. 
To address this question, we propose to
perform an all-sky survey in the 100 MeV region in search for discrete
lines and point-like sources. The working assumption is that DM may annihilate, or
decay, via emission of leptons,  including particles other than
electrons.  e-ASTROGAM will provide unprecedented sensitivity exactly in the energy range where
lines originating from $\mu ^+\mu^-$ annihilation are expected to
emerge. 

\paragraph*{Importance of gamma-ray observations}
Previously, it was suggested that the DM consists of
WIMPs that naturally emerge from
the super-symmetric extension of the Standard model. Such a WIMP
particle was predicted to have a mass of the order of 100
GeV. However, no such particle has been experimentally found and the search
for DM candidates is now being broadened into other
directions.
Recently, the idea of involving a complete hidden sector of new particles
was revitalized. This hidden sector naturally incorporates the DM and 
interacts only through a limited
number of processes with the visible sector, usually through the
so-called mediator, as shown in fig. \ref{fig:mediator}. 
\begin{figure}[h]
\centering
  \includegraphics[width=7cm]{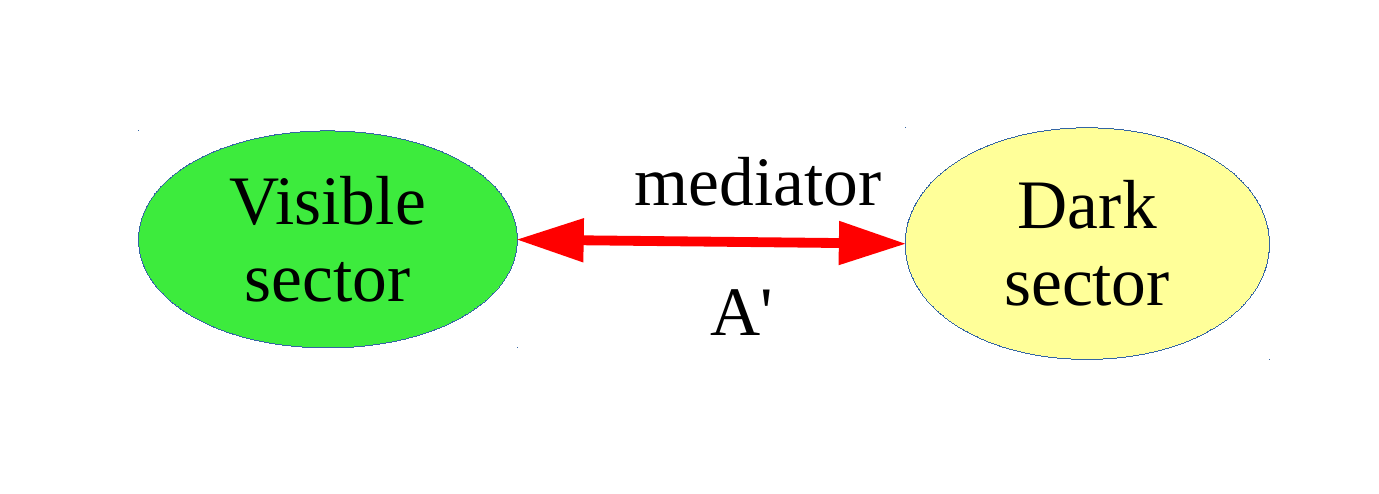}
\caption{A connection between the visible and the hidden sector
  through a vector mediator.}
\label{fig:mediator}
\end{figure}
Even though neither the nature of the DM particle(s) ($\chi$) nor the mechanism
that generates it are known, there are indirect experimental evidences
suggesting that $\chi$ is indeed a weakly interacting particle. Given
that, it is natural to assume that the annihilation, and/or its decay,
will involve leptons, as shown in Fig.\,\ref{fig:dark-sector}. 
These can be electrons and positrons, but also
muons, which can be generated via annihilation $\chi+\chi \to
\mu^++\mu^-$ and/or decay $\chi \to  \mu^+ +\mu^-$. It should be
noted, however, that a similar scenario is not forbidden for the $\tau$
particles either, but the cross-section for formation of a two-tauon
bound state is negligible, and hence, the observation of a signature
of true taonium is considered to be less likely \cite{arX}. The advantage of
using muonium annihilation lines for the search of DM
particles is that the muon mass is much larger than the $e^\pm$ and,
hence, the expected signal will be cleaner.  
\begin{figure}[t]
\centering
  \includegraphics[width=5cm]{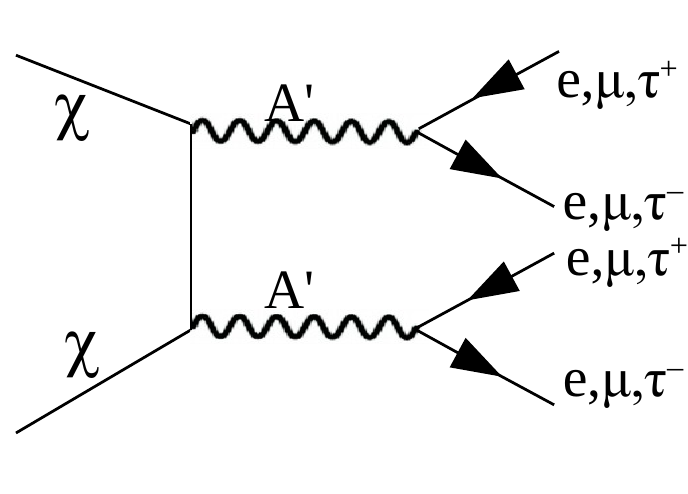}
\caption{\small Feynman diagram for DM annihilation into lepton
  final states.}
\label{fig:dark-sector}
\end{figure}
The simplest effective interaction that can be used to describe the process is:
\begin{equation}
L \sim g'q'\bar{\psi} (\gamma_\mu + \alpha '\gamma_\mu\gamma^5)\psi A'^\mu \ ,
\end{equation}
where $A'$ is the mediator between the Dark and Visible sectors. 
Here
$\psi$ is the leptonic field and $g'$ is the new interaction coupling constant. 
Usually $\alpha _a'=0$. 
The charges, $q_i$, are in general free parameters and 
for the some of the  flavours might vanish - $q_i \rightarrow 0$. 
The branching ratios for $A'\to e^+ + e^-$,
$A'\to\mu ^++\mu^-$, and other competing processes at higher energies are
given in ref. \cite{bib:dpr}. 
There is a threshold of 1022 keV for $e^\pm$
creation and of 210 MeV for $\mu ^\pm$ creation. At higher energies
other channels are enabled. In most of the studied scenarios, it is
also assumed that the mediator 
decays with the same strength to 
different lepton-anti-lepton pairs. But this may prove not to  be true
due to the lepton non-universality, which may lead to an enhancement
of creation of $(\mu ^+, \mu^-)$ pairs via the annihilation reaction $\chi+\chi \to
\mu^++\mu^-$. New experimental results on the muon magnetic moment
\cite{Be06} and the proton radius\cite{Ca15, Ba17, codata}, indeed,
seem to support a different behavior of electrons and muons
with respect to the weak interaction. The $g_\mu-2$ anomaly may be related to a new weakly interacting particle, which lies 
outside the Standard Model, and which would be the best candidate for
the DM $\chi$ particle. 

An all-sky mapping of the 511-keV line was already performed
and it is considered to be among the major achievements  
in \g-ray astronomy. But the origin of the positrons in the
Galaxy is still debated. They can be generated in different processes --
from nuclear reactions and decays, through BH evaporation, to
decay and/or annihilation of DM particles. Hence, it is
difficult to disentangle the processes leading to DM creation. The key
to the problem may lie in the possible complementary channels. 
The other two types of electrically charged leptons in the Standard Model, 
which can annihilate into photons, 
are the muons $\mu$ and tauons $\tau$ with
masses $M_\mu=105.6$ MeV and $M_\tau$=1777 MeV, respectively \cite{bib:pdg}.  
It is worth noting that in
contrast to the electrons and positrons, the muons and the tauons can
not be produced in radioactive decays of atomic nuclei, owing to their
superior masses. As such, the maps based on the $\mu^++\mu^-$ and/or
$\tau^++\tau^-$ annihilation peaks can provide a cleaner signal and a
new information about the sites of enhanced DM concentration which
would be complementary to the data obtained from the 511-keV surveys. 

Further, the leptons can be created not only via processes involving
DM particles such as $\chi + \chi \to l^++l^-$, but in high energy
astrophysical environments a significant number of them can also be
produced via the $\gamma+\gamma \rightarrow l^- + l^+$ and $e^- + e^+
\rightarrow l^- + l^+$ reactions.  However, the muons created in these
high-energy environments have energies much higher than the ionization
energy ($E_{ion}\approx$1.4 keV) of the true muonium \cite{arX} and,
hence, only a small fraction of pairs with energies less then
$E_{ion}$ will form a bound system. The muonium has two states,
depending on the particles spin orientation. These are para- and
orto-muonium. The para-muonioum predominantly decays via two-photon
annihilation, while the orto-muonium -- via electron-positron
annihilation. The energy released in the two-photon annihilation is
E=105.66 MeV \cite{arX}. This is well inside the energy range of e-ASTROGAM
optimal sensitivity. The detection of the muonium annihilation gamma
rays will provide an opportunity to study their production mechanism
or at least to put constraints to the model predictions.  

\paragraph*{Expected results with e-ASTROGAM}
The advantage of using unstable leptons, rather than using electrons,
for tracing DM particles is in their finite lifetime. The tauons
have a lifetime of $2.9\times 10^{-13}$\,s., while the muons have
lifetimes of $2.2\,\mu$s. Their finite lifetimes provide an unique
opportunity for mapping of DM regions with an enhanced
precision. Thus, for example, DM particles with masses of the order of  
$M_{\chi} = 100$ GeV can either annihilate or decay into
muons. For $\gamma_{rel} \sim 1000$ the estimated mean free path of these muons, 
before they decay is of the order of 1000 km. Thus the muons with DM 
origin populate only very close proximity around the dense DM clouds.
This feature provides an excellent instrument for
mapping of regions of DM particles. Before annihilating both $\mu^+$ and 
$\mu^-$ have to be slowed down 
by the medium through ionization losses. This requires the presence  
of high density matter, which had clusterized around the DM clouds. 
Given that the $\mu^+\mu^-$
annihilation could happen only close to their production site, such
processes could provide a higher precision all-sky maps of the DM
distribution in our Galaxy/Universe. 

\begin{figure}[t]
\centering
  \includegraphics[width=10cm]{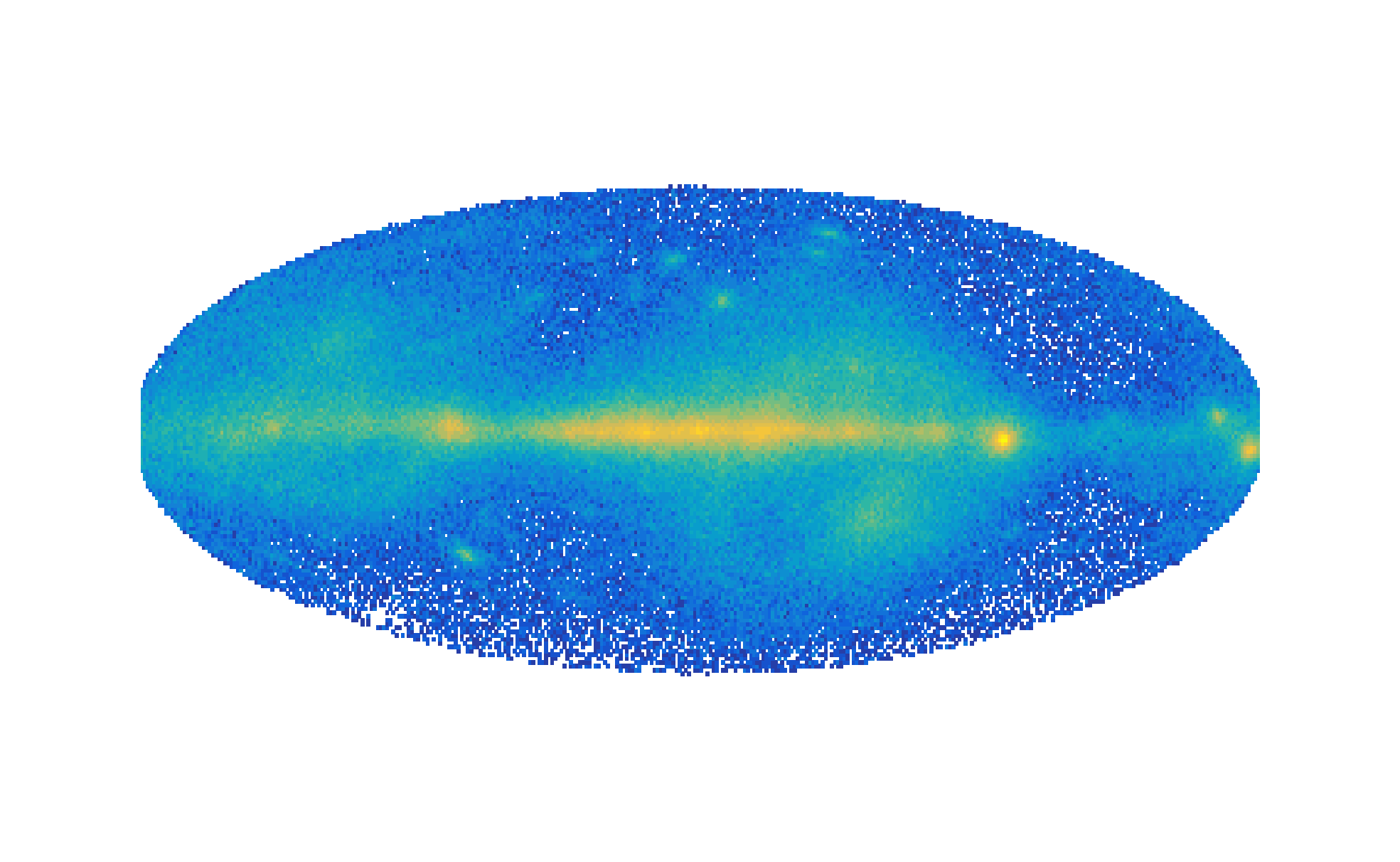}
\caption{All-sky map of O(100 MeV) emission from early FERMI data. This is 
towards the edge of FERMI energy range, where e-ASTROGAM is expected to have 
five-fold superior sensitivity.}
\label{fig:map}
\end{figure}

A preliminary map of the emission in the 100 MeV - 110 MeV region is
shown in Fig.\,\ref{fig:map}. Although the Galactic plane, Vela, Crab, and Geminga
pulsars are clearly visible, the angular resolution limits the
possible observation of weak point-like sources. 
The e-ASTROGAM will have 3 to 5-fold better angular resolution which
will enhance the signal to noise ratio significantly. 
This study will also allow to estimate the 
$\chi+\chi \to \mu^++\mu^-$ branching ratio which would also have an
impact on the understanding of the $g_\mu-2$ anomaly and the nature of
the weak interaction(s). 
Due to low cross-section, the process of muon annihilation into two photons 
has not been observed experimentally so far. On the other side, some astrophysical 
environments where regions with large abundance of DM can provide unique 
opportunity for the observation of such exotic channels.
e-ASTROGAM, being superior than its predecessors in the O(100 MeV) region, 
will be capable of addressing these long standing questions
by directly detecting some of the most exotic particle reactions, or at least 
to put constraints on the production rates of muons and tauons.

\paragraph*{Acknowledgements}
This work is partially supported by the Bulgarian National Science Fund under
contract DN18/17.

\subsection[Particle dark matter searches via angular cross-correlations\\
\noindent
\textit{\small{S. Camera, N. Fornengo, M. Regis}}]{Particle dark matter searches via angular cross-correlations}
\label{s:fp:camera}
\paragraph*{Science questions}
The nature of DM is still an unsolved mystery: its particle physics interpretation is a quite natural option, but a clear and unequivocal signal due to its particle physics nature is nonetheless missing. From the host of investigations of the last two decades, it is now clear that the expected signals have to be extremely weak. Moreover, they need to be isolated from overwhelming and complex astrophysical backgrounds that mask the expected DM signal both morphologically and in terms of spectral features. 
This makes individual DM targets (clusters, galaxies, galaxy satellites or subhaloes) difficult to be detected, although contributing to a (possibly) large cumulative unresolved component. DM constitutes the backbone of all cosmic structures and DM haloes represent, collectively, a potential source of DM decay or annihilation signals. While isotropic at first order, this signal emission reflects the fluctuations of the underlying DM distribution: statistical investigations of maps of large portions of the sky can therefore be a powerful technique that can potentially help in separating the DM signal from the astrophysical backgrounds.
Even if the radiation originating from DM annihilations or decays in a single halo is too faint to be detected, their cumulative signal and its spatial coherence could be.

The non-gravitational signal associated to decay is proportional to the DM density, while the DM annihilation signal is proportional to the density squared; in both cases the emission is peaked at low redshift, say $z<0.3$. The redshift distribution gives a handle to separate DM signals from more mundane astrophysical processes that typically trace the star formation history and peak at higher redshifts. An effective way to  filter out any \g-ray signal that is not associated to DM-dominated structures or that is originated at high redshift is to cross-correlate the \g-ray radiation field with {\it bona fide} low-redshift DM tracers \cite[]{Camera:2012cj,Fornengo:2014cya,Camera:2014rja,2015ApJS..217...15X,Branchini:2016glc,Troster:2016sgf}. Note that this technique has the potential to bring redshift information to the \g-signal, otherwise not available. To perform a measurement of the angular cross-correlation between the \g-ray background and the large scale structure distribution in the Universe with significant statistics, we need surveys with large sky coverage and (at least) sub-degree angular resolution for both the gravitational and \g-ray measurements.

\paragraph*{Importance of gamma-ray observations}
The e-ASTROGAM mission offers intriguing prospects for the identification of \g-ray signals induced by particle DM.
This is true in particular for DM candidates having the peak of the \g-ray emission in the range from sub-MeV up to about 1 GeV. In this range of energy, e-ASTROGAM is superior to the Fermi-LAT satellite in performing the cross-correlation analysis mentioned above. Indeed, not only e-ASTROGAM increases sensitivity and extends the energy range covered by Fermi-LAT, but it also improves the angular resolution, a property of the detector which is very relevant when performing angular correlation studies.
In the following, we illustrate the e-ASTROGAM capability in the specific and yet very relevant framework of light weakly interacting massive particles (WIMP). Indeed, for a WIMP DM candidate annihilating into quarks, the peak of the \g-ray emission occurs at about 1/20 of the DM mass. Therefore, a candidate with mass below $\lesssim$ 20 GeV can be efficiently constrained (or detected) with observations of sub-GeV photons. 
On the other hand, the cross-correlation analysis is not limited to WIMPs. A similar approach can be adopted with e-ASTROGAM to study MeV DM (emitting \g-rays in the MeV range), such as self-interacting DM, `cannibal' DM , strongly-interacting DM, and axion-like-particles.
DM candidates annihilating into leptonic final states or charged pions through s-waves can be strongly constrained by CMB experiments~\cite{Slatyer2016}. For p-wave annihilation and, in general, for DM candidates with prompt \g-ray emission, the constraints derived from \g-rays are instead found to be the strongest~\cite{Bartels:2017dpb}.
The technique proposed here, involving angular cross-correlation of extragalactic DM, have been already proven to provide bounds comparable to local probes (such as dwarf spheroidal galaxies and the GC) for WIMP DM~\cite{Regis:2015zka}. This applies also to MeV DM since the term dependent on particle properties can be (roughly, at first approximation) factorized in the computation of the signals.
Note also that the capability of the cross-correlation analysis will especially benefit from the tremendous improvement expected from cosmological surveys in the next decade, and thus not only from the progresses on gamma-ray detectors.

\paragraph*{Expected results with e-ASTROGAM}
To assess the potential of e-ASTROGAM for the cross-correlation studies, we adopt a Fisher matrix technique to obtain forecasts for the angular cross-correlation signal of DM \g-ray emission with two gravitational tracers of the DM distribution in the Universe, namely cosmic shear and galaxy number counts, as they will be measured by a Stage IV DETF experiment such as, for instance, the Euclid-like satellite [see e.g.\cite{Laureijs:2011gra},\cite{Amendola:2012ys},\cite{Amendola:2016saw}].
\begin{figure}[t]
\centering
\includegraphics[height=0.46\textwidth]{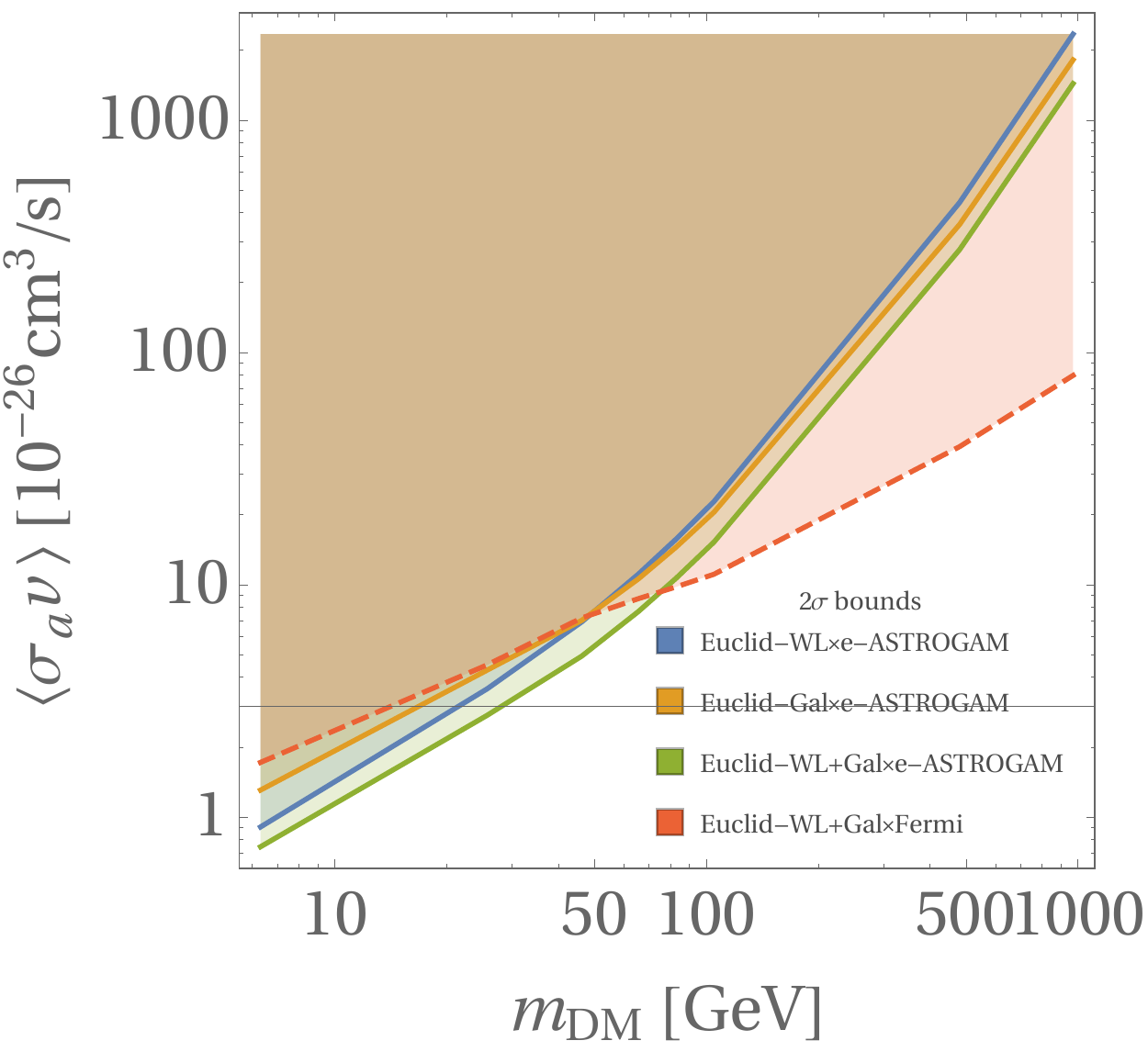}
\includegraphics[height=0.45\textwidth]{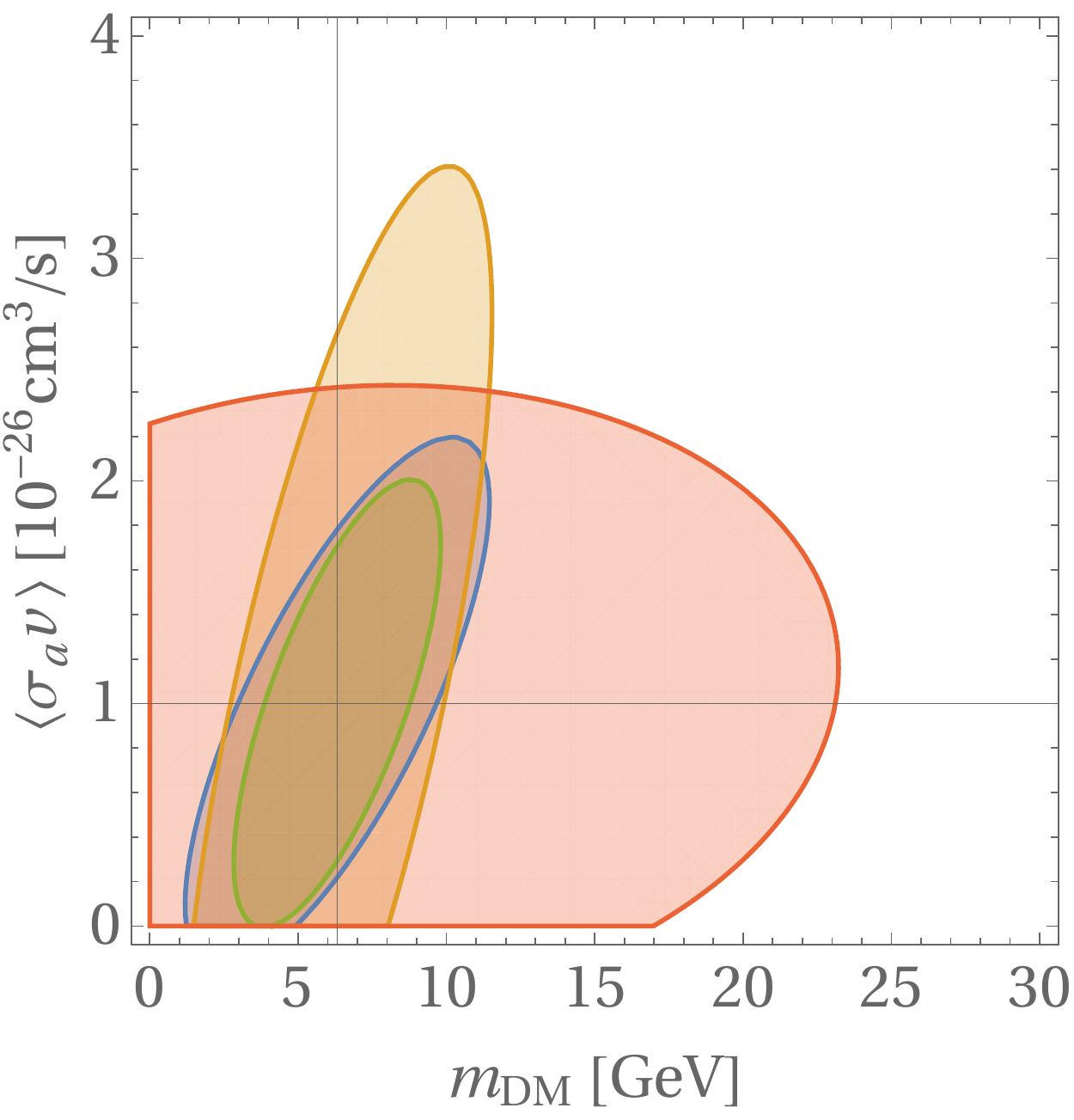}
\caption{ \small Left panel: Expected 2$\sigma$ bounds on the DM annihilation cross section, $\langle\sigma_av\rangle$, versus its mass for a DM candidate annihilating into bottom quarks. The blue line is derived for the cross-correlation of e-ASTROGAM with Euclid weak-lensing cosmic shear, the yellow instead considers the Euclid galaxy clustering, and green line is their combination. The dashed red line shows the bound expected with Euclid data (weak lensing together with galaxy clustering) combined with Fermi-LAT 10 year data dating {\it Right panel:} Expected 1$\sigma$ joint marginal error contours on WIMP parameters for e-ASTROGAM \g-ray data cross-correlated with Euclid (same colour code as in left panel). We chose a WIMP candidate with fiducial DM mass of $\sim$6 GeV and $\langle\sigma_av\rangle=10^{-26}\,{\rm cm^3/s}$.}
\label{fig:bounds}
\end{figure}

The \g-ray background used for this analysis is assumed to be dominated by blazars and is modeled by extrapolating the \g-ray luminosity function that fit Fermi-LAT observations, cross-checking that the derived emission in the sub-GeV range can accommodate Comptel measurements. For details about the computation of the angular power spectrum, the choice of the cosmological parameters, and the DM properties, see Ref.~\cite{Troster:2016sgf}. The performance of e-ASTROGAM is reported in Chapter \ref{intro}. For the sake of simplicity and to be definite, we focus here on the pair production regime for DM annihilation and we consider \g-ray energies above 50 MeV. The Compton domain is very relevant for MeV DM and will be considered in future extensions of this forecast. For the specifications of the Stage IV DETF Euclid-like experiment, we follow Ref.~\cite{Camera:2014rja}.

Bounds on the DM annihilation cross section versus its mass are reported in Fig.~\ref{fig:bounds} (left panel) for a DM candidate annihilating into bottom quarks.
The blue line shows the constraint considering cross-correlation of e-ASTROGAM with Euclid cosmic shear, yellow with Euclid galaxy clustering, while green is their combination. In Fig.~\ref{fig:bounds} (right panel) we show the capability of e-ASTROGAM in reconstructing the microphysics  properties of the DM particle in the case of a positive detection of the cross-correlation signal, under the hypothesis that the fiducial DM mass is about 6 GeV and the annihilation cross section is $10^{-26}\,{\rm cm^3/s}$, i.e. a factor of three below the so-called natural scale for a thermal relic. Fig. \ref{fig:bounds} shows that prospects for e-ASTROGAM in the cross-correlation channel are quite interesting and could lead to relevant limits in a wide portion of the DM parameter space, especially for light DM particles. At the same time, Fig.~\ref{fig:bounds} compares the expectations for e-ASTROGAM with those for the combination of Euclid with a 10-year Fermi-LAT data-taking. Both the bounds (left panel) and the parameter reconstruction capabilities (right panel) clearly show how e-ASTROGAM can play a relevant role for light WIMPs. The results of this exploratory analysis show that the sensitivity of e-ASTROGAM is expected to be even more relevant for DM particles in the MeV-GeV range, and motivates further detailed investigation.
\subsection[Axion-like particles and MeV space gamma-ray detectors\\
\noindent
\textit{\small{A. De Angelis, G. Galanti, M. Roncadelli, F. Tavecchio}}]{Axion-like particles and MeV space gamma-ray detectors}
\label{s:fp:deangelis}
\paragraph*{Science questions}
ALPs are neutral and very light pseudo-scalar bosons $a$~\cite{alp1}. They are predicted by many extensions of the Standard Model, especially by those based on superstrings. They couple to two photons and their interaction Lagrangian is
\begin{equation}
\label{t1}
{\cal L}_{\rm ALP} = \frac{1}{2} \, \partial^{\mu} a \, \partial_{\mu} a - \, \frac{1}{2} \, m^2 \, a^2 + g_{a \gamma} \, a \, {\bf E} \cdot {\bf B}~,
\end{equation}
where ${\bf E}$ and ${\bf B}$ are the electric and magnetic components of the field strength $F^{\mu \nu}$. They are similar to axions  but at variance with them the 2-photon coupling $g_{a \gamma}$ is {\it unrelated} to the ALP mass $m$. The Feynman diagram of the 2-photon ALP interaction is shown in the left panel of Fig.~\ref{immagine3(2)}.
\begin{figure}[h]
\centering
\includegraphics[width=0.27\textwidth]{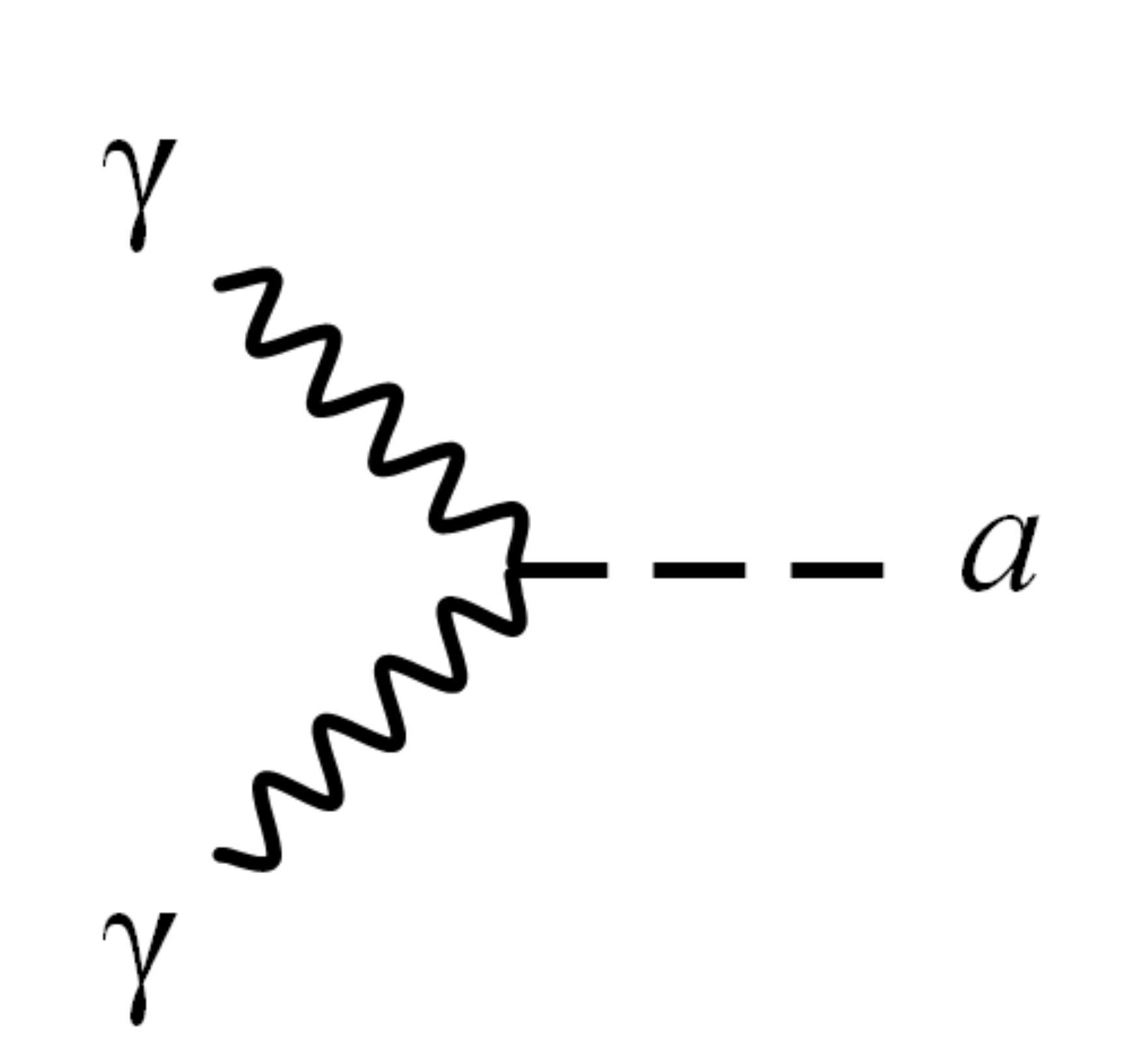} 
\includegraphics[width=0.37\textwidth]{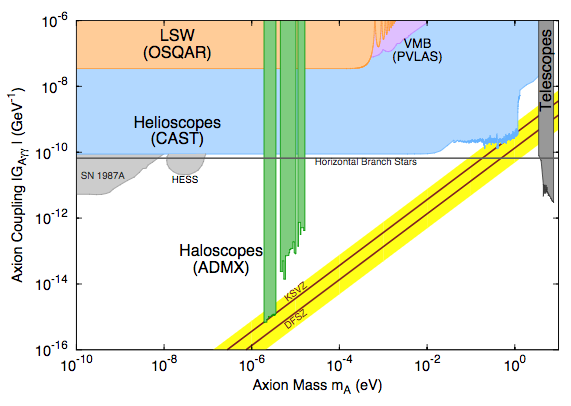} 
\caption{\small \label{immagine3(2)} Left: Photon-photon-ALP vertex. Right: experimental limits on ALPs -- the yellow line indicates standard axions.}
\end{figure}
Present limits (Fig. \ref{immagine3(2)}, right) come mostly from the {(indirect)} non-observation of ALPs  produced in the core of stars (like the Sun) through the Primakoff process in the Coulomb field ${\bf E}$ of ionized matter, illustrated in the left part of Fig.~\ref{fey1}. The CAST experiment at CERN was looking at the Sun and found nothing, thereby deriving $g_{a \gamma} < 0.66 \times 10^{- 10} \, {\rm GeV}^{- 1}$ for $m < 0.02 \, {\rm eV}$~\cite{cast}.  

\paragraph*{Importance of gamma-ray observations}
Let us consider a monochromatic photon beam and assume that an external magnetic field ${\bf B}$ is present (in stars the r\^ole of ${\bf E}$ and ${\bf B}$ is interchanged). Then $\gamma \to a$ conversions occur, as shown in the left part of Fig.~\ref{fey1}, but also the process $a \to \gamma$ takes place, as in right part of Fig.~\ref{fey1}: hence photon-ALP {\it oscillations} $\gamma \leftrightarrow a$ can occur. They can change the intensity of a gamma-ray signal, both increasing and decreasing it~\cite{noi}.
\begin{figure}[h]
\centering
\includegraphics[width=0.5\textwidth]{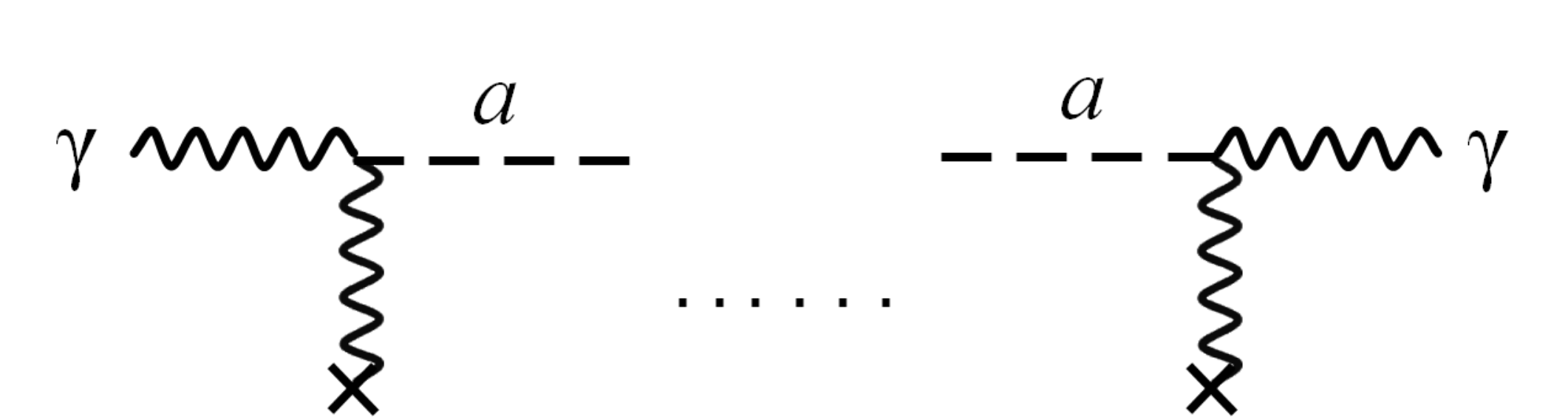}
\caption{\small \label{fey1} Left: $\gamma \to a$ conversion in the external magnetic field ${\bf B}$ (in stars the 
r\^ole of ${\bf E}$ and ${\bf B}$ is interchanged). Right: inverse process $a \to \gamma$.}
\end{figure}
\paragraph*{Expected results with e-ASTROGAM - Evidencing the distortion of a gamma-ray signal}
Suppose that a distant source emits a $\gamma/a$ beam of energy $E$ in the range $0.3 \, {\rm MeV} - 3 \, {\rm GeV}$ which propagates along the $y$ direction reaching us. Consider now the simplest possible case, where no photon absorption takes place and $\bf B$ is homogeneous. Taking ${\bf B}$ along the $z$-axis, we have (see for example~\cite{noi} for a review of the assumptions and the details of the calculations)
\begin{equation} 
P_{\gamma \to a} (E; 0, y) = \left(\frac{g_{a \gamma} \, B}{\Delta_{\rm osc}} \right)^2 \, {\rm sin}^2 \left( \frac{\Delta_{\rm osc} \, y}{2} \right)~, \ \ \ \  {\Delta}_{\rm osc} \equiv \left[\left(\frac{m^2 - {\omega}_{\rm pl}^2}{2 E}  \right)^2 + \bigl(g_{a \gamma} \, B \bigr)^2 \right]^{1/2}~,
\label{a16ghA}
\end{equation}
where ${\omega}_{\rm pl}$ is the plasma frequency of the medium. Defining  $E_* \equiv | m^2 - {\omega}^2_{\rm pl} |/(2 \, g_{a \gamma} \, B)$, one has $P_{\gamma \to a} (E; 0, y) = 0$ for $E \ll E_*$, $P_{\gamma \to a} (E; 0, y)$ rapidly oscillates with $E$ for $E \sim E_*$ -- this is the {\it weak-mixing regime} -- while $P_{\gamma \to a} (E; 0, y)$ is maximal and independent of $m$ and $E$ for $E \gg E_*$ ({\it strong-mixing regime}). The extragalactic magnetic field ${\bf B}$ is usually modeled as a domain-like structure with coherence length $L_{\rm dom} = (1 - 10) \, {\rm Mpc}$, $B = (0.1 - 1) \, {\rm nG}$, and the ${\bf B}$ direction changing randomly among domains. The ${\bf B}$ structure enhances  oscillations around $E_*$ (Fig.~\ref{marco}). 

\begin{figure}[h]
\centering
\includegraphics[width=0.35\textwidth]{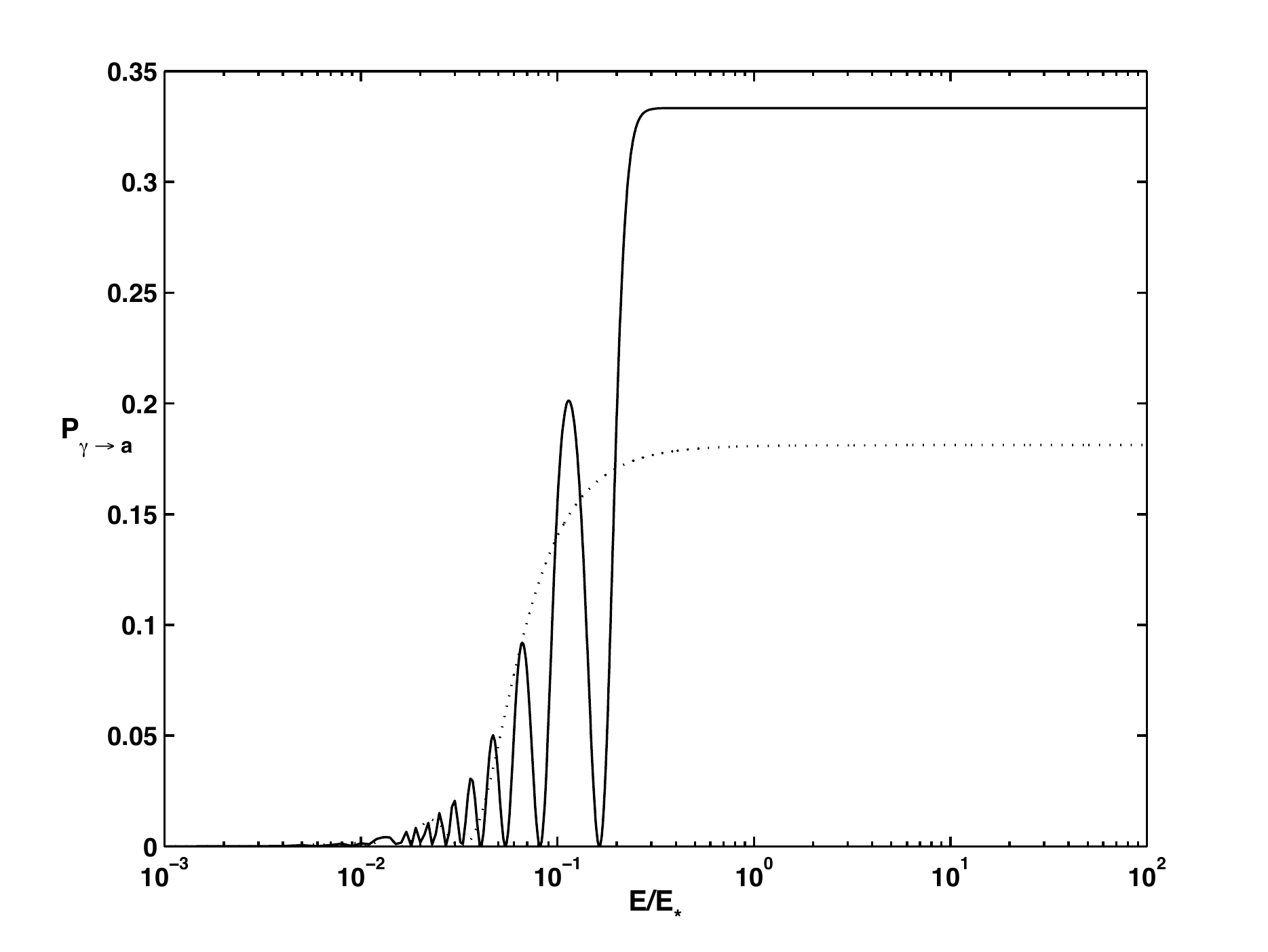}
\caption{\small \label{marco} Oscillatory behavior around $E_*$ for $g_{a \gamma} = 0.33 \times 10^{- 10} \, {\rm GeV}^{- 1}$, 
${\rm B} = 0.5 \, {\rm nG}$ and $N = 200$ magnetic domains.}
\end{figure}

On top of the oscillatory behavior we also have a feature in the energy spectrum followed by a dimming of the signal al lower energies \cite{dmr2008}: the production of ALPs implies  a reduced photon flux. It can be shown that for $N \gg 1$ magnetic domains, the two photon polarization states and the single ALP state undergo equipartition, so that the signal becomes dimmer by a factor of 2/3. 

In addition, the coupling $g_{a \gamma} \, a \, {\bf E} \cdot {\bf B}$ acts as a {\it polarizer}. Photons $\gamma_{\perp}$ with linear polarization orthogonal to the plane defined by ${\bf k}$ and ${\bf B}$ do not mix with $a$, but only photons $\gamma_{\parallel}$ with linear polarization parallel to that plane do \cite{mpz1986}. Two distinct phenomena come about: {\it birefringence}, namely the change of a linear polarization into an elliptical one with the major axis parallel to the initial polarization, and {\it dichroism}, namely a selective conversion $\gamma \to a$ which causes the ellipse's major axis to be misaligned with respect to the initial polarization. Thus, the measure of the polarization of radiation with known initial polarization provides additional information to discriminate an ALP from other possible effects. Actually, we do {\it not need} to know the initial polarization by employing a simple trick. Because when one does not measure the polarization one has to sum over the two final photon polarizations -- while when one does measure it no sum is performed -- the signal has to be 
{\it twice as large} when the polarization is not measure as compared with the case in which the polarization is measured.
 What is the mass range of the ALP that can be probed by e-ASTROGAM? As far as the polarization effect is concerned it is 
of course maximal in the strong mixing regime ($E \gg E_*$) but it is present also in the weak mixing regime ($E \sim E_*$), while the spectral feature shows up only in the weak mixing regime. So, what is required is that $E_*$ falls inside the energy range of e-ASTROGAM. Neglecting ${\omega}_{\rm pl}$ and recalling the definition of $E_*$ we get (regardless of $N$)
\begin{equation} 
0.3 \, {\rm MeV} < \frac{m^2}{2 \, g_{a \gamma} \, B} < 3 \, {\rm GeV}
\label{a16ghAC}
\end{equation}
and by employing the parametrizations $g_{a \gamma} = \alpha \, 10^{- 10} \, {\rm GeV}^{- 1}$ and $B = \beta \, {\rm nG}$, 
Eq. (\ref{a16ghAC}) becomes
\begin{equation} 
1.08 \times 10^{- 12} \, (\alpha \beta)^{1/2} \, {\rm eV} < m < 1.08 \times 10^{- 10} \, (\alpha \beta)^{1/2} \, {\rm eV}
\label{a16ghAE}
\end{equation}
By taking e.g. $g_{a \gamma} = 0.33 \times 10^{- 10} \, {\rm GeV}^{- 1}$ and $B = 0.5 \, {\rm nG}$ one has $0.44 \times 10^{- 12} \, {\rm eV} < m < 0.44 \times 10^{- 10} \, {\rm eV}$. 

\paragraph*{Expected results with e-ASTROGAM - Prompt gamma-ray signal from Type II supernovae} ALPs can be produced at the centre of core-collapse (Type II) supernovae soon after the bounce (when also the neutrino burst is produced) by the Primakoff effect and reconverted to photons of the same energy during their travel in the Milky Way. The arrival time of these photons would be the same as for neutrinos, thus providing a clear-cut signature.

Integrating over the explosion time, which is of the order of 10 \,s, the Authors of~\cite{meyer2017} find that the ALP spectrum can be parametrized by a power law with exponential cutoff,
\begin{equation}
\frac{dN_a}{dE} = C \left(\frac{g_{a\gamma}}{10^{-11}{\rm{GeV}}^{-1}}\right)^2
\left(\frac{E}{E_0}\right)^\beta \exp\left( -\frac{(\beta + 1) E}{E_0}\right) \,\, 
\label{eq:time-int-spec}
\end{equation}
where for a progenitor mass of $10 \, M_{\odot}$, $C$, $E_0$ and $\beta$ are $5.32 \times 10^{50} \, {\rm MeV}^{-1}$, $94 \, {\rm MeV}$, and $2.12$, respectively, while for a progenitor mass of $18 \, M_{\odot}$ they are $9.31 \times 10^{50} \, {\rm MeV}^{-1}$, $102 \, {\rm MeV}$, and $2.25$, respectively.  
%
\begin{figure}[h]
\centering
\includegraphics[width=0.4\textwidth]{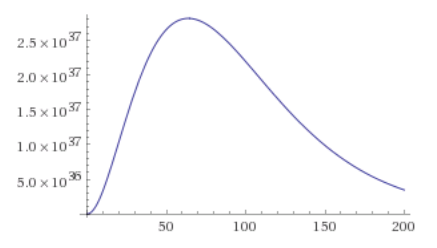}
\includegraphics[width=0.4\textwidth]{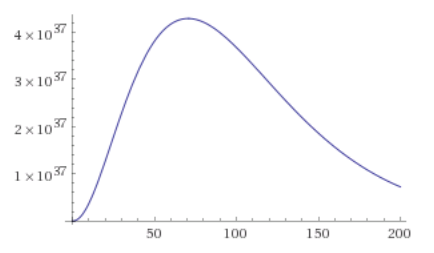}
\caption{\label{fig:alpsn} The differential axion rate from the supernova, $dN_a/dE\,({\rm{GeV}}^{-1})$, for a SN of 10 (left) and 18 (right) solar masses. The abscissa is in MeV.}
\end{figure}

The ALP energy spectrum -- which corresponds to the \g-ray energy spectrum after reconversion -- is shown in Fig. \ref{fig:alpsn}. The bulk is below $\sim \, 100 \, {\rm MeV}$, which shows the potential of e-ASTROGAM for a possible detection. Indeed, e-ASTROGAM has a sensitivity better than Fermi/LAT and can access to much smaller mass/coupling values than dedicated laboratory experiments.
\subsection[Search for signatures of primordial black holes\\
\noindent
\textit{\small{M. Doro, J. Rico, D. Malyshev}}]{Search for signatures of primordial black holes}
\label{s:fp:doro}
\paragraph*{Science questions}
A large number of theories predict the formation of BHs in the
primordial Universe, according to diverse mechanisms: from the
collapse of local overdensities, to that of domain walls, cosmic
strings, etc. Such formation scenarios are reviewed in
\cite{Carr:2005}. Many of these theories predict Primodial Black Holes (PBH) to have formed in
a narrow time period, and therefore having practically a very narrow mass distribution\footnote{However, wider mass distributions is not completely ruled out, see e.g.~\cite{Garcia-Bellido:2017}}. Depending on the formation epoch the mass may vary from few grams to millions of solar masses. 
Non-observation of PBHs of particular masses can constrain cosmological models on small angular scales, 
which are not accessible in CMB observations 
\cite{Josan:2010cj, Linde:2012bt}.  
PBHs radiate particles via the Hawking mechanism~\cite{Hawking:1974},
thus losing mass over time, and accordingly increasing their
temperature following the law $T_{BH} =
\left(8\pi\,G\,M_{BH}\right)^{-1}$. However, specially for larger mass
BH, the possibility of accretion of material could have altered the
above simple evolution formula. In the non-accretion scenario, the Hawking mechanisms predict that, as
the temperature increases, BHs will finally evaporate, where the lapse
time to evaporation is given by: $\tau \sim G^2\,M^3_{BH} \hbar^{-1} c^{-4}$. This allows to make the
straightforward estimation that all PBH of mass smaller than
$10^{14}$~g ($10^{-19}~M_\odot$) would be evaporated today. Small mass PBHs can affect the cosmological observables, such as CMB spectrum or BBN, while larger mass PBHs can be observable with current observations. 
The instantaneous gamma-ray rate for different BH temperatures is shown in \autoref{fig:spectra:doro}. The spectra have two components: the primary component from direct Hawking mechanism, and the secondary component from the decay of hadrons produced by fragmentation of primary quarks and gluons, and by the decay of gauge bosons. The spectrum of secondary photons~\cite{MacGibbon:90} peaks around $E_\gamma=68$~MeV, independent of the BH temperature, because it is dominated by the 2\g-decay of soft neutral pions.
It is  thus clear that
instruments sensitive to the gamma-ray energy band in the $\sim$10~MeV--1GeV range such as e-ASTROGAM, can provide very deep
insights into the questions, in some scenarios providing the strongest
bounds for PBH in the mass range around $10^{14-15}$~g. 
It should also be noted that for $M_{PBH}>10^{15}$~g, their lifetime
exceeds that of the Universe, and therefore PBH could constitute part of the DM (lighter PBH may still have a cosmological role, e.g. in
altering BBN, being involved in baryogenesis, etc.). 
When particles from the Hawing radiation
are injected into the Universe, they are normally too scarce to
significantly alter the energy budget of the Universe or the CMB
number of photons, however, they heat up and ionize the gas, therefore
altering the optical depth of the CMB photons. This provides strong
cosmological bounds~\cite{Poulin:2017,Clark:2017}. Competitive or
stronger bounds can be found from the MeV diffuse component of the extragalactic gamma-ray background (EGB)
\cite{Carr:2010} and from the Galactic diffuse emission \cite{Lehoucq:2009ge}.
\begin{figure}[th]
  \centering 
  \includegraphics[width=0.6\linewidth]{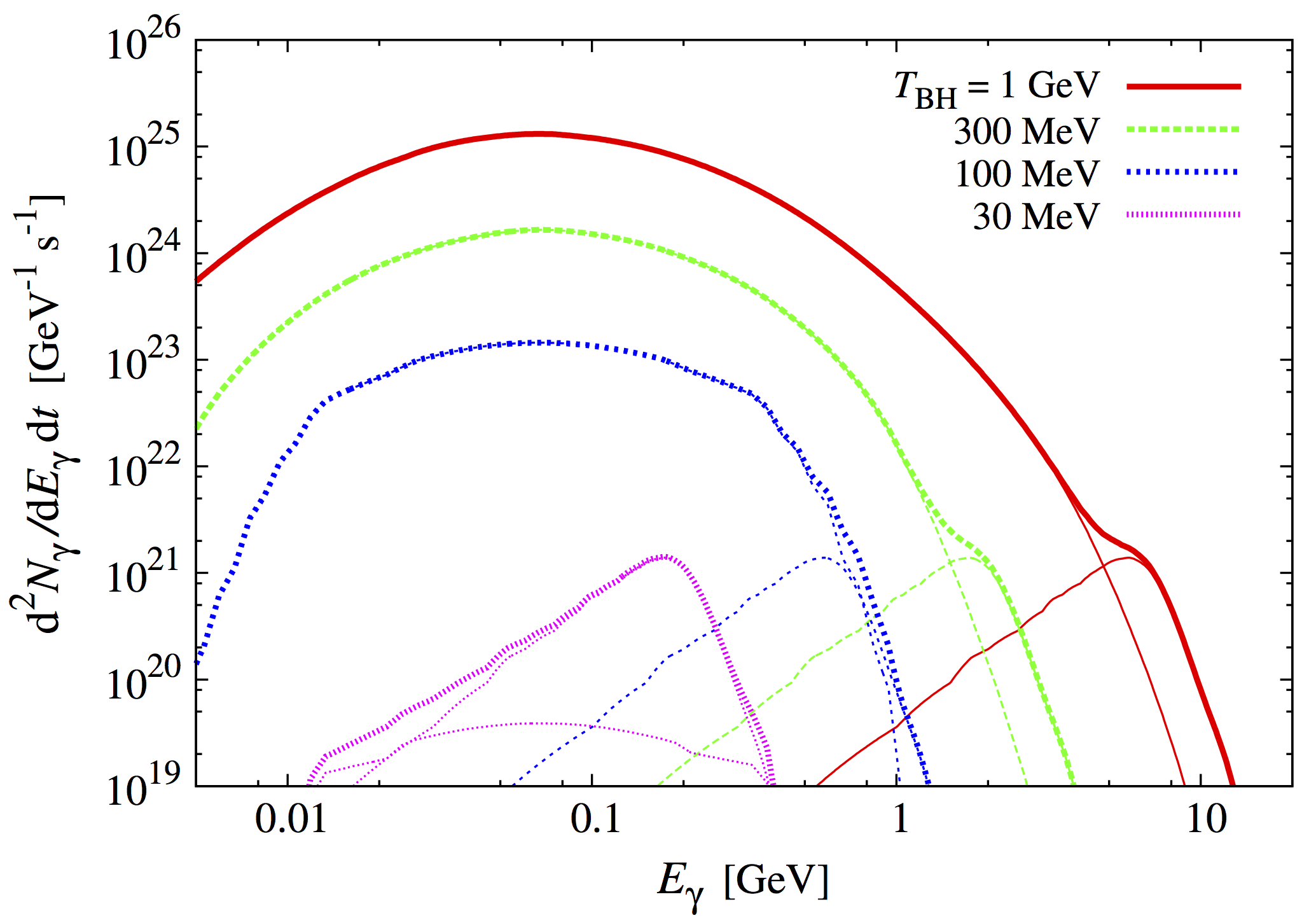}
  \caption{\small \label{fig:spectra:doro} Gamma-ray instantaneous rate for BH at different temperatures. For each temperature, the curve with the peak to the right (left) represents the primary (secondary) component and the thick curve denotes their sum. The figure is a reproduction of Fig. 1 of Ref.~\cite{Carr:2010}.}
  \end{figure}
Not only PBH could constitute part of the DM component,
but their detection could be of utmost interest to understand the
presence and distribution of such elusive objects in the Universe.
PBHs are one of the predictions of general relativity and detection of PBHs would be a spectacular confirmation of quantum field theory in vicinity of BH. The radio telescopes are also approaching the resolution to be able to observe directly the horizon of nearest SMBHs and
specific instruments to observe it are also utilized~\cite{EHT}.
\paragraph*{Importance of gamma-ray observations}
Photons (and other particles) are radiated from BH at any time in its
history, following the Hawking mechanism. In this section, we
concentrate on high energy photon emission. The photon emission is computed
in~\cite{Carr:2010}. For PBHs in the mass range  $10^{15-17}$g,
the peak intensity occurs at $\sim 1--30$,MeV. This means that all PBHs of
those mass, either already evaporated, or close to evaporation, could
have injected a large amount of MeV radiation into the Universe. This
would now be seen as an unresolved contribution into the MeV component
of the EGB, in which the e-ASTROGAM
satellite would be uniquely competitive. 
Besides the continuous (in time) emission from the radiating BH, there
is also the possibility to directly observe the very last final phase of the
BH life, when the BH explodes and vanishes. All non-accreting BH are expected to go into this final
destiny, and the energy and time scales of this phase is governed only
by the mass (or temperature) scale. During
the last phases, a small loss of mass reduces rapidly the BH
lifetime. In comparison to indirect searches like those performed using the EGB, direct searches of the PBHs evaporation bursts are sensitive to the local (sub-kpc scale) PBH distribution. It has been appreciated for a long time \cite{Carr:2010} that by strictly considering Standard Model processes, the likelihood of detecting the final explosive phase of PBH evaporations is very low. However, the physics of the QCD phase transition is still uncertain and the prospects of detecting explosions would be improved in less conventional particle physics models~\cite{EGRET:1994}. For instance, it has been argued that the formation of a fireball at the QCD temperature could explain some of the short-period GRBs (i.e. those with duration less than 100 ms) \cite{Cline:92}.

\paragraph*{Expected results with e-ASTROGAM}
COMPTEL and EGRET data constrained the PBH density using EGB (Ref.~\cite{EGRET:1994,Weidenspointner}).
Some of these constraints are shown
in \autoref{fig:limits}, together with Planck limits~\cite{Clark:2017} and
femtolensing~\cite{Carr:2016}. Planck provides the strongest
constraint on the abundance of PBHs for masses $\sim 10^{15-16}$~g, while
the EGB dominates for smaller masses. These studies used the
EGB as bound the contribution of PBHs, i.e. they were requiring the
integrated MeV contribution not to be larger than that of the measured
EGB. However, the authors themselves claim the interest of actually
considering (fractions of) the EGB as a signal of PBHs. This is an
exciting possibility because the origin of this MeV gamma-ray
background is yet uncertain~\cite{Fornasa:2015qua}.
In Ref.~\cite{Ruiz-Lapuente:2016}, the authors claims that standard astrophysical contributions cannot explain the whole diffuse MeV contribution, see in particular their figure 13. There is therefore space for additional contribution, and PBHs could contribute to some of this missing flux. 
In conclusion, the improved sensitivity of e-ASTROGAM in the MeV range will  allow to use the diffuse MeV component of the EGB to put possibly the strongest constraints on the PBH number density for masses in the range of $10^{15-17}$g. Considering the EGB limits in \autoref{fig:limits} are obtained 	assuming 100\% of the background produced by PBHs, e-ASTROGAM bounds are expected to improve these results.
\begin{figure}[t!]
  \centering 
  \includegraphics[width=0.6\linewidth]{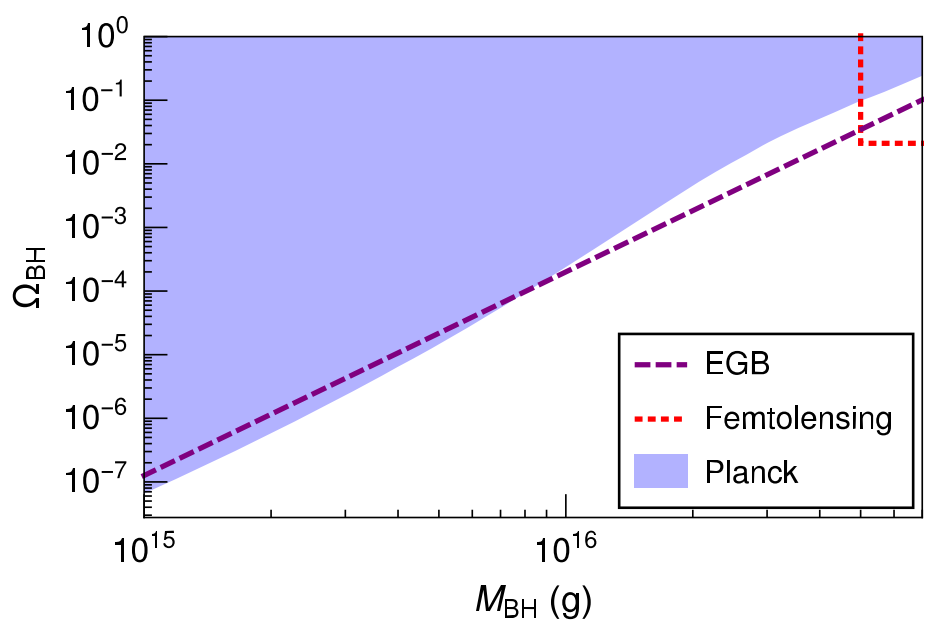}
  \caption{\small \label{fig:limits} Compilation of PBH density bounds in the range $10^{15-17}$g. The figure is a reproduction of Fig. 6 of~\cite{Clark:2017}.}
  \end{figure}
\subsection[Superradiant black holes as particle detectors for very light bosons\\
\noindent
\textit{\small{S. Ciprini}}]{Superradiant black holes as particle detectors for very light bosons}
\label{s:fp:ciprini}
\paragraph*{Science questions}
Dark matter may be a mixture of supersymmetric and axion-like particle (ALP) candidates. The hypothetical pseudo-scalar axion, originally invoked by Peccei and Quinn to elegantly solve the strong-CP problem, is nowadays a strong case for new physics, and a first representative of the Weakly Interacting Slim/Sub-eV Particles (WISPs). WISPs, motivated by string-theory extensions of the Standard Model (SM), are practicable cold DM candidates (e.g., ALPs, hidden-sector/dark photons HPs, millicharged particles).
There is consensus on the viability of gamma-ray spectral observations of cosmological beacons like AGN motivated by the ALPs vs SM-photon mixing during the propagation in intergalactic/Galactic magnetic fields. The dawn of the multi-messenger and gravitational wave (GW) astronomy era, with its revival for BH studies, however offers us a further opportunity: the astrophysical BHs superradiance. BHs can be natural, self-tuned, particle detectors for axions and WISPs (``axionic hairy'' BH configurations).

BH-superradiance could produce nearly monochromatic (resolvable or stochastic)
GW signatures for dense bosonic fields with tiny masses (sub-eV to $10^{-33}$ eV), ascribed to axions, ALPs, HPs, massive gravitons \cite{arvanitaki11,arvanitaki17,brito15,brito17}, and complementing for haloscope (microwave resonant cavity), nuclear magnetic resonance, photon regeneration, helioscope (sun telescopes) experiments like ADMX-HF, CASPEr, ALPS-II/III, CAST, IAXO. Superradiant scattering occurs in, classical and quantum, dissipative systems (dominated by viscosity, friction, turbulence, radiative cooling, tidal acceleration, self-critical cascades). Thermodynamic studies of rotational superradiance in General Relativity (GR) for spinning (Kerr) BHs, established connections to BH-area theorem, Penrose mechanism, tidal forces and event horizon dissipation, with energy/charge/angular momentum extracted from the vacuum and leading to its quantum version (the BH evaporation, \cite{hawking75}).

Spinning BHs are unstable against massive scalar fields. Particles/wave packets trapped in orbit and scattering off the BH ergoregion gain speed/amplitude continuously, by extracting momentum and energy. In superradiant condition $\omega < m \Omega_{H}$ (with $\omega$, $\Omega_{H}$ the wave mode and horizon angular velocities, $m$ the perturbation azimuthal number) ``instabilities'' may naturally grow \cite{press72}.
The multiplication of photons and axions in hairy BHs by superradiance results in an evolutionary BH spin-down on the Regge plane (Fig. 1), through cyclical bosonic field growths/decays, cloud collapses, bosenovae. BH spin measurements are, however, still poor for the heaviest BHs and the OJ 287 exception is anyway tied to the strong assumption of a binary system \cite{valtonen16}.

Superradiance connects strong-field GR, GWs, physics beyond the SM and WISPs, superconductors/fluids, holographic quantum models, SM photon astrophysics, providing, among the other, the following predictions.
1) The formation of quasi-bound states for the QCD axion in the BH ambient (when
$\lambda_{\mathrm{Compton}} \simeq R_{\mathrm{Schwarzschild}}$) with the gravitational potential barrier acting as a mirror in an effective potential (the ``gravitational atom'', Fig. 1). This produces spontaneous atomic excitations (level transitions) with superradiant scalar modes excited by floating (non-decaying) orbits (not possible in classical GR), and subject to annihilations to gravitons, producing GW lines (frequency $f=m \Omega_{H}/\pi$).
2) Amplification of radiation towards strong instabilities (``BH bomb''), induced by massive bosonic fields/condensates, or Reissner-Nordstrom anti-de Sitter (AdS) boundaries, or magnetic fields and nonlinear interactions, and very small novae-like bursts, with collapsing bosonic clouds under the attractive self-interactions of axions (``bosenova'' implosions/explosions).
3) A test of GR in strong field for Kerr spacetime geometry, and applications to Kerr-Newman (charged) BHs, and to any scalar-tensor theory, beyond the GR admitting BH solutions, in addition to new ``hairy'' (e.g. Proca field) BH solutions.
4) Analogies and models of gravity that can be directly devised and experimentally studied in the laboratory \cite{torres17}.
5) Acceleration of ultra-relativistic jets through the rotational energy of the magnetized BH,  described by the Blandford-Znajek mechanism, and upper limits on the BH spin and magnetic field.
6) Plasma mirror interactions, of interest for primordial BHs passed through a cosmological epoch when the mean gas density allowed superradiant instability to be effective (``axionic-BH cosmophysics'').
7) More consequences like stimulated decay lasing (of interest for fast radio bursts, FRBs, X-ray/gamma-ray flashes), GWs scattering produced in BH-NS binaries, the Chandrasekhar-Friedman-Schutz instability of spinning NSs, or the Einstein-Skyrme scenario for chiral hair BHs with topological defects coupled to gravity (the Skyrme solitons). A very open question is whether some of the phenomena introduced above, have CRs and EM counterparts, with photons escaping to infinity and gamma-ray fluences sufficient for a detection that can support beyond-SM GWs signals.

%
\vspace{-0.3cm}
\begin{figure}[ttt!!]\label{fig1superradiance}
\resizebox{\hsize}{!}{ %
\includegraphics[width=10.0cm,angle=0]{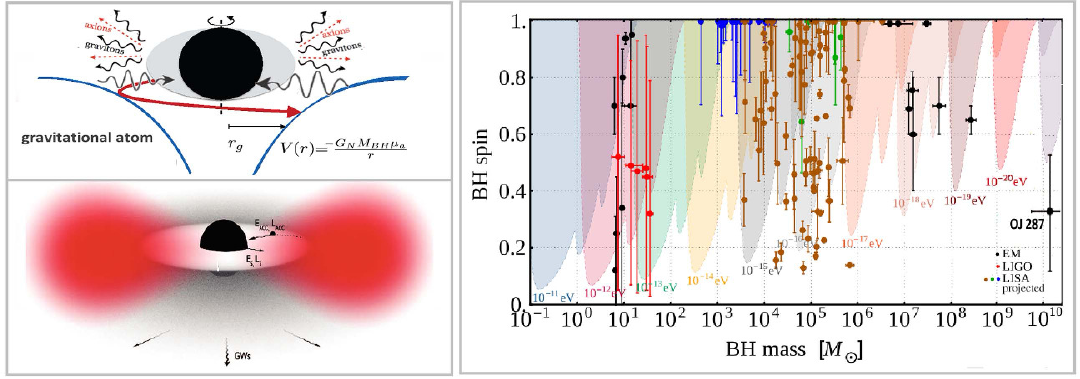}
}%
\vspace{-0.3cm}
\caption{\small{ Left: cartoon of the bosonic gravitational atom (axion, red, cloud bound to a Kerr BH). $E$,$L$ is the energy/angular momentum lost through superradiant scalar waves and GWs (S) or transported in the accretion disc (ACC). The material is in free fall after the innermost stable circular orbit (ISCO). Adapted from \cite{brito15}.
Right:Exclusion regions in the Regge plane for a massive boson field. Black points: BH spin from e.m. estimates (X-ray $K_{\alpha}$ or ISCO-based fits). The OJ 287 value is obtained trough optical variability clocking \cite{valtonen16}.
Red data: LIGO detections GW150914, GW151226, GW170104. Blue/green/brown data: projected LISA. Adapted from \cite{brito17}.}}
\vspace{-0.3cm}%
\end{figure}
\normalsize

\paragraph*{Importance of gamma-ray observations}
GR foresees that rotation increases significantly the density of bound DM clouds close to the BH. The collisional Penrose process (superradiance is its wave analog) is able to reach efficiencies $>600\%$ for rapidly spinning Kerr BHs \cite{schnittman15}. This means that high-energy tails in the gamma-ray spectra of annihilating DM may be observable in gas-poor normal/giant passive elliptical galaxies ($\gtrsim 10^{6}$ M$_{\odot}$ BHs), despite their merging history. This is alternative to nearby dwarf spheroidal galaxy targets. Relic intermediate-mass BHs ($10{^2}-10^{5}$ M$_{\odot}$) in our Galaxy, could be other sources, boosting the annihilation signal.

Superradiant magnetized BHs may have a detectable EM signal joined to the GWs,
considering the QCD axion field and SM photon field coupling via the Chern-Simons interaction (quantum GR supergravity). The conversion rate may be too slow \cite{arvanitaki11}, except during bosenovae. Accretion fluctuations could be resonantly excited by perturbations to the disc by axion/ALPs condensates giving variability signatures. Detectable anomalous gamma-ray lines by nuclear disintegration close to the BH-horizon can be activated by the bosonic axion field.

The inverse Compton scattering on a massive Kerr BH ergosphere (Penrose Comptonization) of the big blue bump photons in AGN, should produce sub-MeV/MeV gamma-ray photons escaping to infinity \cite{williams95}. Evacuated cavities in the plasma around a BH, could instead act as a mirror to confine superradiant waves, leading to an EM version of the ``BH bomb''. BH-superradiance is also able to extract pure EM energy in presence of thin, conducting, accretion discs formed by a neutron star remnant. Highly spinning BHs could behave as a ``sparkplug'', initiating the ultra-relativistic fireball process in GRBs, or also producing some types of FRBs or hypothetical gamma-ray flashes.
Superradiant views for GRBs could be developed and tested thanks to MeV gamma-ray flux/polarization data. Dense axion/ALPs clouds with significant stimulated axion decay into $\gamma \gamma$ pair can originate bright gamma-ray laser pulses and bursts.
Lasing events may be induced by other bosons: the $\pi^{0}$ may power ($\lesssim 50$MeV energy) laser pulses in ($<10^{-18}$ M$_{\odot}$) high-spin primordial BHs, evaporating now at our current epoch.

Axion vs SM-photon mixing is able to create vacuum birefringence and dichroism, qualitatively similar to those in QED magnetized vacuum \cite{mpz1986}. This induces a rotation of the polarization plane of a linearly polarized monochromatic beam, influencing the polarization of a strongly magnetized, broad band gamma-ray source, like a GRB. The inhomogeneous axionic, optically-active, medium affects gamma-ray photon light paths leading to a polarization-dependent bending. Twofold source image splitting and measurable time-delays for the polarized/unpolarized gamma-ray flux, would be an unambiguous signature of superradiant axionic-hairy BHs. Finally some studies on the observations of gamma-rays associated with the bosenova explosions, and radio waves from axion-photon mixing in the Galaxy were recently introduced.

\paragraph*{Expected results with e-ASTROGAM}
$0.3$ MeV - $3$ GeV gamma-ray signatures associated to ultra-light axion/ALP clouds amplification in highly spinning BHs, to Penrose mechanism, to gravitational atom configurations, BH bombs, bosenovae, laser impulses, can be probed through a large field of view, unprecedented sensitivity, space telescope like e-ASTROGAM. First astrophysical hints of axions/ALPs existence have, perhaps, already been seen in the anomalous (excessive) energy cooling of white dwarfs, intermediate-mass stars in the horizontal branch (HB) and red giant (RGs) phases, SNR Cas A neutron star, and the anomalous cosmic transparency for VHE gamma rays. This means axions are coupled directly to electrons, with atomic axio-recombination/de-excitation/bremsstrahlung and Compton scattering.

Gamma-ray polarization measurements of GRBs emission by e-ASTROGAM are expected to constraint the axion-photon coupling constant, while time-delays from polarized/unpolarized fluxes, can shed light on superradiance phenomenon and axions/ALPs hairy BHs. Nearby core-collapse SNe explosions (also a Galactic SN), besides the rich MeV nucleosynthesis physics menu, can
emit ALPs from nucleon-nucleon bremsstrahlung, and can produce a MeV gamma-ray bosenova, considering the high matter density close to the BH horizon shortly after the SN explosion. Strong-nuclear and EM instabilities have enough time to be important during bosenova events.

In strong-field and curved spacetime next to the BH horizon, the collisional, gravito-magnetic, Penrose Comptonization of the optical-UV thermal photons from the disc, can produce a broad spectral component in AGN, typically in the $0.02-12$ MeV band for a $10^{8}$ M$_{\odot}$ BH \cite{williams95}.

The axion/ALPs coupling to gluons and nucleons produces a mixing between ALPs and $\pi^{0}$, through QCD istanton effects (non-perturbative fluctuations of the gluon field). The effective value of the QCD-sector, CP-violating, $\theta$-parameter may become of order one inside the bosonic condensate, affecting the pion mass. SM nuclei are therefore destabilized in the accretion disc towards de-excitation and disintegration. Such fascinating and mostly unexplored new-physics can be probed by e-ASTROGAM searching for exotic gamma-ray lines and signals from unstable anomalous isotopes and nuclear decay products. e-ASTROGAM will discover also new jetted AGN at high redshifts ($z>4$) hosting the most massive BHs. MeV gamma-ray observations and future progresses in numerical simulations hence will allow us to understand the challenging story of BH growth and the new physics (axions/ALPs) might eventually be required to explain the data.

\subsection[Search for matter-antimatter annihilation for testing baryogenesis models\\
\noindent
\textit{\small{C. Bambi, A. D. Dolgov}}]{Search for matter-antimatter annihilation for testing baryogenesis models}
\label{s:fp:bambi}
\paragraph*{Science questions}
The local Universe is clearly matter dominated and the small amount of observed antimatter can be explained as of secondary origin, produced in space by collisions of high energy particles. However, we do not know the origin of this matter-antimatter asymmetry. Matter and antimatter have quite similar properties. The observed asymmetry cannot be explained as due to the Universe initial conditions, because any initial asymmetry would be washed out during inflation and therefore the observed asymmetry today had to be generated by some mechanism after inflation~\cite{school}. The matter-antimatter asymmetry cannot be explained within the Standard model of particle physics, and it is thus one of the cosmological indications of new physics.

We can distinguish three possible scenarios of matter-antimatter asymmetry~\cite{scenarios}:
\begin{enumerate}
\item The asymmetry is spatially constant and the Universe is matter dominated.
\item The Universe is globally baryo-symmetric and there are domains of matter and antimatter.
\item The Universe has a non-vanishing baryonic charge, but the asymmetry is not spatially constant. In particular, there may exist lumps of antimatter in a matter dominated Universe.
\end{enumerate}

Different baryogengesis models can predict any of the scenarios~1-3. Most models proposed in the literature belong to the first class~\cite{school}, but at present there is neither experimental nor observational evidence in favor of one model over another, because they operate at so high energies that it is difficult or impossible to test them in laboratories on Earth. The scenario~2 seems to be observationally excluded, or at least the size of the domain where we live should be larger than the visible Universe~\cite{crg}. The scenario~3 can have interesting phenomenological implications, in particular the existence of antimatter objects in our neighborhood and the observation of matter-antimatter annihilation~\cite{paper1,paper2}.

\paragraph*{Importance of gamma-ray observations}
An unambiguous proof of the existence of primordial antimatter would be the observation of sufficiently heavy anti-nuclei, starting from $^4$He ({\it direct search}). Indeed anti-deuterium can be created in energetic CR reactions, while the probability of production of heavier anti-nuclei is negligible. For example the expected flux of the secondary produced  anti-$^4$He is $ 10^{-15}$/m$^2$/s/sr/(GeV/n) \cite{antinucl1,antinucl2}, i.e. 17 orders of magnitude below the observed flux of normal helium. For the time being, there are only upper bounds on the flux of cosmic anti-$^4$He. The best published limit is by BESS, $\bar {\rm He}^4 / {\rm He}^4 < 3\times 10^{-7}$, though an order of magnitude more stringent bound is expected from PAMELA and another order of magnitude improvement may be achieved by AMS. None of that is yet reported.

A complementary direction for the search of primordial antimatter in the Universe is through the identification of EM radiation produced by matter-antimatter annihilation ({\it indirect search}). In particular, we may expect an excess of $\sim 100$~MeV photons from proton-antiproton annihilation and a 0.511~MeV line from electron-positron annihilation at low energies. Current data provide an upper bound on the possible antimatter abundance in galaxies. If we consider the possibility of the existence of anti-stars, observations require that the ratio between the number of anti-stars and stars is not more than about $10^{-6}$~\cite{steigman}.

The current constraints on the abundance of antimatter become much weaker in the case of anti-matter compact objects~\cite{paper2}. Efficient mechanisms of cosmological production of antimatter lumps were studied in Refs.~\cite{compact1,compact2}. Such antimatter objects would be compact and distributed over large volume (e.g. galactic halos) rather than concentrated in galactic disks, two ingredients that make antimatter objects much more difficult to observe. The phenomenology of such baryogengesis scenarios is discussed in Refs.~\cite{paper1,paper2}, where bounds from current observations are also derived. These antimatter objects may also represent the cosmological DM and therefore baryogengesis models predicting lumps of antimatter can potentially explain both the matter-antimatter asymmetry in the Universe and the origin of DM~\cite{compact2,compact1}, especially if such antimatter objects predominantly form primordial BHs~\cite{BDPP}.

The model of antimatter formation considered in Refs.~\cite{paper1,paper2}  allows for abundant antimatter in the Galaxy but it is difficult to present a reasonable limit on its density because it strongly depends upon the types of the antimatter objects. Some other scenarios of cosmological antimatter creation are discussed in~\cite{KRS} and references therein.
Constraints on the abundance of antimatter from Fermi-LAT are reported in~\cite{pvb14}. Observations require that the antistar to star ratio in the local Galactic neighborhood is less than $4 \cdot 10^{-5}$. The fraction of antimatter in the interstellar medium in the Galaxy and in nearby galaxies (Andromeda, Large Magellanic Cloud, Small Magellanic Cloud) is constrained to be less than $\sim 10^{-16}$. The fraction of antimatter in the medium between galaxy clusters is constrained to be less than $\sim 10^{-8}$.

\paragraph*{Expected results with e-ASTROGAM}
Indirect search for antimatter requires the observation of \g-rays from 0.5 to about 100~MeV, corresponding to the energy range between the electron and the pion masses. Current data at 0.5-1~MeV are from \INTEGRAL, at 1-30~MeV from COMPTEL/CGRO, and at 30-100~MeV from EGRET/CGRO and Fermi-LAT. Generally speaking, e-ASTROGAM will be able to measure lower fluxes and thus improve current bounds on the abundance of anti-objects in our Galaxy and in the whole Universe. e-ASTROGAM can measure fluxes two orders of magnitude smaller than previous missions at energies below $30$~MeV, and about an orders of magnitude smaller than EGRET and Fermi-LAT at 30-100~MeV. Recently, the AMS experiment has detected 4-5~candidate anti-helium-3 particles~\cite{anti-he3}. If confirmed, this would strongly suggest the existence of lumps of primordial antimatter in the contemporary Universe. e-ASTROGAM could investigate such a possibility with a complementary approach, looking for the annihilation signal of such antimatter lumps.
\subsection[Search of gamma-ray coherence effects\\
\noindent
\textit{\small{M. Mariotti, E. Prandini, R. Rando, R. L\'opez-Coto, M. Mallamaci, A. De Angelis}}]{Search of gamma-ray coherence effects}
\label{s:fp:mariotti}
\paragraph*{Science questions}
Coherence effects are taking place by interference of two or more identical photons and can be revealed by searching for coincidence events of a pair (or more) of photons coming from the same source. More than fifty years ago, Hanbury-Brown and Twiss (HBT) \cite{HBT} discovered photon bunching  in the visible light emitted by a chaotic source \cite{Scully} and stimulated the development of modern quantum optics \cite{Glauber}. In a similar manner, coherence at higher energies can be seen for compact and distant sources by searching for a distinct pairs of photons coming at the same time, from the same direction but shifted by some lateral separation. High energy photons can be naturally produced in coherent manner by stimulated radiative process, like ``free electron" laser mechanism \cite{Desy}. \\ We propose to search for event coincidences with a high time resolution detector, in order to understand if stimulated radiative processes are taking place in astrophysical sources. Coherence effects can also take place during the propagation from the source to the observer. If detected, the lateral separation distribution can give precise information on the size of the emitting region.
\paragraph*{Importance of gamma-ray observations}
Interference of EM radiation is a powerful tool for astronomy and gives the opportunity to study in great detail astrophysical sources with an incredible angular resolution, that depends on the ratio between wavelength  and lateral separation of the telescopes.
The angular resolution is improved by increasing the distance of the telescopes or decreasing the wavelength. Unfortunately the degradation of the photon phase, in the visible range traveling in the atmosphere, makes very difficult the use of phase interferometry for short wavelengths. In the 1960s HBT demonstrated however that it was possible to perform stellar interferometry \cite{HBT} by means of the second order coherence, also known as intensity interferometry. 
The HBT effect, seen on visible light, can be extended to an arbitrary pair of quantum mechanically indistinguishable particles\cite{Jeltes}\cite{Schellekens}, including X rays and \g-rays. The effect manifests itself with an enhanced probability to detect two of them simultaneously and nearby with respect to a pair of classically distinguishable particles. 
\begin{figure}
\includegraphics[trim= 0cm -10cm 0cm 12cm, width=0.48\textwidth]{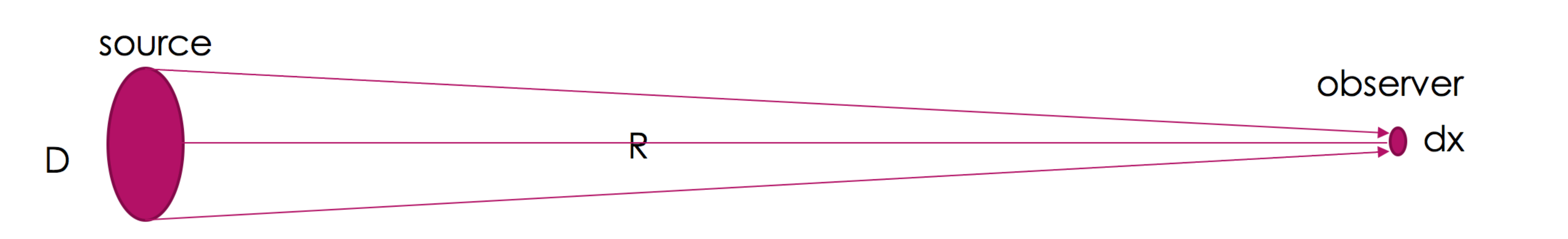}
\includegraphics[width=0.58\textwidth]{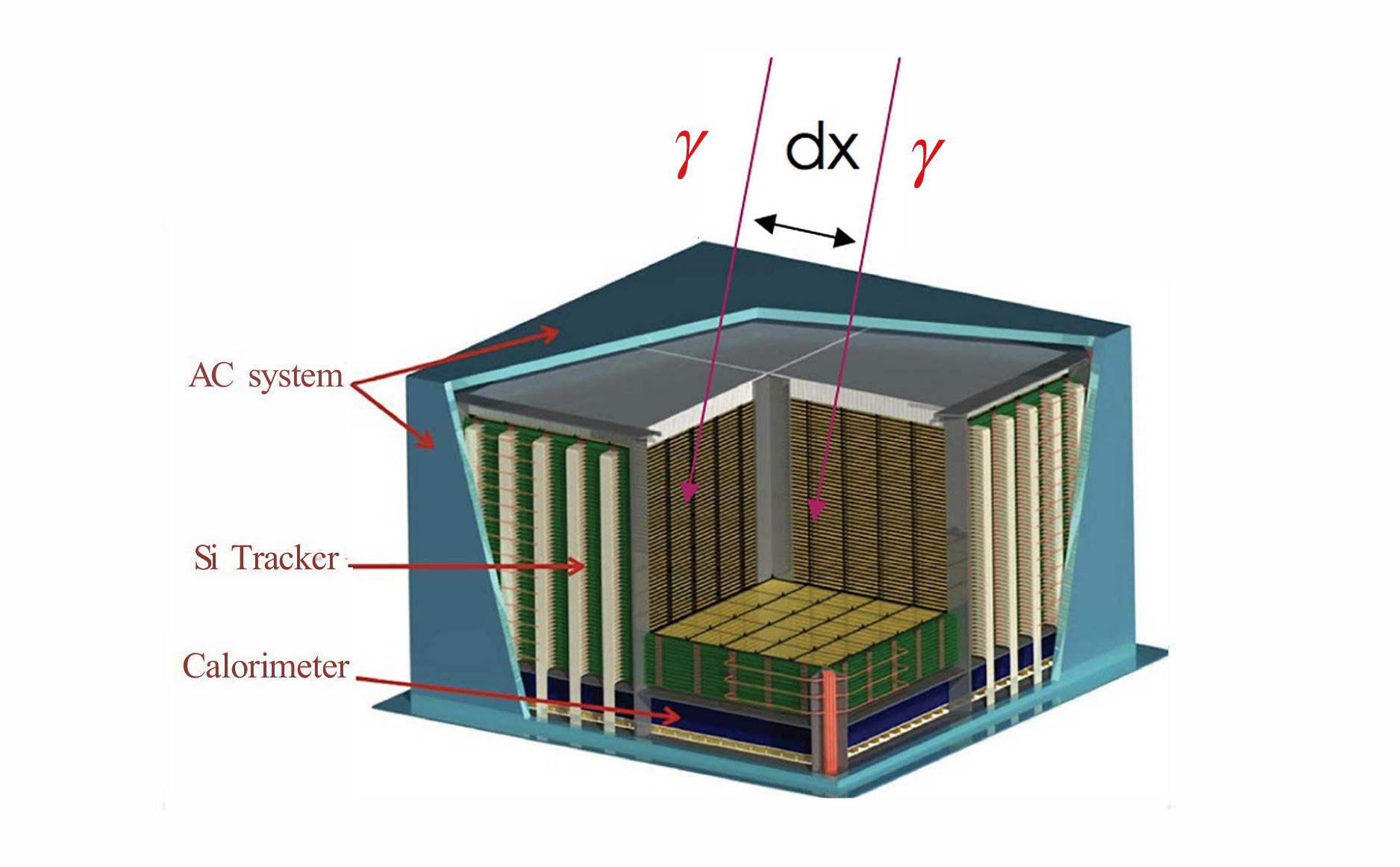}
\caption{Left: Schematic relation between the source emitting region D and the observer, separated by a distance R; $\mathrm{dx}$ is the photon separation in the observer frame. If $\mathrm{dx} < \lambda \mathrm{R/D}$, the pair of bosons are in the same state, and interference might occur. Right: A pair of coherent photon events impinging the detector, separated by a distance $\mathrm{dx}$}.
\label{fig1:mariotti}
\end{figure}
Clearly, this phenomenon is not a property of the source(s) since the emission can be completely independent. It is solely the indistinguishability of quantum mechanical amplitudes at each detector that makes this effect possible.
The compactness of the source and the large distance from the observer can cause this effect at higher energy. The relation 
$$
\mathrm{dx} \leq \lambda \mathrm{R/D}
$$
holds where $\mathrm{dx}$ is the lateral separation of photons at the observer frame, $\lambda$ is the wavelength, R is the distance of the source and D is the size of the emitting region. The relation $\mathrm{dx} \leq \lambda \mathrm{R/D}$ can be satisfied for short wavelengths in the case of a distant compact object resulting in a lateral displacement $\mathrm{dx}$ at the observer within the dimension of the detector and its spatial resolution.
Just to give an example, by taking a star with a diameter like our Sun, at light year ($ly$) distance and photons of visible light of $\lambda$ = 500 nm, we have
$$
\mathrm{R/D} = 6.7\times10^6 / \mathrm{ly} 
$$        
$$
\mathrm{dx}\leq \lambda \mathrm{R/D} = 3.3 \,\mathrm{m/ly}
$$
which means that lateral coherence effects, of enhanced coincidence rate, are taking place only if the two detectors are in within 3.3 m for 1 ly or 330 m in the case of a star at a distance of 100 ly.
If we repeat the same exercise for a compact source, like a pulsar, taking as example the Crab, at a distance of R=$6.78\times10^{16}$ km, and the emitting region size is 20 km or 100 km we can expect coherence effect taking place for displacement:
$$ 
\mathrm{dx} \leq 4.2\ \mathrm{cm}\, (100 \ \mathrm{GeV}\mathrm{,} \ 20 \ \mathrm{km})
$$

$$
\mathrm{dx} \leq 21\  \mathrm{cm}\, (1\,\mathrm{GeV}, 100\  \mathrm{km})
$$ 
Energy and emitting region size scale linearly with the lateral coherence length $\mathrm{dx}$. 
A priori is not possible to know exactly the emitting region size, nor if stimulated coherence emission is taking place (in this latter case the chance probability to detect the effect is statistically much larger), however the above examples are telling us that e-ASTROGAM will have the right size to explore this possible fundamental physics effect.
What has to be implemented in the final design of the detector is a proper trigger and reconstruction algorithm that will not discard a priori a "double gamma" event of the same energy same direction and same arrival time. 
In principle {\it Fermi}-LAT would be capable of observing this effect as well, but the trigger and background filters were not designed to accept such double-track events.
Terrestrial gamma-ray flashes (TGF) are an example of multiple-photon events studied with the \Fermi \cite{Briggs2010}. In this case, though, the GBM data are used to identify the TGF flare, while the raw LAT data are inspected to confirm the presence of multiple tracks with plausible terrestrial origin in the flagged time window ($\leq1$ ms). A special instrument configuration optimized for TGF measurement is available for dedicated, Earth-pointed runs. Of course this is not a viable approach for our science case. 

\paragraph*{Expected results with e-ASTROGAM}
With a proper trigger and DAQ design, able to accept also pair of photons events,  e-ASTROGAM will be an unique instrument to test the HBT effect at energies in the MeV-GeV domain. Although statistically limited for chaotic sources, HBT bouncing effect can give very detailed information on the source emitting region, and it is expected to happen in compact sources.

The HBT bouncing effect is taking place both for chaotic and coherent sources. We do not know a priori if coherent emission processes are taking place in astrophysical sources, however, in case of coherent emission, it will be much easier to detect the HBT effect due to the abundance of native identical photons.
There are several mechanism that can produce coherent gamma-ray emission in astrophysical sources. As example of a possible mechanism, coherent X radiation can be produced in the laboratory \cite{Desy} by stimulated emission of relativistic electrons through a periodically varying magnetic field. This shows that such processes are available in principle. Some promising astrophysical process might arise, for instance, from the interaction of a collimated beam of relativistic electrons and magnetic field \cite{Harwit}.
If the emission is natively coherent in the source, in other words if gamma-rays are emitted as stimulated radiation, the photon bunching effect will be stronger and much easier to reveal with  e-ASTROGAM and, per se, a great discovery.

\newpage
\section[Explosive nucleosynthesis and chemical evolution of the Galaxy\\
\noindent
\textnormal{\small\textnormal{Convenors:} \textit{J. Isern, M. Leising, V. Tatischeff}}
]{Explosive nucleosynthesis and chemical evolution of the Galaxy}
Exploding stars play a very important role in astrophysics since they inject important amounts of kinetic energy and newly synthesized chemical elements into the interstellar medium in such a way that they completely shape the chemical evolution of galaxies. Furthermore, the ``pyrotechnical'' effects associated with such outbursts can be so bright and regular that they can be used to measure distances at the cosmological scale. For instance, Type Ia SNe (SNIa) allowed the discovery that the Universe was expanding in an accelerated way.  

The majority of the outbursts are associated  with instabilities of electron degenerate structures in single stars (core collapse and electron capture supernovae) or when they accrete matter from a companion in a close binary system (SNIa and classical novae, for instance). Systematic research on transient events  have revealed a surprising variety of outbursts that goes from ``Ca-rich'' transients, placed in the gap between Type Ia SNe and novae, Type Iax, ``02es-like'' SNe, ``super-Chandrasekhar'' SNe in the domain of the so-called thermonuclear SNe, to, e.g., Type IIn, Type In, and so-called ``impostors'' in the domain of core collapse of massive stars.

Many of these events, if not all, imply the activation of thermonuclear burning shells that synthesize new isotopes, some of them radioactive. As the ejecta  expand, more and more photons avoid thermalization and escape, such that they can be used as a diagnostic tool. Each one of the different explosion scenarios leads to differences in the intrinsic properties of the ejecta, like the density and velocity profiles, and the nature and distribution of the radioactive material synthesized. This translates into differences in the light curves and line widths of the expected \g-ray emission. Therefore, the observation with \g-rays becomes a privileged diagnostic tool with respect to other measurements thanks to the penetration power of high energy photons and the association of \g-lines to specific isotopes created by the explosion.  
\begin{table}
\begin{center}
\caption{\small Star-produced radioisotopes relevant to \g-ray line astronomy
\label{tab:radioisotopes}}
\begin{tabular}{|c|c|c|c|c|}
\hline
{\bf Isotope} & {\bf Production site$^{\rm a}$} & {\bf Decay chain$^{\rm b}$} & {\bf Half-life$^{\rm c}$} &
{\bf \g-ray energy (keV)} \\
  & & &
  & {\bf and intensity$^{\rm d}$} \\\hline
$^7$Be & Nova & $^7$Be~$\stackrel{\epsilon}{\longrightarrow}$~$^7$Li*
& 53.2~d & 478~(0.10) \\\hline
$^{56}$Ni & SNIa, CCSN &
$^{56}$Ni~$\stackrel{\epsilon}{\longrightarrow}$~$^{56}$Co*
& 6.075~d & 158~(0.99), 812~(0.86) \\
& & $^{56}$Co~$\stackrel{\epsilon(0.81)}{\longrightarrow}$~$^{56}$Fe*
& 77.2~d & {847}~(1), {1238}~(0.66) \\\hline
$^{57}$Ni & SNIa, CCSN &
$^{57}$Ni~$\stackrel{\epsilon(0.56)}{\longrightarrow}$~$^{57}$Co*
& 1.48~d & 1378~(0.82) \\
& & $^{57}$Co~$\stackrel{\epsilon}{\longrightarrow}$~$^{57}$Fe*
& 272~d & {122}~(0.86), {136}~(0.11) \\\hline
$^{22}$Na & Nova &
$^{22}$Na~$\stackrel{\beta^+(0.90)}{\longrightarrow}$~$^{22}$Ne*
& 2.60~y & {1275}~(1) \\\hline
$^{44}$Ti & CCSN, SNIa &
$^{44}$Ti~$\stackrel{\epsilon}{\longrightarrow}$~$^{44}$Sc*
& 60.0~y & {68}~(0.93), {78}~(0.96) \\
& & $^{44}$Sc~$\stackrel{\beta^+(0.94)}{\longrightarrow}$~$^{44}$Ca*
& 3.97~h & {1157}~(1) \\\hline
$^{26}$Al & CCSN, WR &
$^{26}$Al~$\stackrel{\beta^+(0.82)}{\longrightarrow}$~$^{26}$Mg*
& 7.2$\cdot$10$^5$~y & {1809}~(1) \\
 & AGB, Nova & & & \\\hline
$^{60}$Fe & CCSN &
$^{60}$Fe~$\stackrel{\beta^-}{\longrightarrow}$~$^{60}$Co*
& 2.6$\cdot$10$^6$~y & 59~(0.02) \\
& & $^{60}$Co~$\stackrel{\beta^-}{\longrightarrow}$~$^{60}$Ni*
& 5.27~y & 1173~(1), 1332~(1) \\\hline
\end{tabular}
\begin{minipage}{0.92\linewidth}
{\vspace{1 mm} \small
{$^{\rm a}$ Sites which are believed to produce observable
\g-ray line emission. Nova: classical nova; SNIa: thermonuclear
SN (type Ia); CCSN: core-collapse SN; WR: Wolf-Rayet star;
AGB: asymptotic giant branch star.}

{$^{\rm b}$ $\epsilon$: orbital electron capture. When an
isotope decays by a combination of $\epsilon$ and $\beta^+$ emission, only 
the most probable decay mode is given, with the corresponding fraction in 
parenthesis.}

{$^{\rm c}$ Half-lifes of the isotopes decaying by 
$\epsilon$ are for the neutral atoms.}

{$^{\rm d}$ Number of photons emitted in the \g-ray
line per radioactive decay.}
}
\end{minipage}
\end{center}
\end{table}
Table~\ref{tab:radioisotopes} displays the main detectable \g-ray line emissions expected in several nucleosynthesis events (see Ref.~\cite{die11} and references therein). The radioisotopes with a relatively short lifetime can be used to directly characterize the individual explosion events or the first stages of the remnant, while the long-lived radioactivites, i.e., with lifetimes much longer than the characteristic time between events, will produce a diffuse emission resulting from the superposition of many sources that can provide information on stellar nucleosynthesis, but also on the physical conditions and dynamics of the Galactic interstellar medium (see, e.g., \cite{kra15}).

It is important to distinguish here between guaranteed and opportunity observations. By guaranteed, we understand observations that can be predicted with enough anticipation and with the certitude that they can be included into the ordinary mission scheduling. Three examples of guaranteed observations would be:
\begin{enumerate}
\item Measurement of the total mass of $^{56}$Ni/$^{56}$Co ejected by SNIa. This value is fundamental to calibrate the Phillips \cite{phi93} relation and the yield of synthesized Fe. The explosion time and location are not known a priori, but thanks to the sensitivity of e-ASTROGAM, it is expected that about a dozen of SNIa will occur at a distance smaller than 35 Mpc in three years of mission. The observations will have to be performed around 50--100 days after the explosion, when all the SN properties (subtype, luminosity,...) will already be known.
\item Clumping degree of core-collapse SNRs as a diagnostic of internal asymmetries. This property can be obtained from the radioactive emission of the $^{44}$Ti/$^{44}$Sc chain. The sensitivity of e-ASTROGAM would allow the detection of this emission in all young Galactic SNRs and in the remnant of SN1987A.
\item Mapping of the positron annihilation radiation and the long-lived radioactivities $^{26}$Al and $^{60}$Fe. The huge increase in sensitivity of e-ASTROGAM compared to current \g-ray missions should allow the building of detailed maps of these Galactic diffuse emissions, which will shed a new light on nucleosynthesis in massive stars, SNe and novae, as well as on the structure and dynamics of the Galaxy. Individual objects (e.g., SNRs) should also be detected in these lines.
\end{enumerate} 
Given the explosive nature of the events considered here, the majority of the observations will belong to the category of Targets of Opportunity (ToO). The information and the relevance of the observation will depend on the distance of the events. Two examples would be:
\begin{enumerate}
\item Novae. The sensitivity of e-ASTROGAM would allow the detection of the $^{22}$Na (1275 keV) line to a distance large enough to observe about one nova per year, but that of the $^{7}$Be (478 keV) line demands a shorter distance and is thus uncertain during the three years of nominal mission duration. Therefore the results that can be obtained from every individual event will depend not only on the nature of the event, but also on the distance.
\item Type Ia and Core-collapse SNe. The detection of the early \g-ray emission before the maximum optical light in the SNIa case and the determination of the amount of $^{56}$Ni ejected by CCSN would be fundamental to understanding the nature of the progenitor in the first case and of the explosion mechanism in both cases. Given the sensitivity of e-ASTROGAM it is foreseen to detect these details to a distance of about ten Mpc, which ensures the detection of several events during the entire mission and opens the possibility of comparing SN subtypes. 
\end{enumerate}
The observation of ToOs is unpredictable, but extremely rewarding if successful, and exploding stars and related phenomena are within this category. It is important to realize that the increased sensitivity of e-ASTROGAM guarantees that a 
significant number of events will be observed in an effective way.
\subsection[Thermonuclear supernovae (SN Ia)\\
\noindent
\textit{\small{E. Churazov, R. Diehl, J. Isern, V. Tatischeff}}
]{Thermonuclear supernovae (SN Ia)}
\paragraph*{Science questions}
SN Ia are the outcome of a thermonuclear burning front that sweeps a carbon/oxygen white dwarf (WD) in a close binary system. But exactly how the ignition conditions are obtained, and on which WDs, and more so how the thermonuclear runaway proceeds through the WD and turns it into a variety of isotopes that are ejected, all this is subject to considerable debate (see, e.g., \cite{hil00,hi13} and references therein). It seems that several candidate evolutionary channels may all contribute, from the \emph{double degenerate} variant of merging WD binaries disrupting one of the dwarfs through tidal forces or a hard collision, to a variety of \emph{single degenerate} models where accretion of material from a companion star may lead to either the WD reaching the critical Chandrasekhar mass stability limit, or be ignited earlier through a surface explosion from a helium flash.  

Such uncertainties are troublesome for cosmology since the use of SN Ia as standard candles depends on an empirical relationship between the shape and the maximum of the light curve \cite{phi93}. Although useful up to now, in view of the development of \textit{precision cosmology}, a better, astrophysically supported understanding of thermonuclear SNe, as well as their evolutionary effects at large distances and low metallicities, are mandatory. The brightness-decline relation \cite{phi93} is closely related to the mass of synthesized $^{56}$Ni, and factors like the progenitor evolution, ignition density, flame propagation, mixing during the burning, completeness of burning in outer, expanding regions, all lead to different amounts of $^{56}$Ni.

Furthermore, SNIa are the main producers of iron peak elements and understanding the rate at which these elements are injected into the interstellar medium is fundamental to understand the chemical evolution of the Galaxy.
\paragraph*{Importance of \g-ray observations}
The mass of $^{56}$Ni synthesized in the explosion is measured directly through \g-ray lines. On the other hand, radiation transport from radioactivity to optical light and their spectra depend on complex atomic line transitions in the expanding SN, as well as on the total mass burned, the amount and distribution of radioactive nickel and intermediate mass elements, all of which must combine in quite a tight way to reproduce the observations \cite{woo07,ker14}. 

With SN~2014J, for the first time a SN Ia occurred close enough for current generation \g-ray telescopes, at 3.5 Mpc in the starburst galaxy M82. \INTEGRAL data could detect the long awaited \g-ray signatures of the thermonuclear runaway. The lines of  $^{56}$Co (life time of $\sim$111 days) at 847 and 1238 keV, consistent in flux and broadening with the predictions of a canonical Chandrasekhar WD explosion model, have been unambiguously detected \cite{chur14,chur15,diehl15}, as well as a Thomson-scattered continuum and positron annihilation emission. Even although overall significance of the signal was just above $10\,\sigma$, good constraints were obtained on the mass of radioactive material and the expansion velocity of the ejecta. Moreover, possible signatures of the radioactive  $^{56}$Ni (mean lifetime $\sim$8.8 days) lines have been reported  \cite{diehl14,isern16}, albeit at lower statistical significance. The presence of such lines in the spectrum, if confirmed, would suggest either a surface explosion or very special morphology of the runaway in contrast to the conventional model. Clearly, the glimpse offered by SN~2014J observations with \INTEGRAL underline the importance of \g-ray line diagnostics in these systems and emphasize that more and better observations hold the key to a deeper understanding of how the thermonuclear
explosion of a WD star unfolds. 

The presence of a bump in the early light curve of SN2016jhr, as recently reported by \cite{jia17}, provides additional support to the idea that SNe triggered by the ignition of an outer helium layer or cap do occur in nature. However, such lightcurve bumps remain ambiguous in their nature, as they can also be produced by other mechanisms like the interaction of SN ejecta with circumstellar material or internal shocks \cite{noe17}, while characteristic \g-ray lines from the $^{56}$Ni decay chain are unambiguous tracers of the underlying physical process. 
\paragraph*{Expected results with e-ASTROGAM}
\begin{center}
\begin{figure}
\includegraphics[trim= 0.5cm 6cm 1cm 12cm, width=0.98\textwidth,clip=t,angle=0.,scale=0.52]{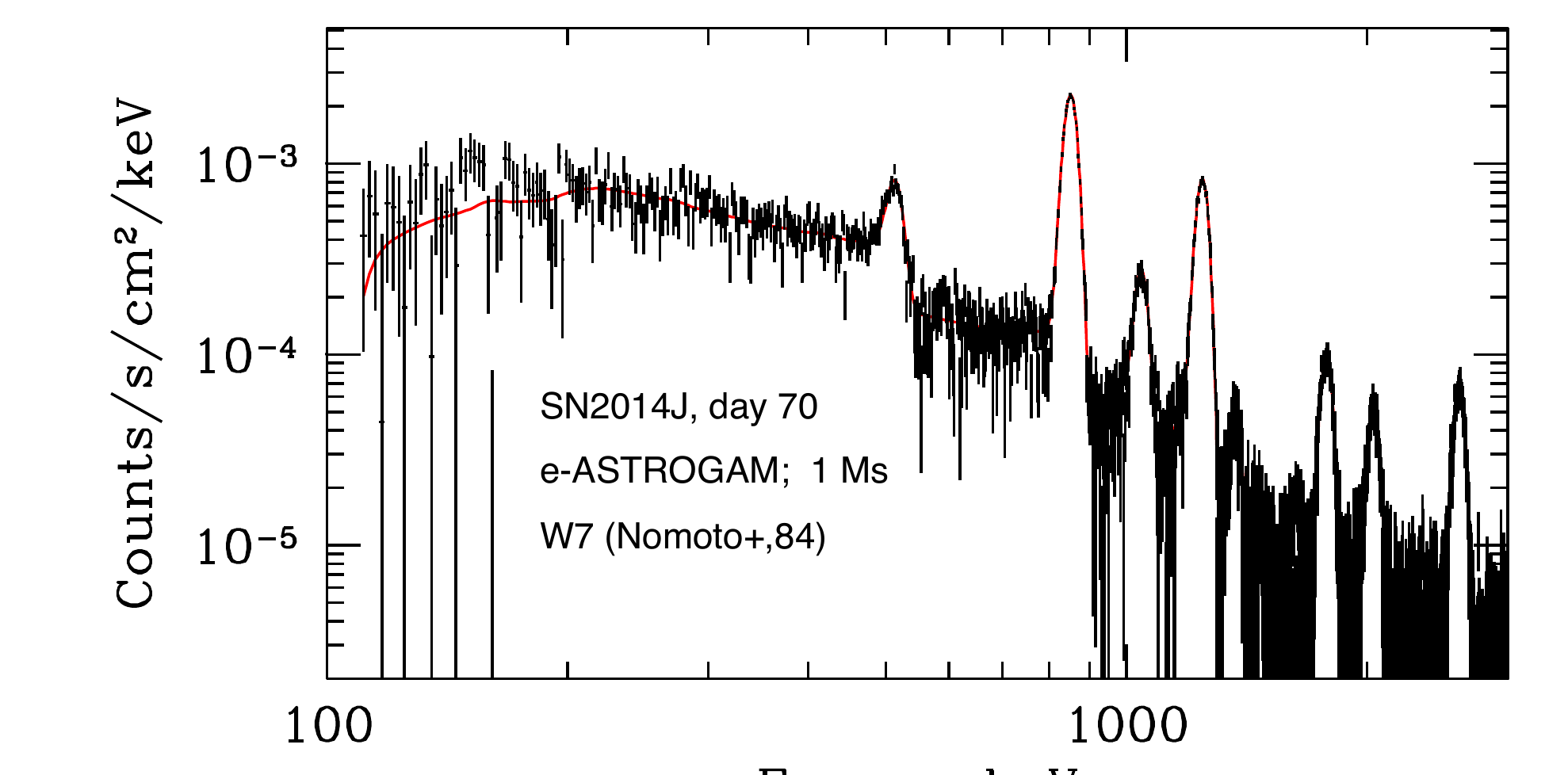}
\includegraphics[trim= 0.5cm 6cm 1cm 12cm, width=0.98\textwidth,clip=t,angle=0.,scale=0.52]{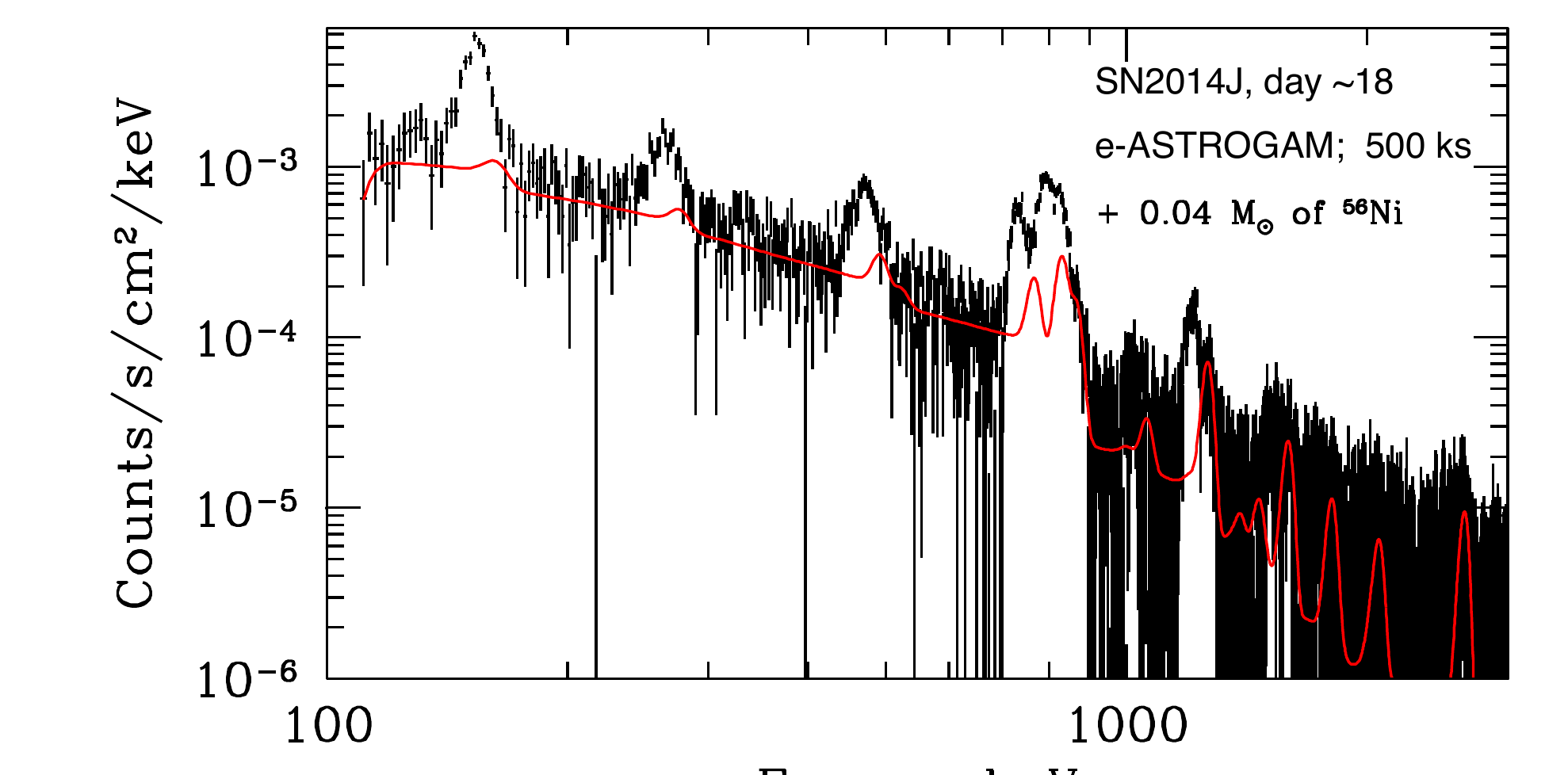}
\caption{\small Left: Simulated background-subtracted spectrum of a SN event like SN~2014J (distance of 3.5 Mpc; 70 days after explosion; 1 Ms exposure). The spectrum is dominated by $^{56}$Co lines, scattered continuum and annihilation emission. The W7 model \cite{nomoto84} is used in the simulations \cite{chur14}.  Such event would be detected at $\sim300\,\sigma$ by e-ASTROGAM.  Right:  Simulated spectrum for a model with 0.04~$M_\odot$ of radioactive $^{56}$Ni located outside the main ejecta (3.5 Mpc; 500 ks exposure centered at day 17.5 after explosion; 3Dball model from \cite{isern16}). Extremely bright lines of $^{56}$Ni at 158 and 810 keV are clearly detected. Such lines would not be seen if all $^{56}$Ni is confined within the ejecta (see red curve that shows a canonical model).  
\label{SNIa:fig1}
}
\end{figure}
\end{center}
e-ASTROGAM will achieve a major gain in sensitivity compared to \INTEGRAL for the main \g-ray lines arising from $^{56}$Ni and $^{56}$Co decays (see Figs.~\ref{SNIa:fig1} and \ref{SNIa:fig2}) allowing for events like SN~2014J the exquisitely accurate (at percent level) measurements of the  Ni mass, the mass of the progenitor and the expansion velocity, easily differentiating between major astrophysical scenarios. For instance, the presence of an envelope of 0.2~$M_\odot$ around a canonical Chandrasekhar WD (such envelope might appear due to the merger of WDs), would be detected at $50\,\sigma$ level in 1~Ms observation. Moreover, e-ASTROGAM will be able to i) detect \g-rays up to 600-700 days after the explosion, when ejecta are essentially transparent to \g-rays, ii) measure with 3-10\% accuracy the annihilation rate of positrons produced during $^{56}$Co decay up to 300 days, iii) verify the presence of even very small ($\sim 2 \times 10^{-3}~M_\odot$) amount of $^{56}$Ni at the surface of the remnant (see Fig.\ref{SNIa:fig1}, right panel) and iv) monitor the emergence of scattered continuum during early phase of the ejecta expansion (Fig.\ref{SNIa:fig2}).
These data will allow us to probe the explosion mechanism in detail, and compare with astrophysical models for each event to better understand the progenitor system(s) and the thermonuclear explosion process. We also note that for a truly nearby type Ia SN, e.g., in M31 or in the Milky Way,  even more ambitious diagnostics will be possible, including a search for asymmetry in the ejecta by using  e-ASTROGAM polarization capabilities or measuring the rate of positrons escape \cite{chur17}.  

Events like SN~2014J are, of course, rare. However, with the e-ASTROGAM sensitivity, the observatory will be able to detect such SN up to a distance of 35 Mpc at $10\,\sigma$ level, i.e., corresponding to the \INTEGRAL detection of SN~2014J at 3.5 Mpc after few Ms integration time. In this volume, that includes, in particular, the Virgo cluster of galaxies, one can expect about 10 type Ia SN explosions in 3 years of nominal mission lifetime. Such sample would allow for a clean and fundamental test of the Phillips relation for a dominant population of Branch Normal type Ia SNe. Moreover, about 30\% of SN~Ia's are peculiar and e-ASTROGAM has an excellent chance to detect few of those, especially bright ones, like SN~1991T, due to the Malmquist bias. Therefore, even without relying on ``lucky'' events like SN~2014J, e-ASTROGAM will be able not only to elucidate the nature of the Phillips relation, but also to study the departures from it.
Overall, e-ASTROGAM will provide a decisive reference set of data on type Ia SNe, addressing questions ranging from the progenitor system(s) and the physics of thermonuclear runaway in WDs to the use of type Ia SNe for cosmology.
\begin{SCfigure}
\centering
\includegraphics[width=0.65\textwidth]{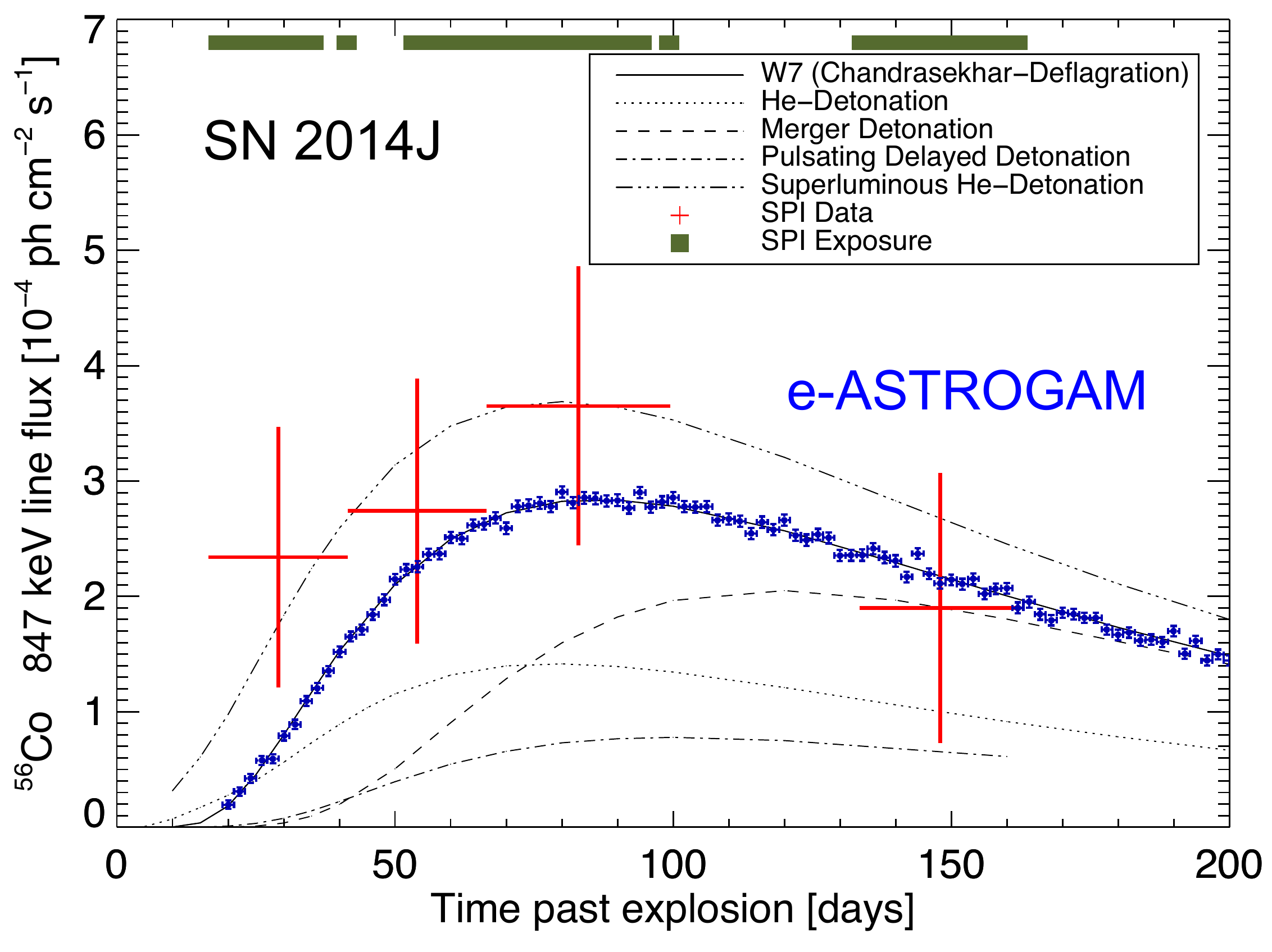}
\caption{\small Light curve of the 847 keV line from $^{56}$Co decay in SN~2014J. \INTEGRAL data (adapted from Fig.~4 in Ref.~\cite{diehl15}, red data points) are compared to various models of Type Ia SN \cite{the14}. A simulation of the e-ASTROGAM response to a time evolution of the 847 keV line such as in the W7 model \cite{nomoto84} shows that the sensitivity improvement by e-ASTROGAM (blue points) will lead to a much better understanding of the SN progenitor system and explosion mechanism.}
\label{SNIa:fig2}
\vspace{-10pt}
\end{SCfigure}
\subsection[Core-collapse supernovae\\
\noindent
\textit{\small{J. Isern, M. Leising, R. Diehl, V. Tatischeff}}
]{Core-collapse supernovae}
\paragraph*{Science questions}
Stars more massive than 11--12 $M_\odot$ develop a massive Si--Fe core that progressively grows until it becomes unstable and collapses to form a compact object (NS or BH), giving rise to a gravitational SN (a core collapse supernova -- CCSN). Stars in the mass-range 10--11 $M_\odot$ develop an O--Ne core that grows until it collapses to form a NS as a consequence of the electron captures on oxygen (an electron capture supernova --ECSN) ~\cite{doh17}. The outcome of such instability is the formation of a proto-NS that can lead to the formation of a NS or a BH after accretion of enough matter. The phenomenological properties of the explosion not only depend on the amount of the energy deposited but also on the structure and chemical composition of the envelope at the moment of the explosion which, in turn, depends on the initial mass and metallicity of the star, and the presence of a companion in the case of close binary systems. To these factors it is necessary to add the influence of rotation and magnetic fields. Just as an example of the existing uncertainties, it is necessary to mention the presence of residual amounts of C in the inner regions of ONe degenerate cores that could completely change the present picture of ECSN~\cite{doh17}. 

The theory of core collapse, which involves hydrodynamics and  shock  physics,  radiative  transfer,  nuclear  physics,  neutrino physics,  particle physics, statistical  physics  and  thermodynamics, gravitational physics, and convection theory, is still not well understood in terms of an astrophysical model (e.g. Refs.~\cite{woo05,jan12,bur13}). The main goal is to explain a tremendous variety of core collapse events, e.g. the peculiarities of the Crab nebula and pulsar, the distribution of the elements in events like Cassiopeia A (Cas~A), collapsars that appear as \g-ray burst sources and produce stellar mass BHs, superluminous SNe that may be powered entirely differently by magnetar rotational energy, or pair instability SNe that create huge amounts of radioactive $^{56}$Ni. This requires quantitative explanations of a number of observation facts, such as \cite{bur13} (i) the relative proportions of stellar BHs and NSs, (ii) the mass distribution of the residual NSs, (iii) the high average proper motion speeds of radio pulsars (the fastest population of stars in the Galaxy), (iv) the observed morphologies of SN explosions and spatial distributions of the ejecta, as well as (v) the measured nucleosynthetic yields as a function of stellar progenitor.

The different scenarios and models that have been advanced to account for these explosive events predict the synthesis of many radioactive isotopes that can be used as a diagnostic tool and can provide key information to understand the phenomenon.
\paragraph*{Importance of \g-ray observations}
The main detectable \g-ray line emissions from radioactive nucleosynthesis products of CCSNe result from the decay chains (i) $^{56}$Ni ($T_{1/2}=6.075$~d) $\rightarrow$ $^{56}$Co ($T_{1/2}=77.2$~d) $\rightarrow$ $^{56}$Fe, (ii) $^{57}$Ni ($T_{1/2}=35.6$~h) $\rightarrow$ $^{57}$Co ($T_{1/2}=271.7$~d) $\rightarrow$ $^{57}$Fe, and (iii) $^{44}$Ti ($T_{1/2}=60.0$~y) $\rightarrow$ $^{44}$Sc ($T_{1/2}=3.97$~h) $\rightarrow$ $^{44}$Ca, as well as from the long-lived radioisotopes $^{26}$Al ($T_{1/2}=7.17\times 10^5$~y) and $^{60}$Fe ($T_{1/2}=2.62\times 10^6$~y). While these two last radioactive isotopes have half-lives that are much longer than the characteristic timescale between two explosive events, $\sim 75$~y for CCSNe \cite{die06,die16}, such that they produce diffuse \g-ray line emissions resulting from the superposition of numerous Galactic sources (see contribution ``Diffuse \g-ray line emissions'' in this White Book), they can also be measured in individual nearby sources, such as the Vela SNR.  Isotopes like  $^{44}$Ti, $^{56}$Co, and $^{56}$Ni can be detected in individual CCSNe, and this is one of the more direct ways to extract information on the inner processes triggering the explosion near the newly forming compact stellar remnant (e.g., \cite{gre14}). Other observables, such as the optical light curve and thermal X-ray emission from shocked-heated gas, are more indirect, and mostly reflect interactions within the envelope, or with circumstellar, pre-explosively ejected, or ambient gas. 

The power of \g-ray observations to study the physics of core collapse is exemplified with the observations of SN~1987A and Cas~A. Thus, the early appearance and measured profiles of $^{56}$Co \g-ray lines in SN~1987A \cite{mat88} provided key indications for an asymmetric explosion and the rapid mixing of $^{56}$Ni in the outer ejecta (e.g. \cite{mah88,tue90}). The spatial distribution of $^{44}$Ti in the remnant of Cas~A as revealed by \nustar's observations provides strong evidence of explosion asymmetries caused by the development of low-mode convective instabilities in CCSNe \cite{gre17,gre14}.
\begin{figure}
\centering
\includegraphics[width=0.65\textwidth]{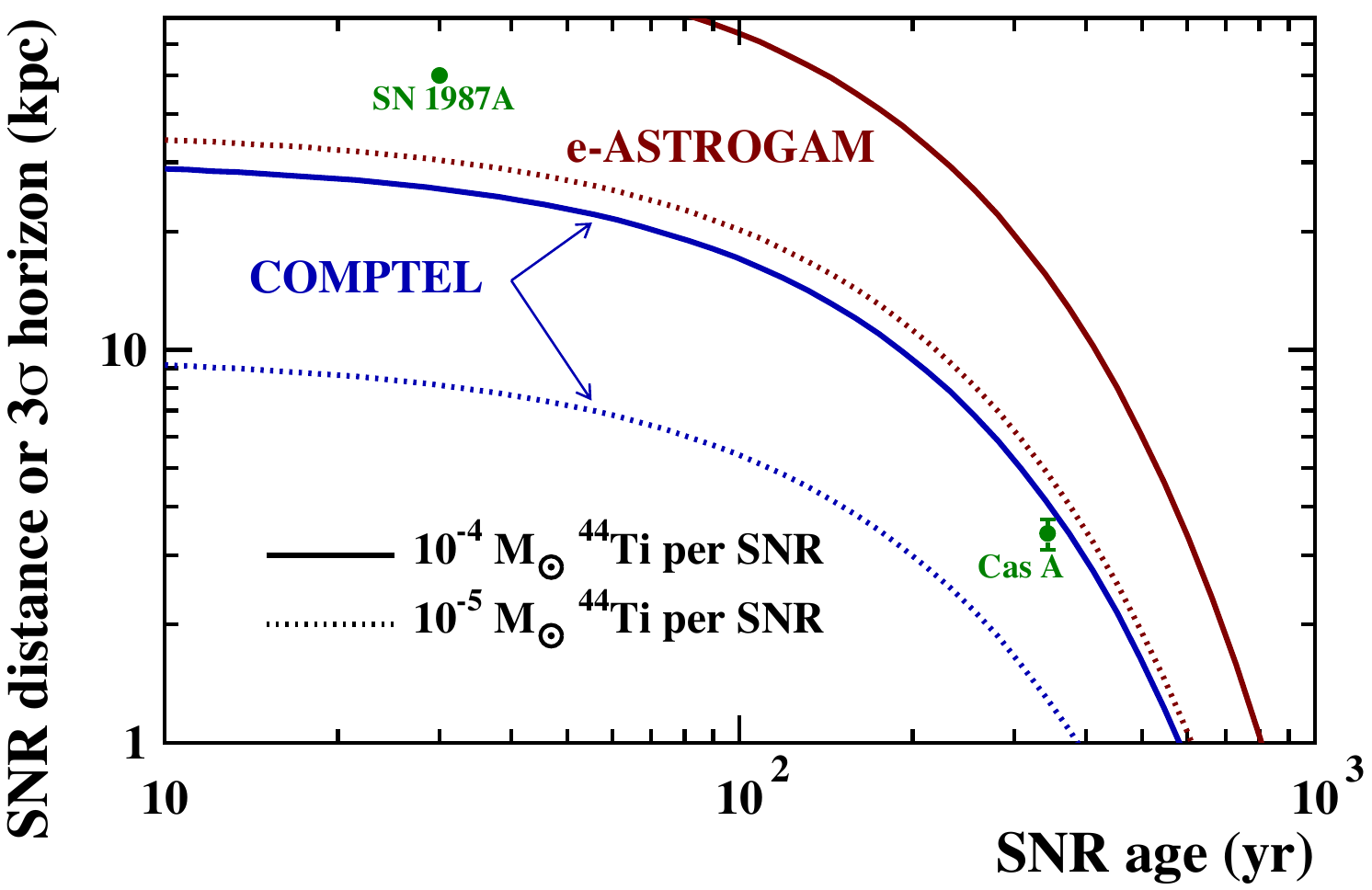}
\caption{\small Horizon of detectability of the $^{44}$Ti line at 1157~keV as a function of SNR age for CGRO/COMPTEL (blue lines) and e-ASTROGAM (red lines). The plotted sensitivities are for an effective source exposure of 1 year (COMPTEL: $9.0 \times 10^{-6}$ ph~cm$^{-2}$~s$^{-1}$; e-ASTROGAM: $6.4 \times 10^{-7}$ ph~cm$^{-2}$~s$^{-1}$), assuming two different yields of $^{44}$Ti production per SNR: $10^{-5}~M_\odot$ (common events; dotted lines) and $10^{-4}~M_\odot$ (Cas~A-like events; solid lines). Also shown are the age and distance of the two CCSNe with detected $^{44}$Ti: SN~1987A and Cas~A.}
\label{fig:ti44}
\end{figure}
\paragraph*{Expected results with e-ASTROGAM}
e-ASTROGAM will achieve a gain in sensitivity for the $^{44}$Ti line at 1157~keV by a factor of 14 compared to CGRO/COMPTEL and 27 compared to \INTEGRAL/SPI (for the same effective time exposure). As illustrated in Fig.~\ref{fig:ti44}, the proposed observatory should detect the radioactive emission from $^{44}$Ti from most of the young (age $\lesssim$ 500 yr) SNRs in the Milky Way, thus uncovering about 10 new, young SNRs in the Galaxy presently hidden in highly obscured clouds. e-ASTROGAM should also detect the youngest SNRs in the Large Magellanic Cloud and will measure precisely the amount of $^{44}$Ti in the remnant of SN~1987A, which is currently disputed in the literature (\cite{gre12,bog15,tsy16}). These observations will give new insights on the physical conditions of nucleosynthesis in the innermost layers of a SN explosion and the dynamics of core collapse near the mass cut. e-ASTROGAM could measure $^{60}$Fe yields in individual SNRs, and should not only measure $^{26}$Al, but also possibly map it in the Vela remnant, discerning whether a fraction of the $^{26}$Al is present in the X-ray emitting shrapnel \cite{asc95}.

e-ASTROGAM should also detect the signatures of $^{56}$Ni and $^{56}$Co decay from several CCSNe in nearby galaxies. The gain in sensitivity for the 847 keV line from $^{56}$Co decay amounts to a factor ranging from 30 to 70 compared to \INTEGRAL/SPI, depending on the width of the \g-ray line (i.e. the velocity dispersion of the ejected radioactive cobalt). The comparison with e-ASTROGAM of \g-ray characteristics of different classes of CCSNe, possibly including the pair instability SNe with their order of magnitude higher $^{56}$Ni production (e.g., \cite{gal09}), will probe potentially large variations in their progenitors and offer a direct view of their central engines. Asymmetries in ejected radioactivity might be reflected in 3--5 times higher line fluxes \cite{hun03}. Rare core collapse events are predicted to have \g-ray line brightnesses orders of magnitude above typical SNe: pair instability and magnetar-powered jet explosions will reveal much larger amounts of $^{56}$Ni. e-ASTROGAM will reach out to a larger part of the nearby universe to constrain the rate of such events, if not detect them.

It is also possible that e-ASTROGAM could identify the long sought site of the r-process production of heavy nuclei. Given the possible very long integration times, relatively long-lived isotopes, such as $^{126}$Sn, in nearby SNRs are the most promising targets \cite{qia98}.

\subsection[Nova explosions\\
\noindent
\textit{\small{M.~Hernanz, J.~Jos\'e, P.~Jean, A.~Coc, V.~Tatischeff, L.~Delgado, G.~Sala, S.~Starrfield, R.~Gehrz, M.~Orio, D.~de~Martino, S.~Balman}}
]{Nova explosions}
\paragraph*{Science questions}
Accreting white dwarfs in close binary systems can explode as novae and/or as SNe Ia. 
Novae are responsible for the enrichment 
of the Galaxy in some species and for the peculiar isotopic signatures found in some presolar grains \cite{JH07}. 
Understanding the origin of the elements in the Galaxy and in the whole Universe is an important topic, 
intimately related to explosive nucleosynthesis and emission of \g-rays. In fact, 
\g-rays directly trace isotopes, whereas observations at other wavelengths give only elemental 
abundances, except some measurements of CO molecular bands in the infrared, where $^{12}$CO and $^{13}$CO can be distinguished, 
thus giving the $^{13}$C/$^{12}$C ratio.

Nova ejecta are enriched in CNO nuclei, as well as in Ne, Na, Mg and even S in 
some cases (see \cite{JH98} and reviews \cite{Gehrz98,Jose16,Starr08,Starr16}). They also produce  
$^{7}$Be, which through electron-capture becomes $^{7}$Li; the role of 
novae in the origin of $^{7}$Li in the Galaxy and the Universe is a hot scientific topic. Also the contribution 
of novae to the galactic content of $^{26}$Al - traced by 1809 keV \g-rays detected since 
long ago - is still not well known. 

There are two main types of binary systems where white dwarfs can accrete matter and subsequently explode as novae. The most common case 
is a cataclysmic variable, where the companion is a main sequence star transferring H-rich matter. In this system, mass transfer 
occurs via Roche lobe overflow, and typical orbital periods range from hours to days. As a consequence of accretion, 
hydrogen burning in degenerate conditions on top of the white dwarf leads to a thermonuclear runaway and a classical nova 
explosion occurs, ejecting $10^{-3}-10^{-7}$M$_{\odot}$ with speeds $10{^2}-10{^3}$ km/s and reaching luminosities $10^5$L$_{\odot}$.
The nova explosion does not disrupt the white dwarf (as occurs in SNe Ia explosions); therefore, after enough mass 
is accreted again from the companion star, a new explosion will occur. The typical recurrence time is $10{^4}$--$10{^5}$ years, although a broader range is not ruled out.
Another scenario where a white dwarf can explode as a nova is a symbiotic binary, where the white dwarf accretes H-rich matter from the stellar 
wind of a red giant companion. Typical orbital periods for these systems are a few 100 days, much larger than in cataclysmic variables. 
This scenario leads to more frequent nova explosions than in 
cataclysmic variables, because of larger accretion rates, with typical recurrence periods smaller than 100 years, and therefore 
more than one outburst can be recorded. Recurrent novae are indeed interesting objects, since the mass of the white dwarf is expected to increase after each eruption - at least in some cases - and thereof they can explode as SNe Ia, when the white dwarf reaches the 
Chandrasekhar mass. 

The symbiotic recurrent nova RS Oph had its last eruption in 2006. It has been identified - based on the analysis of its early X-ray, IR and radio 
emission - as a site of particle acceleration, in the shocks between the ejecta
and the circumstellar matter (red giant companion wind), making it and other novae of this class responsible for
a fraction of the Galactic CRs \cite{TH07}. These ``miniature supernova remnants" are key systems to study the time dependence of 
diffusive shock acceleration of CRs, since their evolution is scaled-down with respect to standard SNRs. 
An important consequence of the production of high-energy particles is that photons with energies higher than about 100 MeV are emitted, 
both via neutral pion decay and \ic processes \cite{HT12}. Importantly enough, five to eight classical and two symbiotic recurrent novae 
have been detected by \fermilat at E$>$100 MeV, since its launch in 2008, confirming theoretical expectations (for symbiotic recurrent 
novae RS Oph-like), but being challenging to interpret for classical novae, where there's not a red giant wind with which the ejecta can interact. Recent studies of internal shocks in the ejecta have started to explain this phenomenon (see for instance \cite{Metz15-Li17}).
In most cases, this emission has been observed early after the explosion, 
around the optical maximum, and for a short period of time \cite{Abdo10,Acker14,Cheung16}. 
\paragraph*{Importance of \g-ray observations}
\g-rays emitted by novae have two very different origins: radioactivity and high energy particle accelerated in diffusive shocks. 
\begin{itemize}
\item
First of all, radioactive isotopes in the ejecta release photons with energies E$\sim$1 MeV. 
This emission has not been detected yet by any space observatory, e.g. CGRO and \INTEGRAL. The potential role of novae as \g-ray emitters related to radioactive nuclei was already pointed out in the 70's \cite{CH74}, referring to electron-positron annihilation 
and $^{22}$Na decays. Seven years later, Clayton \cite{Clayton81} noticed that another \g-ray line could be expected from novae, 
when $^{7}$Be transforms (through an electron capture) to an excited state of $^{7}$Li, which de-excites by emitting a photon of 478 keV.
Thus, two types of \g-ray emission 
related to radioactive nuclei are expected in novae: prompt emission, from electron-positron annihilation (with positrons 
coming from the short-lived $\beta^+$-unstable isotopes $^{13}$N and $^{18}$F), and long-lasting emission, 
from the medium-lived radioactive isotopes $^{7}$Be and $^{22}$Na decays. 
The prompt emission has very short duration (less than 1 day), appears very early (before optical maximum) and consists 
of a continuum (between 20 and 511 keV) and a line at 511 keV \cite{LC87,Gom98}. The origin 
of this emission is e$^+$e$^-$  annihilation and its Comptonization. The long-lasting emission consists of lines at 478 keV (mainly CO novae) 
and 1275 keV (mainly ONe novae), 
lasting around 2 months or 3 years, respectively (CO and ONe refer to the chemical composition of the  
		white dwarf). See Table \ref{tab:radioactivities}, Fig. \ref{nova:fig}, \cite{Gom98} and review \cite{Her08} for details.

Recent detections of radioactive $^7$Be in a few novae in the ultraviolet (see \cite{Tajitsu15-16,Molaro16}), yield an amount 
of ejected $^7$Be larger than the most optimistic 
theoretical values from \cite{JH98}, but anyway, the detectability distances of all the \g-ray lines from novae are still very short, 
around 0.5 kpc with \INTEGRAL/SPI.
\item 
Another way to produce \g-rays in novae is through particle 
acceleration (p and e$^-$), in strong shocks between the ejecta and the surrounding medium - recurrent symbiotic novae - and internal shocks in the ejecta itself - 
in classical novae (see previous section). The high-energy \g-rays originated mainly by neutral pions decay (hadronic origin) and \ic
scattering (leptonic origin) have been detected by \fermilat in a handful of novae \cite{Abdo10,Acker14,Cheung16}. High-energy \g-rays give unique insights on 
the mass ejection processes in novae.
\end{itemize}
\paragraph*{Expected results with e-ASTROGAM}
If the sensitivity of e-ASTROGAM for the nova broad lines at 1275 keV and 478 keV ($\Delta$E (FWHM)$\sim$20 keV and 8 keV, respectively) is  
25 (13 for 478 keV, if we adopt the value for 511 keV) times better than \INTEGRAL/SPI's, then detectability distances would be 5 (3.6) times larger, reaching 3 kpc and 2 kpc. Then it could be expected to detect one nova per year. Detectability distances correspond to model fluxes $1.2\times 10^{-5}$ and 
$10^{-5}$ ph/cm$^2$/s, for the 478 and 1275 keV lines of typical CO and ONe novae, respectively, at d=1 kpc. 

In addition to the direct and unambigous detection for the first time of the radioactive nuclei $^{22}$Na and $^7$Be-$^7$Li in novae (with $^7$Be and $^7$Li now already detected from ground in the near UV and optical, respectively), e-ASTROGAM observations would help to answer some key questions about nova explosions. 
For instance, the mixing between accreted matter (expected to be solar-like) and white dwarf core (CO or ONe) is crucial to understand 
nova explosions, and the amount of $^7$Be and $^{22}$Na directly detected through \g-rays strongly depends on it. The contribution of 
novae to the galactic content of $^7$Li is by itself very relevant and a hot topic;  detection of radioactive $^7$Be with e-ASTROGAM would directly 
lead to the determination of $^7$Li ejected mass: the amount of $^7$Be - and thus $^7$Li - can only be measured unambiguously in the \g-ray range. 

e-ASTROGAM will also help to disentangle the origin of the high-energy \g-ray emission of novae, hadronic or leptonic, thanks to  
its unprecedented sensitivity in the energy range between 10 MeV and 100 MeV (not accessible to \fermilat), together with its excellent coverage 
of the GeV energy range. It is crucial that e-ASTROGAM will be able to observe and detect novae promptly, since this emission appears very early and has relatively short duration.
\begin{figure}[t]
\begin{center}
\includegraphics[width=0.98\textwidth]{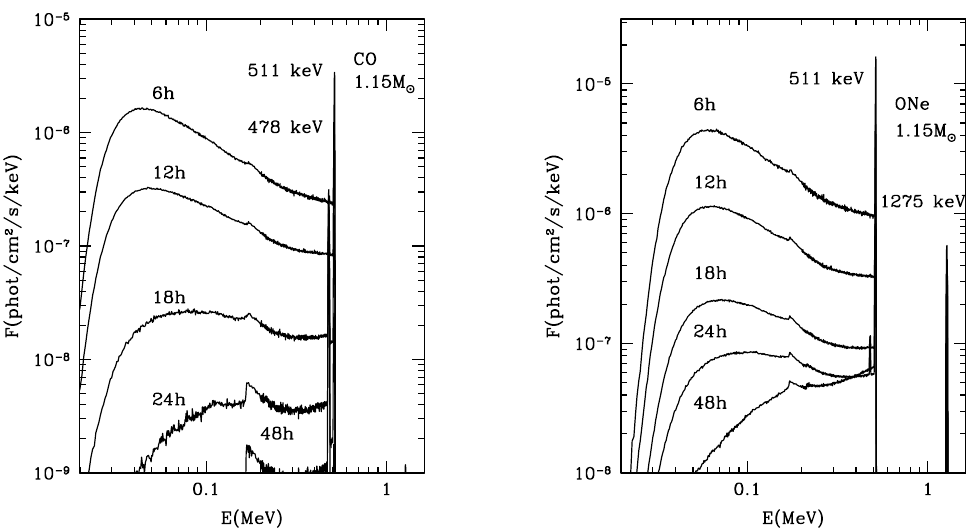}
	\caption{\small Gamma-ray spectra for (left) a CO nova and (right) a ONe nova of 1.15 $M_{\odot}$ at different epochs after the outburst (defined as the peak temperature time) and at the distance of 1 kpc.}
\label{nova:fig}
\end{center} 
\end{figure}

\begin{table}[h]
\caption{\small List of main radioactive isotopes in nova ejecta.}
\label{tab:radioactivities}
\begin{tabular}{ccccc} 
\hline \hline
Isotope   & Lifetime        & Main process  & Type of emission
          & Nova type\\
\hline
$^{13}$N  & 862~s          & $\beta^+$-decay              
                                             & 511~keV line and continuum
          & CO and ONe\\     
$^{18}$F  & 158~min        &  $\beta^+$-decay      
                                              & 511~keV line and continuum
          & CO and ONe\\
$^{7}$Be  & 77~days        & $e^-$-capture      
                                              & 478~keV line 
          & CO\\
$^{22}$Na & 3.75~years  &  $\beta^+$-decay
                                              & 1275 and 511~keV lines  
          & ONe\\
$^{26}$Al & 10$^{6}$~years &  $\beta^+$-decay
                                              & 1809 and 511~keV lines 
          & ONe\\
\hline \hline
\end{tabular}
\end{table}
\subsection[Diffuse \g-ray line emissions\\
\noindent
\textit{\small{R. Diehl, N. Prantzos, V. Tatischeff}}
]{Diffuse \g-ray line emissions}
\paragraph*{Science questions}
Starting from the synthesis of new nuclei by nuclear fusion reactions
within stars and their explosions, the cycle of matter proceeds towards the ejection of metal-enriched and processed stellar gas into
interstellar space. After cooling on the further trajectory, such metal-enriched gas mixes with interstellar gas from other origins and trajectories, to eventually partly ending up in newly-forming stars, closing and starting the cycle of cosmic matter again. 
This cycle includes at least two phases where \g-rays can provide astrophysical and rather direct diagnostics of aspects of cosmic nucleosynthesis: (1) The ejected yields of radioactive by-products of stellar and explosive nucleosynthesis tell us about the conditions of nuclear fusion reactions in those sites, and (2) the tracing of the flow of ejecta over their radioactive lifetimes, which is made possible from radioactive decay \g-rays from longer-lived nuclei because these are independent of thermodynamic conditions or density of gas. Further diagnostics arise (3) from positrons emitted in radioactive decays through their annihilation \g-rays, and (4) from nuclear de-excitation \g-rays caused by CR collisions with ambient-gas nuclei. 
\begin{itemize}
\item[1.]
The yield in specific isotopes from stars and stellar explosions is an important diagnostic of the environmental conditions within such sites. These are occulted and not directly accessible, due to absorption of radiation in massive overlying envelopes. Even \g-rays rarely escape, except for some explosion scenarios. 
Nuclear reactions with their steep temperature sensitivities are excellent probes of the conditions in the nuclear burning regions of cosmic nucleosynthesis. Isotopic yields are the outcome of all nuclear reactions as they are determined by conditions in those inner regions.
Candidate sources are novae from explosive hydrogen burning on the surfaces of white dwarfs composed of C and O or a further-enriched C-O-Ne mixture, the latter from more massive progenitors, SNe in all their variants, and massive stars which experience major mass loss and thus may release nuclearly-processed interior gas (Asymptotic Giant Branch stars and Wolf-Rayet stars). 
When point sources cannot be observed, either due to low individual source yields, or due to superposition from multiple events occurring during a radioactive lifetime span, a diffuse glow of characteristic \g-rays provides a useful signal. 
Specifically, this could be the case for nova-produced $^{22}$Na from
sources within our Galaxy (event rate 30--50 per year in our Galaxy, $^{22}$Na lifetime 3.8 years),
and for SN-produced $^{44}$Ti from nearby galaxies (event rate few per century in M31, $^{44}$Ti lifetime 85 years).
It has been seen already for long-lived \g-ray emitters $^{26}$Al ($\tau$ 1.0~10$^6$ yr) and $^{60}$Fe ($\tau$ 3.8~10$^6$ yr), where many sources along the disk of the Galaxy contribute. 

\par\noindent
Specific science questions include \cite{Meynet17}: Mixing in regions outside of the stellar core; here, stellar rotation, convection, and diffusion from compositional gradients, all interplay in complex ways. The structure of stars in their outer regions is a result of these processes, as they affect the nuclear burning in shell burning regions, which release nuclear binding energy. Further questions address the dynamic environments in stellar explosions. 
These bear the same mixing issues, and in addition non-equilibrium effects such as nuclear-burning fronts and their propagation, the properties of degenerate gas, and neutrino interactions further complicate the burning conditions. 
Again, nuclear-reaction products and their radioactive trace isotopes store such conditions through their production amounts, and carry them outside through the explosion into optically-thin regimes, where decay \g-rays can be observed. 
In nova explosions, science questions are the compositional mix of accreted and underlying white dwarf material within the hydrogen burning region and the propagation of the ignition flame. 
In thermonuclear SNe \cite{hi13}, the ignition of carbon fusion and its flame dynamics is a fundamental issue, then again the flame propagation and how nuclear burning in degenerate cores might be frozen out from nuclear statistical equilibrium as the flame reaches lower-density regions further outside, and lifts degeneracy. 
In the case of core collapses \cite{Burrows13}, electron capture in hot massive cores of massive stars removes pressure and initiates core collapse for the less-massive of the massive stars, while more massive cores collapse under gravity once the nuclear fuel has all been processed to its most stable form of iron. 
The collapse then leads to formation of a NS, and the intense neutrino emission upon its formation and neutronisation of matter may trigger the SN explosion, or not. 
Further collapse towards a BH may occur. In all collapses, vigorous convective flows onto and away from the central compact object include nuclear statistical equilibrium and freezing out from that. The resulting isotopic compositions carry the conditions of such nuclear processing as a memory into the ejecta, specifically for the long-lived $^{44}$Ti,  $^{26}$Al and $^{60}$Fe. 

\par\noindent
Neutrino-induced processing of $^{26}$Al and $^{60}$Fe, in addition to some modest explosive-burning amounts, reflect the conditions within the SN explosion. These two isotopes are plausibly assumed to be mainly produced by the same
sources, by massive stars \cite{Chieffi13}. Therefore, in the measurement of the isotopic ratio $^{60}$Fe/$^{26}$Al any unknowns about source
distances and location will cancel: This ratio is a valuable diagnostic for the internal structure of
massive stars. Although integrated over a large population of such sources, it serves
to test our overall validity of models for the internal structure of massive stars as it evolves over their lifetime,
ending in a core collapse SN. 
$^{26}$Al is mainly produced in the hydrogen burning stages, i.e. the main sequence phase, and O-Ne shell burning
in the late evolution. $^{60}$Fe, on the other hand, is produced in shell He and C burning phases, in the later evolution.
Each of these late shell burning regions is expelled only in the final SN.
 $^{26}$Al from the main sequence phase may be ejected also by Wolf Rayet winds for very massive
stars that evolve rather rapidly within several million years. As massive stars occur in groups, the integrated radioactive
emission from such regions and these two isotopes provides a global test of models of massive star evolution.
Additionally, for individual massive star groups where a steady state has not been reached, this isotopic ratio
encodes the age of the group, as the Wolf-Rayet wind and SN contributions relate to different stellar masses and ages.
\item[2.]
The cyle of matter includes a phase where the how hot nucleosynthesis ejecta cool down and are propagated
towards new star formation. This is often ignored, as it is particularly hard to constrain through observations.
This recycling time scale depends on the structure of the dynamic interstellar medium. 
It could be rather long, up to 10$^7$--10$^8$~years, and thus exceed stellar evolution times which are of the order of tens of Myr,
SNRs can be seen over time scales of few 10$^5$ yr at most. Long-lived radioactive
\g-ray emitters $^{26}$Al ($\tau$ 1.0~10$^6$ yr) and $^{60}$Fe ($\tau$ 3.8~10$^6$ yr) can
trace mixing processes of ejecta into the next generation of star formation over
much longer time. Among others, this provides observational constraints on molecular-cloud lifetime and models
for stimulated/triggered star formation.
On the global, Galactic scale, superbubbles have been proposed to be key structures in the transport of fresh ejecta towards new
star forming regions, from \INTEGRAL/SPI data for the $^{26}$Al line. Mapping Galactic $^{26}$Al \g-ray emission
can thus trace ejecta flows into and through such superbubbles, and compare their connections
to specific star forming regions with their massive-star groups at their respective ages.
\item[3.]
Radioactivity from proton-rich nuclei generally produces positrons, and contributions 
from many such nucleosynthesis sources would integrate to a diffuse emission 
from annihilation \g-rays with the 511 keV line being most prominent. 
Novae and SNe contribute through $^{13}$N, $^{18}$F, $^{56}$Ni, $^{44}$Ti, for example, with 
characteristic radioactive lifetimes from hours to a century, and also other positron sources are known
to exist and add to such diffuse emission. The science question here is a discrimination among the 
different candidate sources.
\item[4.]
Nuclear de-excitation \g-rays reflect CR fluxes as they collide with ambient interstellar gas and are energetic enough to excite nuclear levels \cite{Benhabilez13}. 
The characteristic \g-ray lines provide most-direct measurements of the flux of CRs at several tens of MeVs, which cannot be observed in direct CR measurements as they are deflected by interstellar magnetic fields. One of the unsolved science questions on the origin of CRs could thus be answered by a first 
observation of characteristic \g-ray lines from young SNRs, which generally are considered plausible CR acceleration environments.
\end{itemize} 
\paragraph*{Expected results with e-ASTROGAM}
\begin{figure}[t]
\begin{center}
\includegraphics[width=0.6\textwidth]{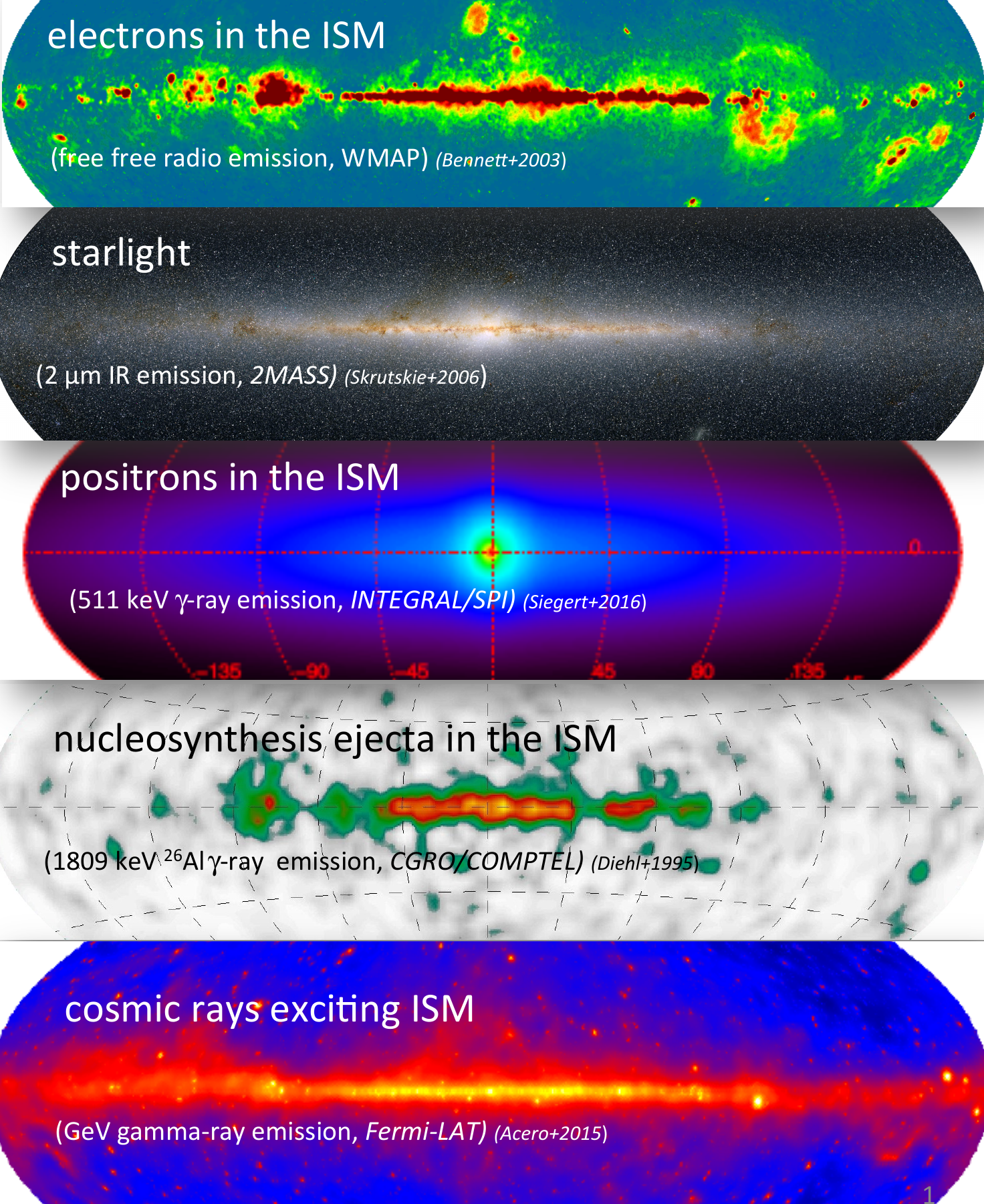}
	\caption{\small The diffuse emissions of our Galaxy across several astronomical bands: e-ASTROGAM will explore the link between starlight (second image from top) and CRs (top and bottom). The current-best images of positron annihilation (3rd from top) and $^{26}$Al radioactivity (4th from top) \g-rays illustrate that this link is not straightforward, and e-ASTROGAM will uncover more detail about the astrophysical links and processes. (Image composed by R. Diehl, from observations with WMAP, 2MASS, \INTEGRAL, CGRO, and \Fermi; Refs. \cite{3FGL,Bennett2003,Skrutskie2006,sie16b,Diehl95})}
\label{diffuse:fig}
\end{center}
\end{figure}
With its huge increase in sensitivity, e-ASTROGAM
will provide a detailed view of the morphology of this emission, with high precision
measurements of the line flux from many regions of the Galaxy (see Fig.~\ref{diffuse:fig}). This will advance the current state of such observational constraints \cite{Diehl13}. For example,
e-ASTROGAM will observe the $^{26}$Al radioactivity from dozens of nearby ($\leq$kpc)
stellar objects and associations. In particular, it will measure precisely the amount
of $^{26}$Al ejected by the Wolf-Rayet star WR11 in the $\gamma^2$-Velorum binary system
(expected line flux is $\sim10^{-5}$ ph~cm$^{-2}$s${-1}$), thus providing a unique calibration of
the $^{26}$Al production during the Wolf-Rayet phase of a massive star. e-ASTROGAM
has also the capability of detecting $^{26}$Al emission from the LMC (expected line flux
of $\sim10^{-6}$ ph~cm$^{-2}$s${-1}$), thus providing new insight into stellar nucleosynthesis
outside the Milky Way.
For the first time, e-ASTROGAM will provide the sensitivity needed to establish
the Galactic $^{60}$Fe emission and build an accurate map of the $^{60}$Fe flux in the Milky
Way, enabling its comparison with the $^{26}$Al map to gain insight into the stellar progenitors
of both radioisotopes. In particular, measuring \g-ray line ratios for specific
massive-star groups will constrain $^{60}$Fe production in massive stars beyond 25--40~M$_{\odot}$,
which directly relates to stellar rotation and uncertain convective-layer evolution in
massive star interiors.
Marginally-bright diffuse radioactivity may arise from nova explosions, due to their higher frequency of occurrence at about 30~yr$^{-1}$ in the Galaxy, ejecting into their surroundings $\beta^+$ emitters such as $^{13}$N, $^{18}$F, and long-lived $^{22}$Na. 
These $\beta^+$ decays inject positrons into interstellar space. e-ASTROGAM will provide a detailed map of annihilation \g-rays, also imaging faint regions near candidate sources along the disk of the Galaxy and in star forming regions. This will allow discrimination of nucleosynthesis contribution of positrons from several other types of sources that are expected to contribute positrons as well, from a variety of electron-positron pair plasma creation scenarios as well as from CR hadronic interactions. 
Sources of CR accelerations will be directly probed with e-ASTROGAM through interactions with ambient interstellar gas, which leads to characteristic nuclear lines, such as from $^{12*}$C at 4430~keV. 
\subsection[Galactic positron annihilation radiation\\
\noindent
\textit{\small{N.~Prantzos, P.~Jean, J.~Kn\"odlseder, P.~von~Ballmoos, T.~Siegert, R.~Diehl, J.~Isern, V.~Tatischeff}}
]{Galactic positron annihilation radiation}
\paragraph*{Science questions}
The 511 keV emission from electron-positron annihilation in the Galaxy is the brightest \g-ray line in the sky, and the first ever detected from outside the solar system \cite{joh72,lev78}. It is produced by the annihilation of a few 10$^{43}$ positrons with electrons of the interstellar medium (ISM) every second (flaring stars could also contribute to the observed annihilation radiation from the Galactic bulge, according to Ref.~\cite{bis17}). Despite more than 40 years of intense observational and theoretical investigation, the origin of annihilating positrons remains a mystery. The emission is strongly concentrated toward the Galactic bulge. The bulge/disk ratio appeared higher than observed in any other wavelength \cite{kno05,wei08}, but with increasing sensitivity the disk emission emerges more clearly, perhaps as an extended, low brightness thick disk \cite{sie16b}. This bulge/disk ratio is believed to point towards source types, hence requires better observations.

High-resolution spectroscopy of the annihilation radiation provides information on the nature of the environment in which the positrons annihilate. Measurements with \INTEGRAL/SPI of the shape of the 511~keV line and positronium fraction in the bulge are consistent with positron annihilation in a mixture of warm ($T \sim 8000$~K) neutral and ionized phases of the interstellar medium \cite{chu05,jea06}.
\begin{figure}[t]
	\includegraphics*[scale=0.38]{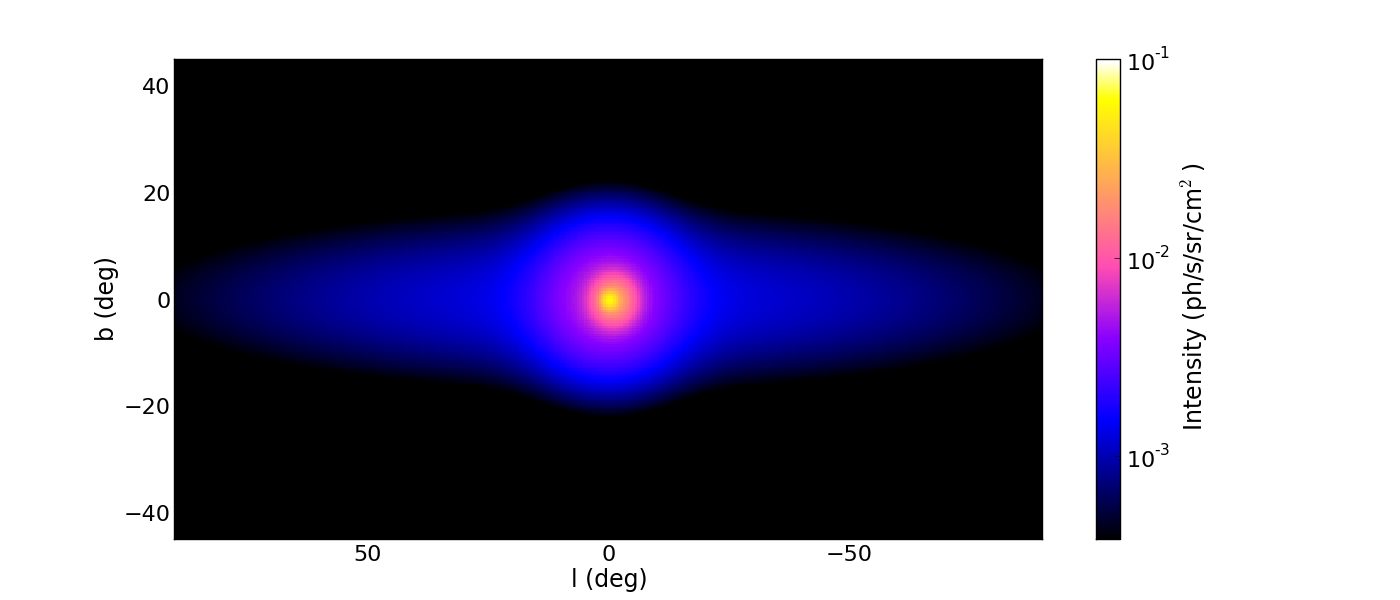}
	\includegraphics*[scale=0.38]{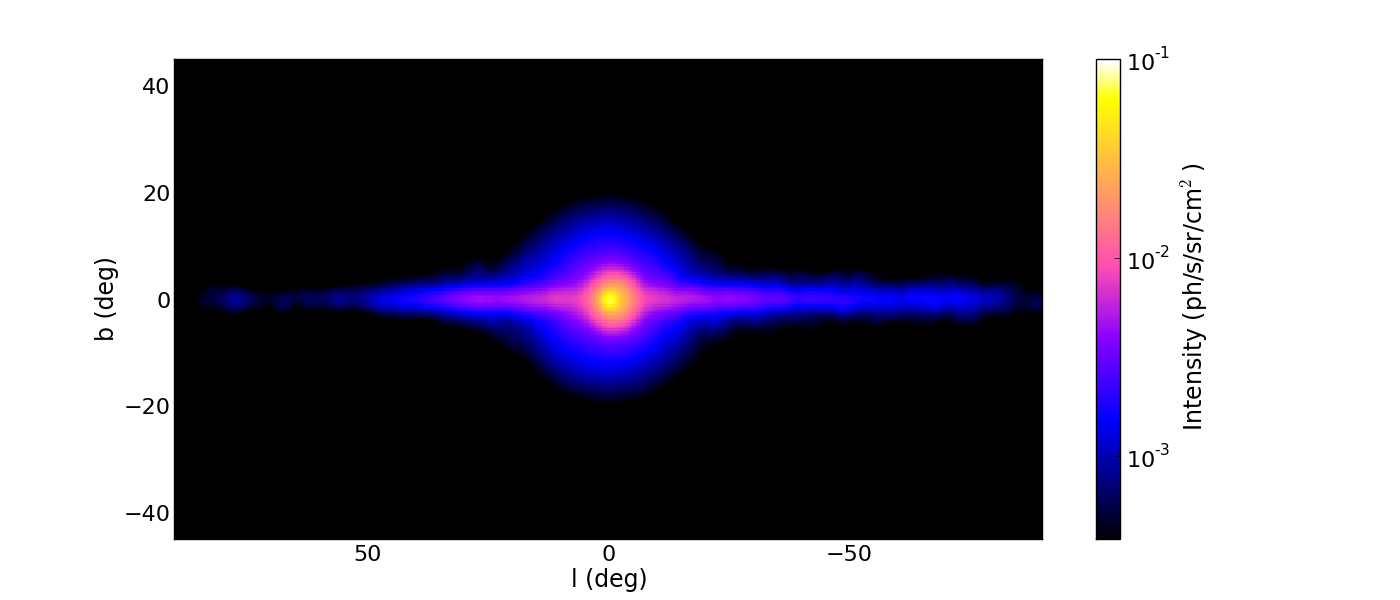}
	\caption{\small Simulated e-ASTROGAM observations of the 511 keV emission, obtained for an exposure of 1 year of the inner Galactic region, assuming (top) the model proposed by \cite{sie16b} with a thick disk and (bottom) a model with the disk of \cite{alex14} and the point source and bulge components of \cite{ski14}. \label{fig:511keVimages}}
\end{figure}
It remains unclear what are the main sources of positrons: SNe of gravitational and thermonuclear types, or pulsars, X-ray binaries and microquasars, or more ``exotic'', such as self-interacting dark matter particles or the Galactic supermassive BH which, appears inactive today but may have been a transient positron injector. The latest proposed positron source is a rare type of thermonuclear SNe known as SN~1991bg-like (resulting from the merger of a CO white dwarf and a He white dwarf) that could cause both the strength and morphology of the Galactic positron annihilation signal as well as the origin of $^{44}$Ca \cite{cro17}. However, a major issue in all studies is that positrons may propagate for several 10$^5$ yr far away from their sources before annihilating --~depending on still poorly understood properties of Galactic ISM and magnetic fields~-- making it difficult to infer positron sources from the observed \g-ray emission (see Ref.~\cite{pra11} and references therein). Understanding the Galactic 511~keV emission constitutes a major challenge for modern astro-particle and astro-physics.
\par\noindent
Progress in the field requires advances in several directions:
\begin{itemize}
\item {\it Observations of 511 keV emission}: What is the true spatial distribution of the emission? How far do the spheroid and disk extend? Can we find support for transport physics or galactic outflows from an extended disk? Is there a distinct central point source, and how concentrated is it \cite{sie16b}? Is the recently-reported \cite{sie16a} broadened positron annihilation emission from the flaring microquasar V404 Cyg really proof of pair plasma near accreting BHs, and can it be confirmed (see dispute by \cite{roq16})? Does the morphology of 511~keV disk emission differ from the one observed at 1.8 MeV, resulting from the decay of radioactive $^{26}$Al, which is a major positron provider in the disk? (Similarity would imply dominance of this source in the disk and that positrons do not travel far away from their sources.)
\item {\it Physics of positron sources}: What is the escaping fraction of positrons from SN~Ia? What is the production yield of $^{44}$Ti in normal, and in SN~1991bg-like thermonuclear SNe? Could these, with $^{26}$Al, explain the Galactic positron production rate? What is the SN~Ia rate in the inner (star forming) and in the outer (inactive) bulge? What are the positron yields, activity time scales, and spatial distribution in the inner Galaxy of X-ray binaries, microquasars and millisecond pulsars? How can the past level of activity of the central supermassive BH be reliably constrained?
\item {\it Studies of positron propagation}: What is the multi-scale morphology of the interstellar medium near positron sources, and how may interstellar turbulences affect the positron transport? What is the large-scale configuration of the Galactic magnetic field? What is the role of particle reacceleration?
\end{itemize}

Those issues are of great interest to a broad community, including researchers in nucleosynthesis and SN physics, compact and accreting objects, the Galactic supermassive BH, as well as CR physics, particle transport in turbulent interstellar plasmas, the large scale galactic magnetic field, and even dark matter research.
\paragraph*{Expected results with e-ASTROGAM}
Figure~\ref{fig:511keVimages} shows two simulated sky maps of the 511 keV intensity distribution observed with e-ASTROGAM: one with a thick Galactic disk as suggested by \cite{sie16b} and another with the disk component of \cite{alex14}, which results from a Monte Carlo modeling of the Galactic propagation of nucleosynthesis positrons produced by the $\beta^+$-decay of $^{26}$Al, $^{44}$Ti, and $^{56}$Ni, and the bulge components of \cite{ski14}. With its large field of view of 46$^\circ$ half width at half maximum (HWHM) at 511 keV, corresponding to a fraction-of-sky coverage in zenith pointing mode of 23\% at any time, e-ASTROGAM will perform a deep Galactic survey of the positron annihilation radiation to search for potential point-like sources, and study in detail the morphology and spectral characteristics (e.g. positronium fraction) of the disk, bulge, and central source emissions. With a predicted point-source sensitivity of $4.1 \times 10^{-6}$ ph~cm$^{-2}$~s$^{-1}$ for the 511~keV line in 1~Ms of integration time, e-ASTROGAM will be able to detect low surface brightness regions outside the Galactic plane and enhanced emission from the inner Bulge, as well as individual star forming regions in the disk, such as the Cygnus region. 

\newpage
\section[Physics of compact objects \\
\noindent
\textnormal{\small\textnormal{Convenors:} \textit{A. Harding, J.M. Paredes}}]{Physics of compact objects}
NS and BHs are the densest objects in the Universe and exhibit a great variety of observational manifestations.  They are observed pulsating and bursting, accreting from a binary companion, interacting with its wind, or even merging with it.  NS are found both in binary systems, often with other compact stars such as white dwarfs or NS, or as isolated sources.  There are many puzzles in the behaviour of NS and in the relationship between their different types.  Observations at MeV energies can uniquely address this and other fundamental questions such as the nature of accretion and the origin and density of pulsar pair plasma.

Explosive magnetars are found with the same outward characteristics, such as surface magnetic field and spin period, as those of more placid rotation-powered pulsars.  We do not understand what as yet hidden property makes them behave so differently, although some rotation-powered pulsars have displayed magnetar-like outbursts.  Models have proposed that the distinguishing property of magnetars and magnetar-like behaviour is a twisted magnetic field structure capable of releasing a power larger than that of dipole spin-down.  Many magnetars have hard non-thermal components extending to at least 100 keV with no observed cutoffs, although one is expected in the MeV band from COMPTEL upper limits.   Detection of such cutoffs as well as their phase-resolved behaviour and polarization with e-ASTROGAM could constrain and probe the magnetar field structure since they are likely caused by attenuation from photon splitting and pair production that is very sensitive to the magnetic field.

Another NS puzzle is the nature of pulsar emission in the 0.1 - 10 MeV band, which was detected from only three rotation-powered pulsars by COMPTEL.  Eighteen pulsars have non-thermal emission detected above 20 keV, and eleven of these have no detected radio or \Fermi pulsations.  Such MeV pulsars appear to have the peaks of their SEDs at MeV energies, so the clues to their nature lie in measurements by more sensitive detectors like e-ASTROGAM.  Many more members of this population could be discovered and such spectral measurements could also reveal the origin of the crucial pair plasma in pulsar magnetospheres.  

A number of pulsars in binary systems are thought to have intra-binary shocks between the pulsar and companion star that can accelerate particles of the pulsar wind to greater than TeV energies.  Gamma-ray binaries, with a young rotation-powered pulsar in orbit around a massive Be star, show orbitally modulated emission at radio, X-ray, GeV and TeV energies.  Models with either inverse-Compton or synchrotron radiation can fit the X-ray to GeV spectrum and better measurements at MeV energies would constrain the mechanism.  Observations of accreting X-ray binaries, that contain either NSNS or BHs, at MeV energies can uncover the emission mechanisms that are operating as well as the role of the jets in these sources.  An exciting possibility is the detection of a 2.2 MeV neutron capture line coming from the inner parts of the accretion disk or from the NS atmosphere, which would be a major discovery and give new constraints on accretion physics and the gravitational redshift at the NS surface, respectively.

Binaries containing millisecond pulsars and low mass companions also show orbitally-modulated X-ray emission from intra-binary shocks and three of these are observed to transition between rotation-powered and accretion-powered states.  e-ASTROGAM observations will fill in the spectral gap from 0.1 - 100 MeV to help us understand the nature of these transitions and the limits to acceleration in the pulsar wind shock. 

Many millisecond pulsars are found in globular clusters and \Fermi has discovered \g-ray emission both from many clusters as well as pulsations from pulsars within some clusters.  The nature of the diffuse X-ray and TeV emission detected from several clusters is presently a mystery and could come from magnetospheric emission or from electron-positron pairs ejected from the pulsars in the cluster.  Detection of the extension of the diffuse X-ray component by e-ASTROGAM can reveal its origin and place strong constraints on the injection rate of pair plasma from millisecond pulsars.
\subsection[Isolated neutron stars and pulsars\\ 
\noindent
\textit{\small{C.~Gouiff\'{e}s, I.~A.~Grenier, A.~Harding, P.~Laurent, R.~Mignani, M.~Lopez~Moya, P.~Saz~Parkinson, A.~Shearer, D.~Torres, L.~Zampieri, S.~Zane}}]
{Isolated neutron stars and pulsars}
\paragraph*{Science questions}
\Fermi revolutionized \g-ray pulsar studies increasing the number of pulsars detected above 100 MeV from 7 with {\it CGRO/EGRET} to about 200 today\footnote{https://confluence.slac.stanford.edu/display/GLAMCOG/Public+List+of+LAT-Detected+Gamma-Ray+Pulsars}. However in the soft  \g-ray region there are only 18 detections above 20 keV and only four have been detected with pulsed emission in the range 1-10 MeV \cite{Kuiper2015}. e-ASTROGAM's sensitivity at 10 MeV is ~100 times better than CGRO/COMPTEL, consequently we would expect a significant number of new pulsar detections at this energy. At lower energy, e-ASTROGAM will study the pulsars and magnetars detected in hard X-rays with \INTEGRAL, \nustar and GBM. This will allow filling the gap in the EM spectrum of these compact objects. Spatial, spectral and temporal data in the e-ASTROGAM energy range will be crucial to better understand the physics, still poorly known, of these sources. Furthermore e-ASTROGAM's polarization sensitivity will enable a unique contribution to pulsar studies. For the first time it should be possible to have a 0.1-10  MeV survey with both pulse shape determination and measurements of phase resolved polarization, both of which will constrain current pulsar emission models and mechanisms.  This mission can play a major role as an alert monitor for the variable sky survey in coordination with all the future radio, infrared, optical and X-rays facilities for the EM domain and the coming neutrino and gravitational observatories.
\begin{figure}[h]
\includegraphics[width=0.6\textwidth]{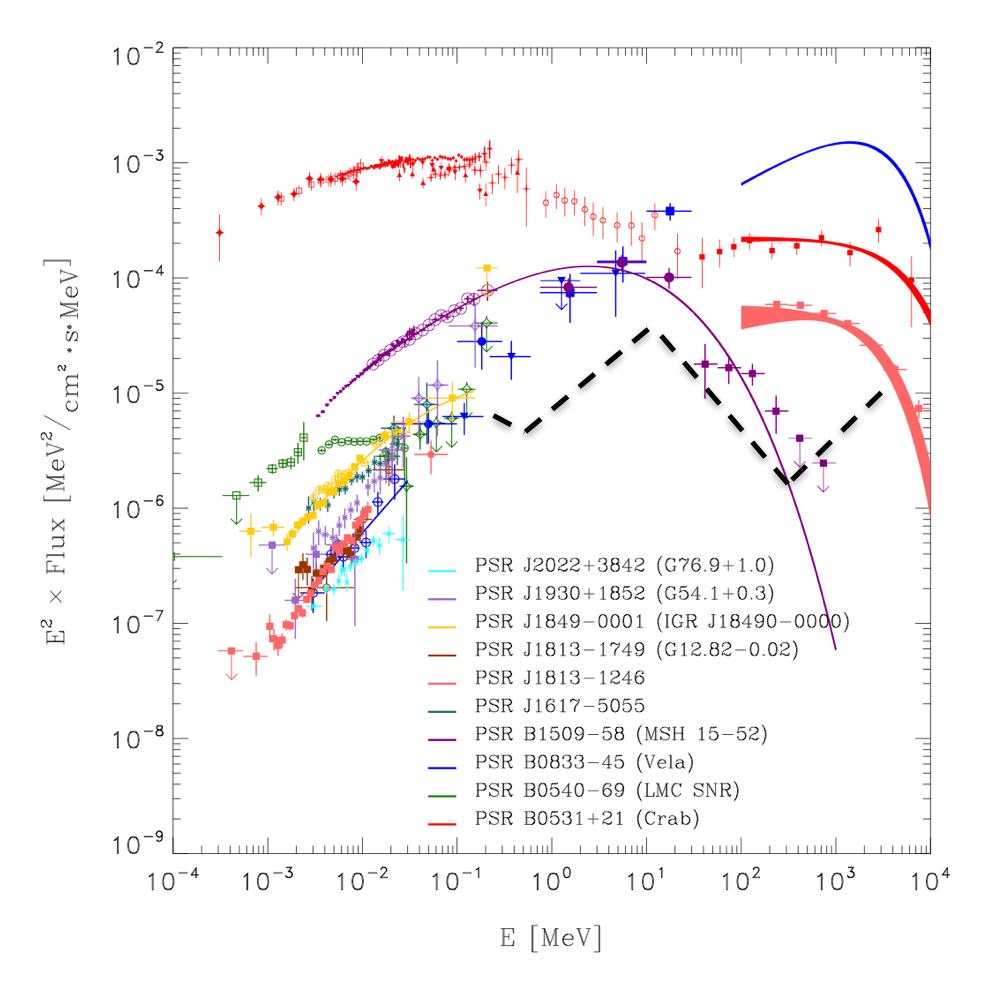}
\includegraphics[width=0.4\textwidth]{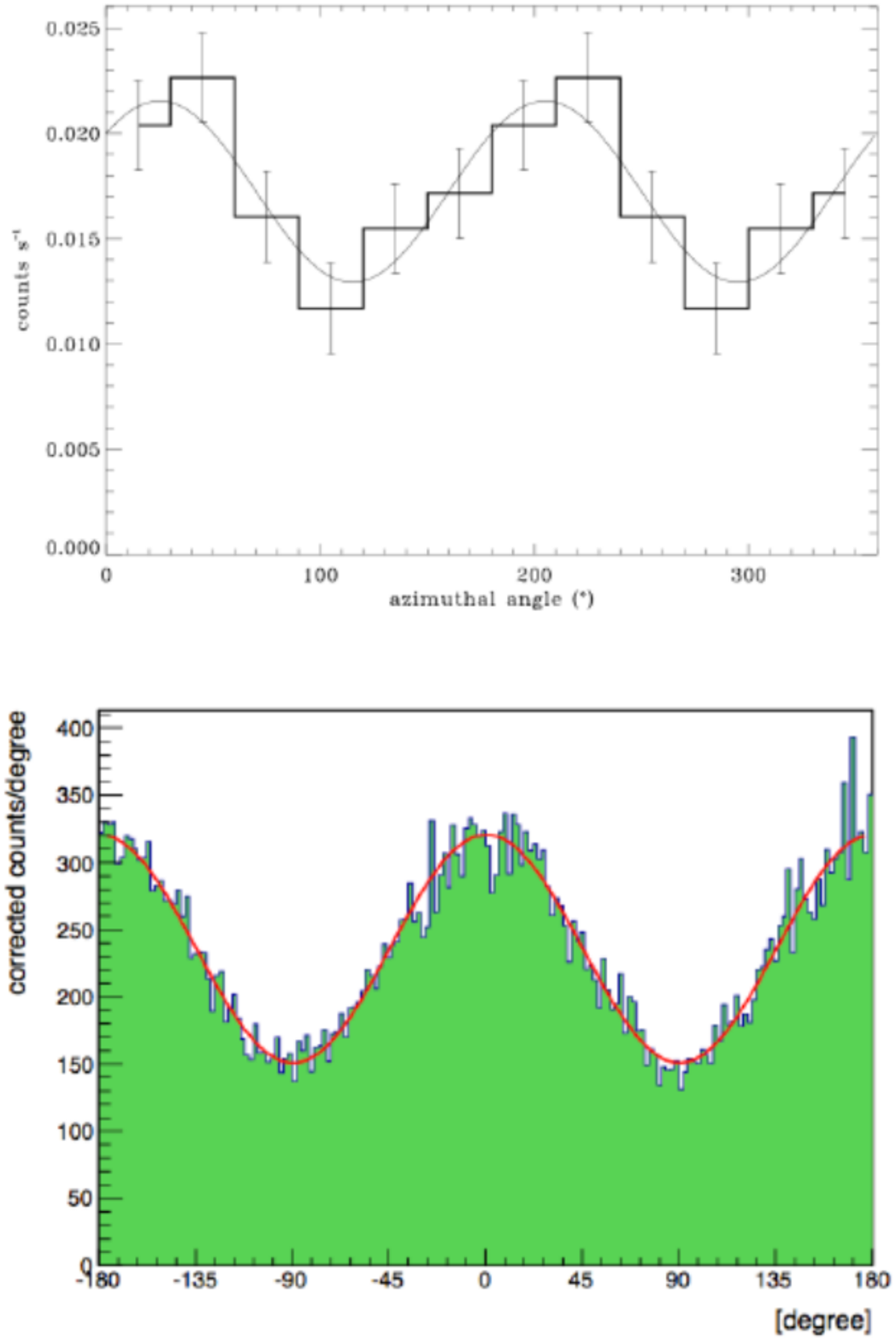}
\caption{\small{Left: SED for a number of soft \g-ray pulsars taken from Kuiper \& Hermsen \cite{Kuiper2015}  with the expected sensitivity for 1 Ms e-ASTROGAM exposure shown (black dashed line). Right: Comparison of the polarization sensitivity of \INTEGRAL-IBIS (top) - observed Crab polarisation from 1 Ms observation (unpublished)  vs e-ASTROGAM (bottom) for 1 Ms observation of a 100\% polarised 10mCrab source in the 0.2 -- 2 MeV range, see chapter \ref{intro}. The spectral fit to the archetypal pulsar PSR B1509-58 is depicted by the purple curve.}}
\label{NS:fig1}
\end{figure}
\paragraph*{Importance of \g-ray observations}
\fermilat has detected many types of \g-ray pulsars: young radio-loud and radio-quiet pulsars \cite{abdo2009,park2010},
millisecond pulsars \cite{Abdo2009_2}, etc. The measured spectral shapes of most \Fermi \g-ray pulsars exhibit
exponential cut-offs in the GeV range.
This has favoured high-altitude models in which the emission originates in the outer magnetosphere, in so-called outer gaps (e.g. \cite{Cheng86,Cheng2000}), or slot gaps \cite{Harding2008}, rather than polar cap models in which super-exponential cutoffs were predicted (e.g. \cite{Rud75}).
However, one class of pulsars that
has hitherto remained relatively elusive are the so-called soft \g-ray pulsars.
Our understanding of soft \g-ray pulsars is in its infancy and limited by the small sample of
objects (see \cite{Kuiper2015}  for a recent review).The majority of these soft \g-ray
pulsars exhibit broad, structured single pulse profiles, and only six have double (or even multiple,
in the case of Vela) pulses. Soft \g-ray pulsars typically have hard power-law spectra in
the hard X-ray band, reaching their maximum luminosities in the MeV range, as opposed to the
GeV range (see Fig.~\ref{NS:fig1}). They tend to be younger and more luminous ($L_{sd} > 4 \times 10^{36}$ ergs/s) than the
overall LAT pulsar population \cite{Kuiper2015} . Only seven soft \g-ray pulsars (as defined by \cite{Kuiper2015} ) have so
far shown pulsed emission detected by the LAT. In fact, PSR B1509-58, the prototypical
 soft \g-ray pulsar, detected in the 1--10 MeV range by COMPTEL \cite{Kuiper1999} and confirmed
by \agile \cite{Pell09}, was particularly challenging to detect with \Fermi \cite{Abdo10_apj} due to its soft spectrum.
Other soft \g-ray pulsars remain undetected by \Fermi (see Table 13 of \cite{Abdo2013}), despite pre-launch
expectations of their detection based on their large spin-down power.
Any soft \g-ray pulsar model must explain why most of them are not seen at GeV energies
by \textit{Fermi}. A possible explanation can be found in \cite{Wang13,Wang14}. 
In a recent work, it has been shown that  all pulsar spectra can be understood in a single theoretical framework \cite{torres18}. 
This model is able to cope with the full range of the multi-frequency spectrum of pulsars. In particular, it can be used as a tool for understanding the X-ray and MeV emission,
and how these connect with measurements at higher energies. 
\begin{figure}[h]
\includegraphics[width=.51\textwidth]{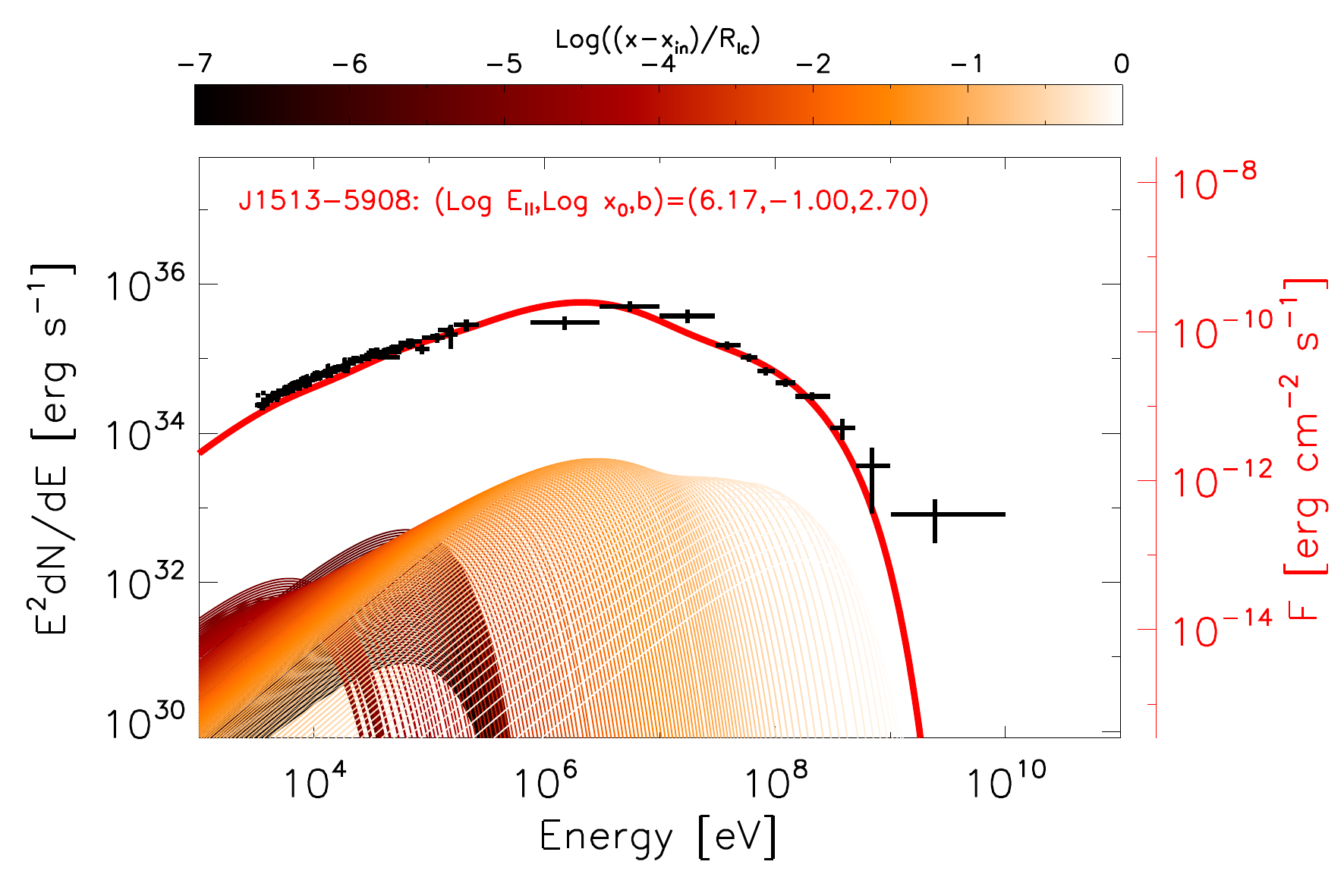}
\includegraphics[width=.51\textwidth]{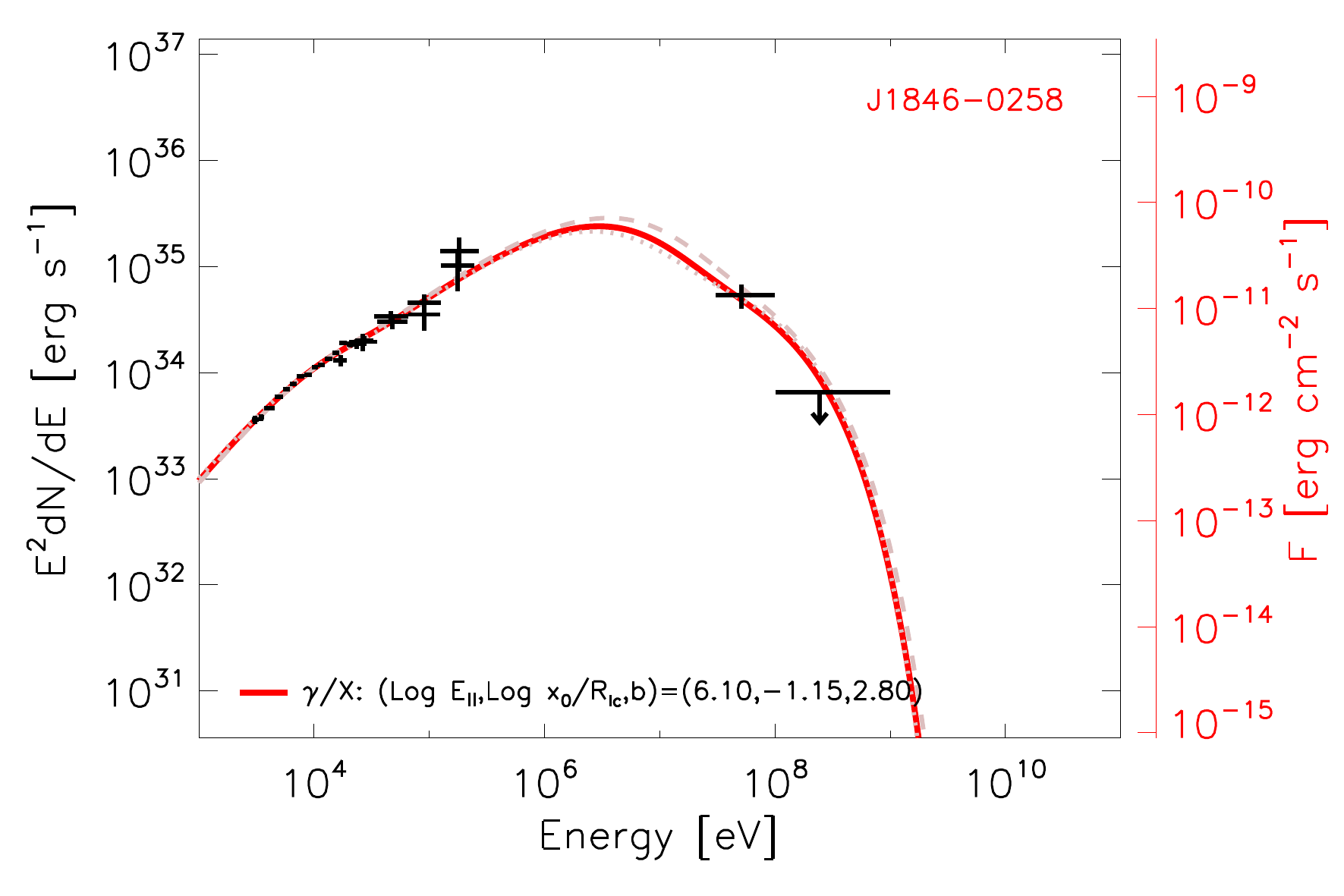}
\includegraphics[width=.51\textwidth]{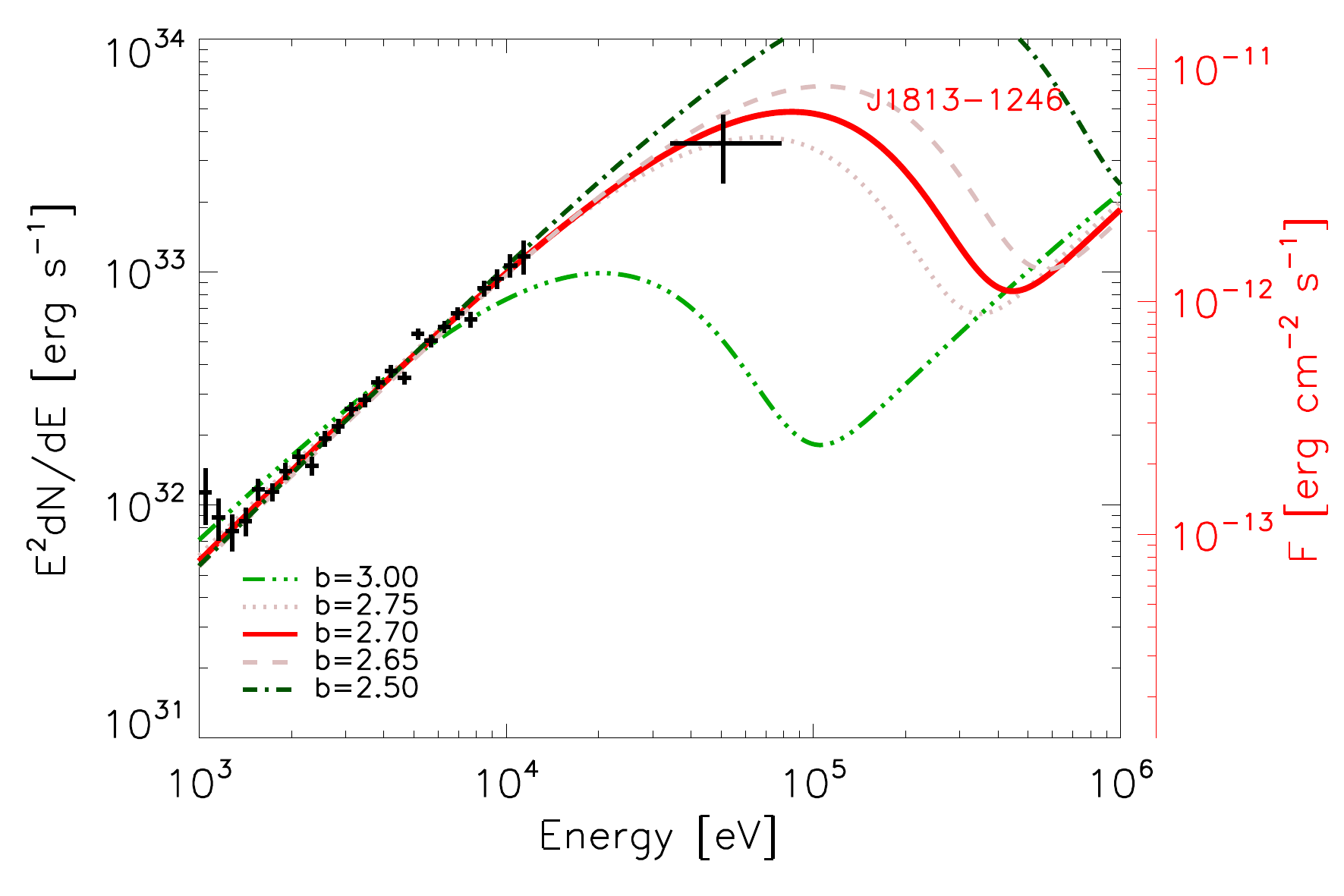} 
\includegraphics[width=.51\textwidth]{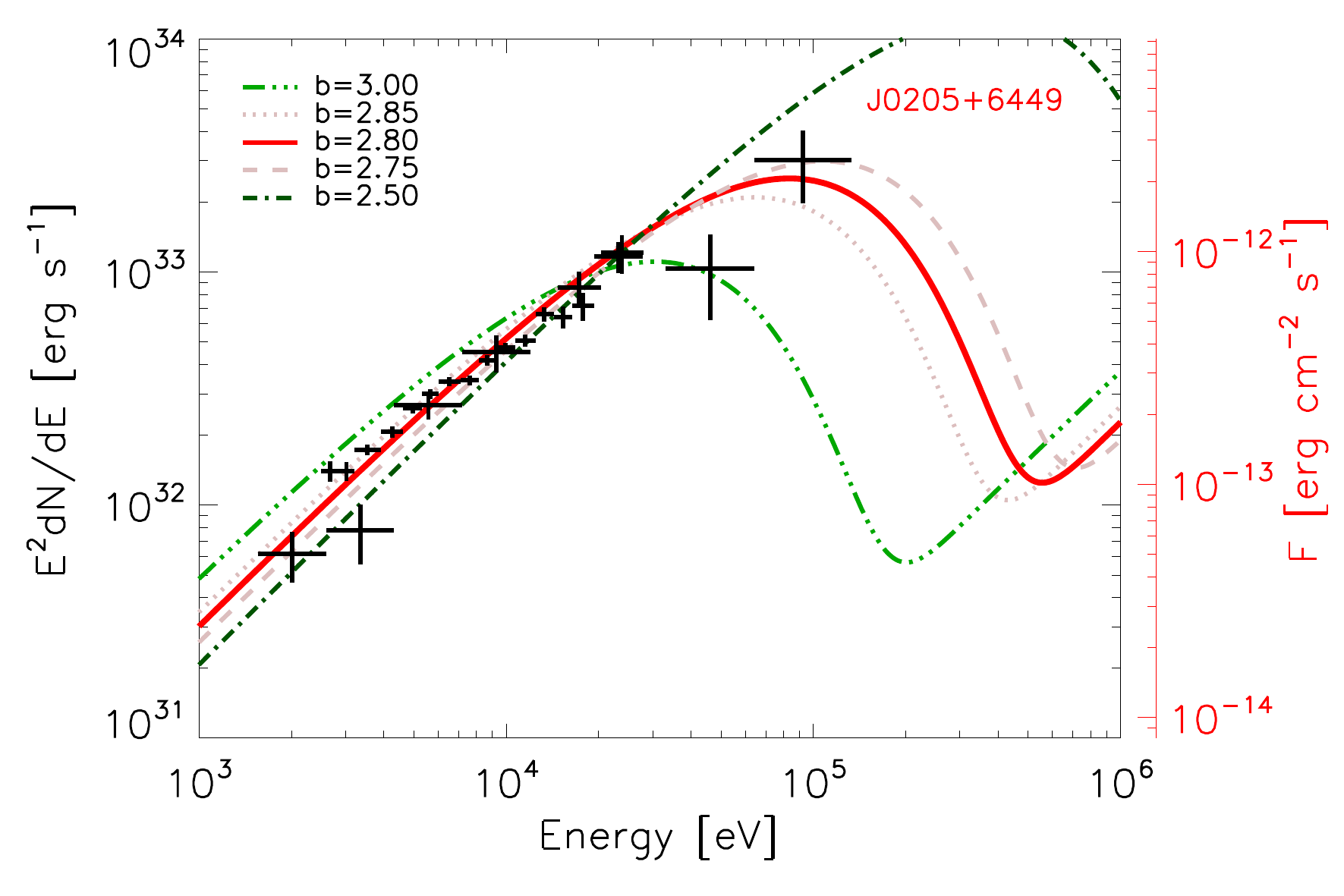}
\caption{\small{Top: Reproduced from corresponding panels in Fig.~1 and 2 of \cite{torres18}. Model fitting for the 2 known MeV pulsars. See text for further comments.
Bottom: Reproduced from corresponding panels of Fig.~2 of the Supplementary Material of \cite{torres18}.
These show examples of the spectral influence of the magnetic gradient ($b$) 
zooming out from X-ray to MeV energies. 
The different curves show the effect on variations on $b$ around the corresponding best-fit one, keeping other parameters fixed at the best-fit solution.}}
\label{MeV_pul:fig}
\end{figure}
Fig. \ref{MeV_pul:fig} shows examples of spectral results, together with the ability for distinguishing among models if observations in the soft \g-ray domain are available.
e-ASTROGAM will enable the detection of new pulsars in the MeV band by carrying out measurements that are difficult or impossible to conduct
with current instruments. \fermilat has detected over 200 pulsars with only 7 out of the 18 known soft \g-ray pulsars  in this sample.  This soft  \g-ray population represents younger pulsars whose spectrum peaks below 100 MeV so they would be hard to detect by \fermilat. From \cite{Kuiper2015}  it is reasonable to expect the number of detected soft \g-ray to at least double with e-ASTROGAM.
\paragraph*{Expected results with e-ASTROGAM}
Unlike traditional \g-ray telescopes, which have relied exclusively on spectroscopy and timing, e-ASTROGAM will also be able to measure the polarization
characteristics of the \g-ray emission from pulsars. Model predictions depend on the pulsar
inclination and viewing angles, which in the best cases are only poorly known. In contrast, the
expected polarization signature differs significantly from one model to another because it is very
sensitive to the EM geometry, and hence to the location of the emitting zones \cite{Cerr2016}.
Nearly all high-energy emission mechanisms can give rise to linearly polarized emission, though the
polarization angle and degree of polarization are highly dependent on the source physics and geometry \cite{Harding2017}. Both synchrotron and curvature radiation produce linearly polarized radiation in which the
angle traces the field direction and the degree of polarization is independent of energy. On the other hand, inverse
Compton scattering produces scattered radiation whose polarization degree depends on energy and
scattering angle. \g-ray polarimetry observations with e-ASTROGAM will thus be crucial to
deliver information on the NS magnetic field and locate the region in the magnetosphere
where the acceleration of particles takes place, as well as to identify different emission mechanisms. As explained in chapter \ref{intro}, e-ASTROGAM will be able to detect  0.7\% polarization from a Crab-like source in 1 Ms; it will be about 100 times more sensitive than \INTEGRAL.
Polarimetry observations in the soft \g-ray regime would be complemented by X-ray (2-10 keV) fluxes detected by the new generation of X-ray polarimetry missions, including the Chinese mission eXTP (enhanced X-ray Timing Polarimetry) a multi-facility X-ray observatory \cite{Zhang2016} with a polarimeter with a time resolution of 100 $\mu$s and NASA's IXPE (Imaging X-ray Polarimeter Explorer) mission \cite{Weis2016}. IXPE's time resolution ( $< 100 \mu$s) would make it an ideal instrument to measure phase-resolved polarization from young and bright X-ray pulsars, decoupling the emission from their surrounding and bright Pulsar Wind Nebulae (PWN). Together with a new generation of optical/infrared polarization instruments designed for the forthcoming 30 m-class telescopes, we will then, for the first time, be able to carry out multi-wavelength polarization studies of pulsars across the entire EM spectrum, from radio to soft \g-rays, including the mm/sub-mm range with ALMA, providing us with unprecedented diagnostic tools to determine key characteristics of pulsar magnetospheres.
\subsection[Transitional Millisecond Pulsars\\
\noindent
\textit{\small{T. Johnson, J. E. Grove, A. Papitto, D. de Martino}}]
{Transitional Millisecond Pulsars}\label{tran_pl}
\paragraph*{Science questions}
Millisecond pulsars (MSPs) are thought to be old, ``recycled'' pulsars, spun up by the transfer of mass and angular momentum from a binary companion \cite{Alpar82}.  The detection of ms pulsations in accretion-powered low-mass X-ray binaries (LMXBs) \cite{Wijnands98} provided early observational support for this scenario.  Dramatic new supporting evidence has come from the recent observation of three MSPs switching in both directions, between rotation-powered pulsar (RPP) and accreting (LMXB) states \cite{Papitto13,Stappers14,Bassa14,Roy15}.  All three transitional MSPs (tMSPs) belong to the class of MSPs known as ``redbacks" binary MSPs with low-mass ($\sim$0.2 to 0.5 M$_{\odot}$), non-degenerate companions (typically G-type stars) and short orbital periods ($\lesssim$ 1 day) \cite{Roberts11}.

Transitions to (from) the LMXB state are accompanied by the disappearance (reappearance) of radio pulsations, a drastic increase (decrease) in high-energy emission -- more than an order of magnitude in X-rays and a factor of a few in \g-rays -- and the appearance (disappearance) of a disk around the pulsar.  Intermittent, accretion-powered X-ray pulsations are detectable in the LMXB state of the three known tMSPs \cite{Papitto13,Papitto15,Archibald15} suggesting episodic accretion.  X-ray emission is detected out to 100 keV with no high or low energy cutoff \cite{Papitto14}.  The $\geq$ 100 MeV emission displays significant spectral curvature, well-described with an exponentially cutoff power-law shape.  The emission mechanism responsible for the enhanced HE emission in the LMXB state is uncertain, and it is unclear what conditions must be met for a system to transition.  Can only redbacks transition? Do all redbacks transition?
\paragraph*{Importance of \g-ray observations}
One of the primary differentiators between models explaining the enhanced HE emission from tMSPs in the LMXB state is whether or not the disk penetrates the pulsar magnetosphere, quenching the RPP emission.  If it does not, the enhanced HE emission is synchrotron X-ray and \ic off UV disk photons $>$ 100 MeV \cite{Takata14}.  If it does, a propeller system is created, and energized electrons emit synchrotron X-rays that then interact with the same electrons to create \ssc $>$ 100 MeV \g-rays \cite{Papitto14}.  Both models can match the X-ray and $>$ 100 MeV spectra reasonably well.  An alternate scenario, discussed in Sec.~\ref{Xbin}, is that the enhanced HE emission during the LMXB state originates from a jet, based on similarities in the X-ray emission properties with microquasars.
Measuring the shape of the spectrum in the MeV range would constrain the physics and conditions in the binary system.  If the disk is outside of the magnetosphere, a slow roll over after a few hundred keV is predicted, turning up at a few tens of MeV as the IC dominates.  In contrast, the propeller model predicts a more gradual transition from synchrotron to SSC dominance.  In the propeller model, the electron energy distribution power-law index can be derived from the X-rays, the maximum Lorentz factor from the $>$ 100 MeV spectrum, and the electron acceleration parameter from the cutoff energy in the few MeV range.  Measuring the spectrum in the MeV range would then leave only the magnetic field strength at the disk-magnetosphere interface not directly constrained in the propeller model.
\begin{figure}
\includegraphics[width=\textwidth]{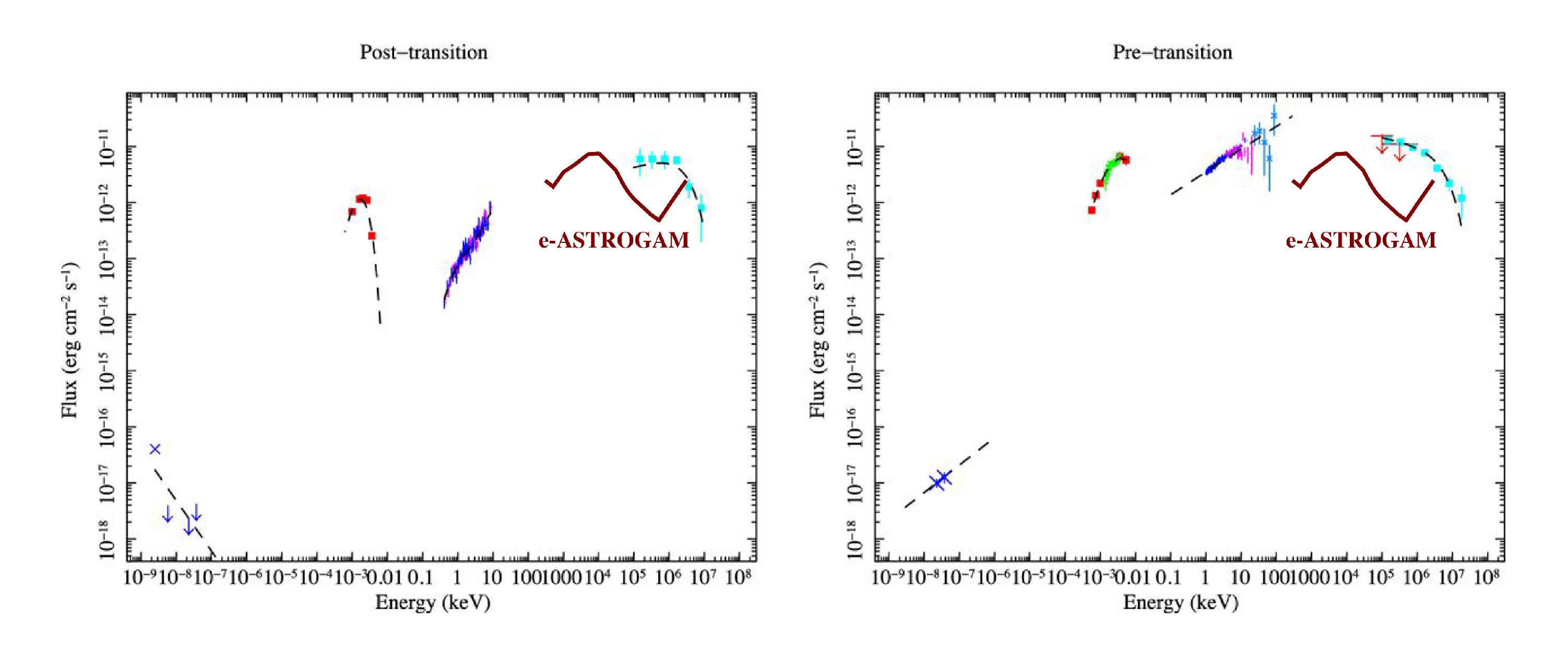}
\caption{\small{Broadband SED of XSS J12270$-$4859/PSR J1227$-$4853 during RPP state (left) and LMXB state (right), taken from \cite{deMartino15}.  The one-year sensitivity of e-ASTROGAM is added to each panel.  This tMSP would be detected with high significance by e-ASTROGAM in either state, and in particular the spectrum from a long-lasting LMXB phase might be detectable across the full science range.}}
\label{TPulsars:fig1}
\end{figure}
\paragraph*{Expected results with e-ASTROGAM}
Theoretical predictions of an energy flux in the MeV domain of few 10$^{-12}$ to few 10$^{-11}$ erg cm$^{-2}$ s$^{-1}$ from tMSPs are within reach of e-ASTROGAM (see Fig.~\ref{TPulsars:fig1}).  While one tMSP transitioned from RPP to LMXB and back over the course of one month \cite{Papitto13}, historical observations of the other two suggests transition time scales of order 10 years.  Thus, tMSPs in the LMXB state may remain there long enough to allow for multiple observations, which can then be stacked.
During the LMXB state, the X-ray emission also varies between low and high ``modes'' with periods of intense flaring, lasting as long as $\sim$45 minutes, in which the luminosity increases by a factor of $\sim$3 \cite{Papitto13,deMartino13,Bogdanov15}.  It is likely that these changes in X-ray flux state, not transitions, are due to changes in which emission mechanism dominates.  If so, the MeV emission could also be variable and at times enhanced above the model predictions discussed previously, and MeV observations triggered by X-ray monitoring could provide a detection on shorter timescales.
\subsection[Magnetars\\
\noindent
\textit{\small{R. Turolla, R. Taverna, S. Zane}}]
{Magnetars}
\paragraph*{Science questions}
Magnetars are ultra-magnetized NS ($B\approx
10^{13}$--$10^{15}$ G) which,  unlike ordinary
radio-pulsars, are powered by their magnetic energy (see e.g \cite{rev15}, for a review).
Observationally identified with two peculiar classes of X-ray
pulsars, the soft \g-ray repeaters (SGRs) and the anomalous X-ray
pulsars (AXPs), their persistent emission has been detected from
the IR/optical range up to the hard X-rays ($\approx 200$ keV)
with the \INTEGRAL satellite (see Fig.~\ref{magnet:fig1}, left). Up to now, only
upper limits at higher energies ($\approx  1$--$10$ MeV) are
available, thanks to old CGRO COMPTEL observations (see again Fig.~\ref{magnet:fig1},
left).
The basic picture for the high-energy magnetar emission involves
the reprocessing of thermal photons emitted by the star surface
through resonant Compton scattering (RCS) onto charges, moving in
a ``twisted''  magnetosphere \cite{tlk02,ntz08}. Many crucial details of the model
are however still unclear. The distribution of the scattering
particles in the velocity space is not completely understood as
yet, nor is the geometry of the  region where currents flow
(the ``j-bundle'' \cite{belo13}). Moreover, a substantial hard X-ray energy emission
is expected from curvature radiation from ultra relativistic
charges accelerated in the external magnetosphere.
\paragraph*{Importance of \g-ray observations}
Observations in the \g-ray range, as those e-ASTROGAM will allow, are key in
addressing the previous issues. Fig.~\ref{magnet:fig1} (right) and Fig.~\ref{magnet:fig2} clearly show
how theoretical spectral predictions (here based on the RCS
scenario)  are substantially different above $\sim  0.5$ MeV,
according to the assumed velocity distribution of the charges, the
geometry of the twisted region (either localized or global) and
the viewing angle.
SGRs and AXPs show somewhat different behaviours at high energies
($\approx 10$--$100$ keV). While the spectrum of the former
steepens, the latter exhibit a spectral upturn. Extrapolating the
hard X-ray flux to the $0.3$--$0.5$ MeV energy band, bright,
persistent magnetar sources are expected to reach fluxes up to
$\sim 10^{-4}\ \rm \mathrm{cts}^{-1}\mathrm{s}^{-1}\mathrm{cm}^{-2}$, as in the case of the AXPs 1RXS
J1708849-4009,  4U 0142+616, 1E 1841-045 and the SGRs 1806-20,
1900+14. At energies $> 1$ MeV, theoretical calculations predict a
steep decline of the flux, in agreement with the upper limits set
by COMPTEL (see Fig.~\ref{magnet:fig1} left).

\begin{figure}
\includegraphics[width=.45\textwidth]{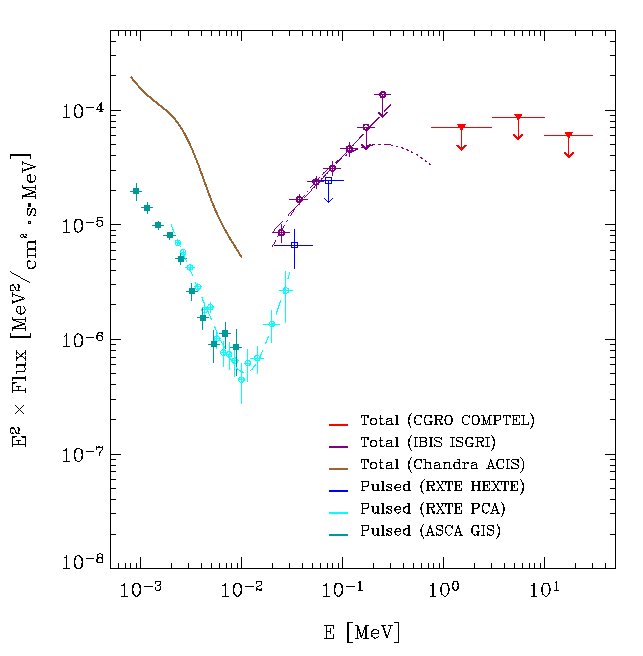}
\includegraphics[width=.45\textwidth]{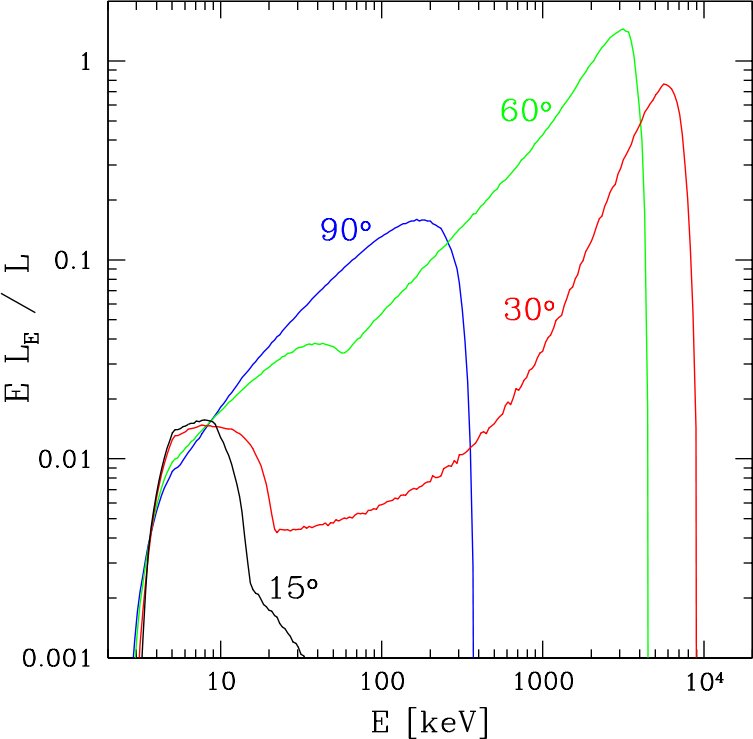}
\caption{\small  Left: The observed SED of the AXP 4U 0142+614 \cite{kui06}.
Right: RCS model spectra  from a localized j-bundle at different viewing angles with respect to the magnetic axis \cite{belo13}.
}
\label{magnet:fig1}
\end{figure}
\begin{figure}
\includegraphics[width=.5\textwidth]{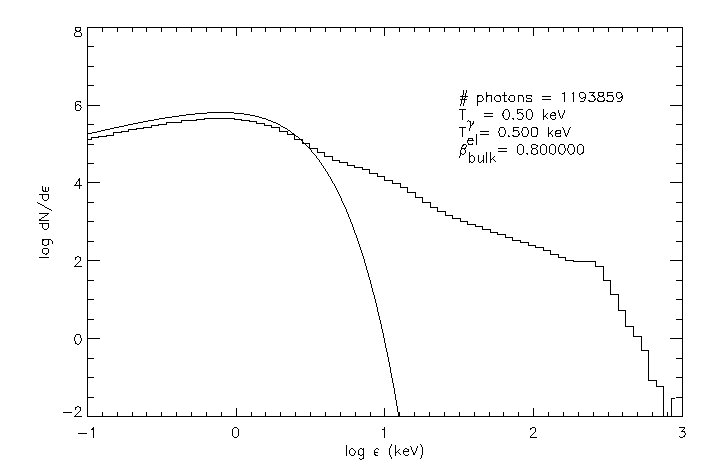}
\includegraphics[width=.5\textwidth]{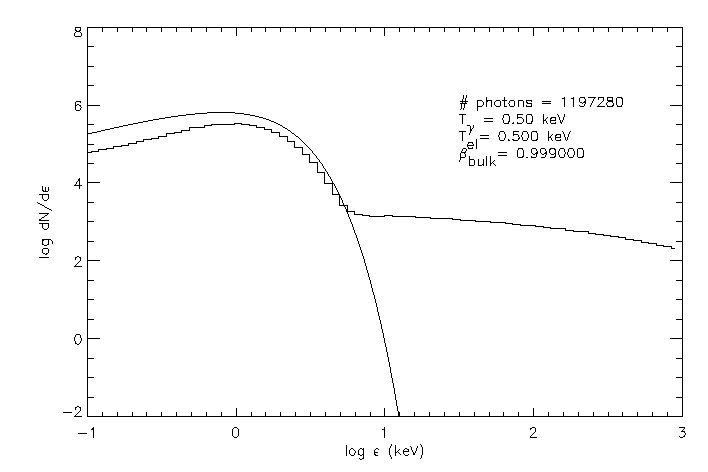}
\caption{\small Monte Carlo simulations of  RCS spectra emerging from a globally twisted magnetosphere for different
values of the bulk electron velocity, $\beta = 0.8$, left, and $\beta = 0.999$, right \cite{zane11}.
}
\label{magnet:fig2}
\end{figure}
\begin{centering}
\footnotesize{
\begin{tabular}{|c|c|c|c|c|}
\hline
  Source & Estimat. flux & e-ASTROGAM  sensitiv.  & Estimat. flux  & e-ASTROGAM sensitiv.\\
 &  $@0.3$ MeV & @$0.3$ MeV &  $@0.5$ MeV  & $@0.5$ MeV\\
 & (cts/s/cm$^2$) &(cts/s/cm$^2$) & (cts/s/cm$^2$) &(cts/s/cm$^2$)\\
  \hline
  &  & & &  \\
  RXS  J1708 & $12\times 10^{-5}$ & $2.8\times 10^{-5}$ & $7\times 10^{-5}$ & $1.3\times 10^{-5}$ \\
  4U 0142 & $23\times 10^{-5}$& $2.8\times 10^{-5}$ & $14\times 10^{-5}$ & $1.3\times 10^{-5}$ \\
  1E 1841 & $8\times 10^{-5}$ & $2.8\times 10^{-5}$ & $5\times 10^{-5}$ & $1.3\times 10^{-5}$ \\
  1806-20 & $8\times 10^{-5}$ & $2.8\times 10^{-5}$ & $5\times 10^{-5}$ & $1.3\times 10^{-5}$ \\
  1900+14 & $10^{-5}$ & $2.8\times 10^{-5}$ & $0.7\times 10^{-5}$ & $1.3\times 10^{-5}$\\
 &  & & &  \\
  \hline
\end{tabular}\label{magnet:tab1}
}
\end{centering}

\paragraph*{Expected results with e-ASTROGAM}
A flux in the range $10^{-4}$--$10^{-5}\ \rm \mathrm{cts}^{-1}\mathrm{s}^{-1}\mathrm{cm}^{-2}$ is well above e-ASTROGAM sensitivity limit, $< 2.8
\times 10^{-5}\ \rm \mathrm{cts}^{-1}\mathrm{s}^{-1}\mathrm{cm}^{-2}$ above $0.3$ MeV for an exposure time of $1$
Ms. A preliminary assessment of the detectability with e-ASTROGAM
of bright magnetar sources is reported in Table~\ref{magnet:tab1}. With the
exception of SGR 1900+14, which falls below the sensitivity
threshold by a factor $\sim 3$, all the other objects should be
easily detectable with an exposure time of $1$ Ms. We stress that
even the absence of a positive detection would be extremely
valuable in constraining magnetar physics.
Polarization studies of magnetars in the $0.3$--$1$ MeV range will
also be extremely important and ideally complement those carried
out by X-ray polarimetric missions, like IXPE, XIPE and e-XTP.
The flux of magnetars in the $2$--$10$ keV range does not exceed a few
mCrab. e-ASTROGAM Minimum Detectable Polarization is $10\%$ at a flux level of $10$ mCrab with
an exposure time of $30$ Ms. Despite the large polarization fractions expected
from these sources ($\gtrsim 50\%$; \cite{tav14}),
quite long exposure times are needed to obtain a significant
measure of the polarization properties.
\subsection[Probing the plasma origin in pulsar magnetospheres\\
\noindent
\textit{\small{A. Harding, I. A. Grenier, P. Saz Parkinson}}]
{Probing the plasma origin in pulsar magnetospheres}
\paragraph*{Science questions}
NS develop rich magnetospheres, filled with plasma pulled out of the star by the large electric fields induced by the fast rotation of the stellar magnetic field.
Large-scale currents flow out of the stellar polar caps and return back along the separatrix between the open and closed magnetic field lines (Fig.~\ref{fig:mgnsph}). When the magnetic dipole is inclined relative to the rotation axis, the thin current sheet undulates around the star. It is stable up to a distance of order ten times the size of the co-rotating part of the magnetosphere (i.e. the so-called light-cylinder radius) \cite{Kostas12c}. Young pulsars and, surprisingly, old recycled millisecond pulsars emit most of their radiated power in the \g-ray band and the \fermilat has transformed our views on the electrodynamical environment of NS by detecting more than 200 \g-ray pulsars \cite{LATPSR}. Their sharp \g-ray pulses and their SEDs and cut-offs at high energy have revealed that the pulses are produced in thin accelerators in the outer regions of the magnetosphere. These characteristics imply that most of the open magnetosphere is filled with a dense plasma that can efficiently screen the electric fields to produce a force free state.
The modelling of the MHD structure and of the global current circulation has rapidly progressed in the last few years, thanks to sophisticated MHD \cite{Kostas12a,Li12} and PIC simulations \cite{Philippov14,Cerutti16,Kostas17}, and to the interpretation of GeV observations. Yet our understanding of the structure of real, dissipative, pulsar magnetospheres and of their potential acceleration sites remains uncertain. The central question that challenges current theories is the origin of the large space densities of charges that support the magnetospheric currents. Specific regions can retain large electric fields along the magnetic field lines to accelerate primary particles to TeV energies. The latter initiate rich cascades of secondary electron-positron pairs, but where are the primary accelerators? Where do the cascades take place?  Can they supply the large charge flows that power the pulsar wind nebulae? What are the dominant radiation mechanisms for the primary and secondary particles? The comparison of the data recorded at MeV and GeV energies is essential to progress.
\begin{figure}
\centering
\includegraphics[width=0.45\textwidth]{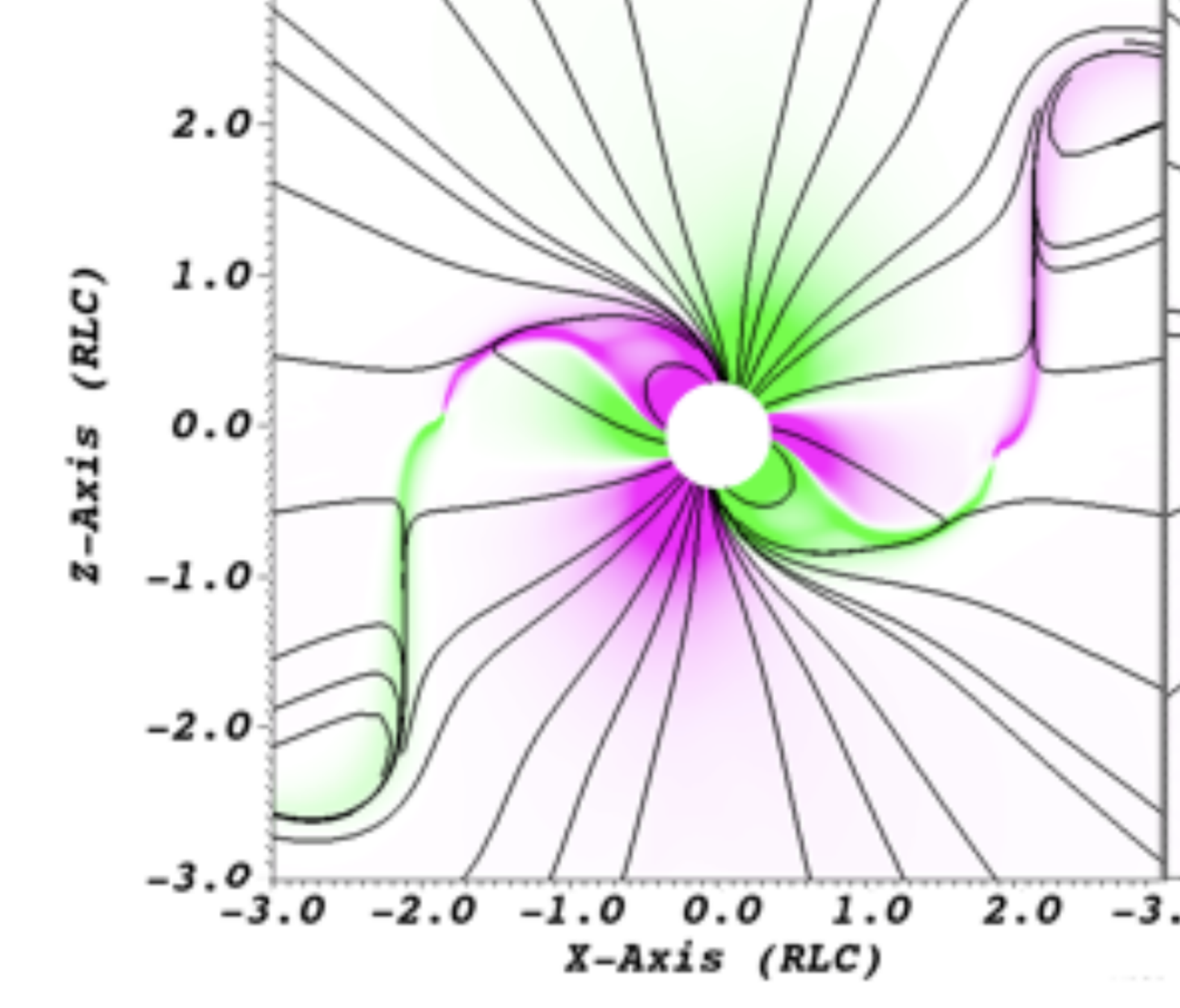}
\caption{\small{Current $J \cdot B$ in a near-force-free magnetosphere.  The pink/green color indicates current along/opposite the local magnetic field direction.  Current flows out of the polar regions and in from the current sheet. From \cite{Brambilla17}.}}
\label{fig:mgnsph}
\end{figure}
\paragraph*{Importance of \g-ray observations}
Our current knowledge suggests that the pulsed GeV emission emanates from particles accelerated in the current sheet, more or less near or beyond the light cylinder. On the other hand, pulsed emission in the MeV band should  relate to `polar' pairs produced  at various altitudes above the polar caps on field lines that do not connect to the current sheet. The spectrum and number density of the bulk of the cascading pairs can be inferred from the SED of the pulsed synchrotron radiation at MeV energies. The peak energy of the SED also yields the maximum energy of the pairs in the cascades.  The MeV data are therefore crucial to constrain how the open magnetosphere manages to be near force free and to produce the outward currents. 
The combination of MeV and GeV information for a population of pulsars with different spin-down powers and different magnetic field strengths near the polar caps and in the current sheet, and with different magnetic obliquities and viewing inclinations, is pivotal to constrain the relative geometries of the primary accelerators and secondary cascades and the beam widths of their respective radiations. In particular, the light curve shapes and the relative phases of the MeV and GeV pulses can inform us on the location of the emission sites and on whether the line-of-sight crosses the thin accelerating regions along the last open field lines and current sheet. 
The polarization data in \g-rays also hold key diagnostics on the radiation processes (synchrotron? curvature? inverse-Compton?) responsible for the pulsed emissions born in the current sheet and in the open magnetosphere. The polarization fraction and polarization angle also bear information on the location of the emitting regions with respect to the light cylinder.
\paragraph*{Expected results with e-ASTROGAM}
The \fermilat data show that the SEDs of pulsars with the largest spin-down powers or largest magnetic-field strengths tend to be very soft, peaking in the 100 keV- 100 MeV band. This trend applies to young and millisecond pulsars indifferently. Eighteen such pulsars are known to exhibit hard X-ray emission that keeps brightening toward the MeV band, but steeply dims or disappears above 100 MeV \cite{Kuiper2015}. Fig.~\ref{pulsar_mag:fig} illustrates that the sensitivity of e-ASTROGAM allows us to easily detect the SED peaks of such energetic young and millisecond pulsars.  We also expect the MeV beam produced by the pairs from cascades on a broad range of field lines to remain detectable over a wide range of viewing angles. \textit{The MeV observations therefore provide the means to uncover a large fraction of the most energetic pulsars present in the Milky Way. }These objects are rare in the radio because the radio beams cover a much smaller solid angle in the sky than the \g-rays, hence the importance of an MeV survey for energetic pulsars. 
A key diagnostic of the origin of the cascade pairs resides in the peak energy, $E_{pk}$, of the synchrotron component seen at MeV energies. It scales as  $\gamma_{\pm}^2 B_{\pm}$, where  $\gamma_{\pm}$ is the maximum Lorentz factor of the pairs and $B_{\pm}$ is the ambient magnetic field strength. Since the bulk of the cascade pairs  should radiate at high altitudes, a characteristic value for the ambient field is the strength $B_{LC}$ at the light cylinder. Primary particles accelerated above the polar caps can produce pairs that acquire pitch angles as they move out to the outer regions through resonant absorption of radio emission \cite{Harding15}. In this case, the maximum energy of the pairs relate to the magnetic field strength near the stellar surface, $B_{NS}$, so the SED peak energy should scale as $E_{pk} \propto B_{NS}\,B_{LC}$. Pair production can also occur in the outer-gap regions \cite{Takata08}. In this case, the SED peak energy depends more strongly on the outer field strength as $E_{pk} \propto B_{LC}^{7/2}$. \textit{Measuring MeV peak energies for a significant sample of pulsars can therefore discriminate between different models and locate the origin of the pair cascades that populate the open magnetosphere. Moreover, because the  MeV luminosity directly relates to the multiplicity of secondary pairs in the cascade, it directly informs us on the amount of plasma that eventually flows into the pulsar wind and termination shock.}
The comparison of the radio-loud and radio-quiet populations of MeV pulsars, as well as the possible correlation between the MeV and radio luminosities, can constrain the origin of the pitch angle of the synchrotron radiating plasma. For instance, these data can discriminate whether polar pairs or current-sheet pairs absorb radio photons as they move out, or whether current-sheet pairs acquire a pitch angle in the reconnecting magnetic field lines inside the current sheet, of if the pitch angles are produced and radiated away in the cascading process itself.
The sensitivity of e-ASTROGAM should allow the detection of large enough samples of MeV and GeV pulsars to perform population studies and explore trends in luminosity, SED shape, and SED peak energy. Based on the latest LAT 4FGL data, we expect to detect over 170 \g-ray pulsars, of which about 50 should be seen below 100 MeV; 20-40\% of them should be millisecond pulsars, depending on energy. The population studies can also bring clues to the origin, possibly geometrical, of the puzzling dichotomy between pulsars seen only at GeV energies (emission from primaries) or only at MeV energies (emission from secondary cascades). 
Finally, we expect curvature radiation from accelerated particles \cite{Harding17} to be much more polarized than synchrotron emission from accelerated particles \cite{Cerutti16} and/or secondaries. An abrupt rise in polarization fraction in the phase-averaged emission, in coincidence with the rise of the GeV emission component, would establish the curvature-radiation origin of the GeV emission. The polarization data can also constrain the altitude of the GeV emission site with respect to the light cylinder and the altitude of the emission zone for polar pairs in the open magnetosphere.  Observations of rotation-powered pulsars with e-ASTROGAM thus holds the promise of constraining the origin(s) and spectrum of the pair plasma that shapes the pulsar magnetosphere, as well as the GeV emission mechanism.
\begin{figure}
\includegraphics[width=0.48\textwidth]{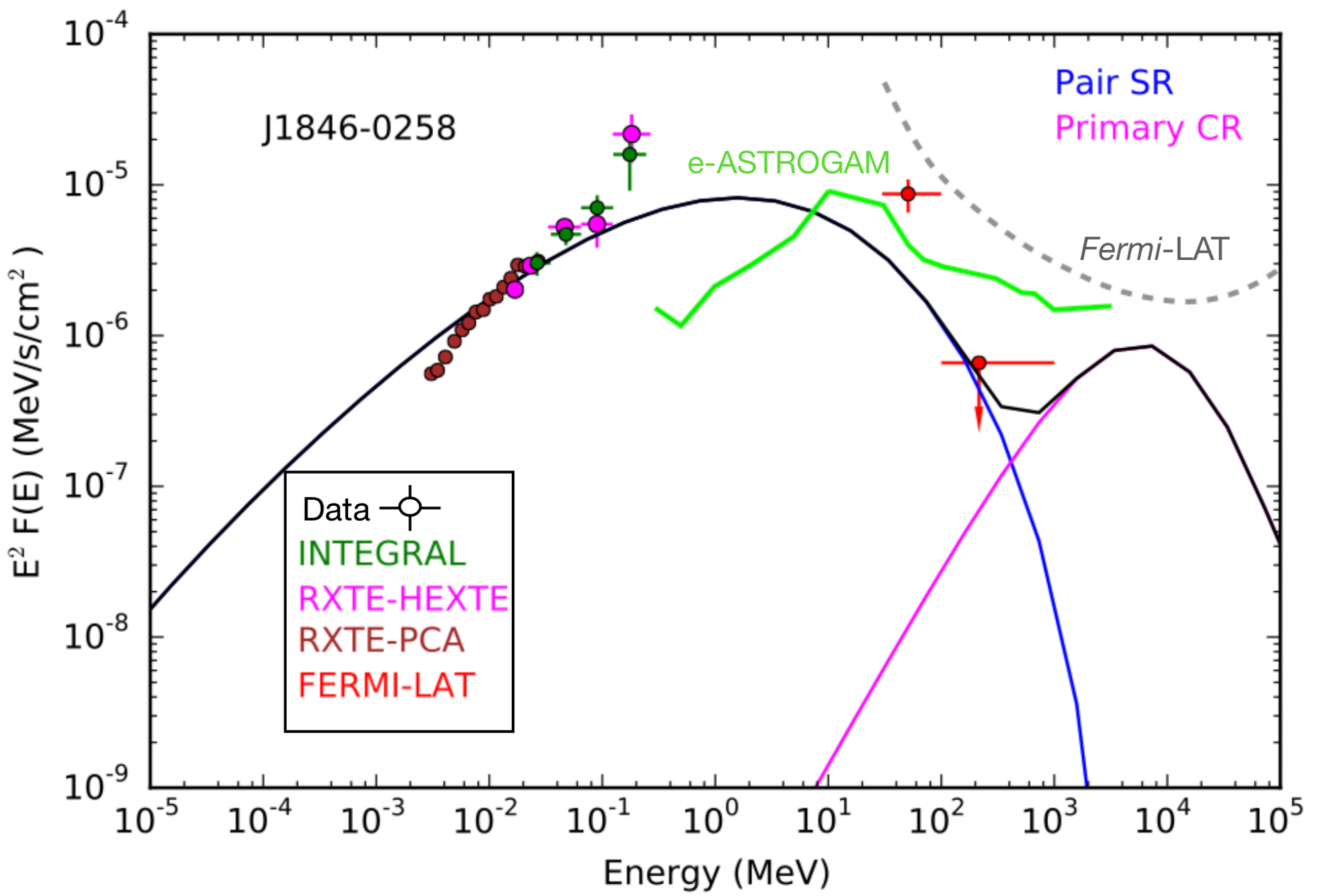}
\includegraphics[width=0.48\textwidth]{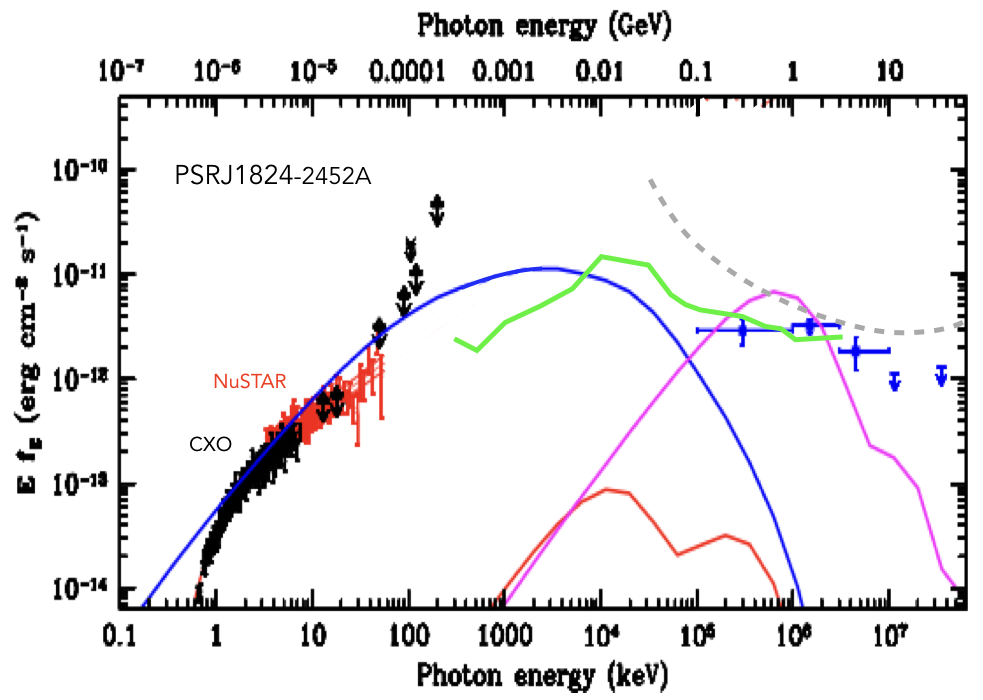}
\caption{\small{SEDs of the young pulsar J1846-0258 (left) and of the millisecond pulsar J1824-2452A (right). The model curves show the expected curvature radiation (magenta) and synchrotron radiation (red)  from the primary particles accelerated in the current sheet, and the synchrotron radiation (blue) from the secondary pairs produced in cascades and radiating in the open magnetosphere. Adapted from \cite{Harding15}, \cite{Harding17} and \cite{Gotthelf17}, with J1846-0258 data from \cite{Kuiper2015} and \cite{Kuiper17}. The sensitivities of e-ASTROGAM (solid green) and of \fermilat (dashed gray) are given for one year of effective exposure in the Galactic disk.}}
\label{pulsar_mag:fig}
\end{figure}
\subsection[Probing the maximum particle energies in pulsar wind nebulae\\
\noindent
\textit{\small{A. Harding, I. A. Grenier, P. Saz Parkinson}}]
{Probing the maximum particle energies in pulsar wind nebulae}
\paragraph*{Science questions}
Pulsar wind nebulae (PWNe) are the manifestation of the particle production by pulsars.  The electron-positron pairs that are produced in cascades in the magnetosphere flow outward and form the pulsar wind that dissipates the spin-down luminosity of the pulsar.  These pairs are accelerated near or in the termination shock, the reverse shock that reacts to the contact outer discontinuity of the nebula with the ISM through pressure balance
\cite{ReesGunn1974}.  Although PWNe are the most numerous Galactic sources detected at TeV energies by Air Cherenkov telescopes, \Fermi has detected relatively few at GeV energies.  PWNe have a multicomponent spectrum consisting of a synchrotron (SR) component in soft \g-rays and an \ic component at higher energies.  The Crab PWN is the brightest and most powerful, its SR component extending to 100 MeV, with the IC component extending to at least 50 TeV.   The Crab nebula differs from the others because the IC emission is synchrotron self-Compton.  The IC components of most other known PWNe  are produced by up-scattering of the ambient soft photon fields.  The sensitivity of \Fermi falls in the valley between the SR and IC components for most other PWNe whose SR spectra extend to lower energies.  The maximum SR photon energy of the Crab PWN, together with the IC spectrum, tells us that the pairs are continuously accelerated to PeV energies \cite{DJH1992}.  A major science question that will be answered by e-ASTROGAM is:
What is the maximum energy of the particles accelerated in PWNe and how does it depend on properties of the pulsar?

\Fermi and \agile discovered surprising flares from the Crab PWN \cite{Abdo2011,BuehlerBlandford2014} with SR photon energies reaching up to 500 MeV (see Fig.~\ref{PWN:fig1}). They require transient particle acceleration to several PeV and violate the 140 MeV diffusive shock acceleration limit \cite{DJ1996}.  The \fermigbm and \swift detected much slower flux variations on year timescales that may be caused by the GeV flares \cite{WH2016}.  e-ASTROGAM, with its wide field of view, will be able to detect possible flares from PWNe with SR cutoffs in the MeV band.  It will also be able to detect flux variations in the  Crab to determine whether the GeV flares produce variability at energies below 100 MeV, with the timescales giving information on the flare location and geometry.  A key question is: what is the very rapid acceleration mechanism that produces the high-energy flare particles.
\paragraph*{Importance of \g-ray observations}
Since the transition between SR and IC components in PWNe spectra falls in the hard X-ray to GeV \g-ray band, \g-ray observations can catch both components.  The maximum steady-state (non-flaring) energy to which particles can be accelerated in a PWN is equal to the voltage across the open field lines, $V_{\rm open} = 6 \times 10^{12}\,B_{12}P^{-2}$ eV, where $B_{12}$ is the pulsar surface magnetic field strength and $P$ is the period.  For the Crab PWN, $V_{\rm open}$ reaches the maximum SR photon energy where SR losses balance acceleration gains (assuming $E<B$.  For most other middle-aged PWNe, $V_{\rm open}$ is much lower and the expected maximum of the SR spectrum, $\epsilon_{SR} \propto V_{\rm open}^2B_s \sim 0.14\, {\rm MeV}\,L_{36}^{6/5}\,[\sigma/(1+\sigma)]^{1/2}\,\tau_{kyr}^{-3/10}$, where $B_s$ is the field strength at the termination shock, $L_{36}$ is the pulsar spin-down luminosity in units of $10^{36}\,\rm erg\,s^{-1}$, $\sigma$ is the wind magnetization and $\tau_{kyr}$ is the pulsar age in kyr.  Therefore, these PWNe should have SR cutoffs visible in the energy range of e-ASTROGAM.

\begin{figure}
\centering
\includegraphics[width=0.8\textwidth]{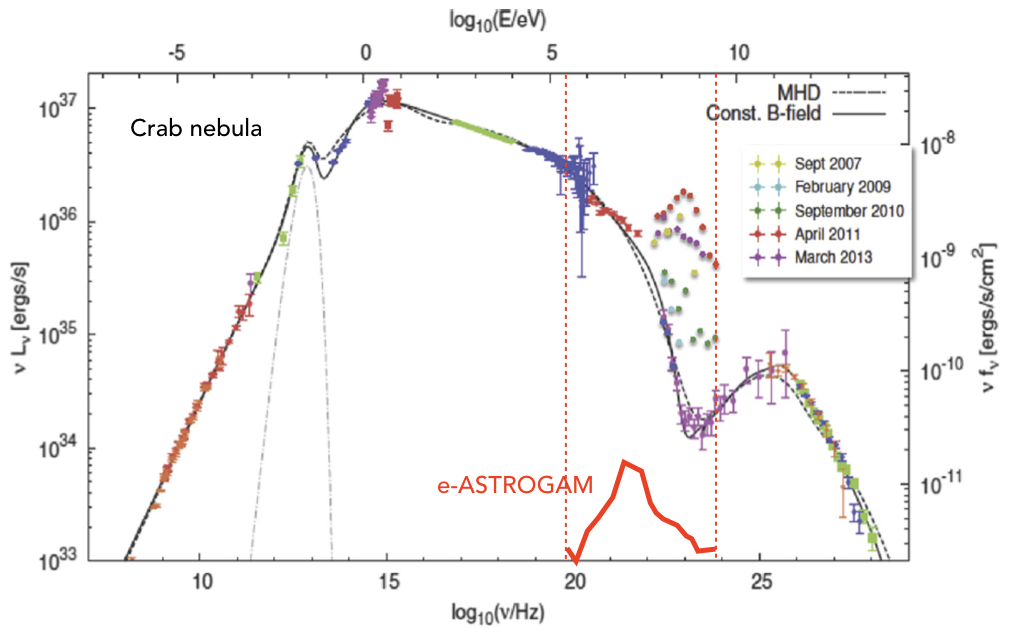}
\caption{\small{SED of the Crab nebula from the radio to VHE \g-rays, also showing the flares. The sensitivity of e-ASTROGAM (red curve) is given for one year of effective exposure in the Galactic disk. Adapted from \cite{BuehlerBlandford2014,Meyer10}.}}
\label{PWN:fig1}
\end{figure}

The detected Crab flares occur near the high-energy cutoff of the SR spectrum, since the maximum-energy particles producing this emission have the fastest SR timescales.
Similarly, flares on longer timescales of months may be expected for other PWNe with SR spectral cutoffs in the MeV band.  If the same processes, such as magnetic reconnection \cite{Cerutti2014}, that are proposed to cause flaring in the Crab PWN are occurring in other PWNe, then we might expect to see flares near the high end of their SR spectra that is not accessible with current telescopes.  Whether the Crab GeV flares are connected with the slower flux variations in the hard X-ray band is currently not clear.  An e-ASTROGAM detection of flux variations of the Crab, which would be on month-long timescales, would confirm whether the variations are due to radiation of flare particles as they lose energy to SR.
\paragraph*{Expected results with e-ASTROGAM}
Fig.~\ref{PWN:fig2} shows examples of two PWNe with measured X-ray and GeV spectra, where the SR cutoff should fall around 1-10 MeV.  It is expected that e-ASTROGAM will detect the SR cutoff in these and other PWNe.  Together with a detected IC component from either e-ASTROGAM or \Fermi, the maximum particle energy can be deduced since the IC spectrum constrains $B_s$, as well as the magnetization of the wind, $\sigma$.  Since most PWNe will have SR spectra with cutoffs in the 100 keV - 10 MeV band, e-ASTROGAM will greatly increase the number of PWN detections at \g-ray energies over the number that \Fermi detected, and will be able constrain the maximum particle energy in a large number of PWN.
Synergy between e-ASTROGAM, Athena and CTA will allow significant progress in understanding how the pulsar spin-down power is transferred to the wind and also how the radiating electron-positron pairs diffuse into the ISM, important in explaining the observed cosmic-ray positron excess \cite{Accardo2014}.  Athena and e-ASTROGAM give information on the maximum particle energy and rapid SR losses in or near the accelerator site.  CTA can image the spatial variations of the spectral losses of the pairs in the wind at tens of pc from the pulsar, thereby mapping the MHD structure of a PWNe.
Finally, detection of flares in older PWNe by e-ASTROGAM would provide valuable information of relativistic reconnection physics, a field still in its infancy \cite{Sironi2016}.
\begin{figure}
\includegraphics[width=0.5\textwidth]{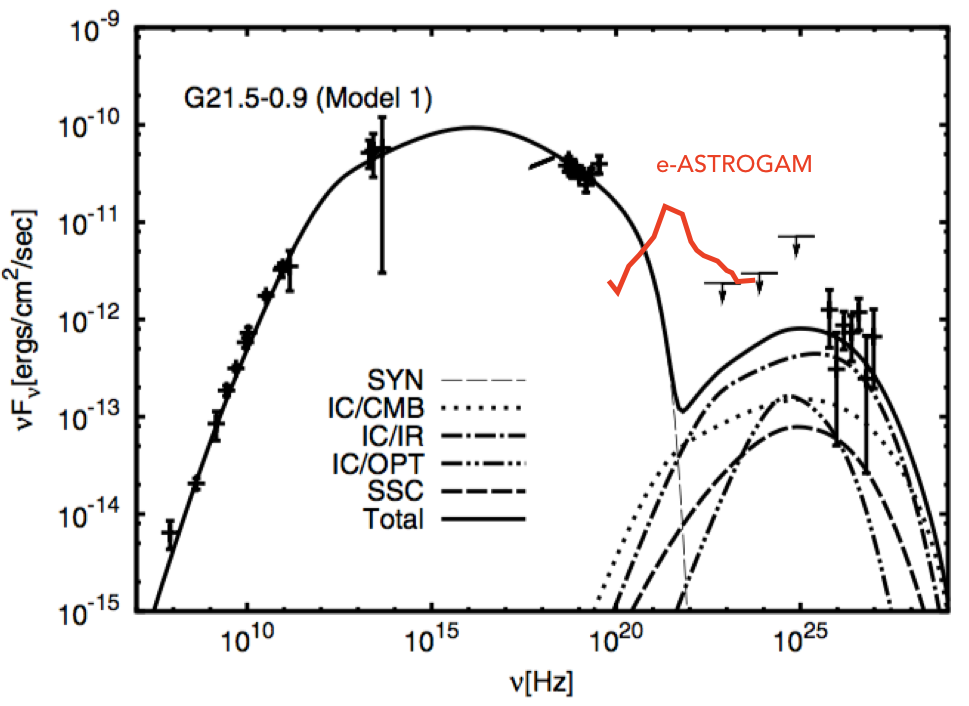}
\includegraphics[width=0.5\textwidth]{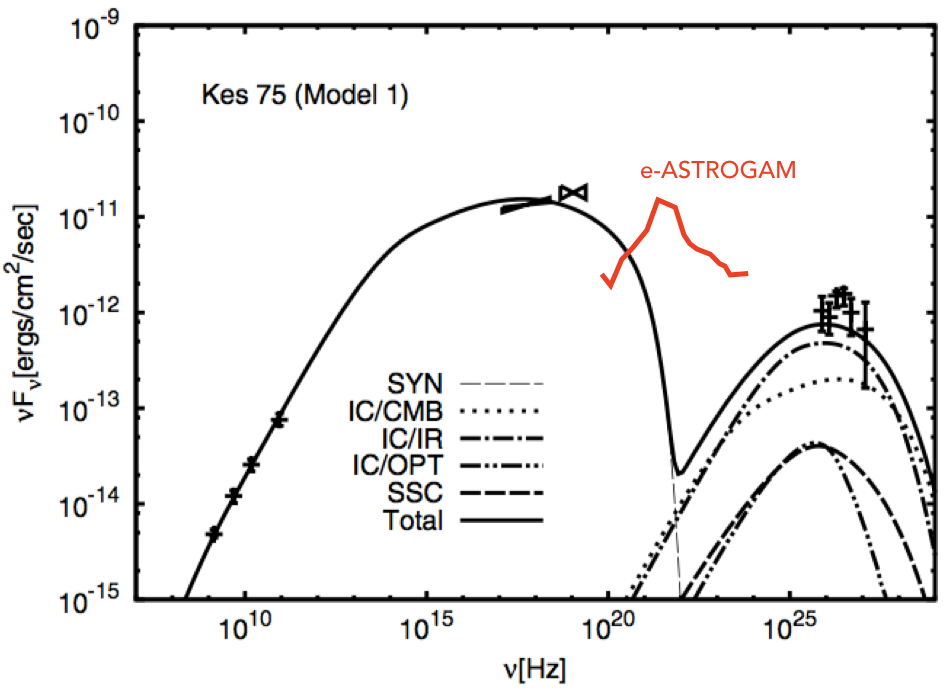}
\caption{\small{Spectra and models for PWN G21.5-0.9 and Kes75 from \cite{TanakaTakahara2011}. The sensitivity of e-ASTROGAM (red curve) is given for one year of effective exposure in the Galactic disk.}}
\label{PWN:fig2}
\end{figure}
\subsection[Gamma-ray binaries\\
\noindent
\textit{\small{J.~M.~Paredes, V.~Bosch-Ramon, D.~de~Martino, A.~Papitto, R.~Walter, \\A.~A.~Zdziarski}}]{Gamma-ray binaries}\label{gbin}
\paragraph*{Science questions}
The nature of \g-ray emission from X-ray binaries presents a number of major puzzles. Generally, that emission can be either powered by accretion onto a compact object or be due to collisions between winds from the binary components \cite{dubus13,dubus15}. Among those sources, the most prominent \g-ray emission is seen from the so-called \g-ray binaries, consisting of a massive star and a compact object, and where \g-rays dominate the SED, peaking above 1 MeV, see Fig.~\ref{LSI}.
There are six \g-ray binaries detected in HE (0.1-100 GeV) or VHE ($>$100 GeV) \g-rays.
 In one case, PSR B1259--63, radio pulsations are detected \cite{joh92}, showing that the compact object is a rotation-powered pulsar, and thus the \g-rays, emitted close to periastron, are likely to be due to interaction between the pulsar and stellar winds \cite{dubus06}, where particles are accelerated at the shock between the winds. No radio pulsations have been found in other cases; although this can be explained by free-free absorption in the stellar wind, no definite proof of the nature of the other sources exist \cite{dubus06}. \fermilat detected  PSR J2032+4127, a new \g-ray  binary that shares many similar characteristics with the previously
unique TeV binary PSR B1259-63. This new source is a long period ($\sim 50$ years) Be binary system hosting a pulsar \cite{lyne15} located
in the vicinity of the first (and yet) unidentified TeV source discovered by HEGRA, TeV J2032+4130. At present, it is not known whether PSR J2032+4127 can produce \g-rays in the star-pulsar wind colliding region. Observations with e-ASTROGAM will be crucial to fully characterize the \g-ray spectrum of the source, and identify a potential
non-magnetospheric component in the MeV-GeV energy band. Perhaps, the best studied \g-ray binary is LS~5039 \cite{par00,aha05b}, which presents, in addition to HE and VHE \g-rays, very strong MeV radiation that is modulated along the orbit. The soft \g-rays seem to naturally follow a synchrotron component coming from $\lesssim 1$~keV and peaking around 30~MeV \cite{col14}.
Generally, satisfactory detailed models explaining both the spectra and orbital modulation of these objects are still missing, largely due to the lack of observations in the MeV range.
\begin{figure}
\centering
\includegraphics[width=0.65\textwidth]{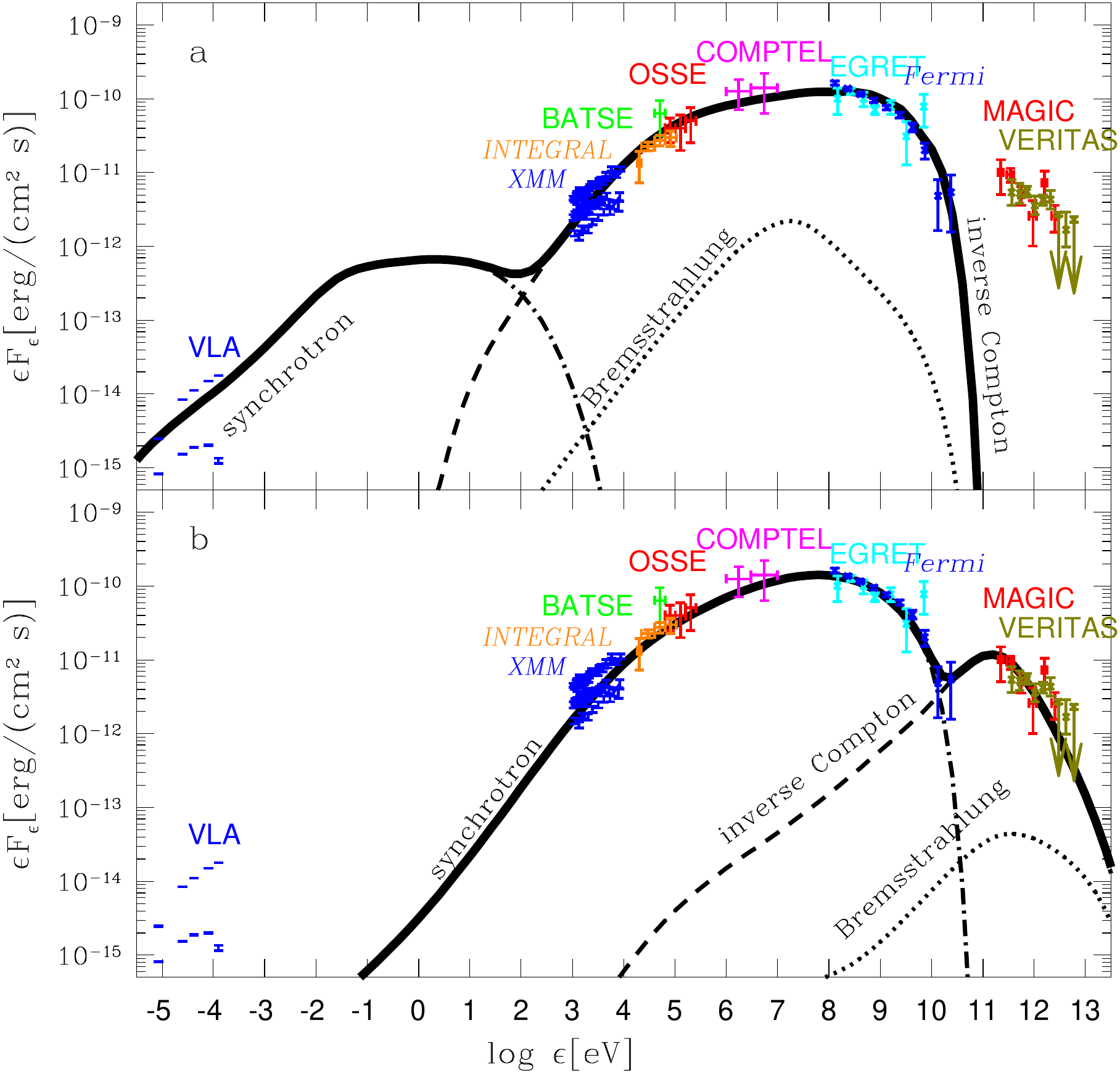}
\caption{\small{The broad-band spectrum of the \g-ray binary LS I +61 303 modelled by two variants of the model of the pulsar-wind/stellar-wind interaction \cite{zdz10}. (a) The model in which the soft and HE \g-rays are dominated by Compton scattering. (b) The model in which the soft and HE \g-rays are dominated by the synchrotron process.}}\label{LSI}
\vspace{-10pt}
\end{figure}
\paragraph*{Importance of \g-ray observations}
Although \fermilat has opened a new discovery space for \g-ray emission from binaries, since their SEDs often peak below $\sim 100$~MeV \cite{col14,col17}, lack of enough coverage and sensitivity at these energies has so far hampered studies of the true nature of the \g-ray emission from these sources. Also, only upper limits were obtained in that range by \agile. Thus, sensitive observations below $\sim 100$~MeV are likely to detect many more of such objects, as the number of \g-ray binaries in the Galaxy is expected to be between $\sim$50 and 200 \cite{dubus17}.
\g-ray binaries have most likely two dominant radiation mechanisms: synchrotron emission, from radio to X-rays/soft \g-rays, and \ic scattering of stellar photons, dominant in the HE and the VHE range \cite{bk09}. The MeV-GeV spectral range is right between the synchrotron and the IC dominance energy ranges \cite{par06}, and is very important to properly understand the physics giving rise to the synchrotron and the IC emission from these objects. Then, \g-ray observations below $\sim 100$~MeV will probe the intersection region, allowing us to distinguish between the two components.
If synchrotron emission is dominant, exploring the MeV-GeV range can allow us to probe extreme particle acceleration. Interestingly, the $\sim$100-MeV synchrotron limit can be exceeded in some cases, as observed in the Crab Nebula. In a \g-ray binary, the observation of a synchrotron component exceeding that limit could unveil important physical information, such as highly relativistic motions, or contamination by a different radiation component.
On the other hand, if IC is dominant, the MeV-GeV range can provide important information related to how non-thermal particles propagate away from the stellar companion, as IC losses are slow for electrons producing MeV photons via IC with stellar photons. The IC process can also probe the geometry of the sources by observing its orbital modulation, related to the varying viewing angle with respect to the binary major axis, which implies changes in the IC emission.
Regardless of the dominant emission process, the MeV-GeV range also permits a careful investigation of the effects of \g-ray absorption and reprocessing on the spectrum, and complements the study of different wind physical conditions in eccentric systems such as the O/Be binaries.
Finally, if \g-ray binaries host a powerful pulsar that powers the non-thermal emission, the MeV photons can interact with the pulsar wind if the latter reaches Lorentz factors of about $10^5-10^6$. This would trigger EM cascades in the pulsar wind that should give rise to strong \g-ray and lower energy radiation, and also strongly modify the wind nature \cite{der12}.
\paragraph*{Expected results with e-ASTROGAM}
COMPTEL data already indicated that \g-ray binaries (in particular LS~5039 \cite{col14} but perhaps most of them \cite{col17}), are powerful MeV emitters. e-ASTROGAM, with its sensitivity in soft \g-rays two orders of magnitude better than that of COMPTEL, will discover many new cases of \g-ray emission from binaries. Its sensitivity will allow the characterization of the orbital light curve and spectral evolution of \g-ray binaries, clearly differentiating the synchrotron and the IC components, probing particle acceleration and \g-ray reprocessing, and potentially revealing pulsar wind physics that can only be probed in this kind of objects.
After the expected launch of e-ASTROGAM, major new facilities from radio to VHE \g-rays, SKA, \textit{Athena\/} and CTA, will also be operational. This will provide an unprecedented opportunity to study particle acceleration, outflows, and wind launching mechanisms in different types of binaries.
\subsection[Gamma-ray emission from accretion-powered X-ray binaries\\
\noindent
\textit{\small{A.~A.~Zdziarski, R.~Walter, V.~Bosch-Ramon, P.~Jean, D.~de~Martino, \\A.~Papitto, J.~M.~Paredes, V.~Tatischeff}}]
{Gamma-ray emission from accretion-powered X-ray binaries}\label{Xbin}
\paragraph*{Science questions}
We consider \g-ray emission from accretion-powered X-ray binaries, excluding the so-called \g-ray binaries (see Sec.~\ref{gbin}), where \g-rays peak above 1 MeV and dominate the SED. Gamma-rays from accretion-powered binaries are usually observed from microquasars, i.e., systems featuring jets. Unambiguous detections of high-energy (HE) \g-rays have only been from high-mass X-ray binaries Cyg X-3 \cite{fermi09} and Cyg X-1 \cite{zanin16,zdz17}. In Cyg X-3, where the nature of the compact object still remains unknown, \g-rays are observed in its soft spectral state, and are strongly orbitally modulated. The \g-ray modulation and spectrum are interpreted as Compton scattering of the blackbody emission of the donor in the jet \cite{dubus10,zdz12a}. However, the models cannot be constrained due to the lack of sensitive observations in the crucial MeV range. In Cyg X-1, HE \g-rays are observed instead only in the hard spectral state (Fig.~\ref{cygx1}), where a compact radio jet is also detected. On the other hand, excess emission below 100 MeV is observed in both hard and soft spectral states (Fig.~\ref{cygx1}), appearing to connect to the high-energy tails observed in soft \g-rays \cite{zdz17}. Another puzzle of Cyg X-1 is the claim of very strong polarization around 1 MeV \cite{jourdain12,rodriguez15}, at face value pointing to synchrotron emission from the jet. This interpretation, however, presents a number of problems.

Interestingly, no HE \g-rays have been detected from low-mass X-ray binaries (LMXBs) containing black-holes, except for a hint of transient emission from V404 Cyg \cite{loh16}. This lack of emission is still not understood. V404 Cyg is also the only object in which an e$^\pm$ annihilation feature has been detected at a relatively high significance \cite{siegert16}. An important science question is how common such emission is and how it can be modelled.

A new type of \g-ray binaries are the so-called transitional ms pulsars, objects showing both rotation-powered and accretion-powered states \cite{Papitto13,Bassa14,archibald09,demartino10} (see Sec.~\ref{tran_pl}). In two sources, transitions between the pulsar and weak accretion states were associated with a power-law-shaped X-ray spectrum with no cut-off up to at least $\sim$100 keV and an increase by up to a factor of a few of the \g-ray flux in the latter \cite{torres17}. The increase of the \g-ray flux has been explained by the interaction of the accretion disk with the pulsar wind \cite{Takata14}
or a propellering magnetosphere \cite{papitto15_apj}. However, the enhanced \g-ray emission was also associated with the appearence of a strong variable radio flux  with a spectral index of $\sim$0 \cite{hill11,deller15}. This behaviour is typical of microquasars, suggesting the possibility that both \g-ray and radio emissions originate in a jet. This would be the first case of steady \g-ray emission from LMXBs during a disk state.

\begin{figure}
\centering
\includegraphics[width=0.65\textwidth]{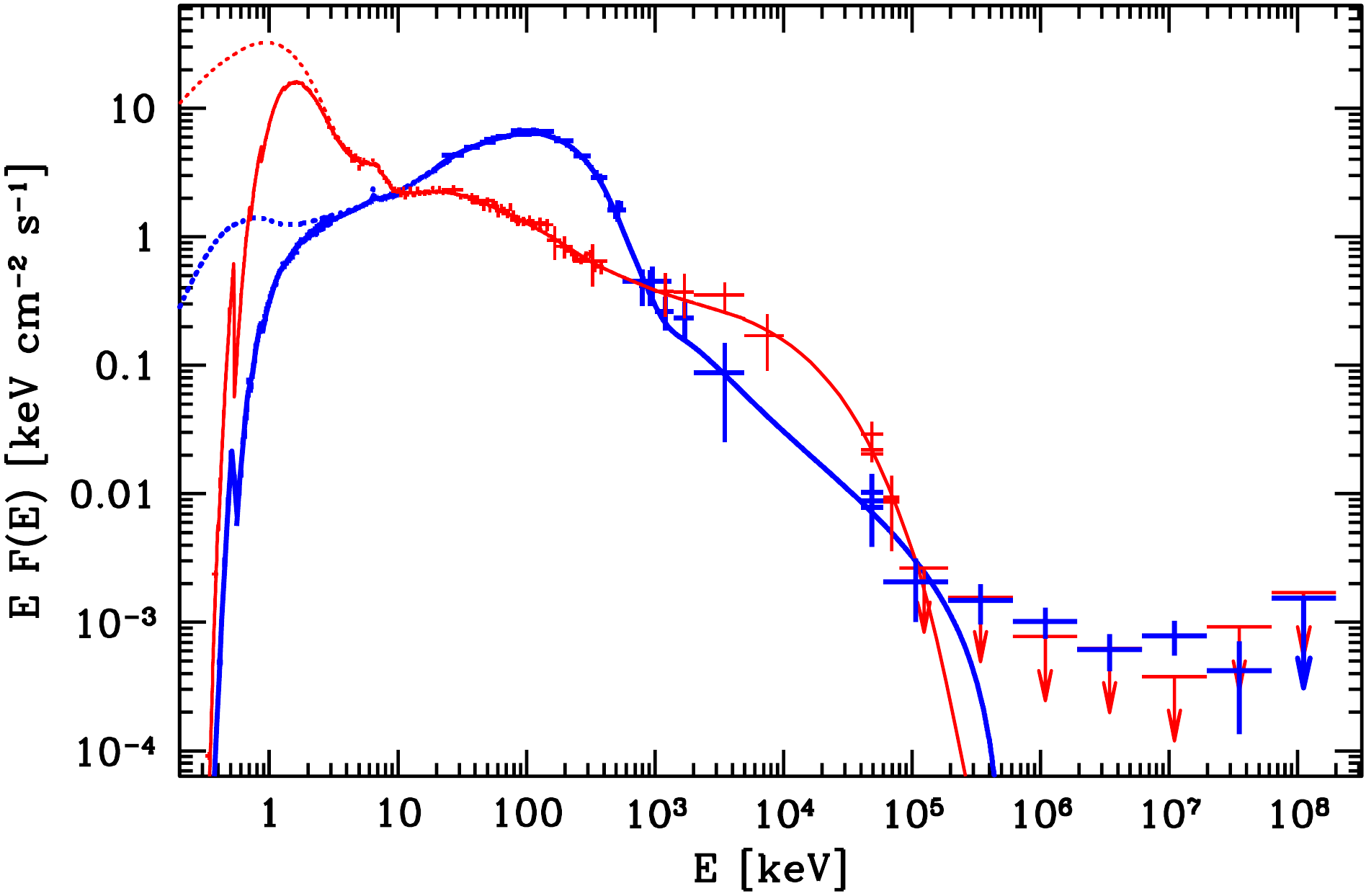}
\caption{\small{Broad-band X-ray/\g-ray spectra for Cyg X-1 in the hard (blue heavy symbols) and soft (red thin symbols) states. The data at $<$ 10 MeV (attenuated by X-ray absorption) are from BeppoSAX and CGRO, while the data at $\geq$40 MeV are from Fermi/LAT. Data are compared to hybrid-Comptonization accretion-flow models \cite{zdz17}. The observed emission above $\sim$100 MeV in the hard state is dominated by the jet. The dotted curves at soft X-rays show the unabsorbed models.}} \label{cygx1}
\vspace{-10pt}
\end{figure}

Finally, the conditions in inner parts of accretion disks can allow neutrons to be produced by spallation of He, at the rate depending on the disk physical state. The neutrons produce 2.2-MeV photons when captured by protons, which can result in a broad line in black-hole accretion disks in the case of fast protons, at the estimated flux of $\sim\!\!10^{-6}$ s$^{-1}$ cm$^{-2}$ at 1 kpc \cite{aharonian84}. If neutron capture takes place in the upper atmosphere of an accreting neutron star, the line will be narrow and gravitationaly redshifted, and its redshift would yield the neutron star mass to radius ratio, and thus a constraint on the equation of state \cite{bildsten93}. Neutrons can also escape the accretion disk and hit the companion star, where they slow down and get captured by ambient protons, resulting in a narrow line \cite{jean01}. The flux in this case depends on many parameters, and a rough estimate is also $\sim\!\!10^{-6}$ s$^{-1}$ cm$^{-2}$ for nearby (1--2 kpc) X-ray binaries \cite{guessoum02}.
\paragraph*{Importance of \g-ray observations}
Although \fermilat has opened a new discovery space for \g-ray emission from binaries, the lack of adequate coverage and sensitivity below 100 MeV has so far hampered investigations of the true nature of their \g-ray emission. Also, sensitive observations in that range are likely to detect many more of such objects.

Detailed modelling of the MeV-range emission will provide the first unambiguous tests of emission mechanisms and help disentangling disk-jet coupling in accretion-powered binaries. The main physical processes contributing to \g-rays in binaries are synchrotron and \ic scattering. The former is, in usual cases, limited to the range of $\lesssim$100 MeV \cite{guilbert83}. Then, \g-ray observations below 100 MeV will probe the intersection region, allowing us to distinguish between the two components. However, the $\sim$100-MeV limit can be exceeded in some cases, as observed in the Crab Nebula, and an observation of a synchrotron component exceeding that in a binary would be of paramount importance. We can also probe the geometry of the sources by observations of their orbital modulation.

If the MeV tail of Cyg X-1 is due to polarized jet synchrotron emission, an intersection of the synchrotron and Compton components is expected below 100 MeV \cite{zdz14}. If, on the other hand, the tail is from Compton scattering by nonthermal electrons in the accretion flow, the intersection will be of the accretion and jet emissions. In the case of Cyg X-3, we observe strong orbital modulation of X-rays up to 100 keV with the minimum at the superior conjunction \cite{zdz12c}, and strong orbital modulation at $>$100 MeV peaking at it \cite{fermi09}. Observations below 100 MeV will allow us to unambiguously distinguish between the jet and accretion components, and e.g., test popular models in which the tail beyond the accretion-disk blackbody peak in the soft states of X-ray binaries is due to jet synchrotron emission. Furthermore, observations of orbital modulation of \g-rays below 100 MeV due to \ic scattering of stellar blackbody photons will allow us a precise determination of both the location of the \g-ray source along the jet and the jet orientation.

Then, detections of the 2.2 MeV line from X-ray binaries would be a major discovery, allowing us to set strong constraints on the physics of their accretion flows.
\paragraph*{Expected results with e-ASTROGAM}
e-ASTROGAM, with its sensitivity in the soft ($<$100 MeV) \g-ray range, 
will discover many new cases of accretion-powered X-ray binaries. Its sensitivity will allow the characterization of their orbital light curves and spectral evolution for the first time down to the soft \g-rays, clearly differentiating the synchrotron/IC and accretion/jet components, probing particle acceleration and \g-ray reprocessing, and potentially revealing jet physics.
\begin{figure}
{\includegraphics[height=0.51\textwidth,angle=-90]{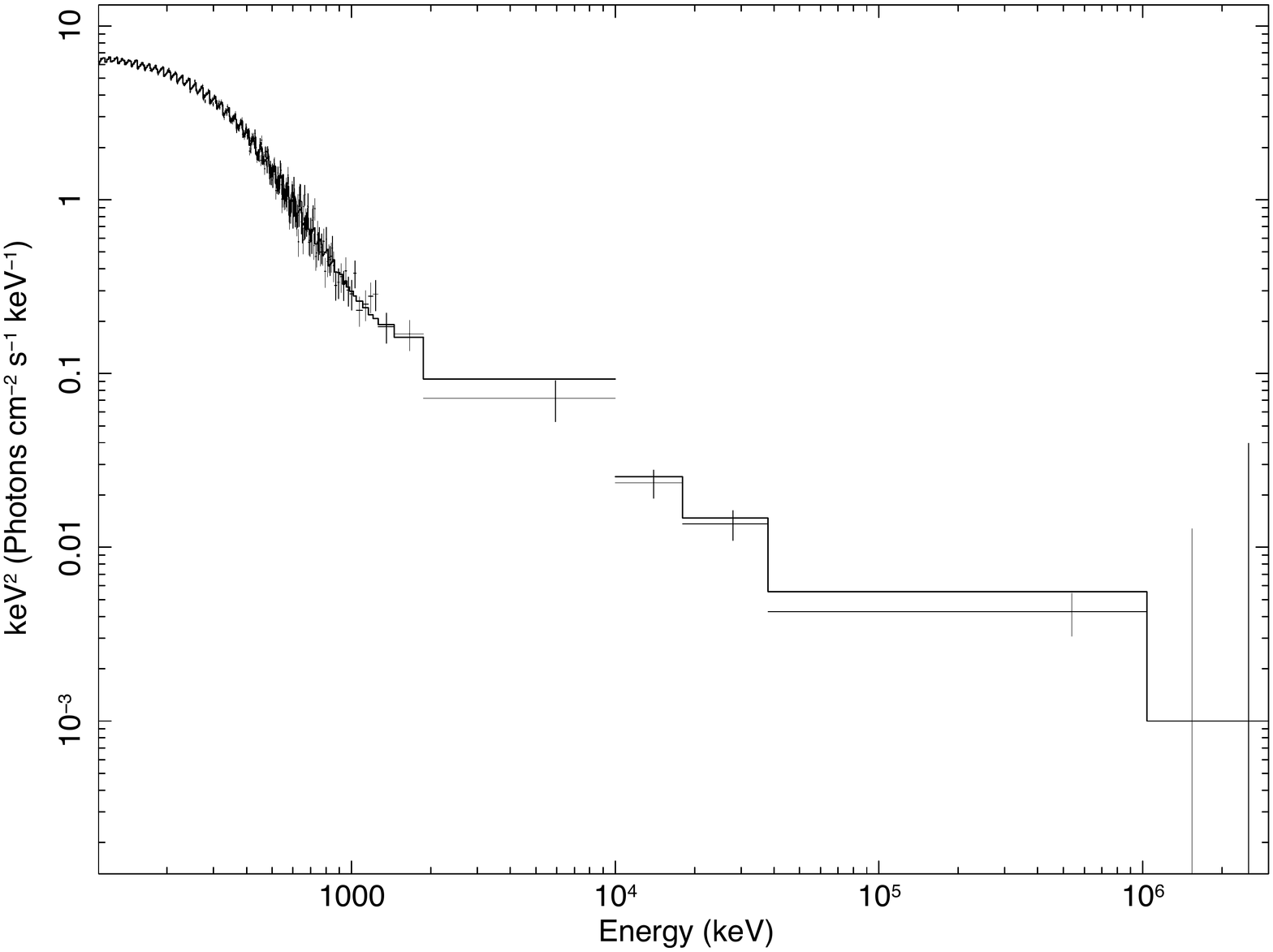}
\includegraphics[height=0.51\textwidth,angle=-90]{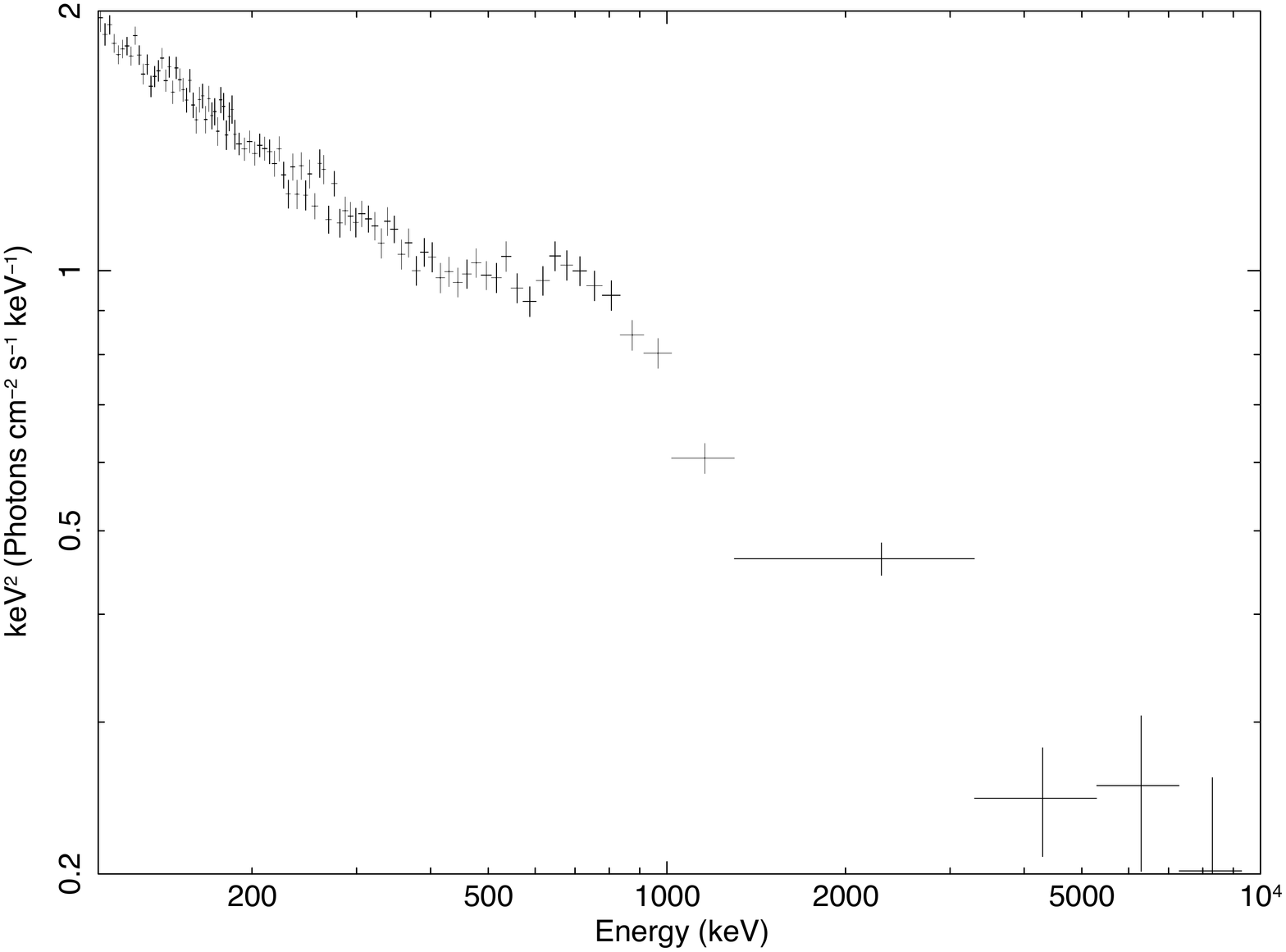}}
\caption{\small{Left: Simulated $10^5$-s e-ASTROGAM average hard-state spectrum of Cyg X-1 assuming contributions from thermal Comptonization in the accretion flow at low energies and from synchrotron emission in a jet at high energies, as in Fig.~\ref{cygx1}. Right: Simulated $10^5$-s spectrum of an X-ray binary with a power law of $\Gamma=2.5$ and a broad line from annihilation of e$^\pm$ \cite{svensson96} at the temperature of $kT\simeq 100$ keV.}}\label{simulated}
\end{figure}
Fig.~\ref{simulated} (left) shows a simulation of a $10^5$-s observation of Cyg X-1 in the hard state. The signal-to-noise ratio of the detection at $>500$ keV is high, 76. We have also found that e-ASTROGAM will be able to detect it in $10^3$ s up to several MeV with the significance similar to that obtained by \INTEGRAL in $2\times 10^6$ s \cite{rodriguez15}. This will allow us to study for the first time correlations between the thermal Comptonization and the high-energy tail components. Given its capability to detect \g-ray polarization, e-ASTROGAM will test the intriguing detection of soft \g-ray polarization in Cyg X-1. In a $10^6$ s exposure, the minimal polarization measurable above 500 keV at 99\% confidence level will be as low as 5\%. We will also be able to search for it in other sources. Fig.~\ref{simulated}(right) demonstrates the sensitivity of e-ASTROGAM to detect e$^\pm$ annihilation lines. At the line equivalent width of 106 keV and the flux an order of magnitude lower than that found in V404 Cyg \cite{siegert16}, the signal-to-noise ratio of the line detection is 32. Finally, the time required for a detection of a broad 2.2 MeV line at the estimated flux is $\sim$4 Ms.
\subsection[Detection of very short Gamma-Ray Bursts in exotic stellar transitions\\
\noindent
\textit{\small{M.~A.~P\'{e}rez Garc\'{i}a, C.~Albertus, M.~Cermen\~{o}, J.~M.~\'{A}lvarez, L.~Roso}}]
{Detection of very short Gamma-Ray Bursts in exotic stellar transitions}
\paragraph*{Science questions}
GRBs are highly energetic phenomena that remain without a definite explanation \cite{ghir}. Their origin is believed to be triggered by cataclysmic events linked to large changes in the internal structure of stellar compact objecs of mass $M\sim 1.5 M_\odot$ and radius $R\sim 12$ km releasing an amount of gravitational energy  $\Delta E \approx G M^2 / {R}\approx 10^{53}-10^{55}$ erg.

One of the possible scenarios where GRBs may be emitted involve NS transitioning to more compact stars. In particular, the possible formation of stars where the quark component may be deconfined out of the the nucleons has been studied in the literature, see for example \cite{alcock:86}. Such a scenario is often referred to as a quark star (QS). In a NS to QS transition, part of the outer stellar crust in the original star can be expelled at relativistic speeds leading to a transient episode of high-energy emission. In those cases the expected duration of the \g-ray signal is much smaller than that typically predicted for short GRBs (SGRBs) at about $\sim 2$ s. The mechanism behind the hipothesized transition is not yet clear but has been considered to be due either to a rise in the central density for slowly rotating old NS or due to the accretion of an exotic dark component \cite{daigne}.
This latter possibility \cite{perez:10, perez:17} links two types of matter (standard and dark) present in our Universe as experimentally determined from complementary indications \cite{bertone} and is another key Physics motivation driving the e-ASTROGAM mission.

One of the key quantities in this SGRB scenario is the isotropic equivalent energy range  $E_{\gamma,iso}\simeq 10^{48}-10^{52}$ erg \cite{nakar:07,berger:07} and the \g-ray signal peak energy expected to arise in the modellization of the (possibly beamed) transient event. This type of SGRBs can occur in any type of galaxy (and location inside) and typically with a time delay above $\sim 10^3-10^5$ yr since the end of the stellar life. As for the local rate it is expected that only a tiny fraction of about $\sim 10^{-3}$ of the SGRBs, $R_{SGRB}\sim (400-1500) \,\rm Gpc ^{-3} yr^{-1}$, is expected to be due to this deconfinement transition.
The possible detection of the associated \g-ray signal from these stellar transitions would be of major importance in our understanding not only of  stellar evolution but also of the interaction of ordinary and dark matter underlaying current particle physics models.
\paragraph*{Importance of \g-ray observations}
The expected properties of SGRBs produced in this scenario are important in the possible identification of the specific central engine that could help discriminate between the possible mechanisms of the underlying  event. The relativistic emission is due to the outer stellar crust ejection with mass $M_\mathrm{ej}=M_\mathrm{ej,-5}  \times 10^{-5}\, M_\odot$ and a width $\Delta = c \Delta t$, where $c$ is the speed of light and $\Delta t=\Delta t_{-6} \times 10^{-6}$, where $\Delta t$ is the time duration in seconds. The initial energy is given by  $E_\mathrm{ej}=f_\mathrm{ej,-3}\times 3.5\times 10^{50}\, \mathrm{erg}$. If this energy is not deposited in a homogeneous way in the expelled crust, the final Lorentz factor $\Gamma$ in the ejecta may not be uniform.  For a thermal acceleration, the saturation to the final Lorentz factor will occur at radius
$R_\mathrm{sat} \simeq \Gamma 
\times 10^7\, M_\mathrm{ej,-5} ^{-1} f_\mathrm{ej,-3}\, \mathrm{cm}$. The ejecta will become transparent to its own radiation at the photospheric radius $R_\mathrm{ph} \simeq \sqrt{\frac{\kappa M_\mathrm{ej}}{4\pi}} \simeq 2\times 10^{13}\, M_\mathrm{ej,-5}^{1/2}\, \mathrm{cm}$. Both radii are estimated based on the fireball model for GRBs \cite{Piran:04}. The internal shock dissipation will occur at a typical radius $R_{sat} \lesssim R_{is}\ll R_{ph}$. We note that an initial free expansion would be followed by a deceleration process in the external medium (two episodes of emission) at a radius $R_{dec}\gg R_{ph}$. The spectrum and duration of the signal depend on the details of the complex crust ejection. However, the duration of the prompt spike, $\Delta t_\mathrm{obs}$,  should be fixed by the intrinsic curvature of the emitting region  and its lateral expansion. This can be written $\Delta t_\mathrm{obs} \simeq \min{\left( \frac{R_\mathrm{ph}}{2\Gamma^2 c};\, \frac{\theta_\mathrm{j}^2 R_\mathrm{ph}}{2 c}\right)}
\simeq \min{\left(M_\mathrm{ej,-5}^2  f_\mathrm{ej,-3}^{-2} ;  \left(\frac{\theta_\mathrm{j}}{3^\circ}\right)^2\right)}\, \times\, 0.8\, M_\mathrm{ej,-5}^{1/2}\, \mathrm{s}$.
Except if the ejection is highly beamed with a beaming angle $\theta_j\sim 2/f^{1/2}_b$, being $f_b$ the beaming factor the minimum is usually given by the first term.
Emission in the \g-ray band require kinetic energies for the outer crust with injected fraction  $f_\mathrm{ej,-3}\sim  (1-10^2)$.
\paragraph*{Expected results with e-ASTROGAM}
e-ASTROGAM will incorporate technology \cite{jm1,jm2}
capable of detecting signals in the energy range $0.3$ MeV$-3$ GeV  as shown in Table~\ref{table:requirements}. As an example, the effective area of e-ASTROGAM at low energies will be about twice than that of SPI and 7.5 times that of COMPTEL (at 1 MeV).The time resolution is expected to be at the sub-ms  level. With this increased accuracy with respect to recent missions such as XMM-Newton or \INTEGRAL \g-ray photons arising from a prompt signal expected in collapse of a dense star are experimentally detectable. The emission of  the relativistic outer crust mass  (with Lorentz factor  $\Gamma>15$)  will allow the detection of specific prompt (sharp) signals beyond the opacity limit.
In Fig.~\ref{vsGRB:fig1} we show the visibility window for the very SGRBs based on the model of P\'erez-Garc\'ia et al \cite{daigne} with the expected performance of e-ASTROGAM. On the left axis the fraction of the energy ejected in the outer crust is shown (solid line) while on the right axis (dashed line) the duration of the expected signal is depicted (in ms) both as a function of the logarithm of the isotropic equivalent energy emitted in the astrophysical event. The duration of the signals is well below the $\sim 1$ s duration, therefore we can refer to them as very SGRBs, i.e. vSGRBs. We have considered an average beaming factor of $f_b\sim 50$. In this scenario the ejected mass $M_\mathrm{ej,-5}\lesssim 10$ for the event energy range considered.
The region above the solid line is where signal peak energies $E_p>300$ keV are thus detectable with e-ASTROGAM.
\begin{figure}
\centering
\includegraphics[width=0.67\textwidth]{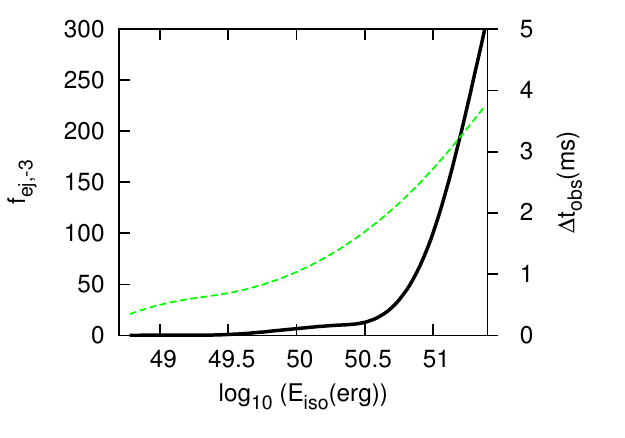}
\caption{\small{Expected e-ASTROGAM visibility window for vSGRBs. Lower energy bound for peak energy at $E_p\sim300$ keV shows the upper region limit (black solid line) where detection with e-ASTROGAM  is possible as a function of the logarithm of the isotropic equivalent energy released in the GRB. The left axis shows the fraction of the energy ejected in the outer crust, while on the right axis (dashed line) the duration of the expected signal is depicted (in ms).}}
\label{vsGRB:fig1}
\vspace{-10pt}
\end{figure}
Thanks to the high sensitivity of e-ASTROGAM, it will be possible to probe the explosion mechanism to improved levels and compare with astrophysical models for each event to better understand the outcome of the transitioning star. In addition, a possible interest for the gravitational wave community (and more generally for the multi-messenger one) is expected to better constrain new physics \cite{gw}.
These events, triggering the emission of vSGRBs, although rare in nature could allow us to discover or find hints, in an astrophysical scenario, of new phases of matter like the deconfined quark matter phase claimed to be first obtained in heavy-ion physics colliders. However, a possible {\it indirect} discovery by e-ASTROGAM seems now at hand.
\subsection[Globular clusters\\
\noindent
\textit{\small{W. Bednarek}}]{Globular clusters}
\paragraph*{Science questions}
Globular clusters (GCs), luminous concentrations of $\sim$10$^5$-10$^6$ low mass stars within the volume of a few parsecs, contain also a large number of compact objects (NS, white dwarfs) which are products of the final evolution of stars with the masses above $\sim$0.8 M$_\odot$. Several GCs have been recently detected by \fermilat at GeV \g-rays \cite{Abdo2009_2,abd10, kon10,tam11}.
The \g-ray emission at TeV energies has been searched with the current Cherenkov telescopes but only detected from the GC Ter 5 \cite{abr11}.
Ter 5 also emits non-thermal diffusive radiation in the $1-7$ keV  energy range \cite{ege10,cla11}.
The origin of the non-thermal X-ray and \g-ray emission is not clear at present.

The GeV \g-ray emission is usually interpreted as a cumulative emission produced in the inner MSP magnetospheres \cite{ven08,ven09}. This scenario is supported by the detection of \g-ray pulsations from two MSPs within GCs, i.e. B1821-24 \cite{fre11} and J1823-3021A \cite{joh13}.
The GeV (and TeV) emission might also originate in the \ic Scattering process of the $e^\pm$ pairs which are injected from the MSP magnetospheres into a dense low energy radiation field present within (and around) GCs \cite{bs07,kop13,che10}.
It is argued that MSPs within GCs can significantly differ from those observed in the Galactic field
 \cite{che10}. They are expected to be frequently captured by the low mass stars in GCs. As a result, their inner magnetic field could have different structure favoring production of a low energy $e^\pm$ plasma.
The $e^\pm$ pairs from MSPs have to pass through a dense radiation field from the GCs (and also from the nearby Galactic disk and the Microwave Background Radiation) producing \g-rays and possibly also diffusive synchrotron radiation \cite{che10}. Their radiation might contribute to the observed \fermilat \g-ray emission. This process can also produce additional emission components at lower energies due to the comptonization of the infrared or the MBR. In fact, in some cases the \g-ray spectra do not show the  characteristic exponential cut-off at a few GeV typical of the MSPs, arguing against the origin within MSP magnetospheres \cite{abd10}.

Some MSPs within GCs are expected to be in an ejector/accretor transition state. A few such systems have been recently discovered. They show enhanced GeV \g-ray emission in the accretor state with respect to that observed in the stationary ejector phase of the MSPs (e.g. PSR J1023+0038 \cite{archibald09})).
Also other high energy components might appear in the hard X-ray
to \g-ray spectrum due to the interaction of the accretion flow with the rotating pulsar magnetosphere
as observed  for example in accreting X-ray binary systems containing NS.
\paragraph*{Importance of \g-ray observations}
Observations of GCs in the hard X-ray to GeV \g-ray energy range (e.g. Ter 5) should allow us to determine the extension of the diffusive, non-thermal X-ray spectrum to energies beyond  those detected by \textit{Chandra}.
Discovery of the hard X-ray emission will provide constraints on:
the magnetic field within the specific GC, the parameters of $e^\pm$ pair plasma injected by the MSPs
(injection rate, maximum energies), the features of relativistic electrons accelerated in the collisions of the MSP winds between themselves or with the winds from the companions stars. The constraints on the injection rate of $e^\pm$ plasma from MSPs will allow us to constrain the models for the non-thermal processes in the inner magnetospheres of the MSPs within GCs.

The detailed studies of the GeV \g-ray morphology of the GCs will allow us to identify the nature of the
discrete sources (ejecting, accreting, transitional MSPs?) or identify processes responsible for this emission.
Discovery of the pulsed \g-ray emission from the many radio MSPs within GCs will support the hypothesis that the observed GeV \g-ray emission originates in this type of compact objects.

The discovery of a new hard X-ray and soft \g-ray features in the non-thermal spectra of GCs will argue for the importance of various radiation processes (or soft radiation fields) as predicted by the IC scattering
model proposed in e.g. \cite{che10}.

Finally, some of the $e^\pm$ pairs from the MSPs can be thermalized in the atmospheres of the companion stars and/or their winds. These $e^\pm$ pairs could annihilate producing a narrow $\sim$0.5 MeV line the intensity of which would allow us to put independent constraints on the $e^\pm$ pair injection rate. 
\paragraph*{Expected results with e-ASTROGAM}
Possible extension of the diffusive synchrotron X-ray emission from Ter 5, observed by
{\it Chandra} in the energy range 1-7 keV \cite{ege10}, to a few hundred keV should be detectable by e-ASTROGAM (see Fig.~\ref{fig:sensitivity}), allowing us to constrain the injection rate of $e^\pm$ pairs by the MSPs and indirectly the MSP models.

A factor of a few better localization of the GeV \g-ray source by e-ASTROGAM (see Fig.~\ref{fig:perf})
should allow us to conclude on the morphology of the emission region, within and/or around the GC, and answer the question whether this emission is related to the distribution of the MSPs within the GC or it has a diffusive nature.

The precise time accuracy of the \g-ray events by the e-ASTROGAM telescope (see Table~\ref{table:requirements} of Chapter~\ref{intro}) will allow to measure the light curves of the MSPs within the GCs. Thus, the fraction
of the GeV \g-ray emission from GCs, which is undoubtedly linked to the MSPs, could be determined.

The e-ASTROGAM mission will have enough sensitivity to detect possible additional components in the \g-ray spectrum due to the Comptonization of different soft radiation fields, such as the infrared emission from the Galactic disk and the microwave background radiation, by a relatively low energy $e^\pm$ pairs (see prediction in Fig.~3 in \cite{che10}).

Finally, the improved sensitivity of e-ASTROGAM (see Table~\ref{table:requirements}) should allow to search for the $e^\pm$ the annihilation feature. It might be produced within a large number of the compact MSP binary systems confined within GCs.  Discovery of such annihilation line should independently constrain the injection rate
of the $e^\pm$ pair plasma from the MSPs within GCs.

\newpage
\section[Solar and Earth science\\
\noindent
\textnormal{\small\textnormal{Convenor:} \textit{F. Longo}}
]{Solar and Earth science}
The same \g-ray emission mechanisms at play in celestial sources can be studied in more detail, even if in different environmental conditions, in local \g-ray sources such as those present in the Solar System. In particular the interactions of CRs with radiation fields and matter, at the Sun and with other Solar System solar bodies, such as the Moon, the acceleration of particles and their emission in the upper atmosphere, the physics of magnetic reconnection and particle acceleration in solar flares are examples of science objectives that e-ASTROGAM will explore by observing \g-rays coming from the Sun, the Moon, the Earth and other bodies in the Solar System.  

\begin{itemize}
\item TGFs are very intense \g-ray emission episodes coming from the upper atmosphere and strongly correlated with lightning activity. They are generally interpreted as \brem high-energy radiation emitted by free electrons in the air, accelerated to relativistic energies by intense electric fields presents in the atmosphere under thunderstorm conditions. The importance of \g-ray observations from space satellites 
flying in Low Earth equatorial orbit, such as e-ASTROGAM, is based on the possibility of detecting TGFs in the tropical regions where the frequency of thunderstorms is higher.  The e-ASTROGAM calorimeter is best suited to monitor all the spectrum of TGF energies allowing a in-depth study of the the atmospheric processes linked to TGFs. e-ASTROGAM will also confirm the possible presence of a high-energy population of TGFs emitting at energies greater than 40 MeV. With its data acquisition system e-ASTROGAM will finally reveal a large number of events, about more than 1000 TGFs per year, with the possibility of multiple sampling of the same thunderstorm regions, providing a huge data set for both the high-energy and the meteorological scientific communities. This topic will be described better in Sec. \ref{Ursi}. 

\item The Moon is one of the brightest sources of high-energy \g-rays in the Solar System. Gamma-rays from the Moon originate in the shower cascades produced by the interactions of Galactic CR nuclei with the lunar surface. The lunar \g-ray emission depends on the fluxes of the primary cosmic-ray nuclei impinging on the Moon and on the mechanisms of their hadronic interactions with the rock composing the lunar surface. In addition to providing a new accurate measurement of the lunar \g-ray spectrum in the MeV-GeV band, e-ASTROGAM data will extend the energy range observed by previous missions towards lower energies. This feature will
provide the unique opportunity to explore possible \g-ray lines in the keV-MeV region, originating from the decays of excited states produced in the interactions of CR nuclei with the lunar rock. Measurements of the \g-ray flux from the Moon also provide a useful tool to study the properties of CRs and
to monitor the solar cycle, since it depends on the primary CR nuclei fluxes, which change with the solar activity.  The lunar \g-ray data at low energies will also represent a powerful tool to monitor the solar modulation and to study the CR spectra impinging on the Moon surface.  This topic will be described better in Sec. \ref{Loparco}, \ref{Loparco2}.

\item The Sun is a known quiescent \g-ray source. Its steady-state \g-ray emission is due to two distinct emissions processes.
A solar disc emission is originated by (i) hadrons interacting with the solar atmosphere and (ii) a spatially extended contribution which comes from \ic interactions of CR electrons on the heliosphere radiation. The observation of both components will allow us a deeper investigation on CRs traveling close to the Sun and provide information on CR propagation in the heliosphere.
close to the Sun and therefore on CR propagation in the heliosphere. CRs in the heliosphere are affected by the solar wind and the magnetic field, which change their spectrum at energies below few tens of GeV/n. The strength of this effect depends on the solar activity. During solar maxima, the solar modulation of CRs is the largest, while during solar minima is the lowest. Being produced by CRs, both \g-ray emission components of the Sun vary as a function of the solar activity. The e-ASTROGAM Point Spread Function will allow us to investigate better the spatial dependence of these two components.  This will enable us to study CR transport in the inner heliosphere, to improve on the models of the solar modulation and the models of CR cascades in the solar atmosphere. This will allow us to trace for the first time the LECR electrons close to the Sun shedding light of the effects of solar modulation in the inner heliosphere. This topic will be described better in Sec. \ref{Orlando}.

\item Solar flares are the most energetic phenomena in the Solar System. These events are often associated with explosive Coronal Mass Ejections (CMEs). The frequency of both flares and CMEs follows the 11-year solar activity cycle, the most intense ones usually occurring during the maximum. What triggers the 
flares is presently not completely understood. Flare energy may be considered to result from reconnecting magnetic fields in the corona. Phenomena similar to solar flares and CMEs are believed to occur at larger scales elsewhere in the Universe. These energetic phenomena from the Sun are therefore the most accessible laboratories for the study of the fundamental physics of transient energy release and efficient particle acceleration in cosmic magnetized plasmas. The \g-ray emission from Solar Flares results from the acceleration of charged particles which then interact with the ambient solar matter in the regions near the footpoints of magnetic field lines. Accelerated electrons mainly produce soft and hard X-rays via non-thermal \brem.   Accelerated protons and ions emit at higher energies: nuclear interactions produce excited and radioactive nuclei, neutrons and pi-mesons. All of these products subsequently are responsible for the \g-ray emission via secondary processes, consisting of nuclear \g-ray lines in the 1-10 MeV range and a continuum spectrum above 100 MeV. The high-energy \g-ray emission light curve can be similar to the one observed in X-rays, lasting for 10-100 s and indicating the acceleration of both ions and electrons from the same solar ambient. This is referred to as "impulsive" phase of the flare. However, some events have been found to have a long-duration \g-ray emission, lasting for several hours after the impulsive phase.  e-ASTROGAM will study the solar flare radiation from 300 keV to 3 GeV, covering a very broad energy range. e-ASTROGAM will then have the opportunity to detect solar flares and to study the evolution in time of the hard-X and \g-radiation from each event, helping in constraining models of acceleration and propagation. e-ASTROGAM will have optimal sensitivity and energy resolution to detect the de-excitations lines from accelerated ions. This will be fundamental to gain insight into the chemical abundances and about the physical conditions where accelerated ions propagate and interact. At higher energies, the spectral analysis performed by e-ASTROGAM will allow disentangling electron \brem and the pion-decay components. A polarised \brem emission in hard X-ray from solar flares is expected if the phase-space distribution of the emitting electrons is anisotropic with important implications for particle acceleration models. e-ASTROGAM's very good angular resolution will localize the source on the solar disk and possible comparisons with location studies in X-rays could give additional information for constraining the emission and acceleration mechanisms. This topic will be described better in Sec. \ref{bissaldi}. 
\end{itemize}

\subsection[Earth: detection of Terrestrial Gamma-ray Flashes\\
\noindent
\textit{\small{A.~Ursi, M.~Tavani, M.~Marisaldi, F.~Fuschino, C.~Labanti, S.~C\'{e}lestin,\\ S.~Dietrich}}
]{Earth: detection of Terrestrial Gamma-ray Flashes}\label{Ursi}
\paragraph*{Science questions}
Terrestrial Gamma-ray Flashes (TGFs) are brief (tens~of~$\mu$s~--~few ms) and intense \g-ray (hundreds of keV~--~tens of MeV) emissions coming from the terrestrial atmosphere ($\sim12-15$~km~a.s.l.), strictly correlated with lightning activity representing the highest-energy natural phenomenon observed on Earth. Representing a crossover between atmospheric physics and high-energy astrophysics, TGFs constitute a really attractive challenge for both scientific fields.
\begin{figure}
 \centering
 \subfigure{\includegraphics[width=14cm]{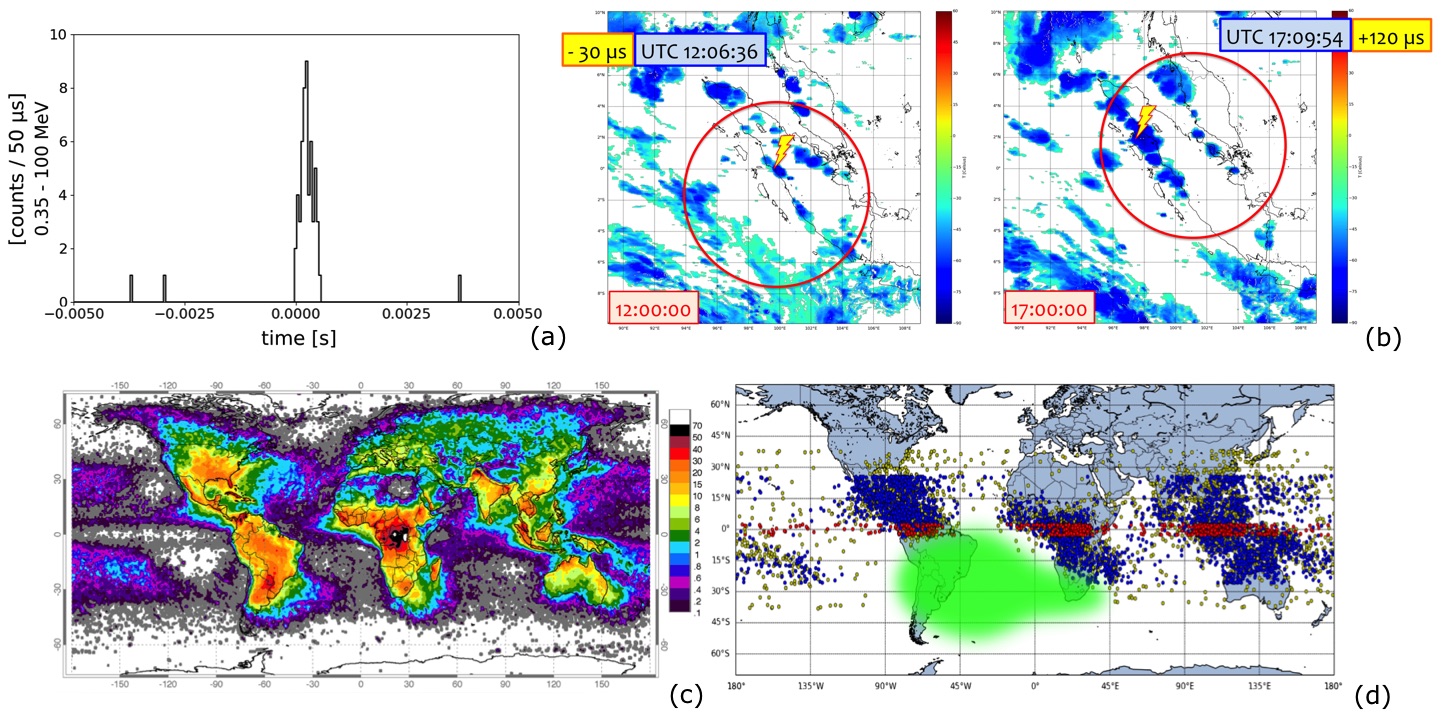}}
 \caption{(a) Light curve of a TGF detected by the \agile MCAL. (b) The annualized distribution of total lightning activity (flashes / km$^2$ / yr), detected by the Lightning Imaging Sensor (LIS) \cite{Christian2003}. (c) World distribution of TGFs detected by RHESSI (yellow), \Fermi (blue) and \agile (red). The green region represents the South Atlantic Anomaly (SAA). (d) Example of two multiple TGFs, with associated WWLLN sferics, detected by the \agile satellite at successive overpasses (within $\sim3$~hours), produced by the same developing storm \cite{Ursi2016}.}
 \label{world}
\end{figure}
The most accepted hypothesis besides their production suggests the upper part of Earth's troposphere behaves as a particle accelerator, under thunderstorm conditions: free electrons in the air, accelerated to relativistic energies by intense electric fields, may produce hard X- and \g-rays via \brem processes on atoms and nuclei in the atmosphere \cite{Gurevich1992, Dwyer2008,Dwyer2012}. Nevertheless, TGFs have also drawn interest, as the significant radiation dose they emit, together with the height at which they occur, have been pointed out as potentially hazardous for aircraft and possible sources of injuries for airlines crews and onboard electronics \cite{Tavani2013,Dwyer2013}.
\paragraph*{Importance of \g-ray observations}
TGFs take place at thundercloud tops and, despite being a completely terrestrial phenomenon, most of the studies about this phenomenon have been carried out using high-energy astrophysics satellites. After the serendipitous discovery in the early 90's by the BATSE experiment \cite{Fishman1994}, wide contributions to their phenomenology have been brought by the NASA Reuven Ramaty High-Energy Solar Spectroscopic Imager (RHESSI) \cite{Smith2005,Grefenstette2009}, \agile \cite{Marisaldi2010a} and \Fermi \cite{Briggs2010}. Moreover, TGFs have also been found within the BeppoSAX (1996-2002) data \cite{Ursi2017b} and have been detected by aircraft \cite{Smith2011} and at ground level both in correlation with natural \cite{Tran2015} and triggered lightning \cite{Dwyer2004,Dwyer2012,Hare2016}.

To date, a wide database including thousands of TGFs for more than 10 years activity is provided by the RHESSI, \agile and \Fermi data \cite{Briggs2010,Grefenstette2009,Marisaldi2010a,Marisaldi2014}. In particular, the \agile satellite produced interesting breakthroughs in the field of TGF science by performing the first imaging of a TGF event exploiting the onboard silicon tracker \cite{Marisaldi2010b}, by investigating the spectrum tail at the highest energies ($>40$~MeV) \cite{Tavani2011a} and by detecting multiple TGFs produced by the same thunderstorm systems taking advantage of meteorological data from geostationary satellites \cite{Ursi2016} (Fig.~\ref{world}(d)).
\paragraph*{Expected results with e-ASTROGAM}
Taking into consideration the heritage provided by the previous TGF-detecting satellites, especially \agile, key points of a suitable TGF detector are represented by a wide energy range, by a high time resolution of the onboard trigger logic timescales (with the possibility of acquiring data in a photon-by-photon mode), as well as by a joint working mode with other onboard instruments (such as a \g-ray imager). Moreover, having a nearly-equatorial satellite orbit plays an important role, in ensuring the monitoring of the tropical regions where most of the lightning activity takes place. Basic contributions and expected results of the e-ASTROGAM mission for what concerns the science of TGFs are listed below.

\begin{itemize}

 \item The strongest point of e-ASTROGAM is the calorimeter, which provides \g-ray data in an energy range (30 keV~--~200 MeV) fully including the typical TGF energies and, in particular, the Compton range (0.3~--~15 MeV) of atmospheric processes linked to TGFs. Such an energy range allows the investigation of the \textbf{TGF high-energy spectral component}, in order to shed light on the nature of the high-energy tail of the TGF spectrum discovered by \agile and the existence of a possible higher-energy TGF population.
 
 \item The calorimeter acquires data in a photon-by-photon mode for triggered events, with a time resolution of 2~$\mu$s (at $3\sigma$): this way, the time and energy binning is limited just by a statistical factor. Moreover, the presence of a \textbf{sub-millisecond trigger logic timescale}, just as for the \agile MCAL, plays a leading role in the detection of very brief events such as TGFs, allowing for revealing a large number of events. Considering the current missions, the e-ASTROGAM calorimeter is expected to detect about $>1000$~TGFs/y, providing a wide database that can be used for scientific purposes.

 \item The calorimeter is an all-sky detector with no imaging capabilities and it is therefore capable of detecting events from every direction, regardless the satellite pointing. Nevertheless, the calorimeter instrument can work alone in a so-called burst mode, or together with the onboard silicon tracker, as a \g-ray imager, in the 0.3~MeV~--3~GeV energy range. This allows to perform \textbf{imaging of TGFs}, reconstructing the incoming direction and geographic position of the TGF source and constraining the \g-ray emission cone.

 \item The e-ASTROGAM satellite will be delivered into \textbf{a near-equatorial orbit} ($\sim2.5^\circ$) that not only guarantees a low and stable charged particle background for the onboard instruments, but also allows for the monitoring of geographic regions with the highest lightning activity on Earth. This strongly increases the chance of detecting TGFs and of revealing multiple TGFs during the same passage and throughout successive overpasses over the same region, providing interesting data for the study of the storm evolution, the associated climatological scenario, and the capability of single storms to produce several TGFs and hence, allowing to refine the production models.

\end{itemize}
\subsection[Gamma-ray spectum of the Moon\\
\noindent
\textit{\small{F. Loparco, M. N. Mazziotta}}
]{Gamma-ray spectum of the Moon}\label{Loparco}
\paragraph*{Science questions}
The Moon is one of the brightest sources of high-energy \g-rays in the Solar System. Gamma-rays from the Moon originate in the shower cascades produced by the interactions of Galactic cosmic-ray (CR) nuclei with
the lunar surface~\cite{morris,egret}. The lunar \g-ray emission depends on the fluxes of the primary cosmic-ray nuclei impinging
on the Moon and on the mechanisms of their hadronic interactions with the rock composing
the lunar surface. The \g-ray energy spectrum of the Moon extends in the energy interval from a few MeV up to a few GeV
and it is well understood, thus making the Moon a useful ``standard candle'' for the calibration of \g-ray telescopes~\cite{abdo2012,ackerman2016}.

As mentioned above, \g-rays emitted from the Moon
are produced after inelastic interactions of charged CRs with the
lunar surface. Assuming
that the CR flux on the lunar surface is spatially isotropic and
indicating with $I_{i}(T)$ the intensity of CRs of the $i$-th species
(in units of $\units{particles~MeV^{-1}~cm^{-2}~sr^{-1}~s^{-1}}$) as a
function of kinetic energy $T$, the rate $\Gamma_{i}(T)$ of CRs of the $i$-th species
(in units of $\units{particles~MeV^{-1}~s^{-1}}$) impinging on the
lunar surface is given by:

\begin{figure}
\begin{center}
\vcenteredhbox{\includegraphics[width=0.33\columnwidth,height=0.21\textheight,clip]{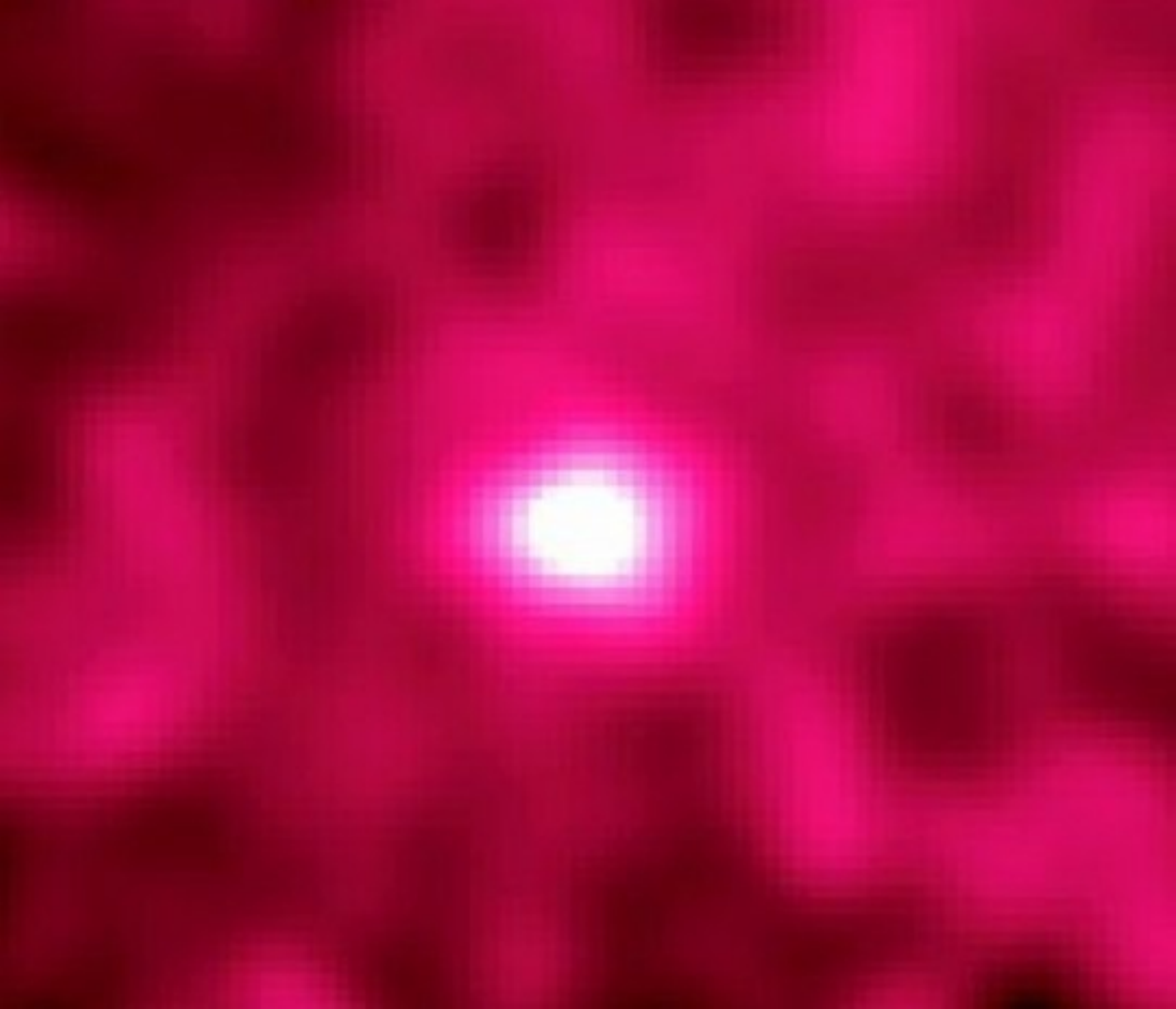}}
\vcenteredhbox{\includegraphics[width=0.65\columnwidth,height=0.3\textheight,clip]{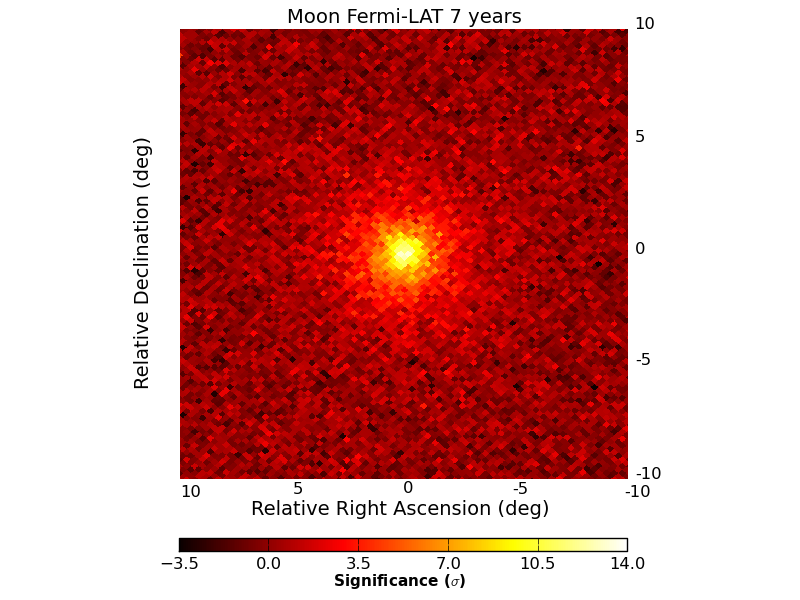}}
\caption{\small Images of the Moon seen by EGRET~\cite{egret} (left) and by the \fermilat~\cite{ackerman2016} (right).
The EGRET plot covers a field of view of roughly $40^\circ$.}
\label{Moon_spec:fig1}
\end{center}
\end{figure}

\begin{equation}
\Gamma_{i}(T) = 4 \pi R_{\leftmoon}^{2} I_{i}(T) \int \cos \theta_{M} d\Omega_{M} =
4 \pi^{2} R_{\leftmoon}^{2} I_{i}(T)
\label{eq:primrate}
\end{equation}
where $R_{\leftmoon}=1737.1\units{km}$ is the radius of the Moon.
In the previous equation we set $d\Omega_{M} = d \cos \theta_{M} d\phi_{M}$,
where $(\theta_{M}, \phi_{M})$ are the zenith and azimuth
angles of CR particles with respect to the lunar surface
($0 < \cos \theta_{M} < 1$ and $0 < \phi_{M} < 2 \pi$).

The differential \g-ray luminosity of the Moon $L_{\gamma}(E_{\gamma})$
(in units of $\units{photons~MeV^{-1}~s^{-1}}$) is given by:
\begin{equation}
L_{\gamma}(E_{\gamma}) = \sum_{i} \int Y_{i}(E_{\gamma} | T) \Gamma_{i}(T) dT 
= 4 \pi^{2} R_{\leftmoon}^{2} \sum_{i} \int Y_{i}(E_{\gamma} | T) I_{i}(T)~dT
\label{eq:luminosity}
\end{equation}
where $Y_{i}(E_{\gamma}|T)$ is the differential gamma-ray yield
(in units of $\units{photons~particle^{-1}~MeV^{-1}}$), i.e. the
number of photons per unit energy produced by a primary particle of the $i$-th species.
The yields $Y_{i}(E_{\gamma}|T)$ depend on the mechanisms of the interactions
of primary CRs with the lunar surface (regolith) and on its composition.

The differential intensity of gamma rays (in units of
$\units{photons~MeV^{-1}~cm^{-2}~sr^{-1}~s^{-1}}$) emitted
from the Moon can be evaluated starting from the differential luminosity and
is given by:

\begin{equation}
I_{\gamma} (E_{\gamma}) = \cfrac{L_{\gamma}(E_{\gamma})}{4\pi^{2} R_{\leftmoon}^{2}}
= \sum_{i} \int Y_{i}(E_{\gamma} | T) I_{i}(T)~dT  
\label{eq:gammaintensity}
\end{equation}
The \g-ray flux observed by a detector at Earth
(in units of $\units{photons~MeV^{-1}~cm^{-2}~s^{-1}}$)
can also be evaluated from the differential luminosity and is given by:
\begin{equation}
\label{eq:moonflux}
\phi_{\gamma}(E_{\gamma}) = \cfrac{L_{\gamma}(E_{\gamma})}{4\pi d^{2}} 
= \cfrac{\pi R_{\leftmoon}^{2}}{d^{2}} I_{\gamma}(E_{\gamma}) 
= \cfrac{\pi R_{\leftmoon}^{2}}{d^{2}} \sum_{i} \int Y_{i}(E_{\gamma} | T) I_{i}(T)~dT  
\end{equation}
where $d$ is the distance between the center of the Moon and the detector.
\begin{figure}
\begin{center}
\includegraphics[width=0.6\columnwidth,height=0.3\textheight,clip]{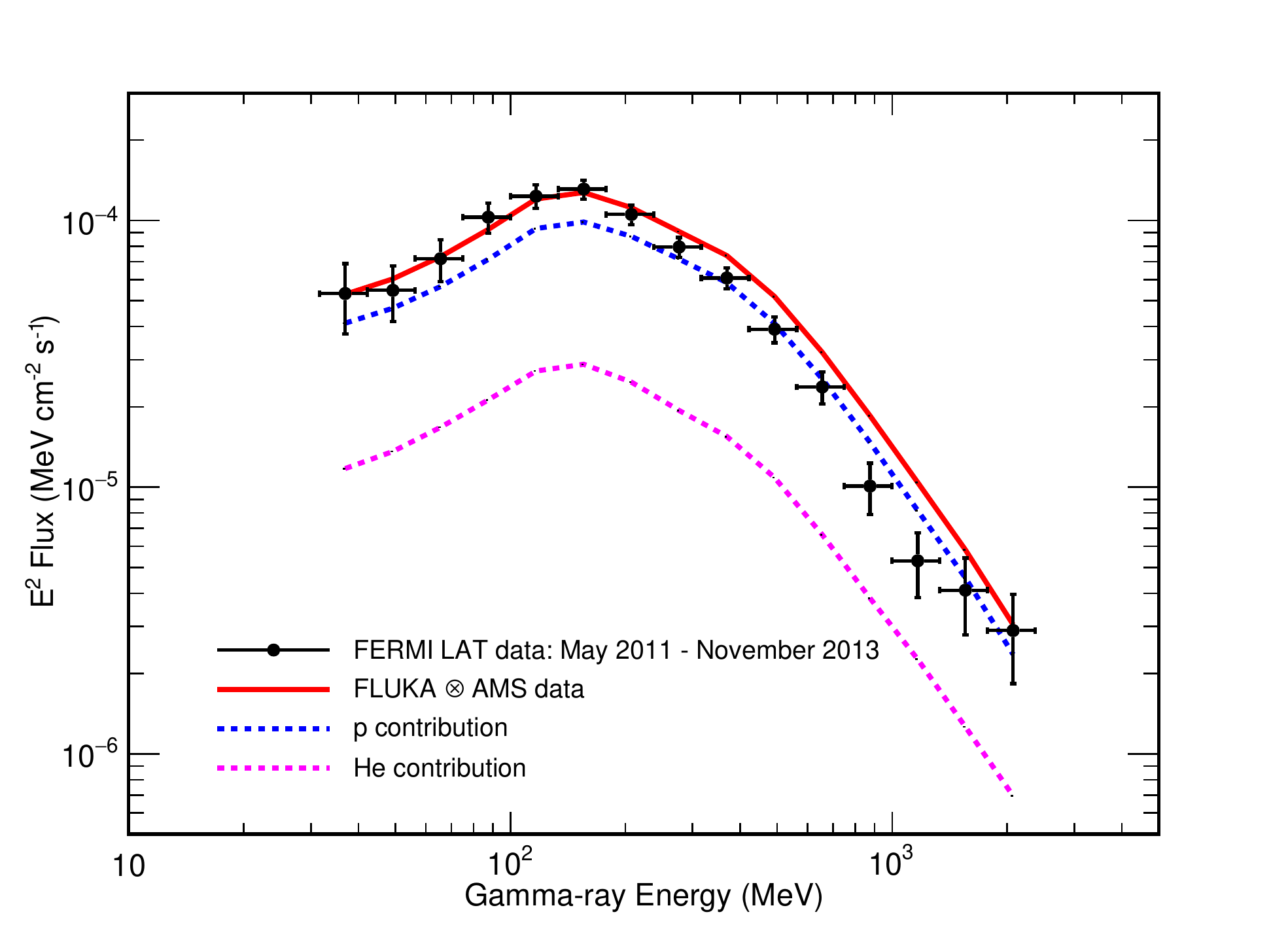}
\end{center}
 \caption{\small Comparison of the lunar \g-ray flux measured by the \fermilat in the period from May 2011 to November 2013~\cite{ackerman2016}
with the predictions obtained by folding the fluxes of cosmic-ray protons and helium nuclei measured by AMS-02~\cite{ams02,amshe}.}
\label{Moon_spec:fig3}
\end{figure}
\paragraph*{Importance of \g-ray observations}
The emission of high-energy \g-rays from the Moon was first observed by the EGRET experiment~\cite{egret}, which operated from 1991 to
2000 onboard the Compton Gamma Ray Observatory. Recently, \fermilat has performed further
measurements~\cite{abdo2012,ackerman2016} of the lunar \g-ray emission, extending the energy range down to 30 MeV with an
improved energy resolution with respect to its predecessor. Fig.~\ref{Moon_spec:fig1} shows the images of the Moon seen by EGRET during eight
exposures in the period 1991-1994 and by the LAT during its first seven years of operation~\cite{ackerman2016}. In addition to the extension of the energy
range with respect to its predecessor, the LAT can also observe the Moon with a better angular resolution.
The direct measurements of the primary proton and helium spectra performed by the AMS-02 experiment~\cite{ams02,amshe} have allowed the \fermilat 
Collaboration to validate their model describing the cosmic-ray interactions with the Moon. Fig.~\ref{Moon_spec:fig3} shows a comparison of the lunar
\g-ray flux measured by the LAT in the same period when AMS-02 performed its measurements of the proton and helium spectra
(May 2011 - November 2013) with the predictions obtained with a full simulation of the interactions of primary CRs with
the lunar surface based on the {\tt FLUKA} code~\cite{battistoni2007,ferrari2005,flukaweb}. The Monte Carlo predictions shown in Fig.~\ref{Moon_spec:fig3}
have been obtained by folding the proton
and helium fluxes measured by AMS-02 with the \g-ray yields predicted by the simulation~\cite{ackerman2016}.
\paragraph*{Expected results with e-ASTROGAM}
The energy range of e-ASTROGAM will cover the whole \g-ray spectrum emitted by the Moon.
In addition to providing a new accurate measurement of the lunar \g-ray spectrum
in the MeV-GeV band, e-ASTROGAM data will extend the energy range observed by the \fermilat
towards lower energies. This feature will provide the unique opportunity to explore
possible \g-ray lines in the keV-MeV region, originating from the decays of excited states
produced in the interactions of CR nuclei with the lunar rock. Moreover, thanks to the better PSF, e-ASTROGAM
will be able to resolve the \g-ray emission from the lunar disk.
\subsection[Cosmic ray studies with the \g-ray emission from the Moon\\
\noindent
\textit{\small{F. Gargano,  F. Loparco, M. N. Mazziotta}}
]{Cosmic ray studies with the \g-ray emission from the Moon}\label{Loparco2}
\paragraph*{Science questions}
The lunar \g-ray emission originates from the hadronic interactions of high-energy CR nuclei with the rock composing
the lunar surface.
Measurements of the \g-ray flux from the Moon also provide a useful tool to study the properties of CRs and to monitor the solar cycle, since it depends
on the primary cosmic-ray nuclei fluxes, which change during the solar cycle.
\paragraph*{Importance of \g-ray observations}
The \fermilat has monitored the time evolution of the lunar \g-ray emission on a 7-year timescale, detecting the expected correlation
with the solar cycle. The left panel of Fig.~\ref{Moon_CR:fig2} shows the time evolution of the \g-ray intensity from the Moon measured by the
LAT above 56, 75, 100 and 178 MeV~\cite{ackerman2016}.
As expected, the \g-ray intensity from the Moon follows the evolution of the solar cycle. This feature is confirmed when looking at the
correlations between the lunar \g-ray intensity and the data from the neutron monitor stations installed in various locations on the Earth.
As an example, in the right panel of Fig.~\ref{Moon_CR:fig2} it is shown a comparison of the lunar \g-ray intensity measured by the LAT with the count rates of the McMurdo
neutron monitor~\cite{bartol}. Furthermore, as the \g-ray threshold energy is increased, the correlation with the solar cycle becomes weaker, as \g-
rays of higher energies are produced by more energetic CRs, which are not affected by the solar modulation.

\begin{figure}[!b]
\begin{center}
\includegraphics[width=0.49\columnwidth,height=0.25\textheight,clip]{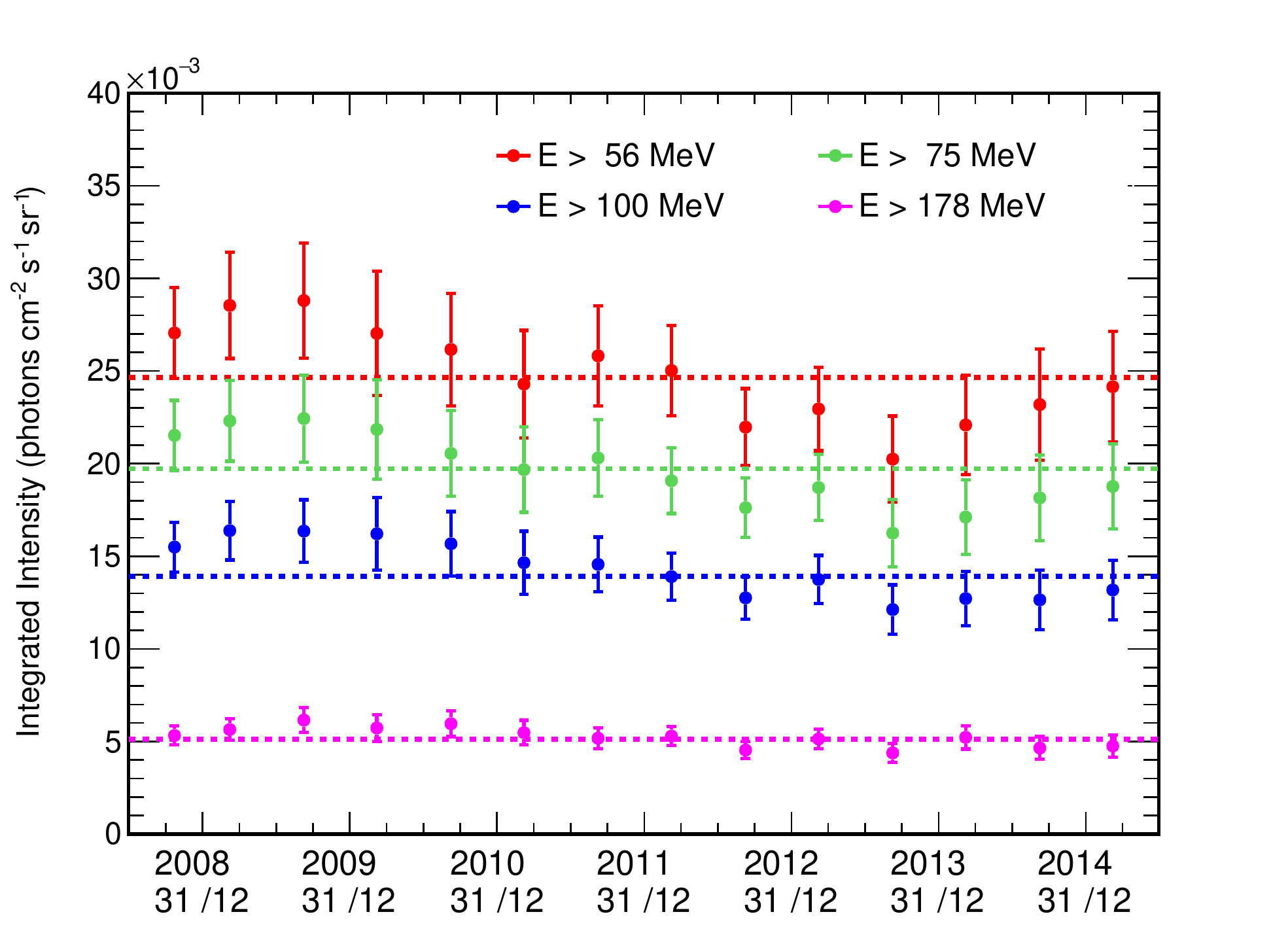}
\includegraphics[width=0.49\columnwidth,height=0.25\textheight,clip]{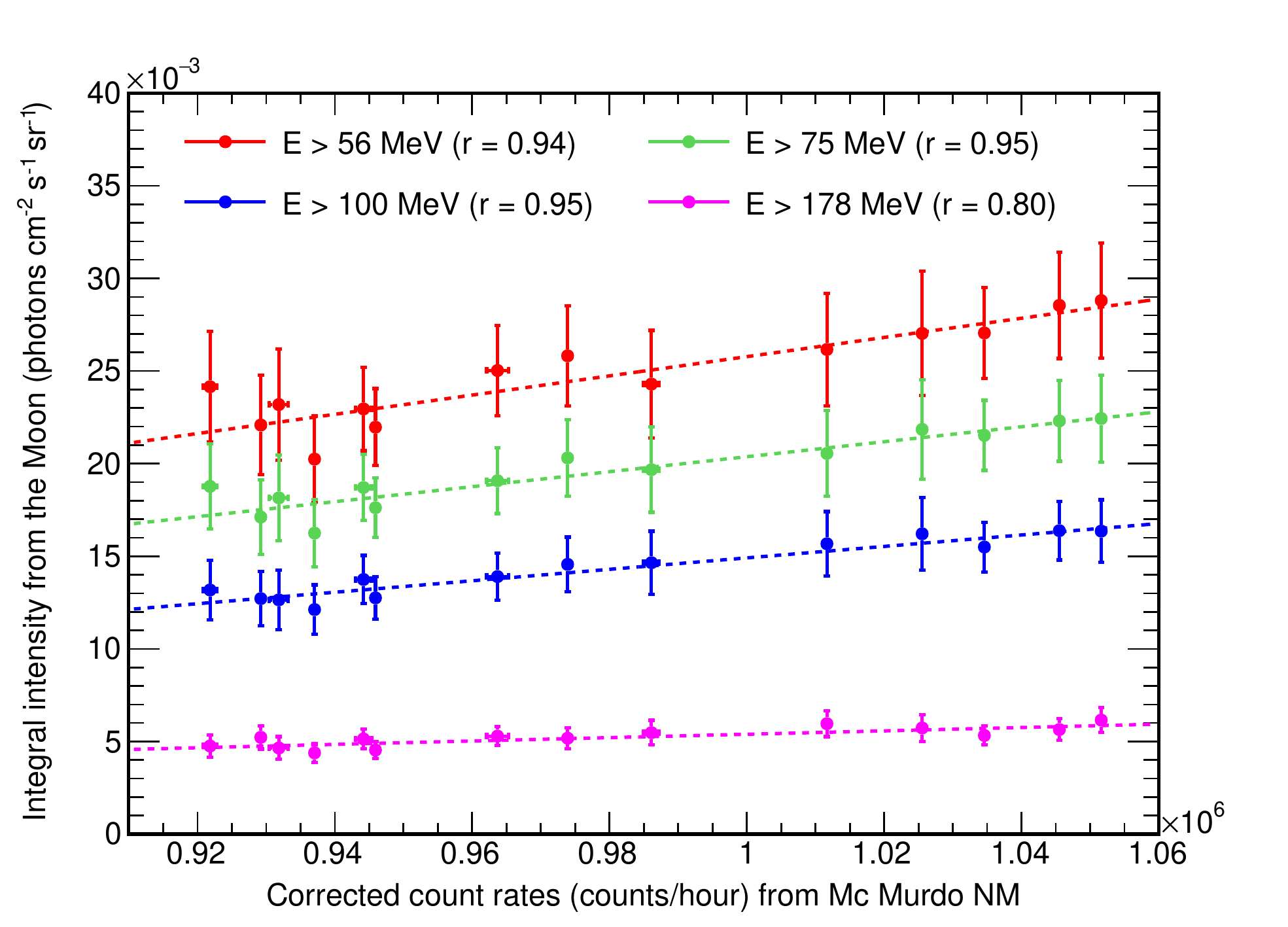}
\end{center}
 \caption{\small Left: time evolution of the lunar \g-ray intensity above 56, 75, 100 and 178 MeV~\cite{ackerman2016}.
Right: study of the correlations
between the lunar \g-ray intensity~\cite{ackerman2016} and the count rates of the McMurdo neutron monitor~\cite{bartol}.
The values reported in brackets are the correlation coefficients.}
\label{Moon_CR:fig2}
\end{figure}

The \fermilat Collaboration has developed a full simulation of the interactions of CR nuclei
with the lunar surface based on the {\tt FLUKA} code~\cite{battistoni2007,ferrari2005,flukaweb}.
Starting from a model for the CR proton and $^{4}$He local interstellar spectra (LIS)~\cite{noipaper}
evaluated using a customized version of the CR propagation code
{\tt DRAGON}~\cite{evoli,gaggero}, the simulation has been used to derive 
derive the solar modulation potential in the framework of the force field approximation from the lunar gamma-ray spectrum
~\cite{gleeson1, gleeson2} and consequently the intensities of
CR protons and $^{4}$He nuclei impinging on the Moon.
The proton and $^{4}$He spectra are shown in the left panel of Fig.~\ref{Moon_CR:fig4},
where they are also compared with the data from direct measurements
performed by AMS-02~\cite{ams02,amshe}  and PAMELA~\cite{pamela} in different epochs. This procedure has also allowed
the \fermilat Collaboration to study the time evolution of the solar modulation potential, as shown in
the right panel of Fig.~\ref{Moon_CR:fig4}.
\begin{figure}[t]
\begin{center}
\includegraphics[width=0.49\columnwidth,height=0.24\textheight,clip]{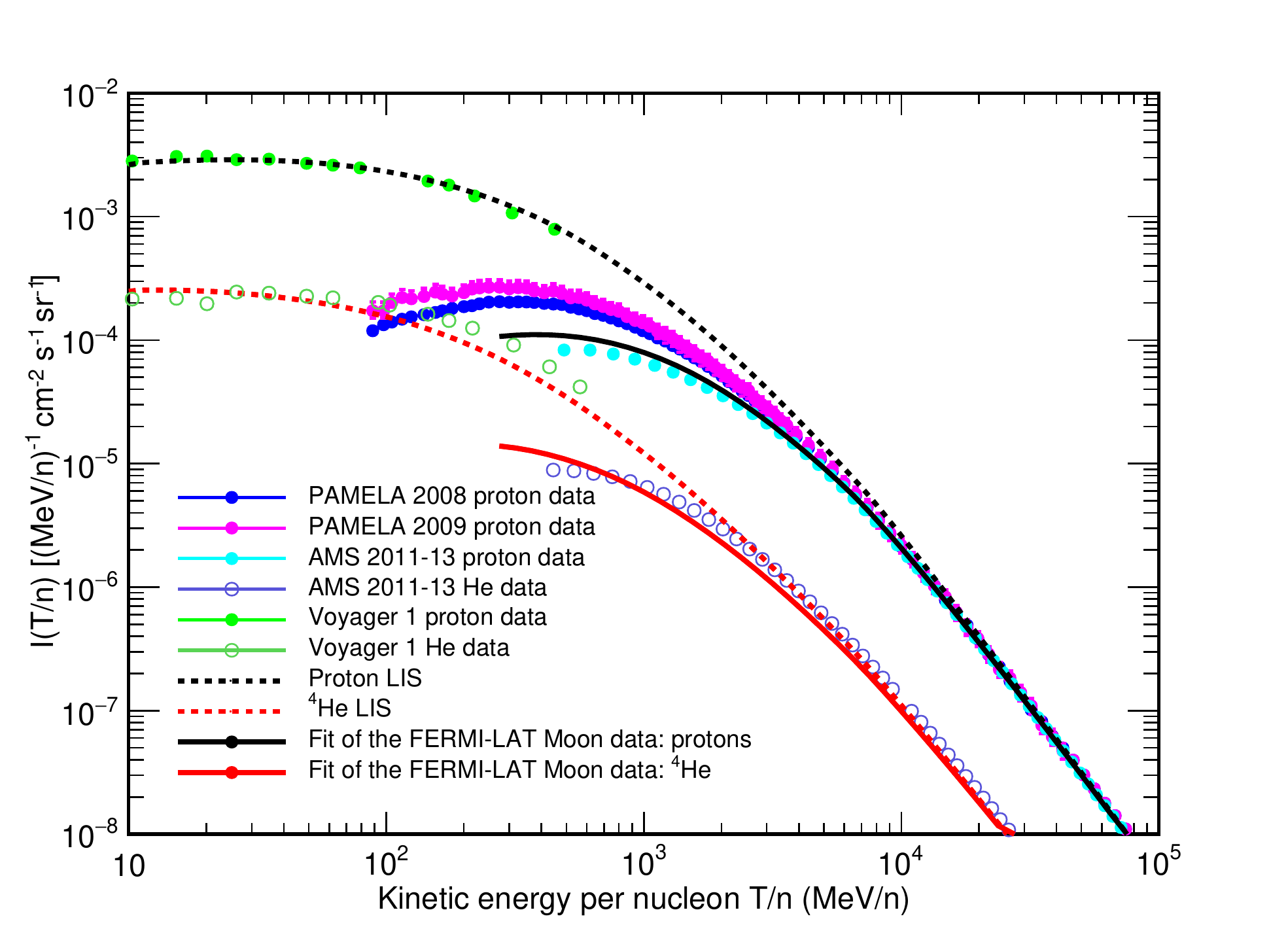}
\includegraphics[width=0.49\columnwidth,height=0.24\textheight,clip]{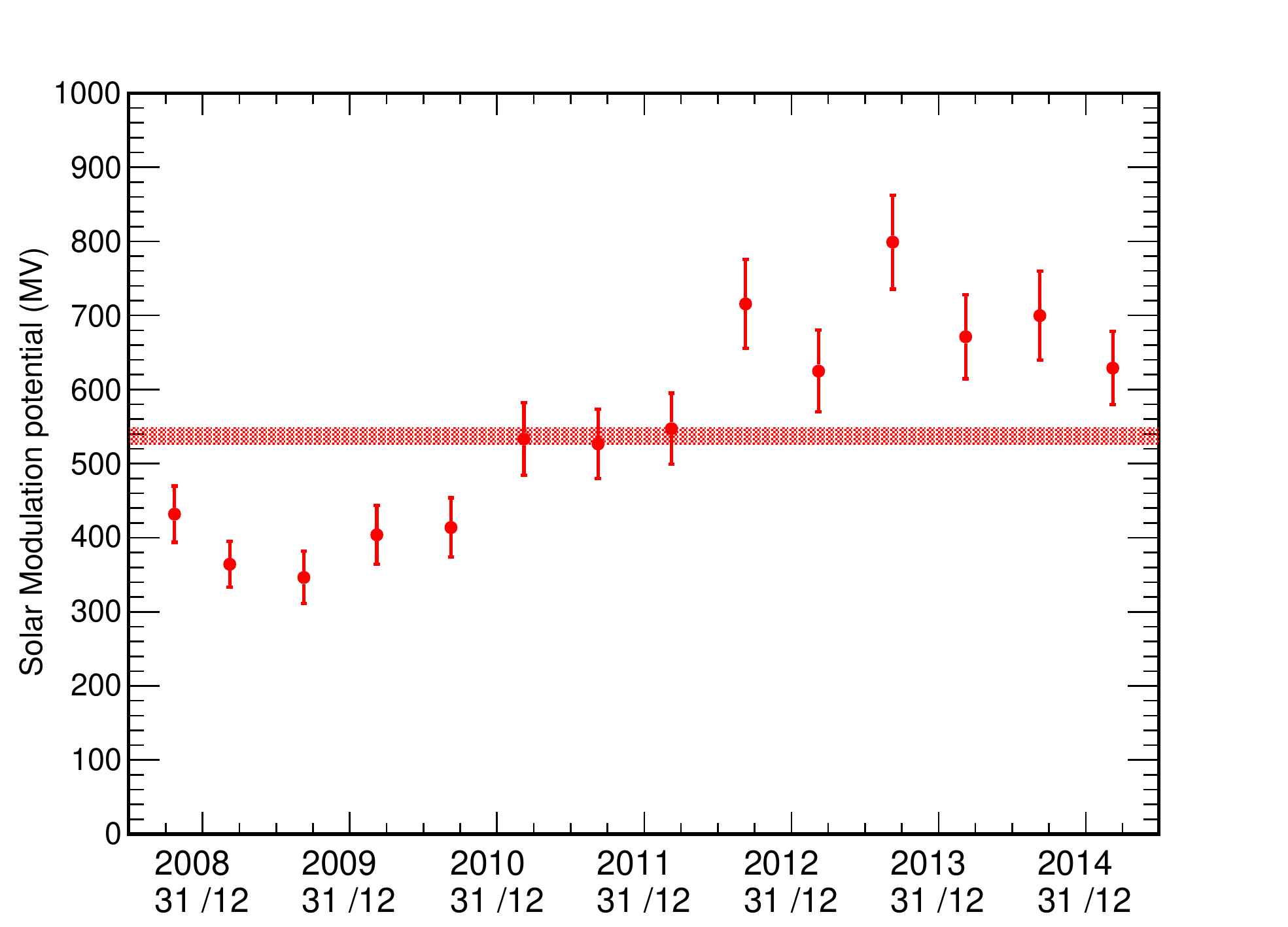}
\end{center}
 \caption{\small Left: CR proton and helium spectra obtained from the best fit
of the \fermilat Moon \g-ray data~\cite{ackerman2016}. The results of the fit (continuous black and red lines)
are compared with the proton measurements taken by PAMELA~\cite{pamela} in
2008 (blue points) and 2009 (purple points) and with the AMS-02~\cite{ams02,amshe}
proton (cyan points) and helium data (violet points). The plot shows also
the proton and helium LIS (dashed black and red lines)
and the Voyager 1 proton (light green points) and helium (dark green)
data~\cite{stone2013}. Right: Time evolution of the solar modulation potential, evaluated from a fit
of the lunar \g-ray emission. The central band corresponds to the average
value of the solar modulation potential during the whole data-taking period.}
\label{Moon_CR:fig4}
\end{figure}
\paragraph*{Expected results with e-ASTROGAM}
The energy range of e-ASTROGAM will cover the whole \g-ray spectrum emitted by the Moon.
e-ASTROGAM data will extend the energy range observed by the \fermilat towards lower energies.
The lunar \g-ray data at low energies will represent a powerful tool to monitor
the solar modulation and to study the CR spectra impinging on the Moon surface.

\subsection[The Sun: a giant lab for cosmic-ray studies\\
\noindent
\textit{\small{E. Orlando, N. Giglietto, M. N. Mazziotta, S. Rain`o, A. Strong}}
]{The Sun: a giant lab for cosmic-ray studies}\label{Orlando}
\paragraph*{Science questions}
The Sun is a known quiescent \g-ray source \cite{Orlando2008, Abdo2011_apj}. Its \g-ray steady-state, characterized by two distinct emissions, is unique for its spatially and spectrally distinct components: 1) disc emission due to pion decay of CR hadrons interacting with the solar atmosphere \cite{Seckel91}; 2) spatially extended emission from \ic scattering of CR electrons on the solar photons of the heliosphere \cite{Moskalenko, Orlando2006}. The latter extends to the whole sky and it is above the background even at large angular distances from the Sun. \\
Observations of the two components of the solar emission allow to gain information on CRs very close to the Sun and on CR propagation in the heliosphere. 
In addition, because CRs are affected by solar modulation, the intensity of both solar emissions is expected to be inversely proportional to the solar activity. This allows to obtain information of CRs close to the Sun as a function of different periods of solar activity. \\
After the discovery of the quiet solar emission in \g-rays with EGRET \cite{Orlando2008}, thanks to \fermilat we can now detect the solar steady state with higher sensitivity and in different periods of solar activity \cite{Abdo2011_apj, Ng}. 
However, at low energy \fermilat has a relatively large PSF that does not allow to disentangle the hadronic disc emission from the leptonic extended emission. This prevents from knowing the CRs and their propagation close to the Sun for those energies where the solar modulation effects are important. Present models of propagation in the inner heliosphere that work well with \fermilat data assume the force field approximation for the modulation of the CRs \cite{gleeson2}. However the reality is more sophisticated\footnote{e.g. see the following code for CR propagation in the heliosphere: www.helmod.org/ and https://github.com/cosmicrays/HELIOPROP}, and this aspect cannot be investigated with the limited PSF and sensitivity of present missions. Even more challenging is that the observed integral flux from the solar disk is found \cite{Abdo2011_apj} to be 7 times higher than predicted by the 'nominal' model of \cite{Seckel91}. This is possibly due to difficulties on the two-component separation, calling for more sensitive observations and better PSF. In addition, observations of the energy range from few hundred MeV to 100 MeV of the Sun, where the solar modulation is very significant, are crucial for understanding low-energy CRs and their propagation in the heliosphere.
\paragraph*{Importance of \g-ray observations}
Our knowledge on CRs at Earth has substantially increased in the recent years thanks to advanced instruments. 
For example, PAMELA \cite{Pamela} launched in
2006, \fermilat \cite{Fermi} in orbit since 2008, and the Alpha Magnetic
Spectrometer-02 (AMS-02) \cite{AMS} working since 2011
have obtained very precise measurements of CRs at Earth. 
Moreover, the data from {\it Voyager~1} \cite{Voyager}, the first human-built instrument leaving the Solar System, made also possible to know the LECRs in the interstellar space. 
Measurements of CRs are also obtained indirectly by looking at the interstellar emission from \g-rays (e.g. \cite{O2017,diffuse1, diffuse2}) to radio-microwave frequencies (e.g. \cite{O2017, Strong2011,Orlando2013mnras}. \\
On the contrary, measurements of the CRs in the inner heliosphere are very difficult, if ever possible. However, an indirect way to probe CRs and their propagation in this region is by looking at the \g-rays from the Sun, and by monitoring its emission components during various solar cycles. 
In more detail, CRs in the heliosphere are affected by the solar wind and the magnetic field, which change their spectrum at energies below few tens of GeV/n. The strength of this effect depends on the solar activity, and it is known as the solar modulation. The solar activity has a period of 22 years, when the Sun changes twice the magnetic field polarity, and it passes through two solar maxima and two solar minima. During solar maxima, the solar modulation of CRs is the largest, while during solar minima is the lowest. As a consequence of being produced by CRs, both \g-ray emission components of the Sun vary as a function of the solar activity. Observations of the two solar components allow us to obtain information of CRs at the Sun.
In addition, observations of the
\ic emission provide information about CR electron spectra
throughout the entire inner heliosphere and
allows  comprehensive  studies  of  the  solar  modulation in this region. \\
The first attempt to detect the disc emission with EGRET data was performed by the EGRET Collaboration \cite{egret} that obtained only an upper limit. An accurate analysis \cite{Orlando2008} of the EGRET data accounting also for the \ic emission component and background sources lead to the first detection of the quiescence \g-ray Sun \cite{Orlando2008}, and to the separation of the disc and the extended \ic components. The flux and spectrum of the two components were found to agree with the expectations. This analysis was performed with data mainly during solar maximum.
During the first two years of the
\Fermi mission
the solar activity has been extremely low, resulting in a high
heliospheric flux of Galactic CRs. Therefore, the CR-induced
quiescent
\g-ray emission from the Sun was expected to be near
its maximum.
The first study with \fermilat data \cite{Abdo2011_apj} allowed to distinguish the two components with higher statistical significance than previously achieved. This analysis was conducted using 18 months of data during low solar activity. Different \ic models have been investigated, yet no best model was found. 
The observed integral flux from the solar disk was found to be ~7 times higher than predicted by the "nominal" model of \cite{Seckel91}. 
A few years ago the solar activity started to increase, allowing us to start studying the evolution of the \g-ray emission for different solar conditions \cite{Ng, Raino} for energies above 100 MeV.
Disentangling the different components and characterizing the sources below 100~MeV with \fermilat is very challenging due to the relatively large PSF and energy dispersion at those energies. Moreover, any analysis below 30~MeV is discouraged\footnote{https://fermi.gsfc.nasa.gov/ssc/data/analysis/LAT\_caveats.html}.
Besides the CR studies, the solar emission need to be accurately modeled in order to properly account for its emission in other studies. Indeed being moving and extended in the sky, the solar emission acts as a confusing source and it should be included in the analyses in a dedicated software as done with the \Fermi Solar Science Tools within the \fermilat Collaboration \cite{SST} that include physically based models of the IC emission \cite{Orlando2013}.
\begin{figure}
\center
\includegraphics[width=0.7\textwidth]{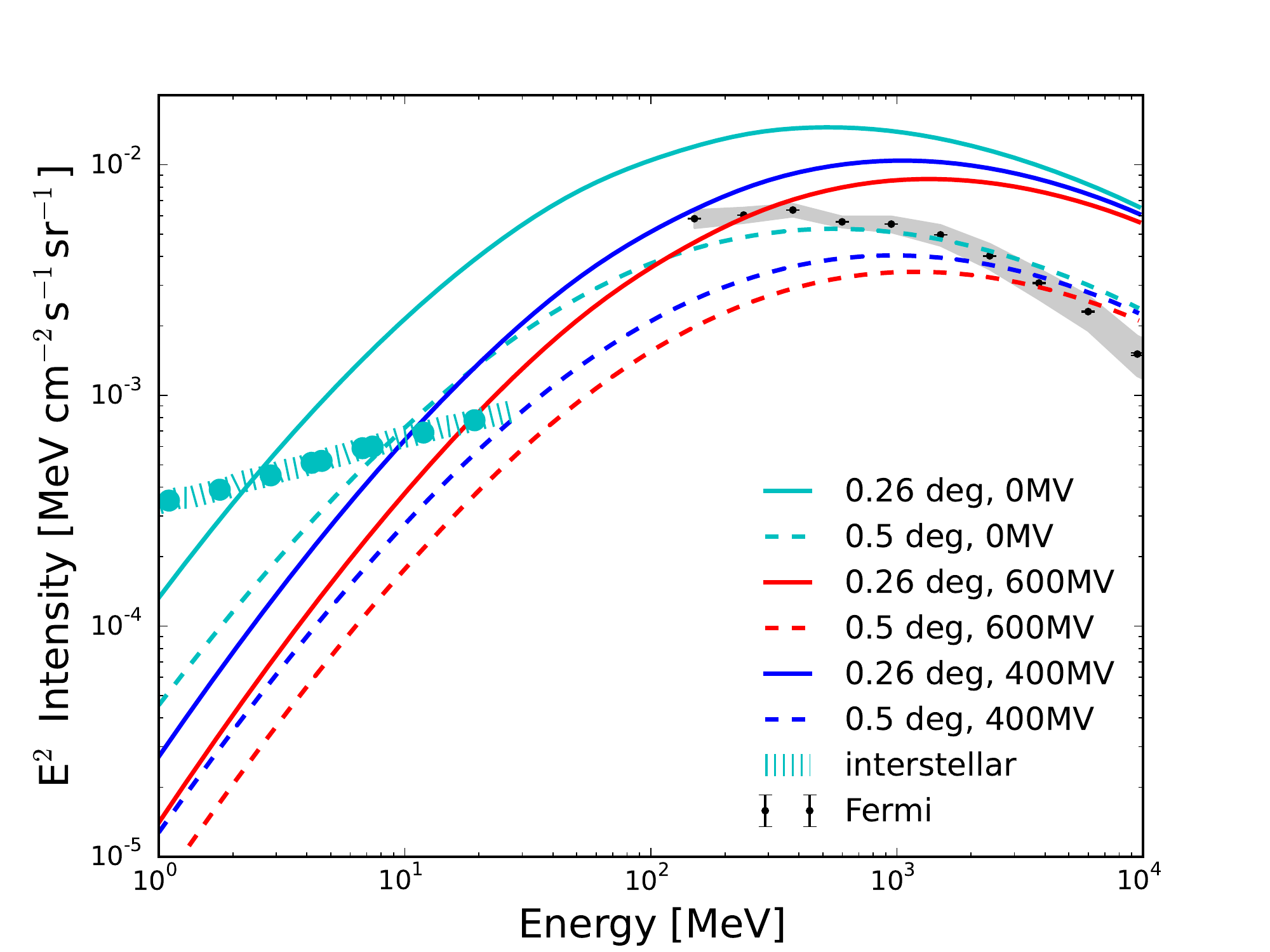}
\caption{Predictions of the intensity of the solar IC emission for the energy range of e-ASTROGAM for various models and various angular distances from the Sun \cite{ICRC2017Sun}. The figure shows also the \fermilat data \cite{diffuse1} and predictions of the interstellar emission at intermediate latitudes for comparison \cite{O2017}. The figure is taken from \cite{ICRC2017Sun}.}
\label{fig_eAstrogam}
\end{figure}
\paragraph*{Expected results with e-ASTROGAM}
The e-ASTROGAM mission will achieve a major gain in sensitivity compared to the COMPTEL missions.
It will also provide improved PSF with respect to \fermilat, which will help in the component separation and angular resolution. This will enable us to study CR transport in the inner heliosphere, to improve on the models of the solar modulation and on the models of CR cascades in the solar atmosphere. \\
As an example, we report here the expected emission due to \ic, as obtained in \cite{ICRC2017Sun}, where we have updated our previous models used in \cite{Abdo2011_apj} to account for the latest more precise AMS-02 CR electron and positrons measurements \cite{AMS_ele}. In that work \cite{ICRC2017Sun}, the StellarICs code \cite{Orlando2013} has been used to extend the predictions down to 1 MeV for various models. Fig.~\ref{fig_eAstrogam} shows these predictions of the \ic component for the entire range of e-ASTROGAM. Solid lines represent the intensity predictions for different solar modulation conditions (0~MV, 400~MV, 600~MV) at 0.26$^\circ$ from the direction of the center of the Sun. As an example, dashed lines represent the same prediction at 0.5$^\circ$ from the direction of the Sun. For more details on the modeling see \cite{ICRC2017Sun}.  The same figure shows also \fermilat data from \cite{diffuse2} at intermediate latitudes and the predictions of the interstellar emission at MeV energies at intermediate latitudes from \cite{O2017}.
In the energy range 1-100 MeV the solar modulation effect is at its maximum, thus allowing to easily distinguish among different models. This will allow tracing for the first time the LECR electrons close to the Sun. \\
In summary, the e-ASTROGAM mission will provide a unique opportunity to monitor the solar emission over the different solar cycles with changes in polarity. Moreover, covering lower energies than \fermilat, e-ASTROGAM will allow accessing the energy range where the solar modulation plays the most important role.  
\subsection[Gamma-ray emission from solar flares\\
\noindent
\textit{\small{E.~Bissaldi, N.~Giglietto, F.~Longo, M.~Mallamaci, S.~Rain\`{o}}}
]{Gamma-ray emission from solar flares}\label{bissaldi}
\paragraph*{Science questions}
Solar flares are the most energetic phenomena in the Solar System. They appear as sudden flashes of light with time scales of minutes to hours, 
releasing up to 10$^{32-33}$ ergs. These events are sometimes associated with explosive blasts of material from the solar corona, i.e. the
CMEs, even if the correlation between the two processes is not clearly established yet.
The frequency of both flares and CMEs follows the 11-year solar activity cycle, the most intense ones usually occurring during the maximum. \\
Solar flares are mainly classified on the basis of the X-ray flux between 0.1 and 0.8 nm measured by 
the Geostationary Operational Satellite Server (GOES). However, during the last few decades, many events have been detected by several 
experiments over a very wide range of energies, going from decameter radio waves to \g-rays beyond 1 GeV, 
hinting at a complex underlying scenario.\\
What triggers the flares and how the Sun releases this energy with such high efficiency is presently not completely understood. 
Flare energy may be considered to result from reconnecting magnetic fields in the corona. 
According to the standard scenario \cite{kopp}, the release of energy derives from accelerating particles, which precipitate 
from the corona to the chromosphere, where they heat the plasma. The hot plasma expands then along the magnetic loop into the corona,
 a process named \textit{evaporation}. This model explains several observations, like the soft and hard X-ray emission, but not all 
 (see for example \cite{kru}). In addition, the acceleration mechanism is not part of the model, being one of the puzzling aspects of the phenomenon. \\
An intriguing counterpart of the solar flares are the so-called Solar Energetic Particles (briefly SEPs), 
a population of charged particles observed in interplanetary space, with energies going from some keV up to GeV.
SEPs can be detected after the solar flares, especially when these are followed by CMEs. 
A key question is whether particles producing the flare radiation and SEPs are accelerated by the same mechanism.\\
Phenomena similar to solar flares and CMEs are believed to occur at larger scales elsewhere in the universe, for example in stellar flares, magnetars, 
young circumstellar disks, SNe shock waves, etc. These energetic phenomena from the Sun are therefore the most accessible laboratories 
for the study of the fundamental physics of transient energy release and efficient particle acceleration in cosmic magnetized plasmas. 
Furthermore, it is worth to study them since they produce the most extreme forms of space weather, like the radiation hazard from the most intense SEP fluxes, 
and the disruption of the heliospheric plasma environment.
\paragraph*{Importance of \g-ray observations}
As outlined above, the solar flares emit radiation with very different energies. 
This results from the acceleration of charged particles which interacts with the ambient solar atmosphere and magnetic fields and typically occurs in the regions near the footpoints of magnetic field lines. 
In particular, accelerated electrons mainly produce X-rays via non-thermal \brem and radio emission via synchrotron mechanism.
On the other hand, accelerated protons and ions come into play emitting at higher energies: nuclear interactions produce 
excited and radioactive nuclei, neutrons and pi-mesons. All of these products subsequently are responsible for 
the \g-ray emission via secondary processes, consisting in nuclear \g-ray lines in the 1-10 MeV range and a continuum spectrum above 100 MeV \cite{murphy87}.  Also accelerated primary electrons, undergoing inelastic scattering, yield \brem radiation 
with a broad energy spectrum extending up to the originating electron energy. \\
Previous \g-ray observations of solar flares were carried out for the first time by the \g-ray spectrometer on board 
of the Solar Maximum Mission (SMM). Later on, many detections were performed by EGRET 
on the Compton Gamma-ray Observatory (CGRO), and in some cases also by RHESSI, 
still operating but mainly designed for hard X-ray energies.
 A review of these \g-ray observations can be found in \cite{chupp}. Besides in \cite{ves}, a compilation of SMM data for 258 \g-ray flares detected above 300 keV is presented.
 Recent observations of solar flares at keV-MeV-GeV energies have been carried out by the two instruments onboard the \Fermi satellite. The secondary instrument, the \fermigbm, consists of two types of detectors, namely the NaI (8-900 keV) and BGO (250 keV - 40 MeV) detectors. GBM triggered on $>$ 1200 solar flares in the hard X-ray band over 9 years. Some of those were also detected in the 1-10 MeV band. However, the BGO energy resolution is not fine enough to perform an accurate line analysis (see next section). \\
The \g-ray emission light curve can be similar to one observed in X-rays, lasting for 10-100 s 
and indicating the acceleration of both ions and electrons from the same solar ambient. 
This is referred to as ''impulsive'' phase of the flare. 
However, some events  have been found to have a long-duration \g-ray emission, lasting for several hours after the impulsive phase \cite{kanbach}. 
In this respect, a relevant number of 
flares detected by  \fermilat above 100 MeV shows this kind of long duration emission \cite{Ack}. Fig.~\ref{fermiflare} displays the temporal evolution of the emission for one of these events. In general, during the extended phase, 
there does not seem to be any other associated radiation,  but most of these flares are associated with fast CMEs and a significant flux of SEPs. The origin of this temporally extended events is not well understood and raised new questions, 
such as the type of radiative process (if hadronic or leptonic), the location of the acceleration (if at the flare site or in the proximity of the CME), the mechanisms of the acceleration \cite{aje}.
Finally, \fermilat has detected an intriguing class of "behind-the-limb" solar flares \cite{ackermann17}, for which one possible explanation is the \g-ray emission by protons in the CME environment.
\begin{figure}
\centering
\includegraphics[width=10cm]{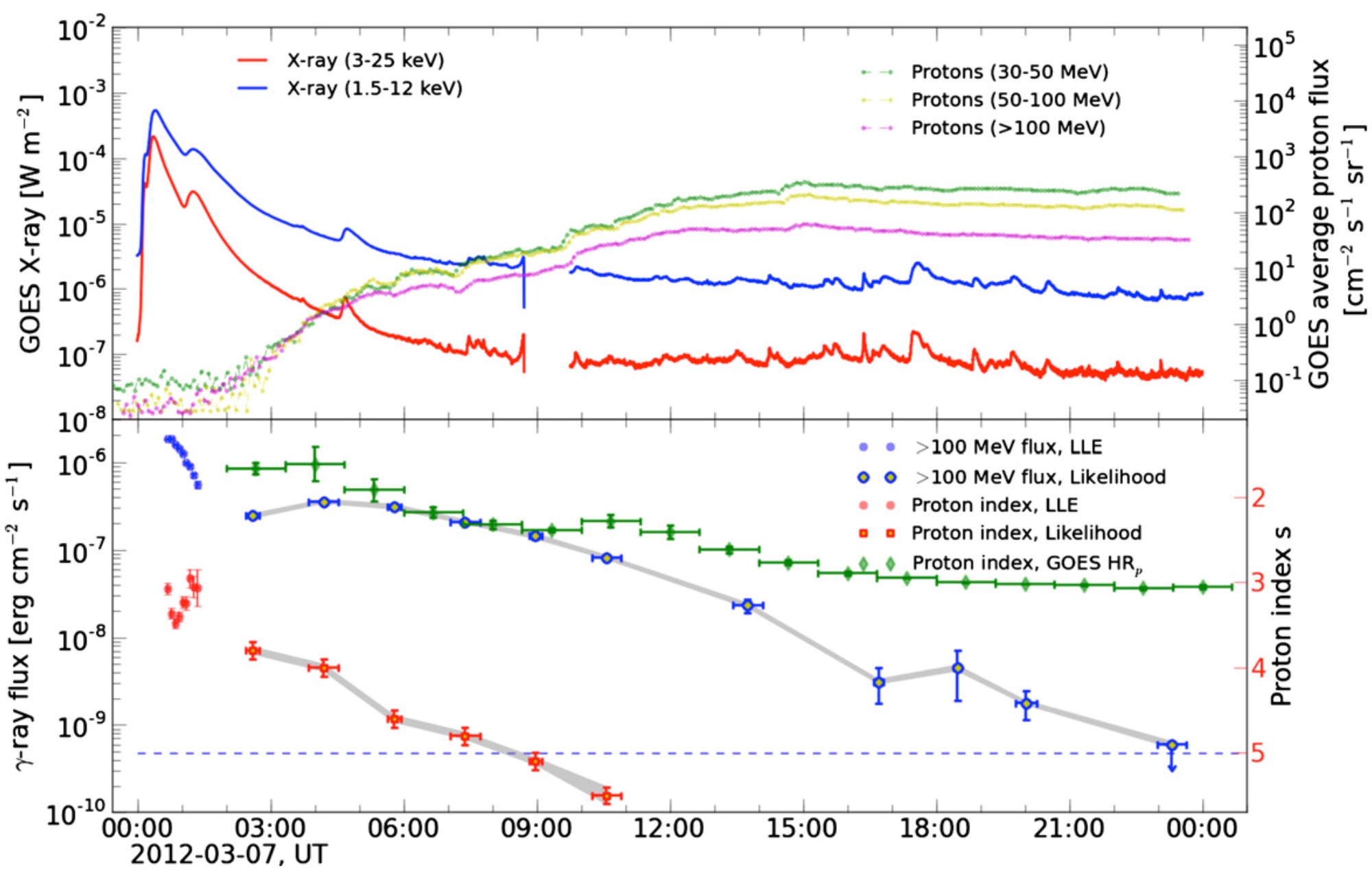}
\caption{\small Temporal evolution of two bright X-class solar flares - 2012 March 7. (Top panel) X-ray emission and 
proton flux detected by GOES satellite. (Bottom panel) Long lasting \g-emission detected by \Fermi. Picture taken from \cite{aje}. 
\label{fermiflare} }
\end{figure}
\paragraph*{Expected results with e-ASTROGAM}
The detections of solar flares by SMM and EGRET, in the past, and more recently by \Fermi
indicate that acceleration of particles in the \g-ray energy range is not such a rare phenomenon, even for more modest events \cite{Ack}.
 e-ASTROGAM will study for the first time the solar flare radiation from 300 keV to 3 GeV, covering therefore a very broad range of energies and complementing information
 collected by dedicated future experiments like the ESA Solar Orbiter \cite{SO}.
In the following, we report the different types of measurements that e-ASTROGAM will be able to perform.
\begin{itemize}
\item{\textbf{Temporal evolution}. During the expected 3 years of operation, e-ASTROGAM will have the opportunity to detect solar flares 
(the number depending on the phase of the solar cycle)  and to study the evolution in time of the \g-radiation from each event. These and  
other observations performed in different energy bands and channels (like SEP fluxes) will give important information about the 
classification of the events between impulsive and long duration events, helping in constraining model of acceleration and propagation \cite{Ack}.} 
\item{\textbf{Energy spectrum.} As already outlined, the \g-ray solar flare spectrum is characterized by a \brem continuum, nuclear lines and pion-decay components. 
e-ASTROGAM will have optimal sensitivity and energy resolution (much better than \fermigbm)  to detect the de-excitations lines from accelerated ions. 
This will be fundamental to gain insight into the chemical abundances and into the physical conditions where accelerated ions propagate and interact \cite{share}. Also, the 0.511 MeV 
and 2.223 MeV lines will be detected. 
It will also be interesting to compare e-ASTROGAM results with  SMM spectroscopic analysis in MeV domain (see for example \cite{murphy}). 
Going at higher energies, the spectral analysis will allow one to distinguish spectroscopically between electron \brem and the pion-decay models. }
\item{\textbf{Photon polarization}. The study of polarization is appealing: \brem emission from solar flares will be polarized if the phase-space distribution of the emitting electrons is anisotropic. 
Polarization measurements therefore provide a direct handle on the extent to which the accelerated electrons are beamed, which, in turn, has important implications for particle acceleration models.
These type of measurements were carried out in X band, while the first and unique measurement of \g-ray polarimetry has been performed in \cite{boggs}, by exploiting RHESSI data between 0.2-1 MeV, but only 
for two solar flares. e-ASTROGAM can be therefore further exploited in this field,  giving unprecedented polarization measurements in the MeV range by means of the Compton interactions in the instrument.}
\item{\textbf{Source localization}. e-ASTROGAM is designed to have an angular resolution of about 0.2$^\circ$ at 1 GeV (a factor 4 better than the \fermilat instrument). 
Even if it will not be able to resolve the details of the \g-ray emission, localizing the source on the solar disk and comparing this with measurements in X-rays 
(from which typically the source region is identified) could give additional information for constraining the emission and acceleration mechanisms.}
\end{itemize}

\newpage
\section{Miscellanea}
\subsection[COMPTEL Data Heritage Project\\
\noindent
\textit{\small{A. Strong, W. Collmar}}
]{COMPTEL Data Heritage Project}
\paragraph*{Science questions}
\fermilat and \agile have provided a very detailed view of the gamma-ray sky in the  range above 100 MeV,
 which in future may extend down to about 30 MeV with the latest \fermilat event analysis techniques ('Pass 8').
Meanwhile we  have a few thousand GeV sources but only about 20 in the 1-30 MeV range from CGRO/COMPTEL.
The new mission e-ASTROGAM is being proposed and the balloon experiment COSI has flown, both promising for the future;
 meanwhile  a long-term  on-going effort to exploit  heritage COMPTEL data is underway at at the
 Max-Planck-Institut f\"ur extraterrestrische Physik
 and the Max-Planck-Institut f\"ur Astrophysik   in   Garching, Germany.
 The new COMPTEL analyses  will be relevant to forecasting to support  e-ASTROGAM science and instrumentation.

The double-Compton telescope COMPTEL flew on the NASA Compton Gamma Ray Observatory (CGRO) satellite from 1991 to 2000,
 and is still the basis of most of our knowledge about the 1-30 MeV sky.
Pending new missions, for the next decade it will still be our most important resource for MeV continuum gamma rays
\footnote{The SPI instrument on \INTEGRAL satellite provides more details on high-resolution
line spectroscopy, in particular the 511 keV positron annihilation line, $^{26}$Al and $^{60}$Fe lines
 but is not very sensitive to continuum emission above an MeV up to now. COMPTEL does not extend down to the 511 keV line.}.

 COMPTEL \cite{schoenfelder1993} consists of an upper layer of 7 liquid-scintillators (D1) and a lower layer (D2) of 14 NaI detectors.
 The energy deposits in D1 and D2 are measured together with the direction of the scattered photon.
 Since only the Compton-scattered photon is measured, the response is basically circles on the sky centred
 on the true photon direction (Compton scattering formula) and broadened by direction and energy measurements.
The full 9 year mission had 341 roughly 2-week observations covering the entire sky, with a field-of-view of about 30$^o$ radius.
 Instrumental background discrimination is obtained with a time-of-flight (TOF) measurement
 and pulse-shape discrimination (PSD).
Instrumental background variations are fitted using a template from high-latitudes
 where the celestial emission is smallest,
 or using a  filtering technique for source detection.
 The main results of COMPTEL were detections and properties of several Galactic and extragalactic sources,
 the $^{44}$Ti radioactive decay line from Cas A,
 mapping of the Galaxy in the 1.8 MeV line of $^{26}$Al,
 and in Galactic continuum emission \cite{strong1999}, cosmic-ray interactions in the interstellar medium
 \cite{strong2011,bouchet2011,grenier2015},
 as well as GRBs and solar flares.
 The source results are collected in \cite{schoenfelder2000}.
A spectrum of the Galactic plane emission from keV to TeV including  COMPTEL and \fermilat is in \cite{grenier2015}.
 For more details on the interstellar emission and the cosmic-ray connection
see the contribution to this White Book by Orlando, Strong and Grenier.

 A more recent result from the continuing analysis of COMPTEL data at MPE is the definitive identification of the
 LS5039 binary via its light-curve \cite{collmar2014} (still however using the earlier data processing).

Several  new developments are completed or in progress for COMPTEL:
 The COMPTEL data analysis system ('COMPASS') was  ported from Sun Solaris to Linux, removing the dependence on the Oracle database.
 New event processing techniques improve the background rejection, and new energy ranges are selected to avoid background lines.
 Time-of-flight (TOF) background rejection has been improved using intra-detector resolution instead of just per detector (TOF-VI vs previous TOF-IV),
 pulse-shape discrimination (PSD) is used with 2-D discrimination using TOF and PSD together.
 The entire COMPTEL event database has been re-processed with the new selections.
 A new source catalogue is being generated with the new event processing.

 The maximum-entropy skymapping method for COMPTEL \cite{strong1999} based on the MEMSYS5 package \cite{skilling1989}
 has been updated to use current state-of-the art
 convolution on the sphere and the HealPix sky projection (uniform pixelization of the sphere),
 and the method has been adapted to modern parallel processing hardware
 so that skymaps can be produced in a short time compared to the large supercomputer requirements of 20 years ago.
 Fig~\ref{skymaps} shows all-sky images in continuum 1--3, 3--10 and 10--30 MeV, and in the 1.8 MeV line of $^{26}$Al.
 using the new maximum entropy algorithm, and data from the first 6 years of the mission and the original processed data.
 The Galactic plane is clearly visible (in continuum mainly interstellar emission from cosmic-ray interactions)
 as well as the principal sources: Crab, Vela pulsar, LS5039, Cyg X-1, 3C273, 3C279, Cen A. The extended feature below the plane at low energies is contamination from earth atmospheric emission.
In future these maps will be updated with the full mission and the new data processing techniques described here. Preliminary maps using the new data processing for the full mission, with the new energy ranges, are shown in \cite{collmar2017}.
In addition, more advanced analysis using Information Field Theory
and the D$^3$PO package \cite{selig2015} is foreseen.
With D$^3$PO the \Fermi gamma-ray sky was reconstructed in nine separate
energy bands. Spatial correlation of the gamma-ray flux was essential to
discriminate the diffuse from the point-source emission and to denoise
and deconvolve the former. Spectral correlations were not exploited. To
also benefit from these, the D$^4$PO code is currently under development
at the MPI for Astrophysics.  This will detect and exploit
spatio-spectral correlation structures of the diffuse emission as well
as correlations in the point source spectra.
\begin{figure}
\includegraphics[width=1\textwidth,clip=t,angle=0.,scale=0.5]{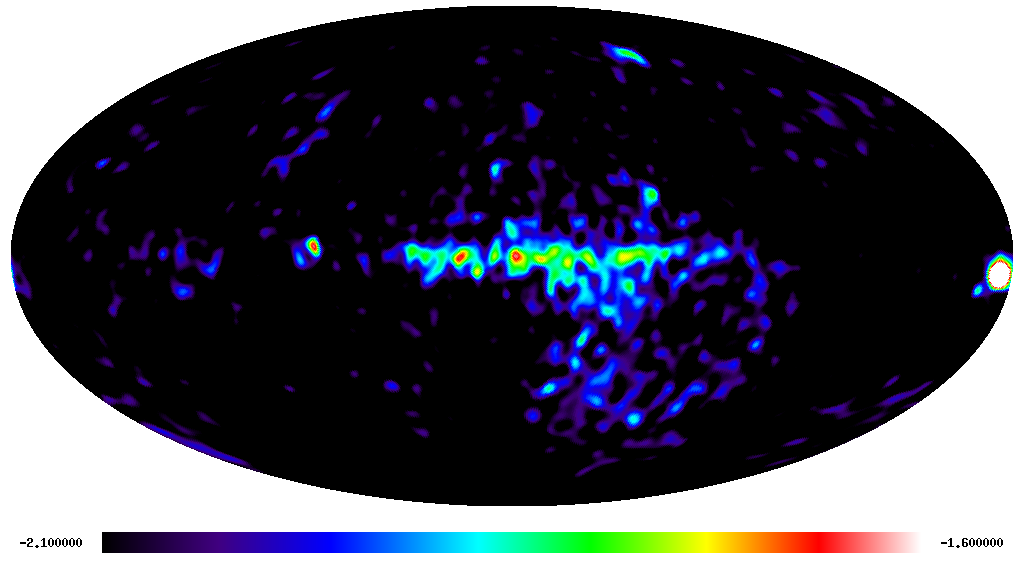}
\includegraphics[width=1\textwidth,clip=t,angle=0.,scale=0.5]{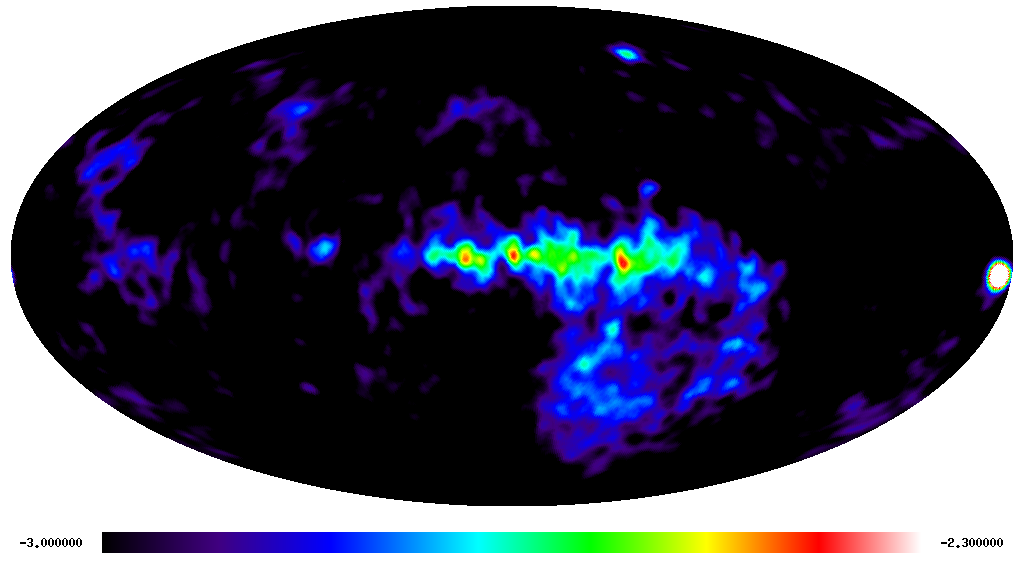}
\includegraphics[width=1\textwidth,clip=t,angle=0.,scale=0.5]{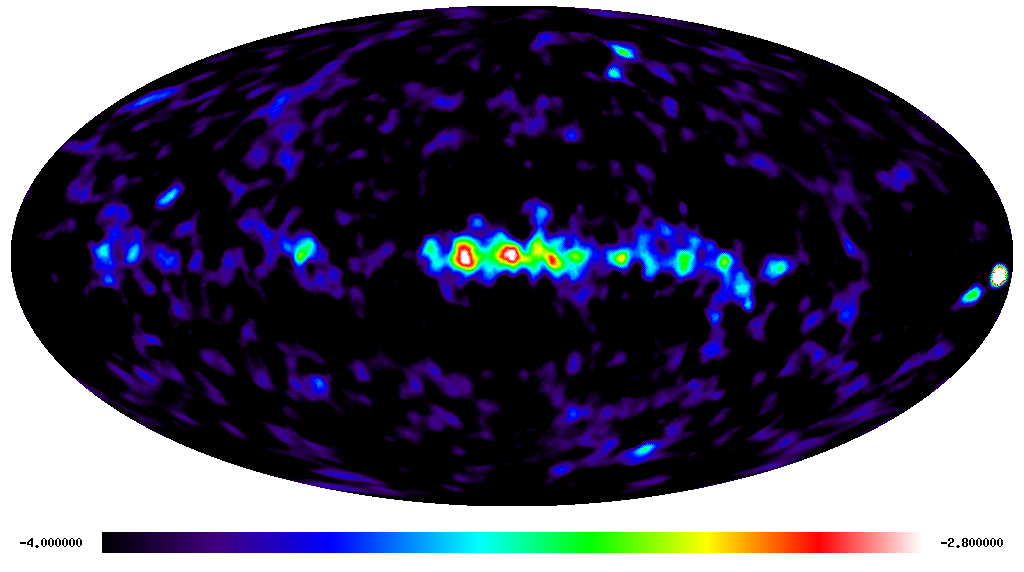}
\includegraphics[width=1\textwidth,clip=t,angle=0.,scale=0.5]{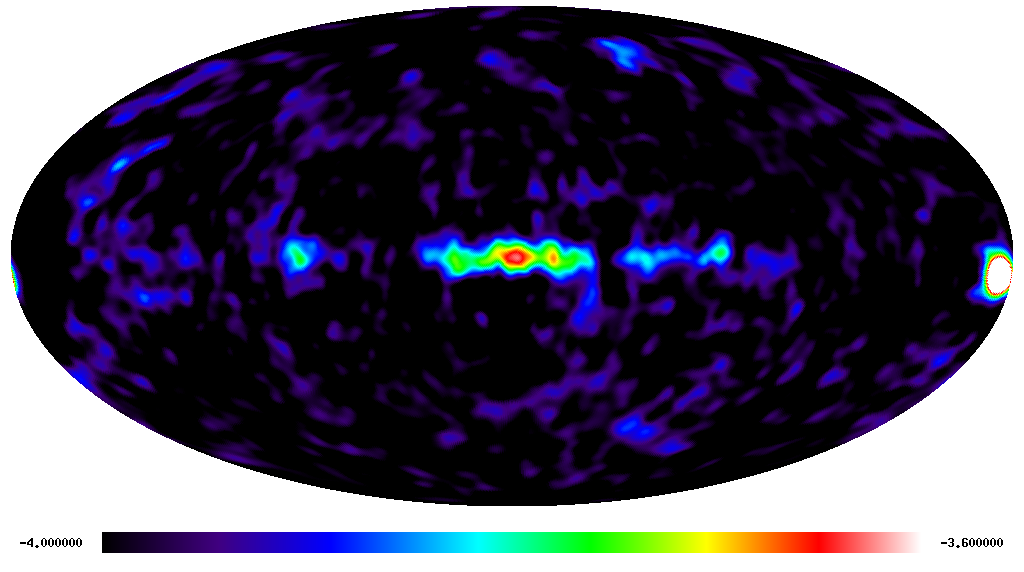}
\caption{\small COMPTEL all-sky images using the current Maximum Entropy implementation. Galactic coordinates, centred on l = 0,  b = 0.
Left to right, top to bottom: 1-3 MeV,  3-10 MeV,  10-30 MeV, 1.8 MeV $^{26}$Al line.
}
\label{skymaps}
\end{figure}
\paragraph*{Acknowledgements --} We thank Martin Reinecke for adapting the Maximum Entropy imaging software as described above, and to Torsten En$\ss$lin and his group at MPA for supporting this project.

\subsection[Cataloguing the MeV sky\\
\noindent
\textit{\small{A.~Dom\'inguez, J.~A.~Barrio, M.~Ajello, M.~L\'opez, B.~Lott, D.~Gasparrini}}
]{Cataloguing the MeV sky}
\paragraph*{Science questions}
The production of source catalogs is a fundamental task of any scientific mission with an instrument that benefits from large field of view (FoV) and high sensitivity such as e-ASTROGAM. According to the Scientific requirements described in Sec. \ref{sec:scirec}, e-ASTROGAM will be designed both with pointing and surveying capabilities. The latter can be activated at any time allowing an optimized all-sky survey.

Source catalogs list sky positions and basic physical properties, which are typically integrated fluxes, photon indices, energy dependent photometry, etc. Other complementary properties, such as redshifts, multiwavelength associations to other catalogs, and source classes, may be included in order to help in the source description and identification. This large collection of high-level data usually is the starting point of many science papers. As an example, the \fermilat catalogs are at the top of the most cited works that have been published by the LAT collaboration (e.g.~ \cite{ackermann13,acero15}).
\paragraph*{Importance of gamma-ray observations}
In the multi-messenger multi-wavelength era that we are entering, cataloguing the sky in the whole electromagnetic spectrum turns out as an indispensable condition for the astronomical community, yet the MeV Universe is still largely uncatalogued. For instance, the sensitivity in the range 100~MeV--500~MeV for the all-sky Third Catalog of \Fermi Sources (3FGL, \cite{acero15}) is 2-3 times worse than what is expected with e-ASTROGAM. The 3FGL catalog contains more than 3,000 sources from the first four years of LAT data and despite its energy threshold, the catalog is more representative of the GeV range. At lower energies of approximately 0.1~MeV and below, there is the Fourth IBIS/ISGRI Soft Gamma-Ray Survey Catalog (IBISCAT4, \cite{bird10}). This catalog provides more than 700 sources both transients and faint persistent objects from the first 5.5 years of data. Also, the \swift-BAT 70 month catalog lists over 1000 sources at similar energies (about 0.1~MeV, \cite{baumgartner13}). Delivering a deep survey of the sky at about 1~MeV will be a major achievement for e-ASTROGAM.
\paragraph*{Expected results with e-ASTROGAM}
As a first step in the construction of an e-ASTROGAM catalog, we are proposing to generate simulated source catalogs based on expected performance of the instrument plus studies of source populations at the lowest energies measured by the \fermilat, and extrapolate them down to MeV energies. This procedure will allow us to estimate the source populations that will be seen by e-ASTROGAM at different exposures and survey strategies. This will help in the surveys optimizations, also between pointing versus survey mode telescope time allocation. In reasonable amounts of telescope exposure, we expect to detect of the order of 1000s sources. These sources wil include blazars, radio galaxies, SNRs, pulsar and PWNe, and likely binary systems, star forming galaxies, lobes of radio galaxies, radio quiet AGN powered by non-thermal electrons in corona, etc. A similar methodology for source population studies has been applied at higher energies to estimate TeV populations expected to be detected by the Cherenkov Telescope Array (CTA, \cite{hassan17}).
Catalogs will be produced from e-ASTROGAM observations providing high-level information to the public. Furthermore, there are plans on building association pipelines for the source identification and classification from information at other wavelengths. It is interesting to find sources with no association or/and class identification (unidentified). The classification of these unIDs sources lead to many interesting scientific possibilities (e.g.~\cite{ackermann12}). Nuclear lines intensities will be also included in the surveys. The e-ASTROGAM catalogs will contain transients as well as persistent sources, thus delivering the most detailed description of the MeV sky for years to come. Monitoring capabilities to constantly look for flares will be explored as well (e.g.~\cite{abdollahi17}). These catalogs will be an essential legacy of the mission.
\subsection[Galactic Center gamma-ray excess: constraining the point source contribution\\
\noindent
\textit{\small{R.~T.~Bartels, K.~Short, C.~Weniger, D.~Malyshev}}
]{Galactic Center gamma-ray excess: constraining the point source contribution}
\paragraph*{Science questions}
The GC is expected to be the brightest source of gamma rays from possible annihilation of DM particles.
An excess of gamma rays, henceforth the Galactic-Center excess (GCE), consistent with DM annihilation in the vicinity of the GC was reported by several groups
\cite{Goodenough:2009gk, 2009arXiv0912.3828V, 2012PhRvD..86h3511A, 2013PDU.....2..118H, Gordon:2013vta, Calore:2014xka, 2016ApJ...819...44A, TheFermi-LAT:2017vmf}.
Apart from DM annihilation, possible explanations of the excess include a population of CR electrons emitted near the GC
and a population of faint but numerous point sources, such as millisecond pulsars (MSPs).
The latter model is supported by various statistical methods, e.g., analysis based on wavelet fluctuations \cite{Bartels2016a},
non-Poissonian template fits \cite{2016PhRvL.116e1103L}, and Monte Carlo reconstruction of Point-Source population near the GC \cite{Fermi-LAT:2017yoi}.
Understanding the origin of the excess is difficult due to significant uncertainties in the diffuse Galactic gamma-ray emission as well as the properties of resolved point sources near the GC.
\begin{figure}
\begin{center}
  \includegraphics[width=0.55\textwidth]{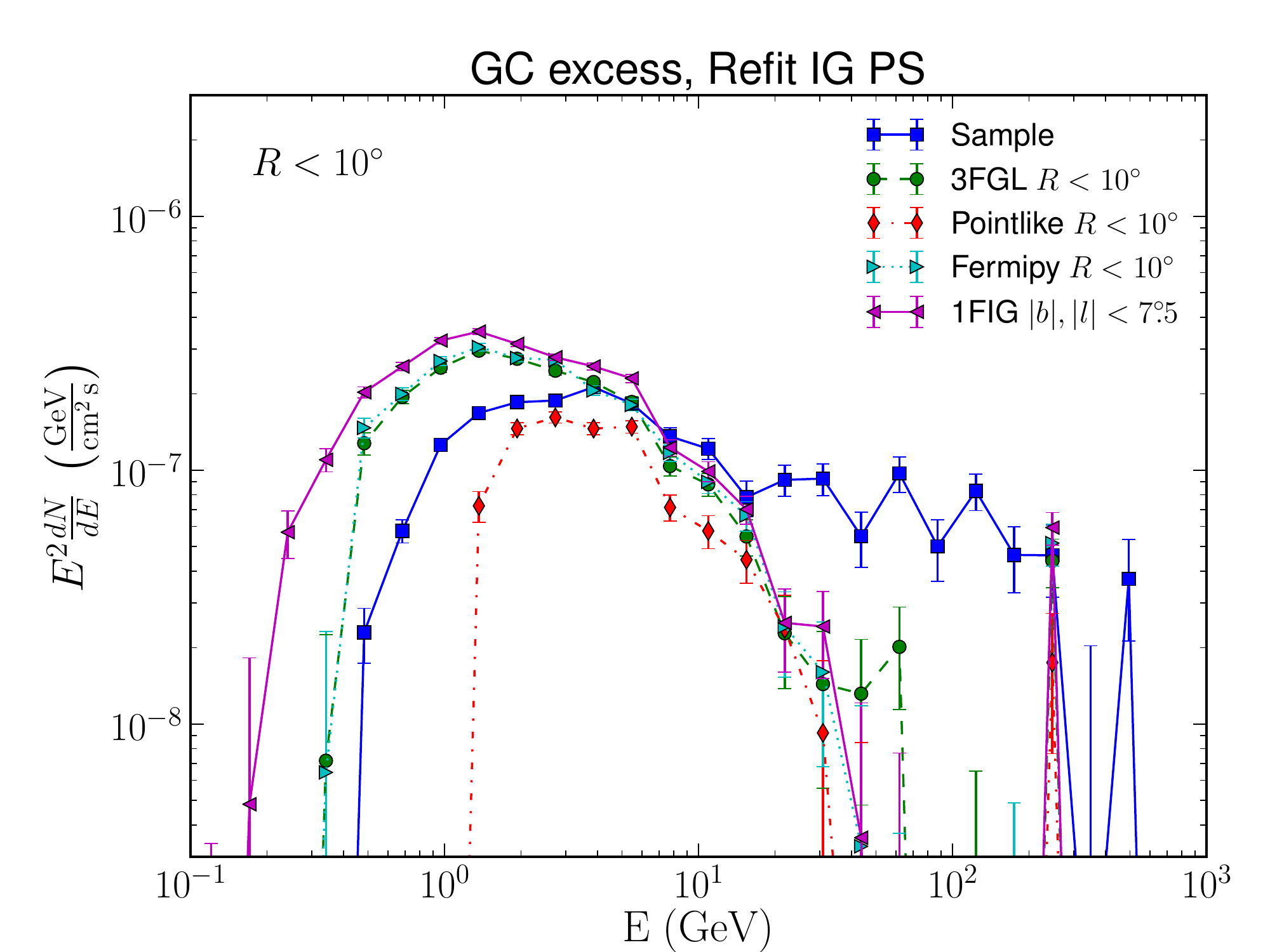}
  \caption{\small Effect on the spectrum of the GC excess from refitting of PS near the GC \cite{TheFermi-LAT:2017vmf}.} \label{fig:PS_modeling}
\end{center}
\end{figure}
In Fig.~\ref{fig:PS_modeling} we illustrate the uncertainty due to resolved PS by showing the effect of refitting PS near the GC found with different PS detection algorithms. One can notice that, at low energies, the changes in the GC excess flux are much larger than the statistical error bars.
This is a manifestation of the fact that our knowledge about the GC excess spectrum is limited by the modeling uncertainties rather than lack of photon statistics.
In part, this is due to relatively poor angular resolution of the \fermilat at energies below 1 GeV.
Improved angular resolution of e-ASTROGAM will help to better separate individual PS and to constrain the spectrum and the morphology of the excess,
which is an important step in distinguishing the MSP hypothesis of the excess from the truly diffuse emission coming from CR electrons or DM annihilation.
Application of statistical methods to the e-ASTROGAM data would even further reduce the uncertainty on the interpretation of the excess.
\paragraph*{Importance of gamma-ray observations}
The gamma-ray GC excess has no clear counterparts in other frequencies, such as radio or X-ray.
This lack of counterparts makes it hard to determine the origin of the excess.
Although proposed future observations with new radio telescopes such as MeerKAT, GBT, VLA, and later SKA have the potential to observe dozens of MSPs in the bulge of the MW,
e.g., if the GC excess is coming from MSPs, then one can expect to detect about 
200 MSPs with SKA when surveying the inner--Galaxy for $\sim 100\mathrm{h}$ \cite{Calore:2015bsx}.
Nonetheless, the gamma-ray observations will remain our main tool to learn about the origin of the excess.
With gamma-ray observations one can either directly search for MSP candidates based on the gamma-ray spectrum \cite{Fermi-LAT:2017yoi}
or use statistical methods to determine the contribution of sub-threshold point sources.
Currently there are about 60 MSP candidates detected in the \fermilat data \cite{Fermi-LAT:2017yoi},
while statistical methods show that all of the excess can be explained with a population of point sources.
Straightforward detection of MSPs in the Inner Galaxy are compromised by large diffuse foregrounds and point source confusion, along with the degrading resolution of \fermilat in the inner few degrees of the GC. Challenges also arise when performing a wavelet analysis in the Inner Galaxy. In principle, there is potential for falsely induced wavelet peaks due to, e.g., mismodeled emission of CRs interacting with interstellar gas. Such concerns can be addressed by a careful analysis of the spatial distribution of peaks, comparison with the expected signal from gas sub-structure only, or a spectral analysis to distinguish between the potential signals. At sub-GeV energies, the poor angular resolution of the \fermilat makes it difficult to constrain the MSP population via both a direct search or statistical analysis using, e.g., wavelet fluctuations analysis.
\paragraph*{Expected results with e-ASTROGAM}
One of the advantages of \agam relative to \fermilat is a better angular resolution at energies below 1 GeV.
Although the statistical sensitivity of \agam around 1 GeV after 5 years of observations is expected to be comparable to \fermilat statistical
sensitivity after 10 years of observations,
the main challenge in analyses near the GC is not the statistical uncertainty,
but rather the source confusion and uncertainties in the diffuse emission modeling.
Thus, it is important to take into account the signal to background ratio (SBR) together with the signal to noise ratio (SNR).
\begin{figure}
\includegraphics[width=13cm]{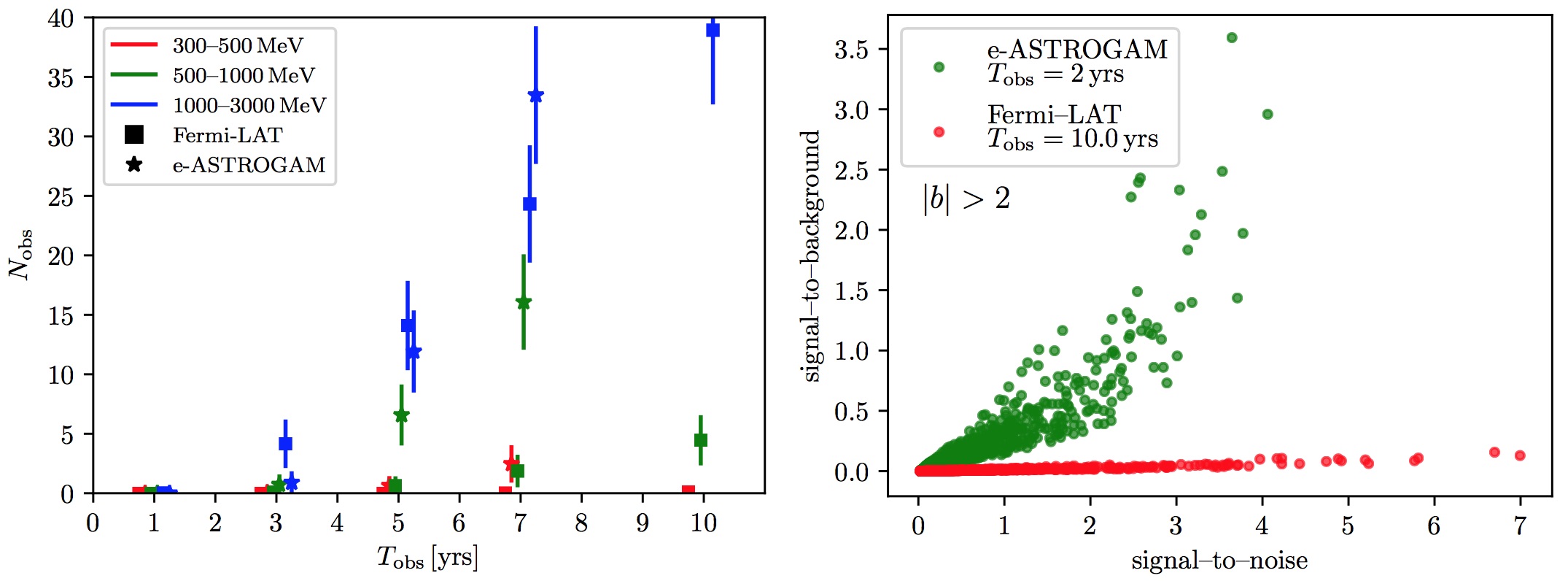}
  \caption{\small Projections for the detectability of the bulge MSP population. Left: Number of sources detectable with $\mathrm{SNR}>5$ as a function of total observation time in different energy ranges and for $|b|>2^\circ$. We show predictions for $T_\mathrm{obs} \in \left[1, 3, 5, 7\right] \mathrm{\,yrs}$ (minor shifts away from these values are for visibility). For \fermilat we also include the 10 year prediction for comparison. Right: Signal--to--noise ratio (SNR) versus signal--to--background ratio (SBR) for a random realization of an MSP population that can reproduce the GCE in the energy range $0.5$--$1\mathrm{\,GeV}$. We compare the current \fermilat sensitivity (red) to that expected for e-ASTROGAM after 2 years of total exposure (green). Due to its larger exposure, the \fermilat can reach a SNR comparable to e-ASTROGAM in this energy range, however, e-ASTROGAM has a superior SBR due to better angular resolution.} \label{fig:agam}
\end{figure}
To estimate the improvement in the source characterization, we simulate a population of MSP-like point sources in the
bulge of the MW.
For the simulation, we are using the best fit distribution of MSPs in the Galactic Bulge as described in \cite{Bartels2016a}, which can explain 100\% of the GCE.
We make predictions for the number of sources detectable by the \fermilat and \agam in the energy ranges $0.3$--$0.5\mathrm{\,GeV}$, $0.5$--$1\mathrm{\,GeV}$ and $1$--$3\mathrm{\,GeV}$. We compute the sensitivity to a point source as a function of sky position by requiring that the signal-to-noise ratio is larger than 5
within the 68\% containment radius of the PSF (Fig.~\ref{fig:agam} left).
In Fig.~\ref{fig:agam} on the right we show the signal--to--noise and signal--to--background ratios for the sources in this population for energy range 0.5 -- 1 GeV.
In this range, the statistical sensitivity of \fermilat is comparable to the sensitivity of \agam, but the signal to background ratio is significantly
better for \agam due to superior angular resolution, which will enable one to better separate the sources from each other and from the diffuse background.
For the wavelet analysis, the number of high significance peaks is expected to be similar for \fermilat and \agam due to similar statistical sensitivity,
but at low significance, the peaks which overlap in \fermilat will be resolved with \agam, which will improve the statistical power of the
wavelet analysis.

The main scientific output of this study will be a better characterization of the GC excess.
If the MSP scenario is disproved, i.e., the excess emission is consistent with truly diffuse component, then the DM interpretation will be still one of the possibilities. If, on the other hand, most of the GC excess emission will be explained by a population of MSPs,
then one will put tighter limits on DM annihilation, which will be competitive or even more constraining for some channels of annihilation than the limits from the dwarf galaxies.

\begin{itemize}
\item  Scientific output:
    \begin{itemize}
        \item Characterization of low-energy component of excess;
        \item Higher resolution of excess emission in individual sources;
        \item Better constraints on DM annihilation.
    \end{itemize}
\end{itemize}
\subsection[Unidentified gamma-ray sources\\
\noindent
\textit{\small{J. M. Paredes, V. Bosch-Ramon, B. Marcote}}
]{Unidentified gamma-ray sources}
\paragraph*{Science questions}
The third EGRET catalogue provided a list of unidentified sources (168 out  271 detected sources) \cite{hartman99}. The discovery of gamma-ray sources with no feasible counterpart at lower energies has been also common in more recent spatial missions such as  \fermilat, or by ground Cherenkov telescopes. Although the fraction of unidentified sources is lower  than in the case of EGRET,  the fraction is still significant, specially in the case of \fermilat \cite{acero15}, in which a large amount of sources have been detected (see Table \ref{Unid:tab1}). In most of the unidentified gamma-ray sources the lower energy part of their spectrum is unknown, and thus e-ASTROGAM will for the first time allow the determination of this part of the spectrum, which can be fundamental for the source identification.
\begin{table}[]
\centering
\caption{\small Fraction of unidentified sources from different gamma-ray instruments.}
\label{Unid:tab1}
\begin{tabular}{llll}
 & Detected & Unidentified &    \\
EGRET & 271 & 168 & 62\%   \\
COMPTEL 1st cat &32  &9  & 28\% \\
 \agile 1st cat & 47 &8  & 17\%  \\
\fermilat 3FGL & 3033 & 1010 & 33\%  \\
 CHERENKOV & 204 &44  &   22\%
\end{tabular}
\end{table}
\paragraph*{Importance of gamma-ray observations}
Observations in the 0.3 MeV--3 GeV range of unidentified sources give valuable information about the unknown spectrum of these sources in the low energy part of the gamma-ray spectrum. On one hand, the majority of the 3FGL Catalog sources (2415 out of 3033) have a power-law spectra (at energies larger than 100 MeV) steeper than E$^{-2}$, and among the unidentified sources, the fraction of them steeper than E$^{-2}$ is 898 out of 1010. This implies that the peak energy output of these sources is below 100 MeV, making them good targets for e-ASTROGAM. On the other hand, the most powerful AGNs peak in the MeV region \cite{ghisellini17}, whereas gamma-ray pulsars typically have spectral peaks in the GeV energy band \cite{thompson04}. Therefore, the knowledge of the MeV--GeV spectrum can already allow one to find possible candidates for the unidentified sources (e.g. AGN vs pulsar), and together with multi-wavelength data, fully reveal the nature of many of them. Finally, it is worth mentioning that there is presently a deficit of blazars in the Galactic plane direction, with estimates of this deficit that amount to $\sim 50--100$ sources \cite{ack_apj_2012}. The high angular resolution of e-ASTROGAM will be of great important to find these missing blazars at low galactic latitudes. Thus, e-ASTROGAM, working in the poorly explored energy range  from 0.3 MeV to 3 GeV, can play a fundamental role in the identification of gamma-ray sources without known counterpart.
\paragraph*{Expected results with e-ASTROGAM}
\begin{figure}
\centering
\includegraphics[width=0.60\textwidth]{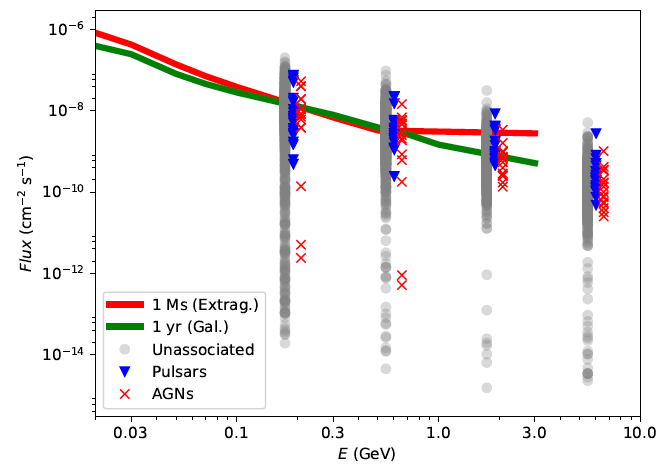}
\caption{\small Flux of the unassociated sources, pulsars and AGNs detected by \fermilat (3FGL) as a function of individual bands. The red and green curves are the e-ASTROGAM sensitivity for different integration times and for the energy range 0.03--3 GeV.}
\label{Unid:fig1}
\end{figure}
\begin{figure}
\includegraphics[trim= 0mm 0cm 0cm 7cm, width=0.5\textwidth,angle=0.,scale=1]{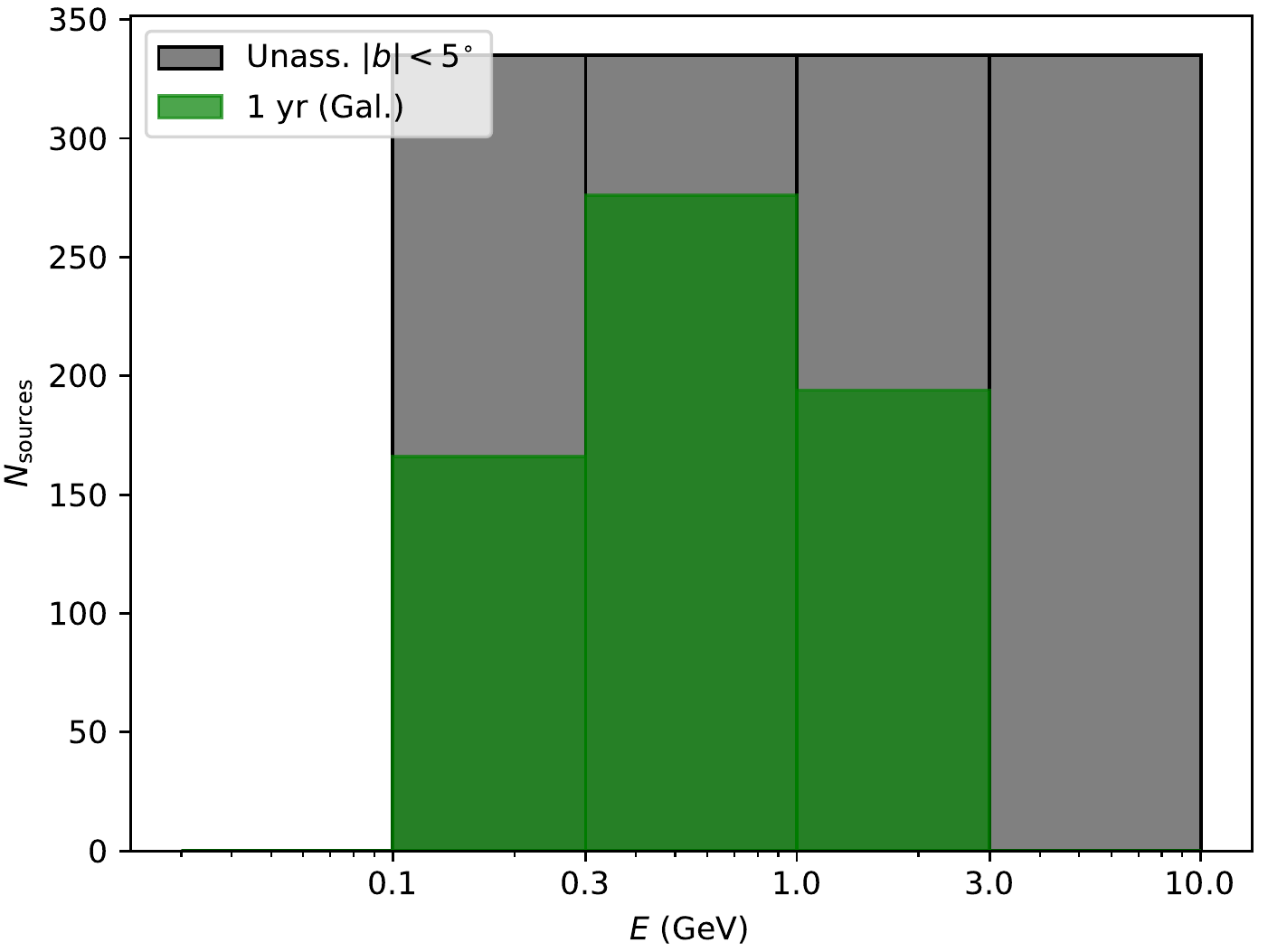}
\includegraphics[trim= 0mm 0cm 0cm 7cm, width=0.5\textwidth,angle=0.,scale=1]{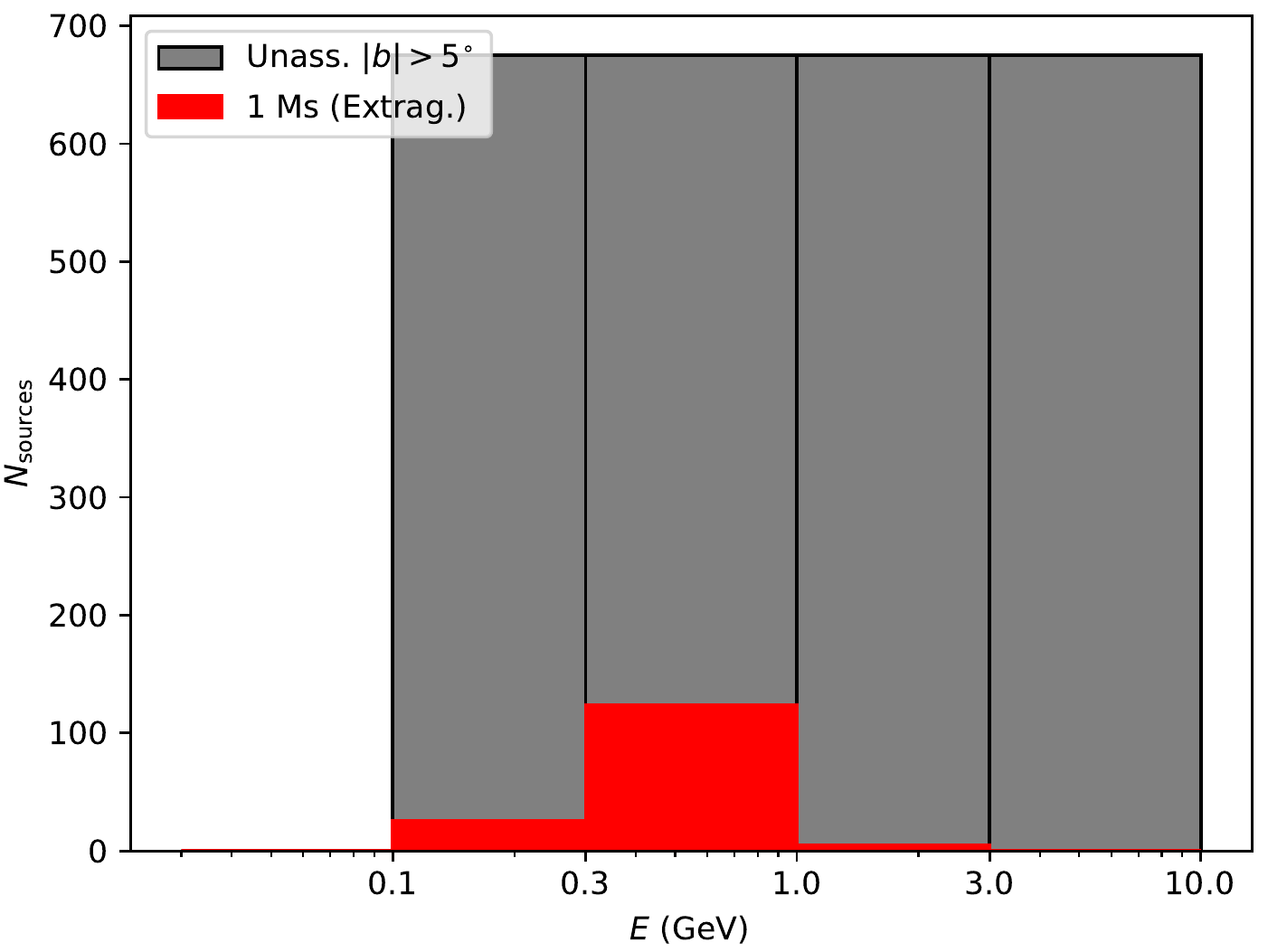}
\caption{\small Left:  Number of unidentified low galactic latitude sources from the 3FGL Catalog that would be detected by e-ASTROGAM for different energy intervals . Right: The same but for high galactic latitude sources.}
\label{Unid:fig2}
\end{figure}
With the e-ASTROGAM sensitivity for different integration times, 1 Ms (Extragalactic case) and 1 year (Galactic case), we have estimated the number of \fermilat unassociated sources, pulsars, and AGNs that could be detected by e-ASTROGAM for each one of the 0.1--0.3, 0.3--1.0, 1--3 GeV energy ranges that are common to  \fermilat and e-ASTROGAM (see Fig.~\ref{Unid:fig1}). In the case of the unassociated sources, we plot in Fig.~\ref{Unid:fig2} two histograms showing how many of them would be detected depending on its galactic latitude ($\abs{b} < 5^{\circ}$ and ($\abs{b} > 5^{\circ}$).  Among the 335 unassociated sources with $\abs{b} < 5^{\circ}$ (Galactic) in the 3FGL Catalog, 166, 276 and 194  of them (50\%, 82\%, and 58\%) would be detected by e-ASTROGAM in the 0.1--0.3, 0.3--1.0, 1--3 GeV energy ranges, respectively.  For the 675 unassociated sources with $\abs{b} > 5^{\circ}$ (Extragalactic), 26, 124 and 5  of them (4\%, 18\%, and 0.7\%) would be detected in the respective energy ranges. Given the steep spectra of many of the sources seen by \fermilat, crucial information at low-energy gamma-rays will be available for those sources detectable above 0.1~GeV for e-ASTROGAM.
\subsection[Fast MeV \g-ray flashes and perspectives on \g-SETI\\
\noindent
\textit{\small{S. Ciprini, C. C Cheung}}
]{Fast MeV \g-ray flashes and perspectives on \g-SETI}
\paragraph*{Science questions}
The rapidly evolving socialization drivers of the globalization development, enabled by the internet networking era and the exponential grow of computing power and data storage, indicate that citizen society (i.e. the participation to the governance by citizens/taxpayers) and  citizen science (scientific data exploration conducted by volunteer individuals) will have a substantial development in the next couple of decades. This is true also for citizen astronomy \cite{marshall2015} based on open data and having non-negligible consequences for space missions and agencies supported by public funds. The hunt for alien worlds and the search for life in the Universe, is a very fascinating topic for citizen astronomy. Citizen astronomers are motivated by being of service to science, as well as by their interest in the subject. A basic question is therefore: how high-energy missions and instruments dedicated to the observation of the \g-ray sky can have a distinctive, and more compelling role in inspiring interest in the wider citizen and public outreach (E\&PO) through the search for signs of extraterrestrial intelligent (ETI) life in newly discovered exoplanets and in extragalactic systems?

As of October 2017, 3691 planets in 2766 planetary systems, with 620 multiple-planet systems are counted \cite{schneider2011}. The future path for the exoplanets science includes advanced searches for Earth-size and super-Earth-size habitable planets, their physical, geological and astrobiology characterization, the search for liquid water and understanding of their atmospheres and, at last, the search for signs of elementary life. Three recent discoveries have substantially impacted the media, enlarging the large public interest for missions motivated partially and primarily to exoplanet detection and characterization (e.g., CoRoT, Kepler, CHEOPS, PLATO, ARIEL, TESS, and also Spitzer, Gaia, JWST, WFIRST, LUVOIR). The Earth-size planet in Proxima Centauri (at 4.2 ly); the first known system of seven Earth-size planets (TRAPPIST-1) at 40 ly; the irregular and unusual optical flux dips, flickering and dimming in the star KIC 8462852 (Tabby's star). This F-type star system KIC 8462852 placed at 1280 ly, is interpreted as a swarm of dusty comet fragments, or a large number of orbiting small masses in tight formation, but more exotic theories are proposed, like patterns and signs of an exo-civilization associated with a construction of Dyson swarm \cite{dyson60,wright16}, that is a popular concept for a Type II extraterrestrial civilization in the Kardashev scale \cite{kardashev64}.

Despite the general scientific skepticism\footnote{
\texttt{``Look for what's detectable, not for what's probable''} (Freeman Dyson 2009).
}
, KIC 8462852 has now been identified as an exceptional target for searches for extraterrestrial intelligence (SETI) signals and transmissions. Historically the SETI initiative have not considered \g-ray energies, therefore a second question is: how a next large \g-ray space telescope characterized by large field of view and improved sensitivity in the MeV band can be useful in the new era of revived scientific and SETI-related projects in the guise of multifrequency time-domain/survey astronomy?
\paragraph*{Importance of gamma-ray observations}
The intriguing fast radio bursts (FRBs) were first discovered in 2007. These are ultrafast radio transients with inferred extragalactic origin based on large dispersion meaures, with typical ms durations and $\sim$GHz flux densities of $\sim$1Jy. There is no concensus explanation for their progenitors \cite{katz16} and they could in fact originate from multiple source populations. Although
there could be thousands of detectable events per day, less than two dozen FRBs have been discovered. An FRB in our Galaxy is predicted
at least every $\sim$300 years -- at $<20$ kpc distances, it would be spectacularly bright with a flux density of 10$^{10}$ Jy, detectable by low-cost radio receivers \cite{maoz17}. Some fraction of FRBs could have a short \g-ray flash following the radio pulse \cite{delaunay16,murase17}, in the hypothesis of a merger of compact binaries (BH/NS).
Short GRBs with temporally-extended emission in hard-X rays and medium-energy \g-rays are expected to be observed in
the 0.2 MeV$-$3 GeV band and this would be important to clarify the nature of FRBs and the related prediction of GW emission
also for some class of FRBs, that could be already detected by advanced LIGO/VIRGO and in future by LISA (launch in 2034). As a remark, supergiant fast X-ray transients, believed to be produced by high mass X-ray stellar binary systems as short, sporadic and bright flares are likely not related to FRBs, but
this do not in principle, exclude that millisecond-duration gamma-ray flashes (MGFs) from other galaxies might exist.

Much more exotic conjectures point out that FRBs could be produced by some activity of extragalactic advanced exo-civilizations (Kardashev II or III types). In some cases (like for the repeating FRB 12110)
they are observed to repeat several times also years later, in agreement with the hypothesis for alien artificial beacons.                                                                                                                                                                                                                                                    When the Fermi Paradox on intelligent civilizations \cite{jones85} was initially proposed, it was thought that planets themselves
were very rare, contrary to the actual evidence that the hundreds of exoplanets found since 1992 are only the tip of the iceberg\footnote{
\texttt{``Alien Worlds Galore''} (M. Cruz \& R. Coontz 2013, introduction to a special issue of Science).\\
\texttt{``If we are alone in the Universe, it sure seems like an awful waste of space.''} (Carl Sagan 1972 paraphrasing Thomas Carlyle
as reported in ``Accepting the Universe'' by John Burroughs, 1920).\\
}
. Some FRBs might originate from radio and coherent beams supplied by stellar energy that would power enormous light sails for
spaceships capable of attaining relativistic speeds \cite{lingam17}. Energetic and engineering constraints both yield similar result
on sail size (comparable to a super-Earth planet) and the optimal powering frequency similar to the detected FRB frequencies. Well observable leakage radiation may well be from the use of power millimeter-wave beaming to transfer energy and accelerate such spaceships, with effective isotropic radiated power of $\sim10^{25}$ erg s$^{-1}$ \cite{benford16}.  \\ Advanced civilizations that have reached a technological singularity (abruptly runaway technological growth) could intentionally transmit a two-millisecond
pulse encoding 10$^{18}$ bits of information \cite{ball96}. GRBs may be also used by civilizations as synchronizers for beamed and short duration SETI transmissions \cite{corbet99}. Considering that civilizations are bathed in optical light,
the absorption/reddening of optical/UV light on the Galactic plane and the terrestrial/solar interference at radio bands,
to send transmissions over the Galaxy is convenient to choose energy bands where the isotropic background and stellar
output is low as like the wide MeV-GeV \g-ray band \cite{arnold13}. Under all the astrophysical/citizen-science/large-public outreach scenarios,
a SETI approach based on \g-ray data (\g-SETI) could be of particular interest, especially if we consider MeV thermonuclear
and matter-antimatter annihilation processes.

Signatures of an advanced exo-civilization in our Galaxy or other galaxies can, more easily, emerge if we observe the sky at MeV/sub-GeV \g-ray energies with respect to other photon frequencies,
and include the following basic categories of technological progenitors.
Artificial objects in central star transiting orbits; Dyson complexes;
deliberate communication signals \cite{arnold13}; directed impulsive beaming for accelerating spacecrafts \cite{benford16,lingam17}; leakage in the electromagnetic spectrum (spectral lines from nuclear fissile waste disposal in stars, tritium leakage, etc.); artifacts such as evidence for energy production/consumption/transportation or for huge space colonies with large-scale industrial engineering, furnaces for antimatter or fusion plants; manipulation of the central star and mining star material (for example the Shklovskii gamma-ray laser mining, or ``graser''); protective blast shields against nearby merging NSs, GRBs or SN; self-destruction of civilizations by global thermonuclear wars and other observational signatures \cite{stevens16}; and the unexpected.
\paragraph*{Expected results with e-ASTROGAM}
Possible artificial and candidate \g-SETI signals from technologically advanced civilizations can be identified by searching for unusual spectral or temporal (dips, periodicy, unusual flickering) features, and with per-photon analysis, using MeV/sub-GeV data obtained by a high-sensitivity and large field of view surveying space telescope like e-ASTROGAM. Beyond the search for possible spectrally/temporally unusual \g-ray signals, the possibly repeating subset of millisecond  \g-ray flashes (MGFs) or other type of mini-bursts are optimal targets for per-photon data studies. It is also of interest to investigate the association/coincidence of multiple \g-ray events with the FRBs \cite{yamasaki16}. The expected number of detectable \g-rays
from a FRB within a direction $\Omega_j$ at redshift $z$ in a MGF search within a time interval $\Delta t$ is given as
$ N_\gamma = \Delta t \int_{\rm 0.2 \ MeV}^{\rm 3 \ GeV} d\epsilon_{\rm obs}
A_{\rm eff}(T_i, \Omega_j) [dF_\gamma(\epsilon_{\rm obs};z, \xi)]/[d\epsilon_{\rm obs}] $,
where $\xi$ is the \g-ray to radio luminosity ratio in the rest-frame of a FRB.
Gamma-ray photon pairs or multiplets within an energy and/or time range, for example with time search windows of
$\Delta t=1,2,5,10$ ms are considered for every reference photon and other
photons are searched, in blind mode, within a spatial distance compatible with per-photon
angular resolution and within $\Delta t$ from the reference photon event.
These simple per-photon analyses have possible implementations
in citizen-science (i.e., within the context of a ``\g-SETI at home'' program or even more interactive platforms). To outline conclusive candidates, Poisson statistics of steady \g-ray fluxes
from bright point sources or diffuse \g-ray background have to be taken into account.
Another example of a possible \g-SETI analysis is to search for spectral, also variable,
annihilation signature in different regions of the \g-ray sky exploiting the excellent
spectral energy resolution of e-ASTROGAM. Such a signal could be produced by artificial $p\overline{p}$ annihilation used for applications requiring portability like
spaceships propulsion. If they exist, this might be detectable in case of solar neighborhood star directions.
An obvious disadvantage of artificial signal searches in \g-rays is the large power output requirements for
exo-engineering, but the history of science teach us that unexpected could be greater than expected.
On the other hand it is time to include \g-rays in SETI and citizen-astronomy frames, especially
considering that the hard-X-ray and soft/medium energy (MeV) sky is
the most promising portion of the electromagnetic spectrum, joined with existing radio and optical-wavebands for such searches.
This also contributes to increase the potential of the e-ASTROGAM mission in terms of public outreach
and synergy with studies of the potentially many, habitable exoplanets
expected to be discovered in the forthcoming decades.


\end{document}